\documentclass[12pt]{thesis}

\pdfoutput=1

\usepackage{amsmath}
\usepackage{amssymb}
\usepackage{graphicx}
\usepackage{natbib}
\usepackage{deluxetable}
\usepackage{lineno}

\def\Fermi{\emph{Fermi}}
\def\FL{\emph{Fermi} LAT}
\def\DEG{$^{\circ}$}
\def\ApJ{\emph{Astrophys. Journal}}
\def\AA{\emph{Astron. \& Astrophys.}}
\def\NIMA{\emph{Nuclear Instrum. and Meth. A}}
\def\ITNS{\emph{IEEE Trans. Nucl. Sci.}}
\def\AsPart{\emph{Astropart. Phys.}}
\def\MNRAS{\emph{Monthly Notices of the Royal Astron. Soc.}}

\begin{document}
\hbox{\ }

\small \normalsize

\begin{center}
\large{{ABSTRACT}} 

\vspace{3em} 

\end{center}
\hspace{-.15in}
\begin{tabular}{ll}
Title of dissertation:    & {\large  CONSTRAINTS ON THE EMISSION}\\
&				{\large  GEOMETRIES OF GAMMA-RAY}\\
&				{\large MILLISECOND PULSARS OBSERVED}\\
&				{\large WITH THE FERMI LARGE AREA} \\
&				{\large TELESCOPE} \\
\ \\
&                          {\large  Tyrel James Johnson, Doctor of Philosophy,} \\
&                          {\large  2011} \\
\ \\
Dissertation directed by: & {\large  Professor Jordan Goodman} \\
&  				{\large	 Department of Physics } \\
&			{\large Dr. Alice Harding} \\
&				{\large NASA Goddard Space Flight Center}\\
\end{tabular}

\vspace{3em}

\large \normalsize

Millisecond pulsars (MSPs) have been established as a class of high-energy ($\geq$0.1 GeV) emitters via the detection of significant pulsed signals, at the radio periods, of many MSPs with the Large Area Telescope (LAT) aboard the \emph{Fermi Gamma-ray Space Telescope}.  The detection of high-energy emission from eight globular clusters (known or suspected of containing many MSPs) that display pulsar-like spectra and detection of several new radio MSPs coincident with unassociated LAT sources further suggests that gamma-ray production must be the rule in MSPs, not the exception.  Most MSP gamma-ray light curves display sharp peaks indicative of thin accelerating gaps, suggesting copious pair-creation in the open volume.  MSP gamma-ray and radio light curves have been simulated using geometric outer-gap (OG), slot-gap/two-pole caustic (TPC), and pair-starved polar cap gamma-ray models and either a hollow-cone beam or altitude-limited, outer-magnetospheric gap radio model, all assuming a vacuum retarded dipolar magnetic field geometry.   A Markov chain Monte Carlo maximum likelihood technique has been developed to find the best-fit model parameters for nineteen MSPs using data from the LAT and various radio observatories.  Confidence contours have been created in pulsar viewing geometry which are compared with constraints from radio and X-ray observations.  Model-derived beaming factors allow for more accurate determinations of gamma-ray luminosities.  The best-fit viewing angles follow a uniform, angular distribution.  The distribution of magnetic inclination angles favors all angles equally, contrary to analyses of non-recycled pulsars, which supports the theory that MSPs have been spun-up via accretion.  There are suggestions that the radio emission should occur nearer the light cylinder.  These results have implications for MSP population simulations and for addressing MSP contributions to diffuse backgrounds.  The likelihood significantly favors one model over another for seven MSPs, with the TPC model largely preferred.  An implied transition in the gamma-ray luminosity versus spin-down power trend is observed but more statistics are necessary to describe it.

\thispagestyle{empty}
\hbox{\ }
\vspace{1in}
\small\normalsize
\begin{center}

\large{{CONSTRAINTS ON THE EMISSION GEOMETRIES OF GAMMA-RAY MILLISECOND PULSARS OBSERVED WITH THE FERMI LARGE AREA TELESCOPE}}\\
\ \\
\ \\
\large{by} \\
\ \\
\large{Tyrel James Johnson}
\ \\
\ \\
\ \\
\ \\
\normalsize
Dissertation submitted to the Faculty of the Graduate School of the \\
University of Maryland, College Park in partial fulfillment \\
of the requirements for the degree of \\
Doctor of Philosophy \\
2011
\end{center}

\vspace{7.5em}

\noindent Advisory Committee: \\
Professor Jordan Goodman, Chair/Advisor\\
Dr. Alice Harding, Co-Advisor\\
Professor Gregory Sullivan\\
Professor M. Coleman Miller\\
Dr. Julie McEnery

\pagestyle{plain}
\pagenumbering{roman}
\setcounter{page}{2}


\renewcommand{\baselinestretch}{2}
\small\normalsize
\hbox{\ }
 
\vspace{-.65in}

\vspace*{\fill}
\begin{center}
\emph{...for my Great-Grandmother Marion who gazed up at the sky in wonder, and for my Grandmother Jan who taught me to always be myself...}
\end{center}
\vspace*{\fill}

\small\normalsize
\hbox{\ }
 
\vspace{-.65in}

\begin{center}
\large{Acknowledgments} 
\end{center} 

\vspace{1ex}

There are many people to whom I owe a debt of gratitude for their part in facilitating the completion of this thesis.  My parents have always supported and encouraged me in my scholarly pursuits and for that I am very grateful.  I also thank all of my family for their understanding these past few months as I've been even worse than usual about keeping in touch and promise to try and do better in the future.

I was fortunate enough to have several helpful and encouraging teachers throughout my grade school years.  In particular, I wish to thank Mr.~Rick Alm who provided me with information about an essay contest which has had a significant impact on my life and career choice.  I also which to thank Mr.~Patrick Karr who taught me most of what I know about writing.  Additionally, I must not forget Mr.~David Huffman who agreed to continue teaching the second semester of senior calculus even though only one student remained in the class.

Several of my undergraduate professors at the University of Idaho were also very helpful in the course of my academic career.  In particular, my undergraduate advisor Dr.~Christine Berven was very helpful in taking on a young college student with little laboratory experience and was always available for a good chat about hunting, fishing, or graduate school.  I owe many thanks to the late Dr.~George Patsakos who taught excellent undergraduate courses in astrophysics and relativity.

I was fortunate enough to be awarded an internship at the Harvard-Smithsonian Center for Astrophysics in the summer of 2004 where I worked under Dr.~Suzanne Romaine and with Mr.~Riccardo Bruni.  This was my first foray into the field of astrophysics, albeit mainly on the instrumentation side, but the skills and experience I acquired were very useful and further cemented the goal of graduate school in my mind.

Dr.~Steven Ritz was my first advisor at the University of Maryland, introduced me to GLAST (now \emph{Fermi}), and set me on the path to my thesis work.  I owe him a great deal of thanks for his patience and willingness to entertain even the silliest of questions.

In March of 2008 I began working on gamma-ray pulsars in anticipation of launch and the planned paper on the Vela pulsar.  It was at this point that I began working with Dr.~Alice Harding at NASA GSFC.  Many thanks are owed her for teaching me about pulsars and gamma-ray emission models, providing many useful comments and suggestions, and being patient through my many coding errors which made the development of the likelihood technique used in this thesis an interesting process.  I also owe her many thanks for reading through rough versions of my thesis chapters and providing useful comments and corrections.

The simulation code used in this thesis was initially developed by Dr.~Jarek Dyks though it has undergone modifications by Dr.~Harding and Dr.~Christo Venter.  I am also grateful to Dr.~Venter for his patience with my many questions and useful suggestions.

To my campus advisor Dr.~Jordan Goodman I owe many thanks as well, especially for being patient with my rather erratic thesis schedule which involved long periods of seemingly little activity followed by many short bursts of emails.

To Dr.~David Thompson and Dr.~Elizabeth Hays I owe many thanks for proofreading early versions of some chapters.  Additionally, I owe many thanks to Dr.~\"{O}zlem \c{C}elik and Dr.~Toby Burnett for their analyses in preparation for the 2FGL and second LAT pulsar catalogs which were used for spectral information in Chapters 7 and 8 of this thesis.  I thank Dr.~Lucas Guillemot for his help with the pulsar timing solutions and comments on sections of Chapter 2.

Dr.~Eric Grove, Dr.~David Smith, and Dr.~Denis Dumora were extremely helpful in regards to the LAT timing verification discussion in Chapter 3.  Dr.~Matthew Kerr has been a very helpful colleague to bounce ideas off of and serve as a reference for statistical analysis.

There are many more people within the \FL{} collaboration to whom thanks are owed, for without the hard work of many individuals (some for more than a decade) this mission would not have come to pass and my thesis work would have been much different.

Last but not least, I must thank my wife Jennifer Johnson for putting up with a much more absent-minded, spacey, and generally distracted husband than usual.  Your wonderful, homecooked meals helped to keep me going through the thesis writing and I hope to somehow return the favor as you prepare for your defense.

\begin{center}
Sincerely,\\
\emph{Tyrel James Johnson}
\end{center}

\renewcommand{\baselinestretch}{1}
\small\normalsize
\tableofcontents
\newpage
\listoftables 
\newpage
\listoffigures 
\newpage
\addcontentsline{toc}{chapter}{List of Abbreviations}

\renewcommand{\baselinestretch}{1}
\small\normalsize
\hbox{\ }

\vspace{-4em}

\begin{center}
\large{List of Abbreviations}
\end{center}

\vspace{3pt}
\begin{tabular}{l l}
ACD & anti-coincidence detector\\
alOG & altitude-limited outer gap\\
alTPC & altitude-limited two-pole caustic\\
BSL & bright source list\\
$c$ & speed of light\\ 
CAL & calorimeter\\
CF & co-rotating frame\\
cm & centimeter\\
CR & curvature radiation\\
CsI & cesium iodide\\
CSPR & calorimeter-seeded pattern recognition\\
CT & classification tree\\
\emph{EGRET} & \emph{Energetic Gamma-Ray Experiment Telescope}\\
eV & electronvolt\\
\emph{Fermi} & \emph{Fermi Gamma-ray Space Telescope}\\
FOV & field of view\\
FWHM & full width at half maximum\\
GeV & gigaelectronvolt\\
GPS & global positioning system\\
HE & high-energy\\
ICS & inverse compton scattering\\
IOF & inertial observer's frame\\
IRF & instrument response function\\
K & Kelvin\\
keV & kiloelectronvolt\\
KF & Kalman filter\\
km & kilometer\\
LAT & large area telescope\\
LRT & likelihood ratio test\\
LMXB & low-mass X-ray binary\\
MC & Monte Carlo\\
MCMC & Markov chain Monte Carlo\\
MeV & megaelectronvolt\\
MHD & magneto-hydrodynamic\\
mm & millimeter\\
MSP & millisecond pulsar\\
M$_{\odot}$ & solar mass = 1.99$\times10^{33}$ g\\
$\mu$m & micrometer\\
OG & outer gap\\
PC & polar cap\\
PTC & Pulsar Timing Consortium\\
PSF & point-spread function\\
PSPC & pair-starved polar cap\\
PSR & pulsar\\
\end{tabular}

\newpage
\renewcommand{\baselinestretch}{1}
\small\normalsize
\hbox{\ }

\vspace{-4em}

\begin{center}
\large{List of Abbreviations (cont'd)}
\end{center}

\vspace{3pt}
\begin{tabular}{l l}
PWN & pulsar wind nebula\\
RFM & radius-to-frequency mapping\\
ROI & region of interest\\
RVM & rotating vector model\\
sr & steradian\\
SSD & silicon strip detector\\
ST & Science Tool\\
TeV & teraelectronvolt\\
TKR & tracker\\
TPC & two-pole caustic\\
TS & test statistic\\
V & volt\\
VHE & very high-energy\\
\end{tabular}

\newpage
\setlength{\parskip}{0em}
\small\normalsize
\setcounter{page}{1}
\pagenumbering{arabic}
\renewcommand{\thechapter}{1}

\chapter{\bf Gamma-ray Astronomy}\label{ch1}
Astronomy began many centuries ago with curious minds looking up to the sky and wondering ``Why?''.  They surveyed the skies first with the naked eye and then with crude telescopes, always striving to learn more about humankind's place in the grand play that is life.  Much has been learned about the universe since that time and ``telescope'' technologies have improved greatly.  However, we continue to gaze at the sky, looking ever deeper and asking the same questions, but now with more subtext.

\begin{figure}
\begin{center}
\includegraphics[width=1.0\textwidth]{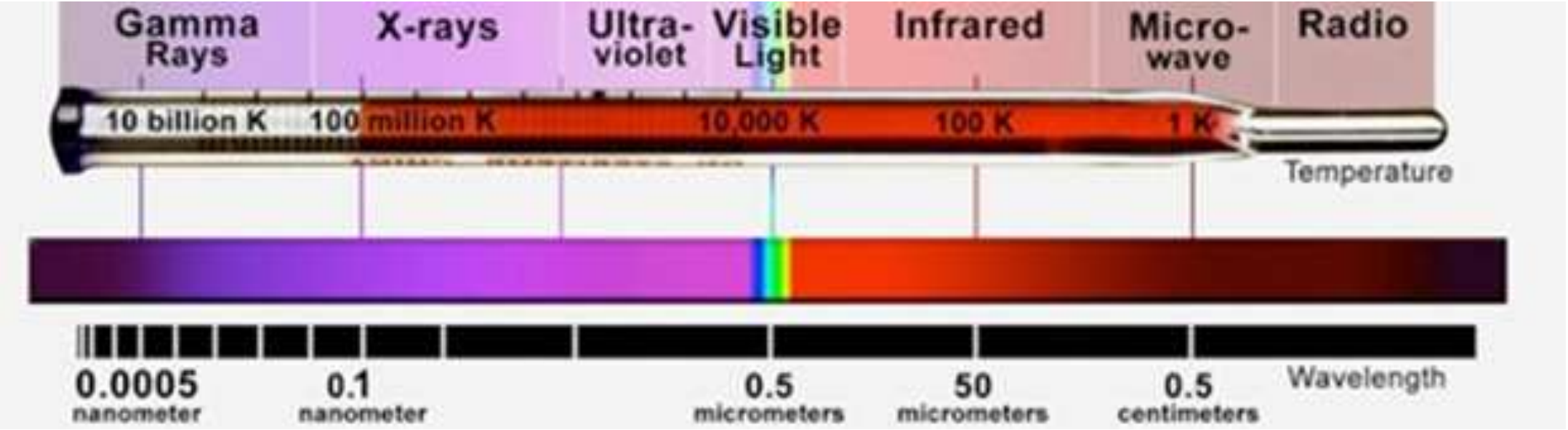}
\end{center}
\small\normalsize
\begin{quote}
\caption[The electromagnetic spectrum]{The electromagnetic spectrum ranging from the shortest wavelengths (gamma rays) to the longest (radio).  Equivalent temperatures for each wavelength are also shown.  \emph{Figure Credit: NASA/CXC} \label{ch1emspec}}
\end{quote}
\end{figure}

Prior to $\sim$100 years ago, astronomy consisted of observations using what is known as ``visible light'' consisting of the small portion of the electromagnetic spectrum (see Fig.~\ref{ch1emspec}\footnote{Adapted from http://chandra.harvard.edu/art/color/colorspace.html}) to which the human eye is sensitive.

\small\normalsize

Early in the twentieth century astronomy began to encompass wavelengths outside the visible band.  Once it was possible to put satellites in space the extent of the electromagnetic spectrum through which humans were observing the universe broadened greatly.  Modern day astronomy encompasses observations of astronomical sources using not only light but also particles (e.g., cosmic rays, neutrinos).  Note that searches for gravitational radiation from astrophysical sources are also underway though no signal has yet been detected.

\section{Electromagnetic Radiation}\label{ch1EandM}
Visible light is the form of electromagnetic radiation to which people are most accustomed.  However, as shown in Fig~\ref{ch1emspec}, this accounts for but a small fraction of the electromagnetic spectrum.

On the macroscopic scale, light can be described as a transverse wave traveling at speed $c\ =\ \lambda\nu$, with wavelength $\lambda$ and frequency $\nu$.  The speed of a wave in vacuum is constant; thus, waves with higher frequencies must have shorter wavelengths and vice versa.  The energy of a wave is $E\ =\ h\nu\ =\ hc/\lambda$, where $h\ \approx\ 6.626\times10^{-27}$ erg s is Planck's constant.

When the frequency of light increases to the point that the wavelength becomes comparable to the size of atoms or smaller, light can act as both a wave and a particle.  In this regime, interactions of light must be treated using quantum electrodynamics in which light is quantized into individual photons.

Electromagnetic waves are governed by Maxwell's Equations (given in Eqs.~\ref{ch1M1},~\ref{ch1M2},~\ref{ch1M3}, and~\ref{ch1M4}) for electric field $\vec{E}$, magnetic field $\vec{B}$, charge density $\rho_{q}$, and current $\vec{J}$.
\begin{equation}\label{ch1M1}
\nabla\cdot\vec{E}\ =\ 4\pi\rho_{q}
\end{equation}
\begin{equation}\label{ch1M2}
\nabla\cdot\vec{B}\ =\ 0
\end{equation}
\begin{equation}\label{ch1M3}
\nabla\times\vec{E}\ =\ -\frac{1}{c}\frac{\partial\vec{B}}{\partial t}
\end{equation}
\begin{equation}\label{ch1M4}
\nabla\times\vec{B}\ =\ \frac{4\pi}{c}\vec{J}+\frac{1}{c}\frac{\partial\vec{E}}{\partial t}
\end{equation}

In the case of no charge or current these reduce to simply Eqs.~\ref{ch1M3} and~\ref{ch1NoJ}.
\begin{equation}\label{ch1NoJ}
\nabla\times\vec{B}\ =\ \frac{1}{c}\frac{\partial\vec{E}}{\partial t}
\end{equation}

Therefore, a time-varying electric field will produce a time-varying magnetic field, and so on, which leads to a self-propagating electromagnetic wave.  The cross product results in $\vec{B}\ \perp\ \vec{E}$ with propagation in a direction perpendicular to both.  This direction is given by the Poynting vector $\vec{S}\ =\ (c/4\pi)\ \vec{E}\times\vec{B}$ which also gives the flux of energy carried by the wave.

The term ``gamma rays'' applies to photons with the shortest wavelengths and thus the highest energies.  This form of light is generally referred to in terms of photon energy as opposed to frequency (as is done for radio waves) or wavelength (as is done for visible light).  As such, it is useful to define a unit of energy known as the electron volt (eV) which is the amount of energy it takes to accelerate an electron through a potential difference of 1 V.  To provide a frame of reference, 1 eV $\approx$ 1.602$\times10^{-19}$ J.

There is no single definition which defines when a photon stops being an X-ray and becomes a gamma ray; however, for the purposes of this thesis a gamma ray shall be taken to mean a photon with energy $\gtrsim$ 1 MeV (where an MeV is a megaelectronvolt = $10^{6}$ eV).  In particular, this discussion will focus on high-energy (HE) gamma rays, those with energy $\geq$ 0.1 GeV (where a GeV is a gigaelectronvolt = $10^{9}$ eV).

\citet{Morrison58} first predicted that astrophysical sources were capable of producing gamma rays at detectable levels.  However, gamma-ray astronomy did not truly begin until the 1960-70's with missions such as Explorer XI, which observed the first gamma rays from astrophysical sources \citep{EXOXI}; SAS-2, which discovered discrete gamma-ray point sources and the diffuse background \citep{SAS2}; and COS-B, which pushed the detectable photon energies up to $\sim$3 GeV and detected even more point sources \citep{Swanenburg81}.

Prior to 2008, the study of HE astrophysics culminated in the launch of the \emph{Compton Gamma-Ray Observatory}.  This observatory consisted of four instruments.  Of particular interest for HE astrophysics is the \emph{Energetic Gamma-Ray Experiment Telescope} (\emph{EGRET}; see Thompson et al., 1993 for instrument details) which operated from 1991 to 2000.  Among the principle achievements of the \emph{EGRET} detector was the discovery of 271 HE point sources, most of which were not firmly identified with objects of known gamma-ray emitting source classes detected at other wavelengths.  Of the identified \emph{EGRET} sources most were active galactic nuclei (AGN) though a handful were pulsars.  \emph{EGRET} also mapped the diffuse emission from the Milky Way and discovered that gamma-ray bursts were capable of producing photons with GeV energies \citep{Hurley94}.  Production of such energetic gamma rays can only occur in the most extreme environments in the universe; thus, HE observations probe regimes of physics which are not easily accessible on Earth.

\section{Thermal Radiation}\label{ch1Therm}
An astrophysical object with a temperature T will emit photons via thermal radiation.  Thermal radiation is defined to be ``radiation emitted by matter in thermal equilibrium'' \citep{RL79}.  The spectrum of thermal radiation is found to follow Planck's Law, Eq.~\ref{ch1therm}, where $k_{B}$ is Boltzmann's constant $\approx$ 8.6$\times10^{-5}$ eV K$^{-1}$.
\begin{equation}\label{ch1therm}
B_{E}\ =\ \frac{2E^{3}}{(hc)^{2}}\frac{1}{\exp\lbrace E/k_{B}\rm T\rbrace -1}
\end{equation}

\begin{figure}[h]
\begin{center}
\includegraphics[width=0.7\textwidth]{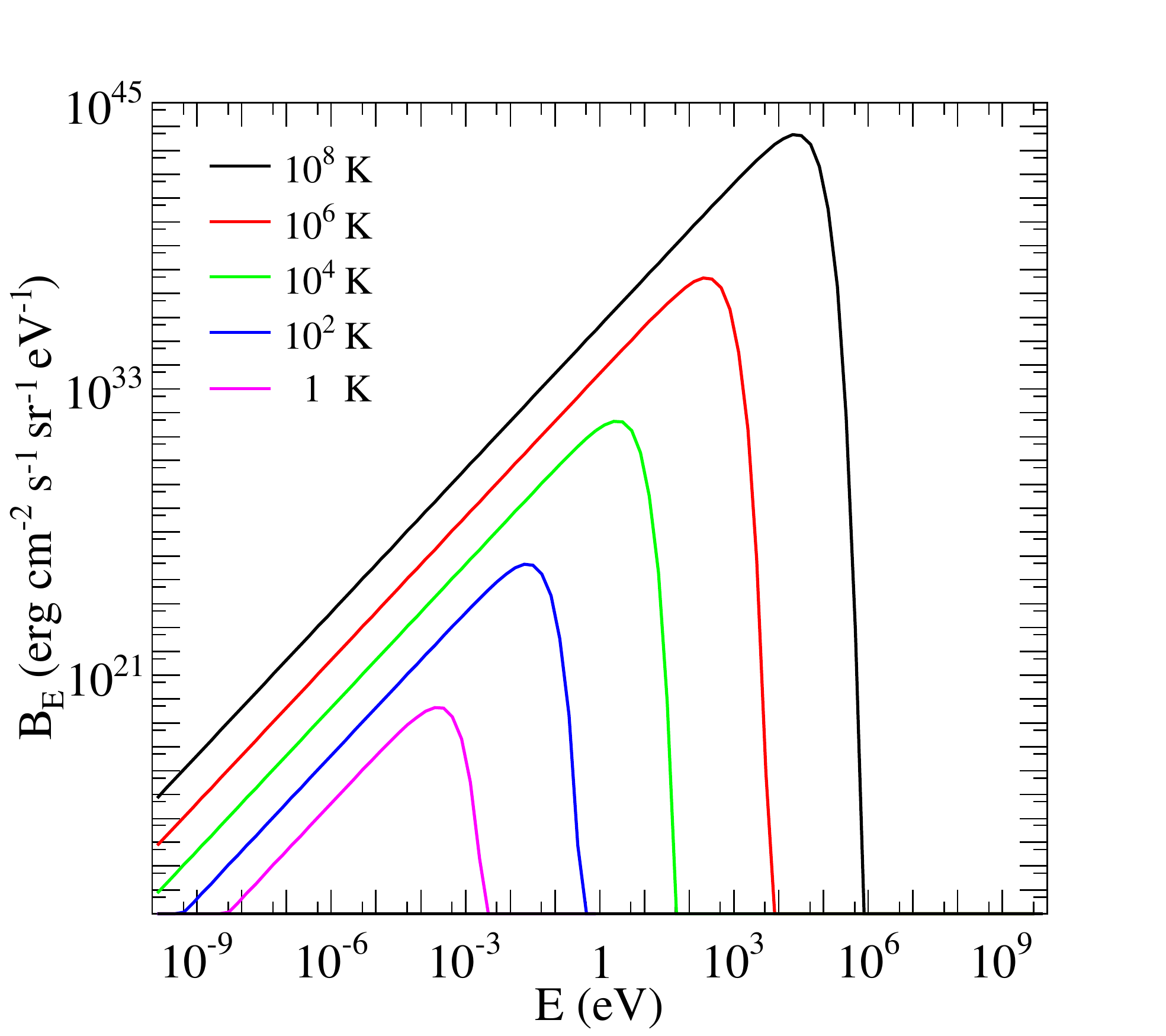}
\end{center}
\small\normalsize
\begin{quote}
\caption[Thermal spectra for different temperatures]{Thermal spectra, Eq.~\ref{ch1therm} for different temperatures as indicated.\label{ch1BBspec}}
\end{quote}
\end{figure}

For a given observed energy, $E$, Eq.~\ref{ch1therm} describes the rate at which energy is emitted into a given solid angle and area.  The shape of this spectrum is shown in Fig.~\ref{ch1BBspec} for different values of T.

\small\normalsize

Wien's Law is apparent in Fig.~\ref{ch1BBspec}; namely, for energies $E\ \gg k_{B}$T the exponential dominates and the spectrum goes like $\exp\lbrace -E/k_{B}\rm T\rbrace$, cutting off very sharply.  Note also that even for temperatures of $10^{8}$ K the spectrum peaks and falls off before 1 GeV.  Production of thermal gamma rays would require extremely high temperatures and any astrophysical body which was that hot would be unstable.  Thus, the radiative processes which produce gamma rays must be non-thermal.  In particular, production of HE gamma-rays generally, but not always, requires acceleration of charged particles to relativistic energies.  Note that while gamma rays must be non-thermal in nature not all non-thermal radiation need be gamma rays.

Many astrophysical sources do emit thermal radiation (often in addition to non-thermal radiation).  This facilitates the definition of a brightness temperature ($T_{b}$).  For a source with measured specific intensity $I_{\nu}$ at frequency $\nu$ the brightness temperature is the temperature for which Eq.~\ref{ch1therm} (with $E$ replaced by $h\nu$) returns $I_{\nu}$.  This is the temperature at which a blackbody would be if it were to have the same specific intensity at the same frequency.  A blackbody is defined as an object which absorbs all incident radiation and re-emits it with the characteristic thermal spectrum \citep{BOB}.

The brightness temperature can often be used to infer physical properties of the source, but only if the emission truly is thermal.  For instance, pulsars are measured to have brightness temperatures at radio wavelengths which imply unphysical temperatures.  For instance, giant pulses from the Crab pulsar imply $T_{b}\ \gtrsim\ 10^{35}$ K \citep{hdbk}.

\section{Non-thermal Emission Processes}\label{ch1emit}
A charged particle which is accelerated will produce electromagnetic radiation.  For a particle with charge $q$ experiencing an acceleration of magnitude $a$ such that the velocity of the charge is much, much less in magnitude than the speed of light (i.e., the particle is non-relativistic) the power radiated away is given by Eq.~\ref{ch1Larmor}, which is the Larmor formula \citep{Jackson} with a negative sign indicating that the particle is losing energy.
\begin{equation}\label{ch1Larmor}
P\ =\ -\frac{2}{3} \frac{q^{2}a^{2}}{c^{3}}
\end{equation}

Charged particles can not radiate away more energy than is given to them by the acceleration.  Assuming that an electron were to emit all of the energy it gained from acceleration at once and that it were to emit a 1 GeV gamma ray, then it must be accelerated to an energy of 1 GeV (neglecting the relatively small rest mass of the electron $m_{e}\ =\ 0.511 MeV/c^{2}$).  This implies a Lorentz factor $\gamma\ \approx\ 2000$ which requires a velocity $\sim$99.99997\% the speed of light.

While charged particles do not radiate all of their energy at once and there are more factors to consider, what is clear from the arguments above is that Eq.~\ref{ch1Larmor} will not be applicable to charged particles energetic enough to emit HE gamma rays.  The relativistic equivalent of Eq.~\ref{ch1Larmor} for a particle with velocity $\vec{\beta}$ and acceleration $\vec{a}$ (in units of $c$) is given by Eq.~\ref{ch1RelP}, this is known as the Li\'{e}nard result \citep{Jackson}.  Note that the squared vector quantities indicate that an inner product of the vector with itself should be taken.
\begin{equation}\label{ch1RelP}
P\ =\ -\frac{2}{3} \frac{q^{2}}{c}\gamma^{6} \Big((\vec{a})^{2}-(\vec{\beta}\times\vec{a})^{2}\Big)
\end{equation}

There are three principal forms of non-thermal radiation of importance for HE pulsar emission models.  These are synchrotron radiation, curvature radiation (CR), and inverse compton scattering (ICS).  While both of the former forms of radiation involve particles being accelerated along curved trajectories in the presence of magnetic fields, the details are somewhat different.  The latter  radiation occurs when light interacts with relativistic charged particles (typically electrons or positrons) and gains energy.

Another important form of non-thermal emission occurs when an electron changes direction quickly due to the presence of an electromagnetic field.  This is called bremsstrahlung (or braking) radiation and is important for the methods of detecting gamma rays discussed in Section~\ref{ch1detection}.  Note that bremsstrahlung is important as an emission mechanism from some astrophysical sources but not pulsars.

\subsection{Synchrotron Radiation}\label{ch1SR}
Synchrotron radiation occurs when a charged particle gyrates around a magnetic field line.  The relativistic form of the Lorentz forces exerted on a charged particle of rest mass $m$ moving with a velocity $\vec{\beta}$ in a magnetic field $\vec{B}$ and an electric field $\vec{E}$ are given by Eq.~\ref{ch1bforce} and~\ref{ch1eforce} \citep{RL79}, assuming no radiative losses.
\begin{equation}\label{ch1bforce}
c\frac{d}{dt}(\gamma m\vec{\beta})\ =\ q\vec{\beta}\times\vec{B}
\end{equation}
\begin{equation}\label{ch1eforce}
\frac{d}{dt}(\gamma mc^{2})\ =\ cq\vec{\beta}\cdot\vec{E}
\end{equation}

In the case that $\vec{E}\ =\ \vec{0}$, Eq.~\ref{ch1eforce} implies that $\gamma$ is constant with time which also implies that $|\vec{\beta}|$ is constant (using that $\gamma\ =\ (1-\beta^{2})^{-1/2}$).  In the event that $\vec{E}\ \neq\ \vec{0}$ this condition can be satisfied by transforming to a frame where the electric field vanishes.  The following arguments will then apply, the radiation properties can be assessed, and then the proper Lorentz transformations can be applied to evaluate the radiation in the original frame.

By the nature of the cross product in Eq.~\ref{ch1bforce} the component $\vec{\beta}$ parallel to $\vec{B}$ will not change with time; thus, the magnitude of the perpendicular component will be constant as well and only the direction will change following Eq.~\ref{ch1bperp}.
\begin{equation}\label{ch1bperp}
\frac{d}{dt}(\vec{\beta}_{\perp})\ =\ \frac{q}{\gamma mc}\ \vec{\beta}\times\vec{B}
\end{equation}

The solution of Eq.~\ref{ch1bperp} is uniform circular motion with angular frequency $\omega_{s}\ =\ qB/\gamma mc$.  The particle will thus circle around the field line while continuing in the direction of $\vec{\beta}_{\parallel}$. In a time which is short compared to the energy loss time, the particle will follow this helical motion; however, as the particle is experiencing accelerated motion it will radiate and thus the motion will diverge from the simple case considered above.

The total power per unit angular frequency radiated away from an electron undergoing such acceleration is \citep{RL79},
\begin{equation}\label{ch1syncspec}
P(\omega)\ =\ -\frac{\sqrt{3}}{2\pi} \frac{q^{3}B}{mc^{2}}\sin(\chi)F\Big(\frac{\omega}{\omega_{c}}\Big),
\end{equation}
\noindent{}where $\chi$ is the angle between the direction of $\vec{B}$ and $\vec{\beta}$ and $\omega_{c}\ \equiv\ (3/2)\gamma^{3}\omega_{s}\sin(\chi)$ is the critical frequency up to which the spectrum should extend before falling off significantly.  The function $F$ is given by Eq.~\ref{ch1syncF} where $K_{5/3}$ is the modified Bessel function of second order with n = 5/3.
\begin{equation}\label{ch1syncF}
F(x)\ =\ x\int_{x}^{\infty}K_{5/3}(\xi)d\xi
\end{equation}

With $\omega\ \ll\ \omega_{c}$, $F$ goes like $\omega^{1/3}$ and for $\omega\ \gg\ \omega_{c}$, $F$ goes like $\omega^{1/2}$ with an exponential cutoff $\exp\lbrace-\omega/\omega_{c}\rbrace$.

In astrophysical sources the accelerated particles are not monoenergetic.  These sources are often observed to have spectra which can be described by a power law over some energy range.  This implies that the underlying particle population will have a power law energy distribution over some energy range.  The observed spectral index can be related to the power law index of the emitting particles.  In particular, assume the number of particles with energy between $E$ and $E+dE$ can be described by $N(E)\ =\ C(E^{-p})dE$ for some constant $C$.  Following \citet{RL79}, the total power per unit angular frequency will also take the form of a power law in $\omega$ with spectral index $s\ =\ -(p-1)/2$, thus relating the observed index to the distribution of accelerated particles.

The equations given above are only valid for magnetic fields $\ll\ 4\times10^{13}$ G.  As will be seen in Chapter 2, the derived mangetic field strengths near the surface of most pulsars violate this requirement and thus quantum synchrotron radiation formulae must be used (e.g., Sokolov \& Ternov, 1968 and Harding \& Lai, 2006).

\subsection{Curvature Radiation}\label{ch1CR}
In Section~\ref{ch1SR} the radius of curvature ($\rho$) of the motion was the radius of circular motion (i.e., the gyro-radius) determined by the strength of the magnetic field responsible for the acceleration.  Synchrotron radiation can be thought of as a specific case of CR corresponding to a circular motion caused by a magnetic field at angle $\chi$ to the particle velocity.  Following \citep{Jackson}, Eq.~\ref{ch1RelP} can be rewritten as given in Eq.~\ref{ch1curvP} to describe the power emitted away from a particle following a generic curved trajectory.
\begin{equation}\label{ch1curvP}
P\ =\ -\frac{2}{3} \frac{q^{2}c}{\rho^{2}}\beta^{4}\gamma^{4}
\end{equation}

The particle need not be accelerated such that $\rho$ is constant (i.e., the motion does not need to be circular) and thus the emitted CR spectrum will differ from that of synchrotron radiation.  However, at any point along the trajectory the particle can be thought of as undergoing instantaneous circle motion.  Thus, the emitted power will be similar to Eq.~\ref{ch1syncspec} with $\omega_{c}$ replaced by $\omega_{CR}(\rho)\ \equiv\ (3/2)\gamma^{3}c/\rho$ \citep{Jackson} and the $\sin(\chi)$ dependence removed.
 
The CR spectrum is often cast in terms of a critical energy $\epsilon_{CR} =\ h\omega_{CR}/2\pi$ which is rewritten by using $\hbar \ =\ h/2\pi$ and substituting the formula for $\omega_{CR}$ above to give $\epsilon_{CR}\ =\ (3/2)c\hbar\gamma^{3}/\rho$.  The power emitted, at photon energy $\epsilon$, from a single charge as a function instantaneous radius of curvature is
\begin{equation}\label{ch1CRspec}
P_{CR}\ =\ -\sqrt{3} \frac{q^{2}}{\hbar c} \frac{\gamma c}{2\pi\rho}F\Big(\frac{\epsilon}{\epsilon_{CR}}\Big),
\end{equation}
\noindent{}where a factor of $\hbar^{-1}$ has been introduced to Eq.~\ref{ch1syncspec} (changes from per unit angular frequency to per unit energy) and the relation $\omega\ =\ c\beta/\rho$ \citep{Jackson} has been used.  Note that the second term in the right hand side of Eq.~\ref{ch1CRspec} reduces to the fine structure constant ($\alpha_{f}$) when $q\ =\ e$.

HE pulsar emission models posit that the observed gamma rays are from electron CR in the radiation-reaction regime (see Section~\ref{ch1RR}) which occurs at $\gamma_{\rm RR}\ =\ (1.5E_{\parallel}/e)^{1/4}\sqrt{\rho}$ \citep{Venter10}, where $E_{\parallel}$ is the magnitude of the accelerating electric field.  This results in a cutoff energy of
\begin{equation}\label{ch1ERR}
\epsilon_{CR}^{\rm RR}\ \sim\ 4\mathnormal E_{\parallel,4}^{3/4}\sqrt{\rho_{8}}\ \rm GeV,
\end{equation}
\noindent{}where $E_{\parallel,4}\ \equiv E_{\parallel}/10^{4}$ statvolt cm$^{-1}$ and $\rho_{8}\ \equiv\ \rho/10^{8}$ cm.

For both synchrotron radiation and CR from relativistic particles the emission is beamed in the instantaneous, forward direction in a cone of half angle $1/\gamma$.  This will be particularly important for HE pulsar emission models in which particles are accelerated along curved magnetic field lines (see Chapter 2) with high enough Lorentz factors such that $1/\gamma\ \sim\ 0$.  This results in gamma rays being emitted, to good approximation, tangent to the field lines which simplifies the geometric models described in Chapter 5.

\subsection{Inverse Compton Scattering}\label{ch1ICS}
The basic Compton scattering process involves the interaction between a photon and a charged particle with rest mass $m$.  For simplicity, assume that the charged particle is at rest; for the case of a non-accelerated particle this can be achieved using a Lorentz transformation.  Let the initial and final energies of the photon be $\epsilon_{i}$ and $\epsilon_{f}$, respectively.  By accounting for energy and momentum conservation the final photon energy is
\begin{equation}\label{ch1compt}
\epsilon_{f}\ =\ \frac{\epsilon_{i}}{1+(1-\cos(\theta))\frac{\epsilon_{i}}{mc^{2}}}
\end{equation}
\noindent{}where $\theta$ is the angle between the initial and final photon directions.  Note that Eq.~\ref{ch1compt} predicts that $\epsilon_{f}\ \leq \epsilon_{i}$ depending on $\theta$ and the quantity $\epsilon_{i}/mc^{2}$.  Therefore, the particle gains energy from the photon in standard Compton scattering.

Eq.~\ref{ch1compt} is valid in the rest frame of the particle.  In this frame, since the particle is initially at rest it must gain energy from the photon.  ICS occurs when the particle is moving in the lab frame resulting in an increase of the photon energy.

In particular, let the particle be moving with velocity $\vec{\beta}$ in the lab frame where the photon has initial energy $\epsilon_{l}$ such that the angle between the photon direction and $\vec{\beta}$ is $\psi$.  The initial photon energy in the electron rest frame is related to the lab frame value as $\epsilon_{i}\ =\ \epsilon_{l}\gamma(1-\beta\cos(\psi))$ using the relativistic Doppler shift \citep{RL79}.

Even though the particle velocity will change due to the interaction, boosting back to the lab frame will still use the same $\gamma$ and $\beta$ as the initial boost.  Let the angle between the final photon direction, as observed in the lab frame, and initial boost direction be $\eta$.  The final energy of the photon in the lab frame is thus $\epsilon_{L}\ =\ \epsilon_{f}(1+\beta\cos(\eta))$.  For highly relativistic electrons, the total ratio between final and initial photon energies can be of order $\sim\gamma^{2}$ resulting in large energy gains for the photon.

In some astrophysical sources the observed HE gamma rays are thought to be ICS of lower energy optical or X-ray photons off of relativistic electrons.  For ICS of an isotropic photon field  with energy density $U$ off an isotropic electron distribution, all with the velocity of magnitude $\beta$, the total power from ICS is \citep{RL79},
\begin{equation}\label{ch1Pcompt}
P\ =\ -\frac{32\pi}{9}r_{0}^{2}c\gamma^{2}\beta^{2}U,
\end{equation}
\noindent{}note that this assumes each photon and each electron interact through ICS only once.  For a distribution of electrons with some spread in $\beta$, Eq.~\ref{ch1Pcompt} can be integrated over the corresponding number density of Lorentz factors to find the total emitted power.  Similar to synchrotron radiation, the spectrum of ICS radiation from a population of electrons with a power law energy distribution with index $-p$ is also a power law with spectral index $s\ =\ -(p-1)/2$.

As noted by \citet{RL79}, the previous arguments assume that these processes can be treated classically.  This is a valid assumption if $\epsilon_{i}\ \lesssim$ 100 keV but for higher photon energies in the particle rest frame quantum effects must be considered.  In particular, treating photons as discrete particles serves to lower the cross section and reduce the efficiency of ICS.

\section{Bremsstrahlung Radiation}\label{ch1bremss}
Bremsstrahlung radiation occurs when the trajectory of a charged particle is deflected by the electromagnetic field of another charged particle.  For the purposes of Section~\ref{ch1detection}, bremsstrahlung radiation from electrons and positrons traveling in matter and interacting with the electromagnetic field of a positively charged nucleus will be considered.

One way to characterize bremsstrahlung radiation \citep{RL79} is to consider an electron interacting with a nucleus of atomic number $Z$ such that the total charge and mass of the nucleus are $q_{n}\ =\ Ze$ and $m_{n}$, respectively.  Assume that both (or one) of the particles are moving at relativistic speeds but that both are unaccelerated such that it is possible to boost to an inertial frame of reference in which the electron is at rest before the interaction occurs.  If the electric field of the nucleus at rest is $\vec{E}_{Z}$, the Lorentz transformation suggests that the electron will observe the fields given in Eqs.~\ref{ch1Enuc} and~\ref{ch1Bnuc}, where the $\parallel$ and $\perp$ indices refer to velocity direction of the nucleus.  Assuming a large Lorentz factor such that $\gamma\vec{E}_{Z,\perp}\ \gg\ \vec{E}_{Z,\parallel}$ this results in $\vec{E}_{Z}^{\prime}\ \sim\ \vec{B}_{Z}^{\prime}$.
\begin{equation}\label{ch1Enuc}
\vec{E}_{Z}^{\prime}\ =\ \vec{E}_{Z,\parallel}+\gamma\vec{E}_{Z,\perp}
\end{equation}
\begin{equation}\label{ch1Bnuc}
\vec{B}_{Z}^{\prime}\ =\ \gamma c\ \vec{E_{Z}}\times\vec{\beta}
\end{equation}

To the electron, this will appear as an electromagnetic wave off of which it will Compton scatter (see Section~\ref{ch1ICS}).  Note that these are virtual quanta which scatter off of the electron but it is a helpful way of addressing the problem.  For an impact parameter (i.e., transverse direction of closest approach) $b$, the emitted energy per unit frequency as observed in the lab frame is given by Eq.~\ref{ch1bremSpec} \citep{RL79}, where $K_{1}$ is the modified Bessel function of the second kind with n = 1.
\begin{equation}\label{ch1bremSpec}
\frac{dW}{d\omega}\ =\ \frac{8Z^{2}e^{6}}{3\pi b^{2}c^{5}m_{n}^{2}}\Big(\frac{b\omega}{\gamma^{2}c}\Big)^{2}K_{1}\Big(\frac{b\omega}{\gamma^{2}c}\Big)
\end{equation}

The above arguments have been made assuming relativistic particles but note that Eq.~\ref{ch1bremSpec} is only valid for frequencies such that $h\nu\ \ll\ \gamma m_{n}c^{2}$.  For higher energies quantum mechanical corrections must be made.  Additionally, for bremsstrahlung in matter of a given density this must be modified to include the path length of the electron through the material.  To first order this should simply be a matter of incorporating the number of nuclei within a cylinder of radius $b$ centered on the projected path of the electron and integrating for a typical path length.

Note that while the above discussion has dealt with non-thermal emission such a thing as thermal bremsstrahlung emission does exist.  This occurs when the observed emission originates from a population of electrons with a thermal distribution of speeds; thus, the emission is still, at heart, non-thermal.

\subsection{Radiation-Reaction Limit}\label{ch1RR}
Newton's third law of motion says that for every action there is an equal and opposite reaction.  This applies to photon emission just as easily as it does to basic kinematics.  Photons carry momentum of magnitude $p\ =\ h\nu/c$ and conservation of momentum requires that the emitting particle experience a change in momentum equal but opposite to that of the photon.  However, if the particle experiences a change in momentum it must experience a force $\sim\ \Delta p$.  This is known as the radiation-reaction force.

This is important for charges being accelerated because it implies that not only must the accelerating force ($\vec{F}_{acc}$) give energy to the particle but some of the energy must go into countering the radiation-reaction force.  The radiation-force must scale as the emitted photon energy, thus when the photon energy is small relative to $F_{acc}$ the particle continues to gain energy.

However, as the particle continues to accelerate and gain energy the possible energy of emitted photons will grow as well.  At some point these two forces will reach a balance where all of $F_{acc}$ goes into counteracting the radiation-reaction force and the particle continues to emit but ceases gaining energy.  This is known as the radiation-reaction limit.

\section{Absorption Processes}\label{ch1absorb}
While it is important to understand the processes which produce the observed gamma rays, it is also important to know in what ways those same gamma rays can be absorbed.  The reason for this is two fold.  Firstly, the high energies of gamma rays means that they have wavelengths similar to the size scale of atoms.  Thus, gamma rays can not be reflected and focused as they will instead travel through matter where interactions can occur which change their energy and/or convert them to a different form.  Secondly, the environments which lead to HE gamma rays are often complicated and certain absorption processes can affect the observed spectrum; therefore, matching spectral features to known processes provides information about the environments and emission processes at work.

One such process, Compton scattering, has already been discussed in Section~\ref{ch1ICS}.  Another important absoption process is pair conversion, i.e., the creation of an electron-positron pair from one or more gamma rays.  Pair conversion occurs when a gamma ray interacts with an electromagnetic field or another photon.

\subsection{Pair Conversion With One Photon}\label{ch1onephoton}
To create an electron-positron pair using one photon requires that the photon have enough energy to create the rest mass of both particles ($\sim$ 1 MeV).  However, an isolated photon can not spontaneously decay into an electron-positron pair, even if the photon energy is much greater than 2$m_{e}c^{2}$, as momentum and energy can not simultaneously be conserved.

Another way in which to see that this is true is to consider a photon with energy $<\ 2m_{e}c^{2}$ in an inertial frame of reference.  Clearly, this photon can not decay into an electron-positron pair.  Suppose that a Lorentz transformation was used to boost to another frame of reference in which the photon is now observed to have an energy significantly greater than 2$m_{e}c^{2}$.  In this new frame pair creation would be energetically viable.  Assume the photon was to spontaneously create an electron positron pair in the boosted frame.  If an interaction occurs in one valid frame of reference it must occur in all; thus, this interaction would also be observed to occur in the original reference frame where it was not energetically feasible.  Therefore, the interaction can not occur, even in a frame for which the photon energy permits it.

Given a photon of sufficient energy all that is needed is some way to ``soak'' up the excess momentum.  The easiest way to do this is to introduce either an electric and/or magnetic field which provides a virtual photon for the initial photon to interact with.

While the extremely short wavelengths of gamma rays allows them to travel through matter easily they can interact with the electric fields of the constituent nuclei.  The attenuation coefficient from this process in the regime where $E\ \gg\ m_{e}c^{2}$ is given by Eq.~\ref{ch1matatten} \citep{Erber66} where $\lambda_{c}\ \equiv\ h/m_{e}c$ is the Compton wavelength, $N_{0}$ is Avogadro's number, $\rho$ is the material density, $A$ is the atomic weight, and $Z$ is the atomic number.
\begin{equation}\label{ch1matatten}
\vartheta(Z)\ =\ \frac{14}{27}\alpha_{f} N_{0} \Big(\frac{\lambda_{c}}{2\pi}\Big)^{2}\Big(h(Z)-0.640\frac{\rho}{A}(\alpha_{f}Z)^2\Big)
\end{equation}

The function $h(Z)$ in Eq.~\ref{ch1matatten} includes screening effects and Coulomb corrections and is given in Eq.~\ref{ch1hz} \citep{Erber66}.
\begin{equation}\label{ch1hz}
h(Z)\ =\ 6\frac{\rho}{A}(\alpha_{f}Z)^{2}\bigg\lbrace\ln\Big(\frac{183}{Z^{1/3}}\Big)+0.083-1.20(\alpha_{f}Z)^{2}\Big(1-0.86(\alpha_{f}Z)^{2}\Big)\bigg\rbrace
\end{equation}

All of the $Z$ dependence of the attenuation is in $h(Z)$ which for high $Z$ is roughly of order $Z^{6}$.  The attenuation increases only linearly with density and this is offset by the increase in $A$ for higher $Z$ materials, so care should be taken when choosing the optimal material for instruments designed to use this absorption process to detect gamma rays (see Section~\ref{ch1detection}).  Note that in this high-energy regime the attenuation is energy independent but the cross section for the interaction will decrease with energy.

One-photon pair production using just a magnetic field is more difficult as it requires that the ratio $R_{B}\ \equiv\ (E/m_{e}c^{2})(B/B_{cr})\ \gtrsim\ 0.1$ to reach significant transition probabilities \citep{Erber66}, where $B_{cr}\ \equiv\ 4.414\times10^{13}$ G is the quantum critical field.  Note that reaching both sufficiently high photon energies ($E$) and magnetic field strength ($B$) in a laboratory setting on Earth is difficult.

As will be seen in Chapter 2, a gamma-ray pulsar can easily create photons with $E\ \geq\ 100$ MeV near the stellar surface.  Achieving $R_{B}\ \gtrsim$ 0.1 with photons of this energy requires a magnetic field strength $B\ \gtrsim\ 2.2\times10^{10}$ G.  This condition is easily satisfied by many pulsars.  Additionally, pulsars are known to produce photons with GeV energies which lowers the required magnetic field strength by another order of magnitude.

\citet{Erber66} gives the photon attenuation coefficient for this process as,
\begin{equation}\label{ch11gamatten}
\Xi =\ 0.16\frac{\alpha_{f}}{\lambda_{c}}\frac{m_{e}c^{2}}{E}K^{2}_{1/3}\Big(\frac{4}{3}\frac{m_{e}c^{2}}{E}\frac{B_{cr}}{B}\Big)
\end{equation}
\noindent{}which reaches a maximum for fixed $B$ at $E/m_{e}c^{2}\ \approx\ 12(B_{cr}/B)$.  For fixed $E$, on the other hand, Eq.~\ref{ch11gamatten} is a strictly increasing function of $B$ which means that for sufficiently high magnetic field strength the process is efficient for arbitrarily high photon energies.

\subsection{Two-Photon Pair Creation}\label{ch1twophoton}
While one photon, alone, can not produce an electron-positron pair for the reasons discussed above, two photons can provided that the energetics permit it.  In principle, the minimum energy required is 2$m_{e}c^{2}\ \sim\ 1$ MeV.  However, this assumes a head-on collision which produces an electron and positron instantaneously at rest.  In practice, the angle between photon velocities will not be head-on and thus the produced pair must have non-zero velocities to conserve momentum which pushes the energy requirement above 2$m_{e}c^{2}$.

Following \citet{Y97}, the cross section for this interaction is given by Eq.~\ref{ch1twosigma}, where $s$ is the total energy in the center-of-momentum frame and $\psi(s)\ =\ 2m_{e}c^{2}/s$.
\begin{align}\label{ch1twosigma}
\sigma(s)\ =\ &r_{0}^{2}\pi\psi(s)\Big[ (2(1+\psi^{2}(s)-\psi^{4}(s)))\cosh^{-1}(\psi^{-1}(s))
\nonumber \\
&-(1+\psi^{2}(s))\sqrt{(1-\psi^{2}(s))}\Big]
\end{align}

This cross section is zero for $\psi(s)\ >$ 1, reaches a maximum value of $r_{0}^{2}\pi$ for $s\ \approx\ 4.4m_{e}c^{2}$, and then decreases for increasing $s$.

\section{Detection Methods}\label{ch1detection}
As noted previously, focusing gamma rays is an unfeasible prospect due to the ease with which they penetrate matter; however, gamma rays undergo well-understood processes in matter.  In particular, gamma rays will pair produce in a dense, high-Z material and the resulting particles can be tracked.  For photon energies $\gtrsim$ 1 GeV pair production is the dominant process; however, from $\sim$1-30 MeV Compton scattering is the dominant absorption process.

There are two classes of gamma-ray telescopes, those which operate in space and those which operate on the ground.  While both are pair-conversion telescopes the details and sensitivities of the two classes are quite different.  For both classes of gamma-ray telescopes the sensitivities are very dependent on the time scales and backgrounds of interest.  Therefore, while approximate energy ranges are given for each class in the following discussion, the interested reader is referred to the appropriate references for more detailed sensitivity curves.

\subsection{Space-borne Observatories}\label{ch1space}
While the elements which comprise the atmosphere do not have particularly high Z values nor is the atmosphere exceptionally dense, it is rather big.  Thus, the atmosphere is effectively opaque to cosmic gamma rays.  This necessitates getting above the atmosphere to perform gamma-ray astronomy, except at the highest energies (see Section~\ref{ch1ground}).

Space-borne, pair-conversion telescopes provide the converting material and track the resulting particles through the detector.  Notable examples include \emph{EGRET} \citep{DJT93}, \emph{AGILE} \citep{Tavani09}, and the \emph{Fermi Gamma-ray Space Telescope} (see Chapter 3).  These telescopes operate(d) in the energy range from tens of MeV to tens of GeV (and beyond).

The maximum energy to which such instruments are sensitive is limited by the amount of converting material which can be launched from Earth.  These requirements generally limit space-borne observatories to energies $\lesssim$ 300 GeV, beyond that the probability of a gamma ray interacting in the instrument is extremely low.

Observatories in space face many other issues as well.  The technology used to detect gamma rays will also trigger on charged particles.  The signal from such cosmic rays is $10^{4}$-$10^{6}$ times greater than some of the brightest gamma-ray sources which means that background rejection is a difficult and important task.  The background rejection is typically achieved, to lowest order, by surrounding the instrument with a charge sensitive material.

Additionally, space is full of ``junk'' which means that the instruments have to be shielded from micrometeorites and other small space debris.  And once something breaks it is typically not possible to access the instrument and fix it.

\subsection{Ground-based Observatories}\label{ch1ground}
As noted in Section~\ref{ch1space}, gamma rays incident on the atmosphere will pair convert.  The resultant electron and positron will be highly energetic and therefore traveling at nearly the speed of light in vacuum.  The speed of light in air is in fact less than $c$ but note that relativity limits the maximum velocity to $c$, not to the speed of light in the particular medium of travel.  Therefore, it is possible for charged particles in the atmosphere to travel faster than the speed of light in air and, in doing so, they will emit what is know as Cherenkov radiation.

The particles can also interact via bremsstrahlung radiation and emit energetic photons which can, in turn, create another energetic electron-positron pair.  This leads to an electromagnetic shower in which further particles are created until the shower constituents no longer have sufficient energy to produce photons capable of pair production.

For photons with very high energies ($\gtrsim$ 100 GeV, VHE) sufficient numbers of secondary particles and/or Cherenkov radiation will make it to the surface of the Earth to be meaningfully detected.  There are two detection methods for such photons, observatories which image the resulting Cherenkov light and those which measure the secondary particles on the ground.

Prominent examples of telescopes which measure the Cherenkov light are Whipple (e.g., Akerlof et al., 1992), MAGIC (e.g., Colin et al.,2009) , HESS (e.g., Hinton, 2004), and VERITAS (e.g., Holder, 2006).  These observatories operate in the energy range of $\sim$ 100 GeV to tens of TeV.  Examples of observatories which measure the particle showers on the ground are Milagro (e.g., Sullivan et al., 2001) and HAWC (e.g., Gonz\`{a}lez et al., 2008).  Such telescopes are generally sensitive to energies $\sim$ 100 GeV to $\sim$100 TeV.

Ground-based observatories have the advantage over space-borne telescopes of being serviceable and upgradable as well as having much larger effective areas.  However, there are downfalls to using the atmosphere as a converting material.  It is not of constant density or composition and changes over time.  Models of the atmosphere do exist and are used to estimate instrument performance but the systematics must be evaluated carefully.  Additionally, events initiated by cosmic ray electrons incident on the atmosphere will look similar to those from gamma rays.  However, the electron events should be, roughly, isotropic on the sky and thus will not create 'point-like' event excesses.

\section{Conclusions}\label{ch1conc}
Gamma-ray astronomy allows some of the most extreme environments of the universe to be explored and tests physical theories in regimes which are not accessible to laboratories on Earth.  What has not been discussed in this chapter is the fact that multi-wavelength studies are the real key to gamma-ray astrophysics (though one might convincingly argue that such studies are the key to all astrophysics).  The emission processes described in Section~\ref{ch1emit} do not emit only HE photons and understanding how the GeV spectrum connects with observations at lower and higher energies is important for fully characterizing the source.

Multi-wavelength studies are especially important for understanding pulsars which, in some cases, are seen to emit at radio, optical, radio, X-ray, and gamma-ray wavelengths.  Note that pulsed gamma rays from the Crab pulsar have been detected up to $\sim$60 GeV by MAGIC \citep{MAGICCRAB}, making pulsars TeV sources as well.  As will be seen in Chapters 7 and 8, when modeling the emission profiles from pulsars using only radio or gamma-ray data it is possible to arrive at incorrect conclusions.  However, combining information from both wavebands leads to strong tests of emission models.

It is an exciting time in gamma-ray astrophysics.  There are two HE missions currently in orbit, several VHE observatories operational on the ground, and a number of X-ray telescopes operating in space as well.  These observatories are augmented by ground based radio and optical telescopes across the globe.  This has facilitated an unprecedented level of coverage for many sources and led to exciting new discoveries (e.g., Abdo et al., 2010i and 2011a).

Our knowledge of HE sources has grown rapidly in the last few years and more questions have arisen which drive the development of new analysis and detection techniques.  In large part this has been facilitated by the willingness of observatories and individuals to share information and work towards a deeper understanding of astrophysical phenomena.  One can only hope that this collaborative attitude persists and leads to many more exciting discoveries.
\renewcommand{\thechapter}{2}

\chapter{\bf Pulsars}\label{ch2}
The first pulsar was discovered by Jocelyn Bell \citep{HB68} using a radio telescope built at the Mullard Radio Astronomy Observatory to measure interplanetary scintillation from quasars.  The pulsar signal was first interpreted as sporadic interference but the periodicity of the signal (with period of 1.337 s), localization to a specific location on they sky, and lack of measurable parallax suggested the source was, in fact, outside the solar system.  \citet{HB68} posited that these sources could be either white dwarfs or neutron stars (see Section~\ref{ch2NSs}) undergoing radial pulsations and suggested that the $\sim$1 s period argued in favor of a neutron star interpretation.

Soon after the initial discoveries \citet{Gold68} argued that, based on the repetition of fine-detail structure and observed polarization, the pulsar powerhouse was, in fact, a rotating neutron star with a strong magnetic field in which the primary magnetic axis was offset from the spin axis by some angle ($\alpha$) \citep{LS68}.  Linking the periodic nature of the emission to the rotation of the neutron star naturally explained the pulse shapes in terms of a 'lighthouse' effect as the magnetic axis swept across the line of sight of an observer located at an angle ($\zeta$) with respect to the spin axis.  \citet{Gold68} also tied the emission mechanism to the co-rotating plasma in the magnetosphere, predicted that the observed periods should be very slowly increasing, and that there should be more objects with even lower periods.

\citet{Pacini68} cautioned that there existed difficulties to overcome in the model of pulsars as rotating neutron stars given the fact that such objects were thought to be born in supernova explosions (Baade \& Zwicky 1934a,b).  In particular, given the supernova environment, it was not clear if pulsed radio emission could escape and retain a periodic signature.  However, using the Deutsch field \citep{Deutsch55} for a rotating star with misaligned rotation and magnetic dipole axes, \citet{Pacini68} demonstrated that such a source could not only power the observed pulsars but also the Crab nebula.

Only a few months later \citet{SR68} discovered two pulsating sources in the vicinity of the Crab nebula (NP 0527 and NP 0532) but, due to a time-resolution of 50 ms, they were unable to provide more than upper limits on the periods.  \citet{Comella69} confirmed the existence of one pulsating source consistent with the center of the Crab nebula and measured its period to be 33.09 ms.  The other source was found to be a 3.75 s pulsar \citep{ZR69} which had already been localized $\sim1^{\circ}.2$ from the center of the Crab Nebula \citep{Reifenstein69}.

The detection of the Crab and, subsequently, Vela \citep{Large68} pulsars with spin periods of $\sim$33 and 89 ms, respectively, firmly ruled out the white dwarf hypothesis.  The minimum spin period of a white dwarf is $\sim$1 s.  For faster rotation matter at the surface of the star would have to travel at a speed above the stellar escape velocity and thus the white dwarf would fly apart.

\citet{Comella69} also measured the change in spin frequency of the Crab pulsar to be $-16.07$ MHz s$^{-1}$, indicating that the pulsar period was increasing very slowly as predicted by \citet{Gold68}.  Assuming that the period is increasing due to magnetic dipole radiation and using the measured spin-down rate with reasonable neutron star properties the pulsar was estimated to be losing rotational energy at a rate of $\sim7\times10^{38}$ erg s$^{-1}$, nearly exactly what is needed to power the Crab nebula (Finzi \& Wolf, 1969 and Gunn \& Ostriker, 1969).  This finding firmly cemented the theory that pulsars are rapidly-rotating, magnetized neutron stars born in supernova explosions.

\section{An Overview of Pulsar Theory}\label{ch2Over}
Before exploring pulsar emission further, it is necessary to first explore the theory behind the pulsar engine.  A full treatment of neutron stars and evaluation of the equation of state is beyond the scope of this thesis; however, most neutron star models share basic traits from which it is possible to infer many properties when combined with timing measurements and assumptions as to the pulsar magnetic field structure.

When discussing pulsar theory it is important to keep in mind that while some properties can be measured very precisely, such as timing parameters, others can not, such as the mass, radius, and magnetic field; thus, there exists a large degree of uncertainty in the neutron star equation of state.

Neutron star magnetic field strengths can be inferred assuming a dipolar geometry which should dominate far from the star but it is likely that higher-order multipoles play an important role near the star though these fields can not be measured directly.  Radius measurements can be made by modeling observed thermal X-ray emission (e.g., Lattimer \& Prakash, 2001) but this has to assume that the entire polar cap is heated and thus some uncertainties still persist in such measurements.  The masses of some neutron stars in binary systems can be estimated from timing measurements to varying degrees of precision (e.g., Demorest et al., 2010).

\subsection{Neutron Star Basics}\label{ch2NSs}
Early in the twentieth century, stars were thought to have one, universal endpoint in the form of white dwarfs.  Normal stars, such as the sun, are stable against gravitational collapse due to the outwardly directed radiation pressure of photons released during fusion processes occurring in the stellar core.  However, it was realized that a star would eventually be unable to continue to sustain the necessary radiation pressure through fusion and be susceptible to gravitational collapse.  Observations of white dwarf stars had revealed them to be extremely dense, a fact which was greatly puzzling until \citet{Fowler26} applied principles of quantum mechanics to the problem and demonstrated that electron degeneracy pressure played a vital role in sustaining such stars against gravitational collapse.

Eddington argued for the existence of a maximum stellar temperature and pressure which would preclude any 'riotus suggestions' of stars with greater densities (Eddington 1931 and 1933) and thus all stars would die a white dwarf death.  However, Chandrasekhar (1931a,b) showed that electron degeneracy pressure was not sufficient to prevent further gravitational collapse for a star more massive than $\sim$0.91 solarmasses (M$_{\odot}$), this was later refined via a more careful treatment of the equation of state to the well-known \emph{Chandrasekhar Limit} of $\sim1.4$ M$_{\odot}$ \citep{Chandra35}.  While neutron degeneracy pressure prevents the collapse of neutron stars with masses above the \emph{Chandrasekhar Limit} it is insufficient to prevent collapse if the mass is above a few M$_{\odot}$.  This brought back the specter of unchecked gravitational collapse and was met with some resistance (e.g., Eddington 1935) as it implied runaway collapse of massive stars (i.e. black holes).

Neutron stars went in and out of scientific popularity as they were thought to be unstable above a mass of $\sim$0.7 M$_{\odot}$ (e.g., Oppenheimer \& Volkoff 1939) at which point the star would, more naturally, be a white dwarf.  However, \citet{Cameron59} demonstrated that by treating the equation of state more realistically (as opposed to previous authors who assumed a non-interacting Fermi gas) the maximum mass of a neutron star was $\sim$3 M$_{\odot}$ at which point the neutron degeneracy pressure would no longer be sufficient to prevent gravitational collapse.  This refueled speculation that neutron stars were created in supernova explosions as it does not require as much mass to be shed during supernovae of massive stars.  However, as discussed above, the supernova to neutron star link remained somewhat uncertain until the detection of the Crab pulsar.

The exact neutron star equation of state is unknown and many different theories exist but most share a few basic traits.  Neutron star models typically consist of five regions \citep{LP04}: an outer atmosphere, envelope, surface crust, outer core, and inner core.  The envelope and atmosphere contain little mass but should affect the emitted radiation to some extent.  The crust is thought to extend a few km below the surface and consist of nuclei, the mixture of which varies with depth as the density changes.  Near the bottom of the crust, top of the outer core, the matter begins to transition into a neutron superfluid.  The inner and outer core contain the majority of the neutron star mass.  It is expected that the neutrons should be a superfluid in the inner core with protons in a superconducting state.  The exact content of the inner core is not known, but it is possible to have matter in exotic states such as strangeness bearing hyperons or Bose condensates of pions or kaons \citep{LP04}.  Note that the recent mass measurement of 1.97$\pm$0.04 M$_{\odot}$ for PSR J1614$-$2230 \citep{Demorest10} rules out many of the more exotic equations of state.

Models typically predict masses between 1 to 2 M$_{\odot}$ with radii between $\sim$5 to 15 km.  Assuming a sphere with a uniform density, the neutron star moment of inertia is $I\ =\ (2/5)$MR$^{2}$ (for a mass M and radius R) resulting in $I\ \approx\ 10^{45}$ g cm$^{2}$ for a mass of 1.4 M$_{\odot}$ and a radius of 10 km.  For a pulsar with angular frequency $\Omega$ the rotational kinetic energy is $I\ \Omega^{2}/2$.  Given that this value of $I$ assumes a uniform mass distribution, which is clearly not realistic, the value of $10^{45}$ g cm$^{2}$ should be taken as an order of magnitude estimate.

Pulsar mass measurements generally take advantage of general relativistic effects of binary systems, such as measurements of excess time delays in eclipsing systems \citep{Demorest10}.  These measurements tend to be roughly consistent with the assumption of 1.4M$_{\odot}$ though millisecond pulsars (MSPs, see Section~\ref{ch2MSPs}) do seem to have slightly higher masses consistent with accretion from a companion.

\subsection{Pulsar Electrodynamics}\label{ch2edyn}
\citet{Deutsch55} solved for the internal and external electromagnetic field of a perfectly conducting sphere with a dipole magnetic field rotating in vacuo.  The functional form of the magnetic field is given in Chapter 5, assuming a dipole moment of magnitude $\mu$ and an angular frequency of rotation $\Omega$, the power radiated away from a rotating dipole in vacuum is given by Eq.~\ref{ch2dipRad} \citep{Jackson} using a convention of positive values for energy loss to reflect the rate at which energy is incident at an observer exterior to the star.  Note that no emission is predicted if the magnetic and rotation axes are aligned (i.e. $\alpha$ = 0\DEG{}).
\begin{equation}\label{ch2dipRad}
\dot{\rm E}_{\mathnormal dip}\ =\ \frac{2\mu^{2}\Omega^{4}}{3c^{3}}\sin^{2}(\alpha)
\end{equation}

However, \citet{GJ69} demonstrated, for an aligned rotator, that the vacuum condition could not be maintained as charges would be pulled from the surface, by a strong component of the electric field parallel to the magnetic field, resulting in a pair-plasma filling the magnetosphere and disturbing the field structure.  One consequence of their work was the realization that rotational energy is lost even for an aligned rotator as it is ultimately the rotation that drives a wind of particles from the stellar surface which carries energy away.

\citet{GJ69} assumed that the star was a perfect conductor.  They argued that the rotation would induce a surface charge distribution resulting in an electric field satisfying Eq.~\ref{ch2Eint}, where $\vec{B}_{int}$ is the internal, dipole field without rotation and $\vec{r}$ is the position vector at a given point in the star.  This condition is simply the statement that the electric field within an perfect conductor is zero, but note that in the surface charge layer this is not true.
\begin{equation}\label{ch2Eint}
\vec{E}_{int}+\frac{1}{c}\Big(\vec{\Omega}\times\vec{r}\Big)\times\vec{B}_{int}\ =\ \vec{0}
\end{equation}

By solving Laplace's equation \citet{GJ69} showed that the potential at the surface of the star was given by Eq.~\ref{ch2Phisurf} where $B_{surf}$ is the magnitude of the magnetic field at the pole, $P_{2}$ is the Legendre polynomial of degree 2, $\theta$ is the polar angle referenced to the spin axis, and R$_{\rm NS}$ is the neutron star radius. In vacuum, the quantity $\vec{E}_{ext}\cdot\vec{B}_{ext}$ exterior to the star (Eq.~\ref{ch2EdotB}) can be calculated by taking the negative gradient of Eq.~\ref{ch2Phisurf} and using the assumption of a dipole magnetic field.
\begin{equation}\label{ch2Phisurf}
\Phi_{surf} =\ -\frac{B_{surf}\Omega \rm R_{NS}^{5}}{3\mathnormal c r^{3}}P_{2}(\cos(\theta))
\end{equation}
\begin{equation}\label{ch2EdotB}
\vec{E}_{ext}\cdot\vec{B}_{ext}\ =\ -\Big(\frac{\Omega \rm R_{NS}}{\mathnormal c}\Big) \Big(\frac{\rm R_{NS}}{\mathnormal r}\Big)^{7}B_{surf}^{2}\cos^{3}(\theta)
\end{equation}

Note that while rotation will disturb the magnetic field structure from that of a static dipole \citet{Deutsch55} showed that, in vacuo, the magnetic field near the surface of the star ($r\ \ll\ c/\Omega$) approximates that of a static dipole in the co-rotating frame.  Thus, Eq.~\ref{ch2EdotB} is approximately correct near the stellar surface which is of interest here.  In particular, the component of the electric field parallel ($E_{\parallel}$) to the magnetic field at the surface can be estimated using Eq.~\ref{ch2Epar}.  This field must vanish inside the star but change continuously from the surface value to zero through the surface-charge layer.
\begin{equation}\label{ch2Epar}
E_{\parallel}\ =\ \bigg[\frac{\vec{E}_{ext}\cdot\vec{B}_{ext}}{B_{surf}}\bigg]_{r=\rm R_{NS}}\ =\ -\Big(\frac{\Omega R \mathnormal B_{surf}}{c}\Big)\cos^{3}(\theta)
\end{equation}

This electric field will exert a force $F\ =\ qE_{\parallel}$ on charges within the surface layer and, assuming $B_{surf}\ \sim\ 10^{12}$ G as derived in Section~\ref{ch2RadioPSRs}, this force exceeds the gravitational attraction $Gm\rm M_{NS}/R^{2}_{NS}$ by $\sim$5 (2) orders of magnitude for electrons (protons) (where $G$ is Newton's gravitational constant, M$_{\rm NS}$ is the mass of the neutron star, and an angular frequency of 63 rad s$^{-1}$, corresponding to a spin period of $\sim$ 100 ms, has been assumed).  Charges are therefore pulled from the surface of the star, populate the magnetosphere, and co-rotate for cylindrical distances less than the light cylinder radius (R$_{\rm LC}\ \equiv\ \mathnormal c/\Omega$) beyond which co-rotation would require velocities in excess of the speed of light.  The charges screen the accelerating $E_{\parallel}$ by increasing amounts until the net charge number density in the magnetosphere reaches the value in Eq.~\ref{ch2ngj} \citep{GJ69} at which point a force-free state is reached.
\begin{equation}\label{ch2ngj}
\mathfrak{n}_{\rm GJ}\ =\ \frac{\rho_{GJ}}{e}\ =\ -\frac{\vec{\Omega}\cdot\vec{B}}{2\pi c} \frac{1}{(1-(\Omega r/c)^{2}\sin^{2}(\theta))}
\end{equation}

\citet{GJ69} noted that this density applies only in the co-rotating portion of the magnetosphere which is bounded by the last closed field lines.  Particles will be streaming out along field lines in the open region which complicates calculation of the charge density as some return current must be hypothesized in order to avoid depleting the star of charge.  This implies that the open field line region is a prime site for particle acceleration.

\citet{Mestel71} applied similar arguments to the case of a non-aligned rotator by starting from the Deutsch field and similarly demonstrating that the vacuum conditions could not remain satisfied, i.e. charges are pulled from the surface as described above.  However, while similar qualitative arguments could be made an analytic form of the magnetic field was not produced.  \citet{MestelPryce92} attempted to derive the magnetic field structure using a Fourier integral technique with some success; however, they assumed the condition $\vec{E}\cdot\vec{B}$ = 0 for all points inside the light cylinder (not just the closed field line region) and their solution was singular at the equator.  Other attempts to derive an analytic form for the magnetic field structure have met with similar difficulties (e.g., Michel, 1974).  Recent efforts have shifted to magneto-hydrodynamic (MHD) simulations to numerically calculate the structure of a realistic pulsar magnetosphere (e.g., Contopolous et al., 1999; Timokhin, 2006).

\section{Radio Pulsars}\label{ch2RadioPSRs}
\begin{figure}
\begin{center}
\includegraphics[width=1.\textwidth]{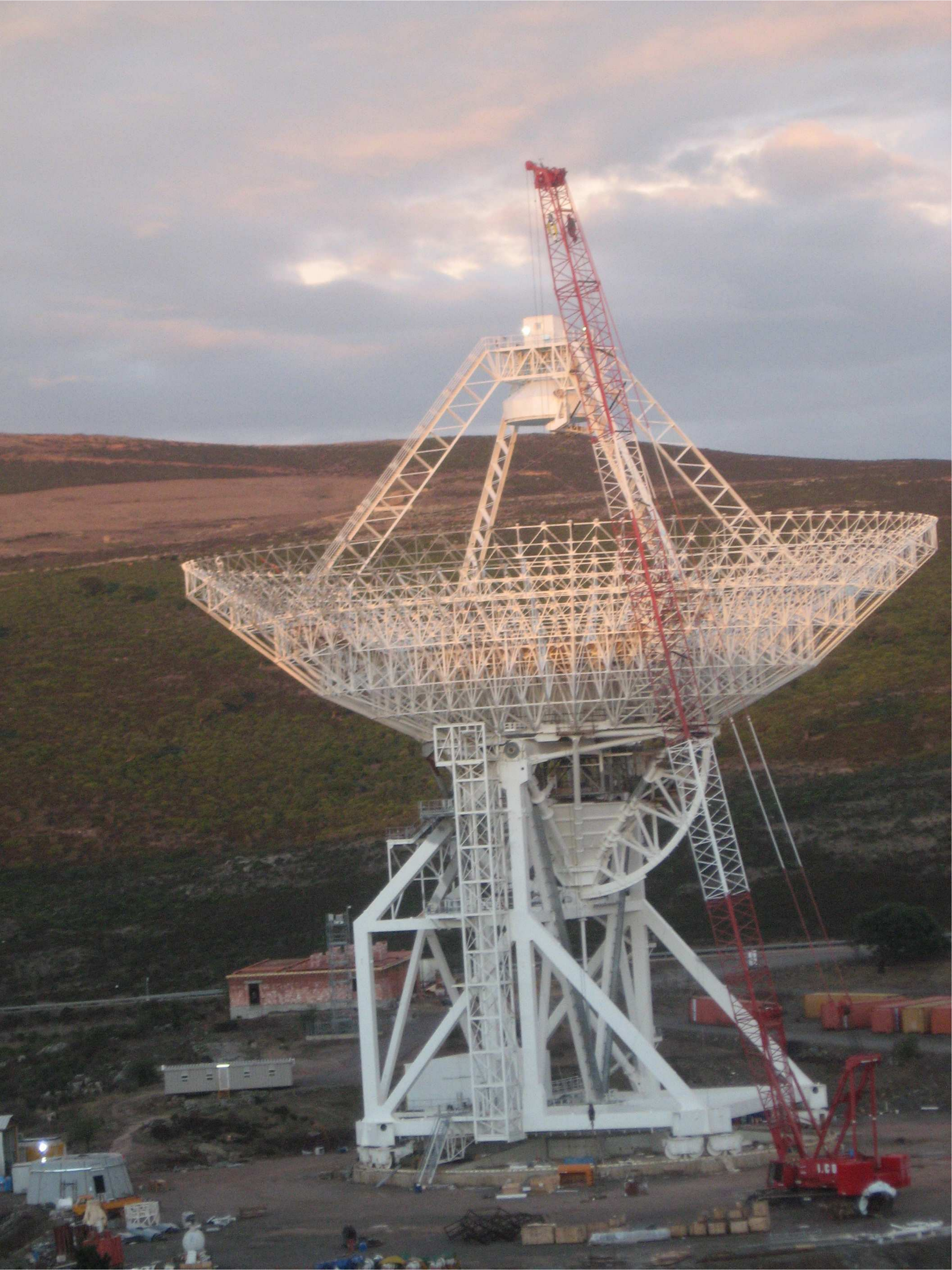}
\end{center}
\small\normalsize
\begin{quote}
\caption[The Sardinia radio telescope under construction]{The Sardinia radio telescope under construction 35 km North of Cagliari, Sardinia, Italy.  \emph{Photo Credit: T. J. Johnson}\label{ch2Sardinia}}
\end{quote}
\end{figure}

The majority of pulsars are observed only at radio wavelengths though many are also observed in the optical, X-ray, and gamma-ray wavebands.  The technology used to find, time, and study pulsars has improved greatly since the initial discovery but such a discussion is beyond the scope of this thesis.  The interested reader is referred to \citet{hdbk} for an informative overview of pulsar astronomy with a focus on radio wavelengths.  As an example of current technology the Sardinia radio telescope, under construction, is shown in Fig.~\ref{ch2Sardinia}.

\small\normalsize

Pulsars are typically weak radio sources with fluxes between $\sim20\ \mu$Jy to 5 Jy (1 Jy $\equiv\ 10^{-26}\ \rm W\ m^{-2}\ Hz^{-1}$) with negative spectral indices \citep{hdbk}.  For comparison, note that the radio flux of the AGN Centaurus A is $\sim$1330 Jy and that of the supernova remnant Cassiopeia A is 2477 Jy.  Therefore, except in sources which are particularly bright or which exhibit giant pulses, such as the Crab pulsar, individual pulses are not observed; rather, average pulse profiles are built up by observing the pulsar over hundreds to thousands of rotations and ``folding'' the data at the pulse period.  The fact that such temporal folding leads to constructive interference of the signal means that pulsar radio emission originates form a coherent process.

\subsection{Pulsar Timing Solutions}\label{ch2time}
Producing a pulsar timing solution requires correcting arrival times to a standard time system which accounts for the motion of the Earth about the sun, several relativistic effects, and for the frequency dependence of scattering in the interstellar medium.  Additionally, the folding of data over many pulse periods to achieve acceptable signal-to-noise level requires the ability to store and handle large amounts of data.

For gamma-ray pulsar studies a timing solution can be used to compute the phase of a given photon taking the fractional part of Eq.~\ref{ch2pphase}, where $t_{0}$ is a reference time for which the phase is known (usually chosen such that $\phi(t_{0})\ =\ 0$) and the rotational frequency is $f$ with the ``dot'' notation for time derivatives.  Note that it is not always possible to measure the second time derivative of a pulsar in which case that term is ignored.
\begin{equation}\label{ch2pphase}
\phi(t)\ =\ \phi(t_{0})+f(t-t_{0})+\frac{1}{2}\dot{f}(t-t_{0})^{2}+\frac{1}{6}\ddot{f}(t-t_{0})^{3}
\end{equation}

The event time $t$ used in Eq.~\ref{ch2pphase} must be transformed to a reference frame which will approximate that of the pulsar to provide sensible phases, i.e., one which can be considered an inertial frame with respect to the pulsar.  The Solar system barycenter (a.k.a center of mass) is taken to be such a reference frame.  For a recorded event time $t$ the equivalent time at the barycenter ($t_{b}$) is calculated using Eq.~\ref{ch2corrtime} \citep{Guillemot09}.
\begin{equation}\label{ch2corrtime}
t_{b}\ =\ t+\Delta_{C}-\Delta_{D}+\Delta_{R,\odot}+\Delta_{E,\odot}-\Delta_{S,\odot}+\Delta_{B}
\end{equation}

Eq.~\ref{ch2corrtime} first corrects $t$ to terrestrial time ($\Delta_{C}$) if necessary and then corrects for the time lag induced by scattering on the interstellar medium ($\Delta_{D}$).  This effect has a 1/$\nu^{2}$ dependence (where $\nu$ is the photon frequency) which can be neglected for gamma-ray pulsar studies.

The light travel time between the observatory position at time $t$ and the Solar system barycenter is corrected for using the term $\Delta_{R,\odot}$.  In the case of space-borne observatories, this last correction requires that the position of the space craft (which is orbiting the Earth) be known precisely at any given time.

The event time then needs to be corrected for the relativistic Shapiro delay ($\Delta_{S,\odot}$) which accounts for space-time deformities near massive objects \citep{Shapiro64}.  The expansion of space-time due to the motion of the Earth in the gravitational potential well of the Solar system is then accounted for ($\Delta_{E,\odot}$).  Finally, for pulsars in binary systems similar corrections must be made to account for the motion of the pulsar and its companion and the associated relativistic effects ($\Delta_{B}$).

Note that the Solar system barycenter is not the only possible reference frame to which the event time can be transformed for pulsar analysis.  The use of barycentric time requires that the pulsar position is well known, and while the positions of radio pulsars are generally very accurate the same is not true for pulsars which have been discovered through their gamma-ray pulsations.  In such cases, incorrect positions will manifest in the timing residuals if the center of the Earth is used as the reference frame.  Thus, for timing pulsars using gamma rays \citet{Ray11} advocate the use of geocentric time and demonstrate its use on pulsars discovered first in gamma rays.

\begin{figure}
\begin{center}
\includegraphics[width=0.75\textwidth]{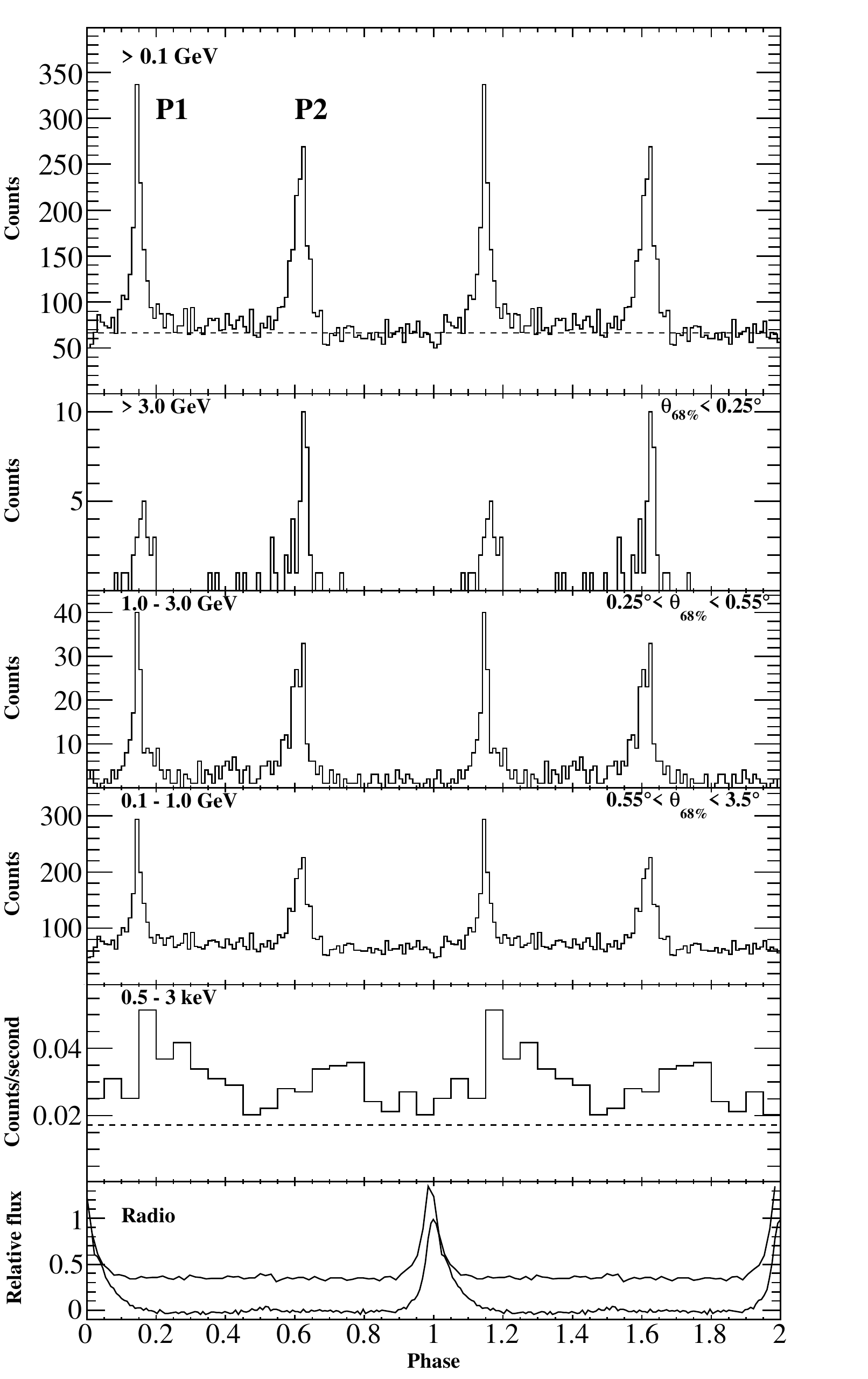}
\end{center}
\small\normalsize
\begin{quote}
\caption[Light curves of PSR J2021+3651]{Multi-band light curves of PSR J2021+3651 shown over two rotation periods.  The top four panels are gamma-ray light curves in different energy bands as marked, the second to lowest panel is the X-ray light curve, and the bottom panel shows the 300 and 1950 MHz radio profiles.  Note the radio structure which occurs right at phase 0 (1) which is difficult to properly characterize when plotted over one rotation period.  Reproduced from \citet{AbdoDF}.\label{ch2exLC}}
\end{quote}
\end{figure}

Once a pulse phase has been assigned to each event light curves are constructed by binning the events in phase.  This is typically done with equal width bins but can be done with variable width bins (see Chapter 4).  Quite often phase 0 is defined to occur at the fiducial point of the radio light curve which is typically at or near the main profile peak.  In order to better assess the structure in such features light curves are typically plotted over two rotation periods (i.e., from phase 0 to 2) as show in Fig.~\ref{ch2exLC}.

\small\normalsize

It is important to note that phases are only calculated between 0 and 1.  When plotting the light curves over two rotation periods values in bins with phases $>$ 1 are simply repeats of the bin values with phases between 0 and 1.  When evaluating the significance of a pulsed detection only the phase values from 0 to 1 are considered.

\subsection{Radio Properties}\label{ch2Rprop}
\begin{figure}
\begin{center}
\includegraphics[width=1.\textwidth]{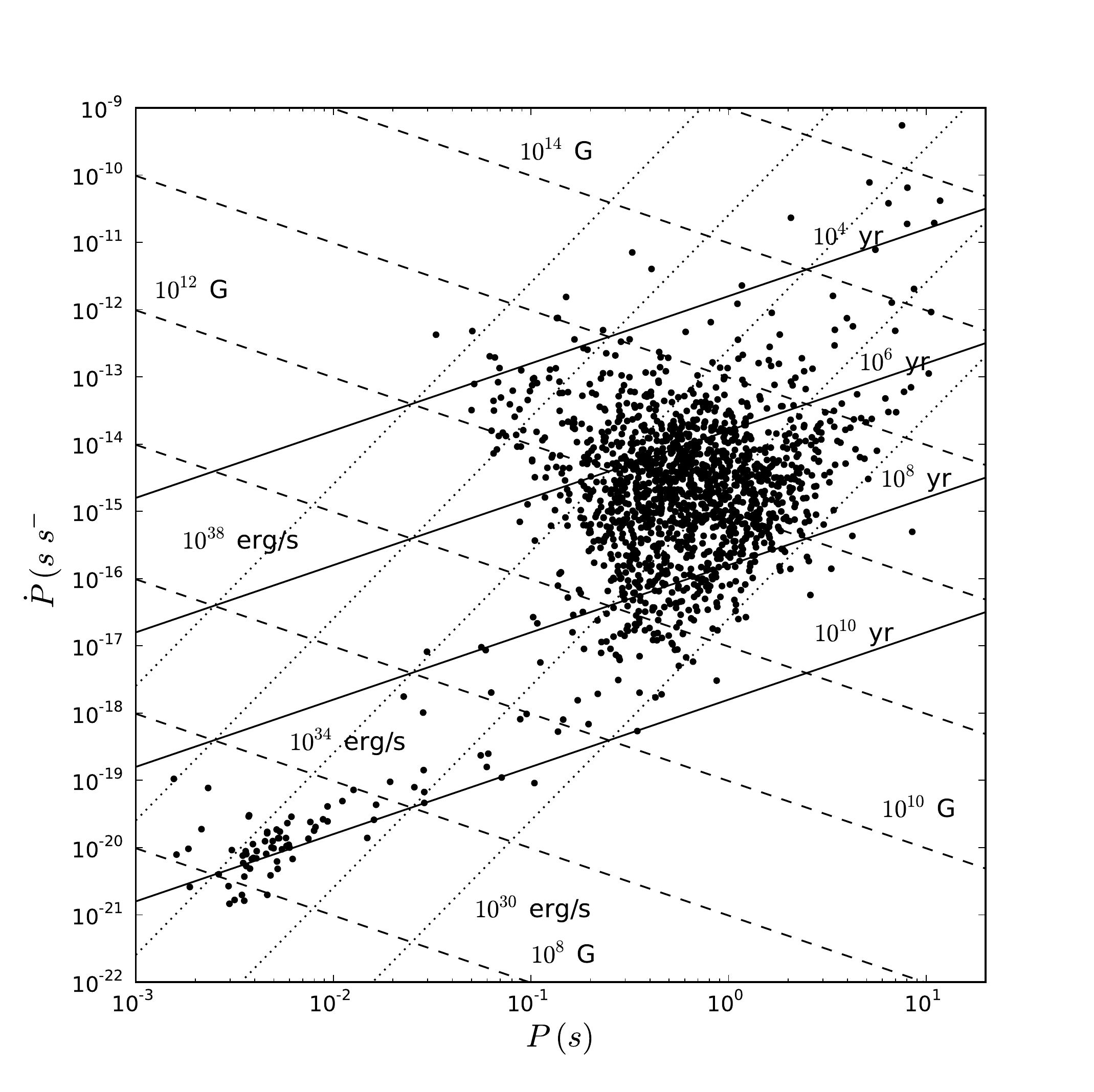}
\end{center}
\small\normalsize
\begin{quote}
\caption[Radio $\dot{\rm P}$-P distribution]{Distribution of measured P and $\dot{\rm P}$ values for known rotation powered pulsars in the ATNF pulsar database \citep{ATNF} as of mid-2009.  Lines of constant magnetic field (dashed), spin-down power (dotted) and characteristic age (solid) are drawn.  Only galactic field MSPs are shown.  Where proper motion measurements exists the $\dot{P}$ values have been corrected for the Shklovskii effect (see Section~\ref{ch2MSPs}).\label{ch2PdotP}}
\end{quote}
\end{figure}

Pulsars have been detected with spin periods (P) and period derivatives ($\dot{\rm P}$) covering a large range of values (see Fig.~\ref{ch2PdotP}) implying the possibility of several pulsar subclasses based on the inferred properties (discussed below).  Note that the $\dot{\rm P}$-P distribution is clearly bimodal, separating into longer period pulsars with relatively high rates of spin down and extremely short period pulsars with much lower spin-down rates.  The latter population will be discussed in more detail in Section~\ref{ch2MSPs}.

\small\normalsize

The exact mechanism behind pulsar radio emission is not known; however, it is generally accepted that for any mechanism to be viable it must be able to produce coherent emission with brightness temperature $\gtrsim\ 10^{20}$ K, be capable of generating highly polarized photons, and be efficient over a large range of frequencies, spin periods, and period derivatives.  Some mechanisms which have been proposed (outlined and discussed in more detail by Lorimer \& Kramer, 2004) are CR from bunches of particles, emission from instabilities in relativistic plasmas, and maser emission.  Even without knowing the exact mechanism responsible for the emission, it is possible to infer the geometry of the emission region from details of the observed pulse profiles at different radio frequencies as discussed in Section~\ref{ch2Beam}.

Many useful pulsar characteristics can be derived using the measured P and $\dot{\rm P}$ values and assuming a vacuum, dipole magnetic field.  The fact that pulsar periods are measured to increase with time suggests that they are losing rotational energy at a rate given by Eq.~\ref{ch2Edrot} in which the time derivative of the angular frequency is $\dot{\Omega}\ =\ -2\pi\dot{\rm P}/P^{2}$.
\begin{equation}\label{ch2Edrot}
\dot{\rm E}_{\mathnormal rot}\ =\ -I\Omega\dot{\Omega}\ =\ 4\pi^{2}I\dot{\rm P}/P^{3}
\end{equation}

If all rotational energy loss is due to magnetic braking then Eqs.~\ref{ch2dipRad} and~\ref{ch2Edrot} can be set equal to each other and used to solve for $\dot{\Omega}$ as given in Eq.~\ref{ch2dotOmega}.
\begin{equation}\label{ch2dotOmega}
\dot{\Omega}\ =\ -\frac{2\mu^{2}\sin^{2}(\alpha)}{3c^{3}I}\Omega^{3}
\end{equation}

The braking index, $n$, is defined by casting Eq.~\ref{ch2dotOmega} in the more generic form of $\dot{\Omega}\ =\ \kappa\Omega^{n}$, which suggests a braking index of 3 for magnetic dipole radiation.  If the pulsar is bright enough that a second derivative on the period (and thus frequency) can be measured then it is possible to solve for the braking index by differentiating and solving for $n\ =\ \Omega\ddot{\Omega}/\dot{\Omega}^{2}$.  Measured values range from $\sim$1.4 to 2.9 \citep{hdbk}, suggesting that other angular momentum loss processes besides magnetic braking are at work.  One such process is a wind of relativistic particles, as predicted by \citet{GJ69}, which leads to a spin-down rate of similar form ($\dot{\Omega}\ \propto\ \Omega^{n}$, with $n$ = 1; Michel, 1969) and disturbing the braking index from the purely dipole radiation value.  While the measurements do suggest $n\ \neq$ 3 it will be assumed that all of the rotational energy loss is from magnetic braking throughout this study.

Eq.~\ref{ch2dotOmega} can be recast in terms of P and $\dot{\rm P}$ using $\Omega\ \equiv\ 2\pi/$P and $\dot{\Omega}\ =\ -2\pi\dot{\rm P}/P^{2}$ (where the latter equation is obtained by simply differentiating the first).  Upon doing so, and consolidating the constant terms into one value $\kappa$, one obtains the expression $\dot{\rm P}\ =\ \kappa P^{2-n}$.  This expression can then be integrated from some initial time $t_{0}$ to some later time $t$ resulting in Eq.~\ref{ch2tau} in which P$_{0}$ is the initial period of the pulsar at time $t_{0}$.  For $n\ =\ 3$ and assuming P$_{0}\ \ll$ P, Eq.~\ref{ch2tau} reduces to the usual formula for the pulsar characteristic age $\tau\ =\ \rm P/2\dot{P}$.
\begin{equation}\label{ch2tau}
t\ =\ \frac{\rm P}{\dot{P} \mathnormal (n-1)} \bigg(1-\Big(\frac{\rm P_{0}}{P}\Big)^{\mathnormal n-1}\bigg)
\end{equation}

As noted by \citet{hdbk}, $\tau$ can provide a reasonable estimate of the pulsar age but the assumption of only magnetic braking and a negligible initial spin period means that it is, at best, an order of magnitude estimate.  In some cases, such as the Crab pulsar, the predicted $\tau$ ($\sim$1240 yr) can be quite close to the true age (957 yr) which is known from records of Chinese astronomers which date supernova explosion in the Crab as occurring in 1054 AD.  In others, such as PSR J0205+6449, estimated ages from supernova remnant associations disagree strongly with derived values of $\tau$.

As discussed in Section~\ref{ch2MSPs}, the theorized evolutionary track of MSPs results in little meaning for $\tau$; however, estimates of MSP ages based on observations of low-mass X-ray binaries are often comparable to the characteristic ages.

The magnetic field strength of a static dipole goes like $\mu/r^{3}$; thus, if one rearranges Eq.~\ref{ch2dotOmega} to solve for $\mu$ and takes $r\ =\ \rm R_{NS}$ the dipolar, surface field strength can be calculated as a function of $I$, P, $\dot{\rm P}$, and $\alpha$ (see Eq.~\ref{ch2Bsurf}).  Following the findings of \citet{Deutsch55}, the dipole approximation should result in a good estimate for the surface field strength.  However, far from the neutron star, in particular near the light cylinder, the field structure is not well known.  With that in mind, the same arguments that lead to Eq.~\ref{ch2Bsurf} also yield the magnetic field strength at the light cylinder (Eq.~\ref{ch2Blc}) which is of more interest for gamma-ray pulsars.  While Eq.~\ref{ch2Blc} can not give the correct field strength the differences in $B_{\rm LC}$ between different pulsars should be in the appropriate direction and thus meaningful trends can be drawn using these values.  Note that in Eq.~\ref{ch2Bsurf} $\alpha$ is usually taken to be 90\DEG{} and this has been assumed in Eq.~\ref{ch2Blc} .
\begin{equation}\label{ch2Bsurf}
B_{surf}\ =\ \frac{1}{\rm R_{NS}^{3}}\frac{1}{2\pi\sin(\alpha)}\sqrt{1.5Ic^{3}\dot{\rm P}\rm P}
\end{equation}
\begin{equation}\label{ch2Blc}
B_{\rm LC}\ =\ 4\pi^{2}\sqrt{\frac{1.5I\dot{\rm P}}{\mathnormal c^{3} \rm P^{5}}}
\end{equation}

All of the above quantities are helpful in looking for trends and distinctions among pulsars under a few key assumptions.  To fully understand the pulsar machine, however, it is necessary to explore the structure and nature of the observed emission in more detail.

\subsubsection{Distance Measurements}\label{ch2dist}
One complication involved in constructing a realistic emission model is deducing the total energy emitted at any given waveband when the distance to the pulsar is not well known.  Estimated pulsar distances, in our Galaxy, span a range from $\sim$100 pc to greater than 10 kpc with a large range of uncertainties.  Neutron stars are not typically bright in the optical waveband and thus, even for nearby pulsars, astrometric parallax measurements are difficult.  As alluded to in Section~\ref{ch2time}, pulsar timing models are very sensitive to the source position and thus a timing parallax and/or a proper motion can be measured for some sources.  The previous methods result in accurate, precise, and generally reliable distance estimates but are not applicable to the majority of known pulsars.  Other than timing, measurement of astrometric parallax with the Very Large Baseline Interferometer is the most promising distance measurement method.

However, the most commonly used method for estimating pulsar distances is to use the dispersion measure (DM) coupled with a model of the Galactic free electron density.  This technique exploits the fact that the interstellar medium, through which the emission must propagate, will result in a delayed arrival time for different radio frequencies (as discussed previously in relation to pulsar timing solutions).  This results in radio photons of different frequencies emitted at the same time (from the pulsar) arriving at different times (with lower frequency emission delayed by a larger amount).  In addition to depending on the frequency of observation, the difference in arrival times will also depend on how much of the interstellar material the signal passed through.  The dispersion in arrival times for different frequencies leads to the DM value (essentially the integrated column density along the line of sight with units of pc cm$^{-3}$, Lorimer \& Kramer, 2004) and the distance can then be inferred if the free electron density is known.  The most commonly used free electron density model is that produced by \citet{NE2001}, typically known as the NE2001 model.

DM-derived distance estimates are useful for gauging the luminosity of a given pulsar but must be treated with caution.  As noted by \citet{Brisken02}, fluctuations in the electron density can lead to uncertainties on DM-derived distances of a factor of $\sim$2.  Such uncertainties get magnified for estimates of pulsar luminosity which go like the distance squared.  One such example is the case of PSR J1939+2134 for which the DM-derived distance is 3.6 kpc (with the NE2001 model) but timing parallax measurements place the pulsar at $\sim$8 kpc \citep{Guillemot11}.

\subsubsection{Beam Structure}\label{ch2Beam}
Different empirical models for the radio beam structure have been constructed based on the general characteristics of observed pulsar light curves (e.g., Rankin, 1993; Story et al., 2007; Lyne \& Manchester, 1988).  These models do not generally require one specific emission mechanism but rather allow for different possibilities while focusing on the geometry of the beam.  The emission is typically assumed to occur in the open field line region where particle acceleration is possible.

A commonly accepted and invoked model is one in which the radio beam consists of a core component along the magnetic dipole axis and/or one or more hollow-cone beams.  In these models, different frequencies are emitted at different altitudes above the stellar surface and each frequency is only emitted at one altitude.  Note that the emission altitudes also depend on P and, more weakly, $\dot{\rm P}$ but even for the fastest MSPs the altitudes are predicted to be $\lesssim$ 30\% of the light cylinder radius, relatively close to the star.  For more details on the hollow-cone beam geometry see Chapter 5.  This prescription naturally explains the observed narrowing of radio peaks with increasing frequency as requiring higher frequencies to be emitted closer to the star which results in narrower beams.

More recently, a new class of radio emission models has re-emerged in which the emission region is further out in the magnetosphere, near the light cylinder, in a fan-like beam (e.g., Manchester, 2005 and Ravi et al., 2010).   These models are motivated by comparisons of gamma-ray and radio pulsar samples which suggest comparable beam sizes, particularly for younger pulsars, and predicts a caustic nature for the radio emission (see Chapter 5 for more details on caustic emission in pulsars).  Note that an early model for radio emission proposed by \citet{Gold69} also assumed that the emission site was near the light cylinder.  In this model gas was pulled from the stellar surface and emitted radio waves just before being flung outside of the light cylinder.

\subsubsection{Polarization}\label{ch2Pol}
Any viable emission mechanism needs to be a coherent process capable of producing highly polarized emission.  The plane of polarization is thought to be tied to the direction of the magnetic field at the emission point.  \citet{RVM} argued that the conal beam model naturally explained the observed polarization angle changes across the pulse phase, usually a characteristic `S' shape, using geometric considerations in what has come to be known as the rotating vector model (RVM).  For any given phase of rotation, the RVM equates the observed angle of linear polarization to the projected angle from the emission point to the dipole axis.  As the cone sweeps across the observer's line of sight the projected angle to the dipole axis changes.  This naturally explained the observed slow variations at peak edges and rapid changes at the center of the peak(s) \citep{hdbk}.

Using the RVM model, it is possible to extract $\alpha$ and $\beta\ \equiv\ \zeta-\alpha$ (where $\zeta$ is the angle between the pulsar spin axis and the observer's line of sight) from an observed pulsar polarization angle pattern.  Such RVM fits are often the only way in which the viewing geometry of a pulsar can be constrained.  If the pulsar wind creates a powerful enough nebula which can be observed in X-rays it is often possible to measure $\zeta$ (e.g., Ng \& Romani, 2008), note that this technique does not work for MSPs which are not observed to power bright nebulae.  Additionally, model dependent constraints on the geometry can be made by modeling pulsar gamma-ray light curves (see Section~\ref{ch2models} and Chapters 5 and 6 for more details).  Note that until quite recently the number of pulsars known to pulse in gamma rays was small and thus this method of constraining the geometry was little used.  Additionally, constraints from the latter method are only valid if one assumes that the underlying gamma-ray emission model used is correct.

\citet{Blaskiewicz91} extended the RVM model to include relativistic effects (see Chapter 5 and Appendix B for more details) such as aberration and time of flight delays.  This proved to be most important for attempting to constrain the geometry of MSPs, most of which exhibit polarization swings which are not easily explained by the RVM model.

\subsection{Millisecond Pulsars}\label{ch2MSPs}
The first pulsar spinning with a millisecond period (PSR B1937+21 a.k.a J1939+2134) was discovered by \citet{Backer82} using the Arecibo radio telescope.  This pulsar was found to have a period of 1.558 ms and a period derivative $<\ 10^{-15}$ s s$^{-1}$.  No evidence was found for a recent supernova event so the hypothesis that this was a very young pulsar with an intrinsically low spin down rate was rejected.  PSR J1939+2134 is an isolated pulsar but \citet{Backer82} noted a nearby H II region with which they argued the MSP was likely associated, suggesting that they were once part of a binary system which has since been disturbed.

Shortly after the first MSP was discovered and it became clear that this was an old system, \citet{Alpar82} proposed the initial version of what has become known as the recycled pulsar model.  The model has evolved over the years but the basic ideas remain the same.  The recycled pulsar model assumes that MSPs are the product of a normal pulsar (where normal, or non-recycled, is taken to mean P$\gtrsim$ 25 ms and $\dot{\rm P}\ \gtrsim\ 10^{-17}$ s s$^{-1}$) in a binary system with a main sequence companion.  The pulsar spins down over millions or billions of years until detectable emission ceases.  The companion enters a giant phase and overflows its Roche lobe. If the pulsar begins accreting mass and angular momentum from the companion the system enters the low-mass X-ray binary (LMXB) phase.  The accretion continues until the neutron star has reached a millisecond spin period at which point the pulsar emission resumes and the accretion disk is blown away leaving a binary MSP system.  \citet{Backus82} had earlier noted the possibility of such occurrences from observations of outliers in the $\dot{\rm P}$-P distribution but did not fully develop the idea.

The recycled pulsar model is supported by several pieces of observational evidence.  Only a few percent of non-recycled pulsars are observed to be in binary systems, likely due to disruption of binary systems from kicks in supernova explosions.  On the other hand, roughly 80\% of MSPs have low-mass binary companions.  Additionally, millisecond period pulsations in X-rays have been observed from LMXBs consistent with the presence of a rapidly rotating neutron star (e.g., Wijnands \& van der Klis, 1998).  More recently, \citet{Archibald09} have discovered a new MSP which, in archival observations, shows evidence for LMXB-type behavior and no pulsations.  This pulsar is thought to represent the missing-link in the LMXB to MSP evolutionary theory.

However, there are still problems with the recycled pulsar model which need to be addressed.  In order for the pulsar to accrete enough material to be spun up to millisecond periods its magnetic field must be relatively low (on the order of $10^{8}$ G).  As can be seen in Fig.~\ref{ch2PdotP}, the majority of non-recycled pulsars have inferred magnetic fields greater than $10^{10}$ G; thus, some mechanism much exist which allows for the pulsar magnetic field to decay with time.  To date, a satisfactory model for magnetic field decay does not exist (e.g., Romani, 1990 and Harding \& Lai, 2006).  Additionally, detections of isolated MSPs (such as the first MSP ever discovered) challenge the model.  Theories have been proposed in which the binary orbit shrinks, due to gravitational radiation, and the companion coalesces with the MSP (e.g., van den Heuvel, 1984).  Observations of black widow MSPs (such as J1959+2048) provide another avenue by which isolated MSPs might form.  These pulsars are observed to have very low mass companions (on the order of the mass of Jupiter) and observations of X-ray emission modulated at the orbital flux suggest that the pulsar wind is ablating the companion (e.g., Stappers et al., 2003).  Given enough time, the companion could be completely ablated away leaving an isolated MSP.

As indicated by Fig.~\ref{ch2PdotP}, many more MSPs have been detected since the initial discovery.  Presently, there are approximately 85 Galactic field MSPs and 147 globular cluster MSPs known though both of these numbers have increased rapidly in recent years.  Globular clusters are old, dense systems in the galaxy that comprise on the order of one million stars.  The density of these systems leads to more binary systems and, potentially, to a larger rate of MSPs.

An important timing effect, known as the ``Shklovskii effect'', in which the measured $\dot{\rm P}$ value of a pulsar is artificially increased due to its proper motion on the sky was proposed by \citet{Shklovskii70}.  For known pulsar proper motions this effect is generally of order $10^{-20}$ s s$^{-1}$ and thus does not strongly affect most non-recycled pulsars.  However, this effect is much more important for MSPs which are found with lower $\dot{\rm P}$ values (see Fig.~\ref{ch2PdotP}) and must be accounted for.

\section{Gamma-ray Pulsars}\label{ch2gPSRs}
\begin{figure}
\begin{center}
\includegraphics[width=1.\textwidth]{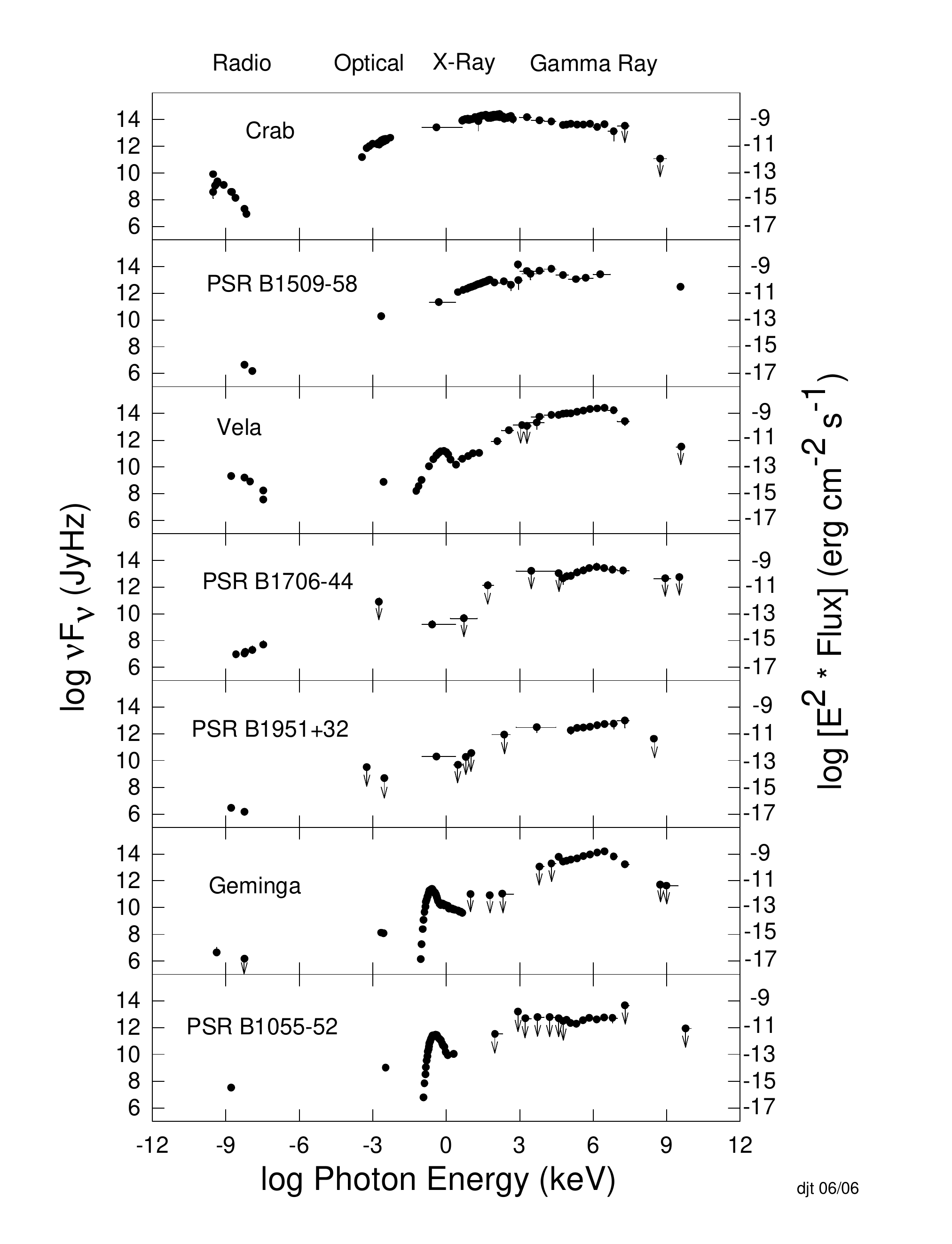}
\end{center}
\small\normalsize
\begin{quote}
\caption[Spectral energy distributions of pulsars detected with the \emph{Compton Gamma-Ray Observatory}]{Spectral energy distributions of pulsars detected by instruments aboard the \emph{Compton Gamma-Ray Observatory} from radio \emph{(top)} to gamma-ray \emph{(bottom)} wavelengths.  \emph{Figure Credit: D. J. Thompson}\label{ch2CGROSpec}}
\end{quote}
\end{figure}

Three years after the discovery of the Crab pulsar gamma-ray pulsations ($>$ 50 MeV) at the radio period were detected with a balloon experiment by \citet{Browning71}.  Gamma-ray pulsations from the Vela pulsar were detected by \citet{DJT75} using data from the SAS-2 satellite.  For nearly two decades Crab and Vela remained the only known gamma-ray pulsars, seeming to be ``special'' objects.  Detecting more gamma-ray pulsars was of extreme interest as the known objects were observed to emit $\sim$1 or 2 orders of magnitude more energy in gamma rays than at any other wavelength (see Fig.~\ref{ch2CGROSpec}).

\small\normalsize

\begin{figure}
\begin{center}
\includegraphics[width=1.\textwidth]{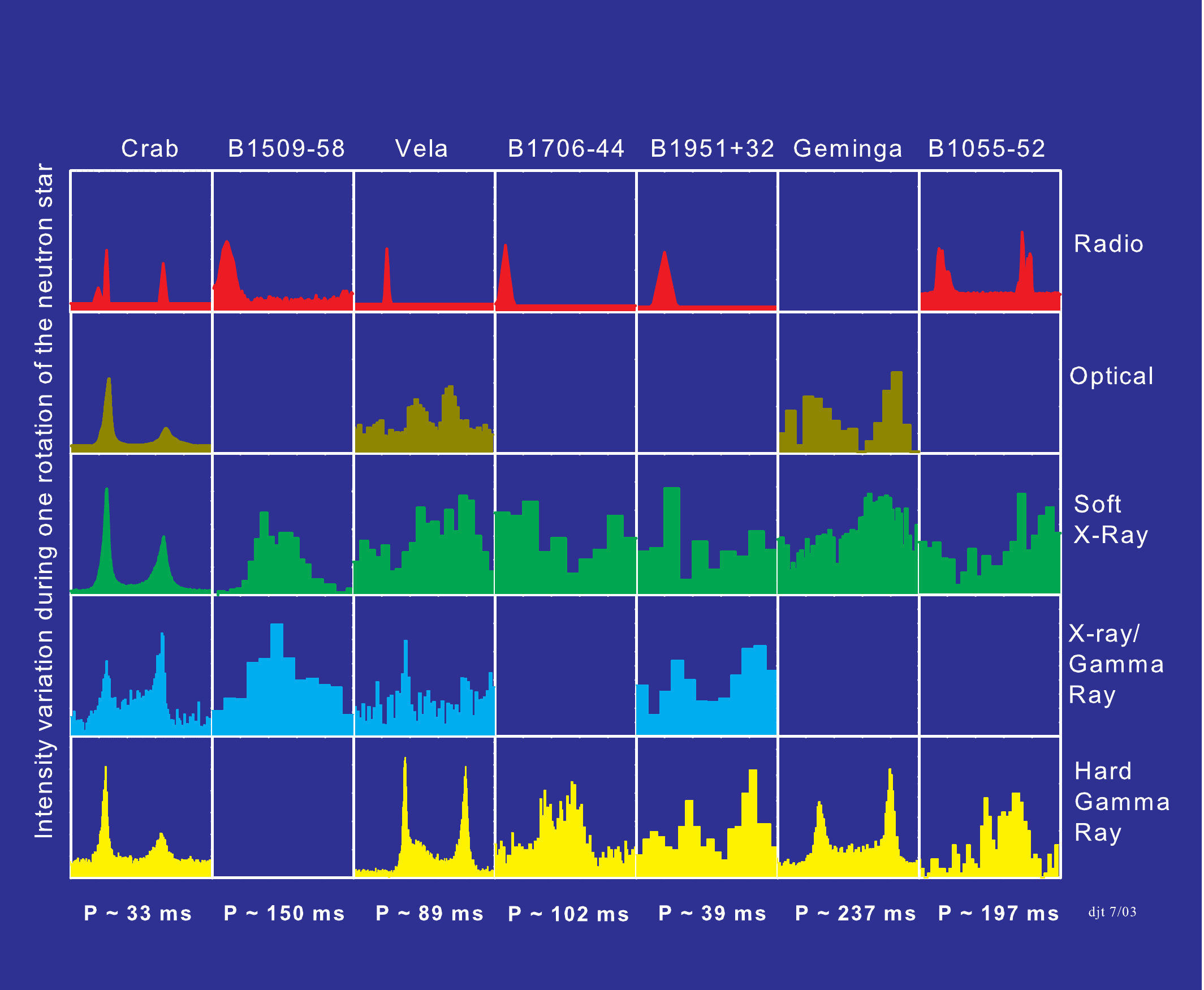}
\end{center}
\small\normalsize
\begin{quote}
\caption[Light curves of pulsars detected with the \emph{Compton Gamma-Ray Observatory}]{Pulsars detected by instruments aboard the \emph{Compton Gamma-Ray Observatory} from radio \emph{(top)} to gamma-ray \emph{(bottom)} wavelengths.  \emph{Figure Credit: D. J. Thompson}\label{ch2CGROPSRS}}
\end{quote}
\end{figure}

One of the main scientific goals of \emph{EGRET} was to search for more gamma-ray pulsars in order to determine if such behavior is common or if Crab and Vela are exceptions.  Two years after launch, the detection of another radio pulsar in gamma rays (J1709-4429) with \emph{EGRET} was announced \citep{DJT92}.  In 1992 \citet{HH92} discovered the period of the Geminga pulsar in X-rays.  This source had long been a mystery of gamma-ray and X-ray astronomy and the X-ray discovery prompted a search and detection of pulsations in the \emph{EGRET} data \citep{Bertsch92}.  Pulsations from two more pulsars (PSR J1057$-$5226 and J1952+3252) were eventually detected with \emph{EGRET} (Fierro et al., 1993 and Ramanamurthy et al., 1995, respectively) bringing the total number of known HE gamma-ray pulsars to six (see Fig.~\ref{ch2CGROPSRS}) with a few plausible candidates which have now been confirmed \citep{AbdoPSRcat}.

While this source sample was still relatively small, it did provide a better perspective on what properties might lead to gamma-ray emission in pulsars.  In particular, it was noted that all of the \emph{EGRET} pulsars were very energetic, occupying the top right of the $\dot{\rm P}$-P diagram.  Additionally, when the gamma-ray luminosity ($L_{\gamma}$) was plotted as a function of the spin-down power, a general trend seemed to appear suggesting $L_{\gamma}\propto\dot{E}^{1/2}$ (e.g., Arons, 1996 and Thompson, 2004).  Such a\\relationship is indicative of a dependence on the open field line voltage; however, it was noted that for pulsars with lower $\dot{E}$ values than the EGRET pulsars this trend would hit the $L_{\gamma}\ =\ \dot{E}$ line and thus some change in either the emission mechanism, the acceleration mechanism, or the efficiency function must occur if such pulsars emit significantly at gamma-ray energies.  Note that this $\dot{E}$ range encompasses most MSPs.

\small\normalsize

\citet{Fierro95} searched \emph{EGRET} data for gamma-rays pulsations from known radio MSPs and calculated 3$\sigma$ pulsed upper limits for nineteen sources.  \citet{Kuiper00} obtained a marginal pulsed detection of the MSP PSR J0218+4232 in the \emph{EGRET} data but a firm detection was complicated by the presence of a near-by and bright gamma-ray AGN.

Following the end of the \emph{Compton Gamma-Ray Observatory's} mission, radio observations of bright, non-variable, unassociated gamma-ray sources from the third \emph{EGRET} catalog (3EG) resulted in the detection of new pulsars (e.g., Keith et al., 2008).  It was suspected, based on positional and energetics arguments, that these pulsars were responsible for the 3EG sources.  However, given the length of time between the \emph{EGRET} observations and the radio discoveries it was not possible to extrapolate the timing solution backwards and successfully detect gamma-ray pulsations.

The field of gamma-ray pulsar physics entered a phase of waiting in which predictions were made for what the new generation of HE missions (\emph{AGILE}, Tavani et al., 2009; and \Fermi{}, formerly \emph{GLAST}, Atwood et al., 2009) might detect but no new gamma-ray pulsars were discovered.  Then, \citet{Halpern08} detected gamma-ray pulsations from PSR J2021+3651 with data from the \emph{AGILE} satellite.  Additionally, \citet{Pellizzoni09} searched \emph{AGILE} data for pulsations from many known radio pulsars and found several promising candidates.  They found 5$\sigma$ pulsations from PSR J2229+6114 with a spin period of 51.6 ms, all other candidates were found with a significance less than 5$\sigma$.  Among their low-significance candidates was PSR J1824$-$2452, an MSP in the globular cluster M28; however, the pulsation was only detectable in $\sim$ 5 days of the data and not over the whole, multi-month observation.  While some, unnamed physical process might be responsible for the disappearance of the signal, the detection has not yet been confirmed and remains suspect.

Thus, shortly after launch of the \emph{Fermi Gamma-ray Space Telescope} (see Chapters 3 and 4 for more details and results) roughly eight pulsars were known to pulse at energies above 100 MeV and it was not clear whether MSPs could emit gamma rays at detectable levels.  The latter uncertainty was, in part, a consequence of lacking strong constraints on the pulsar gamma-ray emission mechanism \citep{Fierro98} to provide a focus for theoretical development.

\section{Pulsar Gamma-ray Emission Models}\label{ch2models}
As discussed in Section~\ref{ch2edyn}, rotation naturally separates the pulsar magnetosphere into zones in which magnetic field lines either close or remain open.  The closed field line region will be in a force-free state and thus particle emission will not be possible.  With this in mind, gamma-ray pulsar emission models have focused on the open field line region as the most probable site for acceleration of particles to energies sufficient for HE gamma-ray production.  The emission models can be grouped into two general categories which are defined primarily by where the particle acceleration is assumed to occur, either at low altitudes above the magnetic polar cap or at high altitudes near the light cylinder.  Basic aspects of typical emission models are presented below as well as a discussion of gamma-ray emission expected from MSPs.

Note that while these models differ on the location in the magnetosphere from which the HE gamma rays originate they all assume that the emission mechanism is CR from electrons accelerated along the magnetic field lines.  In principle, the electrons could emit via synchrotron radiation as well.  However, near the PC primary particles will be constrained to follow the field lines closely and thus the pitch angles required to emit significant amounts of synchrotron photons are unattainable.  As discussed below, secondary particles will be produced at low altitudes which can have larger pitch angles but the energies will be too low to produce HE photons.  Additionally, the magnetic field strength in the outer magnetosphere is too low for electrons to emit HE photons from synchrotron radiation.

\subsection{Polar Cap Model}\label{ch2PC}
The first class of HE pulsar emission models is the polar cap (PC) model which assumes the existence of an accelerating gap just above the magnetic polar cap in the open field line region.

\citet{Sturrock71} first noted that particle acceleration above the polar cap could be important for pulsar emission.  Sturrock's model separated the polar cap into two zones (one from which electrons stream out and one from which protons stream out, following Goldreich \& Julian 1969) and interpreted observations of the Crab pulsar (from radio to gamma-ray wavelengths) as viewing emission only from the proton zone.

\citet{RS75} proposed the existence of a PC gap based on an inability of the pulsar to establish a flow of positively charged particles based on the expected behavior of dense matter in intense magnetic fields.  \citet{Hardee77} showed that near the stellar surface CR could produce photons with energies $\gtrsim$ 1 GeV but argued that it was not clear if the gaps proposed by \citet{RS75} could form (as positive ions should not be bound to the surface strongly enough for fields less than $10^{13}$ G).  However, generation of HE CR photons was not dependent on the gap structure (though the flux and spectrum are).

\citet{HTE78} demonstrated that a spectral cutoff at high energies, different from the CR cutoff, is expected due to one-photon pair production near the stellar surface.  \citet{Harding81} extended this model by tracking the energy losses of emitting primaries and predicted a functional form for the gamma-ray luminosity as a function of period and magnetic field strength and predicting that $L_{\gamma}\ \propto\ \dot{E}^{1/2}$.  These studies neglected the acceleration mechanism and did not consider the effect of secondary particles produced from one-photon pair production.  \citet{DH82} showed that emission from primaries above the acceleration region and from secondaries was important to reproduce the observed HE spectra.  All these considerations have fed into the standard PC emission models (e.g., Daugherty \& Harding 1994,1996; Harding 2009).

Primary particles are injected from the stellar surface and accelerated to a typical height of a few stellar radii.  The observed gamma rays are thought to be from CR of the primaries as they accelerate along the magnetic field lines.  Photons with high enough energies can interact with the strong magnetic field of the pulsar and produce an electron-positron pair.  This mechanism leads to a cascade of particles which screens the accelerating field and sets the maximum acceleration height as follows.

A pair cascade can be initiated by energetic photons from CR or by ICS thermal X-rays from the PC.  A pair cascade is initiated once the CR or ICS photons reach energies $\gtrsim2m_{e}c^{2}$ in the frame where the photon has no momentum parallel to the magnetic field.  These photons will interact with the magnetic field via one-photon pair production.  The height at which pair creation begins denotes the start of the pair-formation front (PFF).

Assuming $\vec{\Omega}\cdot\vec{\mu}$ is such that secondary electrons are accelerated away from the stellar surface, the positrons will decelerate, turn around, and then accelerate back to the stellar surface.  This results in a net negative charge amassing above the PFF.  This charge establishes an electric field which opposes and eventually cancels out the accelerating field.

Charges above the screening layer will flow freely along the field lines and out through the light cylinder.  These charges will be produced with non-zero pitch angles ($\chi$ in Eq.~\ref{ch1syncspec}) to the magnetic field and will thus lose energy rapidly to synchrotron radiation (see Chapter 1).

The spectrum of HE photons which escape from the PC region is expected to be relatively hard (photon index, $\Gamma$, from 1.5-2.0 assuming a $E^{-\Gamma}$ convention) with a spectral cutoff at a few GeV that turns over faster than an exponential due attenuation from one-photon pair production \citep{Harding09}.  The emission pattern is predicted to take on the shape of a hollow cone beam centered on the magnetic axis, similar to what is expected for the radio as described in Section~\ref{ch2Beam}, and thus PC models predict near phase alignment of the radio and gamma-ray profiles.  These models have trouble simultaneously reproducing the widely separated gamma-ray peaks observed in many gamma-ray pulsars without requiring special geometries and artificial flux enhancements near the PC rim \citep{DH96}.  Additionally, recent observations suggest a lack of evidence for faster than exponential cutoffs in known gamma-ray pulsars \citep{AbdoPSRcat}.

\subsection{Slot Gap and Two-pole Caustic Models}\label{ch2SGTPC}
The second class of HE pulsar emission models assumes that the gamma rays come from particles accelerated out to altitudes near the light cylinder.  This class is composed of two main models, the first of which is a natural extension of the PC model.

As discussed above, the altitude of the PFF is determined by the onset of one-photon pair production which requires particles energetic enough to either ICS thermal X-rays or emit CR photons above the pair-production threshold.  This happens sooner for primaries which experience a stronger accelerating field.  This field will be strongest along the magnetic axis and must go to zero at the edges of the open field line region beyond which the magnetospheric charge density is assumed to be that given by \citet{GJ69} and $\vec{E}\cdot\vec{B}$ = 0.  The PFF altitude thus depends on the magnetic colatitude ($\theta_{\mu}$), with the lowest value for $\theta_{\mu}$ = 0\DEG{} and curving upwards to asymptotically approach the surface of last closed field lines.

\citet{AS79} first noted that this led to the existence of a narrow slot gap (SG) near the PC rim which could extend out to high altitudes.  Note that they also found a slot gap which formed along the magnetic axis due to a restriction of particle acceleration to ``favorably'' curved field lines (those which curved toward the rotation axis).  \citet{Arons83} demonstrated that particle acceleration in such SGs was capable of producing HE photons, with emission restricted to favorably curved field lines, by accelerating particles out to high altitudes but was unable to reproduce the observed luminosity of the Crab pulsar.

\citet{HM98} reevaluated PC emission models to include general relativistic effects and the formation of a secondary PFF due to the acceleration of secondary positrons back toward the stellar surface.  They found that a stable accelerating region could be generated with two PFFs if they were due to CR and not ICS.  Their calculations were not restricted to favorably curved field lines which allowed for acceleration above the entire PC.

\citet{MH03} noted that the analysis of \citet{HM98} confirmed the formation of SGs at the PC rim bounded by the surface of last closed field lines.  The predicted gaps were narrower than those of \citet{Arons83} as they recognized that the pair-plasma bordering the gap opposite the last closed field lines would also function as a conducting boundary.  By considering photons produced at the inner edge of the SG from pair cascades (not just primaries) \citet{MH03} demonstrated the viability of this model for HE pulsar emission though they restricted their analysis to low altitudes ($\lesssim$ 5 R$_{\rm NS}$).

\citet{MH04a} extended the SG model to high altitudes, out near the light cylinder, and demonstrated that the particles reach a radiation-reaction limited regime in which energy gained from the accelerating field is balanced by energy lost to CR radiation.  In this SG model emission from primaries occurs throughout the entire gap and is not confined to favorably curved field lines.  They also noted the formation of caustics (see Chapter 5) for emission from particles at 0.1 to 0.7 R$_{\rm LC}$ which naturally explains the sharp, bright peaks observed in gamma-ray light curves.  The magnetic fields at high altitudes are not strong enough to significantly attenuate the emitted spectrum of HE gamma rays; thus, the SG model predicts a spectrum which cuts off only exponentially, following the cutoff in the CR spectrum, as opposed to the faster than exponential cutoff expected from PC emission.

\citet{DR03} developed the (purely geometric) two-pole caustic (TPC) pulsar emission model in order to match observed HE pulsar light curves.  In particular, the TPC model was able to produce  sharp, widely separated peaks which lagged the main radio pulse via the formation of emission caustics.  This model can be considered a geometric realization of the SG model.  The emission is assumed to occur in thin accelerating gaps along the surface of last closed field lines from the PC rim out to high altitudes in the pulsar magnetosphere.  \citet{DR03} followed electrons along curved magnetic field lines (assuming the Deutsch field configuration in the co-rotating frame) where they emit CR photons tangent to the field line.  The photons then travel in a straight line in the inertial observer's frame, after correcting the direction for relativistic aberration and time-of-flight delays (see Chapter 5 for more details).

The most prominent features (i.e., the two bright peaks typically seen in gamma-ray pulsar light curves at the time) of the TPC light curves were found to arise from emission at radial distances $\lesssim$0.75 R$_{\rm LC}$ (in agreement with the findings of Muslimov \& Harding 2004a).  Thus, \citet{DR03} did not follow the emission beyond this distance, presumably for reasons of computation time and also because the magnetic field structure and accelerating field are not as well known near the light cylinder; however, \cite{Venter11} have followed the emission out to larger radial distances and found that there can be significant changes in some of the less prominent features which can be important to matching more recent observations of gamma-ray pulsar light curves \citep{AbdoPSRcat}.

\subsection{Outer-gap Model}\label{ch2OG}
The second type of outer-magnetospheric emission model is the outer-gap (OG) model, the current form of which was first described by \citet{Cheng86a}.  OG models assume that the charge density in the magnetosphere is that given by Eq.~\ref{ch2ngj} everywhere, not just in the co-rotating region.  The sign of the charge distribution changes sign on either side of the null-charge surface (NCS), which is defined by the condition $\vec{\Omega}\cdot\vec{B}$ = 0.  For the case $\vec{\Omega}\cdot\vec{\mu}\ <$ 0, a portion of the open field line region below the NCS but above the closed field line region will have a negative charge distribution.  Electrons will flow out of the light cylinder along these field lines which results in a region of net positive charge near the NCS.  This region will repel positive charges on the other side of the NCS which results in a region void of charge along the surface of last closed field lines where the condition $\vec{E}\cdot\vec{B}$ = 0 is no longer satisfied.  Note that this gap is bounded below by the NCS and thereby distinct from that used in the SG and TPC models.

It is then possible to accelerate electrons in this region which will be confined to move along the curved magnetic field lines and preferentially emit into the open field line region above the NCS.  \citet{Cheng86a} note that these gamma rays could provide the mechanism by which the charge density in the open field volume is maintained and thereby limit the width of the gap.  The observed HE emission is not, in fact, from particles accelerated in the vacuum gap.  Instead, gamma rays from those particles pair produce with thermal X-rays (which penetrate the gap from the stellar surface) near the upper gap boundary and it is emission from these energetic, secondary particles which is seen by an external observer \citep{Cheng86b}.

OG models have been shown to reproduce the observed light curves of gamma-ray pulsars by several authors (e.g., Cheng et al., 1986b; Roman \& Yadigaroglu, 1995).  \citet{Cheng00} used a more detailed, 3D model of the OG regions to calculate phase-resolved spectra.  Similar to the SG and TPC model, spectra from the OG model are expected to cutoff only exponentially.

\subsection{Gamma Rays from MSPs}\label{ch2grayMSPs}
One common feature of OG, SG, and TPC models is the existence of relatively narrow gaps bordering the last closed field line surface.  The creation of these gaps requires that the accelerating field be screened by charged particles in the magnetosphere which requires copious pair-production.  ICS PFFs set in at lower altitudes than those from CR and do not screen the accelerating field \citep{Harding09}, thus CR pair production is the important process for gap creation in outer-magnetospheric emission models.  \citet{HM02} numerically calculated the CR pair-production ``death line'' on the $\dot{\rm P}$-P diagram, any pulsar below this line (corresponding to $\tau\ \sim\ 10^{7}$ yr for non-recycled pulsars and $10^{8}$ yr for MSPs) can not produce pairs in large amounts from CR.  This result suggested that, for nearly all known MSPs, the accelerating field should be unscreened out to high altitudes, near the light cylinder, and led to the development of the pair-starved polar cap (PSPC) emission model (see Section~\ref{ch2PSPC}).

However, some authors (e.g., Zhang \& Cheng, 2003 and Zhang et al., 2007) have developed models for HE emission from MSPs assuming an OG geometry.  If higher order magnetic multipoles are dominant near the surface of MSPs it may still be possible to produce pairs in sufficient quantities to create narrow gaps in the outer magnetosphere.  \citet{ZC03} argued that the observed thermal X-rays from some MSPs argued for the existence of such multipoles.  \citet{Zhang07} refined the MSP OG model to include effects of the viewing geometry and extended the model application to MSPs detected in the globular cluster 47 Tucanae as well as field MSPs.  \citet{Srinivasan90} predicted that even if individual MSPs were not detected in gamma rays the population as a whole might contribute significantly to the observed diffuse emission.  Given recent results on gamma-ray MSPs (see Chapter 4) higher order multipoles may indeed be important in MSPs.

\subsubsection{Pair-starved Polar Cap Model}\label{ch2PSPC}
The PSPC model was first proposed by \citet{MH04b} as a possible gamma-ray emission model for pulsars lying below the CR (and possibly ICS) pair-production death line calculated by \citet{HM02}.  This model assumes that ICS is not efficient at screening the accelerating field and thus particles are accelerated to altitudes near the light cylinder, emitting CR as they go.  \citet{MH04b} demonstrated that it was possible to accelerate particles to sufficiently high energies for production of gamma rays in this model.  \citet{Harding05} investigated the predicted X- and gamma-ray spectrum from such a model and compared it to that of PSR J0218+4232.  They found that the spectrum has contributions from CR, ICS, and synchrotron radiation from cyclotron resonant absorption of radio emission; however, which components are observed depended strongly on the viewing geometry and on the pulsar's relation to the CR death line.  It should be noted that the absence of narrow gaps will, generally, lead to much broader peaks with gamma-ray features leading the radio as demonstrated by \citet{Venter09}.

\section{Conclusions}\label{ch2Conclusion}
Note that while this chapter has been an overview of neutron star and pulsar theory it is by no means exhaustive.  In particular, this discussion has dealt mainly with rotation powered pulsars.  The discovery of accretion powered pulsars is one of the observational pieces of evidence in support of the recycled pulsar model but that topic has been mentioned only briefly.  Additionally, neutron stars present an opportunity to study matter under conditions beyond anything which can be created in a laboratory on Earth and the equations of state are much more varied and complicated than presented here.  Neutron stars have also provided some of the most stringent tests of general relativity (e.g., Taylor \& Weisberg, 1982) and long-term timing of bright, stable millisecond pulsars may provide the first direct detection of gravitational waves (e.g., Manchester, 2011).  In short, pulsars are extremely attractive sources as they allow for the study of a large range of extreme physics.
\renewcommand{\thechapter}{3}

\chapter{\bf The \emph{Fermi Gamma-ray Space Telescope}}\label{ch3}
\begin{figure}
\begin{center}
\includegraphics[width=.9\textwidth]{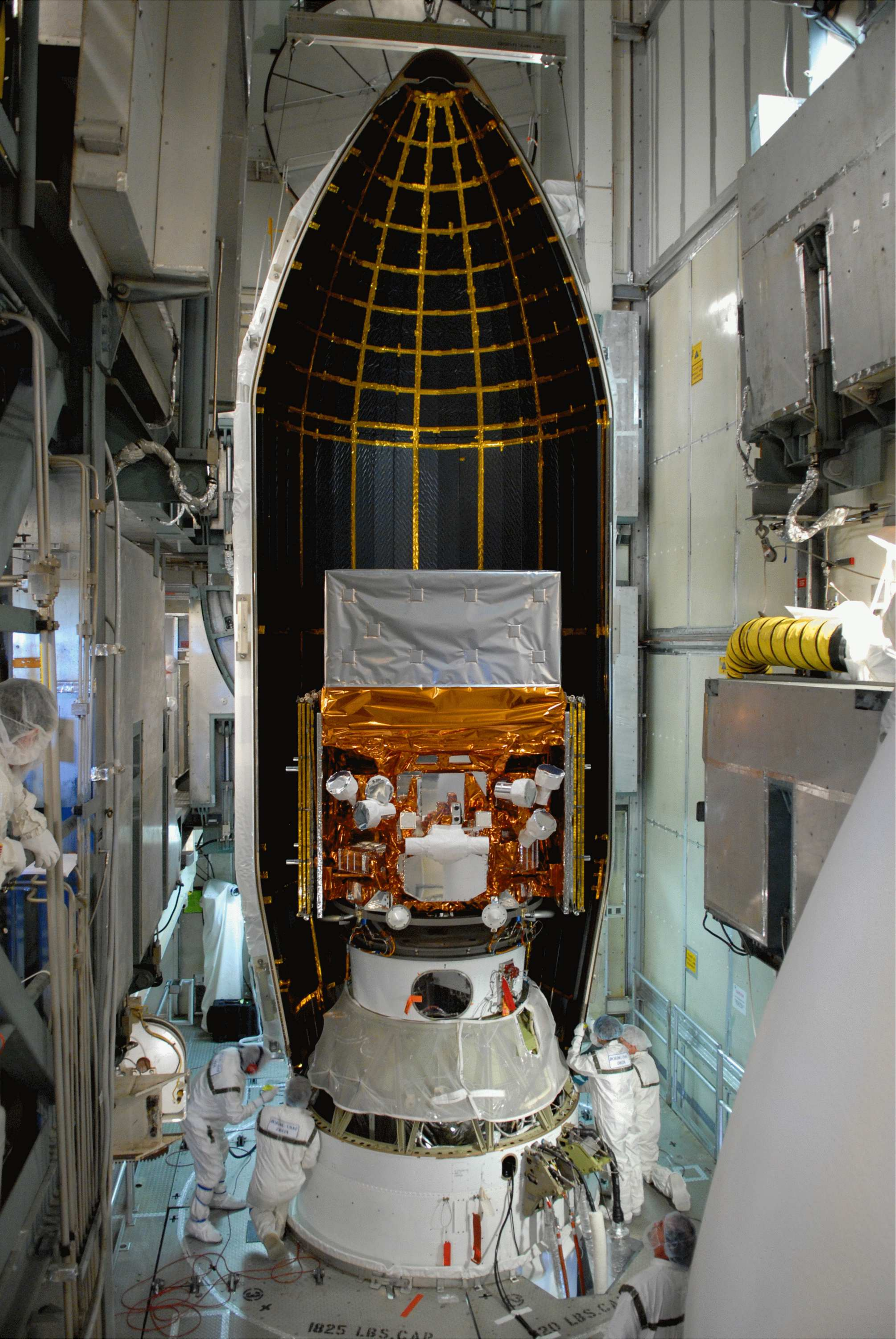}
\end{center}
\small\normalsize
\begin{quote}
\caption[\emph{Fermi} on rocket before fairing closure]{One of the last images of the \Fermi{} observatory.  The spacecraft is in place a top the Delta-II heavy launch vehicle awaiting final fairing closure.  The LAT can be seen atop the spacecraft covered by a kapton, micrometeroid shield.  Gamma-ray Burst Monitor detectors can be seen on the spacecraft, below the LAT, wrapped in white.  \emph{Photo credit: NASA} \label{ch3Fermi}}
\end{quote}
\end{figure}

The \emph{Fermi Gamma-ray Space Telescope} (\Fermi{}, formerly GLAST, see Fig~\ref{ch3Fermi}) was launched from Cape Canaveral, Florida on 11 June 2008, aboard a Delta-II heavy rocket, and successfully reached low-Earth orbit at an altitude of 565 kilometers and an inclination of 25\DEG{}.5 with respect to the equator.  The orbit takes approximately 90 minutes and has a precession period of 55 days.  Data acquisition pauses during passages through the South Atlantic anomaly totalling to a down time of approximately 14.6\%, implying a high duty cycle.

\small\normalsize

There are two instruments aboard \Fermi{}, the Gamma-ray Burst Monitor \citep{Meegan09} and the Large Area Telescope (LAT, Atwood et al., 2009).  \Fermi{} is a joint mission between the National Aeronautics and Space Administration, the United States Department of Energy, and academic institutions in the United States, France, Germany, Italy, Japan, and Sweden.  The mission is planned to last 5 years, with a 2 year minimum requirement and a 10 year goal.  \Fermi{} is nearing 3 years of operation and performing extremely well; the instrumentation consists of radiation hard technology and thus exceeding the 5 year mission lifetime is quite feasible.  For the study presented here only data from the LAT will be considered.
 
\section{The Large Area Telescope}\label{ch3lat}
\begin{figure}[h]
\begin{center}
\includegraphics[width=0.75\textwidth]{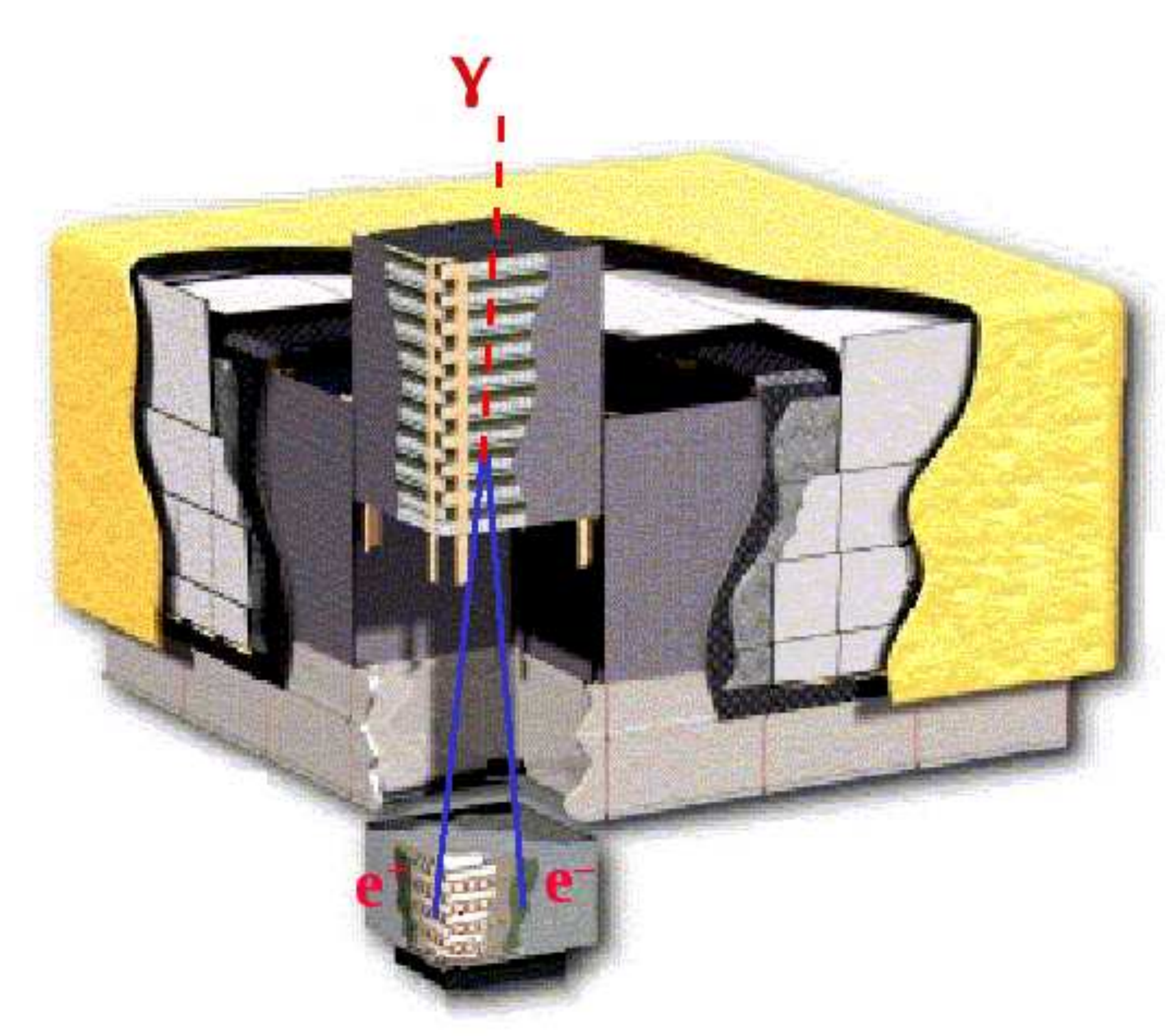}
\end{center}
\small\normalsize
\begin{quote}
\caption[Cutaway cartoon showing a gamma-ray conversion in the LAT]{Cutaway cartoon of the LAT showing an incident gamma-ray (red) converting to an electron-positron pair (blue).  Reproduced from \citet{Atwood09}.\label{ch3Conv}}
\end{quote}
\end{figure}

The LAT is a pair-conversion telescope (see Fig.~\ref{ch3Conv}) consisting of three subsystems: the tracker (TKR; Atwood et al., 2007), for converting incident gamma rays and reconstructing their arrival directions; the calorimeter (CAL; Johnson et al., 2001 and Ferreira et al., 2004), for measuring the energy of the incident gamma ray and assisting in background rejection; and the anti-coincidence detector (ACD; Moiseev et al., 2004 and 2007) for identifying events likely to be due to charged particles and not gamma rays.

\small\normalsize

The initial design for the LAT was proposed by \citet{Atwood94}, motivated by the exciting results from \emph{EGRET}, and has changed very little in realization as the \FL{}.  The modular design of the LAT allowed for feasibility studies to be done with balloon flights (Thompson et al., 2002 and Mizuno et al., 2004) and accelerator beam tests \citep{Couto01}.  The LAT is self-triggering with no expendables and radiation hard technology which has proven effective in HE particle physics experiments.

LAT observations are primarily carried out in a sky-survey mode in which the entire sky is scanned every 2 orbits.  This makes \Fermi{} particularly well suited to the transient nature of the HE gamma-ray sky.  Pointed mode observations are also possible with the LAT, but the increased systematics and loss of exposure for other regions of the sky make this mode less desirable unless given a particularly spectacular occurrence such as a flare of the Crab nebula \citep{AbdoCrabFlare}.  The LAT has a 2.4 sr field of view (FOV) ($\sim$20\% of the sky), a nominal energy range from 0.02 to $>$300 GeV with an energy resolution $<$15\% for events with energies above 0.1 GeV, an on-axis effective area of $\sim$8000 cm$^{2}$ for events with an energy of 1 GeV, and a 68\% point-spread function (PSF) of 0\DEG{}.6 for events with an energy of 1 GeV converting in the front section of the TKR (see Section~\ref{ch3tkr}) and near on-axis.  Among the primary scientific objectives of the LAT mission are: identification of the \emph{EGRET} unidentified sources and diffuse emission, probing particle acceleration mechanisms in astronomical sources, studying the nature of HE transient sources, studying the natue of dark matter, and probing the early universe and cosmic evolution through HE observations of sources with high redshifts.

\subsection{The Tracker}\label{ch3tkr}
The TKR consists of 16 identical towers arranged in a 4$\times$4 square array, each tower is matched with a CAL module below, see Section~\ref{ch3cal}.  The tower and CAL modules are supported by a low-mass aluminum grid structure.  Each tower has 18 x-y tracking planes of single-sided silicon strip detectors (SSDs) and 16 tungsten converting layers.  Solid-state detectors were chosen over gas-based detectors, such as those used in \emph{EGRET}, for several reasons.  The SSDs have no expendables, operate at relatively low voltage ($\sim$100 V), and are self-triggering, reliable, efficient, and robust.  The SSDs used in the LAT, provided by Hammamatsu Photonics in Japan, consist of n type wafers with 384 p-type strips with a separation of 228 $\mu$m.  Four SSDs are bonded in series to form a ladder and each SSD layer consists of 4 ladders and measures $\sim35\times35$ cm$^{2}$.  When a charged particle passes through an SSD layer it will deposit energy through ionization and create electron-hole pairs.  The holes will drift towards the strip and induce a current which is converted to a voltage signal.  With this design, the exact point of passage can not be pinpointed by one SSD layer alone.  However, by placing a second, rotated layer below the first the point of passage can be determined by the intersection of fired strips.

\begin{figure}[h]
\begin{center}
\includegraphics[width=0.6\textwidth]{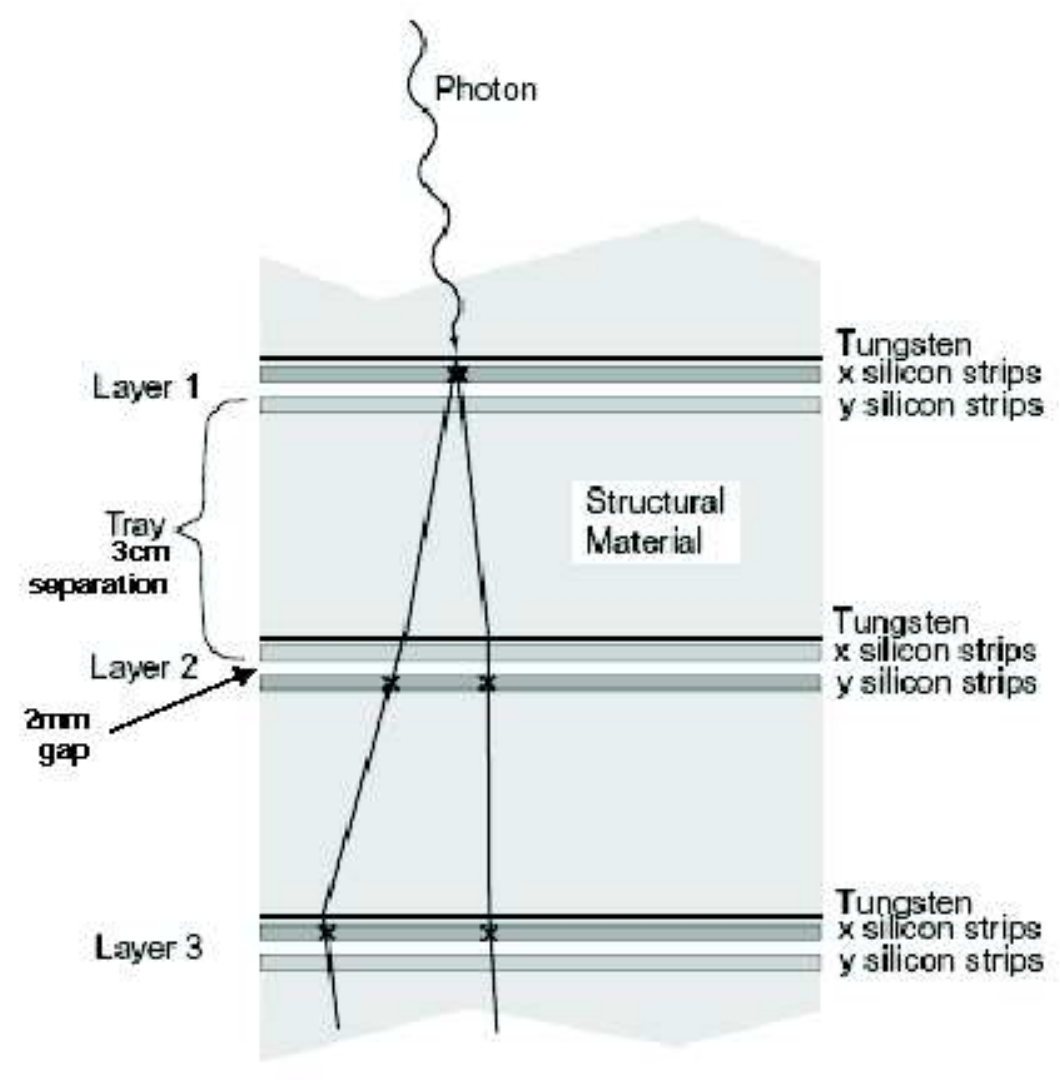}
\end{center}
\small\normalsize
\begin{quote}
\caption[TKR Schematic]{Schematic showing stacking of trays to make x-y tracking layers.  Modified from \citet{Atwood07}.\label{ch3tray}}
\end{quote}
\end{figure}

The towers are composed of 19 trays which house the tracking and converting layers, see Figure~\ref{ch3tray}.  The trays are approximately 3 cm thick and supported by carbon-composite side walls which also serve to conduct heat away from the detectors.  The trays themselves are constructed of a carbon-composite with an aluminum honeycomb core.  Each tray contains 2 SSD layers (except for the top and bottom trays which have only one SSD layer each) and one tungsten converting layer (except for the bottom two trays which have no converters) which is placed immediately before the bottom SSD layer.  The two SSD layers in a given tray measure in the same direction and each tray is rotated by 90\DEG{} with respect to the one above it; thus, an x-y tracking layer is formed from the bottom layer of one tray and the top layer of another.  The z measurement is given by the height of the SSD layer with respect to the bottom of the TKR tower.  The readout electronics (for more details see Baldini et al., 2006) are positioned along the sides of the tower modules to minimize inactive material in the detector FOV, this minimizes missing hits (see Section~\ref{ch3cov}) which increases detection efficiency.  For this same reason, the tungsten foils are positioned such that they only cover the active areas of the SSD layers.

\small\normalsize


In order to more accurately determine the initial photon direction, the x-y tracking planes must be placed as close as possible to the tungsten converter to minimize the effects of multiple scattering.  Multiple scattering limits the PSF at energies near 100 MeV while the separation between SSD strips is the limiting factor for energies above $\sim$10 GeV.  To achieve a PSF of a few degrees at 100 MeV thinner foils are necessary.  However, to provide more effective area at higher energies ($\gtrsim$ 50 GeV) and facilitate meaningful overlap with TeV instruments thicker foils are necessary.   Detailed Monte Carlo (MC) simulations revealed the optimal configuration, given the breadth of LAT science, was to have the first twelve converting layers $\sim$3\% radiation length and the last four $\sim$18\%, these are referred to as the FRONT and BACK of the TKR, respectively.

To trigger the LAT, three consecutive x-y tracking planes must fire, though they need not be in the same tower; therefore, the bottom two x-y tracking planes are not preceded by converting layers as a photon interacting at this point in the TKR would not lead to an event trigger.  In an ideal event, the incident gamma ray converts in one of the tungsten layers as shown in Figure~\ref{ch3tray}; however, gamma rays can interact within any of the other materials which make up the TKR.  As such, the SSD thickness and tower structural material were chosen to be of negligible radiation lengths compared to the tungsten.  Interactions in these materials will still occur; therefore, the LAT team developed a dedicated simulation tool, using the Geant4 particle physics simulation engine (Agostinelli et al., 2003 and Allison et al., 2006), which attempts to take into account all the material of the instrument (including the CAL and ACD subsystems).  Simulations were then used, pre-launch, to develop and validate the event reconstruction and background rejection in a manner which allowed for non-ideal conversions.  The LAT MC simulations will be discussed in more detail in Section~\ref{ch3sims}.

The TKR can measure some of the energy in the shower by counting clusters of adjacent hit strips from electrons and positrons passing through x-y measuring layers and assuming that the number of hit strips is comparable to the energy deposition.  For events with energies $\sim$100 MeV this can amount to a significant fraction of the total energy of the event but becomes negligible at higher energies as the TKR is of finite length.  Therefore, each tower is matched with a CAL module which performs the primary energy measurement and also helps determine the incoming event direction (see Section~\ref{ch3cov}).

\subsection{The Calorimeter}\label{ch3cal}
Each of the 16 CAL modules consists of 96 CsI scintillating crystals.  A scintillating material is one in which atoms are ionized and electrons liberated when a charged particle is incident on the material.  The electrons will be free in the material for a short time (on the order of tens to hundreds of ns) before rejoining with an ionized atom.  To do this, the electron must lose energy, which happens via the emission of a short pulse of light.

The CAL modules each consist of eight, twelve crystal layers.  Each crystal is 2.7 cm $\times$ 2.0 cm $\times$ 32.6 cm, optically isolated from the other crystals, and read out by two photodiodes at each end.  The larger of the two photodiodes, 147 mm$^{2}$, measures energy depositions from 2 MeV to 1.6 GeV while the smaller photodiode, 25 mm$^{2}$, measures energy depositions from 100 MeV to 70 GeV.   At normal incidence, each CAL module is 8.5 radiation lengths deep with gaps and structural material minimized to be $<$16\% of the total mass.

\begin{figure}[h]
\begin{center}
\includegraphics[width=0.6\textwidth]{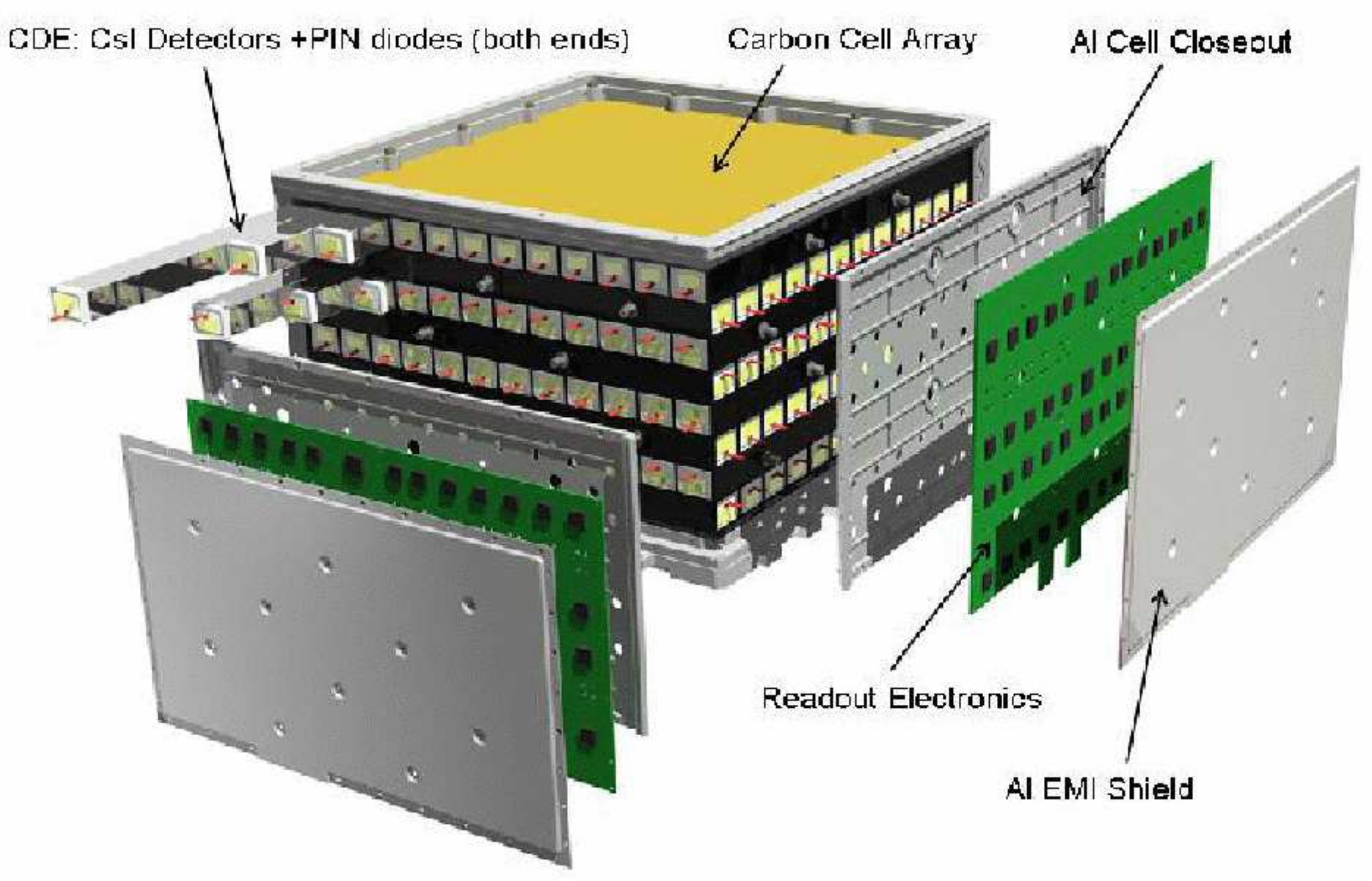}
\end{center}
\small\normalsize
\begin{quote}
\caption[CAL schematic]{Schematic of a CAL module.  Reproduced from \citet{Atwood09}.\label{ch3calSchem}}
\end{quote}
\end{figure}

Similar to the arrangement of x-y planes in the TKR, consecutive CsI layers are rotated by 90\DEG{} (see Fig.~\ref{ch3calSchem}) and the point of passage along a crystal can be estimated by comparing the signal strengths measured at each end, thereby allowing the shower to be imaged.  The latter position measurement is the most accurate, ranging from a few mm for low energy depositions to a fraction of a mm for high energy deposition.  Use of a hodoscopic, segmented CAL has many advantages.

The segmentation allows for precise localizations of the energy centroid (weighted geometrical mean) of electromagnetic showers.  Background rejection is augmented via the identification of showers initiated by particles such as protons, muons, etc., but note that electron initiated-showers will ``look'' the same as those from photons to the CAL.  For events with the highest energies the resulting showers are unlikely to be completely contained in the CAL; however, the segmentation allows for the application of geometric leakage corrections to account for the missing energy.  Additionally,  the CAL can be used to discriminate between downward and upward going tracks, relative to the spacecraft.

\small\normalsize

The initial energy estimate used for track finding (see Section~\ref{ch3latkf}) is simply the sum of all the crystal energies, after a track hypothesis has been fit further refinement of the event energy estimate can be made.  There are three methods used in the LAT to further estimate the event energy.  A parametric method corrects the energy based on the barycenter of the shower and covers the entire energy range of the LAT.  The longitudinal and transverse extents of the shower profile can be fit to account for energy missed due to the finite extent of the CAL but only works for energies greater than 1 GeV.  A maximum likelihood fit correlates the total energy deposition with hits in the TKR but only works below 300 GeV.  Extensive MC simulations and training samples have been used in order to construct logic by which the ``best'' of these three energy estimates is chosen for each event (see Section~\ref{ch3sims}).

\subsection{The Anti-Coincidence Detector}\label{ch3acd}
While in orbit, the LAT is subjected to large fluxes of energetic charged particles, known as cosmic rays, which can also trigger the detectors and could be mistaken for photons.  While hadrons and muons can be differentiated partly based on the signals they leave in the TKR and CAL, signals left by electrons and positrons can not be distinguished from those left by photons.  As such, it was necessary to devise some way by which tracks initiated by charged particles could be separated from those initiated by photons.  This is by no means a new issue for gamma-ray astrophysics and the design of the ACD follows the same basic principles of those systems used in previous HE missions such as \emph{EGRET}.  However, based on insights gained from the experiences of those same missions, the LAT ACD is a much improved design.

\begin{figure}[h]
\begin{center}
\includegraphics[width=0.6\textwidth]{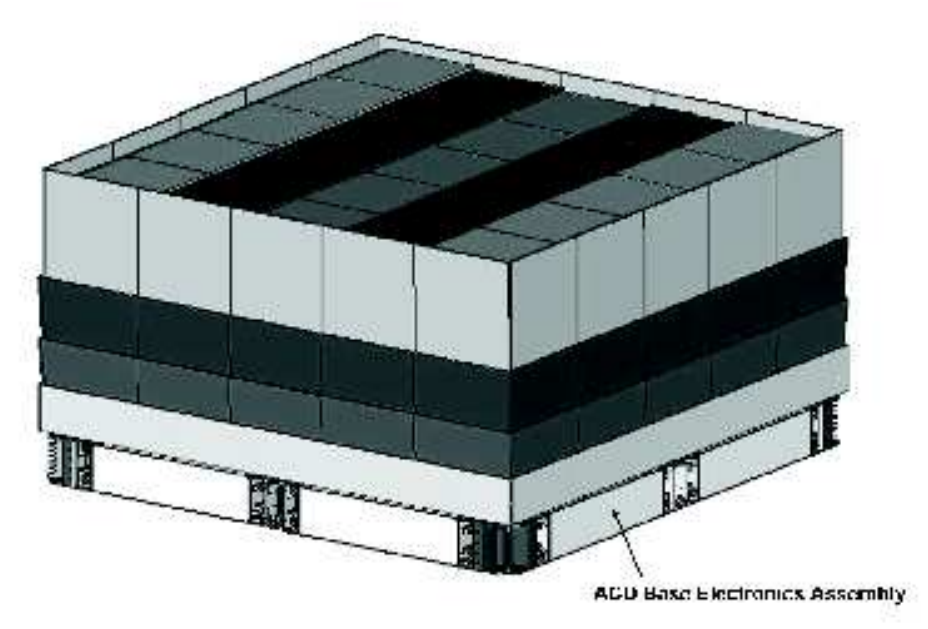}
\end{center}
\small\normalsize
\begin{quote}
\caption[ACD schematic]{Schematic of the ACD tile placement around the LAT.  The top of the LAT is covered by a 5$\times$5 array of tiles.  Each side has 3 rows of 5 tiles with a fourth, bottom row consisting of one long tile.  Reproduced from \citet{Atwood09}.\label{ch3acdSchem}}
\end{quote}
\end{figure}

All anti-coincidence systems use the same principle, put some material before the start of the tracking system with which charged particles will interact but the neutral gamma rays will not.  The most straightforward solution is to surround the detector with a scintillating material which is what has been done with the LAT.  The improvement in the LAT over past missions is that instead of using one, continuous, scintillating detector the LAT ACD consists of 89 plastic, scintillating tiles covering the top and four sides of the instrument, see Figure~\ref{ch3acdSchem}.

\small\normalsize

The ACD tiles are then covered by a micrometeoriod shield to protect the instrument from space debris.  When a charged particle passes through one of the ACD tiles the scintillation light it produces is collected by wavelength shifting fibers embedded in the plastic and read out by two photomultiplier tubes.  Note that to first order the number of liberated electrons, and thus scintillation light produced, depends only on the square of the charge of the incident particle and not its energy.  This means that the ACD is equally sensitive to cosmic rays of low and high energy.

To increase the overall detection efficiency of the ACD, the tiles were made to overlap in one dimension while gaps between tiles in the other dimension were filled with bundles of scintillating fibers.  This results in a detection efficiency, measured pre-launch and verified on-orbit \citep{AbdoOnorb}, exceeding 0.9997.

The ACD was segmented in order to reduce self veto due to backsplash from gamma rays with energies above $\sim$1 GeV.  This effect largely degraded the sensitivity of \emph{EGRET} at the high end of its energy range.  Incoming gamma rays which deposit large amounts of energy in the CAL will produce low-energy particles, mainly 100 to 1000 keV X-rays, which travel upwards in the LAT and can Compton scatter in the ACD tiles leading to a false trigger.  The ACD segmentation allows for such false signals to be easily filtered out.  A crude track reconstruction algorithm approximates the incoming event direction and only tiles within an energy dependent cone (width $\sim\rm{E}^{1/2}$) of this direction can veto the event.  The veto conditions were verified and refined using simulations and data from a beam test conducted at CERN.

\section{Track Reconstruction and Covariance Information}\label{ch3cov}
Reconstructing a particle track through any detector consists of three elements: pattern recognition, track finding, and track fitting.  A complete treatment of these steps can be found in \citet{Fruhwirth00}; a summary of the basic aspects (in a general sense) is given below with a brief description of how tracks are reconstructed in the LAT in Section~\ref{ch3latkf}

\subsection{Pattern Recognition}\label{ch3patrec}
Data in HE physics experiments typically consist of objects, such as hits, clusters, etc.; which have many measurements associated with them.  These measurements lead to a finite number of parameters which describe each object.  It is necessary to have some mechanism with which these objects can be analyzed and patterns searched for.  These patterns can be tracks, images, harmonic signals, etc.

The parameters which describe these objects create a pattern space of dimension $n$, the maximum number of parameters a given object can have.  In general, the pattern space is not a vector space because not every object in the parameter space will have entries defined for all of the possible parameters.  Objects in the parameter space can typically be grouped into different classes, thus creating a classification space.  In principle, objects can belong to multiple classes; however, in HE physics experiments classes do not typically share objects so the problem reduces to that of finding hypersurfaces in parameter space, of dimension $n-1$, which constitute different classes.  These hypersurfaces can be, in physical space for example, clusters of points or vectors which are then useful as starting points for track finding algorithms.

It is often useful to use a training sample to test and verify pattern recognition algorithms and is necessary when no mathematical description of the objects in the parameter space exists.  Training samples are collections of real or simulated data and are particularly helpful for verifying that the covariance matrix $\mathbf {C}$ describing the objects is correct.  If an object $\vec{x}$ consists of separate parameters $x_{i}\in{1,2,...,n}$ and the object comes from a known distribution then the covariance matrix elements can be calculated using Eq.~\ref{ch3Cmat}.
\begin{equation}\label{ch3Cmat}
C_{ij}\ =\ \langle(x_{i}-\langle x\rangle_{i})\times(x_{j}-\langle x\rangle_{j})\rangle\ =\ \sigma_{ij}
\end{equation}

A diagonal element, $\sigma_{ii}$, is called the variance of component $x_{i}$ while an off-diagonal element, $\sigma_{ij}$, is called the covariance of components $x_{i}$ and $x_{j}$ and essentially describes how variation in the $i^{\rm th}$ component affects the $j^{\rm th}$ component.  In Eq.~\ref{ch3Cmat}, $\langle...\rangle$ denotes an expectation value if the events come from a known distribution else it denotes an average from a training sample in which case the variances and covariances are just estimates.

The covariance matrix will, in general, have three independent contributions.  The first such comes from the fact that not all measured parameters are independent, the second from known measurement errors, and the third from random fluctuations which are not known.  These contributions can be combined additively, and therefore separated, due to the fact that they are independent.  When using a training sample generated via some MC method, different effects can be turned on/off to quantify each of these contributions separately.

\subsection{Track Finding}\label{ch3trkfind}
Once pattern recognition has successfully completed, the task of track finding can begin.  The first step in track finding is to separate position measurements into classes.  Members of different classes could have been caused by the same particle but there must exist one class which contains measurements which can not reliably be attributed to any particle.  This class will encompass noise, points skewed due to errors, points excluded by previous track finding steps, etc.

Track finding is, essentially, cluster analysis, where a cluster is defined to be a collection of vectors or points that are close together \citep{Fruhwirth00}.  Clusters are grouped into candidate tracks and fit to a track model.  At this point a decision function must be used to decide whether or not the candidate is a good track.  Often, this method can be sped up by first analyzing a set of measurements, deciding if any are good candidates for the start of a track, and continuing to assemble the track by analyzing more measurements provided, of course, that the first step met with at least one accepted point.  In addition to adding points, it is very important to have an efficient and valid method for removing points.  A track with too many points can result in extra computation time and noise; however, removing points too soon can result in missing good tracks or biasing your method towards tracks of a particular type.  Tracks must also be allowed to share points and points must not be rejected simply because they belonged to a track which was itself rejected.  Points belonging to one or more tracks create vertices and in a detector such as the LAT these are potential sites of pair-production and therefore desirable.  Beyond allowing tracks to share points it is important to have a scheme by which track overlaps are accepted or thrown out as noise.  Such methods can be developed and optimized with appropriate training samples.

\subsection{Track Fitting}\label{ch3trkfit}
After all tracks have been found with high confidence, a track fitting algorithm must be used to determine the parameters of the tracks and group appropriate tracks into primary and secondary vertices.  For experiments in which charged particles are tracked through a detector, this step requires that the amount of material through which the tracks propagate is well known in order to account for the effects of multiple scattering and other energy loss mechanisms.  Precise track models must be known for the different particles involved and the detector response and geometry must be well understood in order to construct these models.  Additionally, it is necessary to have a detailed understanding of and a reliable method for rejecting background events.

It is useful to define the track as a function of the parameters one wishes to determine.  As such, for a track with $m$ parameters one defines an $m$-dimensional measurement space in which to represent the track.  The applicable equations of motions within a detector determine a constraint surface within the measurement space onto which tracks should be mapped.  Experimental errors will cause deviations from the track surface and it is the task of track fitting to meaningfully map position measurements into the measurement space while minimizing the variance from the constraint surface.

\subsection{The Kalman Filter Technique}\label{ch3kf}
\begin{figure}[h]
\begin{center}
\includegraphics[width=0.6\textwidth]{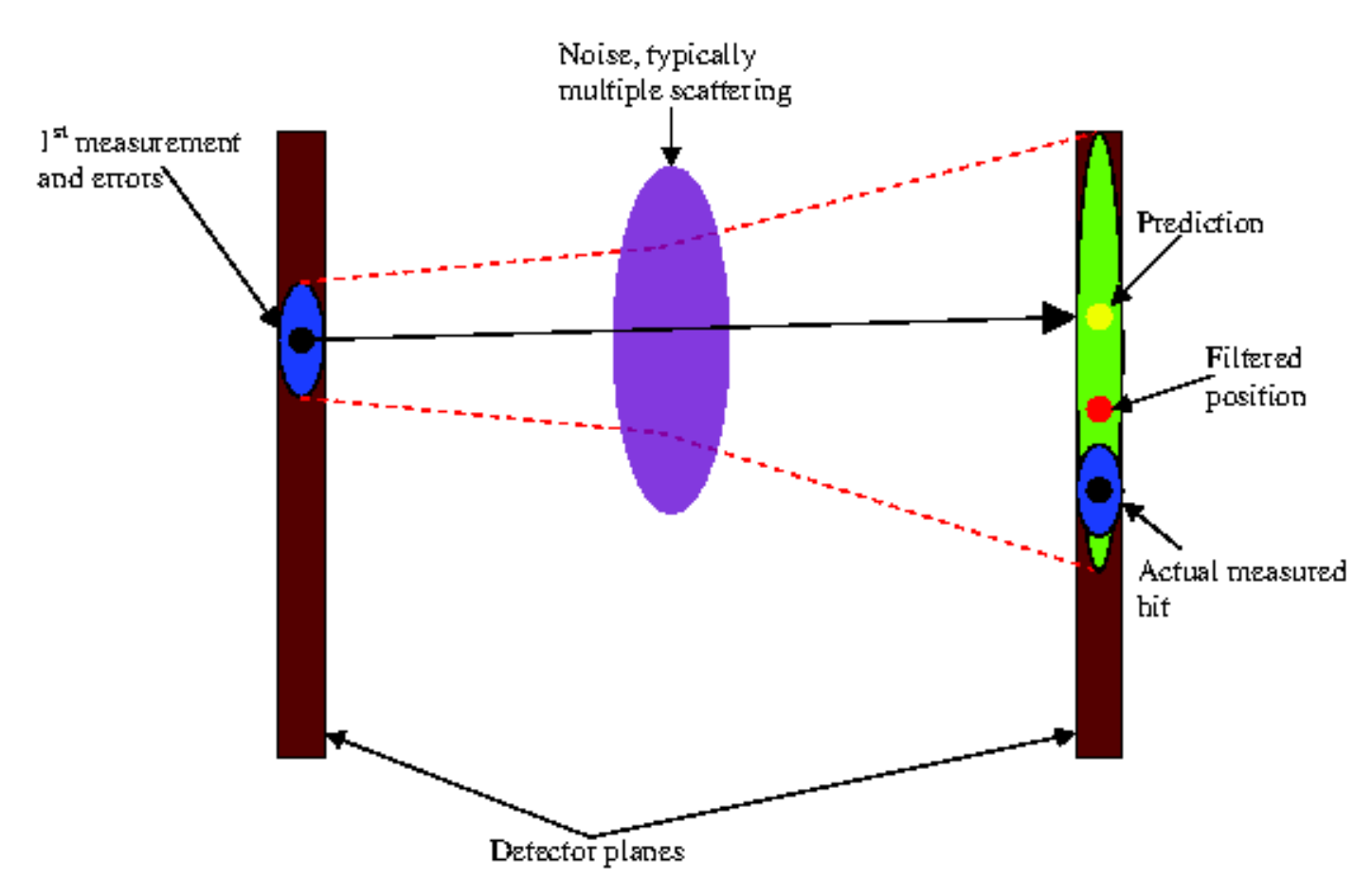}
\end{center}
\small\normalsize
\begin{quote}
\caption[Kalman filter process]{Schematic of the KF process for predicting a track from one detector layer to the next.\label{ch3KFex}}
\end{quote}
\end{figure}

First introduced by \citet{Kalman60}, the Kalman filter (KF) is a technique for estimating the state of a dynamic system.  The KF is a recursive track fitting algorithm which predicts a track forward to the next detector surface and then uses the measurement information to correct the prediction, see Figure~\ref{ch3KFex} .  Once at the end of a track, the filter can be run backwards from the last point to the first in what is known as a smoothing step.  A smoothing step uses the best-fit information from a hit to correct the point(s) before it.  The nature of the KF, using measurements at discrete layers, is particularly well suited to the LAT and the prediction can easily accommodate random noise, such as that introduced by multiple scattering.  Additionally, the effects of multiple scattering in the LAT are more important at energies below 1 GeV and negligible at energies above 100 GeV with a gradual change in importance between.  Such behavior is an important part of the LAT detector response and can be easily incorporated into a KF.

\small\normalsize

The basic mathematical structure of a KF, as outlined below, follows the description in \citet{Fruhwirth00}.  At a given detector surface $k$, the state of the track is described by the vector $\vec{p}_{k}$.  Define a function $\mathbf {f}_{k}$ which, based on the track model used, predicts the behavior of the track between surface $k-1$ and surface $k$.  Then define $\vec{p}_{k}\ \equiv\ \mathbf f_{k}(\vec{p}_{k-1})$.  However, $\mathbf {f}_{k}$ does not consider any of the random noise shown in Figure~\ref{ch3KFex} so this must be introduced separately.  Let $\vec{\delta}_{k}$ represent this noise such that $\langle\vec{\delta}_{k}\rangle\ =\ \vec{0}$ and the effects of the noise on different components of the state vector are described by $\mathbf {P}_{k}$.  The covariance matrix of these errors, $\mathbf {C}(\vec{\delta}_{k})\ \equiv\ \mathbf {Q}_{k}$, is assumed to be known for all $k$.  With the inclusion of this noise, it is now possible to define $\vec{p}_{k}\ \equiv\ \mathbf {f}_{k}(\vec{p}_{k-1})+\mathbf {P}_{k}\vec{\delta}_{k}$.

Ideally, a measurement of the state vector is made at each detector surface.  Each measurement has an error, $\vec{\epsilon}_{k}$, associated with it such that $\langle\vec{\epsilon}_{k}\rangle\ =\ \vec{0}$.  The covariance matrix of these errors, $\mathbf {C}(\vec{\epsilon}_{k})\ \equiv\ \mathbf {V}_{k}$, is also assumed to be known for all $k$.  At each detector surface a function, $\mathbf {h}_{k}$, is defined which maps the state vector onto the measurement and is determined by the specifics of the detector.  This leads to the measurement equation $\vec{m}_{k}\ \equiv\ \mathbf {h}_{k}(\vec{p}_{k})+\vec{\epsilon}_{k}$.

For the KF, it is assumed that $\mathbf h_{k}$ and $\mathbf f_{k}$ are linear functions based on the idea of a linear track from one surface to the next.  In the event that these two functions are non-linear they should be approximated using a first order Taylor expansion.  In particular, it is necessary to calculate the first derivative matrices, $\mathbf H_{k}$ and $\mathbf F_{k}$ respectively, for these functions.  The predicted state vector at surface $k$ using information from layer $k-1$ is then $\vec{p}_{k|k-1}\ \equiv\ \mathbf F_{k}(\vec{p}_{k-1})$.  This is similar to the definition of $\vec{p}_{k}$ above except that now our state vector is a prediction based on the filtered state vector at layer $k-1$.  The filtered step must take the noise into account as well as the actual measured parameter state $\vec{p}_{k}$.  To do this the covariance matrix must also be predicted forward following,
\begin{equation}\label{ch3predCov}
\mathbf C_{k|k-1}\ =\ \mathbf F_{k} \mathbf C_{k-1} \mathbf F_{k}^{T} + \mathbf P_{k} \mathbf Q_{k} \mathbf P_{k}^{T}.
\end{equation}

Then, using a least squares minimization process, the filtered state vector and covariance matrix for the surface $k$ are given by,
\begin{equation}\label{ch3filtState}
\vec{p}_{k}\ =\ \vec{p}_{k|k-1} + \mathbf K_{k}(\vec{m}_{k} - \mathbf H_{k}\vec{p}_{k|k-1})
\end{equation}
\begin{equation}\label{ch3filtCov}
\mathbf C_{k}\ =\ (\mathbf I - \mathbf K_{k} \mathbf H_{k})\mathbf C_{k|k-1}
\end{equation}
\begin{equation}\label{ch3gain}
\mathbf K_{k}\ =\ \mathbf C_{k|k-1} \mathbf H_{k}^{T}(\mathbf V_{k} + \mathbf H_{k} \mathbf C_{k|k-1} \mathbf H_{k}^{T})^{-1}
\end{equation}

where $\mathbf I$ is the identity matrix and $\mathbf K_{k}$ is the gain matrix, Eq.~\ref{ch3gain}, which determines how strongly the predicted values are changed by the inclusion of the measurement information.  As the KF assumes a linear model, it is useful to calculate the residuals and $\chi^{2}$ values as goodness-of-fit measures for each detector surface which are given by Eqs.~\ref{ch3resids} and~\ref{ch3tkrchisq}, respectively.
\begin{equation}\label{ch3resids}
\vec{r}_{k}\ =\ \vec{m}_{k}\ - \mathbf H_{k}\vec{p}_{k|k-1}
\end{equation}
\begin{equation}\label{ch3tkrchisq}
\chi^{2}_{k}\ =\ \vec{r}_{k}^{T}(\mathbf V_{k} - \mathbf H_{k} \mathbf C_{k} \mathbf H_{k}^{T})^{-1} \vec{r}_{k}
\end{equation}

The fact that $\chi^{2}_{k}$ is independent for each $k$ means that the total $\chi^{2}$ and degrees of freedom are sums over the values at each detector surface.  If a smoothing step is run it is possible to calculate a $\chi^{2}$ for the smoothed track but this statistic must be treated carefully as the errors are now correlated \citep{Fruhwirth00}.

\subsection{Implementation of the KF in the LAT}\label{ch3latkf}
Track finding in the LAT follows two procedures which differ mainly in how the candidate first TKR hits (consisting of clusters of fired SSD strips) are selected as described in \citet{Atwood09}.  The first method requires that some energy be deposited in the CAL for a starting guess on the incoming event direction while the second method blindly selects hits.

The first method is known as the \emph{Calorimeter-Seeded Pattern Recognition} (CSPR).  From the energy deposition in the CAL a guess at the event energy is made by summing the signals in all crystals, this energy is given to the KF.  The energy centroid is calculated using a moments analysis and is assumed to lie along the incoming event trajectory.  A starting hit is chosen at random from those in the TKR layer furthest from the CAL with fired strips.  A track hypothesis is generated and fit if a hit can be found in the next layer down which lies along the direction connecting the candidate first hit and the energy centroid.  The KF implementation used in the track accounts for all the mass, including dead material, along the hypothesized track and accounts for possible missing hits in uninstrumented material.  This is done for each possible hit in the layer furthest from the CAL and then moving to subsequent layers until starting hits have been taken from at least two layers and a sufficient quality track has been found.  The KF allows for each hit to be added to the track covariantly as described in Section~\ref{ch3kf} and thus the quality of each hypothesized track can be evaluated by calculating the corresponding $\chi^{2}$ and a ``best'' track can be chosen and the hits belonging to that track are flagged as used and excluded from subsequent track finding steps.  At higher energies ($\gtrsim$ 1 GeV) the process is sped up by restricting the possible starting hits to those within an energy dependent cone (with a half angle which decreases with increasing energy) around a direction provided by the energy centroid (corresponding to the eigenvector with the smallest eigenvalue).

\begin{figure}[h]
\begin{center}
\includegraphics[width=0.6\textwidth]{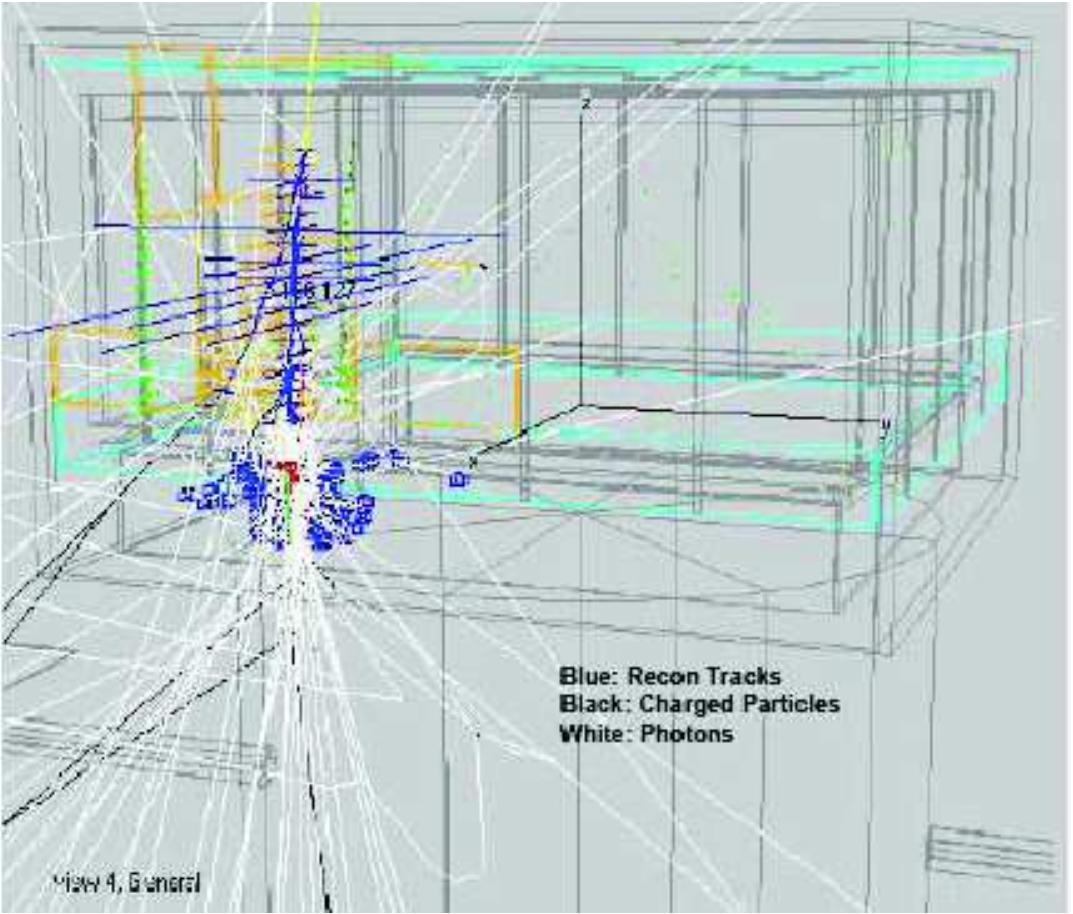}
\end{center}
\small\normalsize
\begin{quote}
\caption[LAT single event display]{Example of the LAT single event display.  The tracks used for reconstruction are shown in blue, with error bars at each TKR layer representing the contribution of that hit to the overall track $\chi^{2}$.  The best direction from the KF is shown as a yellow line.  Fired ACD tiles are outlined in orange while fired TKR strips are green.  CAL crystals with energy depositions are shown as blue boxes beneath the TKR, with the size of the box representing to the amount of energy deposited.\label{ch3Eventex}}
\end{quote}
\end{figure}

The second method for choosing starting track hits is known as the \emph{Blind Search Pattern Recognition}.  This method is used if there is no energy deposition in the CAL (though most analyses require that events have such information) and to search for further tracks after the CSPR has been run and a best track found.  As no information on the event energy is used for this method the KF is given a default energy of 30 MeV.  The starting hit for the candidate track is selected from the layer with fired strips furthest from the CAL (similar to the CSPR) and then a second hit is chosen at random from the next layer down.  If a hit can be found in the subsequent layer which is close to the projected direction from the first two hits a track hypothesis is generated using the same KF implementation as in the CSPR.  It is possible for hits to be shared between tracks but only if the hit corresponds to the start of one track (a possible vertex) or the size of the cluster corresponding to the TKR hit is larger than might be expected for one track.  This procedure stops when either 10 possible tracks have been found or all possibilities have been expended.  For a given event the raw data can be combined with a detailed description of the LAT detector geometry to view the candidate tracks, fired ACD tiles, and CAL crystals with energy deposition as shown in Figure~\ref{ch3Eventex}.

\small\normalsize

At this stage, the best track is used to improve the CAL energy estimate as discussed in Section~\ref{ch3cal}.  The KF is then rerun on the candidate tracks with the new energy estimate but does not change the hits in a given track.  At this point the LAT track finding algorithm groups candidate tracks into vertices.  Starting with the best track, the procedure loops over the other candidate tracks found previously and creates a vertex based on the distance of closest approach between the tracks (at least 6mm).  A $\chi^{2}$ parameter is calculated for each candidate vertex, the best one chosen, and the corresponding track marked as used.  The process then loops over all remaining candidate tracks, flagging those corresponding to the best vertices as used, and placing tracks for which no reasonable vertex can be defined into a vertex with itself such that all candidate tracks are stored as vertices.  The z-coordinate of the vertex, defined with respect to the bottom of the tray closest to the CAL, is determined to be either in the center of the tungsten layer proceeding the first hit, in the center of the SSD layer of the first hit, or in the tray material above the first hit based on the topology of the event.

The nature of the LAT as a pair-production telescope makes finding vertices desirable as the two tracks are expected to correspond to the resultant electron and positron.  However, for some events the resulting electron or positron can radiate via Bremsstrahlung and thus a large portion of the energy is not tracked in the SSD layers, as photons are neutral particles, and two tracks which form the vertex will not point back along the incoming event direction.  This so-called ``neutral energy'' will be reflected in the CAL energy centroid and using that information and the interaction point, as defined by the vertex position, the incoming event direction can be more accurately reconstructed.  As noted by \citet{Atwood09}, combining
\pagebreak

\noindent{}the TKR and CAL information in this manner helps to decrease the non-Gaussian tails of the LAT PSF.

The track of a charged particle can be parametrized by either direction cosines or direction tangents (slopes) and track intercepts.  The LAT track reconstruction software implements a KF which uses the track slopes and intercepts with a filter and smoother step.  The noise is assumed to be from multiple scattering and dominates the covariance at energies below 1 GeV.  This allows for detailed track covariance information to be computed for each event, effectively producing event-by-event errors as described below.

Let $S_{x}\ \equiv\ dx/dz$ and $S_{y}\ \equiv\ dy/dz$ be the x and y slopes, respectively, of a track detected in the LAT.  If the event has only one track then the reconstructed x-, y-, and z-components of the incoming particle direction $\vec{v}$, relative to the spacecraft, are given in Eqs.~\ref{ch3dirx},~\ref{ch3diry}, and~\ref{ch3dirz}, where the negative signs assume that the track is traveling downward in the LAT.
\begin{equation}\label{ch3dirx}
v_{x}\ =\ \frac{-S_{x}}{(1+S_{x}^{2}+S_{y}^{2})^{-1/2}}
\end{equation}
\begin{equation}\label{ch3diry}
v_{x}\ =\ \frac{-S_{y}}{(1+S_{x}^{2}+S_{y}^{2})^{-1/2}}
\end{equation}
\begin{equation}\label{ch3dirz}
v_{x}\ =\ \frac{-1}{(1+S_{x}^{2}+S_{y}^{2})^{-1/2}}
\end{equation}

The slope covariance matrix elements for a track in the LAT are given in Eqs.~\ref{ch3cxx},~\ref{ch3cyy}, and~\ref{ch3cyx} in which $\theta_{ms}$ is the multiple scattering angle which is proportional to the inverse of the track energy, $X$ is the amount of material along the track in units of radiation lengths, $p$ is the particle momentum in MeV, and $\beta c$ is the velocity of the particle.
\begin{equation}\label{ch3cxx}
C_{xx}\ =\ \theta_{ms}^{2}\big[\frac{13.6}{p\beta c}X^{1/2}(1+0.38\ln(X))\big]^2(1+S_{x}^{2})(1+S_{x}^{2}+S_{y}^{2})
\end{equation}
\begin{equation}\label{ch3cyy}
C_{yy}\ =\ \theta_{ms}^{2}\big[\frac{13.6}{p\beta c}X^{1/2}(1+0.38\ln(X))\big]^2(1+S_{y}^{2})(1+S_{x}^{2}+S_{y}^{2})
\end{equation}
\begin{equation}\label{ch3cyx}
C_{xy}\ =\ C_{yx}\ =\ \theta_{ms}^{2}\big[\frac{13.6}{p\beta c}X^{1/2}(1+0.38\ln(X))\big]^2(S_{x}S_{y})(1+S_{x}^{2}+S_{y}^{2})
\end{equation}

Eqs.~\ref{ch3dirx},~\ref{ch3diry}, and~\ref{ch3dirz} can be used to transform from slope space to direction cosine space and from there to instrument, polar coordinates.  These are $\theta\ =\ arccos(-v_{z})$, measured from the normal to the top of the LAT, and $\phi\ =\ arctan(v_{y}/v_{x})$, the azimuthal angle with zero referenced to the positive instrument x-axis.  With these transformations applied to the track covariance matrix it is possible to compute the errors on the $\theta$ and $\phi$ coordinates of the incoming track as well as the covariance of these two parameters.  In principle, this information provides event-by-event errors which could be used to weight tracks.
\begin{equation}\label{ch3sigtheta}
\sigma_{\theta}^{2}\ =\ \cos^{4}(\theta)\big(\cos^{2}(\phi)C_{xx}+2C_{xy}\sin(\phi)\cos(\phi)+\sin^{2}(\phi)C_{yy}\big)
\end{equation}
\begin{equation}\label{ch3sigphi}
\sigma_{phi}^{2}\ =\ \tan^{-2}(\theta)\big(\sin^{2}(\phi)C_{xx}-2C_{xy}\sin(\phi)\cos(\phi)+\cos^{2}(\phi)C_{yy}\big)
\end{equation}
\begin{equation}\label{ch3sigtp}
\sigma_{\theta\phi}\ =\ \sigma_{\phi\theta}\ =\ \frac{\cos^{3}(\theta)}{\sin(\theta)}\Big(\sin(\phi)\cos(\phi)(C_{yy}-C_{xx})+C_{xy}(\cos^{2}(\phi)-\sin^{2}(\phi))\Big)
\end{equation}

The preceding discussion has been a general overview of track finding in the LAT.   However, it should be noted that a good deal of complexity has been smoothed over.  The interested reader is referred to Chapter 8 of \citet{Jones98} which describes, in detail, a KF implementation for a beamtest demonstrating the LAT viability.  Additionally, improvements in the reconstruction (known as ``Pass8'') will add further layers of complexity but should also lead to significant improvements in performance.

\subsection{Using the LAT Track Covariance information}\label{ch3covApp}
Relating $C_{\theta\theta}\ =\ \sigma_{\theta}^{2}$, $C_{\phi\phi}\ =\ \sigma_{\phi}^{2}$, and $C_{\theta\phi}\ =\ \sigma_{\theta\phi}$ and inverting it is possible to construct an error ellipse in $\theta$-$\phi$ space using Eq.~\ref{ch3ellipse}, where the reconstructed direction is ($\theta_{0}$,$\phi_{0}$) and $n\sigma$ is the 'confidence level' of the ellipse.
\begin{equation}\label{ch3ellipse}
(n\sigma)^{2}\ =\ (\theta-\theta_{0})^{2}C_{\theta\theta}^{-1}+2(\theta-\theta_{0})(\phi-\phi_{0})C_{\theta\phi}^{-1}+(\phi-\phi_{0})^{2}C_{\phi\theta}^{-1}
\end{equation}

While the covariance matrix approach does assume Gaussian errors multiple scattering introduces non-Gaussian affects, thus there exists some ambiguity in the definition of $n\sigma$.  Simulated gamma rays of all energies from all directions on the sky (see Section~\ref{ch3sims} for more details, GLAST-release v13r5p3) were used to address the question of how to properly ``size'' these ellipses to achieve 68\% containment.

\begin{figure}[h]
\begin{center}
\includegraphics[width=0.75\textwidth]{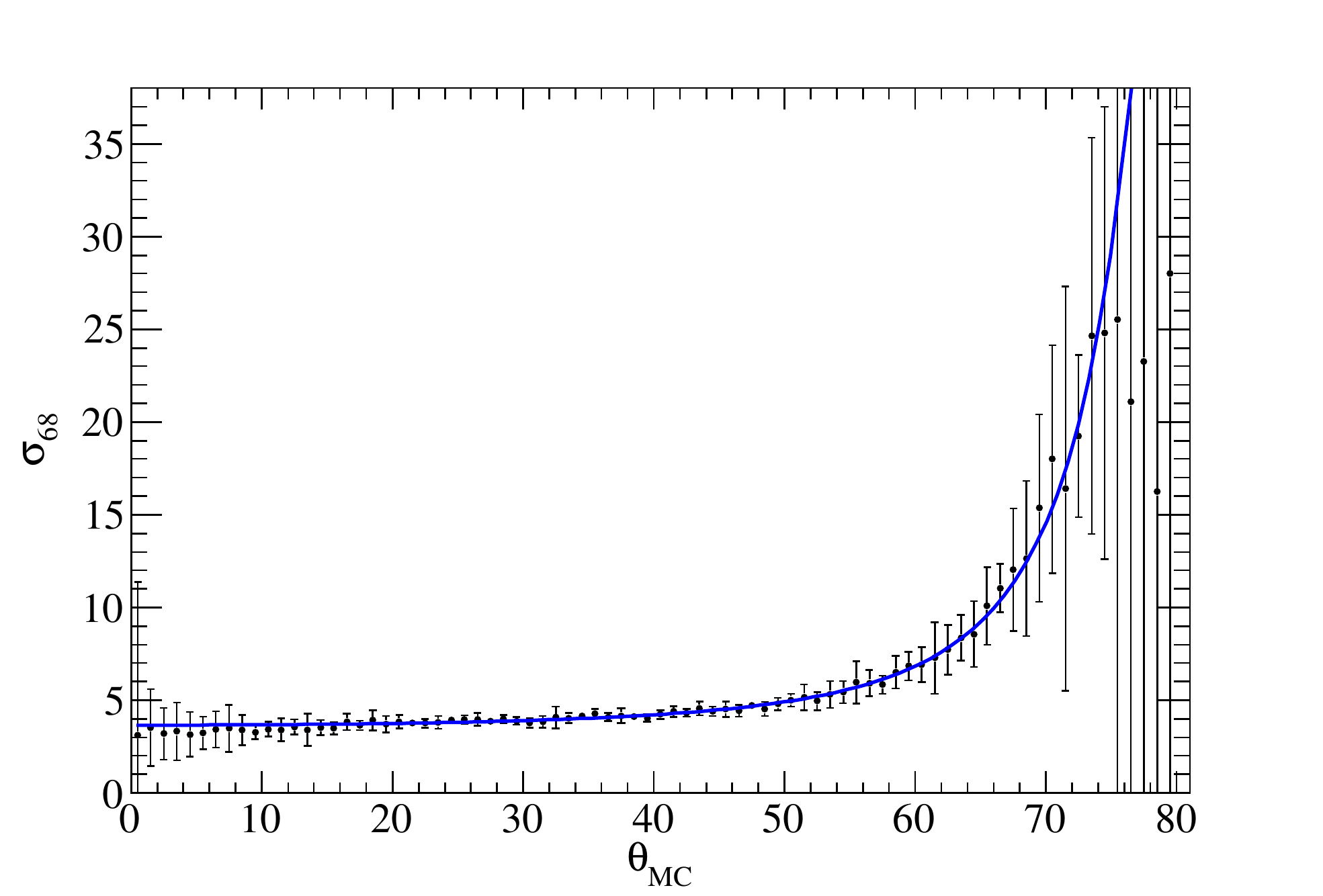}
\end{center}
\small\normalsize
\begin{quote}
\caption[Covariance matrix $\sigma_{68}$ vs. $\theta_{\rm MC}$]{Estimated 68\% containment values verse $\theta_{\rm MC}$.  Blue line is a fit assuming $\cos^{-3}(\theta_{\rm MC})$ dependence as described in the text.\label{ch3thetasig}}
\end{quote}
\end{figure}

Simulated events were sorted into 1\DEG{} bins of simulated instrument polar angle using the MC truth information.  For each bin, the value of Eq.~\ref{ch3ellipse} was calculated substituting $\theta\ =\ \theta_{\rm MC}$ and $\phi\ =\ \phi_{\rm MC}$.  After all events for a given bin have been cycled over the 68\% containment ($\sigma_{68}$) value is estimated, assuming $n$ = 1 in Eq.~\ref{ch3ellipse}. The resulting trend of $\sigma_{68}$ verse $\theta_{\rm MC}$ is shown in Fig.~\ref{ch3thetasig}.

\small\normalsize

The $\sigma_{68}$ values are fairly constant with $\theta_{\rm MC}$ but rise sharply above $\sim$50\DEG{} as expected due to an increased rate of multiple scattering.  The blue line in Fig.~\ref{ch3thetasig} is a fit assuming a functional form $A+B\cos^{3}(\theta_{\rm MC})$ based on $\theta$ dependence of the inverse covariance matrix, the best-fit values were $A\ =\ 3.20\pm0.07$ and $B\ =\ 0.45\pm0.03$.  For this functional form it is assumed that any energy or $\phi$ dependence of $\sigma_{68}$ is absorbed into the value $A$.

\begin{figure}[h]
\begin{center}
\includegraphics[width=0.75\textwidth]{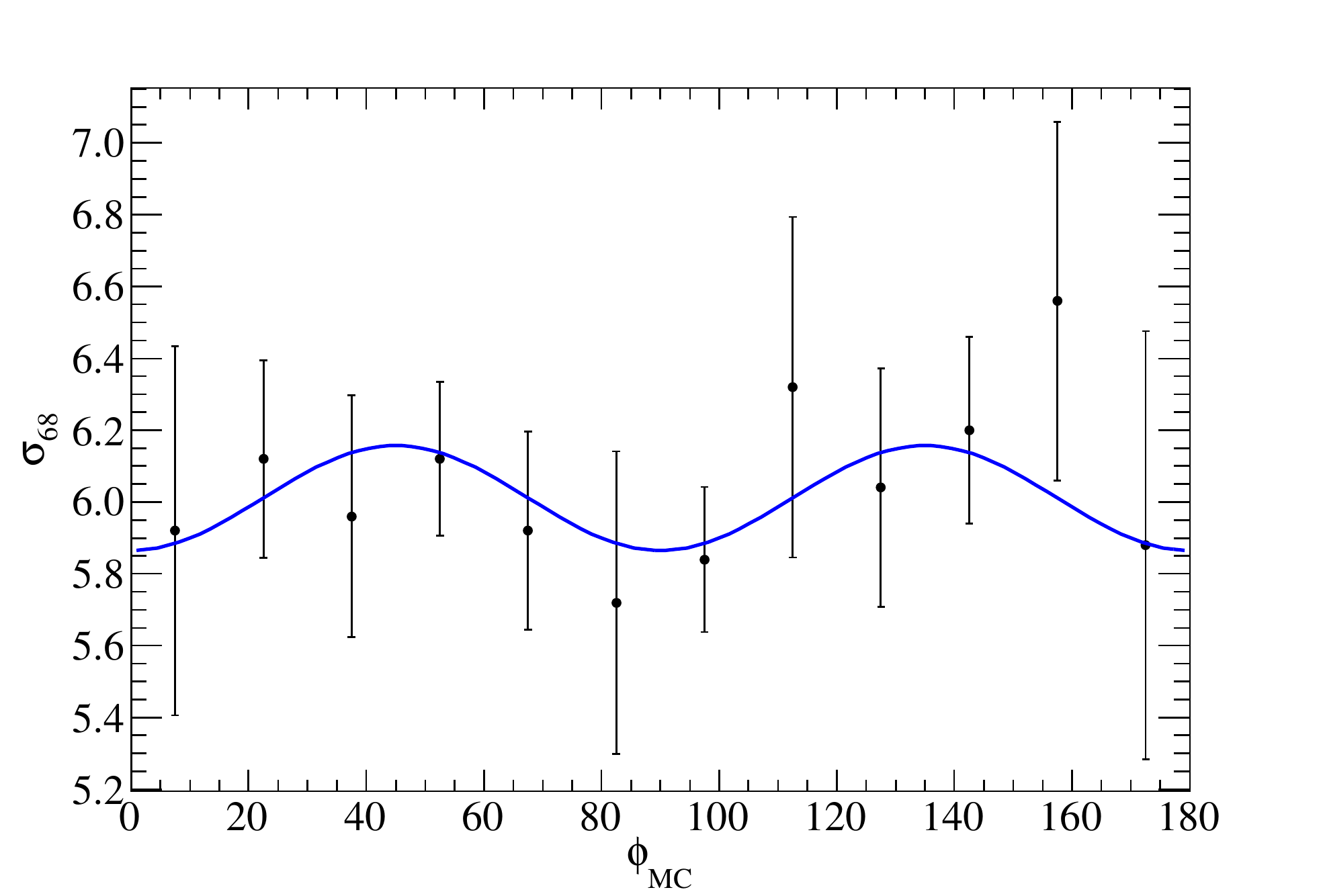}
\end{center}
\small\normalsize
\begin{quote}
\caption[Covariance matrix $\sigma_{68}$ vs. $\phi_{\rm MC}$]{Estimated 68\% containment values verse $\phi_{\rm MC}$.  Blue line is a fit assuming $\cos^{2}(\phi)\sin^{2}(\phi)$ dependence as described in the text.\label{ch3phisig}}
\end{quote}
\end{figure}

In a similar fashion, simulated events are sorted into 15\DEG{} bins of $\phi$ (allows for better statistics in each bin) and $\sigma_{68}$ is estimated as a function of $\phi$.  The results of this analysis are shown in Fig.~\ref{ch3phisig}.

\small\normalsize

The $\sigma_{68}$ values appear to be constant with $\phi_{\rm MC}$ with slight evidence for modulation.  The blue line is a fit assuming a functional form $a+b\cos^{2}(\phi_{\rm MC})\sin^{2}(\phi_{\rm MC})$ which results in $a\ =\ 5.87\pm0.16$ and $b\ =\ 1.17\pm0.98$.

The LAT PSF is not typically parametrized with any dependence on instrument azimuth (see Section~\ref{ch3sims}) though Fig.~\ref{ch3phisig} does predict variation with $\phi$.  However, the uncertainty on $b$ is large and given the assumed functional form of the blue line in Fig.~\ref{ch3phisig} this amounts to, at most, a $\pm$3\% variation in $\sigma_{68}$ with $\phi$.  Thus, neglecting this dependence should have little effect on LAT science results.

While the functional forms of $\theta_{68}$ described above may not be ideal, they do demonstrate that the LAT track covariance information can be used to select events which are reconstructed better.  This could have many applications for LAT science but the development of such an analysis is beyond the scope of this thesis.

It should be noted that currently the LAT response is not characterized using the covariance matrix information as described above.  In part, this is due to the fact that implementation of a fully covariant approach is complicated and characterizing the instrument, as discussed in Section~\ref{ch3sims}, using the average response is simpler.  Additionally, it was found that using the average response satisfied all the mission requirements.

\section{Instrument and Data Simulations}\label{ch3sims}
Throughout the design, construction, and operation of the LAT the instrument response and performance have been evaluated using detailed MC simulations.  These simulations included descriptions of the instrument subsystems and readout electronics as well as realistic models of HE astrophysical sources and charged particle backgrounds expected to be observed by the LAT while in orbit.  The LAT instrument geometry and particle physics interactions are modeled using the Geant4 simulation toolkit.  All instrumented and uninstrumented material is accounted for in the simulation, as well as fine details such as the screw holes in the ACD tiles.  The simulation accounts for interactions of gamma rays and particles in matter and includes a special version of the multiple scattering description in order for the MC to agree more closely with data from beam tests using spare flight hardware.

Classification Trees (CTs) were used to mine the simulations for variables of importance in classifying events with high probability of being gamma rays and those which are most likely charged particle background.  This method allows for selection between the different energy estimates and directions (one track vs. vertex vs. neutral energy).  Additionally, the CT analysis enables the definitions of probabilities that the reconstructed energy is within 1$\sigma$ of the true energy and that the reconstructed direction lies in the core of the PSF and not the non-Gaussian tails (this probability knob is called ``CTBCORE'').  These probabilities can be used to refine the event selections but increasingly restrictive cuts will decrease the effective area.

A realistic model of the charged particle background expected to be experienced by the LAT was used in the simulations to refine the background rejection method using CTs.  This model, described in more detail by \citet{Atwood09}, incorporates data from missions such as AMS \citep{Aguilar02} and BESS \citep{Haino04} for cosmic ray fluxes with dependence on geomagnetic latitude and a reanalysis of EGRET data to model the gamma-ray flux from the limb of the Earth.  This model did not assume time dependent background fluxes which are assumed to be those during solar minimum (when they are at maximum).

The LAT background rejection algorithm needs to be efficient at removing charged particles, as the flux from this background greatly exceeds the average rate of incident, cosmic gamma rays, and this is largely facilitated by segmentation of the ACD (see Section~\ref{ch3acd}).  However, care must be taken not to reject events at the highest energy due to backsplash.  Background events can also be rejected based on the 3D shower distribution in the CAL (with hadronic showers being much broader than those of leptons) which is another advantage of using a hodoscopic calorimeter (see Section~\ref{ch3cal}).  With the MC truth information, the CTs were trained on the simulations in order to optimize the background rejection cuts and define probabilities that a given event was from a charged particle or a gamma ray.

LAT observations cover a broad range of scientific studies which have very different characteristics (i.e., time scales, energy ranges, etc.).  As such, it is necessary to define more than one event class (based on the CT probabilities and other instrument variables) to accommodate different analyses.  For bright, short duration transients the expected background levels are much lower than the source signal and thus cuts can be looser to improve effective area and gamma-ray detection efficiency.  For longer timescale studies the backgrounds become increasingly important and selections must be harsher in order to ensure a cleaner sample of events.  Such considerations were used to define different event classes (currently `TRANSIENT', `SOURCE', `DIFFUSE', and `DATACLEAN') with different intended uses.  Currently, the `TRANSIENT' class is recommended only for analysis of short duration events such as gamma-ray bursts as it has a much higher rate of background events than the other classes.  The `DIFFUSE' event class is recommended for individual source studies while the `DATACLEAN' class is recommended for studies of diffuse emission or as a cross check of an analysis using the `DIFFUSE' class.  The `SOURCE' class was originally designed for point source analysis but, due to unforeseen instrumental backgrounds discussed later, this event class has higher than predicted background levels and is not recommended for use.

\begin{figure}
\begin{center}
\includegraphics[width=1.0\textwidth]{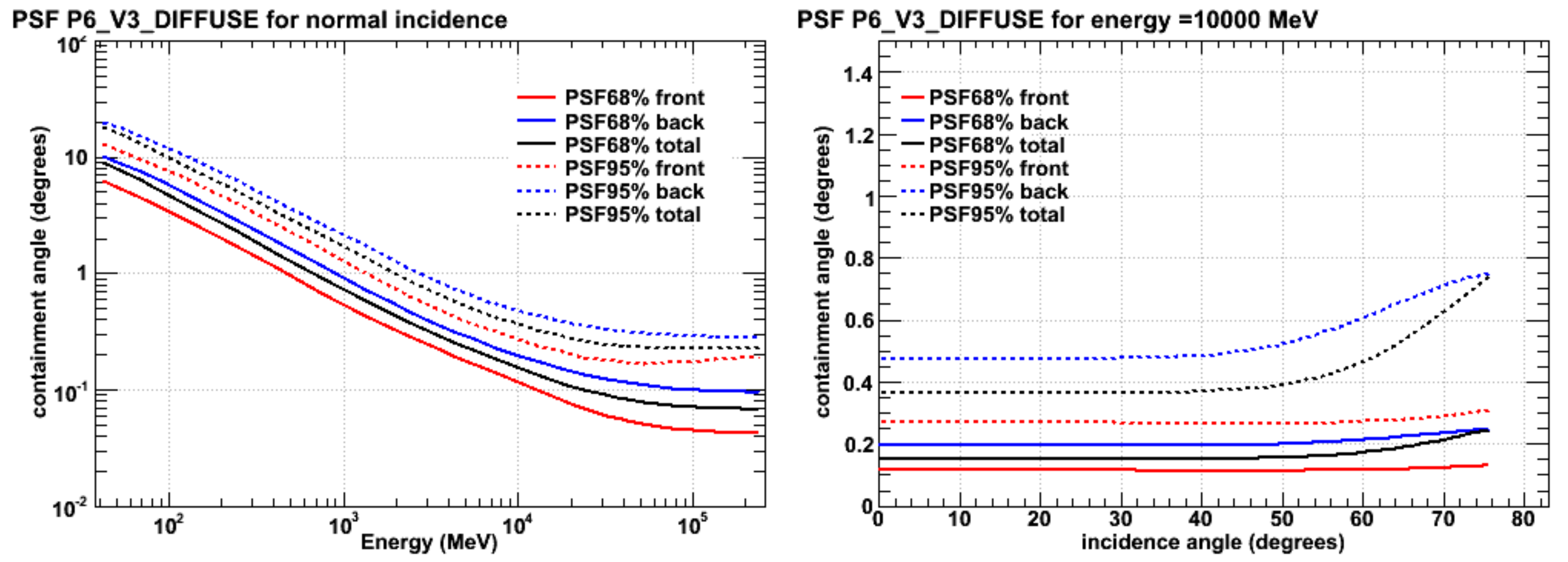}
\end{center}
\small\normalsize
\begin{quote}
\caption[LAT point-spread function]{\emph{(Left)}: PSF vs. energy for near on-axis ($cos(\theta)\ >\ 0.9$) events separated into FRONT, BACK, and total events for 68 and 95\% containment levels for the DIFFUSE event class corresponding to the P6\_V3 IRFs.  \emph{(Right)}: PSF vs. incidence angle ($\theta$) for 10 GeV events.\label{ch3psfplot}}
\end{quote}
\end{figure}

\begin{figure}
\begin{center}
\includegraphics[width=1.0\textwidth]{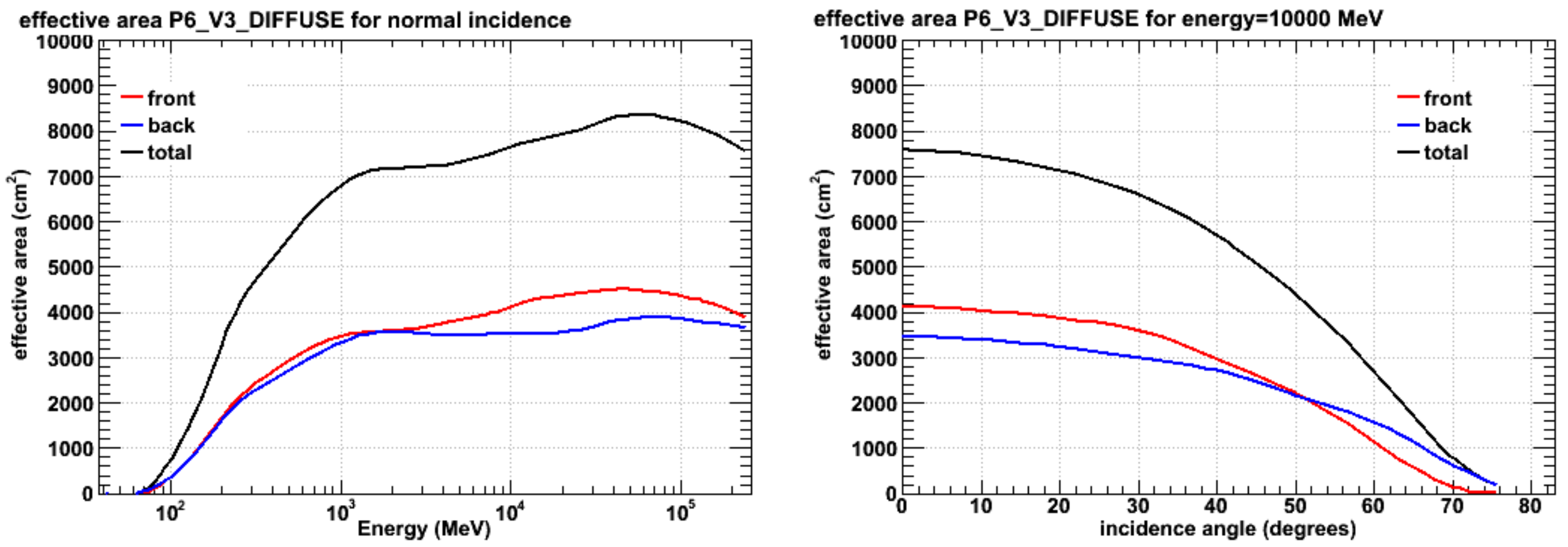}
\end{center}
\small\normalsize
\begin{quote}
\caption[LAT effective area]{\emph{(Left)}: Effective area vs. energy for near on-axis events ($cos(\theta)\ >\ 0.975$) separated into FRONT, BACK, and total events for the DIFFUSE event class corresponding to the P6\_V3 IRFs. \emph{(Right)}: Effective area vs incidence angle ($\theta$) for 10 GeV events.\label{ch3aeffplot}}
\end{quote}
\end{figure}

\begin{figure}
\begin{center}
\includegraphics[width=1.0\textwidth]{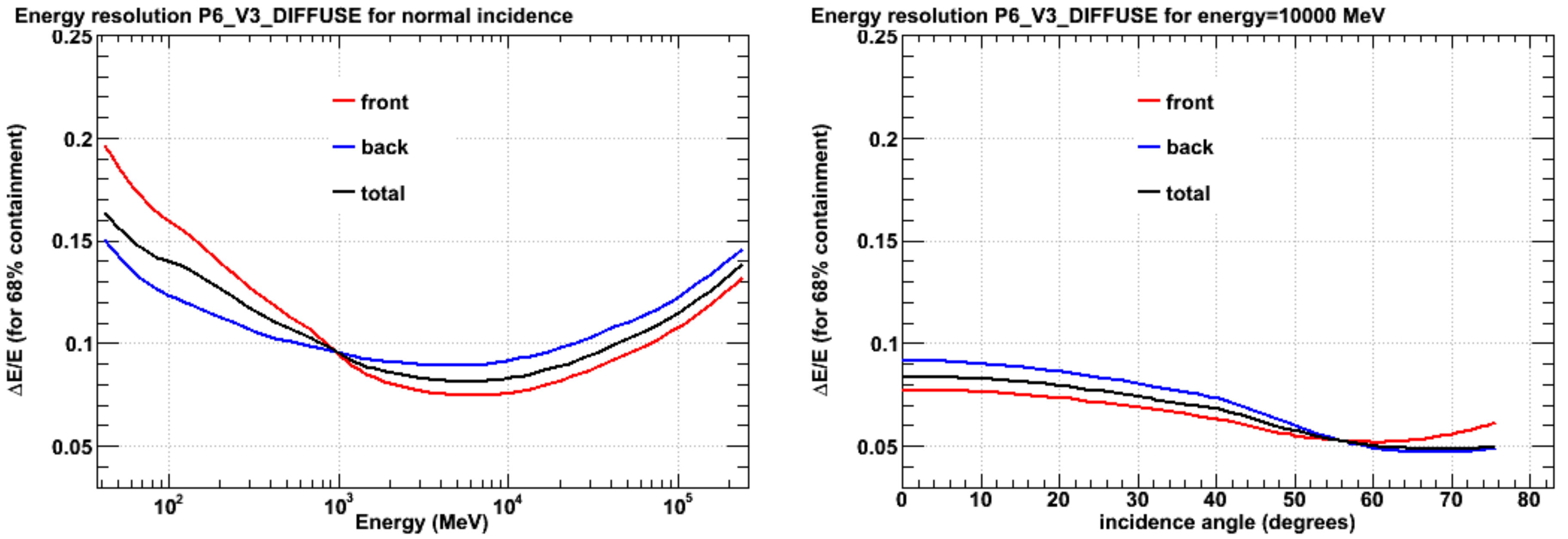}
\end{center}
\small\normalsize
\begin{quote}
\caption[LAT energy resolution]{\emph{(Left)}: Energy resolution (68\% containment level) vs. energy for near on-axis events ($cos(\theta)\ >\ 0.9)$ separated into FRONT, BACK, and total events for the DIFFUSE event class under the P6\_V3 IRFs.  \emph{(Right)}: Energy resolution vs. incidence angle ($\theta$) for 10 GeV events.\label{ch3edispplot}}
\end{quote}
\end{figure}

For a given event class, the appropriate selections are applied to the simulated data which is then binned up in energy, conversion location in the LAT (i.e., FRONT vs. BACK), and $\cos(\theta)$.  The binned, simulated events are then used to derive the instrument response functions (IRFs) as functions of these variables.  These include the PSF (see Fig.~\ref{ch3psfplot}), the effective area (see Fig.~\ref{ch3aeffplot}), and the energy resolution (see Fig.~\ref{ch3edispplot}).  Figs.~\ref{ch3psfplot},~\ref{ch3aeffplot}, and~\ref{ch3edispplot} are available at http://www-glast.slac.stanford.edu/software/IS/glast\_lat\_performance.htm.  Early in the mission, the IRFs (tagged as P6\_V1) were completely derived from simulations and it is a testament to the detail of MC that they worked so well as to allow important science results to be published soon after launch (e.g., Abdo et al., 2008 and 2009b).

\small\normalsize

As the mission continues, these IRFs are expected to be updated to reflect how the real instrument performs.  The first such correction came about in the form of P6\_V3 IRFs which applied corrections to account for slightly lower efficiency due to event pileup \citep{Rando09}.  A second IRF update (P6\_V11) is expected to be released soon which will use a PSF derived from on-orbit observations of bright point sources and applies effective area corrections accounting for MC and data discrepancies.

\small\normalsize

\small\normalsize

\section{LAT Timing Verification}\label{ch3psr}
Events in the LAT are given time stamps derived from a hardware/software system that receives pulse-per-second time hacks from the global positioning system (GPS) receiver on the Fermi spacecraft \citep{AbdoOnorb}.  The LAT absolute timing was verified pre-launch using atmospheric muon data while the LAT CAL modules were being integrated with the TKR at the Naval Research Laboratory in Washington D.C. \citep{Smith06}.

A muon telescope consisting of two plastic scintillating tiles, each read out by two photomultiplier tubes, was placed next to and below the LAT in order to facilitate detection of coincident events.  One data run of thirty minutes was taken in which $\sim$880,000 events were recorded in the LAT and $\sim$10,000 in the muon telescope.  Muons leave straight tracks in the TKR, thus the reconstructed LAT tracks can be extrapolated down to the level of the muon telescope.  Only the $\sim$300,000 events for which the extrapolated track was outside of the LAT at the level of the muon telescope were kept.

Time coincidences between the LAT and muon telescope were found to have a peak time difference of 24 ms and 3.7 ms width.  The observed width was found to be due to a 2.0083 $\mu$s clock drift, the root mean square variance was reduced to 700 ns upon correcting for this drift.  A nearly sinusoidal, systematic drift of $\pm$1 $\mu$s was found and removed from the data, resulting in a normal distribution with 160 ns width \citep{Smith07}.

More muon data was acquired (consisting of eight, thirty minute runs) after the LAT had been integrated with the spacecraft at General Dynamics in Arizona \citep{Smith07}.  In the new runs with the spacecraft locked to the GPS, a drift of 3.4 $\mu$s s$^{-1}$ was observed in coincident events.  This led to time differences which went from 0 to $-1$ ms over $\sim$290 s at which point the GPS lock resulted in a wrap around effect leading to time differences of 0 s again.  In new runs without GPS lock the drift proceeded unchecked resulting in time differences on the order of 10 ms or more.  Such behavior would have had an adverse affect on LAT pulsar science (more details below) if this issue had not been identified and the spacecraft lost GPS lock for a significant amount of time.

This issue was fixed by General Dynamics through a software update which, along with other software improvements, led to the agreement between LAT timestamps and those from the GPS clock to within 0.3 $\mu$s as quoted by \citet{AbdoOnorb}.

\begin{figure}
\begin{center}
\includegraphics[width=1.0\textwidth]{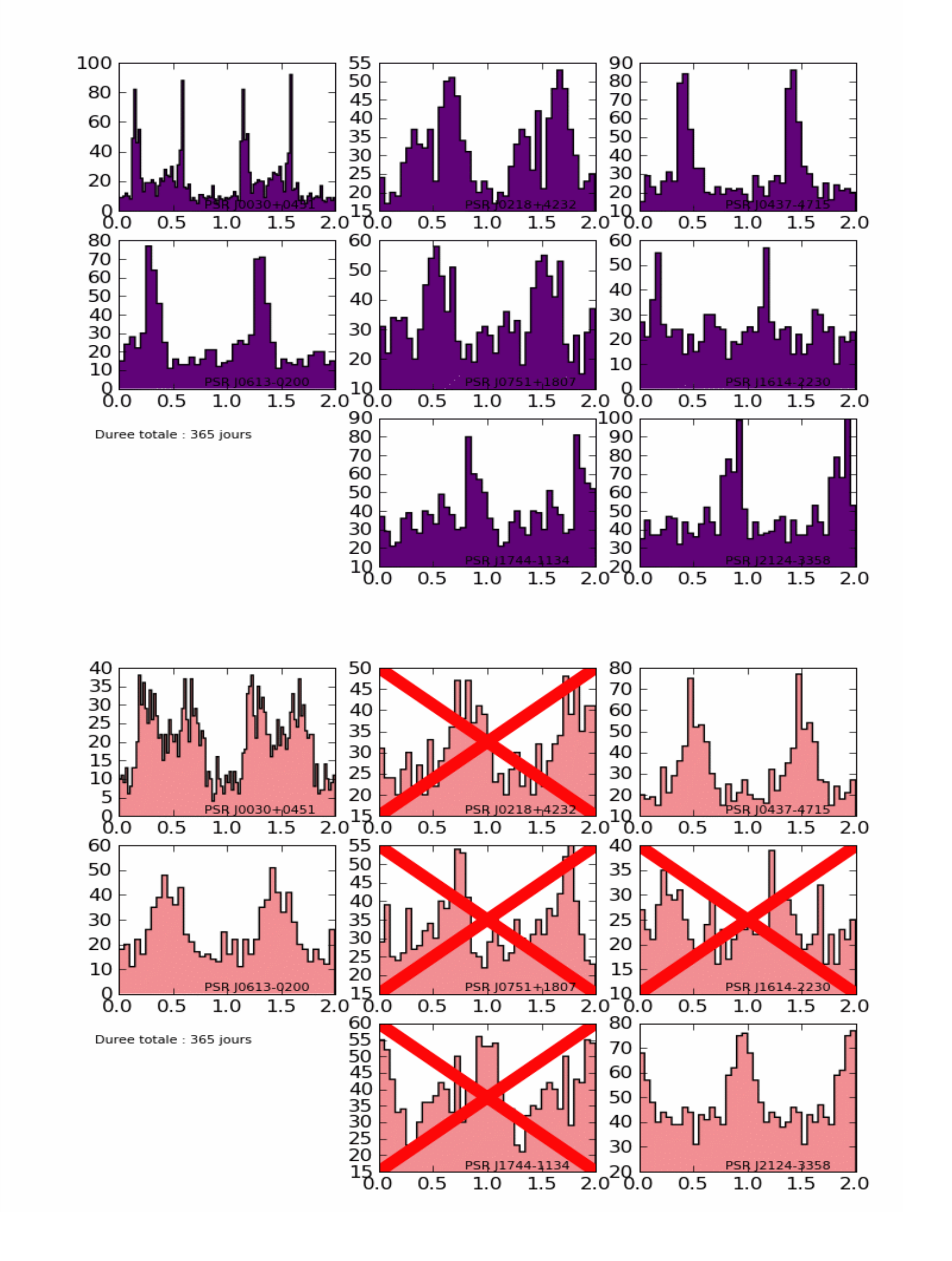}
\end{center}
\small\normalsize
\begin{quote}
\caption[Comparison of MSP light curves with and without drift removed]{\emph{(Top):} Light curves of the first eight LAT-detected MSPs using one year of data.  \emph{(Bottom:)} Light curves of the first eight LAT-detected MSPs using one year of data and modified event arrival times with the $-1$ ms drift added in, crossed out MSPs are those for which a significant detection would not have been claimed.  \emph{Figure credits: D. Dumora}\label{ch3Denis}}
\end{quote}
\end{figure}

As a demonstration of how important verifying the absolute timing is, D. Dumora at CENBG in France produced modified arrival times for one year of LAT data in which the $-1$ ms drift was added as if it had not been removed.  Then, pulse phases were calculated for the first eight MSPs detected with the LAT (Abdo et al., 2009g, and Chapter 4) using the real and modified arrival times.  The resulting phase plots are shown in Fig.~\ref{ch3Denis}.

\small\normalsize

The implications of Fig.~\ref{ch3Denis} are very startling, not only would half of the MSPs not have been significantly detected after one year (roughly 1.5 times the data span used by Abdo et al., 2009g) but the profiles of the other four would have been artificially widened, completely changing the light curve shape and leading to incorrect conclusions regarding the likely emission models.

\begin{figure}
\begin{center}
\includegraphics[width=1.0\textwidth]{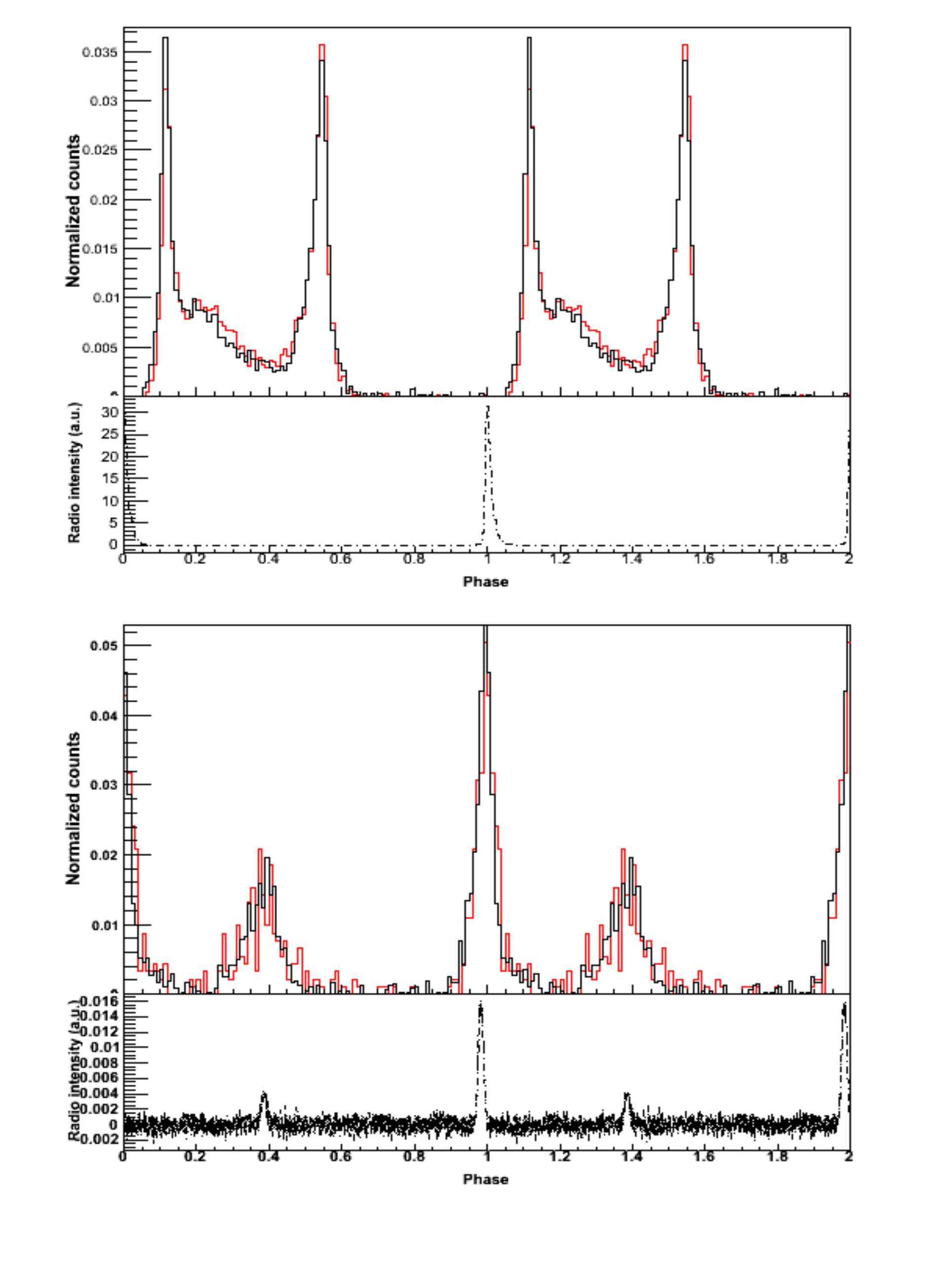}
\end{center}
\small\normalsize
\begin{quote}
\caption[\emph{EGRET} and early LAT light curves of the Vela and Crab pulsars]{\emph{EGRET} light curves (black) of the Vela and Crab pulsars from \citet{Fierro98} compared with LAT light curves (red) using $\sim$3 weeks of early calibration data.  \emph{Figure credits: L. Guillemot}\label{ch3VC}}
\end{quote}
\end{figure}

As an additional check of the LAT timing, data during the calibration phase of the mission was used to produce light curves for the six high-confidence, \emph{EGRET} pulsars.  The LAT light curves were then compared against those from \citet{Fierro98} as shown for the Vela and Crab pulsars in Fig.~\ref{ch3VC}, LAT light curves in red with \emph{EGRET} in black.

The timing solution used for Vela was produced by the Parkes radio observatory in Australia while that for the Crab came from the Nan\c{c}ay radio telescope in France.  The LAT light curves show peak widths, separation, and offsets from the radio in good agreement with those from \emph{EGRET}.  While this comparison can not be used to test absolute timing accuracy, it does demonstrate that the LAT timestamps are at least as precise as those of \emph{EGRET}.  Additionally, as noted by \citet{AbdoOnorb} the gamma-ray peak width of PSR J0030+0451 is $<$100 $\mu$s (see Chapter 4 for more details and Abdo et al., 2009d,g) which suggests that the LAT timestamps are very stable with time.

\small\normalsize

\section{Spectral Analysis of LAT Data}\label{ch3specAn}
Spectral analysis of LAT data is done using a maximum likelihood technique similar to that used for analysis of \emph{EGRET} data \citep{Mattox96}.  The likelihood requires knowing the detector response (as given by the IRFs described in Section~\ref{ch3sims}) and having a model for each source in the region of interest (ROI).  The LAT STs implement both a binned and unbinned likelihood analysis as the \emph{gtlike} tool and the \emph{pyLikelihood} python module. The unbinned analysis treats each event separately whereas the binned analysis sorts the events into bins based on reconstructed direction on the sky and energy.  The unbinned analysis is more exact and is less affected by ignoring the energy dispersion but is also much more memory intensive.  Binning the events reduces the run time but can lead to incorrect results if there are not sufficient counts in each bin.  Therefore, data sets spanning a small range of times or with few events (as may be the case for analyses $\gtrsim$ 100 GeV) should be analyzed with unbinned likelihood while longer, and larger, data sets should use binned likelihood.  However, unbinned analysis has been shown to be unstable for analysis of extended sources and thus binned analysis should be used in those cases (e.g., Abdo et al., 2010e) which generally require longer datasets to properly evaluate the source extent.

The following is a description of the likelihood function for unbinned analysis summarized from Chiang (2002a,b,c,\&d), for binned analysis the integrals are replaced by interpolations over the spatial and energy bins.  The likelihood model is given by Eq.~\ref{ch3lMod} and describes the expected distribution of events as a function of time, reconstructed energy, and direction on the sky.
\begin{equation}\label{ch3lMod}
M(E^{\prime},\hat{p}^{\prime},t)\ =\ \int dE d\hat{p} R(E^{\prime},\hat{p}^{\prime},t;E,\hat{p})S(E,\hat{p})
\end{equation}

The function $R(E^{\prime},\hat{p}^{\prime},t;E,\hat{p})$ defines the total instrument response (encompassed by the IRFs described in Section~\ref{ch3sims}) and $S(E,\hat{p})$ is the source model (including all point and diffuse sources) where $E$ and $\hat{p}$ are the true event energy and direction and $E^{\prime}$ and $\hat{p}^{\prime}$ are the reconstructed values.  Ideally, the total response would include energy dispersion but that is not included in the LAT likelihood analysis; however, the best-fit parameters from a simulated source have been shown to reliably match the input values when neglecting this aspect of the response \citep{Chiang1}.  Additionally, the time dependence of the total response doesn't only depend on when the event was incident on the LAT but also on the position of the spacecraft and instrument mode.

From the likelihood model, the predicted number of events can be calculated using Eq.~\ref{ch3Npred}.  Combining Eqs.~\ref{ch3lMod} and~\ref{ch3Npred} one can construct the logarithm of the Poisson likelihood (Eq.~\ref{ch3Like}) by summing over the individual events with index $j$.  The best-fit model parameters are derived by minimizing the negative of Eq.~\ref{ch3Like} \citep{Chiang2}.
\begin{equation}\label{ch3Npred}
N_{\rm{pred}}\ =\ \int dE^{\prime} d\hat{p}^{\prime} dt M(E^{\prime},\hat{p}^{\prime},t)
\end{equation}
\begin{equation}\label{ch3Like}
log(\mathcal{L})\ =\ \sum_{j}log(M(E^{\prime}_{j},\hat{p}_{j}^{\prime},t_{j}))-N_{\rm pred}
\end{equation}

When preparing for spectral analysis of LAT data, event selections are made on reconstructed energy and position on the sky, defining a circular ROI for unbinned analysis and a square ROI for binned analysis.  Selections are also made on event zenith angle in order to reduce the gamma-ray signal from the limb of the Earth \citep{AbdoAlbedo}.  Good time intervals are then selected based on the data quality and LAT operations mode\footnote{For more details on LAT event selection see http://fermi.gsfc.nasa.gov/ssc/data/analysis/scitools/}.  It is then necessary to calculate total live time and exposure as a function of energy and position on the sky.  These maps need to encompass a region larger than the ROI (typically $\sim$10\DEG{} beyond the ROI limits) to account for the fact that the LAT PSF at low energies (near 100 MeV) is quite broad (see Fig.~\ref{ch3psfplot}) and thus some events may come from sources outside your ROI.  This also means that the source model must include point sources outside the ROI but note that the spectral parameters of these sources must remain fixed.

The maximum likelihood value is not, by itself, a goodness-of-fit test and therefore not able to comment on how well a given model describes a given source.  The ratio of likelihoods (or more practically the difference in the logs of the likelihoods) between different fits can be used to evaluate the significance of a given source and reject one model in favor of another.  This is commonly known as the likelihood ratio test (LRT) which is used to test a more complicated hypothesis versus a simpler, null hypothesis with likelihoods $\mathcal{L}^{\prime}$ and $\mathcal{L}_{0}$, respectively.  The null hypothesis is rejected if the likelihood ratio ($\Lambda\ \equiv\ \mathcal{L}^{\prime}/\mathcal{L}_{0}$) is greater than 1 while the more complex hypothesis is rejected as unnecessary to explain the data if $\Lambda\ \leq\ 1$.  The LRT requires that the null hypothesis occupy a subset of the parameter space available to the more complex hypothesis, in particular, the LRT requires nested models.

The LRT can be used to evaluate the significance of a source by considering the model without the source as the null hypothesis and the model with the source the more complex hypothesis.  If inclusion of the source improves the global fit then the difference of the log likelihoods (defined to be $\Delta{}\log(\mathcal{L})\ \equiv\ \log(\Lambda)\ \equiv\ \log(\mathcal{L}_{\rm{source}})-\log(\mathcal{L}_{\rm{no\ source}})$) will be positive (corresponding to the LRT condition that $\Lambda\ >$ 1).  The significance of the source can be ascertained using Wilks' theorem \citep{Wilks38} relating the likelihood to a $\chi^{2}$ statistic by defining a test statistic (TS) to be 2$\Delta{}\log(\mathcal{L})\ \sim\ \chi^{2}$. Let the number of free parameters in the source model be $n_{\rm src}$ (analogous to degrees of freedom in a $\chi^{2}$ analysis) then the significance of the source is $\sqrt{2}(\rm{erf}^{-1}(1-\mathnormal P(\rm{TS},\mathnormal n_{\rm src}))$, where $P(\rm{TS},\mathnormal n_{\rm src})$ is the probability of finding a value higher than TS from a $\chi^{2}$ distribution with $n_{\rm src}$ degrees of freedom and erf$^{-1}$ is the inverse error function.

For many astrophysical sources of HE gamma rays the predicted spectra in the LAT energy range can be well described by simple power laws (i.e., the number of photons from a given source per unit area per unit time per unit energy is proportional to E$^{-\Gamma}$ where E represents the photon energy and $\Gamma$ is known as the photon index).  For some sources, such as pulsars and AGN, deviations from simple power laws are predicted and/or observed.  The LRT can be used to evaluate the significance of these deviations.  The null hypothesis is taken to be the power law spectrum while the model with deviations from a power law is the more complex model.  As noted above this does require that the models be nested.

The LRT can be used to discriminate between models but does not give one a feeling for how well the fit model for a given source agrees with the data. The likelihood value returned from the \emph{gtlike} and \emph{pyLikelihood} LAT STs apply to the global fit and provide residuals describing how well the total counts over the entire ROI in a given energy bin agree with the total model counts predicted in the same energy range.  In order to make a similar comparison for a given source it is necessary to perform likelihood fits in smaller energy bins.  While this can be done multiple ways, the approach described here is implemented as the \emph{likeSED} family of python macros (likeSED.py, bdlikeSED.py, extbdlikeSED.py\footnote{The macros and usage notes are available for download\\at http://fermi.gsfc.nasa.gov/ssc/data/analysis/user/}).

First, the likeSED macros create a user supplied number of bins of uniform width in log energy and choose the maximum bin for spectra analysis based on the highest energy event found consistent with the source position within the 95\% confidence radius (as defined by the input IRFs).  The code creates event files corresponding to each energy bin and creates the necessary files for unbinned (likeSED.py) or binned (bdlikeSED.py and extbdlikeSED.py) analysis.  A likelihood analysis over the full energy range is then run to get the error information necessary to make a `bowtie' error contour and calculate center energies for each bin.  For each bin a weighted average is calculted using the full energy range spectral information at 100 points between the bin upper and lower limits to calculate sensible values for the center energy.  This is useful as it places the center of each bin nearer to where the bulk of the energy is expected to be for a given bin.

At this stage the code performs likelihood fits for each energy bin by assuming that the source has a power law spectrum.  The photon index of the source can be fixed or left free as the user desires.  The LAT STs allow for two different representations of a power law spectrum.  The first is a \emph{PowerLaw} model (Eq.~\ref{ch3pl}) with 3 parameters: the prefactor ($N_{0}$, default units cm$^{-2}$ s$^{-1}$ MeV$^{-1}$), photon index ($\Gamma$), and scale ($E_{0}$, held fixed in the fit).  Alternatively, one can use a \emph{PowerLaw2} model (Eq.~\ref{ch3pl2}) in which the normalization parameter is the integral flux and not the prefactor and there are 4 parameters: the integral ($F$, default units cm$^{-2}$ s$^{-1}$), photon index ($\Gamma$), lower limit of integration for $F$ ($E_{\rm min}$, default units of MeV), and upper limit of integration for $F$ ($E_{\rm max}$, default units MeV).
\begin{equation}\label{ch3pl}
\frac{dN}{dE}\ =\ N_{0}\ \Big(\frac{E}{E_{0}}\Big)^{-\Gamma}
\end{equation}
\begin{equation}\label{ch3pl2}
\frac{dN}{dE}\ =\ F\ \frac{\Gamma-1}{E_{\rm{min}}^{-\Gamma+1}-E_{\rm{max}}^{-\Gamma+1}}\ E^{-\Gamma}
\end{equation}

If the \emph{PowerLaw} model is used, the center energy of the bin is used as the scale parameter such that the fit prefactor can be used for the counts spectrum plot.  When using the \emph{PowerLaw2} model the upper and lower limits of integration are taken as the upper and lower bounds of the energy bin.  Additionally, before starting the fit the prefactor and flux arguments are rescaled to be the same order of magnitude as that predicted by the full energy range likelihood analysis.  This is done in order to avoid problems with the likelihood code which arise when the starting point of the normalization parameter is several orders of magnitude too large.  The other free sources in the ROI are treated similarly but this is really only important for relatively bright, nearby sources and is mainly done for completeness in the treatment of sources.

\begin{figure}[h]
\begin{center}
\includegraphics[width=0.75\textwidth]{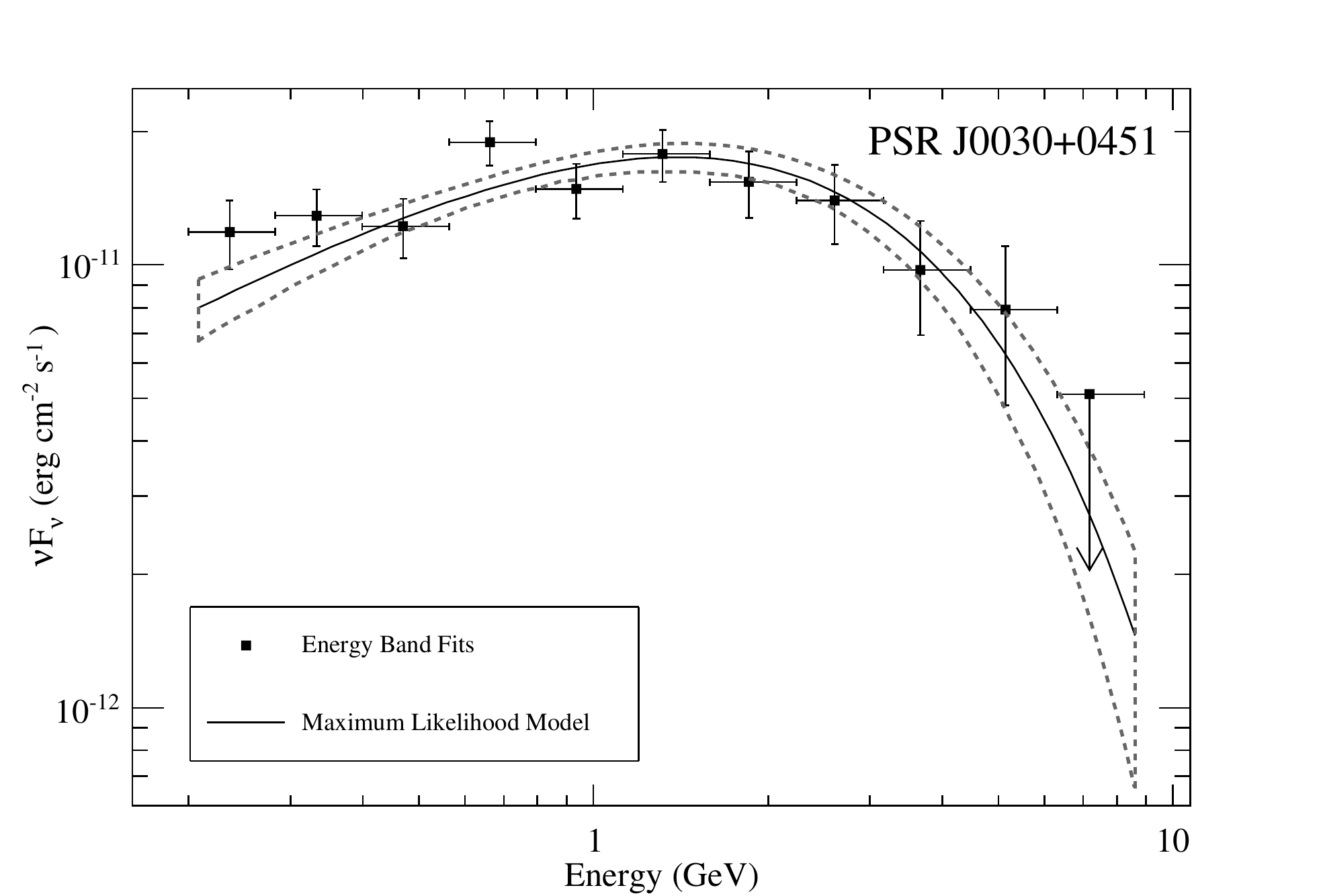}
\end{center}
\small\normalsize
\begin{quote}
\caption[Gamma-ray spectrum of PSR J0030+0451 from likeSED.py]{Spectral plot from the likeSED.py macro for PSR J0030+0451 for the data set used by \citet{AbdoMSPpop}.  The dashed lines are 1$\sigma$ errors on the fit and form the so-called `bowtie' contour.  A \emph{PowerLaw2} model was used for the pulsar and a 95\% confidence level upper limit was caculated if the puslar TS was less than 9 (2.5$\sigma$ significance for 2 free parameters) in a given bin.\label{ch3exmplSpec}}
\end{quote}
\end{figure}

An initial fit with a high tolerance is performed in order to get an estimate of the TS of the free sources in the ROI.  Not all sources will be significant in each energy bin and including sources with TS $\leq$ 0 can adversely affect the fits and often lead to underestimated uncertainties.  Any such sources are removed from the model for that energy bin and then a fit with a much lower tolerance is performed.  If the source of interest is found with a TS less than a user supplied value a 95\% confidence level upper limit is reported for the normalization parameter.  The fit parameters are collected for each energy bin and used to construct spectral plots (see Fig.~\ref{ch3exmplSpec} as an example).  The code outputs three plots, the counts spectrum (units of cm$^{-2}$ s$^{-1}$ GeV$^{-1}$), a plot of source TS in each energy bin, and a $\nu F_{\nu}$ or $E^{2}dN/dE$ plot (units of erg cm$^{-2}$ s$^{-1}$).  If the \emph{PowerLaw} model is used the first plot is made using the prefactor values from the individual band fits while the points for the third plot are made by multiplying each point in the counts spectrum by the square of the center energy of the given bin.  When the \emph{PowerLaw2} model is used the points for the first plot are made by setting the right hand sides of Eqs.~\ref{ch3pl} and~\ref{ch3pl2} equal and solving for $N_{0}$ while the points for the third plot are made by integrating the flux from individual bin fits over the energy bin.

\small\normalsize

In Fig.~\ref{ch3exmplSpec} the maximum likelihood model was obtained from a fit to the entire energy range (solid black line) in which the spectrum of PSR J0030+0451 was modeled as an exponentially cutoff power law, Eq.~\ref{ch3ECO}, with b $\equiv$ 1.  This is the functional form predicted from curvature radiation (see Chapter 2) which is used to fit the spectra of known gamma-ray pulsars.  The exponential index b allows for the possibility of a super- (or sub-) exponential cutoff as may be expected in some pulsars (see Chapters 2 and 4 for more details).
\begin{equation}\label{ch3ECO}
\frac{dN}{dE}\ =\ N_{0}\ \Big(\frac{E}{E_{0}}\Big)^{-\Gamma} \exp\bigg\lbrace -\Big(\frac{E}{E_{C}}\Big)^{\rm b}\bigg\rbrace
\end{equation}

The prefactor, photon index, and scale parameters are defined the same as for Eq.~\ref{ch3pl}.  In addition to the exponential index b, the cutoff energy $E_{C}$ is introduced as an additional parameter.

This should not be considered a model independent method as it does assume that, for small enough bin sizes, the source spectrum can be represented by a power law; the diffuse backgrounds are fit using the standard Galactic and isotropic models; and the choice of the center energy ranges are affected by the full energy range fit (while this is not expected to bias the results of the fits the user can specify custom energy bins and bin centers).  Additionally, flux points from these macros are meant to be a visual check on the best-fit model from the full maximum likelihood analysis and to point out potential discrepancies between the data and best-fit model.  As such, these points should not be used to derive best-fit spectral parameters.

\section{Conclusions}\label{ch3conc}
The LAT is an amazing instrument which has been performing excellently since launch.  The current success of the mission is the result of the hard work of many scientists, engineers, an exceptional flight operations team, and many others.  The merging of the astronomical and high-energy physics community created a unique collaboration which has produced highly cited papers on pulsars, AGN, dark matter limits, gamma-ray bursts, cosmic-ray electrons, and many more scientific topics.

The foresight of the collaboration members in validating the absolute timestamps of the LAT before launch resulted in pulsar science being feasible with early calibration data (e.g., Abdo et al., 2008).  Such accurate time tagging is also of great importance in gamma-ray burst studies, but that is beyond the scope of this thesis.
\renewcommand{\thechapter}{4}

\chapter{\bf LAT Pulsar Science}\label{ch4}
The discovery and characterization of gamma-ray pulsars is one of the main science goals of the \FL{}.  Experience from \emph{EGRET} had shown that the inherent noisiness of young pulsars requires contemporaneous timing solutions, from either radio or X-ray telescopes, in order to get the most significant detections in gamma rays.  As such, the \Fermi{} Pulsar Timing Consortium (PTC) was conceived in order to ensure that timing solutions for 208 of the best gamma-ray pulsar candidates were available when \Fermi{} launched \citep{Smith08}.

\begin{figure}
\begin{center}
\includegraphics[width=1.0\textwidth]{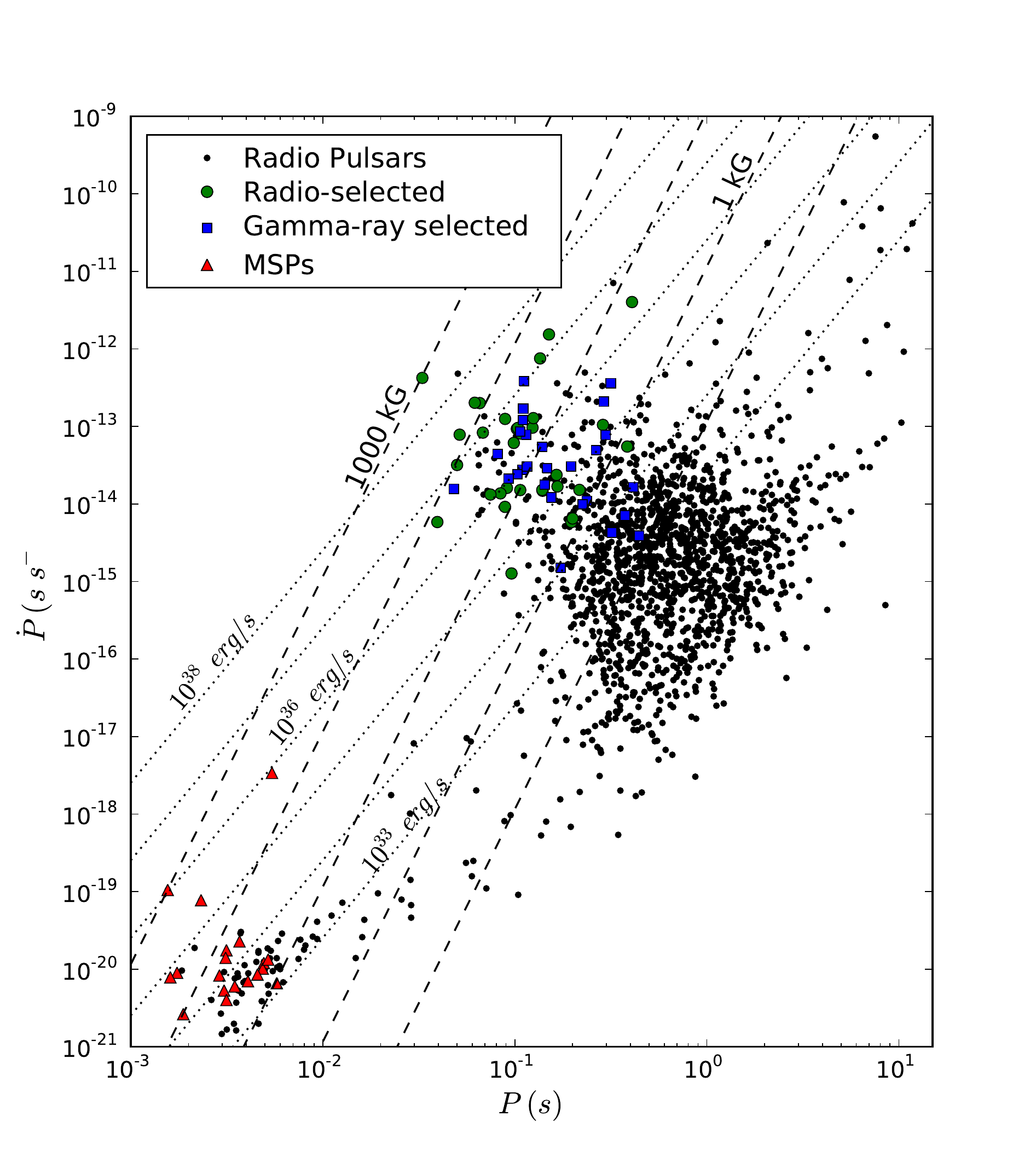}
\end{center}
\small\normalsize
\begin{quote}
\caption[$\dot{\rm P}$-P diagram with gamma-ray pulsars]{$\dot{\rm P}$-P diagram with LAT detected gamma-ray pulsars (note that this plot includes gamma-ray pulsars not in Abdo et al., 2010c).  Red triangles: MSPs; green circles: radio-selected, non-recycled pulsars; blue squares: gamma-ray selected, non-recycled pulsars.  Lines of constant $\dot{E}$ (dotted) and $B_{\rm LC}$ (dashed) are also drawn.  Values of $\dot{P}$ have been corrected for the Shklovskii effect where appropriate.\label{ch4newppdot}}
\end{quote}
\end{figure}

Observations with the \FL{} have increased the number of known gamma-ray pulsars by more than an order of magnitude.  This effort has benefited greatly from the excellent work of the PTC members but also from the ingenuity of \citet{Atwood06} in proposing a time-differencing technique to search gamma-ray data for pulsations which has proved very effective (e.g., Abdo et al., 2008 and 2009e; Saz Parkinson et al., 2010).  Fig.~\ref{ch4newppdot} shows the $\dot{\rm P}$-P distribution for all Galactic field pulsars in the ATNF database as of mid-2009 \citep{ATNF} with LAT detected gamma-ray pulsars and lines of constant $\dot{E}$ and $B_{\rm LC}$.

\small\normalsize

With more than two years of sky-survey data, gamma-ray pulsations from \\$\gtrsim$ 78 pulsars have been detected with the LAT.  These include at least 19 MSPs and 59 non-recycled pulsars.  Of these non-recycled pulsars 32 were known to be radio pulsars prior to the launch of \Fermi{} and 27 were first detected via their X- or gamma-ray pulsations.

\section{Non-recycled Gamma-ray Pulsars}\label{ch4youngPSRs}
Prior to the launch of \Fermi{}, all gamma-ray pulsars known were young ($\tau\ \lesssim$ 1 Myr), non-recycled objects (see Chapter 2 for more details).  While a large number of recycled pulsars (MSPs) have now been seen to pulse in gamma rays they account for only $\sim$1/3 of all gamma-ray pulsars.  The non-recycled gamma-ray pulsars detected by \Fermi{} comprise two subclasses, those detected first in radio observations and those detected first in the X-ray or gamma-ray band, a.k.a radio selected and gamma-ray selected, respectively.

These two subclasses share many traits \citep{AbdoPSRcat} and it is likely that the radio beams of gamma-ray selected pulsars simply miss the Earth.  However, three have now been seen to pulse at radio wavelengths (Camilo et al., 2009 and Abdo et al., 2010b).  It is of interest to note that PSR J1907+0602 is extremely radio-faint and required a 2 hour observation with Arecibo (the largest radio dish in the world) for a $\sim5\sigma$ pulsed detection \citep{Abdo1907} which suggests that some gamma-ray selected pulsars may indeed have geometries for which the radio beams should be seen but may not be due to considerations of distance, energetics, or other issues.  A comparison of the radio and gamma-ray selected pulsars is beyond the scope of this discussion.  Studies of particular radio-selected gamma-ray pulsars are presented below and the basic results should apply equally well to the gamma-ray selected sample.

\subsection{PSR J1028$-$5819}\label{ch4J1028}
PSR J1028$-$5819 was discovered at the Parkes radio telescope shortly before the launch of Fermi \citep{Keith08} in a high-frequency (3.1 GHz) search of the unidentified \emph{EGRET} source 3EG J1027$-$5817.  Measured and derived parameters of PSR J1028$-$5819, from \citet{Keith08}, are given in Table~\ref{ch4J1028vals}, values in parentheses are 1$\sigma$ uncertainties.

\begin{figure}
\begin{center}
\includegraphics[width=0.6\textwidth]{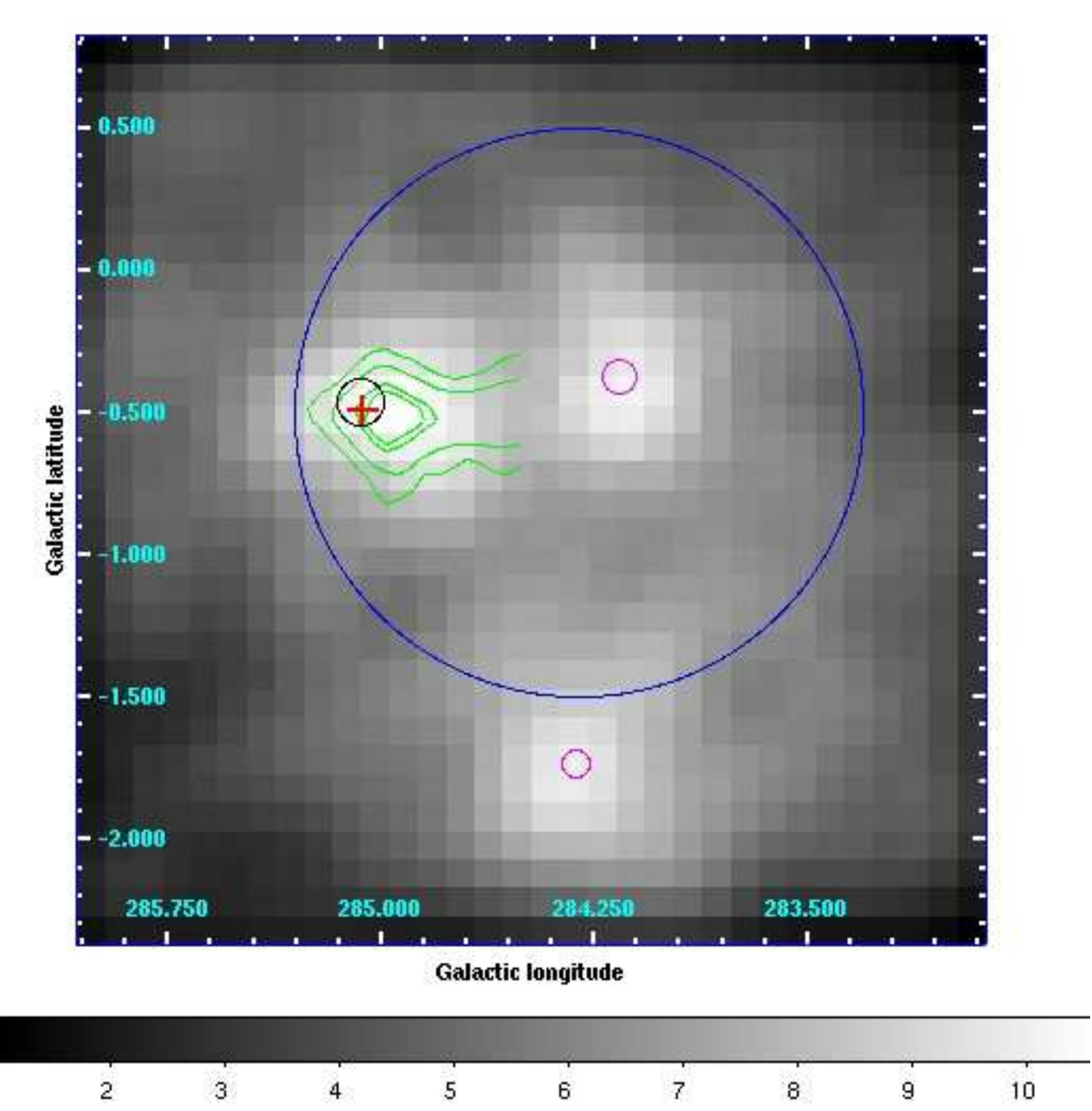}
\end{center}
\small\normalsize
\begin{quote}
\caption[Gamma-ray counts map of the region around PSR J1028$-$5819]{Counts map, $\geq$0.1 GeV, for the same time range as the LAT BSL.  Gaussian smoothing with a kernel of radius $0^{\circ}.3$ has been applied.  The 50, 68, 95, and 99\% confidence level contours for the \emph{EGRET} source are shown in green.  The 95\% confidence level radius is shown in black for 0FGL J1028.6$-$5817 and in magenta for the other two BSL sorces.  Note that the positions have been refined using six months of data.  The error circle of 2CG 284$-$00 is shown in blue.  The red cross marks the radio position of PSR J1028$-$5819.\label{ch4J1028cmap}}
\end{quote}
\end{figure}

The $\dot{E}$ and DM-derived distance made an association with 3EG J1027$-$5817, which was measured to have an integrated flux (0.1-10 GeV) of $6.6\pm0.7\times10^{-7}\ \rm{cm}^{-2}\ \rm{s}^{-1}$ \citep{Hartman99}, very plausible but a detection of pulsed gamma-ray emission was necessary to make an identification.  Using a contemporaneous timing solution, a significant pulsed signal has been detected with the LAT from PSR J1028$-$5819 at a level greater than 10$\sigma$ \citep{AbdoJ1028}.  The pulsar was detected in the LAT bright sources list (BSL, Abdo et al., 2009c) as 0FGL J1028.6$-$5817.  Additionally, the LAT has resolved the region around the pulsar into multiple point sources, Fig.~\ref{ch4J1028cmap}, the combination of which can account for the measured flux of the \emph{EGRET} source and the measured flux of the COS-B source 2CG 284$-$00 \citep{Swanenburg81}.

\small\normalsize

\begin{deluxetable}{l r}
\tablewidth{0pt}
\tablecaption{PSR J1028$-$5819 Timing Parameters}
\startdata
\underline{Measured Parameters}&\\
P (ms) & 91.4032309(14)\\
$\dot{\rm P}$ ($\times10^{-15}$ s s$^{-1}$) & 16.1(8)\\
DM (pc cm$^{-3}$) & 96.525(2)\\
\underline{Derived Parameters}&\\
$\dot{E}$ ($10^{35}$ erg s$^{-1}$) & 8.13\\
Distance (kpc) & 2.3\\
$B_{surf}$ ($10^{12}$ G) & 1.23\\
\enddata\label{ch4J1028vals}
\end{deluxetable}

\small\normalsize

\small\normalsize

\begin{figure}
\begin{center}
\includegraphics[width=1.0\textwidth]{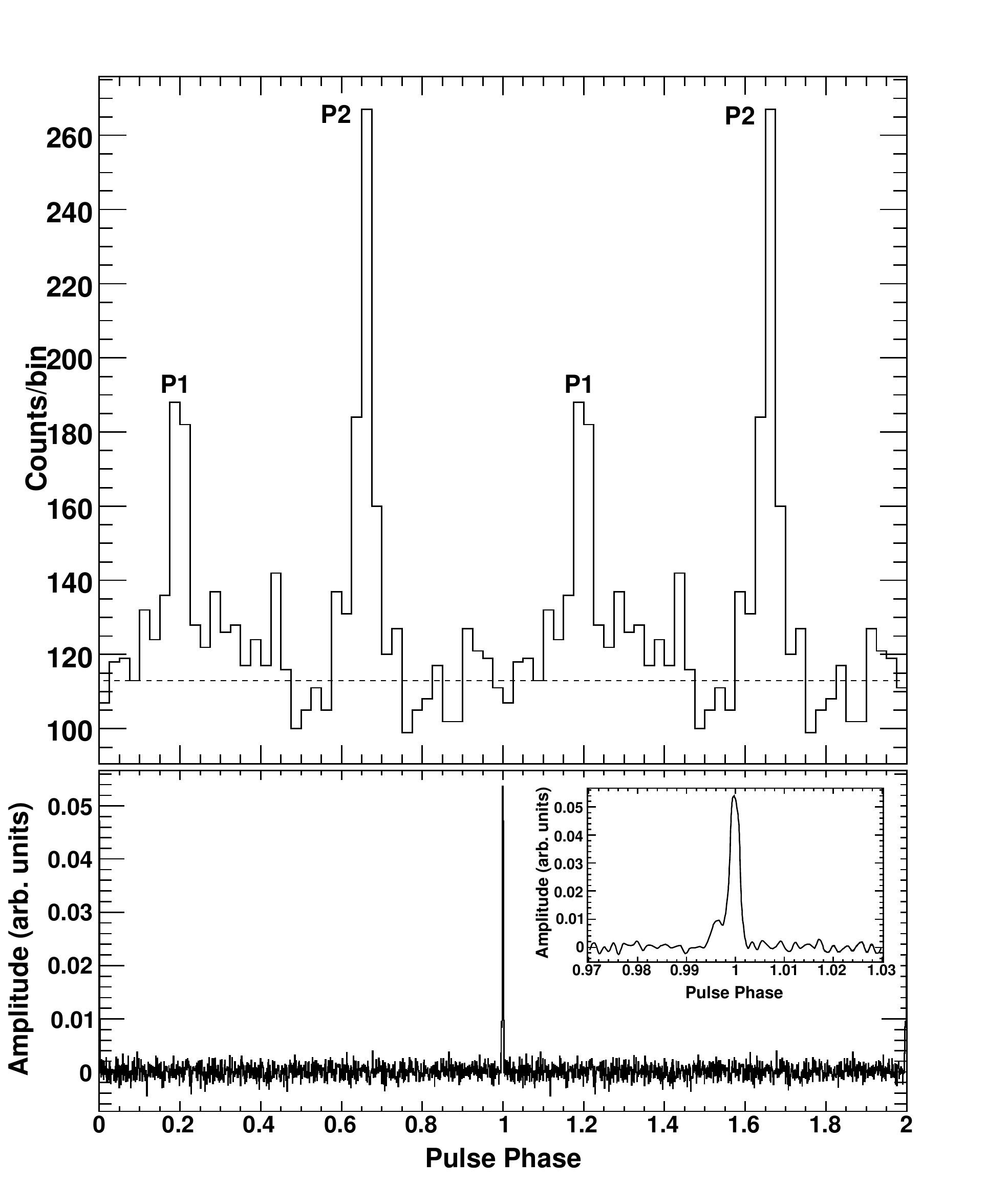}
\end{center}
\small\normalsize
\begin{quote}
\caption[Gamma-ray and radio light curves of PSR J1028$-$5819]{\emph{(Top):} Gamma-ray light curve of PSR J1028$-$5819 in the (0.1-13 GeV) energy band using 40 constant-width bins and shown over two pulse periods.  The horizontal dashed line indicates the estimated background level obtained by fitting the off-peak region as a constant value.  \emph{(Bottom)}: The 1.4 GHz radio pulse profile in arbitrary units.  The inset shows the radio pulse in the phase range 0.97-1.03 with the main peak at phase 1.0 preceded by a smaller, secondary peak at phase $\sim$0.996.  Reproduced from \citet{AbdoJ1028}.\label{ch4J1028lc}}
\end{quote}
\end{figure}

The HE and radio light curves of PSR J1028$-$5819 are shown in Fig.~\ref{ch4J1028lc}.  The gamma-ray light curve was constructed from LAT events spanning the time interval from 30 June to 16 November 2008, belonging to the ``Diffuse" class of photons as defined under the P6\_V1 IRFs (see Chapter 3), have reconstructed energies $\geq$0.1 GeV, and zenith angles$\leq105^{\circ}$.  Additionally, only events with reconstructed directions on the sky within $\theta_{\rm{C}}\times{}(\rm{E}/(100\ \rm{MeV}))^{-0.75}$ of the radio position of the pulsar were used to build the light curve.

\small\normalsize

This energy dependence mimics the 68\% containment radius of the LAT.  For events converting in the FRONT section, $\theta_{\rm{C}}$ was taken to be 3\DEG{} while for those converting in the BACK $\theta_{\rm{C}}$ was taken to be 4\DEG{}.1 (see Chapter 3 for more details on FRONT and BACK converting events).  The events were then phase-folded with the radio timing solution using the \Fermi{} ST \emph{gtptest}.

The gamma-ray light curve of PSR J1028$-$5819 shows two narrow peaks, neither of which are aligned with the narrow radio peaks near phase 0, with the second gamma-ray peak stronger than the first and no significant features observed between the peaks.  Both peaks were fit with Lorentzians on top of a constant background value estimated from a fit of the off-peak region (taken to be phase from 0.8 to 1.0).  The best-fit peak positions ($\phi_{i}$), Lorentzian full width at half max (FWHM) values ($W_{i}$), and peak separation ($\Delta$) are given in Table~\ref{ch4J1028shape}.

\small\normalsize

\begin{deluxetable}{l r}
\tablewidth{0pt}
\tablecaption{PSR J1028$-$5819 Light Curve Parameters}
\startdata
$\phi_{1}$ & 0.200$\pm$0.003\\
$W_{1}$ & 0.040$\pm$0.011\\
$\phi_{2}$ & 0.661$\pm$0.002\\
$W_{2}$ & 0.034$\pm$0.007\\
$\Delta$ & 0.460$\pm$0.002\\
\enddata\label{ch4J1028shape}
\end{deluxetable}

\small\normalsize

The value of $\phi_{1}$ in Table~\ref{ch4J1028shape} is 0.01 greater than that in the first LAT pulsar catalog while the value of $\phi_{2}$ is the same \citep{AbdoPSRcat}.  Note that the pulsar catalog used 6 months of sky-survey data and a different energy-dependent ROI which may account for the discrepancy beyond statistical uncertainties.

\begin{figure}
\begin{center}
\includegraphics[width=1.0\textwidth]{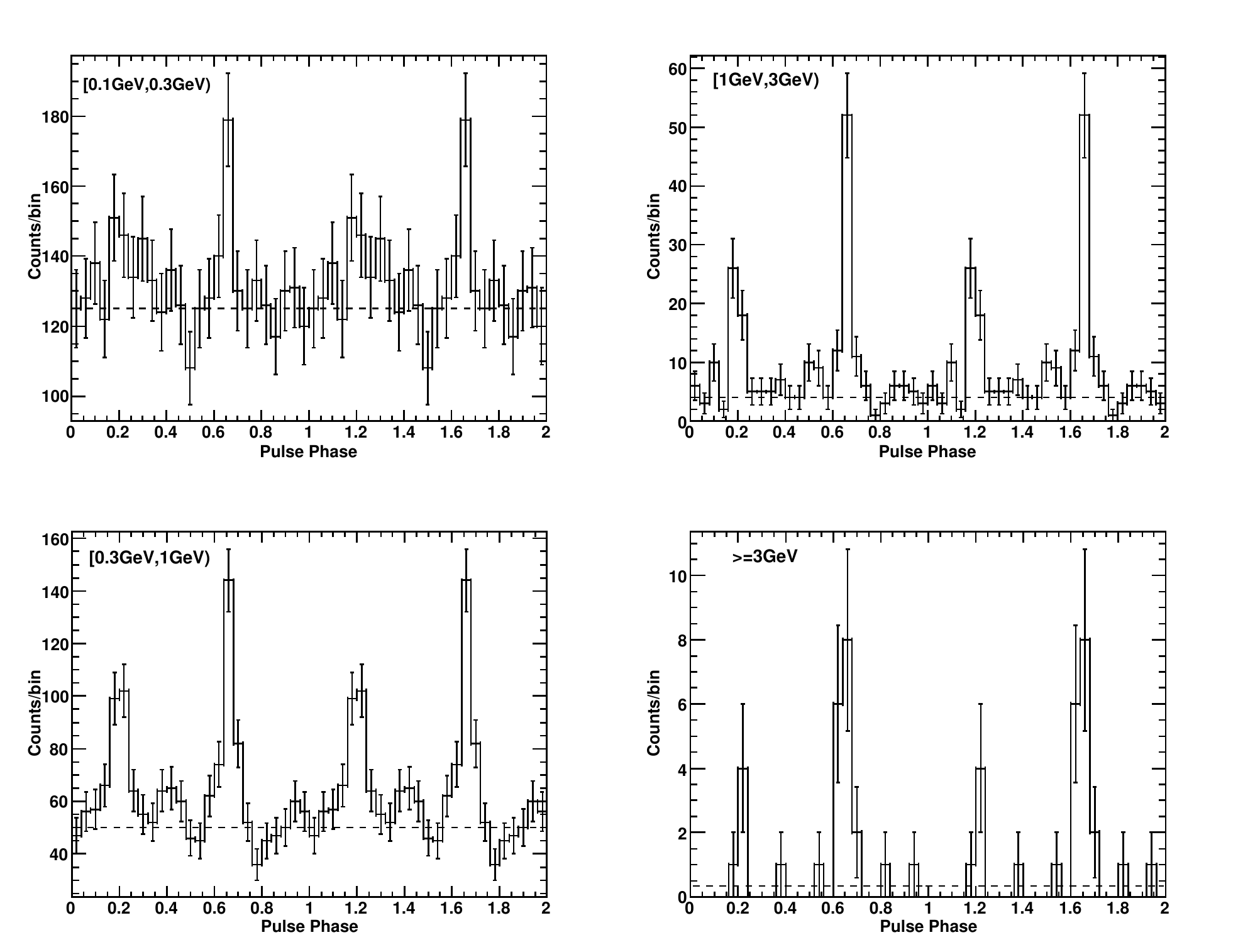}
\end{center}
\small\normalsize
\begin{quote}
\caption[Gamma-ray light curves of PSR J1028$-$5819 in different energy bands]{Light curves of PSR J1028$-$5819 in four energy bands (as labeled) in constant width bins of size 0.04 in phase, shown over two rotations.  Error bars are statistical.  The horizontal dashed lines indicate the estimated background level from the off-peak region.  Reproduced from \citet{AbdoJ1028}.\label{ch4J1028enelc}}
\end{quote}
\end{figure}

In order to explore the energy dependence of the gamma-ray light curve of PSR J1028$-$5819, the events were divided into four energy bins defined as shown in Fig.~\ref{ch4J1028enelc}.  The highest energy event found consistent with the pulsar position within the energy-dependent ROI described above was 13 GeV.

\small\normalsize

\begin{figure}
\begin{center}
\includegraphics[width=0.75\textwidth]{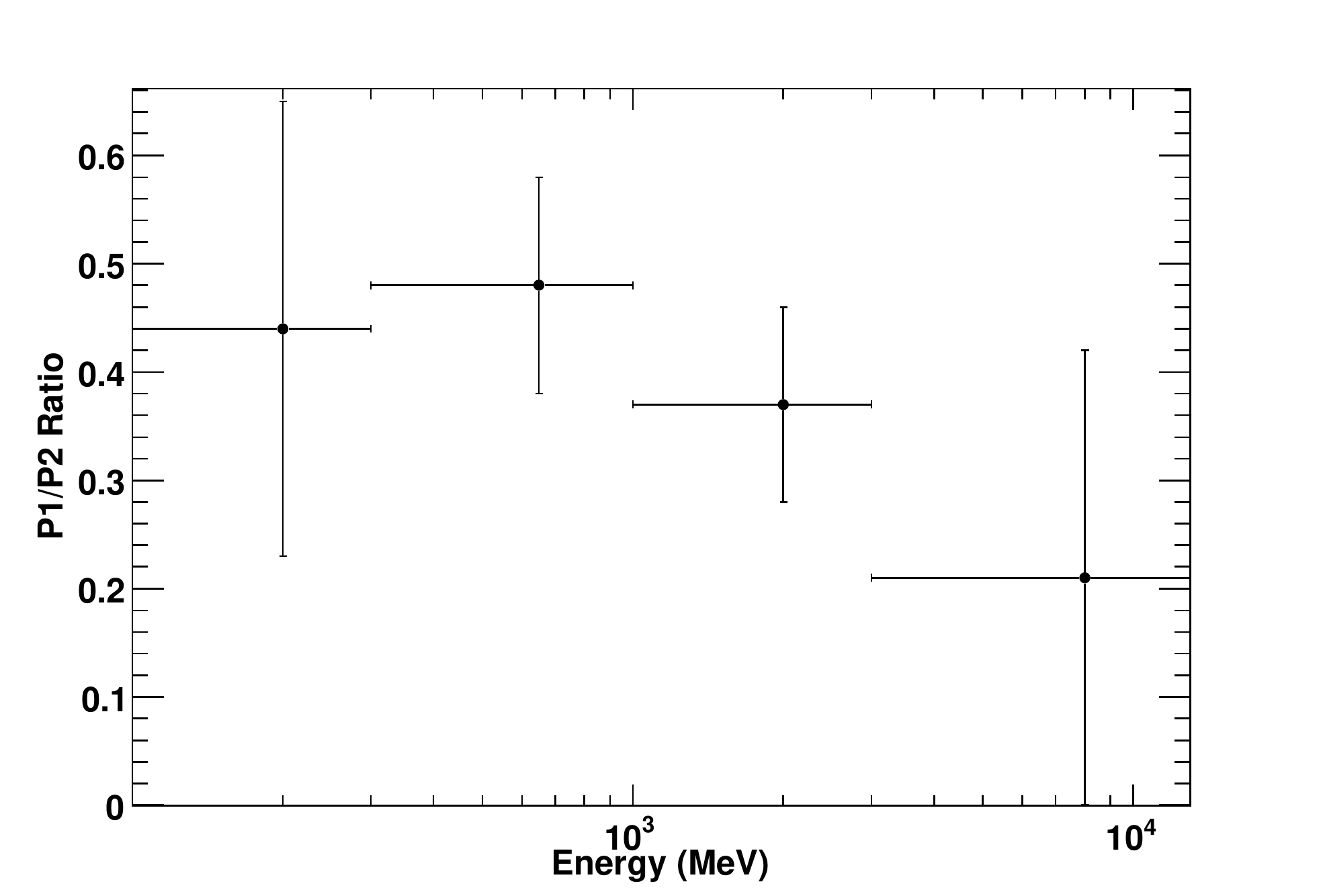}
\end{center}
\small\normalsize
\begin{quote}
\caption[Peak ratio vs. energy of PSR J1028$-$5819]{Ratio of background subtracted counts in the first and second gamma-ray peaks in the same energy bands as Fig.~\ref{ch4J1028enelc}.\label{ch4J1028pr}}
\end{quote}
\end{figure}

The widths of the first and second gamma-ray peaks show only 2.19$\sigma$ and 1.63$\sigma$ deviations from a constant value, respectively, demonstrating that there is no evidence for peak width evolution with energy.  The ratio of counts in the first peak to the second for each energy bin are shown in Fig.~\ref{ch4J1028pr}.  Given the narrowness of the peaks, with Lorentzian widths less than or comparable to the bin size, the counts in each peak were taken to be the counts in the bin corresponding to the fit peak position minus the background estimate from the off-peak region.

\small\normalsize

A fit to these ratios of a constant value yields a reduced $\chi^{2}$ value of 0.54 for 3 degrees of freedom, which implies no evolution of the peak ratio with energy, contrary to what is observed in the Vela pulsar (Abdo et al., 2009b and 2010f); however, the error bars in Fig.~\ref{ch4J1028pr} are quite large and more data may reveal a significant trend.

Using the energy-dependent ROI described above, the significance of the HE light curve of PSR J1028$-$5819 was evaluated for different low energy thresholds.  It was found that using only events $\geq$4 GeV yielded a pulsed detection of 3.5$\sigma$ while using only events $\geq$5 GeV yielded a pulsed detection of only 1.8$\sigma$, suggesting that there is evidence for pulsed emission from up to $\sim$4 GeV.  Eq. 1 of \citet{Baring04} can be inverted to give a lower bound on the gamma-ray emission altitude (Eq.~\ref{ch4lowBound}).  For the lower bound calculation $\epsilon_{max}$ is the maximum pulsed emission energy, $B_{12}$ is the derived pulsar magnetic field in units of $10^{12}$ G, P is the pulsar spin period, and R$_{\ast}$ is the radius of the neutron star.
\begin{equation}\label{ch4lowBound}
r \gtrsim (\frac{\epsilon_{max}\ B_{12}}{1.76\ \rm GeV})^{2/7}\ P^{-1/7}\ R_{\ast}
\end{equation}

Using $\epsilon_{max}$ = 4 GeV, $B_{12}$ = 1.23, and P $\approx$ 0.091 s yields a lower limit of $1.8\ \rm{R}_{\ast}$ for the gamma-ray emission altitude.  While this is not a particularly strong constraint it does preclude emission very near the stellar surface.

Gamma-ray spectral analysis of PSR J1028$-$5819 was done following the procedures described in Chapter 3.  The same selection criteria used to construct the gamma-ray light curve was used for spectral analysis except that all events with reconstructed directions within 15\DEG{} of the pulsar radio position were included.

All point sources, from a preliminary version of the BSL, found above the background with a TS $\geq$ 25  and within 15\DEG{} of the pulsar radio position were included in the model of the region.  The diffuse emission from the Milky Way was modeled using the GALPROP run designation 54\_59Xvarh7S \citep{Strong04}.  The extragalactic diffuse and residual instrument background were modeled as an isotropic power law.

\begin{figure}
\begin{center}
\includegraphics[width=0.75\textwidth]{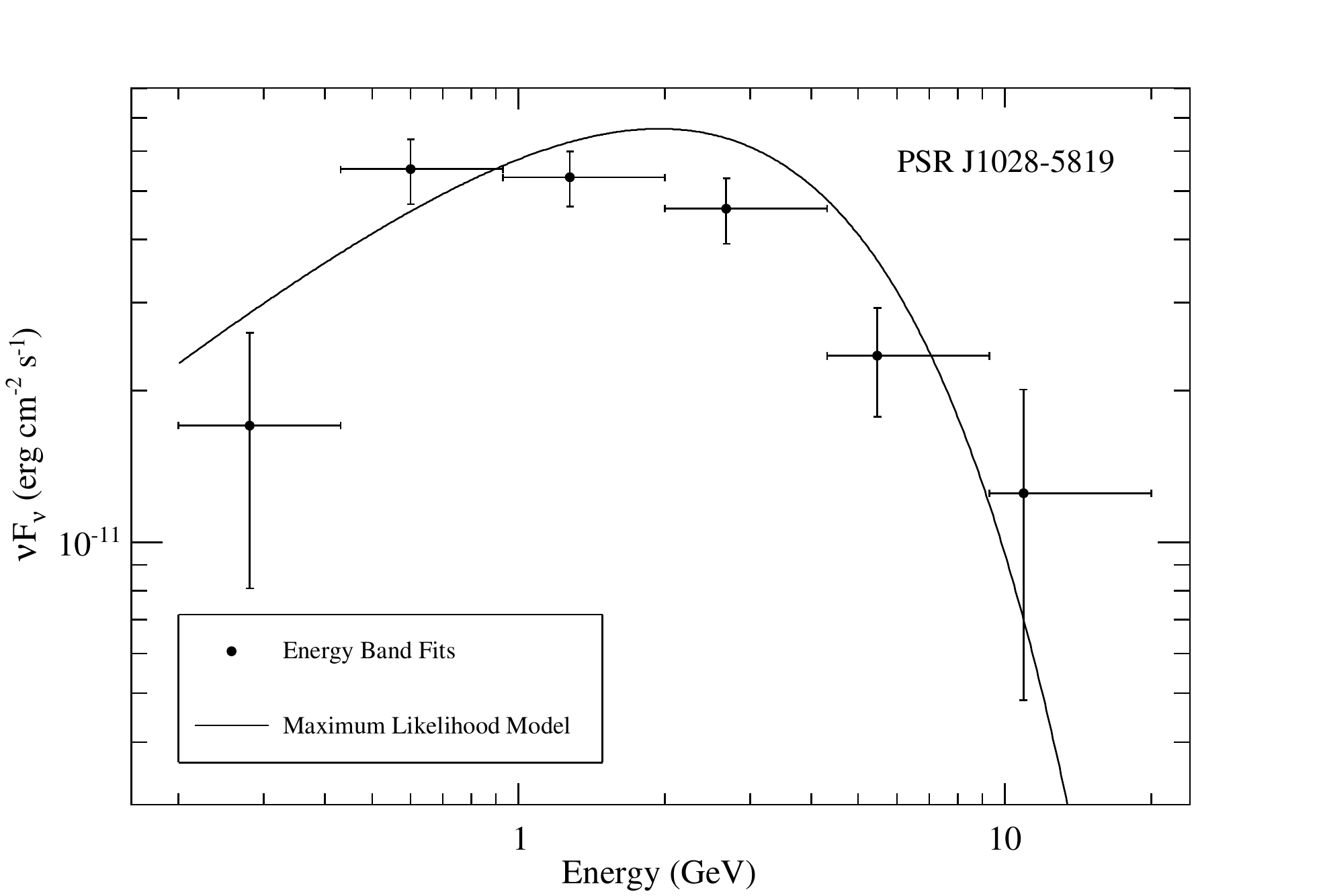}
\end{center}
\small\normalsize
\begin{quote}
\caption[Gamma-ray spectrum of PSR J1028$-$5819]{Gamma-ray spectrum of PSR J1028$-$5819, points derived from individual energy band fits as described in Chapter 3.  Error bars are statistical only.\label{ch4J1028spec}}
\end{quote}
\end{figure}

The gamma-ray spectrum of PSR J1028$-$5819 was modeled as an exponentially cutoff power law, Eq.~\ref{ch3ECO}, with b$\equiv$1.  The other point sources in the region of interest were modeled with power law spectra.  The normalization, photon index, and cutoff energy of PSR J1028$-$5819 were left free in the fit as well as the normalizations and indices of the diffuse sources and nearby point sources.  The best-fit spectral parameters are given in Table~\ref{ch4J1028spectrum}, the first uncertainties are statistical while the second are systematic.  The systematic uncertainties are from early photon-selection efficiency estimates derived by comparing event selections in the on- and off-pulse regions of the Vela pulsar light curve \citep{AbdoVela}.  The gamma-ray spectrum of PSR J1028$-$5819 is shown in Fig.~\ref{ch4J1028spec}.

\small\normalsize

\begin{deluxetable}{l r}
\tablewidth{0pt}
\tablecaption{PSR J1028$-$5819 Spectral Parameters}
\startdata
$\Gamma$ & 1.22$\pm$0.20$\pm$0.12\\
$E_{C}$ (GeV) & 2.5$\pm$0.6$\pm$0.5\\
Flux (0.1-30 GeV) ($10^{-7}$ cm$^{-2}$ s$^{-1}$) & 1.62$\pm$0.27$\pm$0.32\\
Energy Flux (0.1-30 GeV) ($10^{-10}$ erg cm$^{-2}$ s$^{-1}$) & 1.78$\pm$0.15$\pm$0.35\\
\enddata\label{ch4J1028spectrum}
\end{deluxetable}

\small\normalsize

\small\normalsize

The energy band fits in Fig.~\ref{ch4J1028spec} agree reasonably well with the best-fit model though some deviations are observed around 1 GeV.  This may be to due inaccuracies in the Galactic diffuse model used and the fact that the shape of the combined extragalactic diffuse and residual instrument backgrounds is not well described by a single power law.  Nevertheless, the spectral parameters in Table~\ref{ch4J1028spectrum} agree with those of \citet{AbdoPSRcat} within the quoted uncertainties and the noted differences between the P6\_V1 IRFs and the P6\_V3 IRFs \citep{Rando09}.

Using the gamma-ray energy flux  ($G$) and DM-derived distance ($d$) the total gamma-ray luminosity of PSR J1028$-$5819 can be calculated as $L_{\gamma}\ =\ 4\pi$f$_{\Omega}d^{2}G$ = 1.1$\times10^{35}$ f$_{\Omega}$ erg s$^{-1}$, where the pulsar moment of inertia is assumed to be $10^{45}$ g cm$^{2}$ and f$_{\Omega}$ is a beaming factor which depends on the viewing geometry typically of order 1 (see Chapter 5 for more details).  From this luminosity, the efficiency with which PSR J1028$-$5819 converts spin down energy into gamma rays can be calculated as $\eta_{\gamma}\ \equiv\ L_{\gamma}/\dot{E}$ = 0.13 f$_{\Omega}$.

To date, no geometrical constraints on the viewing geometry of PSR J1028$-$5819 have been published.  However, using the gamma-ray pulsar ``Atlas'' of \citet{Watters09} it is possible to constrain the geometry through the characteristics of the HE light curve alone.  To construct this ``Atlas'', a large, random population of pulsars was simulated using geometrical TPC, OG, and PC emission models (see Chapter 2 for more details) and characteristics of the resulting HE light curves were tabulated.  These characteristics were then binned in $\alpha$, $\zeta$, peak multiplicity, and gamma-ray efficiency for the different emission models.  \citet{Watters09} also accounted for HE pulsars which would not be visible in the radio waveband due to geometric constraints.

Given that $\Delta$ for PSR J1028$-$5819 is known and the maximum energy of pulsed HE emission suggests outer-magnetospheric emission, f$_{\Omega}$ can be estimated as $\sim$1 and the middle column of Fig. 5 of the ``Atlas'', which assumes a gamma-ray efficiency of 0.1 in good agreement with the estimate from LAT observations, can be used to constrain the geometry.  This suggests $\alpha\in[70^{\circ},90^{\circ}]$, $\zeta\in[75^{\circ},80^{\circ}$], and f$_{\Omega}\ \sim$ 1.1 for the OG model with $\alpha\in[65^{\circ},80^{\circ}]$, $\zeta\in[60^{\circ},80^{\circ}$], and f$_{\Omega}\ \sim$ 0.9-1.0 for the TPC model.  Comparing these constraints against future polarimetric radio observations and/or X-ray images would serve as a useful test of, and possible discriminator for, the different emission models.

\subsection{Redefining Gamma-ray Pulsar Science with Vela}\label{ch4Vela}
The Vela pulsar was discovered in radio observations by \citet{Large68}.  \citet{DJT74} first discovered a gamma-ray point source consistent with the radio location of the Vela pulsar with SAS-2.  Gamma-ray pulsations were soon detected from the Vela pulsar using a radio timing solution \citep{DJT75}.  In addition to being the second known gamma-ray pulsar (after the Crab), the Vela pulsar is also the brightest, non-variable point source in the HE gamma-ray sky and is, thus, the canonical first target of any HE observatory and testing ground for new pulsar analysis methods.

\subsubsection{Early Observations}\label{ch4VI}
A series of pointed observations of the Vela pulsar were carried out during the commissioning phase of the LAT \citep{AbdoOnorb}.  These observations were undertaken not only to verify and categorize the instrument response during pointed mode but also to increase the number of events from the Vela pulsar to promptly address the nature of its gamma-ray spectrum.  As discussed in Chapter 2, low- and high-altitude emission models predict a different shapes for the gamma-ray spectrum above a few GeV as a result of one-photon pair production; however, observations with the \emph{EGRET} instrument were unable to measure the spectrum with enough precision to differentiate between models with high significance \citep{Fierro98}.

LAT data from the commissioning period and the first 40 days of sky-survey data were used to fit the spectrum of the Vela pulsar \citep{AbdoVela}.  Events were required to belong to the ``Diffuse'' class, as defined under the P6\_V1 IRFs; have reconstructed energies $\geq$0.1 GeV; zenith angles $\leq105^{\circ}$; and reconstructed directions within 15\DEG{} of the Vela pulsar radio position.  The Galactic diffuse emission was modeled with the same GALPROP model used in Section~\ref{ch4J1028} while the extragalactic diffuse and residual instrument backgrounds were modeled as an isotropic power law.

The spectral parameters of the background sources were left free in the fits.  The spectrum of the Vela pulsar was modeled as an exponentially cutoff power law, Eq.~\ref{ch3ECO}, with b$\equiv$1 and b free in order to use the LRT (see Chapter 3) to statistically test for the presence of a super exponentially cutoff power law.  The best-fit spectral parameters are given in Table~\ref{ch4VISpec}, first uncertainties are statistical while second are systematic.

A super exponential cutoff power law model (assuming b = 2) was rejected at the 16.5$\sigma$ level, using the LRT described in  3, if only statistical uncertainties are considered.  When systematic uncertainties are taken into account their is still 0.29\% chance of incorrectly rejecting the b = 2 model.  This suggested that, while low-altitude emission was unlikely, it could not be conclusively ruled out simply due to large systematic uncertainties in the early observations.

\small\normalsize

\begin{deluxetable}{l r}
\tablewidth{0pt}
\tablecaption{Early Vela Spectral Parameters}
\startdata
$\Gamma$ & 1.1$\pm$0.01$\pm$0.07\\
$E_{C}$ (GeV) & 2.86$\pm$0.09$\pm$0.17\\
Energy Flux (0.1-10 GeV) ($10^{-9}$ erg cm$^{-2}$ s$^{-1}$) & 7.87$\pm$0.33$\pm$1.57\\
\enddata\label{ch4VISpec}
\end{deluxetable}

\small\normalsize

\begin{figure}
\begin{center}
\includegraphics[width=0.75\textwidth]{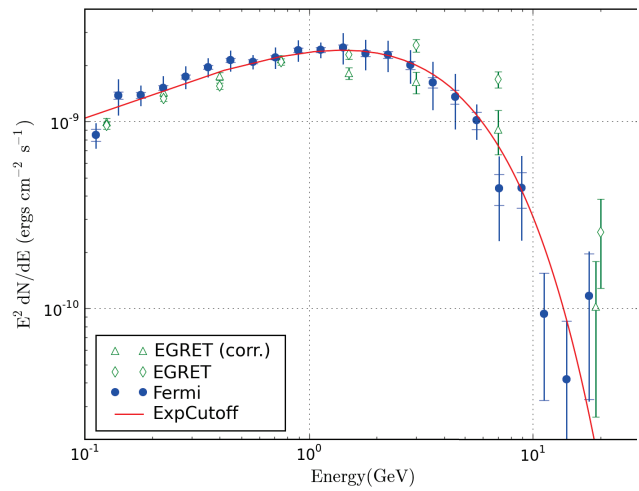}
\end{center}
\small\normalsize
\begin{quote}
\caption[Early LAT gamma-ray spectrum of the Vela pulsar]{Gamma-ray spectrum of the Vela pulsar using early observations described in the text.\label{ch4VelaIspec}}
\end{quote}
\end{figure}

The best-fit spectral model of the Vela pulsar (with b $\equiv$ 1) is shown in Fig.~\ref{ch4VelaIspec} with points derived from an independent, binned analysis tool know as \emph{ptlike}.  The \emph{EGRET} flux points of \citet{Kanbach94} are shown as well for comparison, this was the first indication that the so-called \emph{EGRET} GeV excess was most likely an instrumental effect \citep{AbdoDiffuse}.  ``Corrected'' \emph{EGRET} points, using the analysis of \citet{Stecker08}, are also plotted.

\small\normalsize

However, it should be noted that the best-fit value of b was, in fact, 0.88$\pm$0.04$^{+0.24}_{-0.52}$ which is not explained by gamma-ray emission models which assume CR but is consistent with a value of 1 when all uncertainties are considered.  These early LAT observations did not have sufficient statistics to allow for phase-resolved spectroscopy in fine enough phase bins to be meaningful.

\subsubsection{Vela After Eleven Months}\label{ch4VII}
Sky-survey mode allows the LAT to continuously monitor the entire gamma-ray sky.  Thus, the Vela pulsar is viewed several times every day.  This fact was exploited to create a LAT-only timing solution, phase-aligned with the radio profile from the Parkes Radio Telescope, with 63 $\mu$s residuals using the techniques described in \citet{Ray11}.

With 11 months of sky-survey data it was possible to divide the light curve of the Vela pulsar into 101 variable-width phase bins for phase-resolved spectral analysis \citep{AbdoVII}.  Events were selected from a $20^{\circ}\times20^{\circ}$ region centered on the radio position of the Vela pulsar.  Note that the square ROI was necessary to perform a binned maximum likelihood analysis.  The events were required to belong to the ``Diffuse'' class of events as defined under the P6\_V3 IRFs.  Further selections were made to accept only events with reconstructed energies between 0.1 and 100 GeV and zenith angles less than 105\DEG{}.

\begin{figure}
\begin{center}
\includegraphics[width=1.0\textwidth]{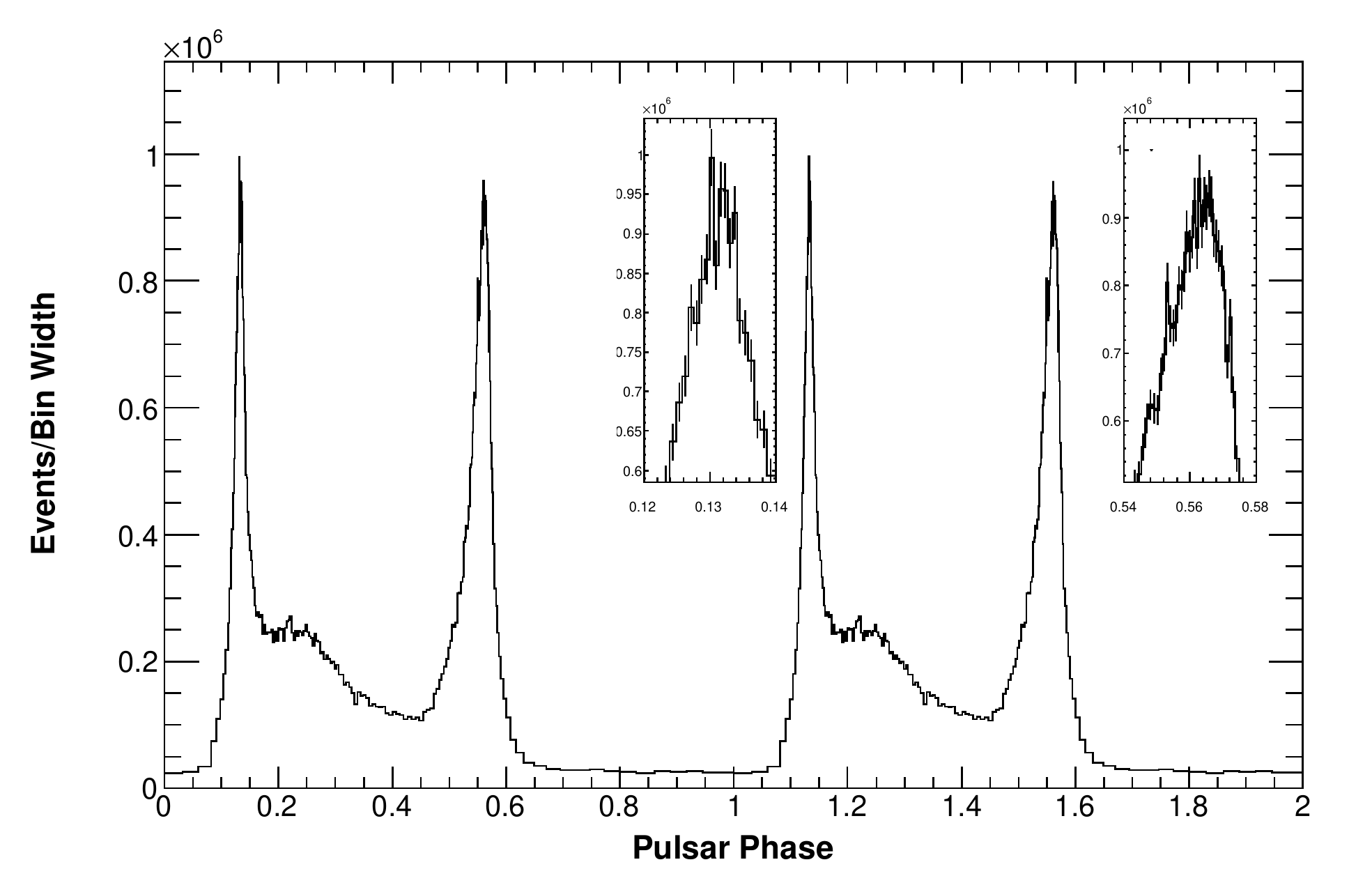}
\end{center}
\small\normalsize
\begin{quote}
\caption[Gamma-ray light curve of the Vela pulsar with 11 months of LAT data]{Light curve of the Vela pulsar using 11-months of sky-survey data in variable-width phase bins with 750 counts each using a gamma-ray derived timing solution, the radio peak is at phase 0.  Compare to the light curves in Fig.~\ref{ch3VC}.  The insets show zoom-ins of the main peaks to demonstrate the fine-scale structure.  Reproduced from \citet{AbdoVII}. \label{ch4VIIlc}}
\end{quote}
\end{figure}

The LAT science tool \emph{gtmktime} was used to exclude times when the rocking angle of the LAT exceeded 50\DEG{} and when the limb of the Earth intruded upon the ROI.  The 11-month light curve of the Vela pulsar is shown in Fig~\ref{ch4VIIlc} with zoom-ins of the main peaks to show the error bars and fine-scale structure.   To build this light curve events were required to have reconstructed directions within $max\lbrace 1.6-3\log_{10}(E/1 \rm GeV),1.3\rbrace$ (units of degrees) of the pulsar radio position and energies $\geq$ 0.03 GeV.

\small\normalsize

The phase bins for the light curve were constructed to each contain 750 events using the energy-dependent ROI described above, the content of each bin is inversely proportional to the width.  This ``fixed-count'' binning scheme serves to bring out fine-detail structure in pulsar light curves but note that it does require a relatively low background level and high statistics.

A binned maximum likelihood analysis was used to fit the phase-averaged spectrum of the Vela pulsar.  All point sources found above the background with a TS of $\geq$ 25 in a preliminary version of the first LAT source catalog (1FGL; Abdo et al., 2010h) within 15\DEG{} of the pulsar were included in the model of the region.  The Galactic diffuse gamma-ray emission was modeled using the \emph{gll\_iem\_v02.fits} mapcube while the isotropic diffuse and residual instrument backgrounds were modeled using the \emph{isotropic\_iem\_v02.txt} template.

The Vela X PWN was included in the model as an extended disc of radius 0\DEG{}.88, for a full analysis of the observed HE gamma-ray emission from this PWN, using the same data set, see \cite{AbdoVX}.  All point sources and the Vela X PWN were modeled with power law spectra while the point source at the position of the Vela pulsar was modeled with an exponentially cutoff power law, Eq.~\ref{ch3ECO}, with b $\equiv\ 1$ and b free.

Examination of the light curve suggests that the pulsar has a significant pulsed fraction with emission from near phase zero until almost 0.8.  Therefore, the off-pulse region was chosen to be $\phi\in(0.8,1.0]$ and spectral analysis was performed in this region, without the pulsar in the model, and with all source normalizations and indices free.

\begin{figure}
\begin{center}
\includegraphics[width=1.0\textwidth]{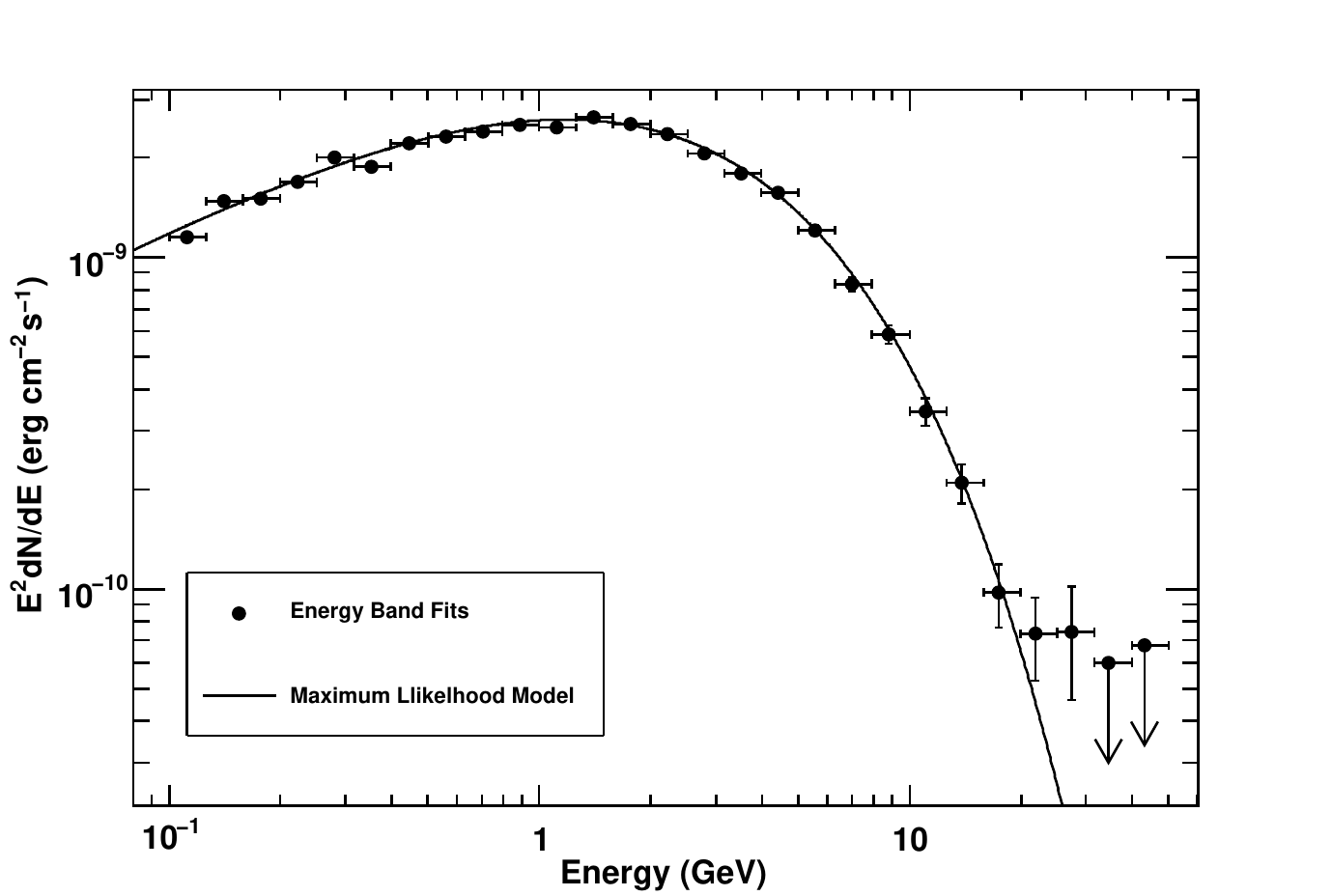}
\end{center}
\small\normalsize
\begin{quote}
\caption[Gamma-ray spectrum of the Vela pulsar with 11 months of LAT data]{Phase-averaged gamma-ray spectrum of the Vela pulsar with 11 months of sky-survey data.  Solid line corresponds to the best-fit spectral parameters given in Table~\ref{ch4VIIvals}, individual flux points are derived as described in Chapter 3.  Reproduced from \citet{AbdoVII}.\label{ch4VIIspec}}
\end{quote}
\end{figure}

The off-pulse results were used as a starting point for a phase-averaged analysis (using $\phi\in[0,1.0]$) in which the parameters of all point sources more than 5\DEG{} from the pulsar and the index of the Vela X PWN were held fixed.  Using the likelihood ratio test, a fit with b free is preferred at the 11$\sigma$ level.  As was found by \citet{AbdoVela}, the best fit value of b was less than 1, see Table~\ref{ch4VIIvals} and Fig.~\ref{ch4VIIspec}.  This is likely due to the superposition of many spectral components with b $\equiv$ 1 and varying values of $E_{C}$ and $\Gamma$ through the pulse.

\small\normalsize

\small\normalsize

\begin{deluxetable}{l r}
\tablewidth{0pt}
\tablecaption{Vela Phase-Averaged Spectral Parameters with Eleven Months}
\startdata
$\Gamma$ & $1.37\pm0.03^{+0.07}_{-0.03}$\\
$E_{C}$ (GeV) & $1.31\pm0.18^{+1.0}_{-0.5}$\\
b & $0.68\pm0.03^{+0.18}_{-0.10}$\\
Flux (0.1-100 GeV) ($10^{-5}\ \rm{cm}^{-2}\ \rm{s}^{-1}$) & $1.07\pm0.01\pm0.03$\\
Energy Flux (0.1-100 GeV) ($10^{-9}$ erg cm${-2}$ s$^{-1}$) & 8.86$\pm$0.05$\pm$0.18\\
\enddata\label{ch4VIIvals}
\end{deluxetable}

\small\normalsize

In order to verify this hypothesis, the LAT ST \emph{gtobssim} and the simulation tool \emph{PulsarSpectrum} \citep{Razzano09} were used to simulate a pulsar source with the period and light curve of the Vela pulsar and varying spectra through the pulse.  The simulation included all other sources in the region model used for spectral analysis and used the real spacecraft file.

A binned maximum likelihood analysis of the simulated data resulted in a best-fit spectrum with b less than 1 as well.  A similar simulation with no spectral variation across the pulse was created and spectral analysis of that simulation did not show preference for a value of b different than 1.  The occurrence of b $<$ 1 spectra in LAT pulsars and a detailed study of the phase-dependent simulation can be found in \citet{Celik11}.

Phase-resolved spectral analysis of the Vela pulsar was carried out using phase bins containing 1500 events each, using the energy-dependent ROI described previously, resulting in 101 phase bins.  For spectral analysis in each phase bin, the photon indices of all other sources in the ROI were kept fixed as well as the normalizations of point sources greater than 5\DEG{} from the pulsar.

\begin{figure}
\begin{center}
\includegraphics[width=1.0\textwidth]{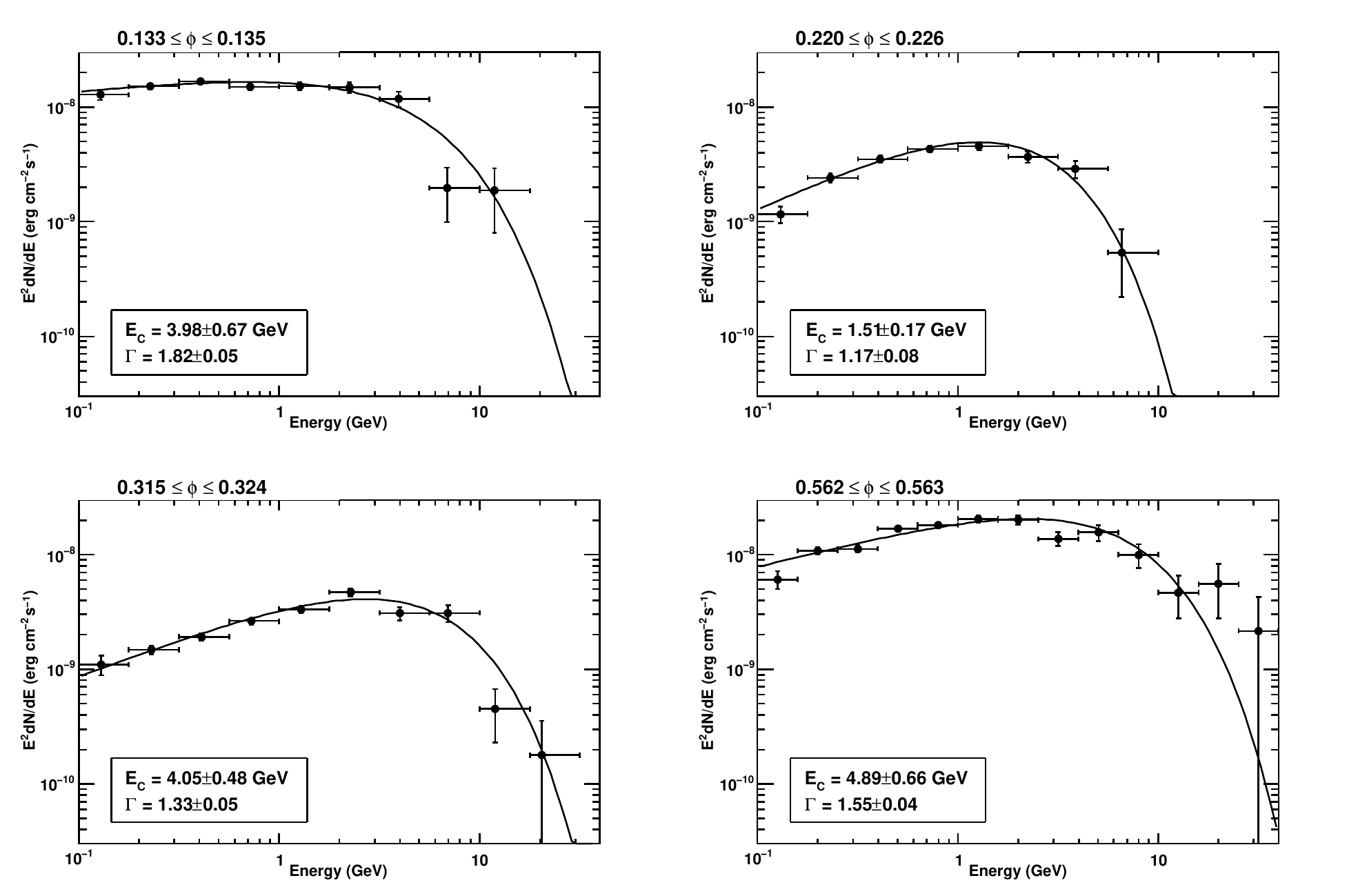}
\end{center}
\small\normalsize
\begin{quote}
\caption[Gamma-ray spectrum of the Vela pulsar in representative phase bins]{Gamma-ray spectrum of the Vela pulsar in the indicated phase bins with b $\equiv$ 1 and best-fit cutoff energies and photon indices as indicated.  Reproduced from \citet{AbdoVII}.\label{ch4VIIspecs}}
\end{quote}
\end{figure}

The pulsar was modeled with a simple exponentially cutoff power law in each phase bin.  The analysis was also performed with b $\equiv$ 2 and b free in order to asses how well the simple cutoff described the data in each bin.  On average, the b free fits were preferred over the b $\equiv$ 2 fits at the 3$\sigma$ level.  For the simple exponentially cutoff models, the b free fits were only preferred at the 1.5$\sigma$ level, on average, indicating that b $\equiv$ 1 is sufficient across the pulse phase as can be seen in Fig.~\ref{ch4VIIspecs} which shows the gamma-ray spectra of the Vela pulsar in four representative phase bins.

\begin{figure}
\begin{center}
\includegraphics[width=1.0\textwidth]{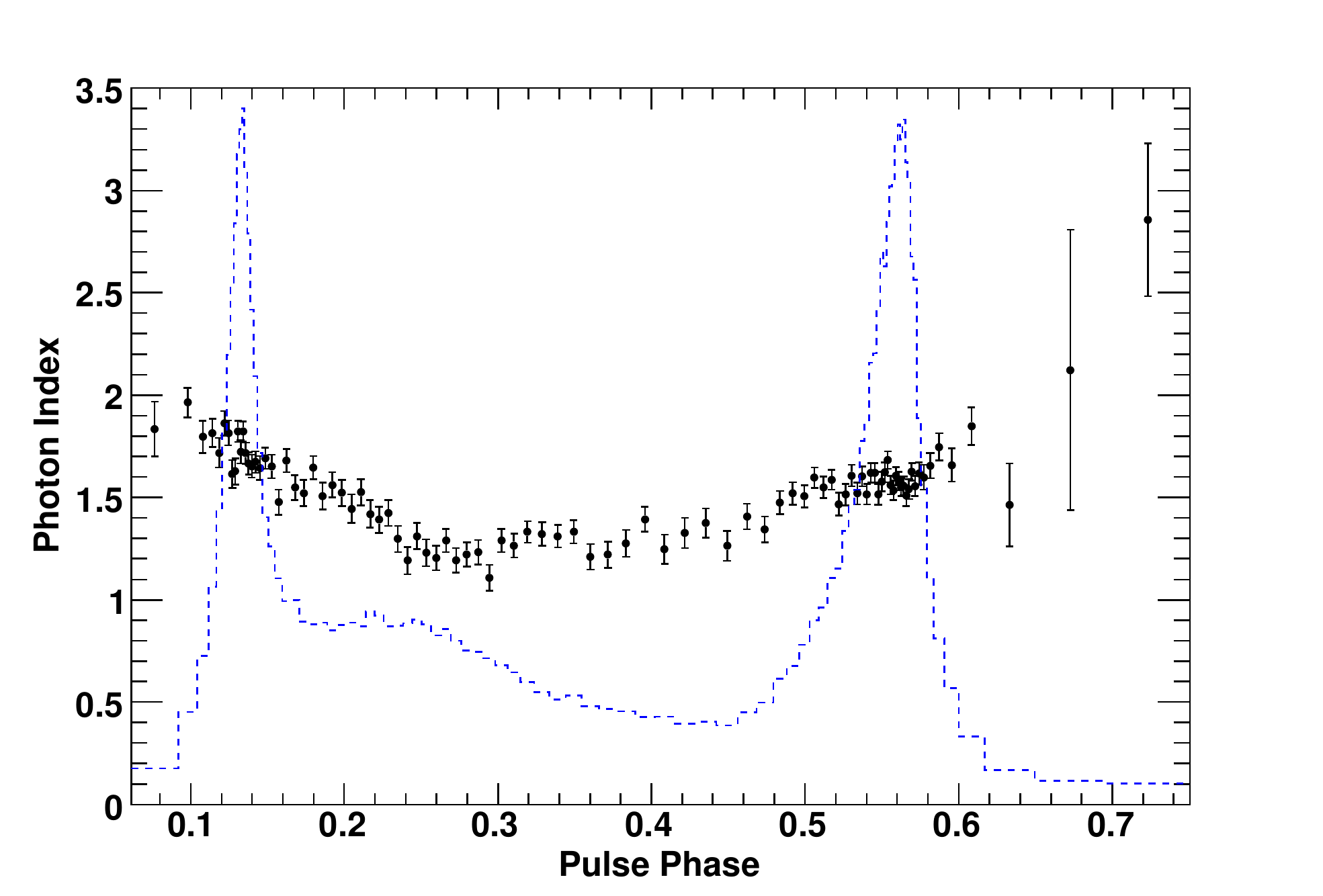}
\end{center}
\small\normalsize
\begin{quote}
\caption[Photon index vs. phase for the Vela pulsar]{Best-fit photon index ($\Gamma$) for the phase-resolved analysis, errors are statistical only, results are only shown for bins in which the pulsar was found above the background with a TS $\geq$25.  Reproduced from \citet{AbdoVII}.\label{ch4VIIgamph}}
\end{quote}
\end{figure}

\begin{figure}
\begin{center}
\includegraphics[width=1.0\textwidth]{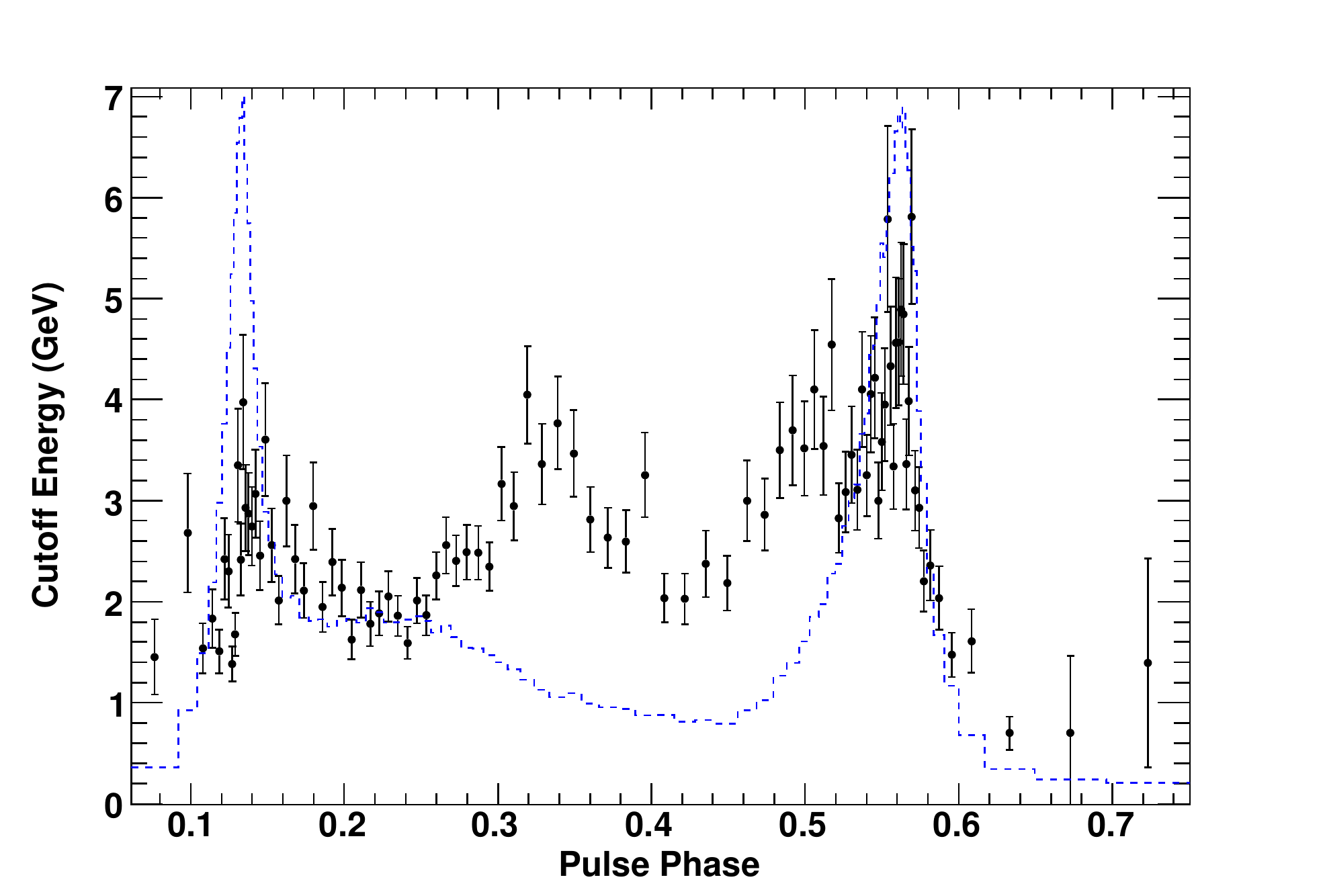}
\end{center}
\small\normalsize
\begin{quote}
\caption[Cutoff energy vs. phase for the Vela pulsar]{Best-fit cutoff energy (E$_{\rm C}$) for the phase-resolved analysis, errors are statistical only, results are only shown for bins in which the pulsar was found above the background with a TS $\geq$25.  Reproduced from \citet{AbdoVII}.\label{ch4VIIecph}}
\end{quote}
\end{figure}

Significant variation is observed in both the cutoff energy and photon index across the phase as can be seen in Figs.~\ref{ch4VIIgamph} and~\ref{ch4VIIecph}.  To evaluate the significance of the observed variations in $E_{C}$ and $\Gamma$ with phase, the analysis was repeated on the b $\equiv$ 1 simulation described above.  The simulation results suggested that point-to-point variations of 0.6 GeV in $E_{C}$ and 0.05 in $\Gamma$ should be expected from the fitting technique alone.  As such, point-to-point variations less than these values can not be considered significant; however, while random fluctuations around the phase-averaged values were observed in analysis of the simulation the systematic trends observed in the data were not reproduced.

\small\normalsize

\small\normalsize

\small\normalsize

\begin{figure}
\begin{center}
\includegraphics[width=0.75\textwidth]{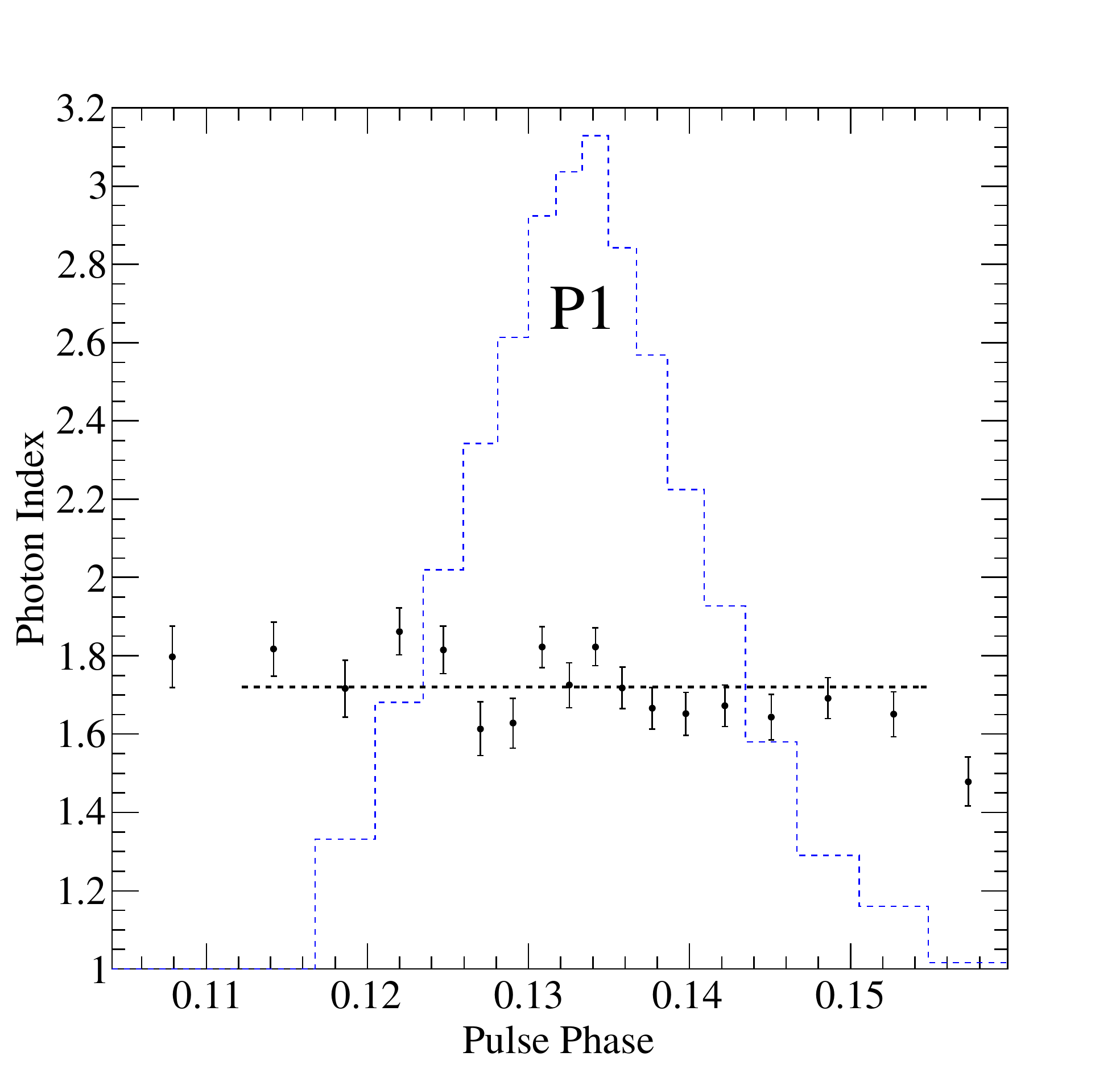}
\end{center}
\small\normalsize
\begin{quote}
\caption[Photon index vs. phase through the first peak of the Vela pulsar]{Photon index with phase for the first peak, dashed line represents the best-fit to these points of $1.71\pm0.01$.\label{ch4Gamp1}}
\end{quote}
\end{figure}

\begin{figure}
\begin{center}
\includegraphics[width=0.75\textwidth]{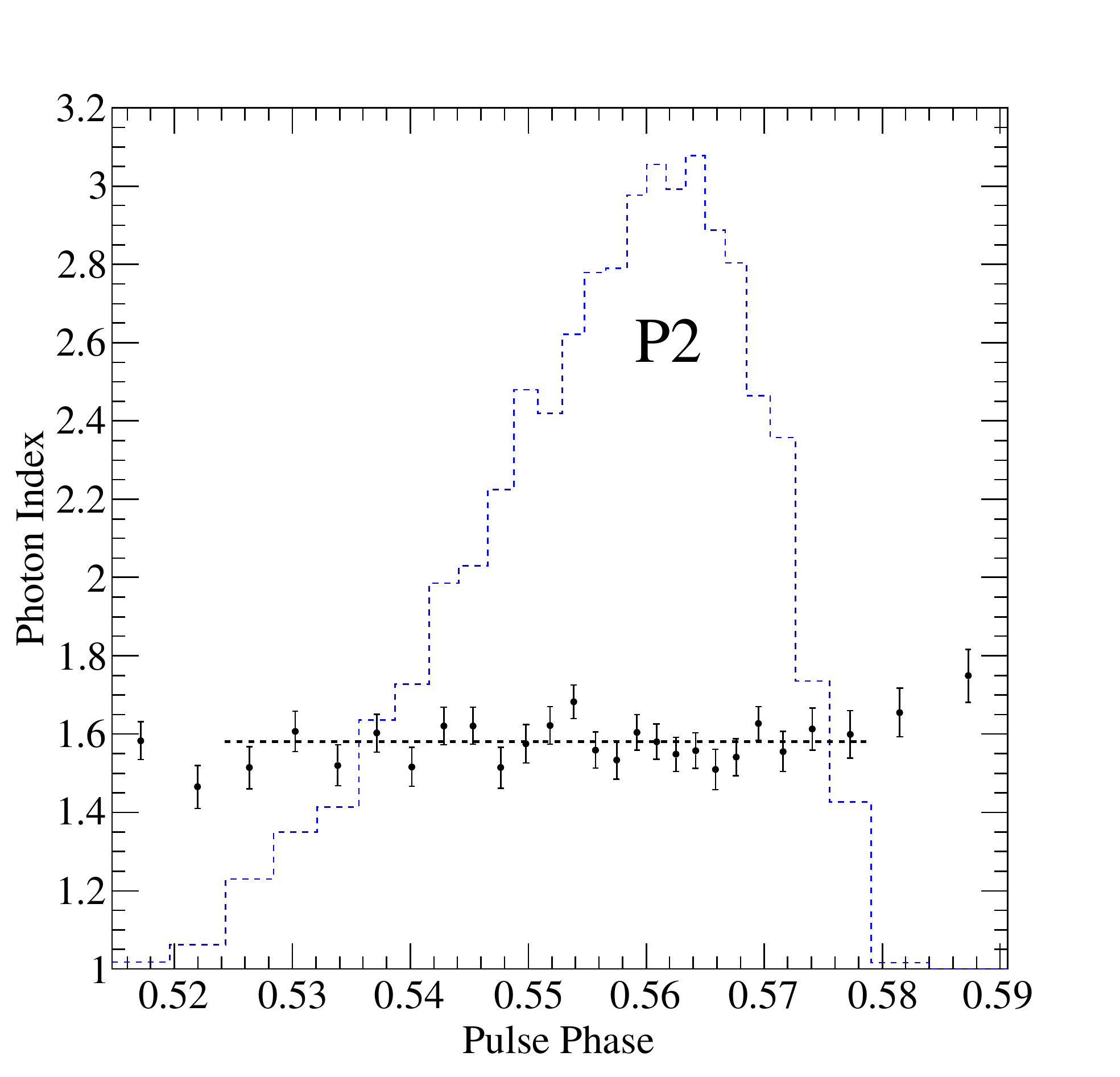}
\end{center}
\small\normalsize
\begin{quote}
\caption[Photon index vs. phase through the second peak of the Vela pulsar]{Photon index with phase for the second peak, dashed line represents the best-fit to these points of $1.58\pm0.01$.\label{ch4Gamp2}}
\end{quote}
\end{figure}

These results confirm that the hardest emission is observed between the two main peaks as suggested by analysis of \emph{EGRET} data \citep{Fierro98}; however, their observation that the photon index changes rapidly through the peaks is not confirmed.  In fact, $\Gamma$ is found to be consistent with constant values of $1.72\pm0.01$ and $1.58\pm0.01$ for the first and second peaks, respectively (see Figs.~\ref{ch4Gamp1} and~\ref{ch4Gamp2}).  This discrepancy is likely due to a lack of statistics above a few GeV in the \emph{EGRET} data which precluded the use of an exponentially cutoff spectral model and thus put all spectral changes into the photon index.

\small\normalsize

\small\normalsize

\begin{figure}
\begin{center}
\includegraphics[width=0.75\textwidth]{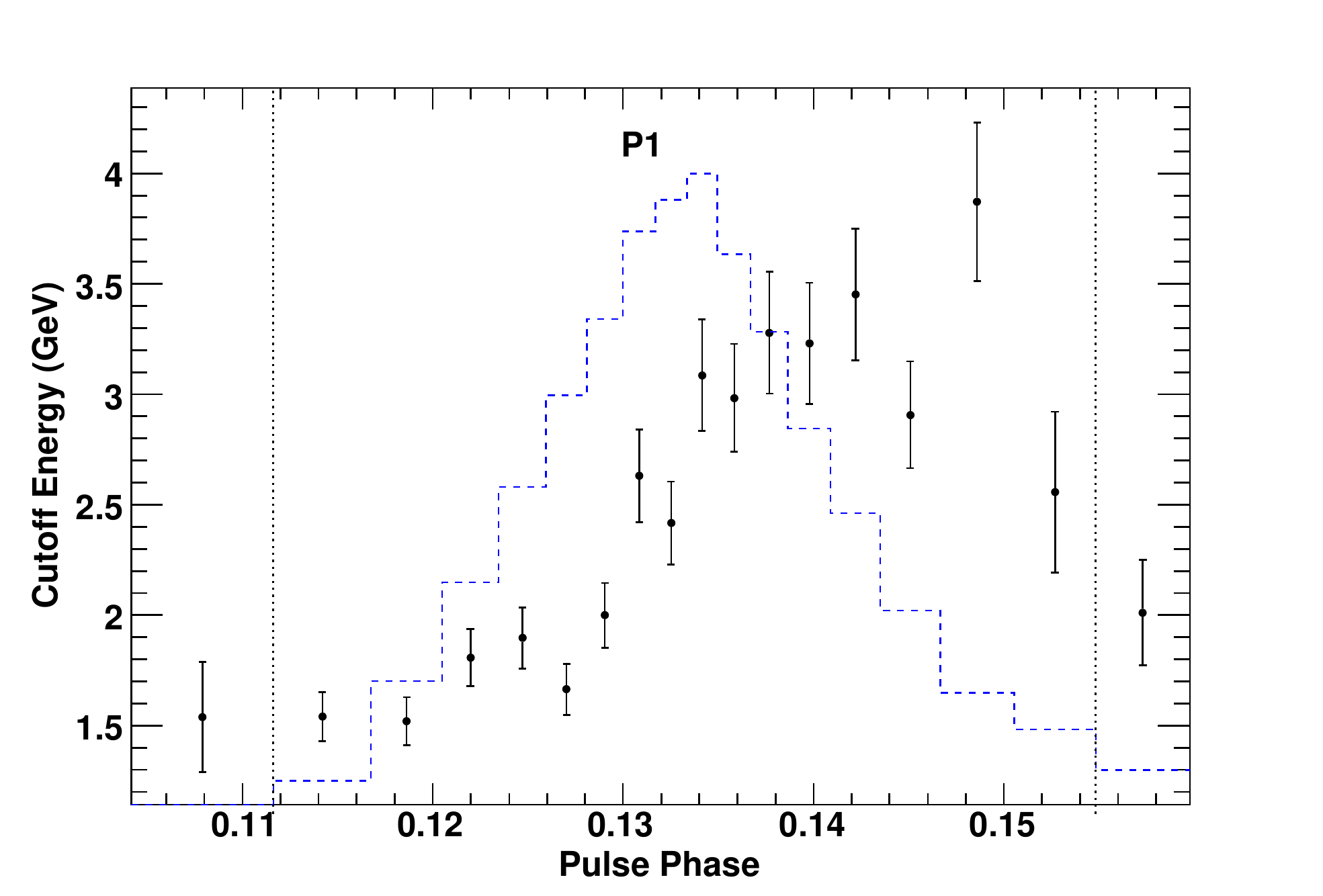}
\end{center}
\small\normalsize
\begin{quote}
\caption[Cutoff energy vs. phase for the first peak of the Vela pulsar]{Cutoff energy with phase for the first peak with $\Gamma$ held fixed to $1.71\pm0.01$ for the phase range defined by the vertical dashed lines.  Reproduced from \citet{AbdoVII}.\label{ch4VIIecp1}}
\end{quote}
\end{figure}

\begin{figure}
\begin{center}
\includegraphics[width=0.75\textwidth]{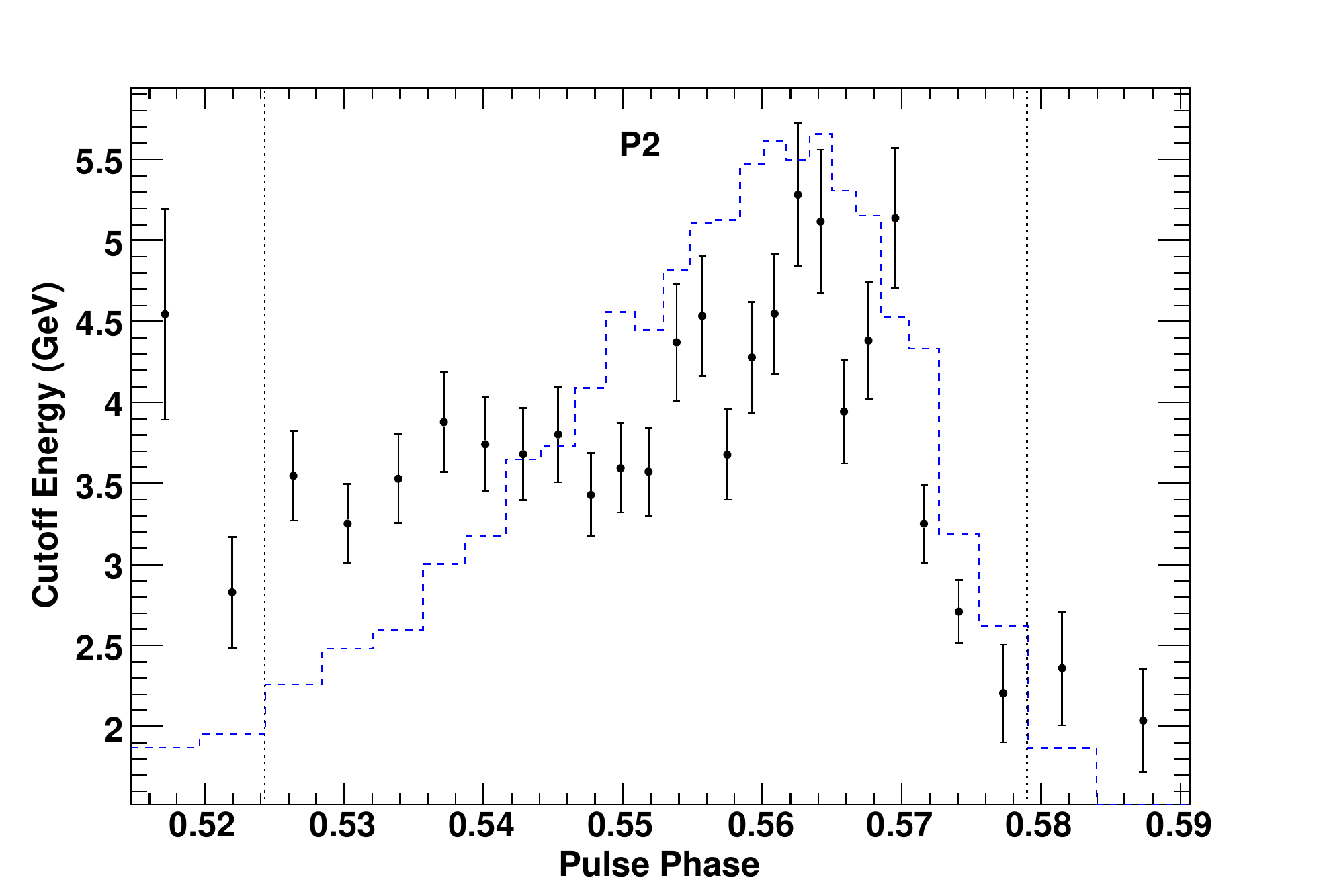}
\end{center}
\small\normalsize
\begin{quote}
\caption[Cutoff energy vs. phase for the second peak of the Vela pulsar]{Cutoff energy with phase for the second peak with $\Gamma$ held fixed to $1.58\pm0.01$ for the phase range defined by the vertical dashed lines.  Reproduced from \citet{AbdoVII}.\label{ch4VIIecp2}}
\end{quote}
\end{figure}

The cutoff energy rises sharply through the main peaks and, surprisingly, between the peaks as well, following the change in position of the third peak with energy (Abdo et al., 2009b and 2010f).  To better evaluate the behavior of the cutoff energy through the peaks the analysis was repeated with the photon index of the pulsar fixed to the constant values found above for $\phi\in[0.112,0.155]$ and $\phi\in[0.524,0.579]$, corresponding to the first and second peaks, respectively.  The resulting trends in cutoff energy are shown in Figs.~\ref{ch4VIIecp1} and~\ref{ch4VIIecp2}.  The cutoff energy is observed to rise smoothly through both peaks with maxima near the start of the trailing edges.

\small\normalsize

\small\normalsize

\citet{AbdoVII} noted that the models give similar CR cutoff energies ($E_{\rm CR}$), Eq.~\ref{ch4Ecr} (units of $mc^{2}$, $\lambda_{c}$ is the electron Compton wavelength), ranging from 1 to 5 GeV, consistent with what is observed in Vela and other gamma-ray pulsars \citep{AbdoPSRcat}.  Note that in outer-magnetospheric emission models the accelerating field $E_{\parallel}$ depends on the magnetic field at the light cylinder and the width of the accelerating gap.
\begin{equation}\label{ch4Ecr}
E_{\rm CR}\ \mathnormal =\ 0.32\lambda_{c}\Big(\frac{E_{\parallel}}{e}\Big)^{\frac{3}{4}}\rho^{\frac{1}{2}}_{c}
\end{equation}
 
It was noted that the cutoff energy depends on the local field line radius of curvature ($\rho_{c}$).  If emission across the pulse originates from regions with different ranges of emission radii then the phase-resolved spectroscopy should map out the emission altitude.  Large variations of $\rho_{c}$, and thus $E_{\rm CR}$, with phase are expected in the models and mapping the minimum $\rho_{c}$, using the geometric models, can produce trends similar to what is seen in Fig.~\ref{ch4VIIecph}.  However, full radiation models will be needed to match all of the observed features (e.g., Du et al., 2011).

\section{Millisecond Pulsars and Gamma rays}\label{ch4gMSPs}
Prior to the launch of \Fermi{} there were no firm detections of pulsed gamma rays from any MSPs.  As discussed in Chapter 2, there were two claims of gamma-ray pulsations from MSPs prior to the launch of \Fermi{}.  However, both detections had only marginal significance and were complicated by other factors.

\subsection{The First LAT Gamma-ray Millisecond Pulsar}\label{ch4J0030}
Shortly after launch pulsations from many new, non-recycled gamma-ray pulsars were detected and it was becoming clear that outer-magnetospheric emission models were favored.  However, it was unclear if MSPs could produce gamma rays via the same models (see Chapter 2 for more details) so it was not a foregone conclusion that MSPs were bright gamma-ray emitters as well.

\begin{figure}
\begin{center}
\includegraphics[width=0.6\textwidth]{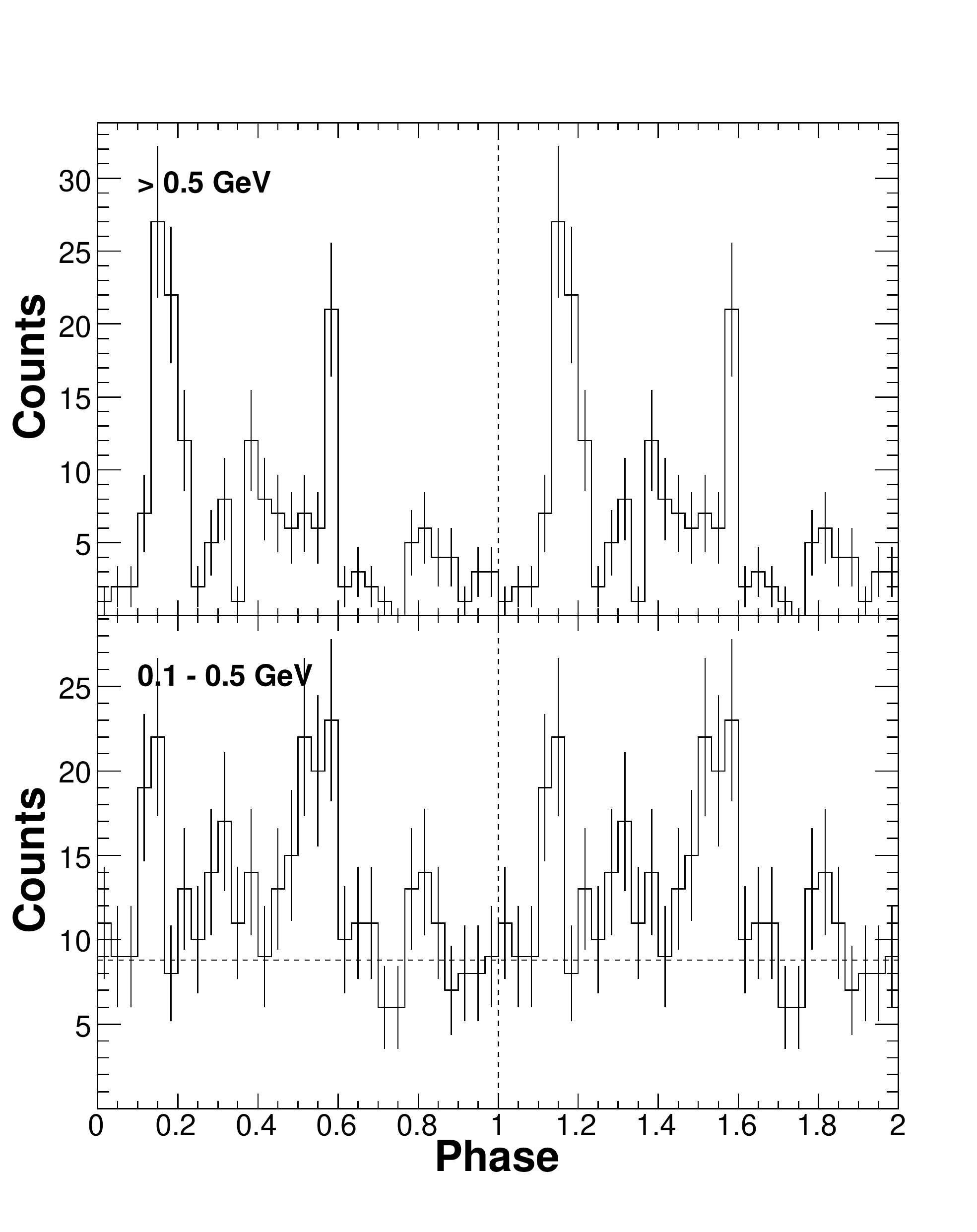}
\end{center}
\small\normalsize
\begin{quote}
\caption[Early gamma-ray light curve of PSR J0030+0451]{Gamma-ray light curve of PSR J0030+0451 in two energy bands, as indicated, sing the first three months of LAT sky-survey data.  Reproduced from \citet{AbdoJ0030}.\label{ch4J0030lc}}
\end{quote}
\end{figure}

With just 1 month of LAT sky-survey data a significant pulsed signal could be seen from the MSP J0030+0451, demonstrating that MSPs could be bright sources of gamma rays.  \citet{AbdoJ0030} presented the first analysis of this pulsar using the 3 months of data.  The HE light curve of PSR J0030+0451, in two energy bands, is shown in Fig.~\ref{ch4J0030lc}.

\small\normalsize

The HE light curve of PSR J0030+0451 is very reminiscent of those observed from known young gamma-ray pulsars.  The first and second peaks were found to be very narrow with Lorentzian FWHM values of 0.07$\pm$0.01 and 0.08$\pm$0.02, respectively.  Such small peak widths are expected of outer-magnetospheric models with narrow accelerating gaps due to copious screening of the accelerating field which was not expected to occur in MSPs.

\begin{figure}
\begin{center}
\includegraphics[width=0.75\textwidth]{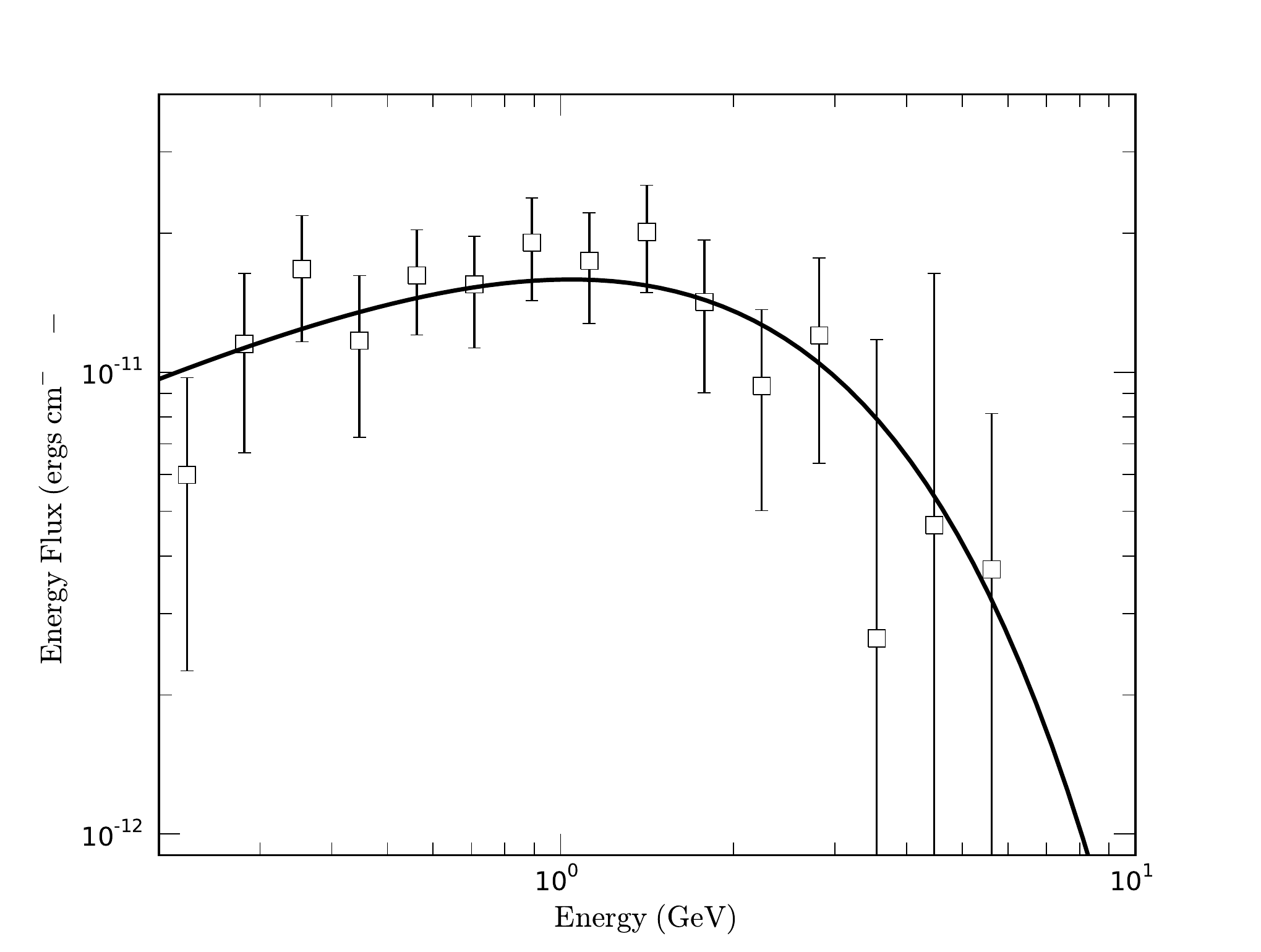}
\end{center}
\small\normalsize
\begin{quote}
\caption[Early gamma-ray spectrum of PSR J0030+0451]{Gamma-ray spectrum of PSR J0030+0451, solid line is the \emph{gtlike} maximum likelihood model while the open squares are flux points derived with \emph{ptlike}.  Reproduced from \citet{AbdoJ0030}.\label{ch4J0030spec}}
\end{quote}
\end{figure}

\citet{AbdoJ0030} selected events with reconstructed energies $\geq$ 0.2 GeV and from a 15\DEG{} radius ROI, centered on the pulsar radio position, for spectral analysis. The best-fit spectral parameters of PSR J0030+0451, given in Table~\ref{ch4J0030vals}, are fairly typical.  Fig.~\ref{ch4J0030spec} shows the best-fit gamma-ray spectrum derived using the LAT science tool \emph{gtlike} with flux points derived using the independent tool \emph{ptlike}.

\small\normalsize

The large amount of screening implied by the sharp gamma-ray peaks of J0030+0451 may suggest that higher-order, magnetic multipoles are more important in MSPs; the typical MSP mass may be closer to $\sim$2 M$_{\odot}$, in agreement with the recycled pulsar model; or that some other processes, such as offset dipoles \citep{HM11}, are in play.

\small\normalsize

\begin{deluxetable}{l r}
\tablewidth{0pt}
\tablecaption{PSR J0030+0451 Spectral Parameters}
\startdata
$\Gamma$ & $1.4\pm0.2\pm0.2$\\
E$_{C}$ (GeV) & $1.7\pm0.4\pm0.5$\\
b & 1 (fixed)\\
Flux (0.1-100 GeV) ($10^{-8}\ \rm{cm}^{-2}\ \rm{s}^{-1}$) & $6.76\pm1.05\pm1.35$\\
Energy Flux (0.1-100 GeV) ($10^{-11}$ erg cm${-2}$ s$^{-1}$) & $4.91\pm0.45\pm0.98$\\
\enddata\label{ch4J0030vals}
\end{deluxetable}

\small\normalsize

\subsection{A Population of Gamma-ray Millisecond Pulsars}\label{ch4MSPpop}
\begin{figure}
\begin{center}
\includegraphics[width=1.\textwidth]{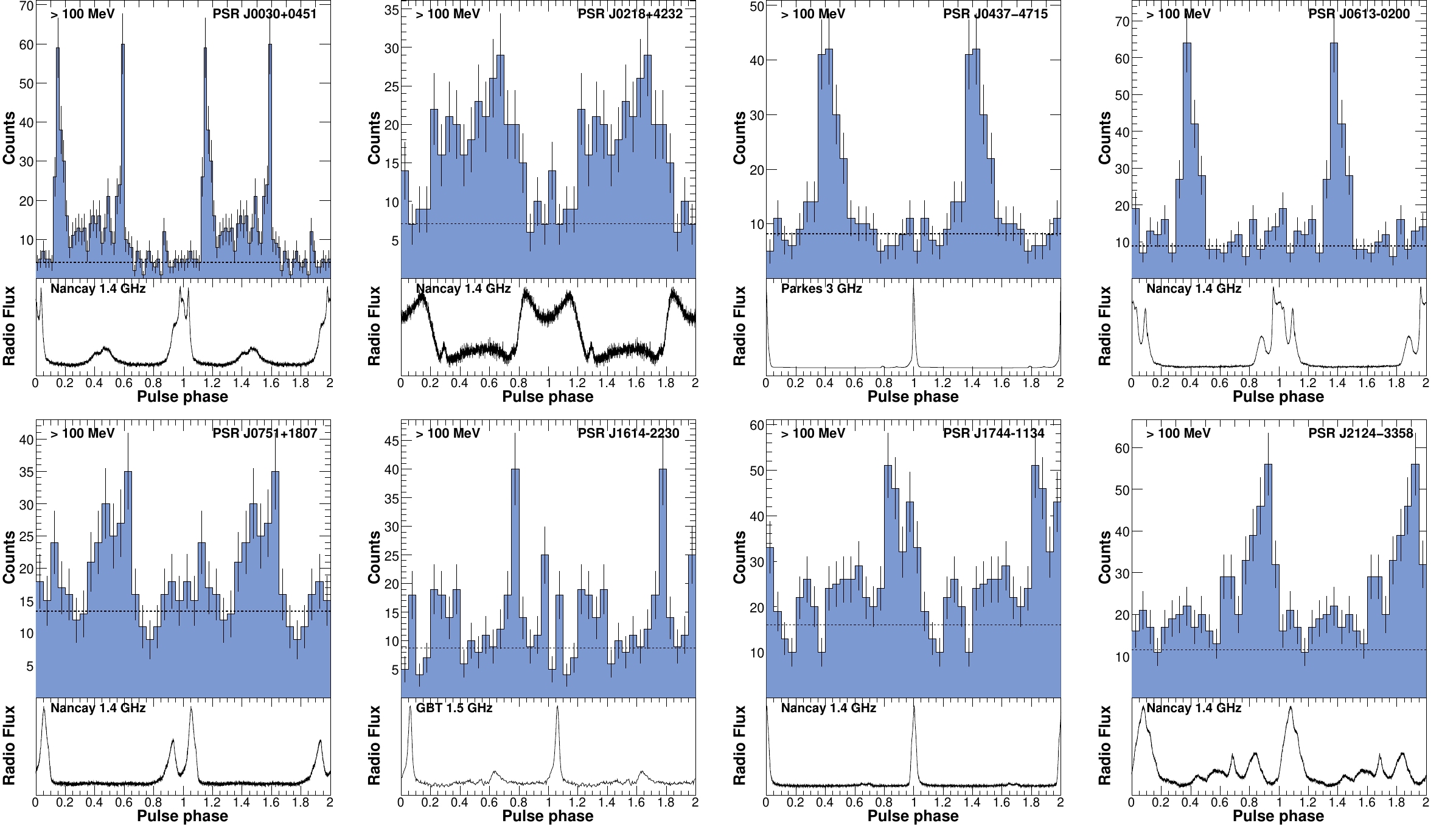}
\end{center}
\small\normalsize
\begin{quote}
\caption[Gamma-ray and radio light curves of the first 8 gamma-ray MSPs detected with the LAT]{Gamma-ray ($\geq$0.1 GeV) and radio light curves of the first 8 LAT detected MSPs, ordered in increasing right ascension from top left to bottom right.  Modified from \citet{AbdoMSPpop}.\label{ch4MSPlcs}}
\end{quote}
\end{figure}

PSR J0030+0451 was the first MSP to be detected with the LAT largely due to the fact that it is located far away from the plane of the Galaxy (Galactic latitude of $-57^{\circ}.6$) in a region of low background.  As more data accumulated in sky-survey mode, significant pulsed gamma-ray signals were seen from more MSPs.  In fact, with $\sim$8.5 months of LAT data significant gamma-ray pulsations were detected from 8 MSPs \citep{AbdoMSPpop}, including PSR J0030+0451 and confirming the marginal detection of PSR J0218+4232 in \emph{EGRET} data by \citet{Kuiper00}.  The $\geq$ 0.1 GeV and radio light curves of these MSPs are shown in Fig.~\ref{ch4MSPlcs}.

\small\normalsize

\citet{AbdoMSPpop} found that the measured spectral parameters of these 8 MSPs were very similar to non-recycled, gamma-ray pulsars.  The cutoff energy of PSR J0218+4232 is somewhat higher than most pulsars but the spectral analysis is complicated by the proximity of a bright, gamma-ray blazar and the uncertainty is large.

Of the first 8 LAT detected MSPs, 7 have sharp, single or double-peaked HE light curves while the remaining MSP (PSR J0218+4232) has (possibly) two, closely-spaced peaks.  These results confirm the implications from the initial detection of PSR J0030+0451 that MSP HE light curves suggest that the observed gamma rays are produced in narrow accelerating gaps near the light cylinder.

\citet{Venter09} modeled the gamma-ray and radio light curves of these MSPs using OG, TPC, and PSPC models (see Chapters 2 and 5 for more details) for the gamma-ray light curves and a hollow-cone beam for the radio profiles.  It was found that six of the MSPs could be well fit by TPC and OG models while only two (PSRs J1744$-$1134 and J2124$-$3358) were well fit by the PSPC model suggesting pair-starved magnetospheres.

\subsection{The Case of PSR J0034$-$0534}\label{ch4J0034}
\citet{AbdoMSPpop} noted that significant point source signals were detected positionally coincident with five additional radio MSPs.  The MSPs were the only plausible, known counterparts for these gamma-ray point sources but significant pulsed signals were not observed.  The 1.88 ms pulsar J0034$-$0534 was one of those five MSPs.

\begin{figure}
\begin{center}
\includegraphics[width=0.6\textwidth]{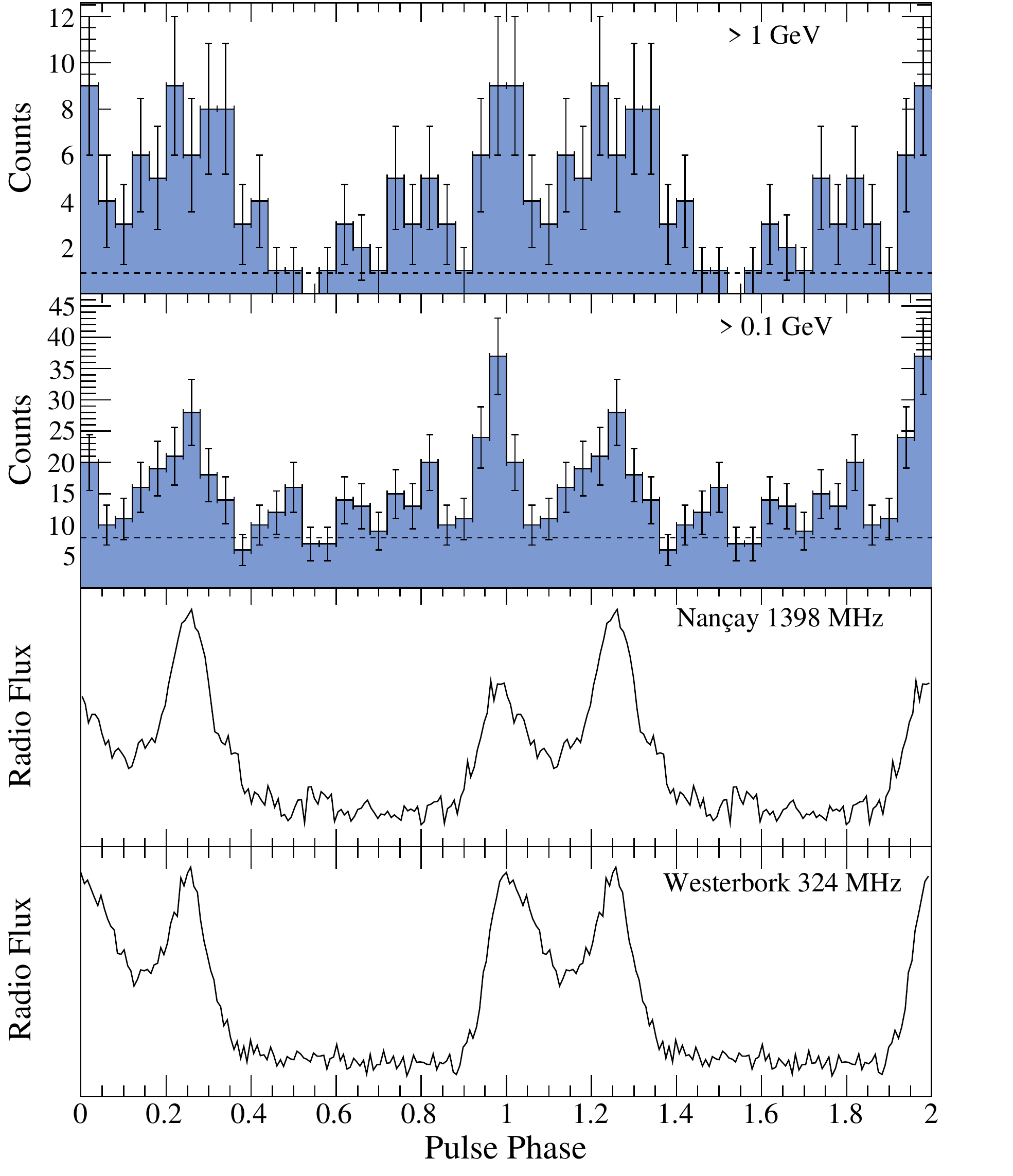}
\end{center}
\small\normalsize
\begin{quote}
\caption[Gamma-ray and radio light curves of PSR J0034$-$0534]{Pulse profiles of PSR J0034$-$0534.  Top panels show LAT data in indicated energy ranges using a circular ROI of 0\DEG{}.8 radius.  The bottom panels show the Nan\c{c}ay and Westerbork radio profiles at indicated observing frequencies.  Reproduced from \citet{AbdoJ0034}.\label{ch4J0034LCs}}
\end{quote}
\end{figure}

With $\sim$13 months of LAT sky-survey data, significant pulsations were detected from PSR J0034$-$0534 \citep{AbdoJ0034}.  The HE light curve of this MSP, Fig.~\ref{ch4J0034LCs}, displays two peaks which are nearly aligned with those observed in the radio profile.

\small\normalsize

The gamma-ray light curve of PSR J0034$-$0534 was fit with two Lorentzians plus a constant offset, estimated from simulations, which resulted in a reduced $\chi^{2}\ \sim$ 1.4 indicating a good fit.  The best-fit light curve parameters are given in Table~\ref{ch4J0034shape}.  The lags between gamma-ray and radio peaks ($\delta_{i}$) were evaluated by assuming the first and second radio peaks to be at phases of 0.0 and 0.258, respectively.  The lag in the second peak is consistent with zero but not the first; however, given the FWHM of the first gamma-ray peak the radio and gamma-ray peak positions are fairly consistent.

\small\normalsize

\begin{deluxetable}{l r}
\tablewidth{0pt}
\tablecaption{PSR J0034$-$0534 Light Curve Parameters}
\startdata
$\phi_{1}$ & -0.027$\pm$0.008\\
$W_{1}$ & 0.066$\pm$0.019\\
$\delta_{1}$ & -0.027$\pm$0.008\\
$\phi_{2}$ & 0.247$\pm$0.013\\
$W_{2}$ & 0.106$\pm$0.038\\
$\delta_{2}$ & 0.011$\pm$0.013\\
$\Delta$ & 0.274$\pm$0.015\\
\enddata\label{ch4J0034shape}
\end{deluxetable}

\small\normalsize

Before the detection of PSR J0034$-$0534, only the Crab pulsar was known to have gamma-ray and radio peaks aligned in phase.  As such, the gamma-ray and radio light curves were modeled in a manner similar to what was done by \citet{Harding08} for the Crab pulsar.  In particular, the gamma-ray light curve was modeled using standard TPC and OG models while the radio emission was modeled as being extended in altitude and co-located with the gamma-ray emission region.  These altitude-limited models have been explored in more detail by \citet{Venter11} (see Chapter 5 for more details as well).

Aligned profiles are also expected with low-altitude emission models and such emission in an MSP is not expected to result in a super exponential cutoff as the derived, dipolar surface magnetic fields of MSPs are too low to cause significant attenuation due to one-photon pair production.

However, attempts to model the light curves with standard low-altitude models were unsuccessful at reproducing the observations \citep{AbdoJ0034}.  \citet{Venter11} have demonstrated that low-altitude SG models can reproduce the observed light curves and compared the model predictions with those from the altitude-limited models.  No model could be conclusively ruled out but outer-magnetospheric emission was somewhat favored.

\begin{figure}
\begin{center}
\includegraphics[width=0.75\textwidth]{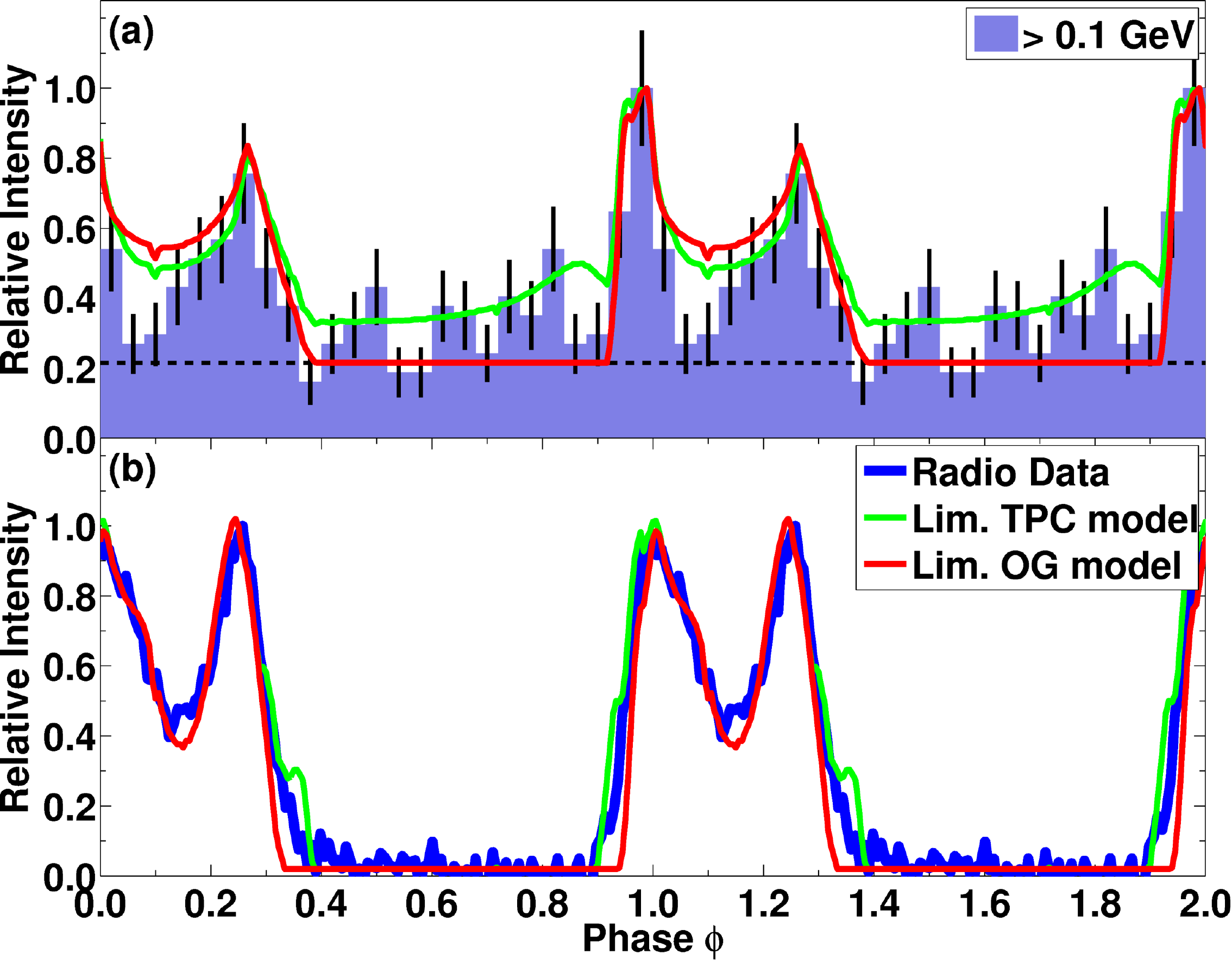}\end{center}
\small\normalsize
\begin{quote}
\caption[Observed and modeled light curves of PSR J0034$-$0534]{Data and model light curves of PSR J0034$-$0534.  The models are for the geometry and altitude ranges given in the text.  The models were matched to the $\geq$ 0.1 GeV and 324 MHz profiles.   Reproduced from \citet{AbdoJ0034}.\label{ch4J0034mods}}
\end{quote}
\end{figure}

Good solutions were found for PSR J0034$-$0534 using the altitude-limited models with $\alpha\ =\ 30^{\circ}$, $\zeta\ =\ 70^{\circ}$, and gap-width of 0.05 (normalized to the polar cap radius) for both TPC and OG models.  The gamma-ray emission region extended from the stellar surface to 0.9 R$_{\rm LC}$ while the radio emission regions spanned from 0.6 to 0.8R$_{\rm LC}$.  For the OG models the emission was always constrained to be above the NCS.   Both model light curves are shown plotted against the data in Fig.~\ref{ch4J0034mods}.  With the statistics of \citet{AbdoJ0034} it was not possible to discriminate between the models.

\small\normalsize

The gamma-ray spectrum of PSR J0034$-$0534 was fit using an unbinned maximum likelihood method, as implemented in the \Fermi{} science tools pyLikelihood python module.  Events were selected from 4 August 2008 to 10 September 2009 which had reconstructed energies from 0.1 to 100 GeV, zenith angles $\leq$ 105\DEG{}, and sky directions within 10\DEG{} of the MSP radio position.  The events were required to belong to the ``Diffuse'' class as defined under the P6\_V3 IRFs.  The \Fermi{} science tool \emph{gtmktime} was used to exclude time periods when the rocking angle of the instrument exceeded 52\DEG{} and when the Earth's limb infringed upon the ROI.

All sources found above the background with a TS of at least 25 in a preliminary version of the 1FGL catalog and within 15\DEG{} of the pulsar were included in the model region.  The spectral parameters of those sources $>$ 10\DEG{} from the pulsar were held fixed to the catalog values while the parameters of the other sources were left free.  The Galactic diffuse gamma-ray emission was modeled using the \emph{gll\_iem\_v02.fits} map cube while the isotropic diffuse and residual instrument backgrounds were modeled jointly using the \emph{isotropic\_iem\_v02.txt} template.  The spectrum of PSR J0034$-$0534 was modeled with both a power law (Eq.~\ref{ch3pl}) and a simple exponentially cutoff power law in separate fits.  The cutoff spectrum is preferred over the power law at the 4.5$\sigma$ level using the LRT.   The best-fit spectral parameters for the simple exponentially cutoff power law are given in Table~\ref{ch4J0034Spect}.

\small\normalsize

\begin{deluxetable}{l r}
\tablewidth{0pt}
\tablecaption{PSR J0034$-$0534 Spectral Parameters}
\startdata
$\Gamma$ & $1.5\pm0.2\pm0.1$\\
E$_{C}$ (GeV) & $1.7\pm0.6\pm0.1$\\
b & 1 (fixed)\\
Flux (0.1-100 GeV) ($10^{-8}\ \rm{cm}^{-2}\ \rm{s}^{-1}$) & $2.7\pm0.5\pm0.4$\\
Energy Flux (0.1-100 GeV) ($10^{-11}$ erg cm${-2}$ s$^{-1}$) & $1.9\pm0.2\pm0.1$\\
\enddata\label{ch4J0034Spect}
\end{deluxetable}

\small\normalsize

\begin{figure}
\begin{center}
\includegraphics[width=0.75\textwidth]{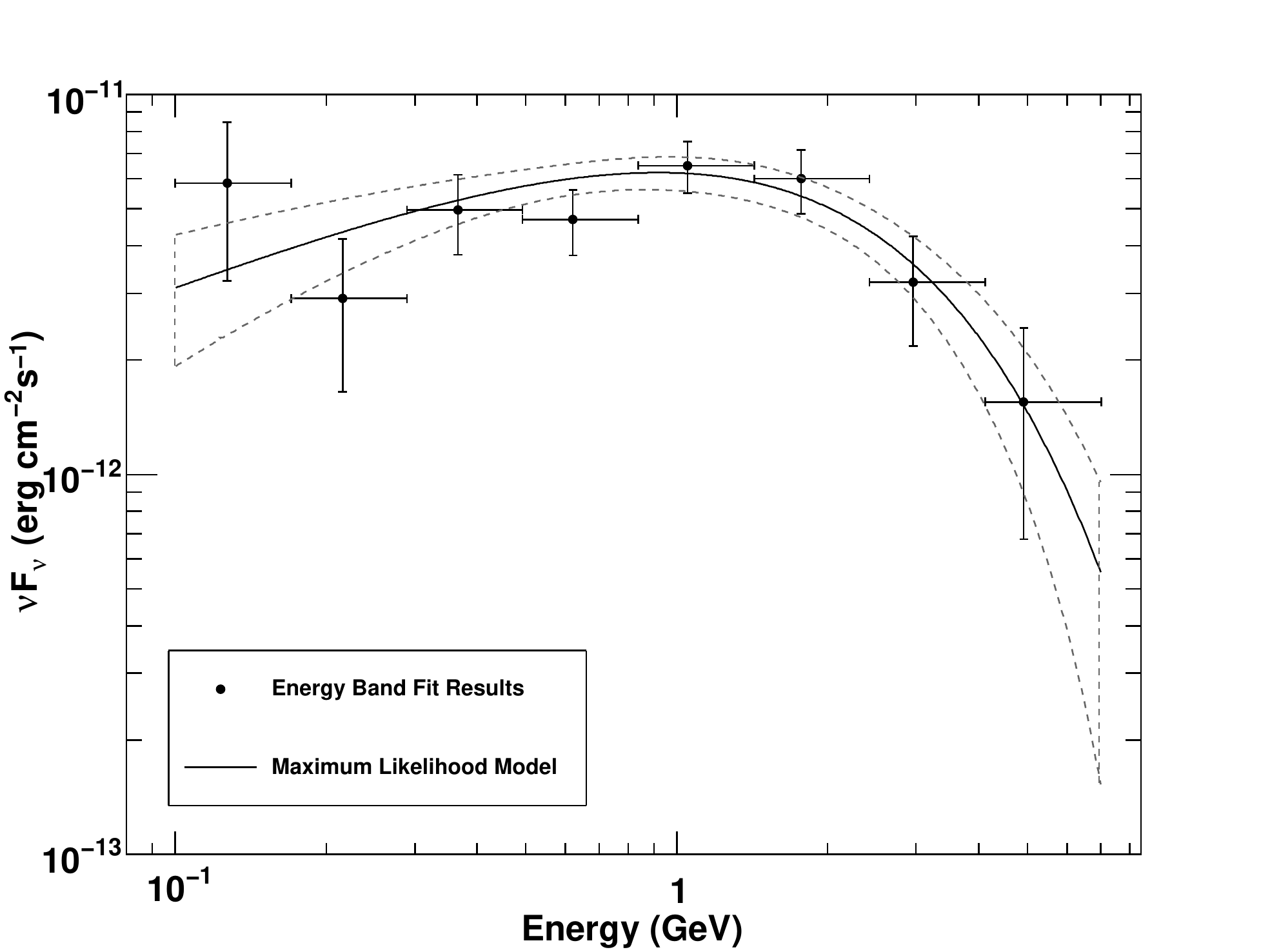}\end{center}
\small\normalsize
\begin{quote}
\caption[Gamma-ray spectrum of PSR J0034$-$0534]{Gamma-ray spectrum of PSR J0034$-$0534, maximum likelihood model is a simple exponentially cutoff power law.  Data points are from individual energy band fits in which the pulsar was modeled with a power law spectrum, 95\% confidence level upper limts were calculated for those energy bands in which the pulsar was found with $<\ 2\sigma$ significance.  Reproduced from \citet{AbdoJ0034}.\label{ch4J0034spec}}
\end{quote}
\end{figure}

With the possibility of near-surface emission for this MSP a separate fit was done in which the b parameter of the exponentially cutoff power law model was left free.  This fit returned a value of b not statistically different from 1 and not favored over the b $\equiv$ 1 model by the LRT.  The gamma-ray spectrum of PSR J0034$-$0534, with b $\equiv$ 1, is shown in Fig.~\ref{ch4J0034spec}.

\small\normalsize

\section{Conclusions}\label{ch4conc}
The LAT has proved to be an excellent instrument for pulsar science and has already facilitated many exciting discoveries.  In addition to establishing single MSPs as HE emitters, LAT observations have detected emission coincident with the location of at least eight globular clusters (Abdo et al., 2009f and 2010j) which display the characteristic pulsar HE spectrum.  Globular clusters are thought to contain many MSPs and, thus, this emission has been interpreted as the combined emission from MSPs which have not been detected individually.  Such combined MSP emission may also have implications for the Galactic and extragalactic gamma-ray backgrounds (e.g., Malyshev et al., 2010).

Of the 1451 sources in the 1FGL catalog 630 could not be reliably associated with a member of any known gamma-ray source class \citep{Abdo1FGL}.  Many of these unassociated sources were found to have non-variable fluxes and display significant spectral curvature, both traits of known gamma-ray pulsars. As such, intensive X-ray and radio campaigns were initiated to search the error circles of these sources for new pulsars.

To date, these searches have been highly successful, especially at finding new radio MSPs.  Some of these pulsars have been seen to pulse in gamma-rays as well (e.g., Ransom et al., 2011; Cognard et al., 2011; Keith et al., 2011), thereby firmly identifying the corresponding 1 FGL sources.  The growing population of gamma-ray MSPs provides a unique opportunity to study HE pulsar emission at lower spin-down energies and thereby understand the evolution of gamma-ray pulsar luminosity.
\renewcommand{\thechapter}{5}

\chapter{\bf MSP Light Curve Modeling}\label{ch5}
There have been many attempts to model pulsar light curves using geometric representations of the magnetic field structure.  Most studies, to date, have assumed a vacuum magnetosphere (e.g., Morini, 1983; Romani \& Yadigaroglu, 1995; Dyks \& Rudak, 2003; Harding et al., 2008; Venter et al., 2009 and 2011).  However, as \citet{GJ69} have shown the vacuum condition can not be maintained as charges would be pulled from the stellar surface and fill the magnetosphere with a pair plasma.

MHD simulations have been used to calculate the pulsar magnetic field structure in the presence of a pair plasma (e.g., Contopolous et al., 1999; Timokhin, 2006).  \citet{BS10b} have used a ``force-free'' MHD solution for the pulsar magnetosphere, in which inertial forces of the plasma are assumed to be negligible, to model gamma-ray pulsar light curves.   This force-free solution can not describe the true magnetic field either as, by its very nature, particle acceleration is not permitted and thus no gamma rays would be produced.

The retarded vacuum dipole field has been used for the simulations described in this chapter and the results presented in Chapter 7.  In principle, the true pulsar magnetosphere lies somewhere between these two extremes and predicted geometries from fits to simulations generated with both field structures should be compared and used to gauge which one lies closer to the truth and the models adjusted accordingly.

\section{The Retarded Dipole Vacuum Field}\label{ch5bfield}
\citet{Deutsch55} first calculated the retarded magnetic field of a star rotating in vacuum which would, if static, possess a magnetic dipole.  This was in order to explain observations of periodic changes in the magnetic field strengths of A stars.  These objects are white, main sequence stars in which the Balmer A0 emission lines are strongest \citep{BOB}.  A stars have spin periods on the order of days, masses typically around a few M$_{\odot}$, and are typically a few solar diameters across.

The star was modeled as a perfectly conducting and rigidly rotating sphere with a co-rotating magnetic field which was symmetric about an axis inclined to the axis of rotation.  Only the field exterior to the star (and the interior field right at the surface for boundary matching conditions) is of interest for the purposes of the geometric pulsar light curve models discussed in this Chapter.

Assuming a static dipole field configuration at the surface and using boundary matching conditions, \citet{Deutsch55} solved for the retarded field structure and calculated the rates at which energy and angular momentum are radiated away from the star.  He also found that for altitudes much less than R$_{\rm LC}$ the field lines rotated approximately rigidly with the star and thus the field structure approximated that of a static dipole.

For a dipole moment $\vec{\mu}\ =\ \mu\hat{z}$, spin frequency $\Omega$, $r\ =\ (x^{2}+y^{2}+z^{2})^{1/2}$ ,and $r_{n}\ = r/\rm R_{LC}$ the retarded dipole magnetic field is given by Eqs.~\ref{ch5Bretx},~\ref{ch5Brety}, and~\ref{ch5Bretz} \citep{DH04}.  The static dipole field is obtained by setting $r_{n}\ =\ 0$.
\begin{align}\label{ch5Bretx}
B_{ret,x}\ =\ &\frac{\mu}{r^{5}}\bigg(3xz\cos(\alpha)+\sin(\alpha)\Big\lbrace[(3x^{2}-r^{2})+3xyr_{n}+(r^{2}-x^{2})r_{n}^{2}]\cos(\Omega t-r_{n}) \nonumber \\
&+[3xy-(3x^{2}-r^{2})r_{n}-xyr_{n}^{2}]\sin(\Omega t-r_{n})\Big\rbrace\bigg)
\end{align}
\begin{align}\label{ch5Brety}
B_{ret,y}\ =\ &\frac{\mu}{r^{5}}\bigg(3yz\cos(\alpha)+\sin(\alpha)\Big\lbrace[3xy+(3y^{2}-r^{2})r_{n}-xyr_{n}^{2}]\cos(\Omega t-r_{n}) \nonumber \\
&+[(3y^{2}-r^{2})-3xyr_{n}+(r^{2}-y^{2})r_{n}^{2}]\sin(\Omega t-r_{n})\Big\rbrace\bigg)
\end{align}
\begin{align}\label{ch5Bretz}
B_{ret,z}\ =\ &\frac{\mu}{r^{5}}\bigg((3z^{2}-r^{2})\cos(\alpha)+\sin(\alpha)\Big\lbrace(3xz+3yzr_{n}-xzr_{n}^{2})\cos(\Omega t-r_{n}) \nonumber \\
&+(3yz-3xzr_{n}-yzr_{n}^2)\sin(\Omega t-r_{n})\Big\rbrace\bigg)
\end{align}

The properties of A stars are quite different from those of neutron stars but the results are applicable to both.  As discussed in Chapter 2, the Deutsch field was first applied to pulsars by \citet{Pacini68} who used it to show that a highly-magnetized, rotating neutron star could explain the pulsar phenomenon.  However, it was many years later before the Deutsch field structure was first used to model pulsar light curves \citet{RY95}.  Previous modeling attempts had employed a static shape dipole which should be a good approximation for very low altitude emission ($r\ \ll\ \rm R_{LC}$) but, as discussed below, will not be valid in the outer magnetosphere where the bulk of pulsar HE emission is now thought to originate.

\section{Special Relativistic Effects}\label{ch5SR}
In order to reproduce the bright, sharp peaks observed in pulsar HE light curves it is necessary to consider the special relativistic effects of aberration, time-of-flight delays, and rotational sweepback of magnetic field lines.

\citet{DH04} investigated the effects of magnetic field sweepback in the context of geometric radio emission models and found that, while the polarization angle curve is affected only weakly, the open field line region is shifted back significantly with respect to the direction of rotation.  This can lead to a phase shift of the same order or greater as the time-of-flight and aberration effects which lead to emission caustics.  Note that this effect is automatically included through the use of the retarded dipole magnetic field.

Photons radiated by particles traveling along curved magnetic field lines will be emitted in a cone of opening angle $1/\gamma$, where $\gamma$ is the particle Lorentz factor.  For highly relativistic particles this angle is approximately zero and the photon emission direction is nearly tangent to the field line.  Additionally, due to the finite speed of light, photons emitted in the same direction but at diffrent altitudes will be observed at different phases.

\citet{Morini83} first applied relativistic aberration and time-of-flight delays to a model of the optical and HE emission from the Vela pulsar.  He found that these two effects nearly cancel out the differences in phase which would otherwise be observed for photons emitted at different altitudes on the trailing edge of the PC.

These effects naturally lead to the formation of emission caustics where photons originating at different altitudes arrive closely spaced in phase creating bright peaks in the observed light curve.

\subsection{Magnetic Field Sweepback}\label{ch5sweep}

The sweepback effect simply refers to the rotationally induced distortion of the magnetic field lines from the static dipole form.  This is a direct consequence of the fact that magnetic field lines have momentum and thus can not rotate rigidly with the star.

Early studies of this effect (e.g., Shitov, 1983) focused on the field lines near the stellar surface (such that $r\ \ll\ \rm R_{LC}$) where radio emission is thought to occur.  In particular, these studies assessed how rotational deflections of the field direction would affect polarization profiles expected from the RVM.

\citet{Shitov83} found that the sweepback resulted in a deflection of order $r_{n}^{3}$ by relating magnetic torques with pulsar spin down.  This will lead to a lag in the expected polarization curve \citep{Shitov85} but the effect will be much less than that found by incorporating relativistic effects to the RVM which goes like $r_{n}$ \citep{Blaskiewicz91}.  An important effect for HE emission (from the outer magnetosphere) noted by \citet{Shitov85} is that the open field line volume is shifted backward with respect to the magnetic PC.

\citet{DH04} used the vacuum retarded dipole magnetic field solution (Eqs.~\ref{ch5Bretx},~\ref{ch5Brety}, and~\ref{ch5Bretz}), including the aberration and time-of-flight effects discussed below, to address the sweepback effect in more detail.  They found that the retarded field is disturbed from the static configuration, in the near-field region, by angle of order $r_{n}^{2}$, though along the dipole axis of an orthogonal rotator the deflection is of order $r_{n}^{3}$ which is roughly in agreement with \citet{Shitov83}.

\citet{DH04} defined the open field line region following the prescription of \citet{Dyks04} (for more details see Section~\ref{ch5simLCs}).  This approach accounts for the fact that the shape of the open field line region is affected strongly by the configuration of the magnetic field near the light cylinder.  They demonstrated that the open volume is displaced backwards (shifted to later phase) by an amount of order $r_{n}^{1/2}$.  Note that while it is important to include this effect the relativistic corrections discussed below will dominate at high altitudes where the observed HE emission is thought to originate.

\subsection{Relativistic Aberration}\label{ch5ab}
Photons are emitted in a direction tangent to the magnetic field in the co-rotating frame (CF).  To assess the observed phase of this emission the direction must then be transformed to an inertial observer's frame (IOF) via a Lorentz transformation as follows.

Let the emission direction in the CF be $\hat{k}^{\prime}$ with coordinate system ($t^{\prime}$,$x^{\prime}$,$y^{\prime}$,$z^{\prime}$) such that $\vec\Omega\ =\ \Omega\hat{z}^{\prime}$.  Let the IOF coordinate system be ($t$,$x$,$y$,$z$) such that $z\ = z^{\prime}$.

Assuming that the magnetic field structure described in Section~\ref{ch5bfield} does not change with time in the CF, it is possible to evaluate the emitted photon direction in the CF and IOF at any time $t$; therefore, let the direction be evaluated at time $t$ = 0 such 
that the $\vec{\Omega}$-$\vec{\mu}$ plane coincides with the $x$-$z$ plane.

This choice of $t$ allows for phase 0 to be assigned to those photons emitted from the center of the star with IOF directions in the $y$ = 0 plane.  With this definition and $\vec{\Omega}$ along the $z$-axis phases are negative for positive $y$ and positive for negative $y$.  Note that no emission is actually observed from the center of the star, this choice serves to simplify the time-of-flight delay equation described in Section~\ref{ch5tof}.

Consider the point of emission to be, instantaneously, an inertial frame of reference moving with velocity $\vec{\beta}_{\Omega}$ with respect to the IOF.  For an emission point $\vec{r}$, referenced from the center of the neutron star, the co-rotation velocity (in units of $c$) is given by Eq.~\ref{ch5beta}, where $\zeta$ is the polar angle with respect to the rotation axis (in the IOF) and $\hat{\beta}_{\Omega}$ is a unit vector in the direction of $\vec{\beta}_{\Omega}$.
\begin{equation}\label{ch5beta}
\vec{\beta}_{\Omega}\ =\ \frac{\vec{\Omega}\times\vec{r}}{c}\ =\ \frac{\Omega r}{c}\sin(\zeta)\hat{\beta}_{\Omega}\ =\ r_{n}\sin(\zeta)\hat{\beta}_{\Omega}
\end{equation}

In principle, the co-rotation velocity should be defined using the polar angle in the CF; however, \citet{Dyks04} showed that to first order in $r/\rm R_{LC}$ the polar angles in the CF and IOF are identical.

The emission direction in the IOF is given by Eq.~\ref{ch5abform} (derivation in Appendix~\ref{appB}).  To lowest order in $r_{n}$, the aberration goes like $\vec{\beta}_{\Omega}$ the magnitude of which grows linearly with $r_{n}$.
\begin{equation}\label{ch5abform}
\hat{k}\ =\ \frac{\hat{k}^{\prime}+\Big(\gamma+(\gamma-1)\frac{\vec{\beta}_{\Omega}\cdot\hat{k}^{\prime}}{\beta_{\Omega}^{2}}\Big)\vec{\beta}_{\Omega}}{\gamma(1+\vec{\beta}_{\Omega}\cdot\hat{k}^{\prime})}
\end{equation}

For photons emitted on the leading edge of the PC $\hat{k}^{\prime}$ has a component parallel to $\vec{\beta}_{\Omega}$ while those from the trailing edge have a component anti-parallel.  On the leading edge this results in a spreading out, in phase, of emission as observed in the IOF while emission from the trailing edge is seen to bunch in phase.

\subsection{Time-of-flight Delays}\label{ch5tof}
As noted above, zero phase is referenced to emission from the center of the star with $\hat{k}\cdot\hat{y}$ = 0.  Due to the finite speed of light, photons emitted at higher altitudes will arrive at the observer earlier in phase.

The difference in light travel times between a fictitious photon emitted from the center of the star and one emitted with the same $\hat{k}$ is given by $\Delta t\ =\ r\cdot\hat{k}/c$, where the dot product reflects the fact that only the distance (in the CF) along the line of sight to the observer contributes to the time delay.  This corresponds to a backward shift in phase of $-\Omega\Delta t\ =\ -\Omega r\cdot\hat{k}/c\ =\ -r_{n}\hat{r}\cdot\hat{k}$.

On the leading edge of the PC emission at higher altitudes is already shifted to earlier phases by the aberration effect; including the time-of-flight delay spreads emission out even further.  For photons on the trailing edge of the PC time-of-flight delays combine with aberration to bunch the emission even more closely in phase.

Using the conventions above with negative phase in the region of positive $y$, the observed photon phase can be calculated using Eq.~\ref{ch5finphase}.
\begin{equation}\label{ch5finphase}
\phi_{obs}\ =\ -(\tan^{-1}\Big(\frac{\hat{k}_{y}}{\hat{k}_{x}}\Big)+r_{n}\hat{r}\cdot\hat{k})
\end{equation}

The first term on the right hand side calculates the phase of the photon if it were from the center of the star, incorporating aberration, while the second term adds the time-of-flight delay.

\section{Simulating Light Curves}\label{ch5simLCs}
To simulate MSP radio and gamma-ray light curves the radius and mass of the neutron star are assumed to be $10^{6}$ cm and 1.4 M$_{\odot}$, respectively.  The spin period, magnetic inclination angle, and period derivative are parameters of the simulation (as well as an observation frequency for the hollow-cone beam radio models).

The next step is to find the rim of the PC via bisection in magnetic polar angle ($\theta_{\mu}$) following the prescription outlined in \citet{Dyks04}.  The PC rim is defined as the contour on the stellar surface from which those magnetic field lines which lie on the surface of last closed field lines originate.  Field lines from this contour will have $\hat{B}_{\rm LC}\cdot\hat{\rho}\ =\ 0$, where $\hat{B}_{\rm LC}$ is the direction of the field line at the light cylinder and $\hat{\rho}$ is the usual unit vector in the direction of the cylindrical radius coordinate $\rho$.

For a fixed value of magnetic azimuth ($\phi_{\mu}$) the magnetic field line originating from the stellar surface with $\theta_{\mu}$ = 0.7 $\Theta_{\rm PC}$ (where $\Theta_{\rm PC} \approx (\Omega R_{NS}/\mathnormal c)^{1/2}$ is the PC opening angle) is traced out to the light cylinder via Runge-Kutta integration.  Thus, if the current field line closes before the light cylinder a new field line is traced from the surface with $\theta_{\mu}$ smaller than the last step.  If the current field line is open at the light cylinder a new field line is traced from the surface with $\theta_{\mu}$ greater than the last step.

This continues until a field line is found which has $\hat{B}_{\rm LC}\cdot\hat{\rho}\ =\ 0$.  A step is made in $\phi_{\mu}$ (using a step size of 0\DEG{}.5) and the process is repeated starting at the value of $\theta_{\mu}$ corresponding to the rim for the previous $\phi_{\mu}$.

For $\phi_{\mu}$ $\sim$115\DEG{} many authors (e.g., Romani \& Yadigaroglu, 1995) found a deformity of the PC rim, due to distortion from magnetic field sweep back, for which the rim had three values of $\theta_{\mu}$ for a given $\phi_{\mu}$.  Thus, a jump discontinuity was introduced and no field lines originating from this region of the rim were used.

\begin{figure}[h]
\begin{center}
\includegraphics[width=.75\textwidth]{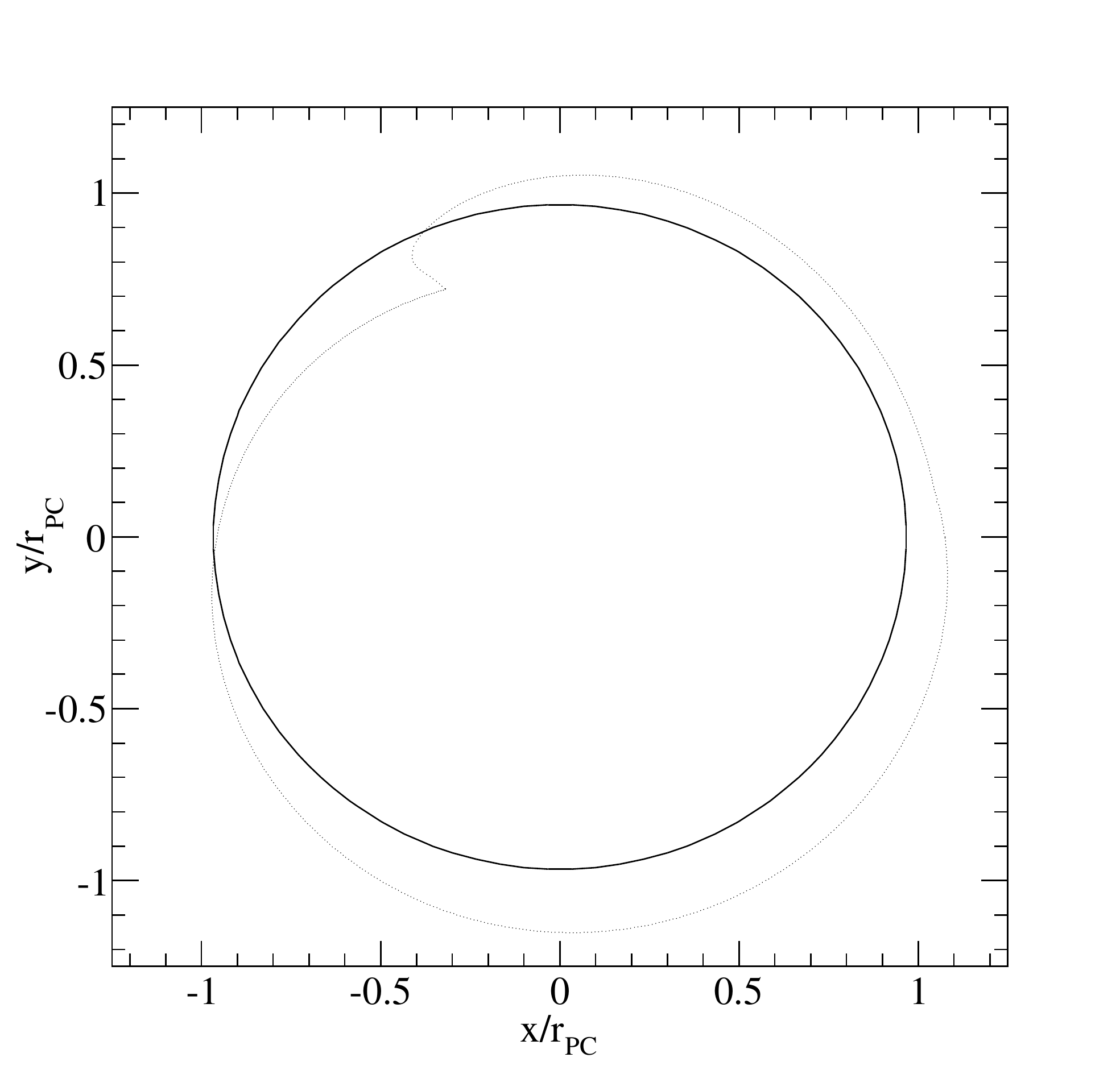}
\end{center}
\small\normalsize
\begin{quote}
\caption[Distorted polar cap rim contour]{Points used to define the PC rim projected onto a 2D space with $\vec{\mu}$ coming out of the page at (0,0).  The coordinates have been normalized to the PC rim, a circle of radius $r_{\rm PC}$ centered on the magnetic pole is shown for contrast.  This rim corresponds to a simulation with $\alpha$ = 45\DEG{} and spin period = 2.5 ms.\label{ch5rim}}
\end{quote}
\end{figure}

However, \citet{Dyks04} applied bisection in $\phi_{\mu}$ to trace the rim in this region which led them to discover that the distortion manifested as a ``notch'' in the rim (see Fig.~\ref{ch5rim}).  They also discovered that ignoring the field lines originating from the notch resulted in a large fraction of the open-field line volume being empty for moderate inclination angles.  Such a situation is highly undesirable for the generation of HE pulsar light curves.

\small\normalsize

The simulations in this study follow the prescription of \citet{Dyks04} in the vicinity of the notch when tracing the PC rim.  For small steps in $\phi_{\mu}$ the location of the rim in $\theta_{\mu}$ should change very little.  Thus, bisection in $\phi_{\mu}$ is used when a field line has not been found with $\hat{B}_{\rm LC}\cdot\hat{\rho}\ =\ 0$ and the trial $\theta_{\mu}$ has moved more than 1.5\% of the PC angle from the previous step.  This algorithm proceeds similarly to that described for bisection in $\theta_{\mu}$ until the rim is found.  Once clear of the notch the remainder of the rim is found using bisection in $\theta_{\mu}$.

Once the full rim has been found open volume coordinates $r_{ovc}$ and $l_{ovc}$ are defined as in \citet{Dyks04}.  The $r_{ovc}$ coordinate is defined to be 1 at the rim and 0 at the magnetic pole, in analog to $\theta_{\mu}/\Theta_{\rm PC}$.  The $l_{ovc}$ coordinate is defined as arc length along a contour of constant $r_{ovc}$, defined to be zero at $\phi_{\mu}$ = 0 and increasing in the same direction.  While $l_{ovc}$ is very similar to magnetic azimuth note that it follows the shape of the PC rim and thus is not a monotonic function of $\phi_{\mu}$ in the region of the notch.

Self-similar rings are defined on the surface of the star in open volume coordinates between a specified $r_{ovc}^{min}$ and $r_{ovc}^{max}$, with a spacing between rings of $\delta_{ring}$ = 0.005 (units of $r_{ovc}$).  The rings are divided into equal area segments by taking constant steps in $l_{ovc}$, resulting in rings with lower values of $r_{ovc}$ (inner rings) having fewer segments. Each segment is assumed to contain one magnetic field line along which an electron will be followed and emitted photons collected.

Electrons are distributed uniformly over the PC, one per ring segment, which allows for the area of each surface element to be approximated as $dS\ \approx\ \pi r_{\rm PC}/\mathnormal N_{e,tot}$ (Eq. 20 of \citet{Venter09}), where $N_{e,tot}$ is the total number of electrons (dependent on the number of rings and ring segments) and $r_{\rm PC}\ =\ (\Omega R_{NS}^{3}/\mathnormal c)^{-1/2}$ is the approximate radius of the PC.

Starting at the outermost ring and moving inwards, electrons are followed along magnetic field lines in the CF.  For the gamma-ray emission models, the electrons are assumed to emit via CR with a uniform emissivity along the field lines.  The phase of an emitted photon is calculated accounting for aberration and time-of-flight delays and the corresponding phase bin content is incremented by 1.

In reality, the emissivity should decrease far from the surface of the star.  To reflect this fact, the emission is only followed out to a specified radial distance (R$_{max}$), essentially implementing a step function from uniform to zero emissivity.

In principle, pulsed emission is possible from particles co-rotating with the star out to the edge of the light cylinder.  However, the magnetic field structure is not well known near the light cylinder and thus the emission is only followed to a cylindrical distance of 0.95 R$_{\rm LC}$.

For generation of radio light curves with a hollow-cone beam structure (see Section~\ref{ch5radio}), the field lines are traced to one specified altitude where emission occurs.  The phase and colatitude of an emitted photon is calculated and the corresponding phase bin is incremented by a specific flux value.

In order to calculate the observed phase of a photon emitted at a given point on a field line, the direction tangent to the field line in the CF must first be calculated.  This is used as the initial photon direction and then transformed to the IOF, which accounts for relativistic aberration (Eq.~\ref{ch5abform}).  The phase is then calculated from this direction, using the geometric convention described in Section~\ref{ch5SR}, and corrected for time-of-flight delays (see Section~\ref{ch5tof}).

\begin{figure}[h]
\begin{center}
\includegraphics[width=1.\textwidth]{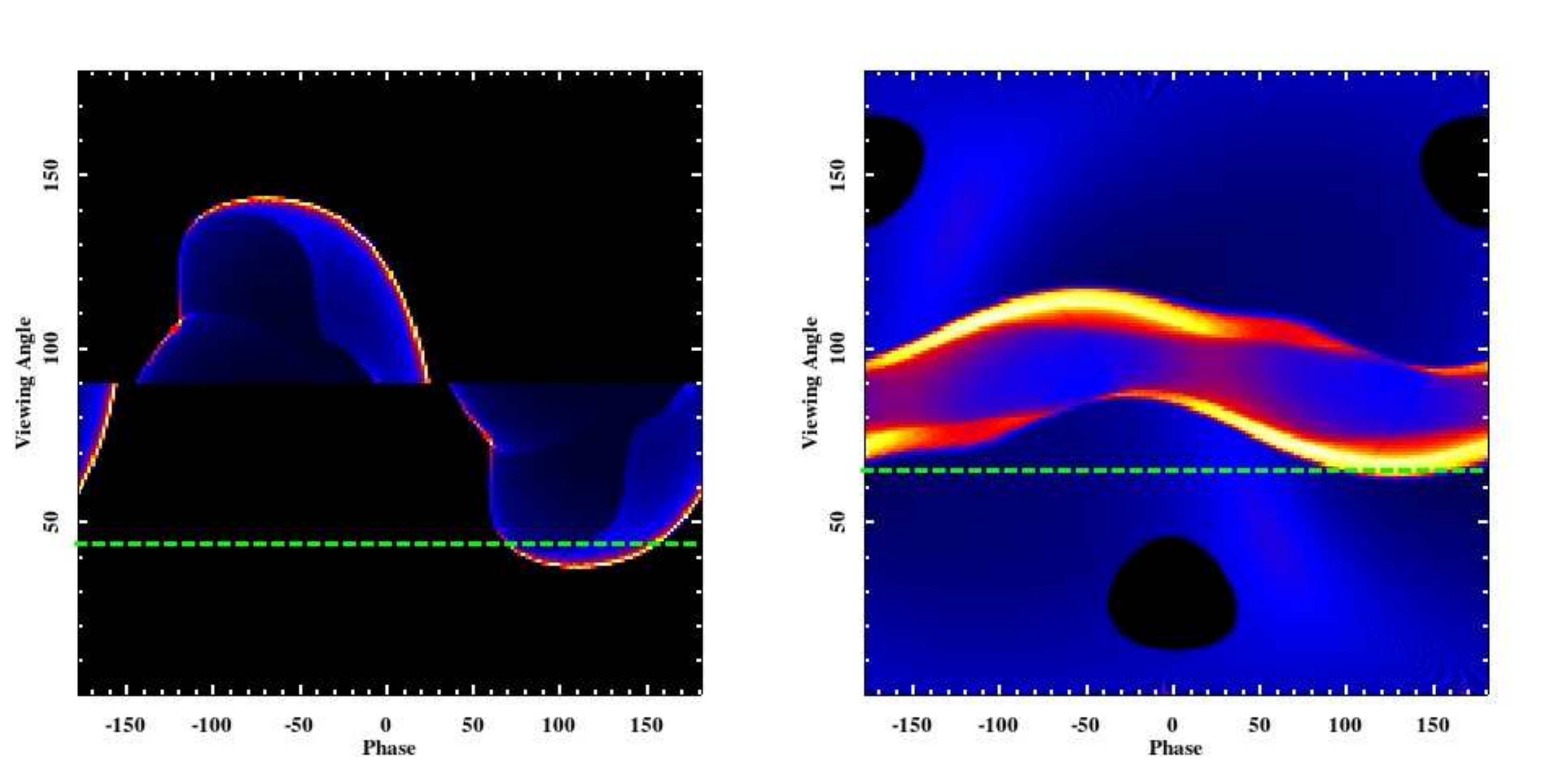}
\end{center}
\small\normalsize
\begin{quote}
\caption[Example of simulated gamma-ray phase plots]{Simulated phase plots for OG (\emph{left}, $\alpha$ = 72\DEG{}) and TPC (\emph{right}, $\alpha$ = 28\DEG{}) gamma-ray emission models with a spin period of 5.5 ms.  The phase plots were chosen to match the best-fit viewing geometries and gap widths for PSR J0437$-$0534 (see Chapter 7).\label{ch5zphi}}
\end{quote}
\end{figure}

The emitted photons are accumulated in bins of colatitude ($\zeta$) and pulse phase, both relative to the rotation axis, with 1\DEG{} resolution in $\zeta$ and 2\DEG{} in phase (see Fig.~\ref{ch5zphi}).  The content of each bin is then divided by the solid angle it subtends resulting in units of photons/primary/solid angle.  This also removes distortions in the phase plots (e.g., Fig.~\ref{ch5zphi}) due to the variation of solid angle with $\sin(\zeta)$.

\small\normalsize

The resulting light curve for a given viewing angle is formed by selecting the proper $\zeta$ bin and plotting the contents in ascending phase.  When comparing the simulated light curves to data the relative, not absolute, magnitudes of each bin are the important values as these define the shape.

Previous studies (e.g., Venter et al., 2009; Abdo et al., 2010d) determined the magnetic field direction in the IOF, used that as the emitted photon direction in the CF, and then applied a Lorentz transformation to calculate the IOF photon direction.  \citet{BS10a} argued that the magnetic field direction should be first transformed to the CF, before calculating the emission direction, for self-consistency.

The simulations used in this study now calculate the magnetic field in the CF using Eq.~\ref{ch5BfLT} \citep{Ohanian01}.  This Lorentz transformation is for the force-free case and the perpendicular and parallel suffixes are referenced to $\vec{\beta}_{\Omega}$.  For details of how the simulation code implements this transformation see Appendix B.
\begin{equation}\label{ch5BfLT}
\vec{B}^{\rm CF}\ =\ \vec{B}_{\parallel}^{IOF} + \gamma^{-1}\vec{B}_{\perp}^{IOF}
\end{equation}

This had been neglected previously because the total effect, after accounting for abberation, is of order $\beta_{\Omega}^{2}\ \ll$ 1 for $r_{n}\ll1$.  However, \citet{BS10a} demonstrated that including the transformation can affect the shape of simulated light curves.

Use of Eq.~\ref{ch5BfLT} before determining the photon CF direction can lead to slightly wider peaks. Additionally, this transformation has the effect of beaming the emission towards the rotational equator which leads to a lack of emission, most pronounced in TPC models, for large $\alpha$ and small $\zeta$ where previous studies would have predicted emission.  However, the relative flux level and phase modulation at these geometries predicted previously are very low.  Thus, this does not impact a region which detected gamma-ray pulsars are expected to populate.

\subsection{Gamma-ray Light Curves}\label{ch5gamma}
Geometric versions of the SG, OG, and PSPC models (see Chapter 2) can be realized by specifying $r_{ovc}^{min}$, $r_{ovc}^{max}$, and the altitudes between which emission is calculated to define accelerating and emitting gaps.  Note that the TPC model \citep{DR03} is taken to be a geometric realization of the SG model for the purposes of this study.

As discussed in Chapter 2, initial studies with the TPC model only followed emission out to radial distances of 0.75 R$_{\rm LC}$.  \citet{DR03} found that the features of known HE light curves (two bright, widely separated peaks) were satisfactorily reproduced without following emission higher in altitude.  However, given the improvement in HE light curve detail afforded by data from the \FL{} and following the findings of \citet{Venter11}, the TPC model used here allows the emission to occur at radial distances $>$ 0.75 R$_{\rm LC}$.

The assumption of uniform emissivity along the magnetic field lines for TPC and OG models means that the rate of emitted photons is proportional to the distance traveled along the field line and no explicit form is necessary for the accelerating electric field.  For the PSPC model the accelerating electric field solution of \citet{Venter09} is used to calculate the rate of emitted photons out to high altitudes in the magnetosphere (see Chapter 2 for more details).

\subsubsection{TPC and OG Realizations}\label{ch5TPCOG}
\begin{figure}[h]
\begin{center}
\includegraphics[width=.9\textwidth]{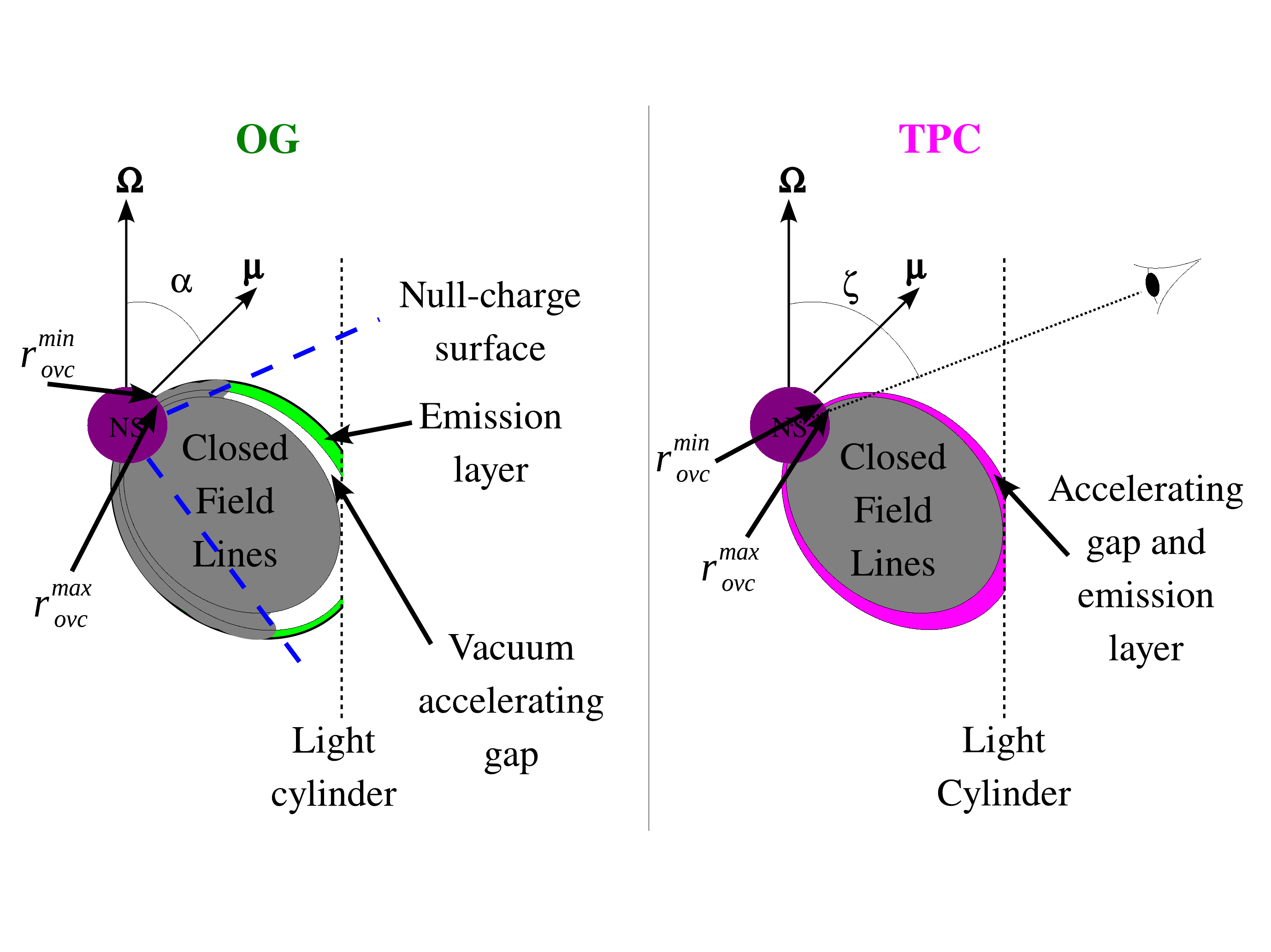}
\end{center}
\small\normalsize
\begin{quote}
\caption[Schematic of the TPC and OG emission geometries]{Schematic of the TPC \emph{(right)} and OG \emph{(left)} emission geometries.  The emission gaps are defined by $r_{ovc}^{min}$ and $r_{ovc}^{max}$ as described in the text, emission is assumed to originate from the green (pink) parts of the gap for the OG (TPC) model.\label{ch5mygeom}}
\end{quote}
\end{figure}

The accelerating gaps and emission regions for the TPC and OG models are defined as follows and shown schematically in Fig.~\ref{ch5mygeom}.

\small\normalsize

For the TPC model, acceleration and emission are assumed to occur within the same, finite-size gap bounded by the surface of last closed field lines (i.e., the PC rim) which leads to $r_{ovc}^{max}\ \equiv\ 1.0$ for all TPC realizations.  The inner boundary of the gap is defined by field lines emerging from a ring on the stellar surface with $r_{ovc}\ =\ r_{ovc}^{min}\ \leq\ 1.0$.  The gap width is therefore defined to be $1-r_{ovc}^{min}$.

Electrons are followed from the stellar surface, originating between $r_{ovc}^{min}$ and $r_{ovc}^{max}$, to R$_{max}$ = 1.2 R$_{\rm LC}$ and emission is collected from points along the magnetic field lines as described previously.

For the OG model, acceleration is assumed to occur in a vacuum gap, bounded on one side by the surface of last closed field lines, with the other boundary within the open field line volume and specified by $r_{ovc}^{max}$.  The accelerating gap width is defined to be $1-r_{ovc}^{max}$.

The emission should then occur in a small layer just inside $r_{ovc}^{max}$.  This inner boundary is specified by $r_{ovc}^{min}$ which is constrained to satisfy $0.5(1-r_{ovc}^{max})\ \geq (r_{ovc}^{max}-r_{ovc}^{min})$ following the findings of \citet{Wang10} that this layer should be small compared to the total gap size.  The width of the emitting layer is defined to be $r_{ovc}^{max}-r_{ovc}^{min}$ and the ``total'' gap width is the sum of these two widths.

Similar to the TPC model, electrons are followed from the surface, originating between $r_{ovc}^{min}$ and $r_{ovc}^{max}$, out to R$_{max}$ = 1.2 R$_{\rm LC}$.  However, no emission is collected from below the NCS.  This boundary is defined by the requirement that $\vec{\Omega}\cdot\vec{B}$ = 0.  Therefore, emission is only collected from points where the $z$-component of the magnetic field is negative.

To model the radio and gamma-ray light curves of phase-aligned MSPs, \citet{AbdoJ0034} introduced the so-called ``altitude-limited'' TPC and OG models (alTPC and alOG, respectively) in which the gamma-ray and radio emission are both assumed to be caustic in nature and come from regions co-located within the magnetosphere.  These models, as well as a low-altitude version of the SG, are developed and discussed in more detail by \citet{Venter11}.

The emission gaps for the gamma-ray alTPC/OG models are defined in the same manner discussed above for the TPC/OG models with one important difference.  For these models R$_{max}$ is a parameter of the simulation but, currently, constrained to be $\geq$ 0.7 R$_{\rm LC}$.  The minimum emission altitude remains fixed at the stellar surface for alTPC models or the NCS for alOG models.

\subsubsection{PSPC Realization}\label{ch5PSPC}
The PSPC model uses the full open field line volume to accelerate electrons and produce gamma rays.  This is implemented by setting $r_{ovc}^{min}$ = 0 and $r_{ovc}^{max}$ = 1, effectively sectioning the entire PC into $l_{ovc}$ rings which are populated with electrons.

The accelerating field ($E_{\parallel}$) should change significantly over the open volume and it is thus necessary to abandon the uniform emissivity assumption for this model and include the functional form of the accelerating electric field.  A detailed description of the method used to accelerate electrons and accumulate curvature radiation emission is given in \citet{Venter09}.  A summary of the method follows.

The forms of $E_{\parallel}$ for low and medium altitudes have been solved by \citet{MH97} and \citet{HM98} and are given, in reduced form, by Eqs.~\ref{ch5E1} and~\ref{ch5E2}, where $\phi_{\rm PC}$ is the azimuthal coordinate referenced to the polar cap at the point of interest and $\xi\ \equiv\ \theta/\Theta_{\rm PC}$.
\begin{equation}\label{ch5E1}
E_{\parallel}^{(1)}\ =\ -\frac{B_{surf}\Omega \rm R_{NS}}{c} (\theta_{0}^{\mathnormal GR})^{2} F_{1}(\Omega,\alpha,\phi_{\rm PC},r,\rm R_{NS},\xi)
\end{equation}
\begin{equation}\label{ch5E2}
E_{\parallel}^{(2)}\ =\ -\frac{B_{surf}\Omega \rm R_{NS}}{c} (\theta_{0}^{\mathnormal GR})^{2} F_{2}(\Omega,\alpha,\phi_{\rm PC},r,\rm R_{NS},\xi)(1-\xi^{2})
\end{equation}

The full functional forms of $F_{1}$ and $F_{2}$ are given in \citet{Venter09}, these terms account for, among other things, General Relativistic inertial frame dragging.

Note that Eq.~\ref{ch5E1} is valid for $r/\rm R_{NS}\ -\ 1\ \ll\ 1$ while Eq.~\ref{ch5E2} is valid for $\theta_{0}^{GR}\ \ll\ r/\rm R_{NS}\ -\ 1\ \ll\ R_{LC}/R_{NS}$, where $\theta_{0}^{GR}\ \equiv\ \theta^{GR}(1)$ and $\theta^{GR}$ given in Eq.~\ref{ch5thetaGR}, which approximates $\Theta_{\rm PC}$.  The function $f(r/\rm R_{NS})$ is generally of order unity and is defined in \citet{MT92}.
\begin{equation}\label{ch5thetaGR}
\theta^{GR} \approx\ \bigg(\frac{\Omega}{c}\frac{r}{f(r/\rm R_{NS})}\bigg)^{1/2}
\end{equation}

The accelerating field solutions given above are not valid out to the light cylinder, therefore it is necessary to use the solution of \cite{MH04b} given in Eq.~\ref{ch5E3} which gives $E_{\parallel}$ near the light cylinder.
\begin{equation}\label{ch5E3}
E_{\parallel}^{(3)}\ \approx\ -\frac{3}{16}\Big(\frac{\rm R_{NS}}{R_{LC}}\Big)^{3} \frac{B_{surf}}{f(1)}F_{3}(\Omega,\alpha,\phi_{\rm PC},r,\rm R_{NS},\xi)
\end{equation}

The full functional form of $F_{3}$ can be found in \citet{Venter09}.  The low altitude solutions $E_{\parallel}^{(1)}$ and $E_{\parallel}^{(2)}$ are matched at an altitude of $r/\rm R_{NS}\ \approx\ 1+0.0123\ P^{-0.333}$ following \citet{Venter08}.  Matching the high altitude solutions $E_{\parallel}^{(2)}$ and $E_{\parallel}^{(3)}$ is done separately for each field line using an altitude which depends on P, $\dot{\rm P}$, $\alpha$, $\xi$, and $\phi_{\rm PC}$ as described in \citet{Venter09}.

Note that the accelerating field is directly proportional to the surface magnetic field.  For the results presented in Chapter 7 all simulations were produced assuming a period derivative of $10^{-20}\ \rm s\ s^{-1}$ which leads to  $B_{surf}\ \approx\ 4.94\ \sqrt{\rm P_{ms}}\ 10^{9}\ G$, where P$_{\rm ms}$ is the pulsar period in units of ms.

Only losses due to CR from primary particles are considered using the transport equation given in Eq.~\ref{ch5trans} (e.g., Daugherty \& Harding, 1996).  The electrons are assumed to be travelling with velocity $\beta\ \approx\ 1$ which ignores the initial acceleration from the stellar surface.
\begin{equation}\label{ch5trans}
\dot{E}_{e}\ =\ ecE_{\parallel}-\frac{2e^{2}c}{3\rho^{2}}\gamma^{4}
\end{equation}

The first term on the right hand side of Eq.~\ref{ch5trans} is the energy gain from the accelerating field (where the appropriate $E_{\parallel}$ is used based on the current particle altitude) while the second term is the energy loss due to CR (see Eq.~\ref{ch1curvP}).

The particle outflow along each field line (Eq.~\ref{ch5dN}) is normalized to the general relativistic counterpart of the Goldreich-Julian charge density $\rho_{e}$ (Eq. 12 of Harding \& Muslimov, 1998 ,with $r\ =\ \rm R_{NS}$) and the surface patch area estimate ($dS$) given in Section~\ref{ch5simLCs} of this thesis.
\begin{equation}\label{ch5dN}
d\dot{N}\ =\ -\frac{\rho_{e}(1,\xi,\phi_{\rm PC})}{e} \mathnormal dS\beta_{0}c
\end{equation}

The particles are assumed to have an initial speed $\beta_{0}c$ which is taken to be such that $\gamma_{0}\ =$ 100.  As \citet{Venter09} note this choice does not strongly affect the results as $\gamma$ reaches vales of $\sim10^{6}$ or more a short distance above the stellar surface.

For each step along a field line the number of CR photons is calculated using Eqs.~\ref{ch5dN} and~\ref{ch1CRspec} (with $q\ =\ e$ and using the instantaneous field line radius of curvature) scaled to an average photon energy of 100 MeV.  The direction and phase of emitted photons are calculated as described in Section~\ref{ch5simLCs}.

\citet{Venter09} found energy balance and near radiation-reaction conditions which suggests that the simulation describes the acceleration process accurately.

\subsection{Radio Light Curves}\label{ch5radio}
For MSPs with significant lags between radio and gamma-ray light curve peaks, the radio emission is modeled using a hollow-cone beam geometry.  This geometry is based on empirical fits of radio profiles by \citet{Rankin93} which suggested that the light curves of many pulsars could be fit with a core and one or more surrounding hollow-cone beams.  The models of \citet{Rankin93} assume that the cone is fully illuminated though some authors have argued for only partially illuminated (or patchy) cones (e.g., Lyne \& Manchester, 1988).

The radio profiles of MSPs with non-aligned profiles discussed in Chapter 7 are modeled as single, hollow-cone beams with no core component.  Several studies have suggested that cone beams dominate in short period pulsars (e.g., Johnston \& Weisberg 2006).  However, as discussed in Chapter 7, there are some cases for which this model is likely not sufficient.

The cone beam geometry used here follows that of \citet{Story07} and \citet{Harding08} (where the two differ the model of Story et al., 2007 has been used).  These models build on the analysis of \citet{Gonthier04} who added frequency dependence to the flux model of \citet{Arzoumanian02}.

The cone flux (in units of mJy sr$^{-1}$) as a function of $\theta_{\mu}$ and radio frequency ($\nu$) is given by Eq.~\ref{ch5Stheta}.
\begin{equation}\label{ch5Stheta}
S_{cone}(\theta_{\mu},\nu)\ =\ F_{cone}(\nu)exp\bigg\lbrace -\frac{(\theta_{\mu}-\bar{\theta})^{2}}{w_{e}^{2}}\bigg\rbrace,
\end{equation}

The angular position of the cone (with respect to $\vec{\mu}$) is $\bar{\theta}\ =\ (1-2.63\delta_{w})\rho_{cone}$.  The width of the cone is $w_{e}\ =\ \delta_{w}\rho_{cone}$, using the parameters $\delta_{w}$ = 0.18 (Harding et al. 2008 and Gonthier et al. 2006) and $\rho_{cone}\ =\ 1^{\circ}.24 \sqrt{r_{\rm KG}/\rm P}$.  The emission is emitted at a single altitude in the magnetosphere given by Eq.~\ref{ch5rKG} (in units of R$_{\rm NS}$) where $\nu_{\rm GHz}$ is the radio frequency in units of GHz \citep{KG03}.
\begin{equation}\label{ch5rKG}
r_{\rm KG}\ =\ 40\bigg(\frac{\rm \dot{P}}{10^{-15}s\ s^{-1}}\bigg)^{0.07}\rm P^{0.3}\nu_{GHz}^{-0.26} .
\end{equation}

The flux per solid angle of the cone ($F_{cone}$) is given by Eq.~\ref{ch5Fcone} for a cone of solid angle $\Omega_{cone}$, assuming a spectral index $\alpha_{cone}\ =\ -1.72$, and a distance of $D\ =\ 1$kpc.  The gamma-ray MSPs detected by the LAT span a distance range from $\sim$0.1 to 8 kpc but the use of 1 kpc in all simulations does not adversely affect the light curve fits presented in Chapter 7 as only the light curve shapes, not the observed fluxes, are compared.
\begin{equation}\label{ch5Fcone}
F_{cone}(\nu)\ =\ \frac{-(1+\alpha_{cone})}{\nu}\bigg(\frac{\nu}{50\ \rm MHz}\bigg)^{\alpha_{\mathnormal cone}+1}\frac{L_{cone}}{\Omega_{cone}D^{2}}
\end{equation}

The cone luminosity ($L_{cone}$) is given by Eq.~\ref{ch5Lcone} (following the P and $\dot{\rm P}$ dependence found by Arzomanian et al., 2002) in which $L_{r}$ is taken to be $1.76\times10^{10}\ \rm P^{-1.05}\ \dot{P}^{0.37}$ mJy kpc$^{2}$ MHz following the findings of \citet{Story07}.
\begin{equation}\label{ch5Lcone}
L_{cone}\ =\ \frac{L_{r}}{1+(r/r_{0})}
\end{equation}

The core-to-cone peak flux ratio ($r$) is 25 P$^{1.3}\nu_{\rm GHz}^{0.9}$ (Gonthier et al., 2006 and Harding et al., 2008; for P $<$ 0.7 s).  The parameter $r_{0}$ (Eq.~\ref{ch5r0}) is the ratio of energy emitted at frequency $\nu$ in the core versus the cone component, where $\Omega_{core}$ is the solid angle of the core component which is assumed to have a spectral index of $\alpha_{core}\ =\ -2.36$ \citep{Story07}.
\begin{equation}\label{ch5r0}
r_{0}\ =\ \frac{\Omega_{cone}r_{\rm KG}}{\Omega_{\mathnormal core}} \frac{\alpha_{core}+1}{\alpha_{cone}+1} \bigg(\frac{\nu}{50 \rm MHz}\bigg)^{\nu_{\mathnormal core}-\nu_{cone}}
\end{equation}

The PC is partitioned into rings between $r_{ovc}^{min}$ = 0.1 and $r_{ovc}^{max}$ = 1.2 as described previously.  The magnetic field line in each footprint on the stellar surface is traced outward until $r_{\rm KG}$ is reached.  At this point the phase and co-latitude of a photon emitted tangent to the field line is calculated.  The corresponding phaseplot bin is then incremented by the radio flux given in Eq.~\ref{ch5Rcone}.
\begin{equation}\label{ch5Rcone}
R_{cone}(\theta_{\mu},r_{ovc})\ =\ S_{cone}(\theta_{\mu},\nu)\Omega_{seg}
\end{equation}

\begin{figure}[h]
\begin{center}
\includegraphics[width=1.\textwidth]{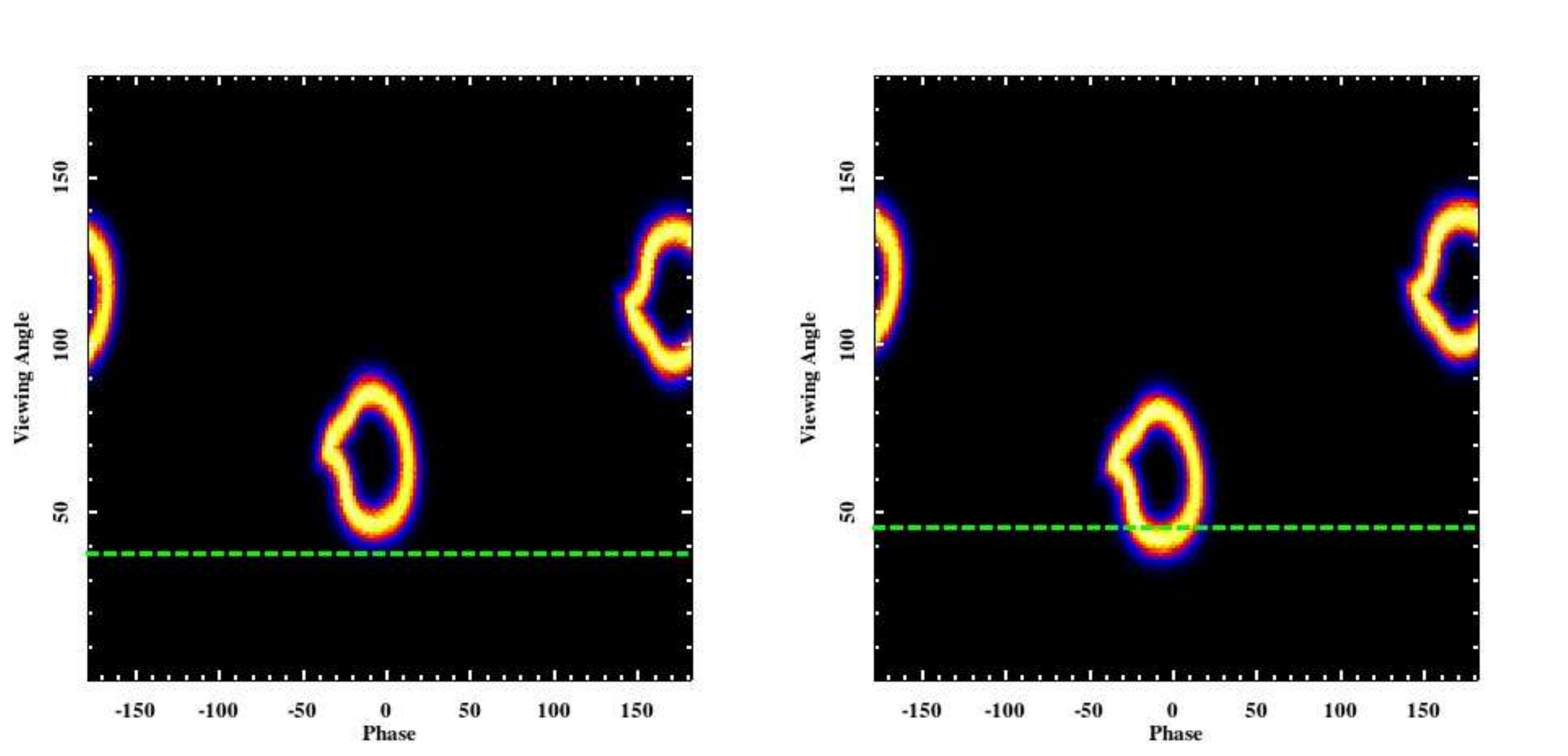}
\end{center}
\small\normalsize
\begin{quote}
\caption[Example simulated radio phase plots]{Simulated radio phase plots assuming a hollow-cone beam geometry and a spin period of 5.5 ms.  The phase plots were chosen to match the best-fit viewing geometries of PSR J2302+4442 with OG (\emph{left}, $\alpha$ = 64\DEG{}) and TPC (\emph{right}, $\alpha$ = 59\DEG{}) gamma-ray emission models (see Chapter 7).\label{ch5radiozphi}}
\end{quote}
\end{figure}

The solid angle of a ring segment is approximated as $\Omega_{seg}\ \equiv\ (d\phi) (d\theta)\ =\ (2\pi\sin(\theta_{\mu})l_{ring})(\delta_{ring}\Theta_{\rm PC}\mathnormal r_{ovc}^{max})$ where $l_{ring}$ is the length of a ring segment in $l_{ovc}$.  An example of the simulated radio emission is shown in Fig.~\ref{ch5radiozphi}.

\small\normalsize

The hollow-cone beam model can match the shape of some MSP radio light curves, at least in part, but for gamma-ray MSPs it is also important to match the observed phase delay between features at both wavelengths.  In particular, assuming gamma-ray emission originating in the outer magnetosphere the hollow-cone beam geometry (with the emission altitude of Kijak \& Gil, 2003) can not produce phase-aligned radio and gamma-ray light curves.

Low-altitude, gamma-ray emission models such as the PC model (see Chapter 2) do predict phase-alignment but, as discussed in \citet{AbdoJ0034}, standard PC models are not able to properly match the observed gamma-ray light curve shapes.  However, low-altitude SG models may be a viable alternative.  These models have been explored in more detail by \citet{Venter11} but are beyond the scope of this study.

The radio light curves of phase-aligned MSPs are simulated using alTPC and alOG models.  The emitting regions are defined in the same manner as those of the gamma-ray alTPC/OG models with one exception.  In the radio alTPC/OG models the minimum altitude of emission (R$_{min}$) is also a free parameter.

In particular, the value of R$_{min}$ can be different from the stellar surface (alTPC) or NCS (alOG) for the radio simulations.  Note that emission in the alOG radio models is still confined to be above the null-charge surface and thus the true minimum emission altitude is a function of $\alpha$, $\theta_{\mu}$, and $\phi_{\mu}$.

Modeling the radio emission as significantly extended in altitude implies that it is of caustic origin.  \citet{Manchester05} and \citet{Ravi10} have also argued that radio pulsar emission should be from further out in the magnetosphere in wide, fan-like beams based on comparisons of gamma-ray and radio pulsar populations.  However, the caustic nature of the emission also affects the polarization properties \citep{Venter11} and thus applicability to all pulsars is unclear.

\subsection{Beaming Correction Factors}\label{ch5beam}
One uncertainty which has plagued gamma-ray pulsar science is the determination of gamma-ray luminosity when it is not known how the emission is distributed across the sky.  In particular, for a pulsar with distance $d$ and observed gamma-ray energy flux $G_{obs}$, the gamma-ray luminosity can be expressed as \citep{Venter08},
\begin{equation}\label{ch5lgam}
L_{\gamma}\ =\ G_{obs}d^{2}\lambda
\end{equation}
\noindent{}where $\lambda\ =\ \varepsilon\bar{\Delta\Omega}/\beta^{obs}$ corrects for the fact that the pulsar emission is not isotropic.  The parameter $\varepsilon$ is defined to be $\beta^{obs}G_{tot}/G_{obs}$, $\beta^{obs}$ is the pulsar duty cycle, and $\bar{\Delta\Omega}$ is the average beaming angle.

Without knowing the details of the emitting geometry this ambiguity can not be resolved.  However, as noted by other authors (e.g., Watters et al., 2009 and Venter et al. 2009) the necessary beaming correction factor can be estimated from geometric simulations such as those described in this chapter.  Following Eq. 4 of \citet{Watters09} the fraction of 4$\pi$ into which the pulsar emission is beamed can be estimated using Eq.~\ref{ch5fomega}.
\begin{equation}\label{ch5fomega}
\rm{f}_{\Omega}(\alpha,\zeta)\ =\ \frac{\iint \mathnormal F_{sim}(\alpha,\zeta^{\prime},\phi)\sin(\zeta^{\prime})d\zeta^{\prime} d\phi}{2\int F_{sim}(\alpha,\zeta,\phi)d\phi}
\end{equation}

The quantity $F_{sim}(\alpha,\zeta,\phi)$ is the simulated flux for a given magnetic inclination, viewing angle, and phase bin.  The numerator sums the emission over the whole sky while the denominator sums only that emission seen at a particular viewing angle.  This correction factor can be connected to Eq.~\ref{ch5lgam} by equating $\lambda\ \approx\ 4\pi\rm{f}_{\Omega}$.

Calculating f$_{\Omega}$ from the simulations described here requires summing over the phase plot bins for a given $\alpha$ (see Fig.~\ref{ch5zphi}).  The numerator in Eq.~\ref{ch5fomega} is calculated as the sum of all bins while the denominator is twice the sum of the phase bins for a given $\zeta$.

\section{Simulations}\label{ch5sims}
TPC, OG, and PSPC simulations have been generated for periods of 1.5, 2.5, 3.5, 4.5, and 5.5 ms while alTPC and alOG simulations have only been generated for a period of 1.5 ms.  Hollow-cone beam radio simulations have been generated for the same five spin periods at frequencies which closely match the observations (values of 300, 800, 1400, and 3000 MHz).

The code which generated the simulations used in this thesis was originally developed by Alice Harding and Joe Daugherty \citep{DH96} to produce PC model light curves using a static dipole field.  Jarek Dyks later modified the code to use the retarded dipole field and to calculate polarization \citep{Dyks04}.  The current version has further been modified to allow for PSPC and altitude-limited models by Christo Venter as well as to incorporate the Lorentz transformation of the IOF magnetic field as detailed in Section~\ref{ch5simLCs} and Appendix B.

The simulations are generated with a resolution of 1\DEG{} in $\alpha$ and with steps of 0.05 in $r_{ovc}^{min}$ and $r_{ovc}^{max}$.  Note that this rather coarse resolution in the $r_{ovc}$ parameters results in most OG models having zero width emission layers.  For the altitude-limited models, steps of 0.1 (in units of R$_{\rm LC}$) have been used for the emission altitudes.  For a period of 1.5 ms R$_{\rm NS}$ = 0.14 R$_{\rm LC}$ which results in one altitude step of 0.06 between of R$_{\rm NS}$ and R$_{max}$ = 0.2.

An overview of the simulated parameter ranges for each model is given in Table~\ref{ch5pars}. For the OG and alOG models, $r_{ovc}^{min}$ can only take on the value of 0.85 if $r_{ovc}^{max}$ = 0.9.

\begin{deluxetable}{l c c c c c}
\tablewidth{0pt}
\rotate
\tablecaption{Simulation Parameter Ranges}
\startdata
\underline{Model} & \underline{$\alpha$ (\DEG{})} & \underline{$r_{ovc}^{min}$ ($\Theta_{\rm PC}$)} & \underline{$r_{ovc}^{max}$ ($\Theta_{\rm PC}$)} & \underline{R$_{min}$ (R$_{\rm LC}$)} & \underline{R$_{max}$ (R$_{\rm LC}$)} \\
TPC & 1-90 & \{1.0,0.95,0.9\} & 1.0 & R$_{\rm NS}$ & 1.2 \\
OG & 1-90 & \{1.0,0.95,0.9,0.85\} & \{1.0,0.95,0.9\} & R$_{NCS}$ & 1.2\\
PSPC & 1-90 & 0.0 & 1.0 & R$_{\rm NS}$ & 1.2\\
alTPC (gamma) & 1-90 & \{1.0,0.95,0.9\} & 1.0 & R$_{\rm NS}$ & \{0.7,0.8,...,1.2\}\\
alOG (gamma) & 1-90 & \{1.0,0.95,0.9,0.85\} & \{1.0,0.95,0.9\} & R$_{NCS}$ & \{0.7,0.8,...,1.2\}\\
alTPC (radio) & 1-90 & \{1.0,0.95,0.9\} & 1.0 & \{R$_{\rm NS}$,0.2,...,0.9\} & \{0.2,0.3,...,1.2\}\\
alOG (radio) & 1-90 & \{1.0,0.95,0.9,0.85\} & \{1.0,0.95,0.9\} & \{R$_{\rm NS}$,0.2,...,0.9\}\tablenotemark{a} & \{0.2,0.3,...,1.2\}\\
\enddata
\tablenotetext{a}{For the radio alOG models R$_{min}$ is taken as a lower limit since the emission is constrained to be above R$_{\rm NCS}$ which is a function of $\alpha$, $\theta_{\mu}$, and $\phi_{\mu}$.}\label{ch5pars}
\end{deluxetable}

Models with infinitely-thin gaps, $r_{ovc}^{min}\ =\ r_{ovc}^{max}\ =\ 1.0$, are unphysical. For any MSP where the best-fit parameters result in zero width (see Chapter 6) this should be taken to mean that the true gap width is less than the current resolution of 0.05.

For a select set of simulation parameters the PC rim was not completely defined.  In particular, for a spin period of 4.5 ms and $\alpha$ = 42\DEG{} and for a spin period of 5.5 ms and $\alpha$ = 42, 45, and 69\DEG{} a field line with $\hat{B}_{\rm LC}\cdot\hat{\rho}\ =\ 0$ could not be found for at least some range of $\theta_{\mu}$ and $\phi_{\mu}$.

However, the procedure was successful if values of $\alpha$ = 41.9, 44.9, and 68\DEG{}.9 were used.  Thus, for all simulations with 4.5 and 5.5 ms periods the latter $\alpha$ values were used to generate the light curves but in reporting best-fit values they are treated as 42, 45, and 69\DEG{}.

\section{Conclusions}\label{ch5conc}
This chapter outlines a method of simulating gamma-ray and radio light curves without the need for a detailed radiation model.  Similar simulations have been used previously to study pulsar light curves and polarization angle sweeps (e.g., Romani \& Yadigaroglu, 1995; Dyks \& Rudak, 2003; Venter et. al, 2009).

This approach exploits the fact that the shape of a pulsar's light curve should reflect the structure of the magnetosphere along the line of sight.  While any energy dependent behavior will require more detailed physical models, much can be inferred about the field structure simply by matching the basic light curve shape.

In particular, the key components to consider when matching observations to geometric models are: peak multiplicity, peak separation, radio-to-gamma lag, and off-peak emission level.

Full radiation models provide detailed information about the emission regions and particle populations (e.g., Du et al., 2011) but are time intensive and must be tailored to each source.  Geometric models can be used to derive the basic properties of many sources which can then be used as inputs to population synthesis studies where specific details of the emission processes are of less importance.

The simulations described in Section~\ref{ch5sims} have finer resolution in $\alpha$ than previous studies which will allow for better comparison of the best-fit geometries with estimates from radio and X-ray observations (see Chapters 6 and 7).
\renewcommand{\thechapter}{6}
\chapter{\bf Likelihood Fitting Method}\label{ch6}
Previous modeling studies of MSP gamma-ray and radio light curves estimated viewing geometries by eye (e.g., Venter et al., 2009 and Abdo et al., 2010d).  This involved scanning through the models for geometries which were known to give the appropriate number of peaks, reproduce the observed radio-to-gamma lag, and match the gamma-ray peak separation (in the case of two-peaked gamma-ray light curves).

While these studies were successful in finding models which matched the observed profiles well such methods are time consuming, provide no means by which one model can be preferred over another, and tend to favor fitting the gamma-ray light curves better than the radio.  Additionally, if one wishes to explore a larger parameter space (i.e. different gap widths, emission altitudes, etc.) it is unclear how the best parameters would be chosen with confidence and the number of light curves to scan grows with each new parameter.  Therefore, it is desirable to create a fitting technique to statistically determine the best-fit parameters for a given emission model.

Maximum likelihood estimation methods are well suited to fitting data using complex models with large parameter spaces and providing estimates of the best-fit parameters.  The likelihood value ($\mathcal L$) is proportional to the probability that a model (with a given set of parameters) accurately describes the data.

The value $\mathcal L$ itself is not a goodness-of-fit measure; however, the ratio of likelihood values for different models (used to fit the same data) can be used to reject one model over another (e.g., the LRT as described in Section~\ref{ch3specAn}).  In practice, it is often easier to cast a maximum likelihood technique as a problem of minimizing $-\log(\mathcal L)$.  As such, the likelihood function and fitting techniques described below will generally deal with these values and not the likelihoods themselves.

While the likelihood function for a given set of model parameters and observed profiles can be written analytically (see Section~\ref{ch6likefunc}) the multi-dimensional likelihood surface itself is quite complex and does not easily lend itself to a functional form.  Markov chain Monte Carlo (MCMC) methods are designed to map out unknown distributions and lend themselves nicely to maximum likelihood problems (e.g., Verde et al., 2003).

\section{Likelihood Function}\label{ch6likefunc}
Poisson likelihood is used to describe experiments in which a certain number of events are expected to be observed in a given amount of time.  In particular, the likelihood of observing $x$ events (for $x$ a nonzero integer) at a rate $\lambda$ is given by Eq.~\ref{ch6Plike}.
\begin{equation}\label{ch6Plike}
\mathcal{L}\ =\ \frac{\lambda^{x}\exp\Big\lbrace -\lambda \Big\rbrace}{x!}
\end{equation}

This likelihood statistic is well suited to gamma-ray pulsar light curves.  With Poisson likelihood, the uncertainty in the data value $x$ is $\pm\sqrt{x}$ (for $x\ \gg\ 1$).

The radio profiles, however, do not consist of integer counts and thus Poisson likelihood is not applicable.  However, a $\chi^{2}$ statistic can be used to fit the radio profiles and turned into a likelihood, Eq.~\ref{ch6chilike}, using Wilks' theorem \citep{Wilks38}.
\begin{equation}\label{ch6chilike}
\rm \Delta\log(\mathcal{L})\ =\ -0.5\Delta\chi^{2}
\end{equation}

When using Eq.~\ref{ch6Plike} the likelihood is maximized by fitting the parameter $\lambda$.  For the gamma-ray light curves, it is assumed that the counts in each phase bin follow Poisson statistics and the value of $\lambda$ for each bin corresponds to the value of the model light curve on top of a constant background.

At first, this seems to be a problem with many variables to fit when fitting a typical light curve with sixty bins.  However, the problem simplifies when one recognizes that the individual model bin values are not independent.  In fact, they must maintain a fixed ratio with respect to one another, a relation which can be exploited to simplify the likelihood maximization to a problem of optimizing one variable (two when including the radio profiles) as shown in Eq.~\ref{ch6grayFunc}.
\begin{equation}\label{ch6grayFunc}
-\log(\mathcal{L}_{\gamma})\ =\ -\log\bigg[\prod_{i=0}^{\rm{N}-1} \frac{(c_{\gamma,i}\lambda_{\psi}+b_{\gamma})^{d_{\gamma,i}}\exp\Big\lbrace-(c_{\gamma,i}\lambda_{\psi}+b_{\gamma})\Big\rbrace}{d_{\gamma,i}!}\bigg]
\end{equation}

For each light curve bin (from 0 to N-1) the $i^{th}$ gamma-ray datum value is represented by $d_{\gamma,i}$, the background estimate by $b_{\gamma}$, and the model value by $c_{\gamma,i}\lambda_{\psi}$ for some reference bin $\psi$ such that $\lambda_{\psi}\ \neq\ 0$ with $c_{\gamma,i}\ \equiv\ \lambda_{i}/\lambda_{\psi}$.

The radio profiles are handled in a similar manner as indicated by Eq.~\ref{ch6radioFunc}.
\begin{equation}\label{ch6radioFunc}
-\log(\mathcal{L}_{\rm{R}})\ =\ \frac{0.5}{\sigma_{\rm R}^{2}} \mathnormal \sum_{i=0}^{M-1}\Big((c_{\rm{R},i}R_{\Psi}+b_{\rm{R}})-d_{\rm{R},i}\Big)^{2}
\end{equation}

For each bin (from 0 to M-1, with M$\geq$N) the $i^{th}$ radio datum value is represented by $d_{\rm{R},i}$, the background estimate by $b_{\rm R}$, the error used for each radio bin is $\sigma_{\rm R}$, and the model value by $c_{\rm{R},i}R_{\Psi}$ for some reference bin $\Psi$ such that $R_{\Psi}\ \neq\ 0$ with $c_{\rm{R},i}\ \equiv\ R_{i}/R_{\Psi}$.

In order to balance the relative contributions from the gamma-ray and radio light curves to the total likelihood ($\mathcal L\ =\ \mathcal L_{\gamma} \mathcal L_{\rm R}$), $\sigma_{\rm R}$ has been chosen to be the average gamma-ray relative uncertainty in the ``on-peak'' region times the maximum value of the radio light curve.  In cases where the radio profile has more bins than the gamma-ray light curve $\sigma_{\rm R}$ is divided by the ratio of the radio to gamma-ray bin numbers.

While this approach does simplify the likelihood maximization, it also requires that the model and observed profiles have matching numbers of bins.  As detailed in Chapter 5, both the radio and gamma-ray model light curves are generated with 180 bins in phase which means that the observed light curves must be binned such that N and M are both integer divisors of 180.  This results in a loss of fine structure for the radio profiles but this would not be significant given the values of $\sigma_{\rm R}$ chosen for this study.

\section{Markov Chain Monte Carlo}\label{ch6MCMC}
As there is no clear functional form for the likelihood surface in the model parameter space, only an equation to calculate the likelihood for a given parameter state, it is not possible to analytically solve for the maximum.  However, as mentioned above, MCMC techniques are designed for problems which involve samples drawn from a distribution which can be calculated at discrete points but for which an analytic expression is not known.

A Markov chain is a series of parameter states which have the Markov property \citep{Markov06}, namely, the probability of the $(i+1)^{th}$ state being added to the chain only depends on the $i^{th}$ state.  An MCMC involves taking random steps in parameter space, based upon a proposal distribution ($P_{\rm step}$), where the step is taken or not based on a specified acceptance criterion.

A commonly used acceptance criterion in MCMC maximum likelihood analysis is the Metropolis-Hastings algorithm \citep{Hastings70} which uses the likelihood ratio of consecutive steps to determine whether or not a step is added to the chain.  In particular,  let the current parameter state in the chain be $\mathbf x$ with likelihood $\mathcal L$.  Let the proposed parameter step be $\mathbf x^{\prime}$ with likelihood $\mathcal L^{\prime}$.  Using the likelihood ratio $\Lambda\ \equiv\ \mathcal L^{\prime}/ \mathcal L$ the step is added to the chain if $\alpha_{\rm MH}$, Eq.~\ref{ch6MH}, is greater than a random number $\epsilon\in[0,1)$.
\begin{equation}\label{ch6MH}
\alpha_{\rm{MH}}\ =\ min\Big(\Lambda\times\frac{P_{\rm step}(\mathbf{x}^{\prime}|\mathbf{x})}{P_{\rm step}(\mathbf{x}|\mathbf{x}^{\prime})},1\Big)
\end{equation}

In Eq.~\ref{ch6MH}, $P_{\rm step}(\mathbf{x}|\mathbf{x}^{\prime})$ denotes to the probability of going from $\mathbf x$ to $\mathbf x^{\prime}$.  In the event that $P_{\rm step}$ is symmetric (i.e., $P_{\rm step}(\mathbf{x}|\mathbf{x}^{\prime})\ =\ P_{\rm step}(\mathbf{x}^{\prime}|\mathbf{x})$) these quantities cancel and this reduces to the Metropolis method \citep{Metropolis53}.

This type of acceptance criterion has the positive aspect that it naturally moves towards maxima of the distribution in question but does provide some probability of taking a step to a state where the distribution has a lower value which allows the chain to more fully explore the parameter space.  For a multi-modal distribution this does run the risk of the chain spending long times in local maxima which can lead to poor mixing (i.e., not fully sampling the parameter space) and low acceptance rates (i.e., rejecting many steps between each accepted step).  Two methods to mitigate this issue and a test for convergence are described below.

Note that it is necessary to undergo an initial burn-in period, which uses the same proposal distribution and acceptance criteria, but for which no steps are added to the chain.  This is done to remove any dependence on the initial parameters.

\subsection{Small-World Chains and Simulated Annealing}\label{ch6improve}
\begin{figure}[h]
\begin{center}
\includegraphics[width=0.75\textwidth]{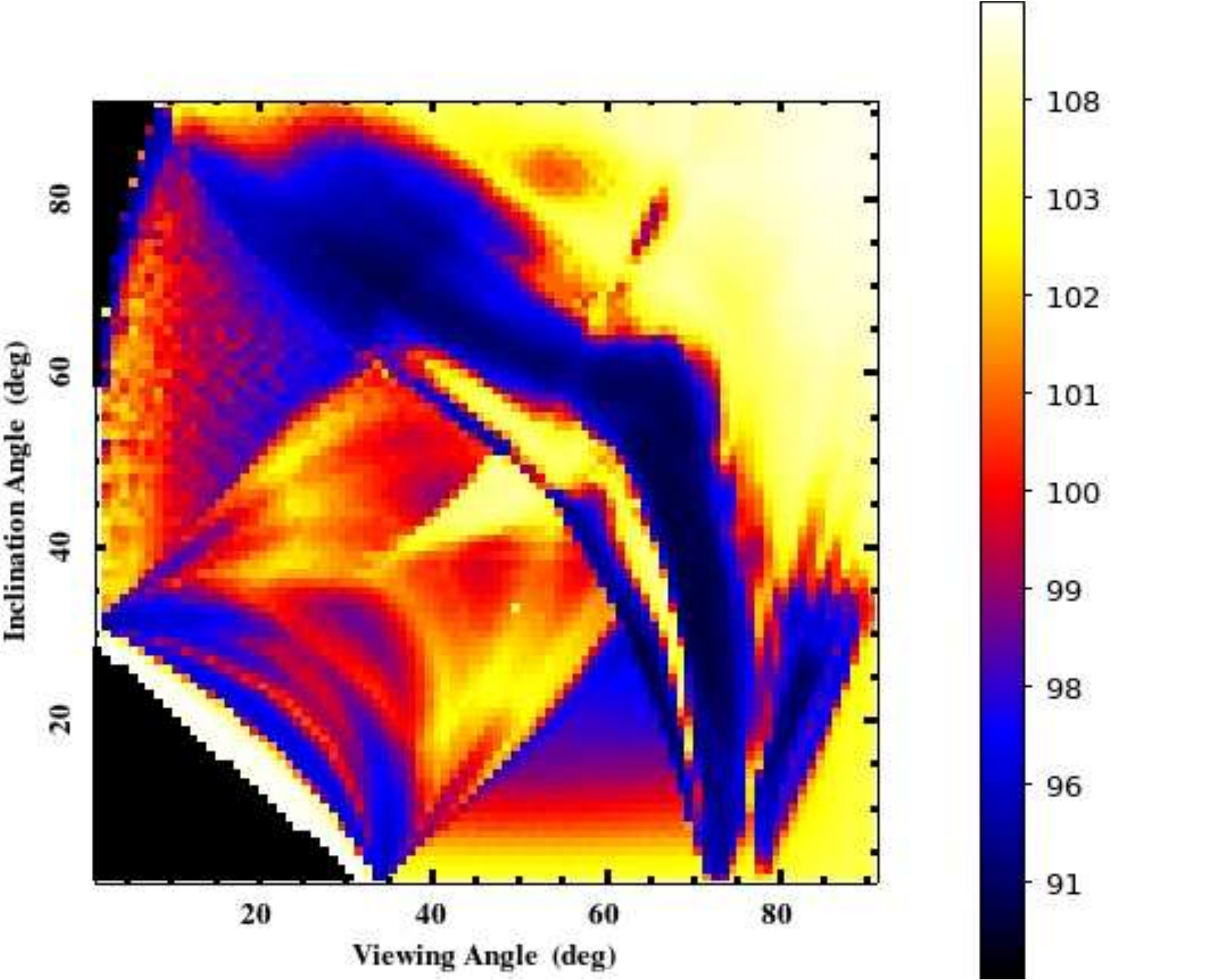}
\end{center}
\small\normalsize
\begin{quote}
\caption[Example likelihood surface]{Example of a likelihood surface for preliminary tests of the likelihood function for PSR J0034$-$0534, gamma-ray fit only, the color scale gives the $-\log(\mathcal L)$ for that geometry assuming a gap width of 0.05.  The black triangles near ($\alpha$,$\zeta$) = (0\DEG{},0\DEG{}) and for $\zeta\ \sim$ 0 and $\alpha\ \gtrsim$ 60\DEG{} represent viewing geometries which result in no visible emission.  Note that there are several local minima.\label{ch6llsurf}}
\end{quote}
\end{figure}

Preliminary profile scans of the $\alpha$-$\zeta$ plane for fixed gap width demonstrated that the likelihood surfaces for the MSP light curve fits can be very multi-modal (e.g., Fig.~\ref{ch6llsurf}); thus, it was clear that steps needed to be taken while designing the MCMC light curve fitting to ensure good acceptance and mixing.  This has been done through the use of small-world chains (Guan et al., 2006 and Guan \& Krone, 2007) and simulated annealing (Marinari \& Parisi, 1992 and Guan \& Krone, 2007).

\small\normalsize

The small-world chains approach involves combining a local step, one which decays exponentially, with a probability $1-s$ and a global step, one which decays as a power law, with a probability $s$.  Such a proposal distribution results in the MCMC occasionally taking a very large step in parameter space which can facilitate moving away from a local maximum.

\begin{figure}[h]
\begin{center}
\includegraphics[width=0.9\textwidth]{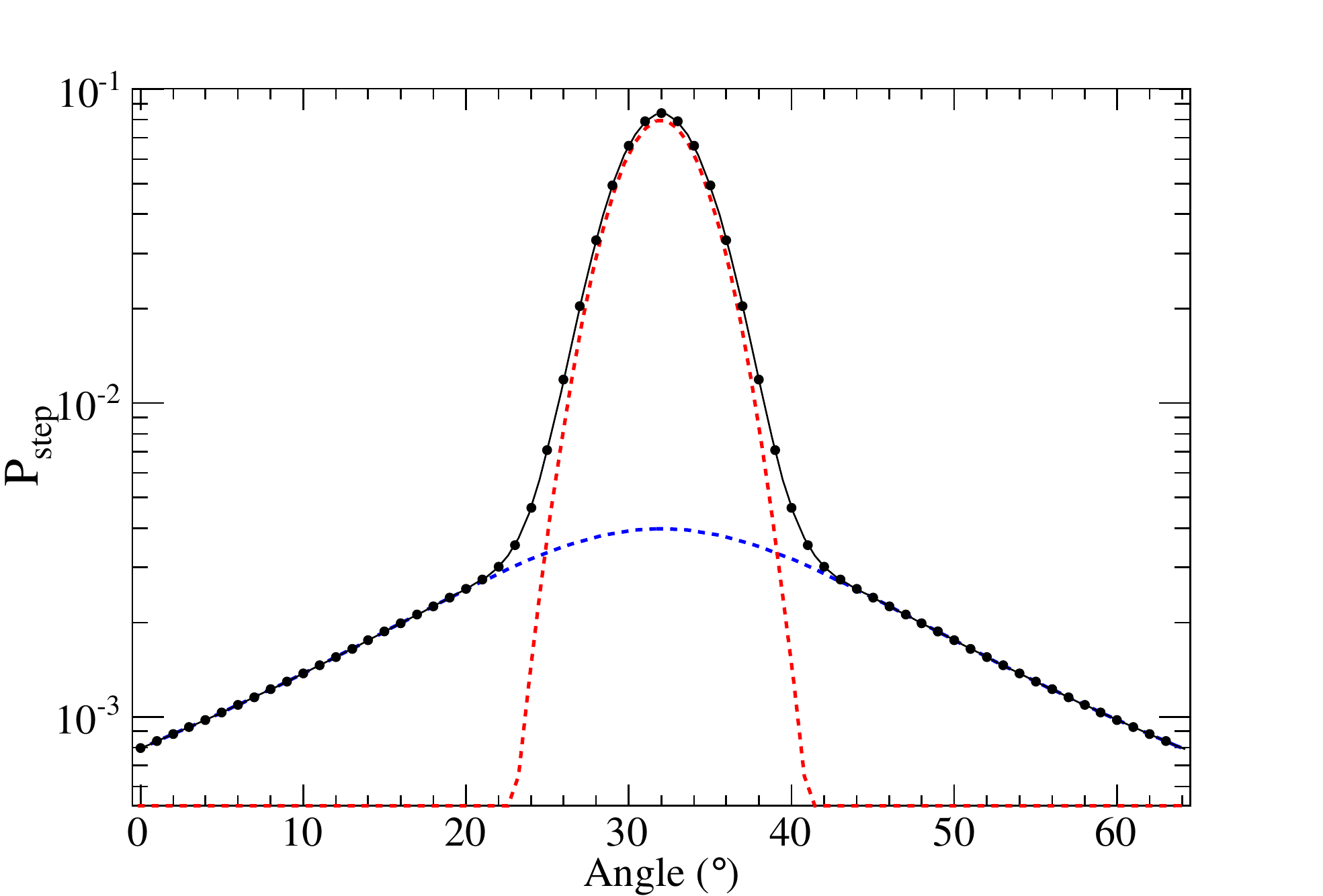}
\end{center}
\small\normalsize
\begin{quote}
\caption[Example small-world probability distribution]{Example of the small-world chains step distribution with $s$=1/5, Gaussian width ($\sigma$) of 4\DEG{}, Lorentzian width ($w$) of 16\DEG, and centered at $x_{0}$ = 32\DEG{}.  The red dashed line is the Gaussian component and the blue dashed line is the Lorentzian.  The solid black line is the combined function as given in Eq.~\ref{ch6Prop} and the black points are from the \emph{scipy.stats.rv\_continuous} realization of this distribution demonstrating the fidelity of the implementation.\label{ch6step}}
\end{quote}
\end{figure}

For the MCMC results presented in Chapter 7, the proposal distribution draws trial $\alpha$ and $\zeta$ parameters using a small-world chain step consisting of a narrow Gaussian combined with a wider Lorentzian, see Eq.~\ref{ch6Prop} and Fig.~\ref{ch6step}.
\small\normalsize
\begin{equation}\label{ch6Prop}
P(x,x_{0})_{\alpha/\zeta,\rm{step}} =\ (1-s) \frac{1}{\sqrt{2\pi\sigma^{2}}} \exp \Big \lbrace \mathnormal{\frac{-(x-x_{0})^{2}}{\sigma^{2}} \Big \rbrace\ +\ s \frac{1}{\pi w} \frac{1}{1+\frac{(x-x_{0})^{2}} {w^{2}}}}
\end{equation}

While a small-world chain step is useful for fully exploring the phase-space, it is not sufficient for cases where the local maxima rise steeply and thus the large steps are rarely accepted due to low likelihood ratios.  In such cases, simulated annealing is a useful method of leveling the distribution in order to make accepting steps more likely by effectively introducing a temperature to the distribution.

Instead of using the distribution in question, $f(\mathbf{x})$, to calculate $\Lambda\ =\ f(\mathbf{x}^{\prime})/f(\mathbf{x})$ in Eq.~\ref{ch6MH} one instead uses $f_{t}(\mathbf{x})\ \equiv\ f(\mathbf{x})^{1/t}$ with $t\in\lbrace1,2,..,n\rbrace$.  For high values of $t$, a hot distribution, the local maxima are smoothed down which makes accepting a new step more likely.

After updating the current step using the modified form of Eq.~\ref{ch6MH}, one then updates the temperature if $\alpha_{\rm{MH},t}$, Eq.~\ref{ch6SAupdt}, is greater than a random number\\ $\epsilon\in[0,1)$.
\begin{equation}\label{ch6SAupdt}
\alpha_{\rm{MH},t}\ =\ min\Big(\frac{f_{t^{\prime}}(\mathbf{x})q(t^{\prime})p(t^{\prime}|t)}{f_{t}(\mathbf{x})q(t)p(t|t^{\prime})},1\Big)
\end{equation}

In Eq.~\ref{ch6SAupdt}, $q(t)$ is the auxiliary probability of temperature $t$ and $p(t|t^{\prime})$ is the probability of transitioning from temperature $t$ to $t^{\prime}$ which is equal to 1 if $t\in\lbrace1,n\rbrace$ and 0.5 otherwise.

\begin{figure}
\begin{center}
\includegraphics[width=1.\textwidth]{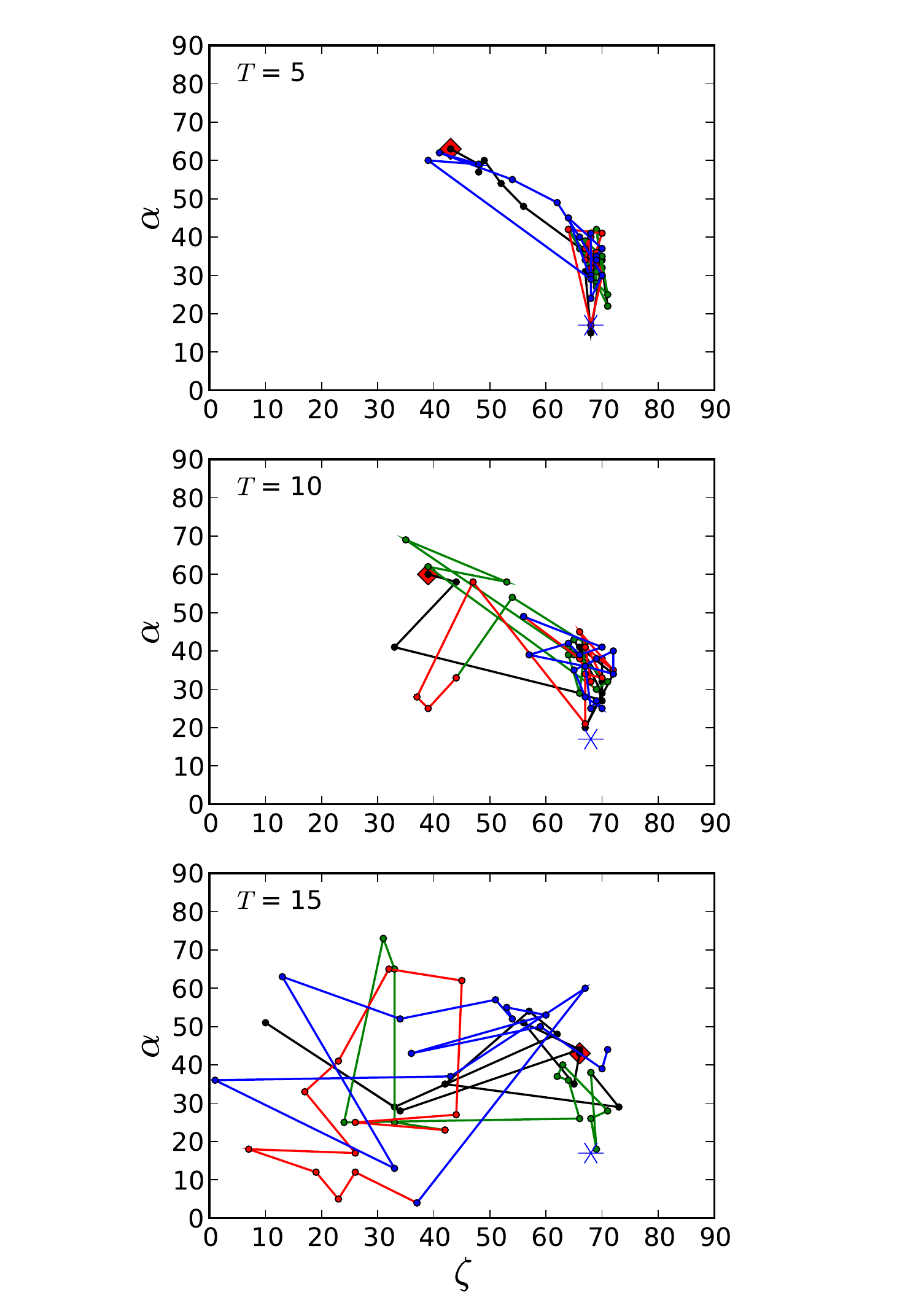}
\end{center}
\small\normalsize
\begin{quote}
\caption[Example step history for different maximum temperatures]{Accepted ($\alpha$,$\zeta$) steps for 3 chains (post burn-in) with different $T$ values.   Chains start at red diamond, steps proceed black to green to red to blue, best-fit is shown as a blue asterisk.  The units of the axes are degrees.\label{ch6Steps}}
\end{quote}
\end{figure}

Example chains for different values of $T$ are shown in Fig.~\ref{ch6Steps}.  Note that all three chains have included steps near the best-fit geometry (blue asterisk) but that the chain with the highest $T$ (bottom) spends little time exploring that region.  The chain with the lowest value of $T$ explores the region near the best-fit very well but does not stray far from it, if this chain were to have started in a local maxima it would take a long time to get out of it.  The middle plot has a more reasonable value of $T$ such that the region around the best-fit is well explored but the chain also samples the surrounding phase-space well.

\small\normalsize

One of the main appeals of choosing the best-fit model via a maximum likelihood analysis is to provide confidence contours in viewing geometry (see Section~\ref{ch6CnC}).  For the light curve fits presented in this thesis confidence contours are generated by marginalizing over the other fit parameters.  However, while simulated annealing does not change the likelihood distribution it does have a significant impact on the confidence contours.  Larger values of $T$ will include parameter states with lower likelihood values more frequently, compared to using a smaller value of $T$, and thus widen the confidence contours.  Therefore, care should be taken when choosing $T$ in order to strike a reasonable balance between having swift computation times and meaningful confidence contours.

\subsection{Convergence and Confidence Contours}\label{ch6CnC}
The light curve fits presented in Chapter 7 are from MCMC analyses consisting of either 8 or 16 chains (the larger number being used for altitude-limited models which have larger parameter spaces) consisting of 12,500 steps each.  This was done in order to fill out the tails of the confidence contours and to ensure that the chains had converged.  Convergence was verified using the criterion of \citet{GR92} as described by \citet{Verde03} and outlined below.

For a given parameter $x$ (i.e., $\alpha$, $\zeta$, etc.), this convergence criterion involves comparing two estimates for the variance of the distribution as represented by the last half of the parameter states in each chain.  Let the number of steps in each chain be $N_{\rm step}$ and the number of chains be $N_{\rm chain}$.

To start, the average value of $x$ within the $j^{th}$ chain is calculated as\\ $\langle x^{j} \rangle\ =\ (N_{\rm step}/2)^{-1}\ \sum_{j= \mathnormal N_{\rm step}/2+1}^{\mathnormal N_{\rm step}} x_{i}$.  Next, the average of the individual chain averages is calculated as $\bar{x}\ =\ N_{\rm chain}^{-1}\ \sum_{j=1}^{N_{\rm chain}}\langle \mathnormal x^{j} \rangle$.

Using the individual $\langle x^{j} \rangle$ values and $\bar{x}$, the variance between chains ($\mathcal B$) and the variance within chains ($\mathcal W$) are calculated using Eqs.~\ref{ch6Bn} and~\ref{ch6W}, respectively.
\begin{equation}\label{ch6Bn}
\mathcal B\ =\ \frac{1}{\mathnormal N_{\rm{chain}}-1} \sum_{j=1}^{N_{\rm{chain}}}(\langle x^{j} \rangle - \bar{x})^{2}
\end{equation}
\begin{equation}\label{ch6W}
\mathcal W\ =\ \frac{1}{\mathnormal N_{\rm{chain}}(N_{\rm{step}}/2 - 1)} \sum_{j=1}^{N_{\rm{chain}}} \bigg\lbrace \sum_{i=(N_{\rm{step}}/2+1)}^{N_{\rm{step}}} (x_{i}^{j}-\langle x^{j} \rangle)^{2} \bigg\rbrace
\end{equation}

If the distribution is stationary, the variance can also be estimated using Eq.~\ref{ch6Rtop}, note that this will be an overestimate if the distribution is not stationary \citep{Verde03}.
\begin{equation}\label{ch6Rtop}
\mathcal V\ =\ \big[(\mathnormal N_{\rm{step}}/2-1)/(N_{\rm{step}}/2)\big]\mathcal W + \mathcal B(1+1/\mathnormal N_{\rm{chain}})
\end{equation}
 
The convergence criteria is the ratio $\mathcal R\ =\ \mathcal V/\mathcal W$.  \citet{GR92} recommend running the chains until $\mathcal R <$ 1.2 while \citet{Verde03} recommend a more conservative upper limit of 1.1.  For the number of chains and steps given above, the MCMC fits presented in Chapter 7 satisfied both convergence criteria with typical variance ratios of $\sim$ 1.0.

\begin{figure}
\begin{center}
\includegraphics[width=1.0\textwidth]{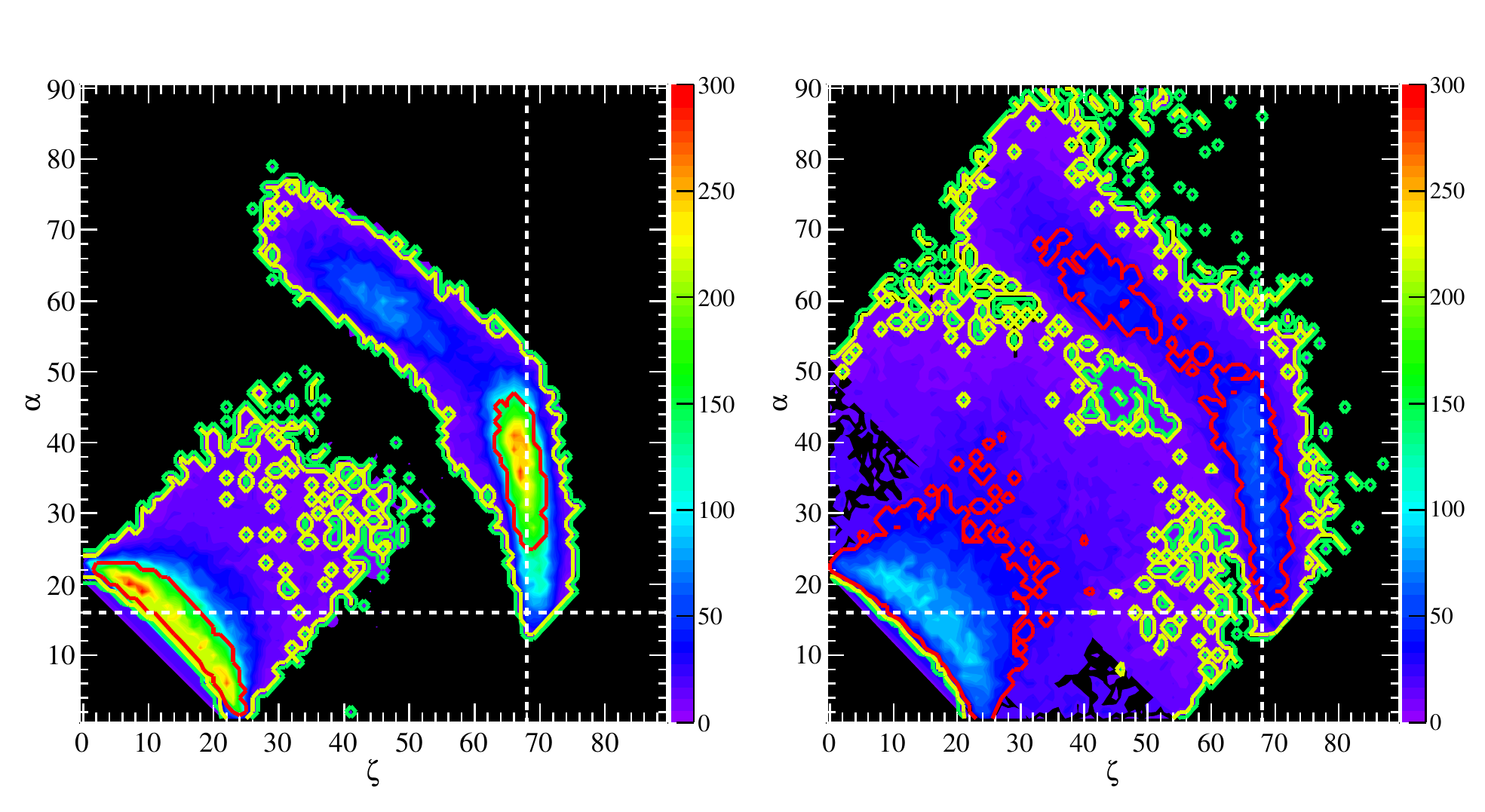}
\end{center}
\small\normalsize
\begin{quote}
\caption[Example marginalized confidence contours with different maximum temperatures]{Marginalized $\alpha$-$\zeta$ confidence contours from a TPC fit to PSR J2017+0603 using the data in \citet{Cognard11}, color scale represents number of times a given ($\zeta$,$\alpha$) pair were in an accepted parameter state in the chain.  Colored contours indicate 39\% confidence levels (red), 68\% (yellow), and 95\% (green).  White dashed lines indicate the best-fit solution.  The fit in the left panel used $T$ = 10 while that in the right panel used $T$ = 20.\label{ch6cont}}
\end{quote}
\end{figure}

Another advantage of the MCMC technique for maximum likelihood estimation is the ability to produce confidence contours in some sub-set of the parameters marginalized over the remaining parameters.  To do this, the ($\alpha,\zeta$) pairs for each step in the chain are collected in a 2-D histogram.  From this histogram, the confidence contours are constructed for 39, 68, and 95\% confidence levels corresponding to 1, 1.5, and 2.5 $\sigma$ confidence levels for 2 degrees of freedom, respectively. Fig.~\ref{ch6cont} presents an example of such marginalized contours for TPC fits to PSR J2017+0603 as described in \citet{Cognard11}.

\small\normalsize

To demonstrate the effect of $T$ on the marginalized contours the left panel of Fig.~\ref{ch6cont} corresponds to a fit with $T$ = 10 while the right panel is from a fit to the same data with $T$ = 20.  The color scales in Fig.~\ref{ch6cont} have been matched in order to demonstrate the difference in time spent near the best-fit solution for different values of $T$.  Note that the best-fit solution did not change between the two fits though a very large value of $T$ may result in the MCMC not finding the same best-fit solution as the chain will spend too much time in parameter states with low likelihood.

There are some instances for which the marginalized confidence contours do not agree well with the best-fit geometry.  In particular, if the best-fit geometry is right on the edge of the allowable parameter space (i.e., going a little ways in one direction leads to one or both profile components not being seen) then trial steps will often be to disallowed states which means they are not accepted.  Thus, the marginalized contours will be highly skewed in the opposite direction.

Additionally, constraining $\alpha\ \leq$ 90\DEG{} and $\zeta\ <$ 90\DEG{} can adversely affect the contours when one, or both, of the parameters is near the boundary.  In particular, if the proposed step is over the boundary the value is reflected back to an allowed state.  This leads to confidence contours which should be symmetric around the best-fit value being skewed such that the peak of the confidence contours is displaced from the best-fit geometry.

\subsection{Emission Altitude Likelihood Profiles}\label{ch6proflike}
In order to address the uncertainty in predicted emission altitudes from fits using the alTPC and alOG models, a profile likelihood method was used.  This method involves finding the lowest $-\log(\mathcal L)$ value in the resulting chains for a given emission altitude and comparing this to the overall minimum.  Differences in $-\log(\mathcal L)$ of 0.5 and 1.92 correspond to 1$\sigma$ and 2$\sigma$ uncertainties, respectively, for 1 degree of freedom.

\begin{figure}
\begin{center}
\includegraphics[width=1.0\textwidth]{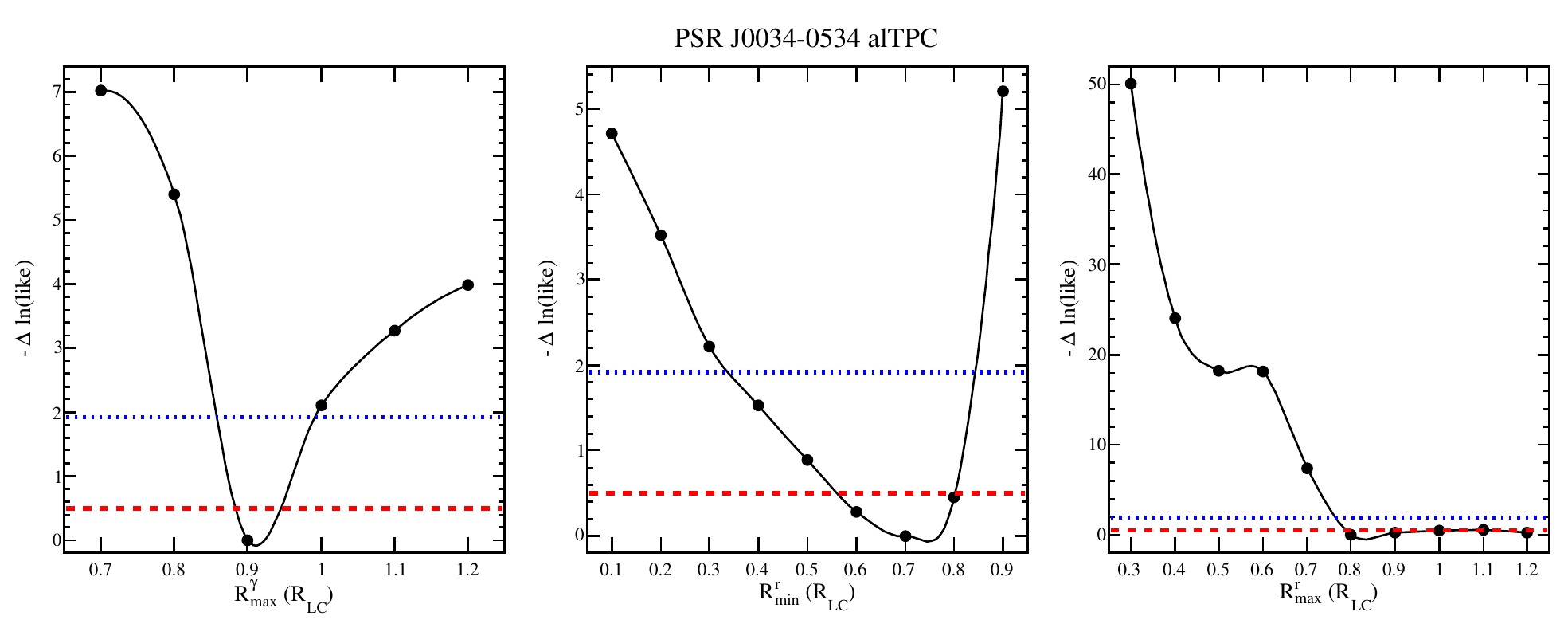}
\end{center}
\small\normalsize
\begin{quote}
\caption[Sample emission altitude likelihood profiles]{Likelihood profiles corresponding to alTPC fits to PSR J0034$-$0534 as described in \citet{Venter11}.  The likelihood profile for R$_{max}^{\gamma}$ is shown in the left panel, for R$_{min}^{\rm R}$ in the middle panel, and for R$_{max}^{\rm R}$ in the right panel.  The solid black lines are only drawn to guide the eye and do not represent the actual likelihood surface between the black points.  Below the red dashed line denotes 1$\sigma$ confidence while below the blue dashed line denotes 2$\sigma$ confidence.\label{ch6likeprof}}
\end{quote}
\end{figure}

Inspection of the likelihood profiles, see Fig.~\ref{ch6likeprof}, implied that most 1$\sigma$ uncertainties (but not all) were $\lesssim$ 0.1 R$_{\rm LC}$.  Therefore, for the best-fit values reported in Chapter 7 the 2$\sigma$ confidence intervals are quoted.  It is clear, however, that finer resolution in emission altitude is warranted.

Using the uncertainties derived from this method does assume that the parameter subspace with the emission altitude in question fixed at the given value has been explored sufficiently to find the overall minimum.  Near the best-fit value this should be the case as, compared to $\alpha$ and $\zeta$, the emission altitudes can take on a relatively small number of values and, as discussed in Section~\ref{ch6impl}, the total number of steps in the final chain is large.

\small\normalsize

\section{Implementation}\label{ch6impl}
The model light curves from simulations described in Chapter 5 are saved as ROOT\footnote{See http://root.cern.ch/drupal/ for documentation.} 1-D histograms with model specific specifiers in the file names in order to facilitate automated access by the MCMC code.  Additionally, the MCMC relies heavily upon the \emph{scipy}\footnote{See http://docs.scipy.org/doc/ for documentation.} python module.

The starting parameter state is chosen at random and the MCMC proceeds without recording steps until a specified number of burn-in steps have been accepted (50 for TPC, OG, and PSPC models and 100 for alTPC and alOG models).  The code finishes when a specified number of steps have been accepted to and recorded in the chain.

The MCMC parameter space for TPC and OG models consists of $\alpha$, $\zeta$, phase shift ($\Phi$), and $r_{ovc}^{min/max}$ with ranges given in Table~\ref{ch5pars}.  While the simulations do calculate $\zeta$ from 0\DEG{} to 180\DEG{}, the fitting restricts this range to be $<$ 90\DEG{} as the generated light curves for $\zeta\ \geq$ 90\DEG{} are identical to those $<$ 90\DEG{} but shifted by 180\DEG{} in phase.   The variable $\Phi$ is the number of bins to shift the model light curves in order to account for the fact that the definition of phase = 0 in the simulation is not, necessarily, the same as that used to produce the observed profiles.  The PSPC MCMC parameter space only consists of $\alpha$, $\zeta$, and $\Phi$.

The alTPC and alOG model parameter space contains that used for the standard TPC and OG models with the addition of R$_{min/max}$ as described in Chapter 5 with ranges given in Table~\ref{ch5pars}.  There are two R$_{min}$ and R$_{max}$ parameters for each fit, one for the radio and one for the gamma rays, and there are similarly two sets of $r_{ovc}^{min}$ and $r_{ovc}^{max}$, allowing for the gamma-ray and radio emission regions to have different gap widths.

The proposal distribution uses small-world steps (Eq.~\ref{ch6Prop}) in $\alpha$ and $\zeta$, both with Gaussian width ($\sigma$) of 4\DEG{} and Lorentzian width ($w$) of 20\DEG{} and with $s$ = 1/5.  The small-world step distribution is made using the \emph{scipy.stats.rv\_continuous} class for which the probability and cumulative distribution functions are supplied.  The black points in Fig.~\ref{ch6step} demonstrate that fidelity of this implementation.  Steps in $\Phi$ are chosen from a Gaussian with $\sigma$ = 4, rounded to the nearest integer to account for the binned nature of the light curves, using the \emph{scipy.random.normal} function.

The $r_{ovc}^{min}$ and $r_{ovc}^{max}$ parameters are chosen at random with a uniform probability.  In particular, for these parameters a list is made of the possible values and one of these is chosen at random using the \emph{scipy.random.randint} function.

The code verifies that the selected combination is valid.  A valid parameter state will have $r_{ovc}^{min}\ \leq\ r_{ovc}^{max}$ with the additional requirement for OG/alOG models that $(r_{ovc}^{max}-r_{ovc}^{min})\ \leq 0.5(1.0-r_{ovc}^{max})$ based on requiring realistic sizes for the emission layer as discussed in Chapters 2 and 5.

This type of random walk proposal distribution can lead to long convergence times as discussed below.  However, given the relatively small number of possibilities for these two parameters with the current, coarse resolution the random walk is not expected to affect the convergence time greatly.

For the alTPC and alOG models, initial analyses used a random walk to choose new R$_{min/max}$ parameters as was done for the $r_{ovc}^{min/max}$ parameters.  However, upon examination of the likelihood profiles for these parameters it was clear that the parameter space near the best-fit values was not being fully explored.  Therefore, a Gaussian step with $\sigma$ = 0.1 R$_{LC}$ was implemented, rounding to the nearest simulated value.

\begin{figure}
\begin{center}
\includegraphics[width=1.0\textwidth]{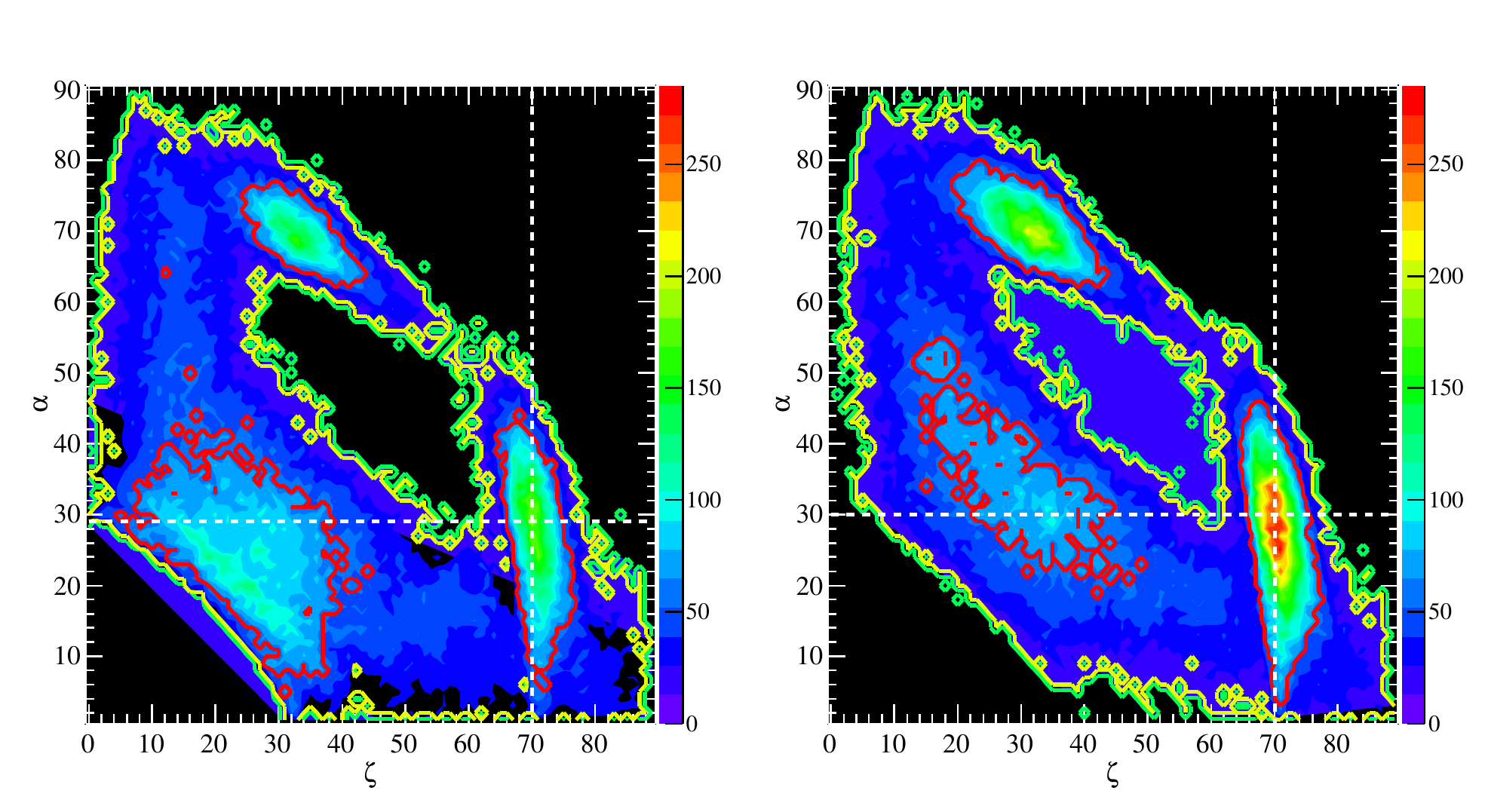}
\end{center}
\small\normalsize
\begin{quote}
\caption[Comparison of marginalized contours with different steps in altitude space]{Marginalized $\alpha$-$\zeta$ confidence contours from a TPC fit to PSR J0034$-$0534 using the same dataset as \citet{PWNcat}, color scale represents number of times a given ($\zeta$,$\alpha$) pair were in an accepted parameter state in the chain.  Colored contours indicate 39\% confidence levels (red), 68\% (yellow), and 95\% (green).  White dashed lines indicate the best-fit solution.  The fit in the left panel used a random walk to choose new emission altitudes while the right panel used a Guassian distribution.\label{ch6altStep}}
\end{quote}
\end{figure}

This was found to not only speed up the convergence and tighten the contours (see Fig.~\ref{ch6altStep}) but also allow the MCMC to find a slightly better solution as the chain could more fully explore the parameter space near the minimum.  Note that the best-fit $\alpha$ in the right panel of Fig.~\ref{ch6altStep} only changed by 1\DEG{} from that in the right panel when going from a random walk to a more local step in R$_{min/max}$ while the radio emission altitudes changed significantly.  Similar to steps in $\Phi$, the R$_{min/max}$ steps are chosen using the \emph{scipy.random.normal} function.

\small\normalsize

After the parameters have all been updated a check is made to ensure that the new step is not exactly the same as the previous step, in the extremely unlikely event that this occurs new trial parameters are chosen.  For each proposed step the model light curves are accessed and a check is made to ensure that neither model is empty, i.e. to make sure that both radio and gamma-ray emission are visible.  If one or both models is empty a new proposal step is chosen.

Once an acceptable proposal step is found, the models are rebinned to match the data (N bins for the gamma-rays and M for the radio such that M $\geq$ N).  The model light curves are then rescaled to integrate to the same values as the data profiles in order to provide a good starting point for optimization of the model normalizations.  A reference bin is chosen for both models ($\lambda_{\psi}$ and $R_{\Psi}$ in Eqs.~\ref{ch6grayFunc} and~\ref{ch6radioFunc}), the models are shifted $\Phi$ bins ($\Phi\times$M/N for the radio), and then the ratios $c_{\gamma,i}$ and $c_{\rm{R},i}$ are calculated.

In order to calculate the best likelihood value for a given parameter state, the model normalizations must be optimized to obtain the best-fit possible to the data.  To do this, the \emph{scipy.optimize.fmin\_l\_bfgs\_b} multi-variate, bound optimizer (Byrd et al., 1995 and Zhu et al., 1997) is used to find the minimum $-\log(\mathcal L)$ with those specific parameters.

The radio and gamma-ray model normalizations are effectively optimized separately as the corresponding contributions to the likelihood are independent.  In order to increase the speed of the optimizer, it is useful to supply a gradient function.  The gamma-ray and radio components of the joint log likelihood gradient are given in Eqs.~\ref{ch6grayDeriv} and~\ref{ch6radioDeriv}, respectively.  For optimization, the variables $\lambda_{\psi}$ and $R_{\Psi}$ are constrained to be $\geq$0 and to increase precision the optimizer inputs ``factr'' and ``pgtol'' are set to 1 and 1e-7, respectively.
\begin{equation}\label{ch6grayDeriv}
\rm \frac{d[log(\mathcal{L}_{\gamma})]}{d\lambda_{\psi}}\ =\ \mathnormal -\sum{}_{i=0}^{\rm{N}-1} \frac{c_{\gamma,i}(d_{\gamma,i}-(c_{\gamma,i}\lambda_{\psi}+b_{\gamma}))}{c_{\gamma,i}\lambda_{\psi}+b_{\gamma}}
\end{equation}
\begin{equation}\label{ch6radioDeriv}
\rm \frac{d[log(\mathcal{L}_{\rm{R}})]}{dR_{\Psi}}\ =\ \mathnormal{} \frac{1}{\sigma_{\rm R}^{2}}\ \sum_{i=0}^{\rm{M}-1} c_{i,\rm{R}}((c_{i,\rm{R}}R_{\Psi}+b_{\rm{R}})-d_{\rm{R}})
\end{equation}

The MCMC continues following the prescription described above and the final chain is output as a FITS file.  The resulting chains are merged, checked for convergence, and the parameter state corresponding to the maximum likelihood is reported.

It is not possible to use the LRT (described in Chapter 3) to evaluate whether or not one model can be preferred over another based on the maximum likelihood values.  The models can not, strictly speaking, be thought of as nested which means that the idea of null and trial hypotheses is ambiguous.

One is left with the option of simply comparing the difference in $-\log{\mathcal L}$ values and making general statements regarding how strongly one model is preferred over another (or not).  In particular, a difference of 10 in $-\log{\mathcal L}$ suggests that there is a difference of $\exp\lbrace10\rbrace\ \sim\ 22\times10^{3}$ between the probabilities that a given model describes the data.  However, this argument does not consider any systematic issues in the fitting technique.

In particular, the TPC and OG models predict very different levels of off-peak emission which means that the estimated background level in the LAT gamma-ray light curves can strongly affect the likelihood values.  A background estimate which is too high will systematically disfavor TPC models while an estimate which is too low will systematically disfavor OG models.  Likelihood fits of the same MSP light curve with gamma-ray background levels different by 1-2 counts/bin can lead to changes in $-\log{\mathcal L}$ for a given model by as much as 5-6.  Therefore, when discussing the fits in Chapter 7, a model is said to be somewhat preferred if the $-\log{\mathcal L}$ is smaller by $\sim15$.  The method by which the gamma-ray background estimates have been determined is discussed in Chapter 7.

\section{Conclusions}\label{ch6conc}
As will be seen in the next chapter, there are many advantages to using maximum likelihood to determine the best-fit parameters from geometric light curve models.  However, there are also some pitfalls.

Use of maximum likelihood allows for the creation of meaningful confidence intervals and, potentially, for one model to be favored over another.  While a trained eye can guess the approximate geometry which will fit the observed light curves it can be difficult to sift through all possibilities when using a resolution of 1\DEG{} in both $\alpha$ and $\zeta$.  Additionally, use of the MCMC allows for the discovery of unexpected solutions.

The likelihood is, by definition, tied to the model used to describe the data.  Therefore, any inadequacies in the model will strongly affect the location of the best-fit geometry.  In particular, the radio light curves of MSPs without aligned profiles are only fit with a hollow-cone beam model.  For those MSPs which display evidence for a core component the likelihood may try to fit part of the cone beam to this component and thus skew the geometry.  Additionally, the emission height of the cone beam is taken from an analysis of the profiles of non-recycled, longer period pulsars \citep{KG03}.  If this model does not apply equally well to MSPs it can be difficult to fit the observed profiles at the true geometry.  Even for MSPs which have no evidence of core emission the observed profiles can be quite complex and may require multiple cone beams.

The convention adopted here for choosing $\sigma_{\rm R}$ serves to mitigate some of the issues outlined above.  In particular, by balancing the likelihood contributions of the gamma-ray and radio light curves the best-fit solutions should be strongly influenced by getting the gamma-ray shape correct.  At the very least, the confidence contours will be widened such that regions which fit the gamma-ray profile well are included with high confidence and thus future constraints from radio and X-ray observations, along with improved radio models, will serve to narrow in on the true geometry.

\renewcommand{\thechapter}{7}

\chapter{\bf Results}\label{ch7}
With $\sim$13 months of data, pulsations from nine gamma-ray MSPs had been firmly detected with the LAT.  With more than one year of additional data and new MSPs detected in radio searches of \FL{} unassociated sources (e.g., Ransom et al., 2011; Cognard et al., 2011; Keith et al., 2011) the number of detected MSPs has grown to $\gtrsim$20, see Table~\ref{ch7MSPtable}.  The light curves of nineteen gamma-ray MSPs have been fit using the simulations described in Chapter 5 and the MCMC maximum likelihood procedure described in Chapter 6.  The results of these fits are discussed below.

\small\normalsize

\begin{deluxetable}{l c c c c}
\tablewidth{0pt}
\tablecaption{LAT Detected Gamma-ray MSPs}
\startdata
\underline{JName} & \underline{P$_{obs}$ (ms)} &\underline{$\dot{\rm P}_{\mathnormal obs}$\tablenotemark{a} ($10^{-20}$ s s$^{-1}$)} & \underline{$\nu_{obs}$ (MHz)} & \underline{H Test Value}\tablenotemark{b}\\ 
J0030+0451 & 4.8655 & 1.0169 & 1400 & 1793.3\\
J0034$-$0534 & 1.8772 & 0.26300 & 324 & 76.7\\
J0218+4232\tablenotemark{c} & 2.3231 & 7.7393 & 1400 & 113.7\\
J0437$-$4715 & 5.7575 & 0.6600 & 3000 & 473.6\\
J0613$-$0200 & 3.0618 & 0.5300 & 1400 & 387.8\\
J0614$-$3329 & 3.1487 & 1.7548 & 820 & 1963.3\\
J0751+1807 & 3.4788 & 0.6000 & 1400 & 111.1\\
J1231$-$1411 & 3.6839 & 2.2799 & 820 & 1719.6\\
J1614$-$2230 & 3.1510 & 0.4000 & 1500 & 72.7\\
J1713+0747 & 4.5700 & 0.8530 & 1400 & 44.8\\
J1744$-$1134 & 4.0745 & 0.7000 & 1400 & 93.7\\
J1823$-$3021A\tablenotemark{d} & 5.4400 & 340.0 & 1400 & 60.5\\
J1902$-$5105 & 1.7424 & 0.9000 & 1400 & 140.2\\
J1939+2134 & 1.5578 & 10.51 & 1400 & 23.1\\
J1959+2048 & 1.6074 & 0.7850 & 300 & 56.3\\
J2017+0603 & 2.8962 & 0.8300 &  1400 & 256.5\\
J2124$-$3358 & 4.9311 & 1.2000 & 1400 & 347.4\\
J2214+3000 & 3.1192 & 1.4011 & 820 & 112.4\\
J2302+4442 & 5.1923 & 1.3300 & 1400 & 429.9\\
\enddata
\tablenotetext{a}{Values have been corrected for the Shklovskii effect.}
\tablenotetext{b}{Values correspond to 2 years of diffuse class events with energies $\geq$ 0.1 GeV and within $0^{\circ}.8$ of the pulsar radio position.}
\tablenotetext{c}{Events taken from within $0^{\circ}.5$ of pulsar radio position.}
\tablenotetext{d}{Events with energies $\geq$ 0.5 GeV used.}\label{ch7MSPtable}
\end{deluxetable}

\small\normalsize

\section{Analysis Setup}\label{ch7setup}
For each of the MSPs in Table~\ref{ch7MSPtable} the gamma-ray light curves were constructed by selecting ``Diffuse'' class events, as defined under the P6 IRFs, from the first 2 years of LAT sky survey having reconstructed directions within 0\DEG{}.8 of the radio position; reconstructed energies from 0.1 to 100 GeV; and zenith angles $\leq$105\DEG{}.  Additionally, the LAT ST \emph{gtmktime} was used to exclude time periods when the rocking angle of the spacecraft exceeded 52\DEG{}, when the DATA\_QUALITY flag was set to zero, or when the LAT was not in standard science mode (i.e., LAT\_CONFIG $\neq$ 1).

The Htest value \citep{deJager89}, derived using the \FL{} ST \emph{gtptest}, for each MSP with these selection criteria is given in column 5 of Table~\ref{ch7MSPtable}.

The timing solutions were provided from radio observatories around the world under the PTC agreement \citep{Smith08}.  The pulse phase for each event was calculated using the radio timing solutions for the corresponding MSP and the TEMPO2\footnote{http://tempo2.sourceforge.net/} \citep{TEMPO2} fermi plug-in written by L.~Guillemot.  This tool applies the necessary time corrections (Eq.~\ref{ch2corrtime}) and then calculates the phase from the input pulsar parameters (eq.~\ref{ch2pphase}).  Note that the phase calculations and time corrections discussed in Chapter 2 were very simplistic and did not account for other fit terms which can be used to remove timing noise (though that is comparatively minimal in MSPs) which do not, strictly speaking, represent physical quantities.  The fermi plug-in accounts for all of these factors, see \citet{Guillemot09} for more detail concerning additional timing fit parameters.

The spin periods and radio frequencies used to generate the simulations to which the MSP light curves were fit are given in Table~\ref{ch7FitParTable}.  The estimated radio and gamma-ray background levels are given in columns 4 and 5.  The radio uncertainties in column 7 were estimated using the prescription described in Chapter 6 and the on-peak intervals given in column 6.

\small\normalsize

\begin{deluxetable}{l c c c c c c}
\tablewidth{0pt}
\rotate
\tablecaption{MSP Simulation and Fitting Parameters}
\startdata
\underline{JName} & \underline{P$_{sim}$ (ms)} & \underline{$\nu_{sim}$\tablenotemark{a} (MHz)} & \underline{b$_{\gamma}$ (counts)} & \underline{b$_{\rm R}$} & \underline{On-Peak} & \underline{$\sigma_{\rm R}$}\\ 
J0030+0451 & 5.0 & 1400 & 359 & 0.0 & [0.12,0.65] & 0.057\\
J0034$-$0534 & 1.5 & NA & 335 & 1.365 & [0.0,0.33]$\cup$[0.9,1.0] & 3.397\\
J0218+4232 & 2.5 & 1400 & 601 & 0.1 & [0.16,0.90] & 0.593\\
J0437$-$4715 & 5.5 & 3000 & 386 & 0 & [0.28,0.56] & 1.399\tablenotemark{b}\\
J0613$-$0200 & 3.5 & 1400 & 1137 & 0 & [0.05,0.45] & 6.765$\times10^{-4}$\\
J0614$-$3329 & 3.5 & 820 & 411 & 0.023 & [0.0,0.32]$\cup$[0.52,0.95] & 0.1195\\
J0751+1807 & 3.5 & 1400 & 433 & 0.0931 & [0.4,0.73] & 0.610\\
J1231$-$1411 & 3.5 & 800 & 498 & 0.0863 & [0.22,0.78] & 0.294\\
J1614$-$2230 & 3.5 & 1400 & 1405 & 0.2 & [0.2,0.84] & 0.248\\
J1713+0747 & 4.5 & 1400 & 812 & 0.024 & [0.14,0.53] & 0.259\\
J1744$-$1134 & 4.5 & 1400 & 2518 & 0.083 & [0.0,0.13]$\cup$[0.43,0.63]$\cup$[0.8,1.0] & 3.652\\
J1823$-$3021A & 1.5 & NA & 1975 & 0.2995 & [0.0,0.06]$\cup$[0.57,0.7]$\cup$[0.9,1.0] & 1.345\\
J1902$-$5105 & 1.5 & NA & 746 & 0.0 & [0.0,0.76] & 0.262 \\
J1939+2134 & 1.5 & NA & 7364 & 0.0023 & [0.0,0.35]$\cup$[0.5,0.77]$\cup$[0.86,1.0] & 0.027\\
J1959+2048 & 1.5 & NA & 1937 & 3.839 & [0.0,0.7] & 12.53\\
J2017+0603 & 2.5 & 1400 & 827 & 0.335 & [0.3,0.7] & 0.6337\\
J2124$-$3358 & 4.5 & 1400 & 537 & 0.0 & [0.72,1.0] & 0.057\\
J2214+3000 & 1.5 & NA & 558 & 0.0268 & [0.0,0.23]$\cup$[0.33,0.7]$\cup$[0.84,1.0] & 0.171\\
J2302+4442 & 5.5 & 1400 & 908 & 0.5494 & [0.27,0.72] & 2.326\\
\enddata
\tablenotetext{a}{For the phase-aligned MSPs the radio frequency does not enter into the simulations.}
\tablenotetext{b}{The radio uncertainty used is 1/3 the normal estimate as the single radio peak is very sharp and the fitting routine did not give it enough significance with larger error estimates.}\label{ch7FitParTable}
\end{deluxetable}

\small\normalsize

As noted at the end of Chapter 6, it is important to properly estimate the gamma-ray background level in order to minimize systematic biases in the light curve fits.  Therefore, the gamma-ray background estimates given in column 4 of Table~\ref{ch7FitParTable} are derived using the LAT ST \emph{gtsrcprob}.

Using a spectral model for the sources and background components in a given ROI, \emph{gtsrcprob} calculates the probability that a given event came from a given source.  For event $i$ observed at time $t$ with reconstructed energy $E_{i}^{\prime}$ and direction on the sky $\hat{p}_{i}^{\prime}$, \emph{gtsrcprob} calculates the probability of that event being associated with source $K$ as $P_{K,i}\ =\ M_{K}/ \sum_{J}M_{J}$, where the index $J$ runs over all sources in the model and $M_{J}$ is the likelihood source model as given in Eq.~\ref{ch3lMod} \citep{Chiang10}.

An enhanced pulsation search technique using such spectrally derived event probabilities has recently been proposed by \citet{Kerr10} to increase the LAT pulsar sensitivity.  The effectiveness of this method has been demonstrated for the MSPs J1939+2134 and J1959+2048 by \citet{Guillemot11}.

To derive the gamma-ray background estimates in Table~\ref{ch7FitParTable}, the event probabilities were calculated using spectral information from a preliminary version of the 2 year \FL{} catalog (Abdo et al., 2011b; 2FGL) using the updated P6\_V11\_DIFFUSE IRFs.  The 2FGL catalog uses source finding and analysis methods similar to those described in \citet{Abdo1FGL} for the 1 year LAT catalog (1FGL) with several improvements.  Among the improvements over the 1FGL catalog is the use of exponentially cutoff power law models for the spectra of known gamma-ray pulsars.  Note that the background estimate for the MSP J1902$-$5105 \citep{Camilo11} comes from an indpendent reanalysis of the source region as gamma-ray pulsations had not yet been detected when the preliminary 2FGL analysis was done.

The P6\_V11\_DIFFUSE IRFs include a corrected description of the LAT PSF using on-orbit observations of bright pulsars and AGN \citep{Roth11}.  This analysis demonstrated that the LAT PSF was larger above $\sim$10 GeV than pre-flight expectations.  The effect is seen both in LAT detected AGN, bright pulsars, and pulsar wind nebulae suggesting that it is truly an instrumental artifact.  Additionally, the P6\_V11\_DIFFUSE IRFs include an updated description of the LAT effective area which corrects for livetime effects and discrepancies between how the CTBCORE (see Chapter 3) probability knob was applied to the data and MC.

Once the event probabilities have been computed, the gamma-ray background level in the given ROI is estimated as $b_{\gamma}\ = N-\sum_{j}^{N}\ P_{\rm MSP,\mathnormal j}$ where the $j$ index runs over the events, $N$ is the total number of events in the ROI, and $P_{\rm MSP,\mathnormal j}$ is the \emph{gtsrcprob} probability that event $j$ comes from the MSP.  This formula uses the sum of the probabilities as a proxy for the source signal ($S_{src}$) and equates $N$ with the total signal ($S_{tot}$) to calculate the background as $S_{tot}-S_{src}$.

In principle, it should be possible to use the full covariance matrix of the spectral fit to calculate an uncertainty on the event probabilities which could then, in turn, be used to calculate an uncertainty on the background estimate.  These values could also be used to calculate uncertainties for probability weighted light curves, following the prescription of \citet{Kerr10}, to be used in the light curve fitting described in Chapter 6.  However, \emph{gtsrcprob} does not yet provide such uncertainty estimates.

Table~\ref{ch7MCMCtable} gives the MCMC parameters used for each MSP including the number of gamma-ray and radio light curve bins, the maximum temperature, the number of burn-in steps, chain steps, and chains.  For those MSPs fit with both TPC and OG (or alTPC and alOG) models the same value of $T$ was used in most cases.  However, the OG (alOG) models were found to occasionally require higher values of $T$ to finish in a reasonable amount of time.  For those cases in which different maximum temperatures were used, the TPC (alTPC) $T$ is given as normal in column 4 of Table~\ref{ch7MCMCtable} and the OG (alOG) $T$ is given in parentheses.

\small\normalsize

\begin{deluxetable}{l c c c c c c}
\tablewidth{0pt}
\tablecaption{MSP MCMC Fitting Details}
\startdata
\underline{JName} & \underline{Nbins$_{\gamma}$} & \underline{Nbins$_{\rm R}$} & \underline{$T$\tablenotemark{a}} & \underline{N$_{\rm chains}$} & \underline{N$_{\rm steps}$} & \underline{N$_{\rm burn}$}\\ 
J0030+0451 & 60 & 60 & 30 & 8 & 12500 & 50\\
J0034$-$0534 & 30 & 30 & 5 & 16 & 12500 & 100\\
J0218+4232 & 30 & 30 & 10 & 8 & 12500 & 50\\
J0437$-$4715 & 60 & 60 & 10 & 8 & 12500 & 50\\
J0613$-$0200 & 60 & 60 & 10(15) & 8 & 12500 & 50\\
J0614$-$3329 & 60 & 60 & 15(20) & 8 & 12500 & 50\\
J0751+1807 & 30 & 60 & 10 & 8 & 12500 & 50\\
J1231$-$1411 & 90 & 90 & 20 & 8 & 12500 & 50\\
J1614$-$2230 & 30 & 60 & 10(15) & 8 & 12500 & 50\\
J1713+0747 & 15 & 30 & 10 & 8 & 12500 & 50\\
J1744$-$1134 & 60 & 60 & 10 & 8 & 12500 & 50\\
J1823$-$3021A & 60 & 60 & 10 & 16 & 12500 & 100\\
J1902$-$5105 & 30 & 30 & 10 & 16 & 12500 & 100\\
J1939+2134 & 30 & 30 & 20 & 16 & 12500 & 100\\
J1959+2048 & 60 & 60 & 25 & 16 & 12500 & 100\\
J2017+0603 & 60 & 60 & 10 & 8 & 12500 & 50\\
J2124$-$3358 & 60 & 60 & 10 & 8 & 12500 & 50\\
J2214+3000 & 30 & 60 & 15(20) & 16 & 12500 & 100\\
J2302+4442 & 60 & 60 & 10 & 8 & 12500 & 50\\
\enddata
\tablenotetext{a}{If two values for $T$ are given the value in parentheses applies to the OG (alOG) model fit.}\label{ch7MCMCtable}
\end{deluxetable}

\small\normalsize

\section{Results}\label{ch7res}

The best-fit parameters for MSPs fit with TPC and OG models are given in Table~\ref{ch7TPCOG}, for MSPs fit with the PSPC model in Table~\ref{ch7PSPC}, and for MSPs fit with the alTPC and alOG models in Table~\ref{ch7alTPCOG}.  The best-fit light curves, marginalized contours, and simulated emission phase plots for all MSPs are given in Appendix~\ref{appA} along with brief historical information.  In Tables~\ref{ch7TPCOG} and~\ref{ch7alTPCOG} the gap widths for each model type are as defined in Chapter 5, values of zero gap width are unphysical and should be taken to mean that the true, best-fit gap width is less than the simulation resolution of 0.05.  When considering the predicted viewing geometries, the reported value of $\zeta\ =\ Z$ should be taken to mean $\zeta\in[Z,Z+1)$ as this reflects the manner in which simulated phase plots are binned.

\small\normalsize

\begin{deluxetable}{l c c c c c c c c}
\tablewidth{0pt}
\tablecaption{TPC and OG Best-Fit Results}
\startdata
\underline{JName} & \underline{$\-log(\mathcal L)$} & \underline{$\alpha$ (\DEG{})} & \underline{$\zeta$ (\DEG{})} & \underline{$\Phi$} & \underline{$w\ (\Theta_{\rm PC})$} & \underline{f$_{\Omega}$} & \underline{$\chi^{2}$} & \underline{$\nu_{dof}$}\\
\underline{TPC:}\\
J0030+0451 & 356.8 & 73$\pm$6 & 57$^{+6}_{-10}$ & 0.050 & 0.05 & 1.17$^{+0.10}_{-0.47}$ & 689.5 & 114\\
J0218+4232 & 115.5 & 24$^{+4}_{-1}$ & 8$^{+6}_{-5}$ & 0.033 & 0.10 & 1.86$^{+0.14}_{-0.40}$ & 68.4 & 54\\
J0437$-$4715 & 206.8 & 28$^{+12}_{-2}$ & 65$^{+2}_{-5}$ & 0.033 & 0.05 & 0.80$^{+0.20}_{-0.15}$ & 247.8 & 114\\
J0613$-$0200 & 231.6 & 57$^{+3}_{-2}$ & 42$\pm$3 & 0.067 & 0.05 & 0.91$\pm$0.20 & 182.4 & 114\\
J0614$-$3329 & 469.4 & 81$^{+3}_{-10}$ & 76$^{+9}_{-4}$ & -0.450 & 0.10 & 1.03$^{+0.07}_{-0.20}$ & 677.2 & 114\\
J0751+1807 & 160.7 & 61$^{+9}_{-30}$ & 73$^{+5}_{-27}$ & -0.233 & 0.00 & 0.74$^{+0.36}_{-0.24}$ & 180.6 & 86\\
J1231$-$1411 & 465.2 & 24$^{+8}_{-4}$ & 69$\pm$1 & 0.078 & 0.00 & 0.50$^{+0.62}_{-0.02}$ & 747.8 & 174\\
J1614$-$2230 & 146.1 & 22$^{+36}_{-2}$ & 71$^{+18}_{-3}$ & 0.100 & 0.00 & 0.52$^{+0.60}_{-0.06}$ & 115.6 & 84\\
J1713+0747 & 56.7 & 21$^{+20}_{-1}$ & 65$^{+12}_{-5}$ & 0.033 & 0.00 & 0.60$^{+0.35}_{-0.13}$ & 25.5 & 41\\
J2017+0603 & 199.3 & 17$^{+38}_{-5}$ & 68$^{+7}_{-10}$ & 0.117 & 0.00 & 0.49$^{+0.25}_{-0.04}$ & 125.8 & 114\\
J2302+4442 & 265.4 & 59$^{+6}_{-4}$ & 46$\pm$7 & 0.183 & 0.00 & 0.96$^{+0.10}_{-0.06}$ & 235.4 & 114\\
\underline{OG:}\\
J0030+0451 & 320.9 & 81$\pm$9 & 66$^{+4}_{-2}$ & 0.000 & 0.00 & 1.09$^{+0.05}_{-0.37}$ & 441.3 & 113\\
J0218+4232 & 199.6 & 76$^{+14}_{-4}$ & 32$^{+12}_{-8}$ & 0.167 & 0.00 & 1.01$^{+0.10}_{-0.14}$ & 243.1 & 53\\
J0437$-$4715 & 228.3 & 72$\pm$4 & 44$^{+8}_{-4}$ & 0.033 & 0.10 & 0.74$^{+0.16}_{-0.04}$ & 274.1 & 113\\
J0613$-$0200 & 248.6 & 63$^{+7}_{-3}$ & 30$^{+4}_{-8}$ & 0.033 & 0.00 & 0.71$^{+0.24}_{-0.06}$ & 210.5 & 113\\
J0614$-$3329 & 634.6 & 46$^{+11}_{-10}$ & 88$\pm$2 & 0.067 & 0.00 & 0.87$\pm$0.10 & 1185.0 & 113\\
J0751+1807 & 155.4 & 61$^{+8}_{-11}$ & 73$^{+6}_{-11}$ & -0.233 & 0.00 & 0.54$^{+0.66}_{-0.10}$ & 163.5 & 85\\
J1231$-$1411 & 487.1 & 82$^{+5}_{-10}$ & 65$^{+1}_{-20}$ & 0.100 & 0.00 & 1.16$^{+0.10}_{-0.30}$ & 536.6 & 173\\
J1614$-$2230 & 142.8 & 29$^{+31}_{-8}$ & 78$^{+12}_{-8}$ & 0.100 & 0.05 & 0.32$^{+0.60}_{-0.06}$ & 115.7 & 83\\
J1713+0747 & 54.9 & 35$^{+9}_{-7}$ & 73$^{+5}_{-9}$ & 0.003 & 0.10 & 0.34$^{+0.66}_{-0.04}$ & 20.6 & 40\\
J2017+0603 & 196.0 & 17$^{+33}_{-5}$ & 68$^{+8}_{-5}$ & 0.117 & 0.00 & 0.30$^{+0.90}_{-0.02}$ & 101.8 & 113\\
J2302+4442 & 266.9 & 64$\pm$5 & 38$^{+7}_{-6}$ & 0.183 & 0.00 & 0.98$^{+0.25}_{-0.07}$ & 229.4 & 113\\
\enddata\label{ch7TPCOG}
\end{deluxetable}

\small\normalsize

\small\normalsize

\begin{deluxetable}{l c c c c c c c}
\tablewidth{0pt}
\tablecaption{PSPC Best-Fit Results}
\startdata
\underline{JName} & \underline{$\-log(\mathcal L)$} & \underline{$\alpha$ (\DEG{})} & \underline{$\zeta$ (\DEG{})} & \underline{$\Phi$} & \underline{f$_{\Omega}$} & \underline{$\chi^{2}$} & \underline{$\nu_{dof}$}\\
J1744$-$1134 & 226.7 & 51$^{+16}_{-19}$ & 85$^{+3}_{-12}$ & 0.033 & 1.04$^{+0.80}_{-0.50}$ & 110.4 & 115\\
J2124$-$3358 & 297.6 & 23$^{+4}_{-7}$ & 20$^{+5}_{-8}$ & -0.017 & 0.50$^{+0.03}_{-0.14}$ & 365.6 & 115\\
\enddata\label{ch7PSPC}
\end{deluxetable}

\small\normalsize

\small\normalsize

\begin{deluxetable}{l c c c c c c c c c c c c}
\tablewidth{0pt}
\tabletypesize{\footnotesize}
\tablecaption{alTPC and alOG Best-Fit Results}
\rotate
\startdata
\underline{JName} & \underline{$\-log(\mathcal L)$} & \underline{$\alpha$ (\DEG{})} & \underline{$\zeta$ (\DEG{})} & \underline{$\Phi$} & \underline{$w_{\gamma}\ (\Theta_{\rm PC})$} &\underline{$w_{\rm R}\ (\Theta_{\rm PC})$} & \underline{R$_{max}^{\gamma} (\rm R_{LC})$} & \underline{R$_{min}^{\rm R}\ (\rm R_{LC})$}& \underline{R$_{max}^{\rm R}\ (\rm R_{LC})$} & \underline{f$_{\Omega}$} & \underline{$\chi^{2}$} & \underline{$\nu_{dof}$}\\
\underline{alTPC:}\\
J0034$-$0534 & 89.0 & 12$^{+31}_{-6}$ & 69$^{+7}_{-4}$ & -0.267 & 0.00 & 0.00 & 0.9$\pm$0.1 & 0.5$^{+0.3}_{-0.2}$ & 1.0$\pm$0.2 & 0.54$^{+0.46}_{-0.01}$ & 39.3 & 50\\
J1823$-$3021A & 175.0 & 52$^{+9}_{-51}$ & 68$^{+20}_{-14}$ & 0.483 & 0.00 & 0.00 & 1.0$\pm$0.1 & 0.4$^{+0.1}_{-0.26}$ & 0.9$\pm$0.1 & 0.87$\pm$0.40 & 108.3 & 110\\
J1902$-$5105 & 126.6 & 44$^{+15}_{-23}$ & 42$^{+17}_{-22}$ & -0.100 & 0.00 & 0.10 & 0.9$^{+0.3}_{-0.1}$ & 0.2$^{+0.2}_{-0.06}$ & 0.8$^{+0.2}_{-0.3}$ & 1.51$^{+0.50}_{-1.00}$ & 89.7 & 50\\
J1939+2134 & 129.5 & 78$^{+12}_{-77}$ & 82$^{+8}_{-32}$ & 0.433 & 0.10 & 0.00 & 1.1$^{+0.1}_{-0.3}$ & 0.7$\pm$0.1 & 1.0$\pm$0.1 & 1.02$^{+0.56}_{-0.60}$ & 35.3 & 50\\
J1959+2048 & 214.5 & 47$^{+9}_{-46}$ & 83$^{+7}_{-31}$ & -0.483 & 0.10 & 0.05 & 1.2$\pm$0.1 & 0.8$\pm$0.1 & 1.1$\pm$0.1 & 0.77$^{+0.90}_{-0.30}$ & 92.0 & 110\\
J2214+3000 & 120.8 & 89$^{+1}_{-13}$ & 14$^{+46}_{-8}$ & 0.140 & 0.10 & 0.10 & 0.8$\pm$0.1 & 0.2$^{+0.3}_{-0.06}$ & 0.6$\pm$0.1 & 1.30$^{+0.30}_{-0.60}$ & 83.4 & 80\\
\underline{alOG:}\\
J0034$-$0534 & 103.4 & 12$^{+32}_{-12}$ & 69$^{+21}_{-4}$ & -0.267 & 0.00 & 0.00 & 0.9$\pm$0.1 & 0.2$^{+0.4}_{-0.06}$ & 1.2$^{+0.1}_{-0.3}$ & 0.31$^{+0.90}_{-0.10}$ & 53.2 & 48\\
J1823$-$3021A & 168.2 & 65$^{+25}_{-65}$ & 63$^{+27}_{-47}$ & 0.467 & 0.00 & 0.00 & 1.1$\pm$0.1 & 0.3$^{+0.1}_{-0.16}$ & 1.1$^{+0.1}_{-0.3}$ & 1.01$^{+0.60}_{-0.20}$ & 83.6 & 108\\
J1902$-$5105 & 127.7 & 15$^{+43}_{-14}$ & 78$^{+12}_{-4}$ & 0.133 & 0.00 & 0.00 & 1.0$\pm$0.1 & 0.2$^{+0.1}_{-0.06}$ & 1.2$^{+0.1}_{-0.3}$ & 0.31$^{+0.90}_{-0.10}$ & 97.2 & 48\\
J1939+2134 & 133.6 & 84$^{+6}_{-83}$ & 84$^{+6}_{-15}$ & 0.433 & 0.05 & 0.00 & 1.2$^{+0.1}_{-0.2}$ & 0.6$\pm$0.1 & 0.9$\pm$0.1 & 1.12$^{+0.10}_{-0.90}$ & 42.4 & 48\\
J1959+2048 & 219.1 & 51$^{+9}_{-50}$ & 84$^{+6}_{-9}$ & -0.467 & 0.00 & 0.00 & 1.1$\pm$0.1 & 0.4$^{+0.5}_{-0.1}$ & 1.0$^{+0.2}_{-0.1}$ & 0.85$^{+0.25}_{-0.45}$ & 101.6 & 108\\
J2214+3000 & 157.7 & 83$^{+7}_{-17}$ & 20$^{+36}_{-10}$ & -0.367 & 0.00 & 0.00 & 1.1$\pm$0.1 & 0.5$\pm$01 & 0.6$^{+0.3}_{-0.1}$ & 0.76$^{+0.10}_{-0.60}$ & 184.9 & 78\\
\enddata\label{ch7alTPCOG}
\end{deluxetable}

\small\normalsize

Error estimates are given for $\alpha$ and $\zeta$ taken from analysis of the marginalized confidence contours in Appendix A.  Note that, in some cases, the 1$\sigma$ confidence regions (39\% probability for 2 degrees of freedom) are not simply connected and, thus, the error estimates provided in Tables~\ref{ch7TPCOG},~\ref{ch7PSPC}, and~\ref{ch7alTPCOG} are somewhat optimistic.

Additionally, as discussed in Chapter 6 there are cases where the best-fit geometry is not encompassed by all of the marginalized contours.  For those cases the viewing geometry errors given are quite large to be conservative but their significance is not well defined.

In all cases, the $\alpha$ and $\zeta$ uncertainties provided in the tables should be taken as approximations and comparisons with radio and X-ray constraints should use the marginalized contours.

For the best-fit emission altitudes in Table~\ref{ch7alTPCOG}, 2$\sigma$ errors are reported using the profile likelihood method described in Chapter 6.  For those best-fit values which are found at the maximum of the allowed range an upper uncertainty of 0.1 is reported but this only reflects the simulation resolution.

Confidence intervals are supplied for the f$_{\Omega}$ estimates in Tables~\ref{ch7TPCOG},~\ref{ch7PSPC}, and~\ref{ch7alTPCOG}.  These are estimated by collecting the f$_{\Omega}$ estimates from all ($\alpha$,$\zeta$) pairs in the 39\% confidence marginalized contours.  Changes in viewing geometry were found to produce a larger spread in f$_{\Omega}$ than changes in either gap width or emission altitude.

No uncertainties are reported for the gap widths in Tables~\ref{ch7TPCOG} and~\ref{ch7alTPCOG} as, at best, these can only be constrained to within $\pm$0.05 and, given the coarse resolution of the current simulations, no meaningful constraints can be derived on these parameters.

Note that none of the uncertainties reported in Tables~\ref{ch7TPCOG},~\ref{ch7PSPC}, and~\ref{ch7alTPCOG} account for potential systematics due to inadequacies in the emission models or issues with the gamma-ray background estimates.

The last two columns of Tables~\ref{ch7TPCOG},~\ref{ch7PSPC}, and~\ref{ch7alTPCOG} give the $\chi^{2}$ values and degrees of freedom ($\nu_{dof}$) for each fit.  The degrees of freedom are calculated by summing the number of gamma-ray and radio bins and then subtracting the number of model parameters including the normalizations of both model light curves.  Note that for each MSP there is at least one fit presented for which the $\chi^{2}$ and $\nu_{dof}$ suggest reasonable agreement between the model and the data (using the $\sigma_{\rm R}$ estimates from Table~\ref{ch7FitParTable}) except for PSR J0614$-$3329 for which neither best-fit model in Table~\ref{ch7TPCOG} results in a reasonable fit.

\begin{figure}[h]
\begin{center}
\includegraphics[width=.6\textwidth]{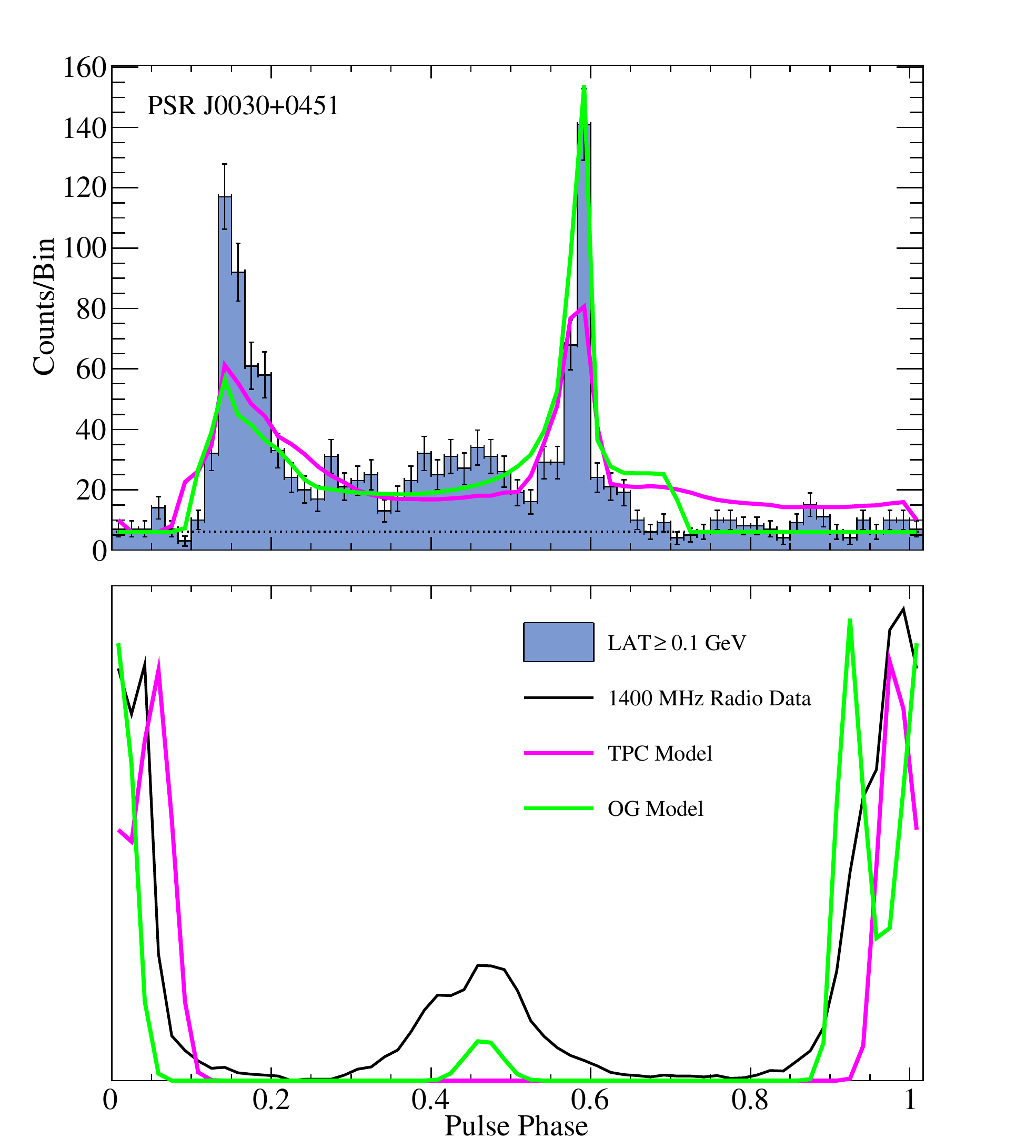}
\end{center}
\small\normalsize
\begin{quote}
\caption[Example light curve fits to PSR J0030+0451 using TPC and OG models]{Gamma-ray (\emph{top}) and radio (\emph{bottom}) data and best-fit model light curves for PSR J0030+0451 using standard TPC (pink) and OG (green) gamma-ray models with a hollow-cone radio beam.\label{ch7J0030LC}}
\end{quote}
\end{figure}

\begin{figure}[h]
\begin{center}
\includegraphics[width=.6\textwidth]{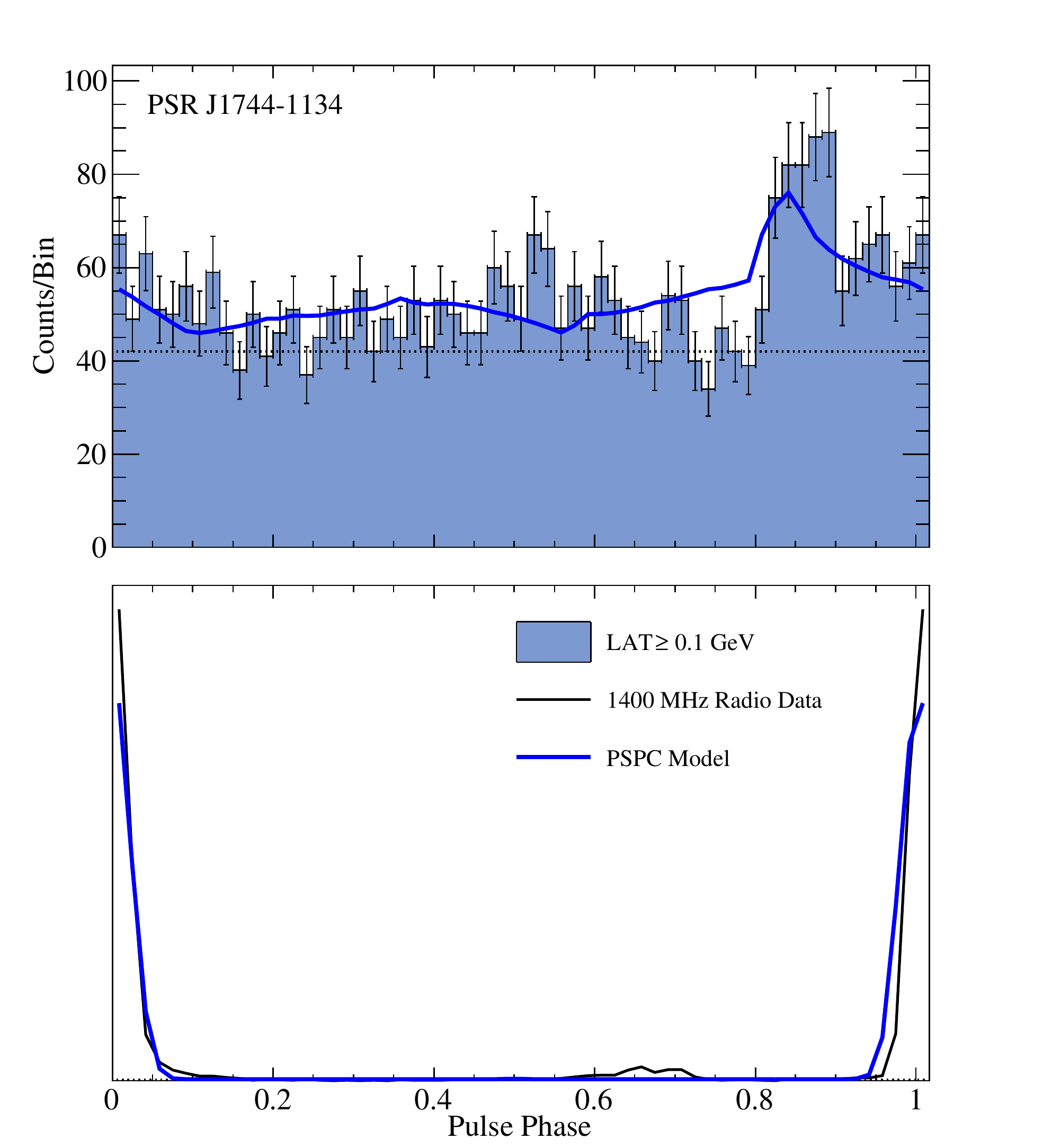}
\end{center}
\small\normalsize
\begin{quote}
\caption[Example light curve fits to PSR J174$-$1134 using PSPC gamma-ray model]{Gamma-ray (\emph{top}) and radio (\emph{bottom}) data and best-fit model light curves for PSR J1744$-$1134 using the PSPC (blue) gamma-ray model with a hollow-cone radio beam.\label{ch7J1744LC}}
\end{quote}
\end{figure}

\begin{figure}[h]
\begin{center}
\includegraphics[width=.6\textwidth]{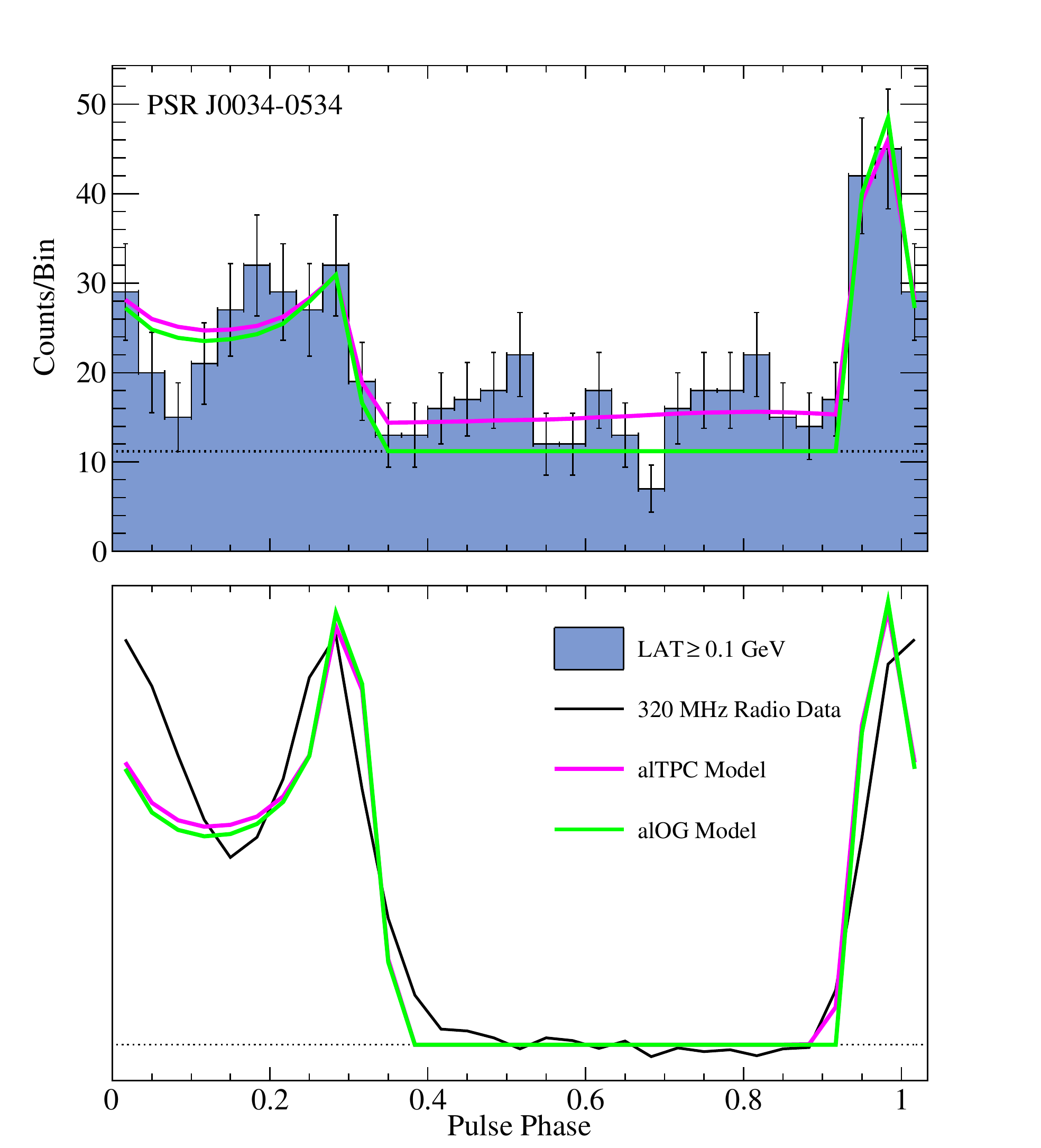}
\end{center}
\small\normalsize
\begin{quote}
\caption[Example light curve fits to PSR J0034$-$0534 using alTPC and alOG models]{Gamma-ray (\emph{top}) and radio (\emph{bottom}) data and best-fit model light curves for PSR J0034$-$0534 using alTPC (pink) and alOG (green) gamma-ray and radio models.\label{ch7J0034LC}}
\end{quote}
\end{figure}

Example fits are shown in Figs.~\ref{ch7J0030LC},~\ref{ch7J1744LC}, and~\ref{ch7J0034LC}.  Fig.~\ref{ch7J0030LC} shows the data and best-fit model light curves for PSR J0030+0451 (see Section~\ref{ch7J0030}) which has been fit with standard TPC and OG models for the gamma rays and a hollow-cone beam for the radio.  Fig.~\ref{ch7J1744LC} shows the data and best-fit model light curves of PSR J1744$-$1134 (see Section~\ref{ch7J1744}) which has been fit with the PSPC model for the gamma rays and a hollow-cone beam for the radio.  Fig.~\ref{ch7J0034LC} shows the data and best-fit model light curves of PSR J0034$-$0534 (see Section~\ref{ch7J0034}) which has been fit with the alTPC and alOG models for both the gamma rays and the radio.

\small\normalsize

\small\normalsize

\small\normalsize

\subsection{Comparison With Previous Geometric Predictions}\label{ch7individual}
The MSPs J0614$-$3329, J1231$-$1411, J2214+3000, J2017+0603, J2302+4442, and J1902$-$5105 have only recently been discovered in radio observations of unassociated \FL{} sources (Ransom et al., 2011; Cognard et al., 2011; and Camilo et al., 2011; respectively) and thus no radio polarization measurements, or geometric constraints from other wavelengths, have yet been published.

Polarization observations exist for many of the MSPs analyzed here but the measurements do not often lend themselves to straightforward and/or constraining interpretations.  Comparison with alternative predictions for the viewing geometries of the MSPs in Table~\ref{ch7MSPtable}, when they exist, are discussed below.

\subsubsection{PSR J0030+0451}\label{ch7J0030}
PSR J0030+0451 was first discovered by \citet{Lommen00} who also attempted to produce RVM fits to the polarization measurements in order to constrain the viewing geometry.  Results of the RVM fits depended greatly on the starting parameters and were, thus, non-unique.

However, based on the presence of a radio interpulse (IP, relatively minor peak separated in phase from the main component by $\sim$180\DEG{} in phase) \citet{Lommen00} argued that the pulsar was either nearly aligned or nearly orthogonal.  They presented RVM calculations for $\alpha$ = 8\DEG{} with $\beta\ \equiv\ \zeta-\alpha$ = 1\DEG{} and $\alpha$ = 62\DEG{} with $\beta$ = 10\DEG{} based on ``reasonable'' agreement with the data and argued that the data suggested a large value of $\alpha$.

\citet{Bogdanov08} modeled the thermal X-ray emission from PSR J0030+0451 assuming a hot polar cap and ruled out the low $\alpha$ solution as it would not lead to the observed, double-peaked X-ray light curve.

\citet{Venter09} modeled the gamma-ray and radio light curves of the first 8 gamma-ray MSPs detected with the LAT, including PSR J0030+0451.  They only fit MSPs using the TPC, OG, and PSPC models with hollow-cone beam radio emission and used simulations with 5\DEG{} resolution in $\alpha$.  Note that their OG model definitions were slightly different from that described in Chapter 5 of this thesis.  Additionally, the dataset they used was $\sim$8.5 months of LAT data which made identification of significant features beyond the main peak(s) difficult.  Their best-fit geometries were picked out by eye, which tends to favor the HE profiles, but were still able to reproduce the observed profiles remarkably well.

For J0030+0451, \citet{Venter09} estimated $(\alpha,\zeta)$ = (70\DEG{},80\DEG{}) for a TPC model with a gap width of 0.10 and $(\alpha,\zeta)$ = (60\DEG{},60\DEG{}) for an OG model with an infinitely-thin gap and emission layer of size 0.05, using the accelerating gap and emission layer definitions given in Chapter 5 of this thesis.

The results presented here agree with the radio and X-ray analysis suggesting a large value of $\alpha$ though the predicted $\beta$ is higher in both fits.  The likelihood favors the OG model, which does predict a radio IP though this is done with a cone beam whereas \citet{Lommen00} argued that a nearly orthogonal geometry would suggest that the observed components were from the core components of both poles.

The results presented here favor the OG model but predict narrower gaps for this model than that used by \citet{Venter09}, a fact which may in part be due to the inclusion of the magnetic field Lorentz transform (see Section~\ref{ch5simLCs}) which can slightly widen the simulated peaks.  The TPC $\alpha$ values agree well but the $\zeta$ values are quite different, likely due to the likelihood seeking to decrease the amount of off-peak emission predicted to agree with observations.  The situation is reversed for the OG fits where the $\zeta$ values agree well but the predicted $\alpha$ values are quite different.  In the latter case no obvious factors motivating either value of $\alpha$ are clear but there may be some difference between the two fits related to the relatively large size of the emission layer used by \citet{Venter09}.

\subsubsection{PSR J0034$-$0534}\label{ch7J0034}
Polarimetric observations of PSR J0034$-$0534 have been reported by \citet{Stairs99} at 410 MHz.  The mean linear and circular polarization amplitudes measured were 0 and 18\%, respectively.  Additionally, no swing of the polarization angle was observed through the peaks.

The radio profile of this MSP has been modeled with the alTPC and alOG models which implies it is caustic in nature.  If this is accurate then the lack of significant polarization could be explained as the caustic mixing of different polarizations at different altitudes (similar to the effects which lead to bright HE peaks) which washes-out the polarization \citep{Dyks04}.

\citet{AbdoJ0034} modeled the gamma-ray and radio light curves of this pulsar using methods similar to \citet{Venter09} and $\sim$13 months of LAT data.  Attempts to match the near-alignment of the HE and radio peaks led them to develop the alTPC and alOG models as standard TPC and OG models with a hollow-cone radio beam could not match the alignment and low-altitude PC models could not reproduce the correct peak shapes.  They found good fits for alTPC and alOG models with gap widths of 0.05 and $(\alpha,\zeta)$ = (30\DEG{},70\DEG{}) for both.  Additionally, both models used $R_{max}^{\gamma}$ = 0.9, $R_{min}^{\rm R}$ = 0.6, and $R_{max}^{\rm R}$ = 0.8 suggesting that the radio emission originates near the light cylinder, co-located with the gamma-ray emission region.

These models were explored in more detail by \citet{Venter11} in which the MCMC maximum likelihood fitting technique described in Chapter 6 was used to fit the light curves of PSR J0034$-$0534 with the $\sim$16 month dataset analyzed by \citet{PWNcat}, this study also included the Lorentz transform of the magnetic field.  They found the same geometry for the alTPC model as \citet{AbdoJ0034} with the exception that the width of the radio emission region was 0.10 and $R_{min}^{\rm R}$ was 0.7.  The best-fit alOG model was found to have  $\zeta$ = 69 \DEG{} and $\alpha$ = 12\DEG{}.  Both emission regions were fit with infinitely thin gaps.  The extent of the gamma-ray emission was the same as that found by \citet{AbdoJ0034} but the radio region was found to be much more extended with $R_{min}^{\rm R}$ = 0.2 and $R_{max}^{\rm R}$ = 1.1.

The results presented here use more data and an improved method of estimating the gamma-ray background level.  The alOG fit agrees very well with that of \citet{Venter11} while the viewing geometry of the alTPC model now agrees with what was found for the alOG model.  The alTPC model is now found with infinitely thin gaps for both gamma-ray and radio profiles with a slightly larger extent of the radio emission region.  This is likely due to the increase in statistics which has allowed for the gamma-ray peak near a phase of 1.0 to become more well-defined and sharper, driving the fit to a lower value of $\alpha$.  The likelihood appears to favor the alTPC model, likely due to the apparent level of off-peak emission, in accord with what was found by \citet{PWNcat}.

\subsubsection{PSR J0218+4232}\label{ch7J0218}
Polarimetric observations and RVM model fits for PSR J0218+4232 have been reported by \citet{Stairs99} at both 410 and 610 MHz.  Application of the RVM model to this pulsar yielded best-fit values of $\alpha$ = 8\DEG{}$\pm$11(15)\DEG{} for the 410(610) MHz data with unconstrained values of $\beta$.

The nearly-aligned geometry is supported by the observation of radio emission across nearly the entire pulse phase, with only $\sim$50\% of the flux in the main peaks.  \citet{Stairs99} comment that the the RVM fits do allow for $\beta\ \sim$90\DEG{} which would make this a good candidate for Shapiro delay observations if confirmed.

\citet{Venter09} found reasonable fits for PSR J0218+4232 using TPC and OG models for the gamma-ray light curve and a hollow-cone beam model for the radio.  Their TPC model used a gap width of 0.10 and a viewing geometry of $(\alpha,\zeta)$ = (60\DEG{},60\DEG{}).  They used an OG model with an infinitely-thin gap and an emission layer of 0.05 for a viewing geometry of $(\alpha,\zeta)$ = (50\DEG{},70\DEG{}).

The TPC model fits reported in Table~\ref{ch7TPCOG} for PSR J0218+4232 agree well with the RVM predictions of \citet{Stairs99} while the OG model values do not.  The best-fit TPC $\alpha$ agrees at the large end of the quoted error range but note that the marginalized confidence contours do suggest a range of $\alpha$ as low as 20\DEG{} for the same value of $\zeta$ and as low as $\alpha\ \sim$8\DEG{} for $\zeta\ \sim$30\DEG{}.  The OG models do not find reasonable fits for low values of $\alpha$ and those geometries which do reproduce the observed gamma-ray profiles can not match the wide radio peaks.

When the HE pulsations are taken into account, which are also quite wide and cover nearly the entire pulse phase, the viability of a large $\beta$ facilitating Shapiro delay measurements seems low and no such measurement has been reported to date.

When compared to the best-fit geometries of \citet{Venter09} the geometries do not agree well but note that they found fits with sharper peaks than are observed and were not able to reproduce the width and separation of the radio peaks.

\subsubsection{PSR J0437$-$4715}\label{ch7J0437}
PSR J0437$-$4715 is the nearest known MSP and has been the subject of much detailed research and speculation concerning its potential HE emission.  \citet{AbdoMSPpop} first reported the detection of significant pulsed gamma-rays from this MSP and found it to be a relatively weak gamma-ray emitter when compared to other, known HE pulsars (see Table 4 of Abdo et al., 2010c).  Nevertheless, this MSP remains a source of interest.

\citet{MJ95} and \citet{Navarro97} presented polarimetric observations of PSR J0437$-$4715 and noted that the radio profile is complex.  Both argued that the observation of a high level of circular polarization near the profile center, which presents a clear sense reversal at the peak, argues for a small value of $\beta$ and is characteristic of emission from a core beam component.

Standard RVM fits of the polarization swing were unsuccessful as the parameters could not be constrained; however, \citet{MJ95} did present RVM curves for $\alpha$ = 145\DEG{} (35\DEG{}) and $\zeta$ = 140\DEG{} (40\DEG{}) which they believed represented the data well but are not fits and thus provide no actual geometric constraints for the pulsar.  More recent observations by \citet{Yan11} have confirmed earlier findings.

\citet{GK97} applied a relativistic version of the RVM model, developed by \citet{Blaskiewicz91}, to the polarization data of \citet{MJ95}.  This version of the RVM model corrects for first order, special relativistic effects due to fast co-rotation.  The prescription of \citet{GK97} was to find values of $\alpha$ and $\zeta$ which reproduced the basic profile characteristics and then calculate the predicted polarization swing which resulted in a predicted geometry of $\alpha$ = 20\DEG{} and $\beta$ = $-4^{\circ}$ though no comments on the uncertainty of these estimates was provided.

\citet{Bogdanov07} modeled the X-ray emission from PSR J0437$-$4715 using a thermal hotspot model.  They were unable to fit the observed pulse profile with a simple dipole but were successful when introducing an offset to the dipole axis.  For an offset dipole $\alpha$ describes the center of the hot PC from which the dipole axis is offset in stellar azimuth and colatitude by $\Delta\phi$ and $\Delta\alpha$, respectively.  Initial fits were highly unconstrained in $\alpha$ and $\zeta$.  By assuming that the spin axis is nearly orthogonal to orbit plane they estimate $\zeta\ \approx\ 42^{\circ}$.  Using this value they find $\alpha\ =$ 25\DEG{}-90\DEG{}, $\Delta\alpha\ =\ -50^{\circ}$-20\DEG{}, and $\Delta\phi\ = -(23^{\circ}$-$14^{\circ})$, with quoted ranges corresponding to 1$\sigma$ confidence intervals.

Using similar TPC and OG models for the HE light curves and a hollow-cone beam for the radio profile, \citet{Venter09} estimated a viewing geometry of $(\alpha,\zeta)$ = (30\DEG{},60\DEG{}).  They used a TPC model with gap width 0.05 and an OG model with an infinitely thin gap and an emission layer of size 0.05.

The best-fit geometries presented here do not agree with the predictions of a small impact parameter but it should be noted that the fits were done assuming a hollow-cone radio beam and the only way to get such a sharp, main component is to clip the outer edge of one cone.  Such an arrangement suggests an impact parameter with absolute value larger than the cone annulus $\bar{\theta}\ \approx\ 16^{\circ}$ as can be seen from the fits in Table~\ref{ch7TPCOG} which give values of 37\DEG{} and -28\DEG{}.  Therefore, future studies should model the radio emission of this MSP with a core component in order to compare with both the predictions in Table~\ref{ch7TPCOG} and those from radio considerations.

It is difficult to compare the predictions of \citet{Bogdanov07} with the results in Table~\ref{ch7TPCOG} as their $\alpha$ values do not correspond to the definition used here.  The OG model finds a value of $\zeta$ in agreement with the orbital inclination and $\alpha$ is easily in agreement with the suggested range of $\alpha\ +\ \Delta\alpha$ though it is not clear if such a comparison is valid as the shapes of HE light curves from pulsars with offset dipoles have not yet been investigated.  The TPC best-fit $\alpha$ is in good agreement with the large range they predict but the best-fit $\zeta$ is in disagreement with the orbital inclination.

The TPC fits presented here agree well with those of \citet{Venter09} but prefer a slightly larger value of $\zeta$ as this leads to a sharper radio peak and a vanishing of the precursor gamma-ray peak in their model.  The geometry of the best-fit OG model presented here differs drastically from what they found but early analysis with an unrealistically large emission layer did find similar fit values and thus the coarse gap resolution may be the cause of the observed discrepancy.

\subsubsection{PSR J0613$-$0200}\label{ch7J0613}
Many authors have reported on polarization measurements for PSR J0613$-$0200 and while all agree it presents a complex profile none completely agree on the polarization characteristics.

In particular, \citet{Xilouris98}, \citet{Stairs99}, and \citet{Ord04} all measure some degree of circular polarization through the main pulse component with clear sense reversal indicating a small impact parameter and a core emission component while disagreeing on the polarization of the rest of the pulse.  \citet{ManHan04}, on the other hand, measure no circular polarization through the pulse and thus do not require a core component for the pulsar.  More recent polarimetric observations by \citet{Yan11} have confirmed earlier findings suggesting a core component and detected a weak trailing component to the pulse with similar polarization properties.

Joint modeling of the gamma-ray and radio light curves by \citet{Venter09} found reasonable fits with $(\alpha,\zeta)$ = (30\DEG{},60\DEG{}) for both TPC and OG models.  The TPC model used a gap width of 0.05 while the OG model had an infinitely-thin gap with an emission layer of size 0.05.

The best-fit results presented here do not agree well with those of \citet{Venter09}, but this is largely driven by the increased gamma-ray statistics and the use of maximum likelihood which puts approximately equal weight to both the gamma-ray and radio profiles.  Note that these fits predict relatively large impact parameters, at odds with the suggestions of a core component, but this is to be expected given the use of a hollow-cone beam for the radio emission model.

\begin{figure}[h]
\begin{center}
\includegraphics[width=0.85\textwidth]{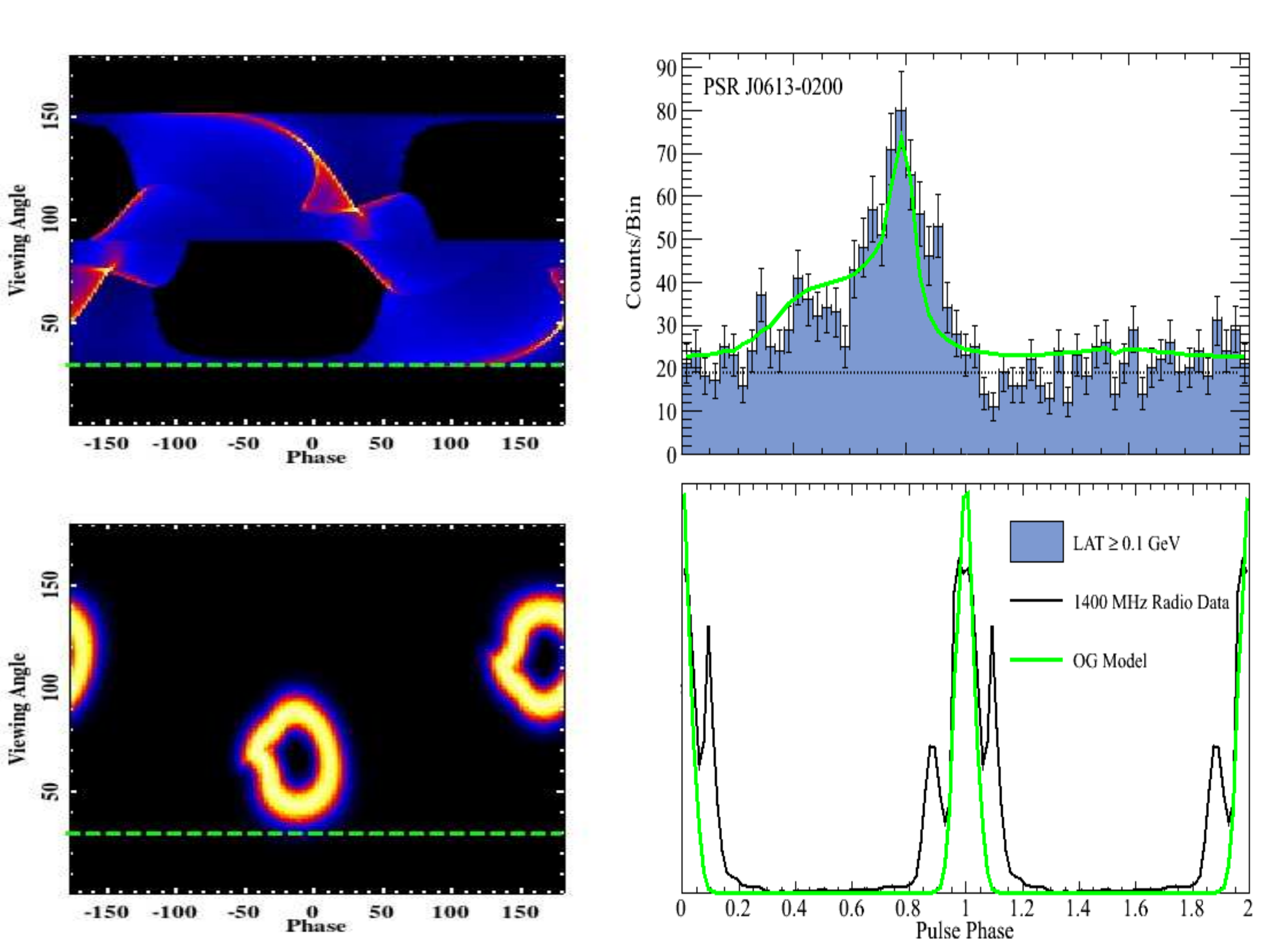}
\end{center}
\small\normalsize
\begin{quote}
\caption[Best-fit OG phase plots for PSR J0613$-$0200 with data and best-fit light curves]{\emph{(Top left):} Gamma-ray phase plot corresponding to the best-fit geometry for PSR J0613$-$0200 using an OG model and $\alpha$ = 63\DEG{} and $\zeta$ = 30\DEG{} (dashed green line), color scale is square root to show fainter features.  \emph{(Bottom left):} Radio phase plot for the same fit, color scale is also square root for consistency.  \emph{(Top right):} Gamma-ray data and best-fit OG light curve for PSR J0613$-$0200, note that the model gamma-ray emission persists across the entire phase.  \emph{(Bottom right):} Radio data and model light curves.\label{ch7J0613phplot}}
\end{quote}
\end{figure}

Note that the best-fit OG model for this MSP occupies a rare part of phase space which predicts emission across the entire pulse (see Fig.~\ref{ch7J0613phplot}).  For large inclination angles ($\gtrsim60^{\circ}$) the full shape of the polar cap reforms on the phase plot and thus a constant, low-level gamma-ray component is predicted for large values of $\beta$.  This is a side effect of two facts 1) the emission is followed out to large radial distance and 2) the radio beams of MSPs are much larger than in younger pulsars for which this viewing geometry would not result in a radio-loud, gamma-ray pulsar.

\small\normalsize

\subsubsection{PSR J0751+1807}\label{ch7J0751}
Polarization measurements have been reported for PSR J0751+1807 by \citet{Xilouris98} though they make no predictions on the viewing geometry.  This MSP exhibited unusual polarization behavior with the linear polarization of the leading peak changing over time while that of the trailing peak was steady.

\citet{Venter09} modeled the gamma-ray and radio profiles of this MSP in the context of TPC and OG models with a hollow-cone beam for the radio emission.  They predict $(\alpha,\zeta)$ = (50\DEG{},50\DEG{}) with a gap width of 0.05 for the TPC model and an infinitely-thin gap with an emission layer of 0.05 for the OG model.

The results presented here predict larger values for both $\alpha$ and $\zeta$ with a viewing angle below the magnetic axis.  While neither of the two results satisfactorily match the gamma-ray profile the model radio peaks of \citet{Venter09} were displaced with respect to the observed profile.  Additionally, the best-fit photon index of this MSP as reported by \citet{AbdoPSRcat} is 0.7, which is quite hard, so fitting the light curve with a slightly higher minimum energy may be warranted.

\subsubsection{PSR J1614$-$2230}\label{ch7J1614}
The predicted geometry for this pulsar from \citet{Venter09} is $(\alpha,\zeta)$ = (40\DEG{},80\DEG{}) for a TPC model with gap width 0.05 and an OG model with an infinitely thin gap and an emission layer of 0.05.  These geometries predict beam-correction factors of f$_{\Omega}\ \sim$ 1, but as noted by \citet{AbdoMSPpop} and \citet{AbdoPSRcat} the measured gamma-ray energy flux predicts an efficiency for this MSP $\gtrsim$ 100\%.

\citet{Demorest10} reported a precise measurement of the Shapiro delay in this system and strongly constrained the mass and orbital inclination angle of this system.  They found a value for the inclination angle of 89\DEG{}.17$\pm$0\DEG{}.02, suggesting that the binary is edge-on.  Assuming that the spin axis of the MSP is perpendicular to the orbital plane, this suggests the same value for $\zeta$.

The best-fit values reported for PSR J1614$-$2230 do not agree with those of \citet{Venter09}, predicting similar values of $\zeta$ but much lower values of $\alpha$.  These geometries are better at matching the main radio peak but miss the radio IP entirely.

\begin{figure}
\begin{center}
\includegraphics[width=.75\textwidth]{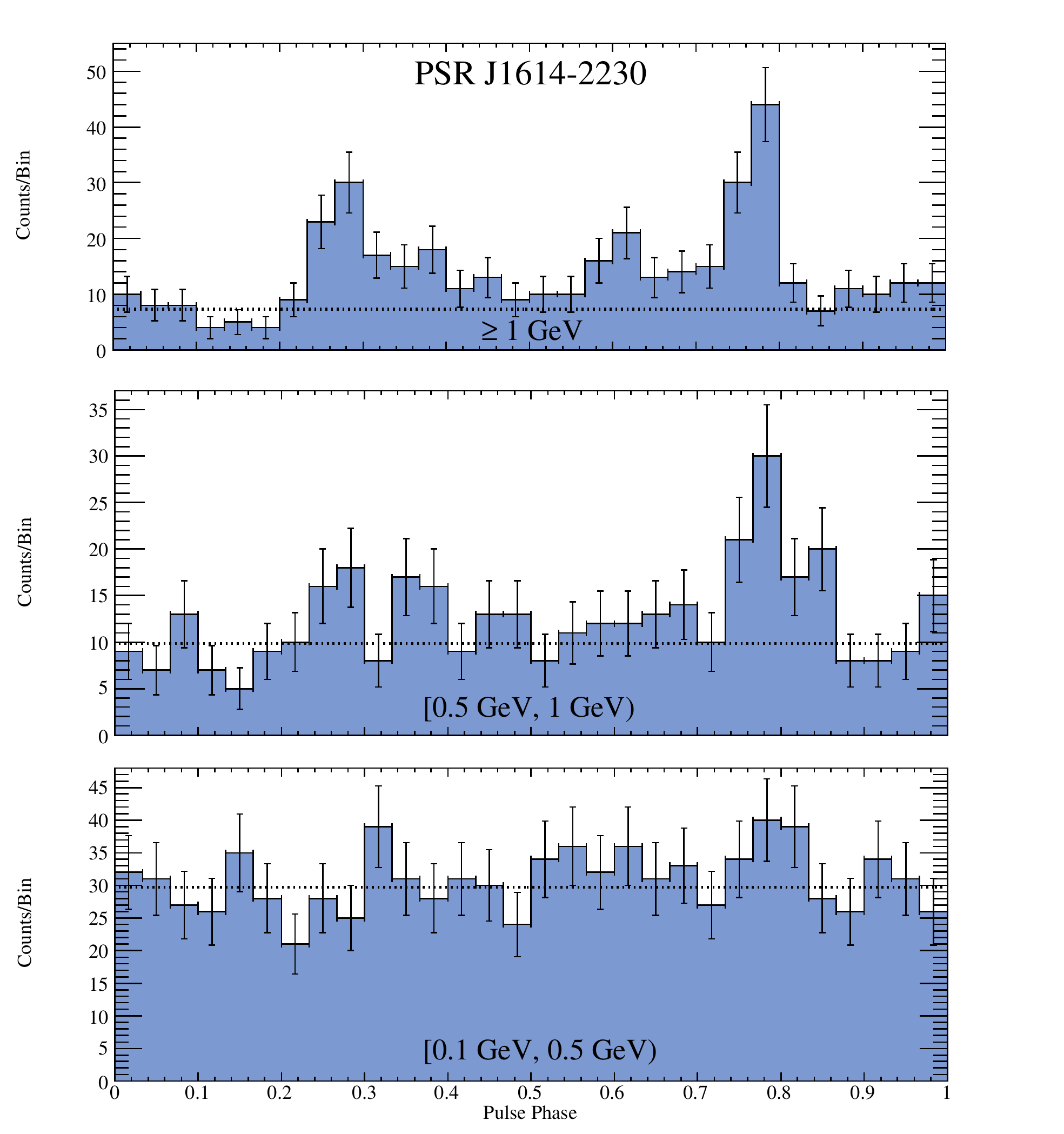}
\end{center}
\small\normalsize
\begin{quote}
\caption[Gamma-ray light curves of PSR J1614$-$2230 in different energy bands]{\emph{(Top):} Gamma-ray light curve of PSR J1614$-$2230 following the selections described in the text but only for events with reconstructed energies $\geq$ 1 GeV. \emph{(Middle):}  The same, but for events with reconstructed energies $\geq$ 0.5 GeV and $<$ 1 GeV.  \emph{(Bottom):} The same, but for events with reconstructed energies $\geq$ 0.1 GeV and $<$ 0.5 GeV.\label{ch7J1614eneLCs}}
\end{quote}
\end{figure}

The predicted height of the gamma-ray peaks do not match those observed, particularly for the second gamma-ray peak, but peak ratios in gamma-ray pulsars are known to vary with energy (see Fig.~\ref{ch7J1614eneLCs}) and such behavior is not captured by these geometric models.  Additionally, the predicted f$_{\Omega}$ values are both $\lesssim$ 0.5 predicting high but reasonable efficiencies.  While this last argument does argue for the geometry predicted here it should be noted that such considerations are not part of the likelihood fitting.

\small\normalsize

\begin{figure}[h]
\begin{center}
\includegraphics[width=.6\textwidth]{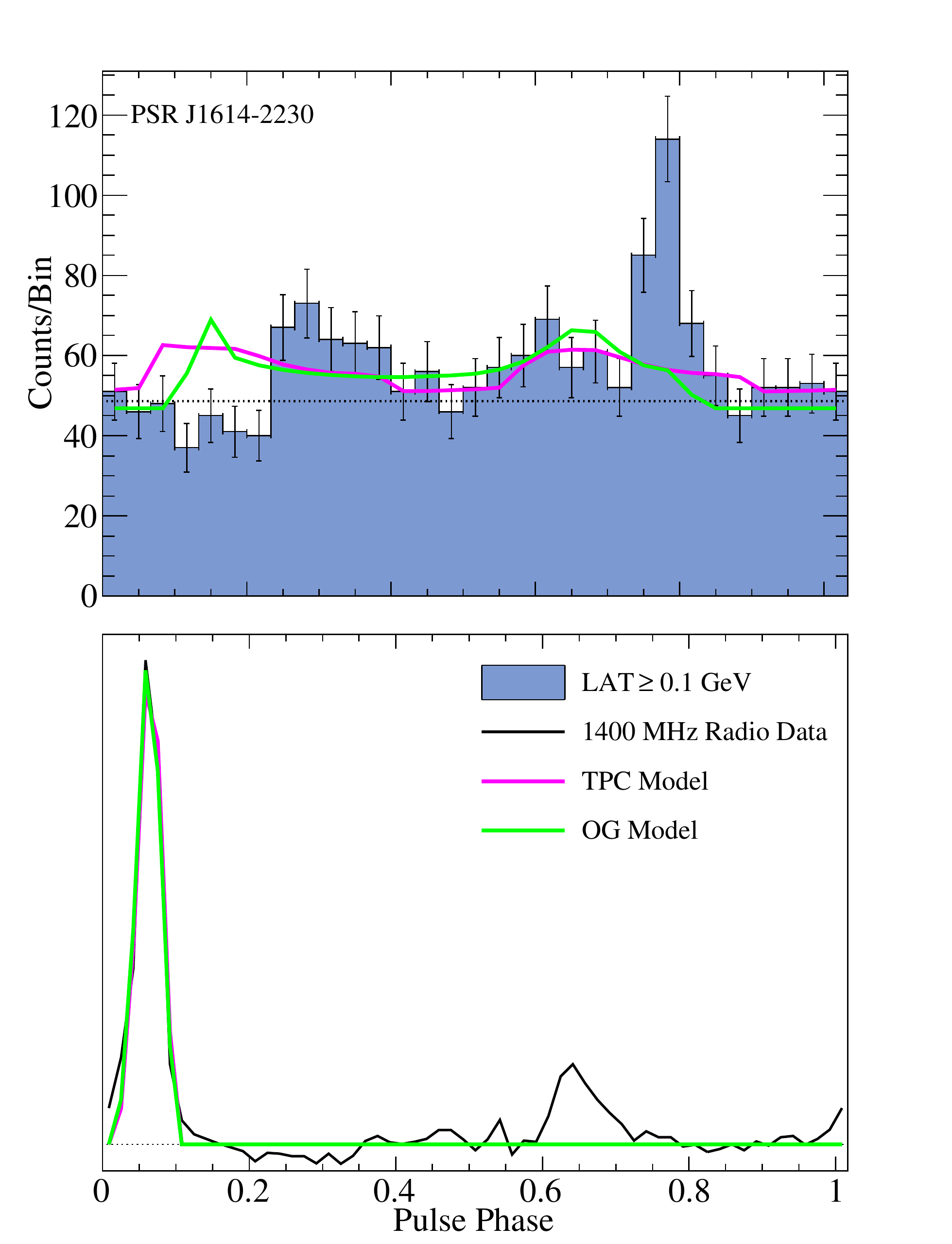}
\end{center}
\small\normalsize
\begin{quote}
\caption[Data and best-fit light curves of PSR J1614$-$2230 for $\zeta\ \geq$ 80\DEG{}]{\emph{(Top):} Gamma-ray data and best-fit model light curves of PSR J1614$-$2230 under the constraint $\zeta\ \geq$ 80\DEG{}, pink is TPC and green is OG. \emph{(Bottom):} The same, but for the radio data and model.\label{ch7highZJ1614}}
\end{quote}
\end{figure}

The best-fit $\zeta$ values of \citet{Venter09} are in better agreement with the constraints of \citet{Demorest10} than the values reported in Table~\ref{ch7TPCOG}.  For both the TPC and OG model the confidence contours do allow for the possibility of solutions with $\zeta$ near 90\DEG{} and $\alpha\ \lesssim$ 40\DEG{}.  Restricting $\zeta$ to be $\geq$ 80\DEG{} and searching the chains for the best-fit results in the light curves shown in Fig.~\ref{ch7highZJ1614}.

\small\normalsize

The TPC model parameters are $(\alpha,\zeta)$ = (40\DEG{},89\DEG{}) with an infinitely thin gap width.  The OG model has parameters  are $(\alpha,\zeta)$ = (39\DEG{},88\DEG{}) with a gap width of 0.05 and an infinitely thin emission layer.  The $-log(\mathcal L)$ values for these TPC and OG models are 157.7 and 157.4, respectively.

For these TPC and OG model geometries f$_{\Omega}$ = 0.89 and 0.59, respectively.  This predicts efficiencies of $\sim$100 \% and $\sim$67\%, for the TPC and OG geometries, respectively.  The predicted efficiency of the OG model is higer than average but still less than 100\%.

The best-fit geometries, with constrained $\zeta$, agree very well with that found by \citet{Demorest10}; however, the model radio light curves are no better than those of the overall maximum likelihood geometry and the gamma-ray models are clearly not good fits.  While the main radio pulse is matched well, the smaller radio peak is not found in any of the fits presented here.  That may suggest that either the emission altitude of the cone beam is too low or that the radio emission from this pulsar should be modeled with a core component.

To date, no polarimetric data has been published for this source to indicate if the core emission component is warranted or not.  However, a core component would suggest a higher inclination angle which could lead to better gamma-ray model light curves and lower predicted efficiencies for both TPC and OG models.

\subsubsection{PSR J1713+0747}\label{ch7J1713}
Several authors have reported polarizations measurements of PSR J1713+0747, all measuring some degree of circular polarization, with sense reversing in some cases, suggestive of a core component (Xilouris et al., 1998; Stairs et al., 1999; Ord et al., 2004; Yan et al., 2011).  No predictions on the viewing geometry of this MSP have been made.

The TPC and OG models, in conjunction with a hollow-cone beam radio model, reproduce the observed light curves remarkably well.  This is at odds with the claims of the main radio component having the qualities of a core emission component but as the gamma-ray statistics are still rather poor it can not be ruled out that a fit with a core radio beam model could also produce a comparable fit.  Such an exercise should be carried out and compared to the hollow-cone beam predictions, especially as further gamma-ray statistics accumulate.

\subsubsection{PSR J1744$-$1134}\label{ch7J1744}
Polarization measurements for PSR J1744$-$1134 all present a flat position angle swing through the peaks which favor the cone beam interpretation.  \citet{Xilouris98} detect only a partially polarized profile but \citet{Stairs99}, \citet{Yan11}, and \citet{Ord04} observe profiles with near complete linear polarization.  Additionally, \citet{Yan11} suggest that $\zeta\ >\ \alpha$ based on the assumption that the polarization angle swing in the main component is a continuation of that in the precursor component.

This MSP was one of only two for which \citet{Venter09} found it necessary to use the PSPC model to explain the profile shape and observed radio-to-gamma lag.  They predict a viewing geometry of $(\alpha,\zeta)$ = (50\DEG{},80\DEG{}) and were able to match both the radio and gamma-ray profiles well.

The best-fit values in Table~\ref{ch7PSPC} for PSR J1744$-$1134 agree quite well with those of \citet{Venter09}.  Both fits predict $\zeta\ >\ \alpha$ as suggested by \citet{Yan11}.

The gamma-ray light curve of this MSP was also fit with standard TPC and OG emission models in order to ensure that the PSPC model was the best-fit.  When using the TPC and OG models, the gamma-ray light curve could be matched well but not for geometries which resulted in good fits to the radio.

\begin{figure}[h]
\begin{center}
\includegraphics[width=1.0\textwidth]{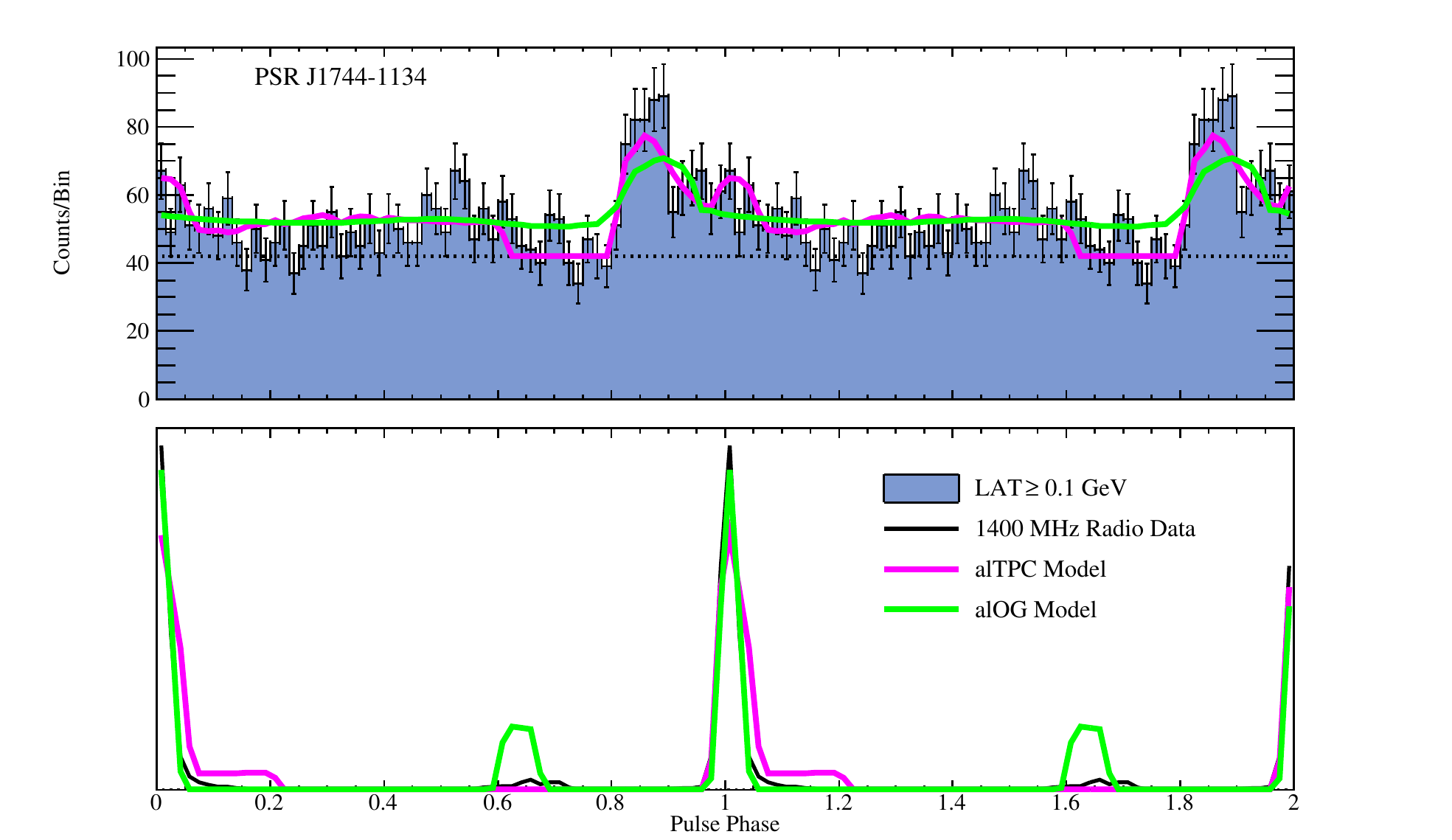}
\end{center}
\small\normalsize
\begin{quote}
\caption[Altitude-limited model fits for PSR J1744$-$1134]{Data and best-fit gamma-ray (\emph{top}) and radio (\emph{bottom}) light curves for PSR J1744$-$1134 using the alTPC (pink) and alOG (green) models.\label{ch7J1744alLCs}}
\end{quote}
\end{figure}

Therefore, this MSP was also fit with the alTPC and alOG models (see Fig.~\ref{ch7J1744alLCs}) both of which returned slightly better likelihood values than the PSPC model.  The high degree of linear polarization is contrary to what is seen in other phase-aligned, gamma-ray MSPs (PSRs J0034$-$0534 and J1959+2134) but is also observed in PSR J1939+2134 (though it seems that this is due to the predicted orthogonal geometry, see Section~\ref{ch7J1939}).  Thus, it is unclear if the altitude-limited models are viable for this MSP and more investigation into the expected polarization properties is warranted.

\small\normalsize

\subsubsection{PSR J1823$-$3021A}\label{ch7J1823}
PSR J1823$-$3021A is the first globular cluster MSP from which significant HE pulsations have been detected \citep{Freire11}.  \citet{Pellizzoni09} have claimed a pulsed detection from PSR B1821$-$24 in the globular cluster M28 using \emph{AGILE}.  However, the significance was low, 4.2$\sigma$; the pulsar was only detectable over 5 days out of their multi-month observation; and the detection has not yet been reconfirmed by either \Fermi{} or \emph{AGILE}.   While some ``strange'' physics can be invoked \citep{Pellizzoni09} to explain the transitory nature of the HE pulsations, without confirmation it is unclear if PSR B1821$-$24 does emit significantly in gamma-rays.

PSR J1823$-$3021A has a measured $\dot{\rm P}$ (see Table~\ref{ch7MSPtable}) on the order of $10^{-18}\ \rm s\ s^{-1}$, 10 to 100 times that of typical MSPs, which results in large derived values of spin-down energy and magnetic field strengths.  However, this MSP is located in the globular cluster NGC 6624 and projected to lie near the cluster core.  Thus, the measured $\dot{\rm P}$ may be increased significantly from the true value via gravitational acceleration along the line-of-sight.

\citet{Freire11} first reported the HE pulsations from this source and attempted to use the LAT observations to place constraints on the intrinsic $\dot{\rm P}$ value.  After fitting the observed HE spectrum and calculating the energy flux it was found that PSR J1823$-$3021A converts rotational energy into gamma rays with an efficiency of $\sim$10\%, typical of other MSPs and suggesting that the measured $\dot{\rm P}$ need not be that strongly affected by gravitational acceleration.

\citet{Stairs99} reported polarization measurements for this MSP at 610 MHz.  Similar to what was found for PSR J0034$-$0534, this pulsar shows no polarization angle swing across the pulse with a mean linear polarization of 0\%.  This is consistent with the radio emission being caustic in nature.

This pulsar is among the class of HE MSPs for which the radio and gamma-ray profiles are nearly-aligned in phase.  Thus, \citet{Freire11} (see their supplementary online material when published) modelled the gamma-ray and radio emission from PSR J1823$-$3021A with the alTPC and alOG models, choosing the best-fit geometries via the MCMC maximum likelihood technique described in Chapter 6.  They found best-fit emission geometries for the alTPC model of $(\alpha,\zeta)$ = (51\DEG{},68\DEG{}) with infinitely thin gaps for both the radio and gamma-ray emission regions with R$_{max}^{\gamma}$ = 1.0 and (R$_{min}^{\rm R}$,R$_{max}^{\rm R}$) = (0.2,0.9).  For the alOG model they found $(\alpha,\zeta)$ = (69\DEG{},75\DEG{}) with a gamma-ray gap width of 0.05, a radio gap width of 0.10, R$_{max}^{\gamma}$ = 1.2, and (R$_{min}^{\rm R}$,R$_{max}^{\rm R}$) = (0.2,0.8).

The results presented in Table~\ref{ch7alTPCOG} for PSR J1823$-$3021A agree well with those of \citet{Freire11} though the altitude extent of the radio emission regions are slightly different and the alOG results presented here have infinitely thin gaps.  Their analysis was done for $\sim$26 months of LAT data and the light curve fits were applied to LAT data $\geq$ 0.5 GeV.  The latter choice was motivated by the use of a simple annular-ring to estimate the background level which seemed to underestimate the low-energy contribution of the pulsar.  The results presented here also fit the $\geq$ 0.5 GeV light curve but this was done to mitigate issues with the normalization of the isotropic diffuse component (fit to a value of $\sim10^{-6}$) which could have effected the \emph{gtsrcprob} derived background estimate at lower energies.

\subsubsection{PSR J1939+2134}\label{ch7J1939}
PSR J1939+2134 is the first MSP ever discovered \citep{Backer82} and has, therefore, occasioned a great deal of scientific study.  Shortly after the initial discovery \citet{Usov83} assessed the viability of detecting this MSP in HE gamma-rays using a PC model, motivated in part by its relatively large spin-down energy loss rate, predicting an efficiency of $\sim$1\%.

Many authors have reported polarimetric observations of PSR J1939+2134 and agree that both peaks show a high degree of linear polarization with flat position angle swings.  \citet{Ashworth83} first reported on the polarimetric properties of this MSP, and noted that while the near 180\DEG{} phase separation of the two peaks argued for an orthogonal geometry the near 90\DEG{} phase separation of the position angles between the two peaks argued against this interpretation as they should be nearly equal.

\citet{Stinebring83} confirmed early polarization measurements and suggested that the existence of orthogonal mode switches, hinted at by depolarization near the inner peak edges, could reconcile the position angle differences with the orthogonal rotator geometry.

\citet{SC83} reported polarization observations to lower frequencies and confirmed earlier findings which suggested a decrease in the average linear polarization percentage in the IP with decreasing frequency.  The lower frequency observations showed nearly equal position angles in each peak, contrary to what is seen at higher frequencies.

Orthogonal mode switching for all frequencies was demonstrated by \citet{TS90}.  However, they also argued that the typical core beam interpretation, as expected in an orthogonal rotator, can not explain the narrowness of the peaks.  \citet{TS90} also noted that the flat position angle swings through the peaks is consistent with the line of sight passing through the magnetic axis only if one assmes the existence of an unresolved 180\DEG{} transition at the center of each peak.

Later polarization measurements (Xilouris et al., 1998; Stairs et al., 1999; Ord et al., 2004; and Yan et al., 2011) confirmed earlier findings.  \citet{Yan11} discovered weak features before both the main pulse and the IP which were not observable in earlier observations with lower resolution.  With flux densities $\sim$0.5\% of that of the main pulse, these feautures would not be significant in the likelihood fits using the value of $\sigma_{\rm R}$ quoted in Table~\ref{ch7FitParTable}.  Additionally, \citet{Yan11} suggest that the orthogonal mode switching observed by other authors is closer to 60\DEG{}, arguing against an orthogonal geometry.

\citet{CS84} extrapolated timing measurements of J1939+2134 at 4 frequencies (from 320 to 1390 MHz) to infinite frequency and found residuals with no apparent frequency dependence.  This suggested that, given their timing uncertainty, the emission altitudes for all frequencies were the same within $\pm$2 km.

HE pulsations from this MSP were first announced by \citet{Guillemot11} who fit the gamma-ray and radio profiles with alTPC and alOG models using the MCMC maximum likelihood technique described in Chapter 6 and $\sim$18 months of LAT survey data.  They also reanalyzed the polarization data of \citet{Stairs99} in order to generate RVM based predictions on the geometry.  They found best-fit parameters for the alTPC model of $(\alpha,\zeta)$ = (75\DEG{},80\DEG{}), $w^{\gamma}$ = 0.10, $w^{\rm R}$ = 0.0, R$_{max}^{\gamma}$ = 1.0, and (R$_{min}^{\rm R}$,R$_{max}^{\rm R}$) = (0.7,0.9).  The RVM analysis yielded $\alpha$ = 89\DEG{} and $\beta\ =\ -3^{\circ}$, a geometry which is well within the 39\% marginalized confidence contours.  Their best-fit alOG parameters were $(\alpha,\zeta)$ = (84\DEG{},84\DEG{}), $w^{\gamma}$ = 0.05, $w^{\rm R}$ = 0.0, R$_{max}^{\gamma}$ = 1.0, and (R$_{min}^{\rm R}$,R$_{max}^{\rm R}$) = (0.6,0.9).

Spectral analysis of PSR J1939+2134 is complicated by the bright, Galactic diffuse emission \citep{Guillemot11} and the fact that the pulsar is at Galactic latitude of $\sim$0\DEG{}.  However, given the larger parallax distance and the observed flux, it is clear that an efficiency greater than 1\% is needed, casting doubt on the model of \citet{Usov83}.  Additionally, given the large PCs of MSPs it is difficult to reproduce the relatively sharp peaks observed in the LAT profile with low-altitude PC models.

The results presented here agree very well with those of \citet{Guillemot11} and both geometry estimates agree with suggestions from the radio data of an orthogonal rotator with a viewing angle looking nearly down the magnetic axis.  Unlike PSR J0034$-$0534 which shows no strong polarization, this MSP displays high levels of linear polarization.  However, for an orthogonal rotator with $\zeta\ \sim$ 90\DEG{} little depolarization is expected in the altitude-limited model \citet{Venter11} which further argues in favor of the best-fit geometry in Table~\ref{ch7alTPCOG}.

While the best-fit results predict that the radio emission occurs over a significant range of altitudes the caustic nature of the altitude-limited models is consistent with the findings of \citet{CS84} if one assumes that all frequencies are emitted over an extended range of altitudes.  The results of light curve fits with the altitude-limited models using radio profiles at different frequencies would confirm this agreement if the best-fit parameters found the same values R$_{min}^{\rm R}$ and R$_{max}^{\rm R}$ for all frequencies.  Additionally, \citet{Guillemot11} were able to use the RVM fits to constrain the emission altitude of the radio to be 0.65 R$_{\rm LC}$ which is consistent with both the alTPC and alOG predictions within the current resolution of 0.10 R$_{\rm LC}$.

\subsubsection{PSR J1959+2048}\label{ch7J1959}
PSR J1959+2134 was the first ``black widow'' MSP ever discovered \citep{Fruchter88}, so called because it is in a binary orbit with an extremely low-mass companion ($\sim$0.022 M$_{\odot}$) which is believed to have been mostly ablated away by the pulsar wind.  This pulsar displays less than 2\% linear polarization though some evidence exists for sign-changing, circular polarization through both peaks \citep{TS90}.  Observations well away from the eclipsing phase have confirmed that the lack of polarization is not due to interactions with the companion.  Additionally, the eclipsing nature of this system and assumptions that the spin axis should have aligned with the orbital angular momentum suggest that the viewing angle of the system is near 90\DEG{}.

\citet{Guillemot11} reported the pulsed detection of PSR J1959+2134 above 0.1 GeV with the LAT.  They also modeled this pulsar with the alTPC and alOG models.  Their best-fit alTPC model parameters were $(\alpha,\zeta)$ = (43\DEG{},44\DEG{}), infinitely thin radio and gamma-ray gap widths, R$_{max}^{\gamma}$ = 0.8, and (R$_{min}^{\rm R}$,R$_{max}^{\rm R}$) = (0.3,0.5).  Their best-fit alOG models parameters were $(\alpha,\zeta)$ = (16\DEG{},82\DEG{}), gamma-ray and radio gap widths of 0.10 each, R$_{max}^{\gamma}$ = 1.20, and (R$_{min}^{\rm R}$,R$_{max}^{\rm R}$) = (0.9,1.1).

The results presented here disagree with those of \citet{Guillemot11}, mainly in the best-fit $\alpha$ and $\zeta$ values.  Note that they used a gamma-ray light curve of 30 bins owing to lower statistics.  The use of 60 gamma-ray bins leads to both fits with $\zeta$ near 90\DEG{}, as opposed to only the alOG model of \citet{Guillemot11} having such a large viewing angle.

Both models have trouble reproducing the HE peak near 0.2 in phase properly.  This may, in part, be due to the appearance of lower-significance structure.  These fits do suggest a large viewing angle for PSR J1959+2048 with moderate inclination angles, with large degrees of depolarization predicted, in line with the observations of $<$ 2\% linear polarization.

\subsubsection{PSRs J2017+0603 and J2302+4442}\label{ch7NancayMSPs}
PSRs J2017+0603 and J2302+4442 were discovered in radio observations of bright, \FL{} unassociated sources \citep{Cognard11}.  The gamma-ray and radio light curves of both MSPs have been fit with OG and TPC models using the MCMC maximum likelihood technique described in Chapter 6.  No polarization measurements have yet been published for either MSP.

The best-fit parameters in Table~\ref{ch7TPCOG} for PSR J2017+0603 and J2302+4442 are the same to within $\pm1^{\circ}$ of the values reported by \citet{Cognard11}.  This is not suprising as the gamma-ray data they used spanned nearly 22 months and their methods of event selection and background estimation are the same as those used here.  The radio profiles of both MSPs are complex with three or more components each, suggesting that use of a more complex radio beam geometry may be necessary.

\subsubsection{PSR J2124$-$3358}\label{ch7J2124}
Polarimetric observations of PSR J2124$-$3358 have been reported by \citet{Ord04} and \citet{ManHan04}.  This MSP displays radio emission across the entire pulse with a complex shape.  All components display some degree of linear polarization with no identifiable core component.

\cite{ManHan04} attempted to fit the position angle swing using the RVM and obtained a best-fit geometry of $(\alpha,\zeta)$ = ($48^{\circ}\pm3^{\circ},67^{\circ}\pm5^{\circ}$) suggesting that emission is visible from both poles.  However, they note that there is a large degree of covariance in the parameters and suggest a more realistic range from $(\alpha,\zeta)$ = (20\DEG{},27\DEG{}) to (60\DEG{},80\DEG{}).  Additionally, the presence of radio emission through the entire pulse argues for emission from one pole and thus a low value of $\alpha$.

\citet{Bogdanov08} modeled the thermal X-ray emission from this MSP but the results were unconstrained in $\alpha$ and $\zeta$.

More recently, \citet{Yan11} have reported new polarization measurements of PSR J2124$-$3358 which agree with earlier observations.  They note that if the linear-polarization position angles between $-0.3$ and $-0.2$ in phase were increased, uniformly, by 90\DEG{} then the total position angle curve would approximately match that expected from the RVM with $\alpha\sim25^{\circ}$ and $\zeta\ >\ \alpha$.  This geometry would be in agreement with the fact that radio emission is observed across almost the entire pulse phase.  However, no mechanism is proposed for how the position angles in this phase range might be shifted from that predicted by the RVM.

PSR J2124$-$3358 was the second MSP for which \citet{Venter09} found it necessary to invoke the PSPC model to match the HE profile and radio-to-gamma lag.  They found a best-fit viewing geometry of ($\alpha,\zeta$) = (40\DEG{},80\DEG{}).

The best-fit results presented in Table~\ref{ch7PSPC} disagree with those of \citet{Venter09}.  Their solution finds a radio profile which is too narrow and a gamma-ray solution which is too broad.  These results also disagree with the best-fit RVM geometry of \citet{ManHan04} but agree well with their broader confidence range.  Note that neither the results of \citet{Venter09} nor those presented here are able to reproduced all of the observed features in the radio profile.

The value of $\alpha$ = 25\DEG{}, suggested by \citet{Yan11}, agrees quite well with the maximum likelihood value but the best-fit $\zeta$ is $<\ \alpha$, contrary to their prediction.  However, the marginalized confidence contours easily permit a solution with similar $\alpha$ and a larger $\zeta$.  Light curve fitting with a more complicated radio emission model is needed to test these predictions further.

\subsubsection{PSR J2214+3000}\label{ch7J2214}
PSR J2214+3000 has been fit with the alTPC and alOG models based on the apparent alignment of the radio and gamma-ray peaks; however, it should be noted that the gamma-ray peaks are very broad which complicates the claim of phase-alignment.  Light curve fits with standard TPC and OG models can not reproduce the observed radio and gamma-ray light curves simultaneously.  If the profiles are not aligned then the radio lag is small with the gamma-ray light curve possibly preceding the radio and thus a PSPC model is possible.

\begin{figure}
\begin{center}
\includegraphics[width=1.0\textwidth]{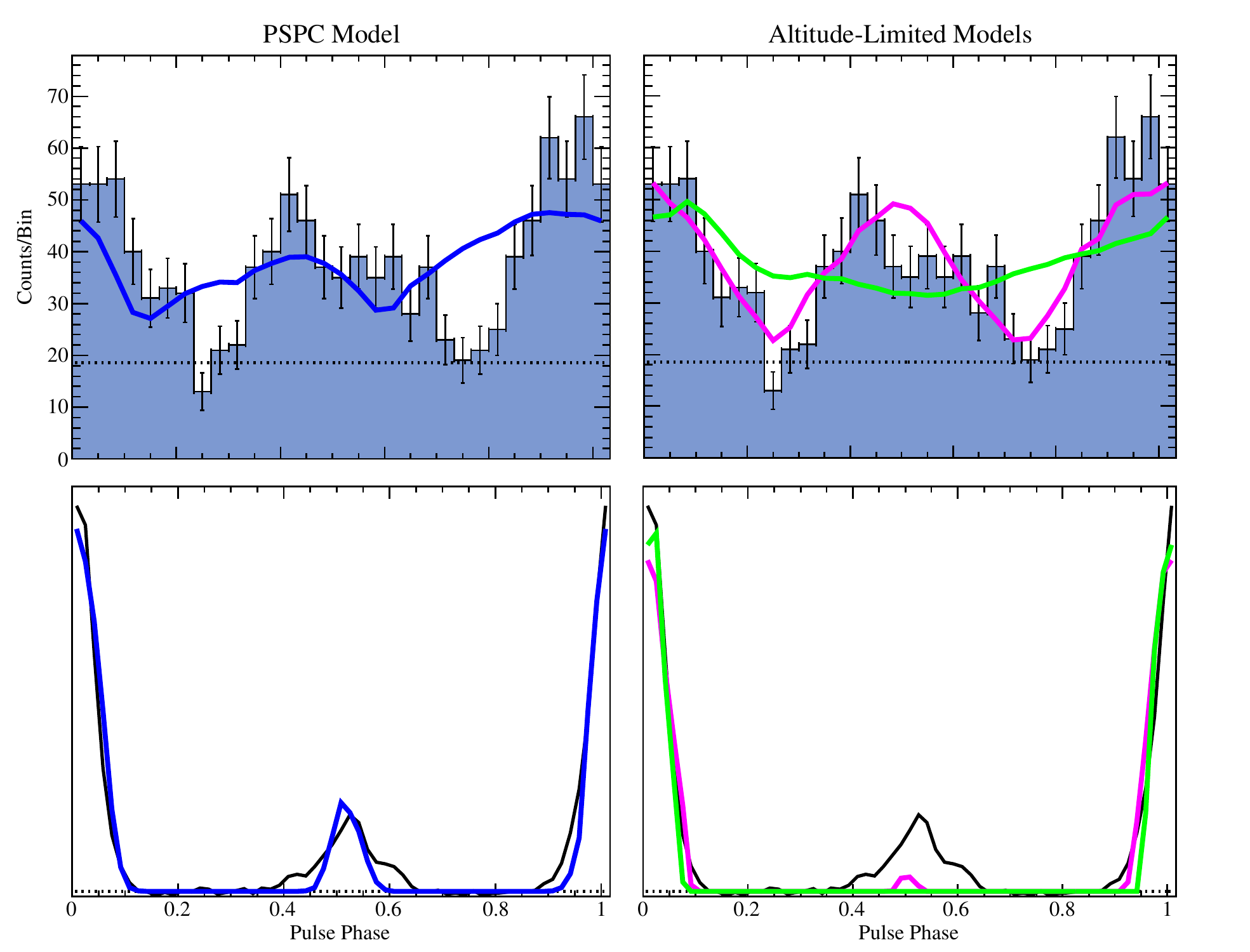}
\end{center}
\small\normalsize
\begin{quote}
\caption[Comparison of PSPC and alTPC fits to PSR J2214+3000]{\emph{(Left):} Observed and best-fit gamma-ray (\emph{top}) and radio (\emph{bottom}) light curves for the PSPC model (blue curves).  \emph{(Right):} Observed and best-fit gamma-ray (\emph{top}) and radio (\emph{bottom}) light curves for the alTPC (pink curves) and alOG (green curves) models.\label{ch7J2214LCs}}
\end{quote}
\end{figure}

Fig.~\ref{ch7J2214LCs} compares fits to J2214+3000 with the altitude-limited and PSPC models.  The PSPC fit has a $-\log(\mathcal{L})$ = 145.8, while this is clearly a better fit than the alOG fitting with the alTPC model decreases the $-\log(\mathcal{L})$ by $\sim15$.  The PSPC model has problems properly matching the gamma-ray peaks but reproduces the radio peaks well.  The alTPC model matches the gamma-ray peaks well and does find two radio peaks at the proper phases.  The minor radio peak in the alTPC model is not reproduced with a high enough intensity; however, the disparity in the alTPC radio peaks is controlled by limiting the value of R$_{max}^{\rm R}$, with finer steps in altitude it may be possible to more accurately reproduce the radio profile.

\small\normalsize

Given the preceding arguments, the alTPC model appears very convincing but the PSPC model can not be ruled out even though the likelihood also prefers the alTPC model.  Polarization measurements would help to further choose between the models.  Little, or no, observed polarization would argue strongly in favor of the alTPC model while a high degree of polarization would cast doubt on radio emission of a caustic nature.  Additionally, if the polarization measurements suggested a core beam, as opposed to the hollow-cone beam used here, that could lead to more favorable PSPC geometries.

\begin{figure}
\begin{center}
\includegraphics[width=1.0\textwidth]{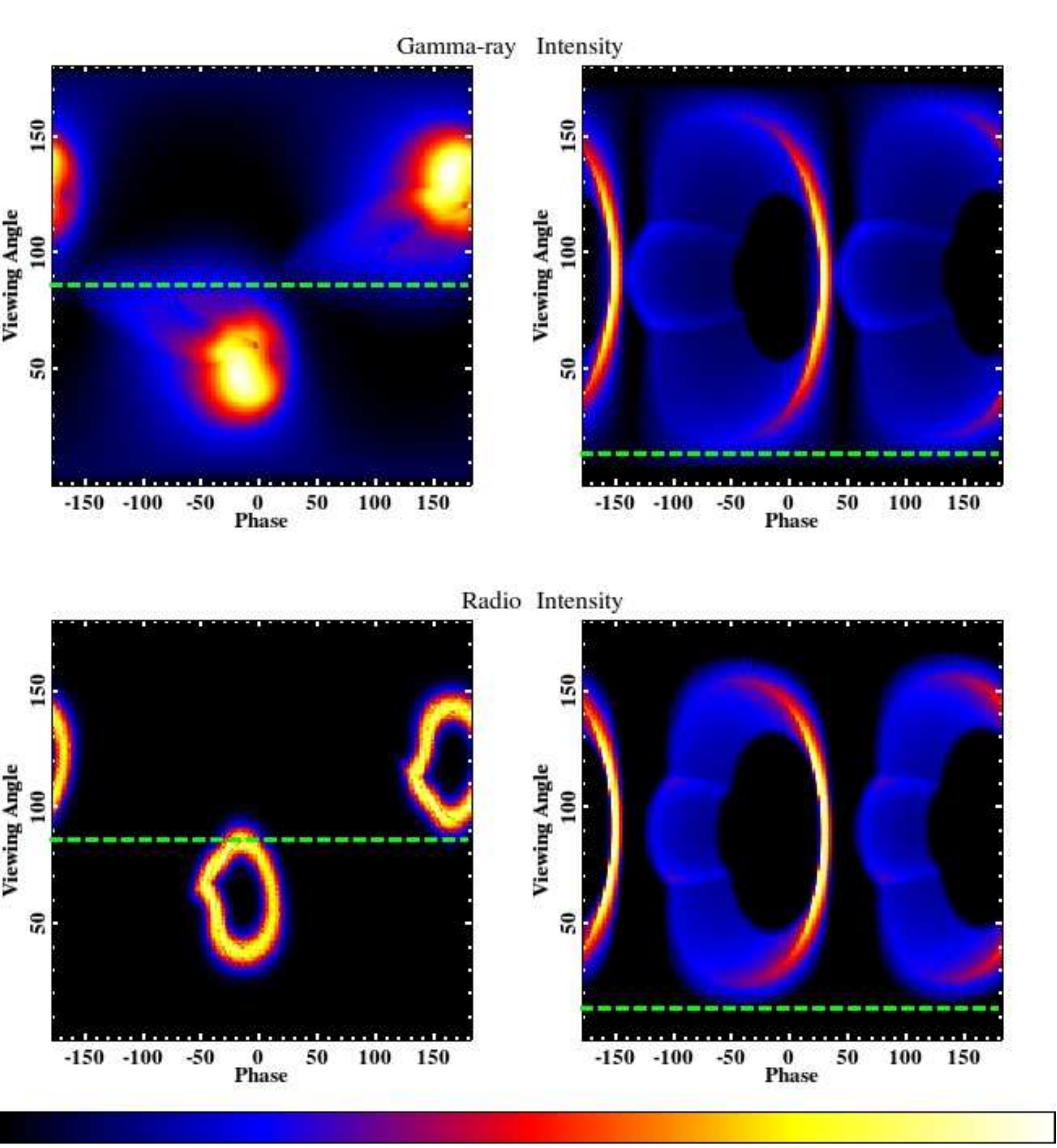}
\end{center}
\small\normalsize
\begin{quote}
\caption[PSPC and alTPC simulated emission phase plots for PSR J2214+3000]{Simulated gamma-ray (\emph{top}) and radio (\emph{bottom}) emission phase plots for PSPC (\emph{left}) and alTPC (\emph{right}) fits to the light curves of PSR J2214+3000.  Green dashed lines represent best-fit $\zeta$ values.\label{ch7J2214ppComp}}
\end{quote}
\end{figure}

By examining the simulated emission phase plots corresponding to the PSPC and alTPC best-fit geometries (Fig.~\ref{ch7J2214ppComp}) it is clear that the observed light curves require that both poles are viewed.  However, the two models achieve this in different ways.  The PSPC model clips both poles as they pass above and below the line of sight.  The alTPC model requires a nearly orthogonal geometry with a low viewing angle such that both poles are above the line of sight.

\small\normalsize

\subsection{Combined Results}\label{ch7combined}
Pulsar parameters derived from timing observations as well as cutoff energies from spectral analyses are given in Table~\ref{ch7Vitals}, distances are taken from \citet{AbdoPSRcat}, \citet{AbdoJ0034}, \citet{Freire11}, \citet{Cognard11}, \citet{Ransom11}, and \citet{Camilo94} and references therein; however, the distance of PSR J1902$-$5105 was derived directly from the NE2001 electron density model \citep{NE2001} using the position and dispersion measure values of \citet{Camilo11}.

\small\normalsize

\begin{deluxetable}{l c c c c}
\tablewidth{0pt}
\tablecaption{Additional MSP Measured and Derived Parameters}
\startdata
\underline{JName} & \underline{$\dot{\rm E}$ ($10^{33}$ erg s$^{-1}$)} & \underline{d (pc)} & \underline{B$_{\rm LC}$ (kG)} & \underline{$E_{C}$ (GeV)}\\ 
J0030+0451 & 3.00 & 300$\pm$90 & 17.8 & 2.5$\pm$0.3\\
J0034$-$0534 & 15.7 & 530$\pm$210 & 98.9 & 2.0$\pm$0.6\\
J0218+4232 & 240.0 & 2700$\pm$600 & 313.1 & 4.7$\pm$1.3\\
J0437$-$4715 & 3.00 & 156$\pm$1 & 13.7 & 2.7$\pm$0.9\\
J0613$-$0200 & 13.0 & 480$\pm$190 & 54.3 & 6.6$\pm$0.2\\
J0614$-$3329 & 22.0 & 1900$\pm$440 & 70.1 & 4.0$\pm$0.4\\
J0751+1807 & 6.00 & 600$\pm$600 & 32.3 & 6.7$\pm$2.9\\
J1231$-$1411 & 18.0 & 400$\pm$51 & 53.9 & 2.9$\pm$0.4\\
J1614$-$2230 & 5.00 & 1270$\pm$390 & 33.7 & 1.7$\pm$0.3\\
J1713+0747 & 3.53 & 1100$\pm$500 & 19.2 & 4.7$\pm$2.6\\
J1744$-$1134 & 4.00 & 357$\pm$43 & 24.0 & 0.7$\pm$0.4\\
J1823$-$3021A & 830.0 & 8400$\pm$600 & 248.6 & 1.4$\pm$0.6\\
J1902$-$5105 & 68.0 & 1164$\pm$208 & 220.3 & 3.0$\pm$0.8\\
J1939+2134 & 1097.6 & 7700$\pm$3800 & 984.7 & 1.2$\pm$0.7\\
J1959+2048 & 74.8 & 2500$\pm$1000 & 249.0 & 1.3$\pm$0.6\\
J2017+0603 & 13.4 & 1560$\pm$160 & 58.6 & 2.9$\pm$0.5\\
J2124$-$3358 & 4.00 & 250$\pm$250 & 18.8 & 1.8$\pm$0.2\\
J2214+3000 & 18.0 & 1500$\pm$189 & 64.1 & 2.4$\pm$0.4\\
J2302+4442 & 3.74 & 1180$\pm$230 & 17.3 & 3.8$\pm$0.7\\
\enddata\label{ch7Vitals}
\end{deluxetable}

\small\normalsize

The $\dot{E}$ values reported in Table~\ref{ch7Vitals} do not account for any possible contribution to the observed $\dot{P}$ values due to Galactic gravitational acceleration along the line of sight ($a_{l}$) which could increase the measured $\dot{P}$ by
\begin{equation}\label{ch7al}
\dot{\rm P}_{\mathnormal acc}\ =\ \frac{a_{l}}{c} \rm P
\end{equation}
\noindent{}from the true value.  Note that Eq.~\ref{ch7al} is an increasing function of P.  The gravitational acceleration can be approximated from Galactic rotational velocity curves by relating $a_{l}\ =\ v_{rot}^{2}/r_{gal}$, where $r_{gal}$ is the distance at the MSP position to the Galactic center.  Note that this implies Eq.~\ref{ch7al} is an increasing function of $v_{rot}$ and a decreasing function of $r_{gal}$, though $v_{rot}$ will be a function of $r_{gal}$ so the overall dependence is unclear.

\citet{Fich89} fit multiple functional forms to the Galactic rotation velocity (valid for $r_{gal}$ from 3 to 17 kpc) curve using measurements of gas in the Galaxy.  They found that the data could either be described by a power law or linear functional form.  The power law functional form is preferred in the inner Galaxy and leads to a maximum $\dot{\rm P}_{\mathnormal acc}\ \sim\ 5\times10^{-23}$ s s$^{-1}$ for a distance of 3 kpc and assuming that the entire acceleration is along the line of sight to Earth.  This is typically 2 to 3 orders of magnitude less than the $\dot{\rm P}$ values in Table~\ref{ch7MSPtable} which suggests that the $\dot{E}$ values in Table~\ref{ch7Vitals} are not greatly affected by Galactic gravitational acceleration.

The cutoff energies (and other spectral parameters) have been taken from spectral analysis\footnote{Credit and many thanks to \"{O}. \c{C}elik (NASA GSFC).} of the same data used to produce the gamma-ray light curves, but using 10\DEG{} ROIs centered on the radio MSP positions, and sources from a preliminary version of the 2FGL catalog \citep{Abdo2FGL} except for the parameters of PSRs J1744$-$1134, J1823$-$3021A, J1939+2134, and J1959+2048.  The first two MSPs encountered errors for unbinned maximum likelihood analysis and thus values from the LAT PSR catalog \citep{AbdoPSRcat} were used for PSR J1744$-$1134 and those from \citet{Freire11} for PSR J1823$-$3021A.  For the latter two MSPs the values from \citet{Guillemot11} have been used.

Measured gamma-ray luminosities (incorporating the predicted beaming factors) and derived efficiencies ($\eta_{\gamma}\ \equiv\ L_{\gamma}/\dot{E}$) for each MSP analyzed in this thesis are presented in Tables~\ref{ch7TPCOGspec},~\ref{ch7PSPCspec}, and~\ref{ch7alTPCOGspec} sorted by which models were used to fit the light curves.

\begin{figure}
\begin{center}
\includegraphics[width=1.\textwidth]{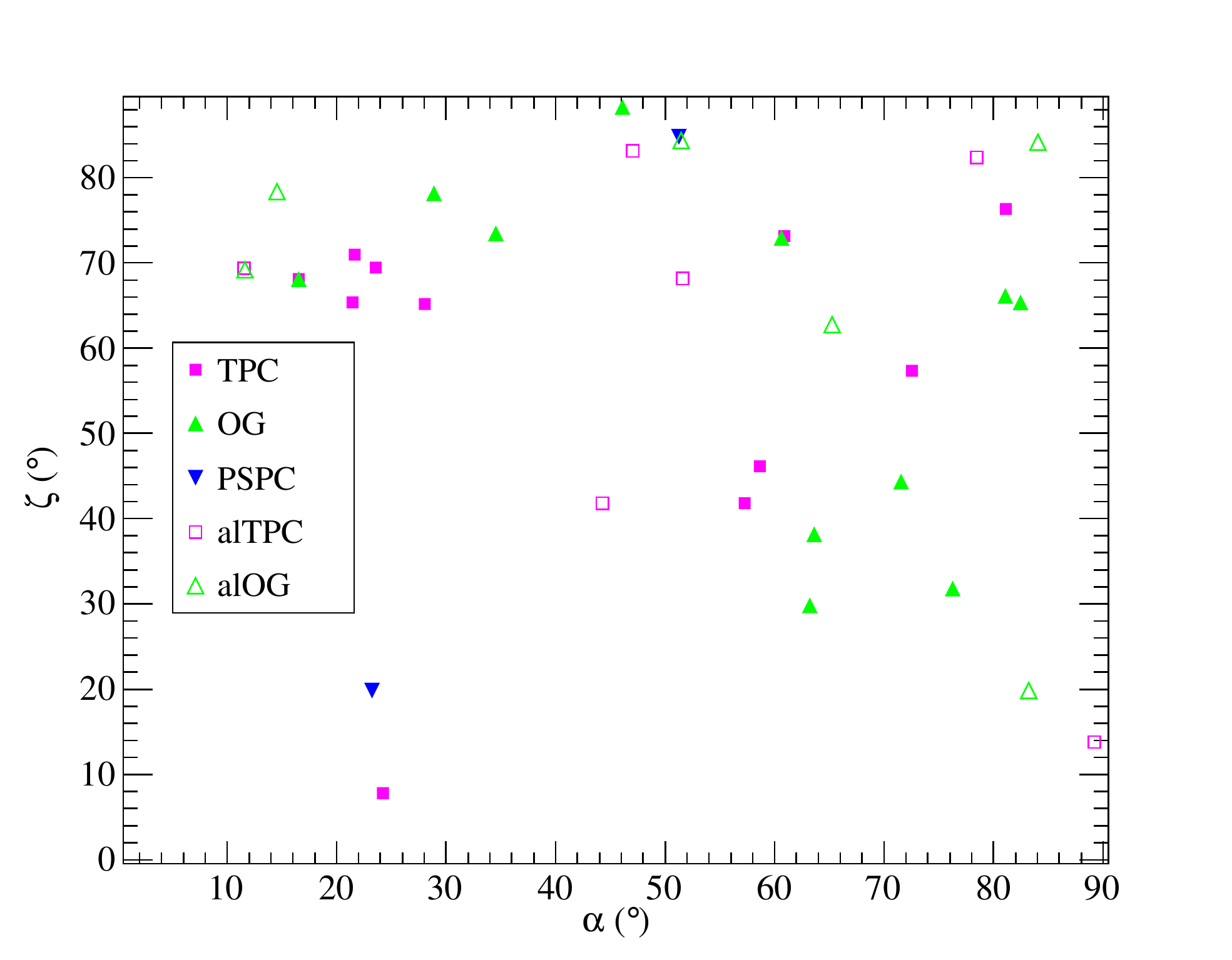}
\end{center}
\small\normalsize
\begin{quote}
\caption[Distribution of all best-fit $\alpha$-$\zeta$ pairs]{Distribution of best-fit $(\alpha,\zeta)$ pairs.  Filled pink squares represent values from TPC model fits, filled green triangles are from OG model fits, and filled blue triangles are from PSPC fits.  The open pink squares are from alTPC fits while open green triangles are from alOG fits.\label{ch7allAZ}}
\end{quote}
\end{figure}

It is of interest to look for possible trends in the best-fit viewing geometries of the MSPs studied here and, in particular, to compare them with similar studies for non-recycled pulsars and evaluate whether or not anything can be said regarding models of pulsar evolution.  The distribution of best-fit $(\alpha,\zeta)$ pairs for the MSPs analyzed in this thesis is shown in Fig.~\ref{ch7allAZ}.

\small\normalsize

\begin{deluxetable}{l c c }
\tablewidth{0pt}
\tablecaption{Luminosities and Efficiencies for MSPs fit with TPC and OG models}
\startdata
\underline{JName} & \underline{$L_{\gamma}\ (10^{33}$ erg s$^{-1})$} & \underline{$\eta_{\gamma}$}\\
\underline{TPC:}\\
J0030+0451 & 0.84$_{-0.61}^{+0.51}$ & 0.28$_{-0.20}^{+0.17}$\\
J0218+4232 & 72.9$_{-36.4}^{+33.3}$ & 0.30$_{-0.15}^{+0.14}$\\
J0437$-$4715 & 0.06$\pm$0.01 & 0.02$_{-0.004}^{+0.005}$\\
J0613$-$0200 & 1.23$\pm$1.02 & 0.09$\pm$0.08\\
J0614$-$3329 & 51.5$_{-26.0}^{+24.3}$ & 2.34$_{-1.18}^{+1.10}$\\
J0751+1807 & 0.58$_{-1.19}^{+1.21}$ & 0.10$\pm$0.20\\
J1231$-$1411 & 1.02$_{-0.26}^{+1.29}$ & 0.06$_{-0.01}^{+0.07}$\\
J1614$-$2230 & 2.94$_{-1.85}^{+3.87}$ & 0.59$_{-0.37}^{+0.77}$\\
J1713+0747 & 0.91$_{-0.87}^{+1.00}$ & 0.26$_{-0.25}^{+0.28}$\\
J2017+0603 & 5.65$_{-1.30}^{+3.14}$ & 0.41$_{-0.10}^{+0.23}$\\
J2302+4442 & 7.36$_{-2.93}^{+3.00}$ & 1.97$_{-0.78}^{+0.80}$\\
\underline{OG:}\\
J0030+0451 & 0.79$_{-0.46}^{+0.54}$ & 0.26$_{-0.16}^{+0.18}$\\
J0218+4232 & 39.3$_{-18.5}^{+18.1}$ & 0.16$\pm$0.08\\
J0437$-$4715 & 0.05$_{-0.005}^{+0.0.01}$ & 0.02$_{-0.002}^{+0.004}$\\
J0613$-$0200 & 0.97$_{-0.7}^{+0.84}$ & 0.07$\pm$0.06\\
J0614$-$3329 & 43.3$\pm$20.8 & 1.97$\pm$0.94\\
J0751+1807 & 0.43$_{-0.86}^{+1.00}$ & 0.07$_{-0.14}^{+0.17}$\\
J1231$-$1411 & 2.36$_{-0.86}^{+0.64}$ & 0.13$_{-0.05}^{+0.04}$\\
J1614$-$2230 & 1.85$_{-1.19}^{+3.61}$ & 0.37$_{-0.24}^{+0.72}$\\
J1713+0747 & 0.52$_{-0.48}^{+1.12}$ & 0.15$_{-0.14}^{+0.32}$\\
J2017+0603 & 3.45$_{-0.78}^{+10.47}$ & 0.26$_{-0.06}^{+0.78}$\\
J2302+4442 & 7.49$_{-3.00}^{+3.51}$ & 2.00$_{-0.80}^{+0.94}$\\
\enddata\label{ch7TPCOGspec}
\end{deluxetable}

\small\normalsize

\small\normalsize

\begin{deluxetable}{l c c }
\tablewidth{0pt}
\tablecaption{Luminosities and Efficiencies for MSPs fit with the PSPC Model}
\startdata
\underline{JName} & \underline{$L_{\gamma}\ (10^{33}$ erg s$^{-1})$} & \underline{$\eta_{\gamma}$}\\
J1744$-$1134 & 0.45$_{-0.26}^{+0.37}$& 0.11$_{-0.06}^{+0.09}$\\
J2124$-$3358 & 0.16$\pm$0.32 & 0.04$\pm$0.08\\
\enddata\label{ch7PSPCspec}
\end{deluxetable}

\small\normalsize

\small\normalsize

\begin{deluxetable}{l c c }
\tablewidth{0pt}
\tablecaption{Luminosities and Efficiencies for MSPs fit with alTPC and alOG models}
\startdata
\underline{JName} & \underline{$L_{\gamma}\ (10^{33}$ erg s$^{-1})$} & \underline{$\eta_{\gamma}$}\\
\underline{alTPC:}\\
J0034$-$0534 & 0.34$_{-0.27}^{+0.40}$ & 0.02$_{-0.02}^{+0.03}$\\
J1823$-$3021A & 80.9$_{-41.6}^{+76.7}$ & 0.10$_{-0.05}^{+0.09}$\\
J1902$-$5105 & 6.44$_{-4.88}^{+3.18}$ & 0.09$_{-0.07}^{+0.05}$\\
J1939+2134 & 213.9$_{-251.5}^{+247.4}$ & 0.19$\pm$0.23\\
J1959+2048 & 12.4$_{-11.5}^{+17.9}$ & 0.17$_{-0.15}^{+0.24}$\\
J2214+3000 & 12.3$_{-6.5}^{+4.3}$ & 0.68$_{-0.36}^{+0.24}$\\
\underline{alOG:}\\
J0034$-$0534 & 0.20$_{-0.17}^{+0.60}$ & 0.01$_{-0.01}^{+0.04}$\\
J1823$-$3021A & 94.3$_{-59.9}^{+28.7}$ & 0.11$_{-0.07}^{+0.03}$\\
J1902$-$5105 & 1.34$_{-0.65}^{+3.87}$ & 0.02$_{-0.01}^{+0.06}$\\
J1939+2134 & 233.4$_{-303.3}^{+238.8}$ & 0.21$_{-0.28}^{+0.22}$\\
J1959+2048 & 13.7$_{-13.6}^{+12.2}$ & 0.18$_{-0.18}^{+0.16}$\\
J2214+3000 & 7.25$_{-6.0}^{+2.1}$ & 0.40$_{-0.33}^{+0.12}$\\
\enddata\label{ch7alTPCOGspec}
\end{deluxetable}

\small\normalsize

\small\normalsize

\begin{figure}[h]
\begin{center}
\includegraphics[width=0.75\textwidth]{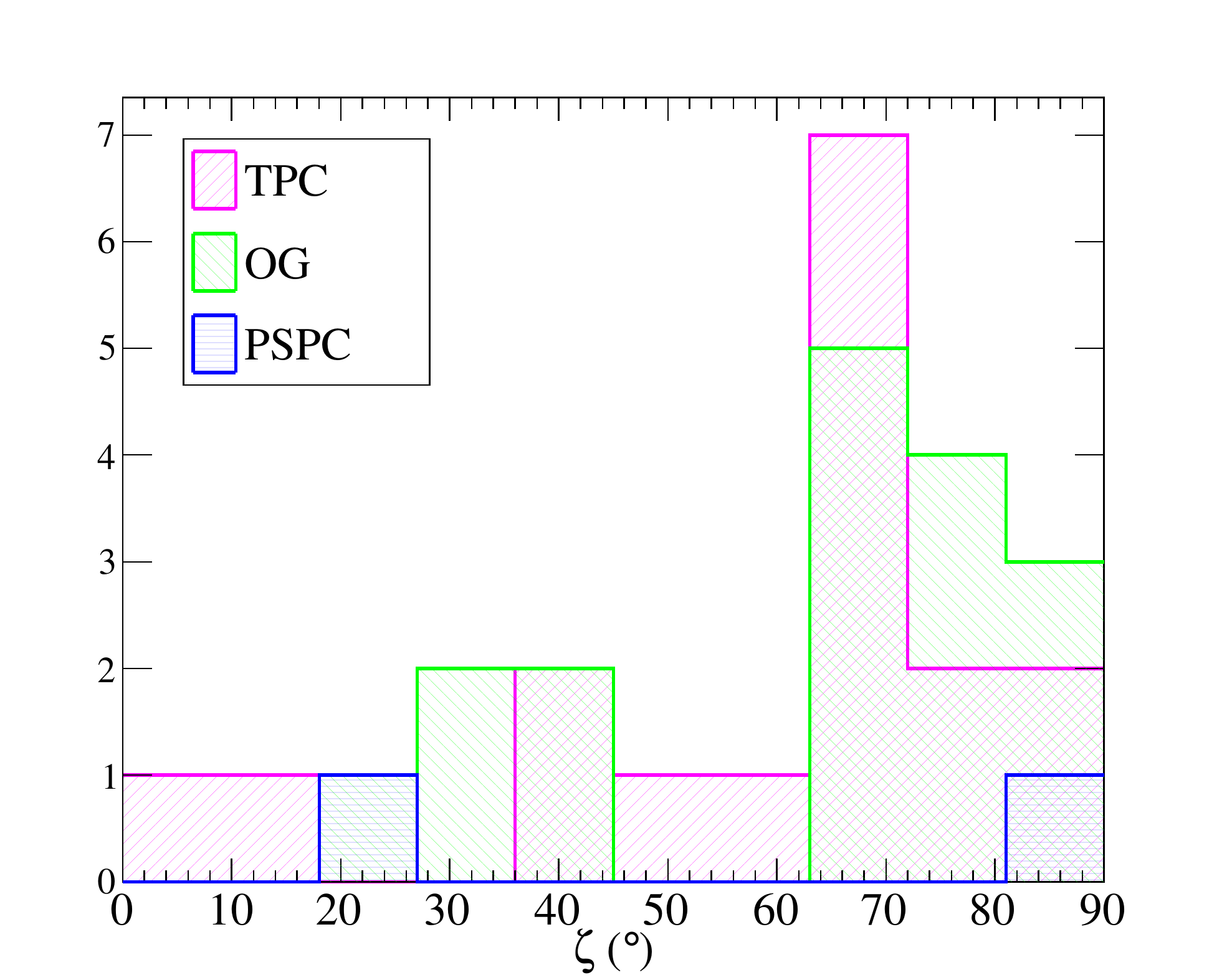}
\end{center}
\small\normalsize
\begin{quote}
\caption[Distribution of all best-fit $\zeta$ values.]{Distribution of best-fit $\zeta$ values.  The pink histogram corresponds to values from TPC fits, the green histogram from OG fits, and the blue histogram from PC fits.\label{ch71Dzs}}
\end{quote}
\end{figure}

The best-fit $\zeta$ values appear to favor larger viewing angles near 90\DEG{} as shown more clearly in Fig.~\ref{ch71Dzs}.  In Fig.~\ref{ch71Dzs} and elsewhere, the labels TPC and OG include both standard and altitude-limited models unless it is specifically stated otherwise.

\small\normalsize

If the pulsar spin axes are assumed to be distributed randomly, with respect to the line of sight to Earth, then the viewing angles should follow a $\sin(\zeta)$ distribution with a higher probability of pulsars being observed at larger $\zeta$.  Additionally, the brightest emission in outer-magnetospheric emission models is typically observed at larger values of $\zeta$ which introduces another selection effect.  This will be revisited in the next chapter after a ``best'' solution for each MSP is chosen.  Note that these findings are in agreement with a similar analysis for non-recycled gamma-ray pulsars (Pierbattista, 2011 and Pierbattista et al., 2011).

\begin{figure}[h]
\begin{center}
\includegraphics[width=0.75\textwidth]{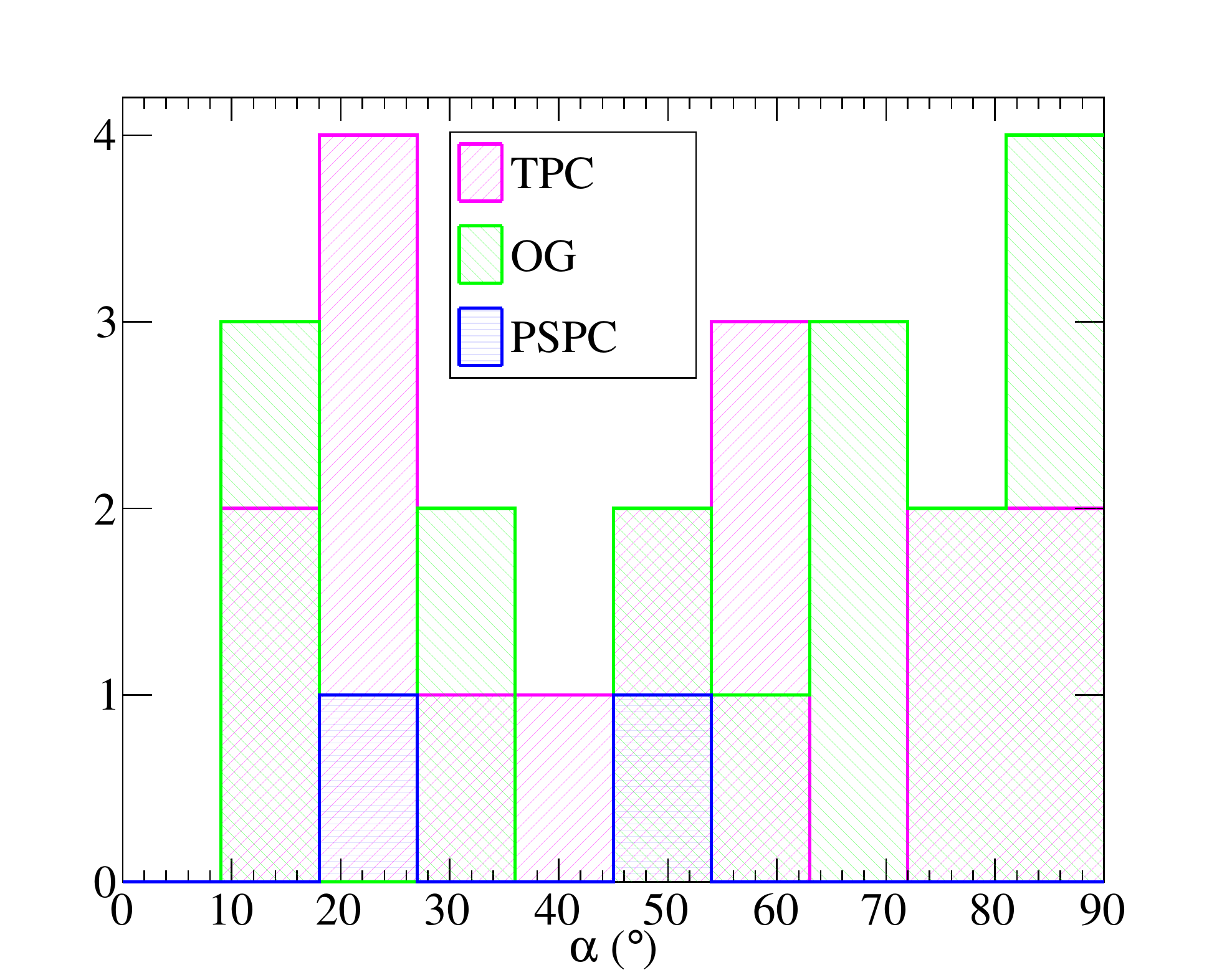}
\end{center}
\small\normalsize
\begin{quote}
\caption[Distribution of all best-fit $\alpha$ values]{Distribution of best-fit $\alpha$ values.  The pink histogram corresponds to values from TPC fits, the green histogram from OG fits, and the blue histogram from PC fits.\label{ch71Das}}
\end{quote}
\end{figure}

The best-fit MSP $\alpha$ values do not follow a uniform, angular distribution (see Fig.~\ref{ch71Das}), instead appearing to favor all angles roughly equally.  This is in contrast to findings for non-recycled pulsars (Pierbattista, 2011 and Pierbattista et al., 2011) which suggest a predominance of large $\alpha$ values.

\small\normalsize

While it might be feasible that the magnetic inclination angles of newly born pulsars should follow a uniform, angular distribution it is not clear that this should remain the case when looking at pulsars of different ages.  In particular, there are many theories as to the evolution of $\alpha$ as a pulsar spins down (e.g., Jones, 1976; Ruderman, 1991).

Some studies suggest that the spin and magnetic axes should align as a pulsar spins down and analysis of non-recycled radio pulsar profiles supports this theory with a suggested alignment timescale of $\sim10^{6}$ yr \citep{Young10}.  Gradual alignment has also been used by \citet{WR11} to explain an observed lack of young, low-$\dot{\rm E}$, radio-loud pulsars which are also observed to be gamma-ray loud.

\citet{Ruderman91} has argued that, based on consideration of stellar crust stresses from superfluid vortices, low magnetic field neutron stars being spun up due to accretion should experience a gradual alignment of spin and dipole axes.  After accretion has stopped, however, the dipole axes of such low magnetic field pulsars spinning down should move away from the spin axis into an orthogonal configuration.

If the model of \citet{Ruderman91} is correct, the distribution of $\alpha$ values in Fig.~\ref{ch71Das} could represent the age of the MSP after accretion stopped.  However, the difficulty with such an interpretation is that it is unclear if all MSPs are spun up to the same extent or if accretion stops earlier in some sources leaving them with an initial $\alpha$ greater than in other sources.  Evidence for inefficiencies in the recycling process has been found via the discovery of a pulsar which is only mildly recycled \citep{Keith09}.

\begin{figure}[h]
\begin{center}
\includegraphics[width=0.75\textwidth]{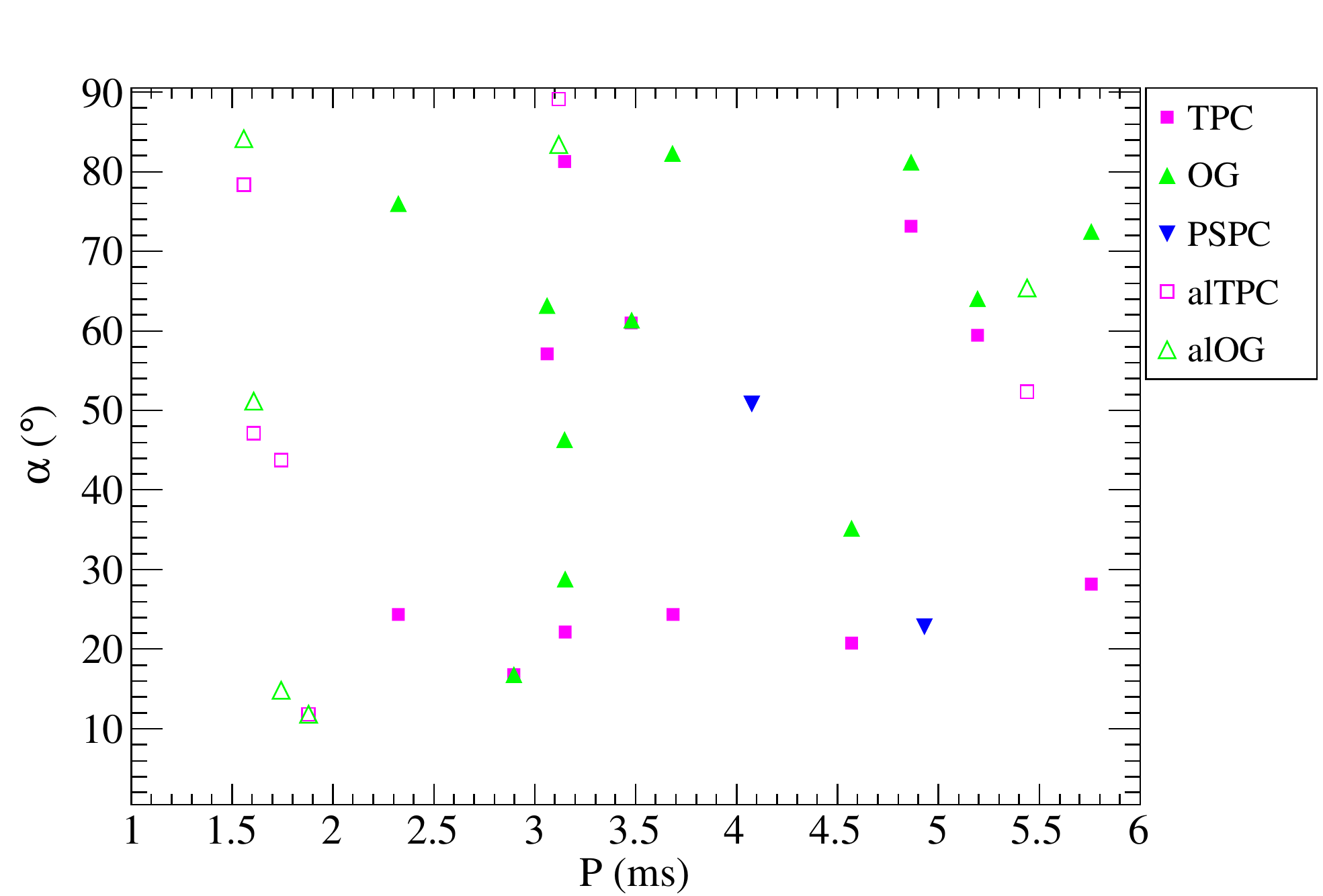}
\end{center}
\small\normalsize
\begin{quote}
\caption[Distribution of all best-fit $\alpha$ values versus pulsar period]{Distribution of best-fit $\alpha$ values versus pulsar period.  Filled pink squares represent values from TPC model fits, filled green triangles are from OG model fits, and filled blue triangles are from PSPC fits.  The open pink squares are from alTPC fits while open green triangles are from alOG fits.\label{ch7avp}}
\end{quote}
\end{figure}

Fig.~\ref{ch7avp} shows the distribution of best-fit $\alpha$ values versus the MSP spin periods.  If the distribution of best-fit magnetic inclination angles in Fig.~\ref{ch71Das} is truly just a reflection of the recycling process one might expect to see a preference for larger values of $\alpha$ for greater spin periods which is clearly not seen.  This casts some doubt on the proposed interpretation of the observed $\alpha$ distribution.

\small\normalsize

\begin{figure}
\begin{center}
\includegraphics[width=0.75\textwidth]{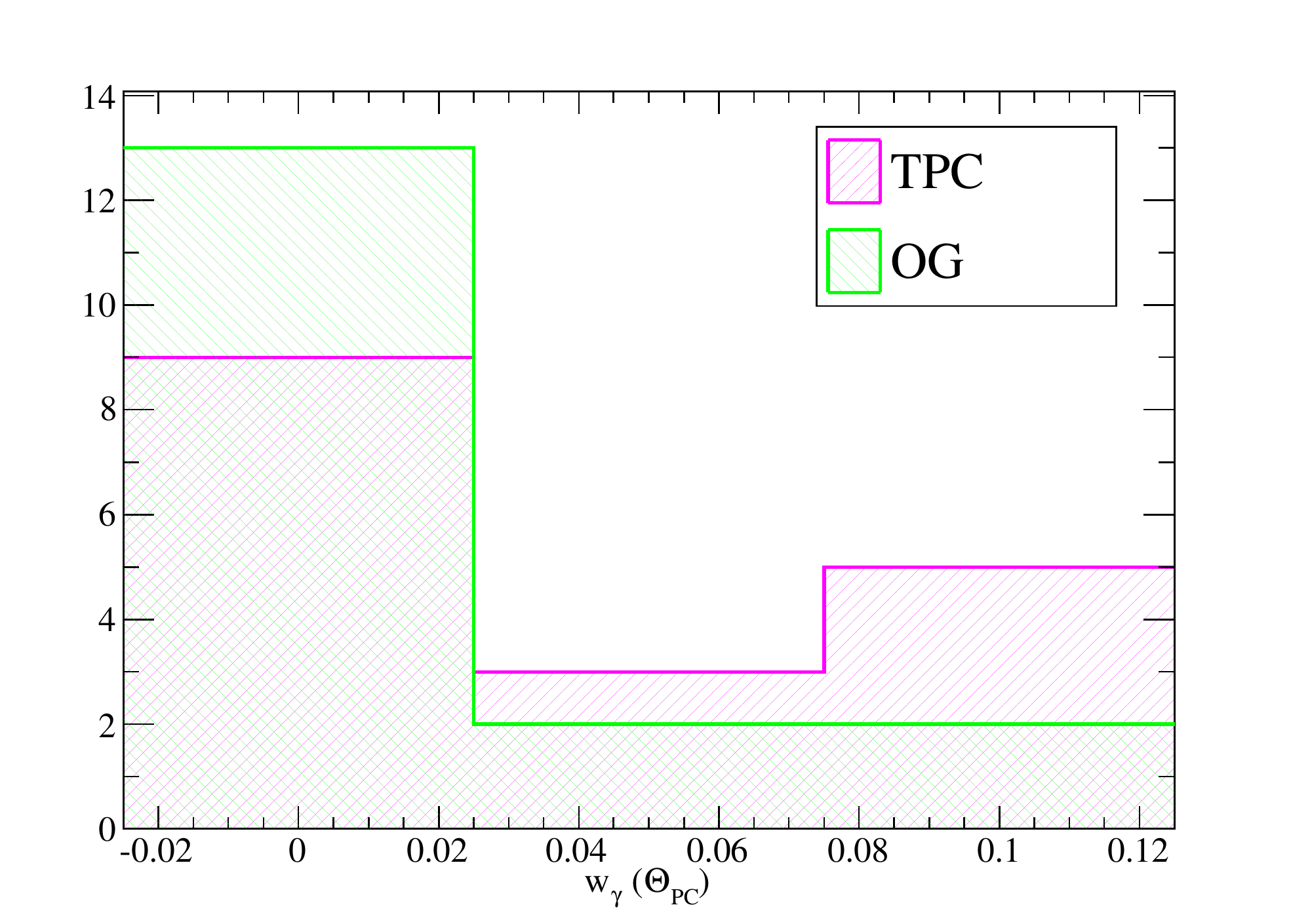}
\end{center}
\small\normalsize
\begin{quote}
\caption[Distribution of all best-fit gamma-ray gap widths]{Distribution of best-fit, gamma-ray gap width values.  The pink histogram corresponds to values from TPC fits and the green histogram from OG fits.\label{ch7wgs}}
\end{quote}
\end{figure}

As noted by \citet{Venter09}, the gamma-ray light curves of LAT detected MSPs are suggestive, in all but two cases, of emission occurring in narrow accelerating gaps.  The fits presented here use simulations with a resolution of 0.05 in gap width and as can be seen in Fig.~\ref{ch7wgs} best-fit gap widths of 0.0 are found for two thirds of the fits.  This suggests that the best-fit gap width for the majority of MSPs is $\lesssim$ 0.05 which further suggests that the accelerating electric fields are highly screened.

\section{Concerning the Absolute Goodness-of-fit}\label{ch7absgood}
It is worth reminding the reader that the likelihood value is not an absolute goodness-of-fit measure.  Statistical tools, such as the LRT described in Chapter 3, can be used to reject one model in favor of another but they do not guarantee that the preferred model accurately describes the data.  Some of the fits in Appendix A match the data reasonably well but others do not; however, all of the fits will be used to infer general MSP properties in Chapter 8.  Thus, those conclusions should be viewed with some caution.

\small\normalsize

Most of the results presented in this thesis make the underlying assumption that the basics of the assumed emission models are correct.  However, the use of geometric emission models and uniform emissivity precludes the ability to match certain light curve features and the vacuum retarded dipole field geometry is known to be unrealistic.  Additionally, details of the hollow-cone beam radio model may not be correct and some MSPs likely require a core component and/or additional cones.  Thus, it is important to know which light curve features are more (or less) dependent on the model details.

Geometric models are able to match the positions of caustic emission peaks rather well as these depend largely on the boundary of the open volume.  While these boundaries will not change in a full radiation model they do depend strongly on the choice of magnetic field geometry.  In particular, a vacuum state can not be maintained everywhere in the pulsar magnetosphere and the presence of an electron-positron pair plasma will affect the shape of the PC rim.

\citet{BS10b} showed that the open volume region is larger and the PC rim more circular in the force-free model compared to the vacuum assumption though they did not quantify the difference or address how different altitudes were effected.  For a larger open volume, the same peak spacing can be achieved (assuming $\alpha,\ \zeta\ <$ 90\DEG{}) by either decreasing $\zeta$ or increasing $\alpha$ though large changes can result in different peak multiplicities and predicted off-peak emission levels.  Lacking a detailed study of the force-free open volume it is not possible to comment on how much the geometries reported here might change.

\begin{figure}[h]
\begin{center}
\includegraphics[width=.6\textwidth]{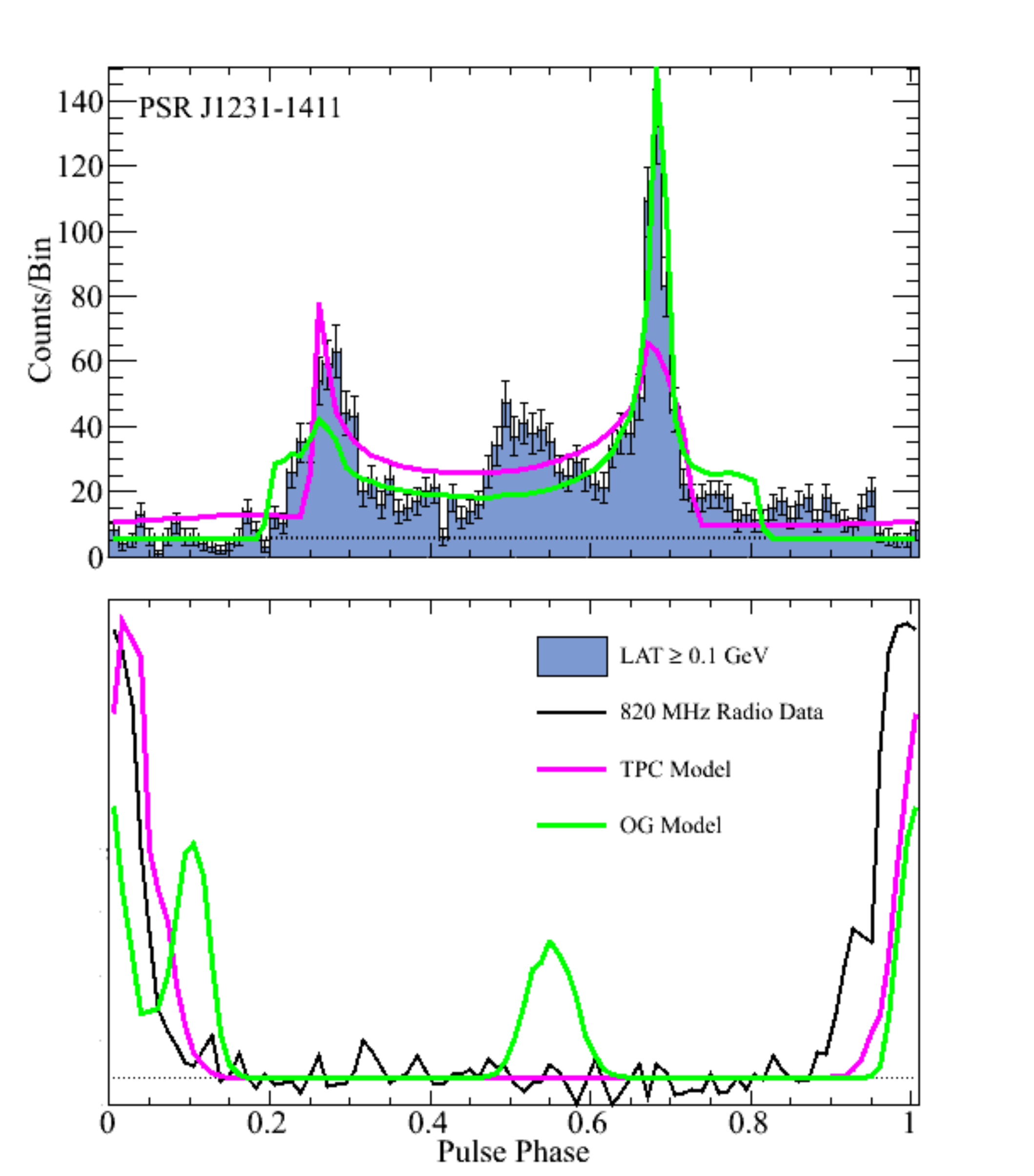}
\end{center}
\small\normalsize
\begin{quote}
\caption[Light curve fits to PSR J1231$-$1411 using TPC and OG models]{Gamma-ray (\emph{top}) and radio (\emph{bottom}) data and best-fit model light curves for PSR J1231$-$1411 using standard TPC (pink) and OG (green) gamma-ray models with a hollow-cone radio beam.\label{ch7J12311ROT}}
\end{quote}
\end{figure}

The previous arguments apply to the main gamma-ray peak(s) and do not apply to features which move significantly in phase with increasing photon energy such as has been observed for the Vela pulsar.  Geometric models can not generally reproduce these features but fitting techniques, such as that described in Chapter 6, will still try to match such features and this can affect the best-fit results.  One example is PSR J1231$-$1411 (best-fit light curves shown in Fig.~\ref{ch7J12311ROT}) for which the two main gamma-ray peaks are fit relatively well but the minor gamma-ray peak near phase 0.5 is not.

\small\normalsize

\begin{figure}
\begin{center}
\includegraphics[width=1.0\textwidth]{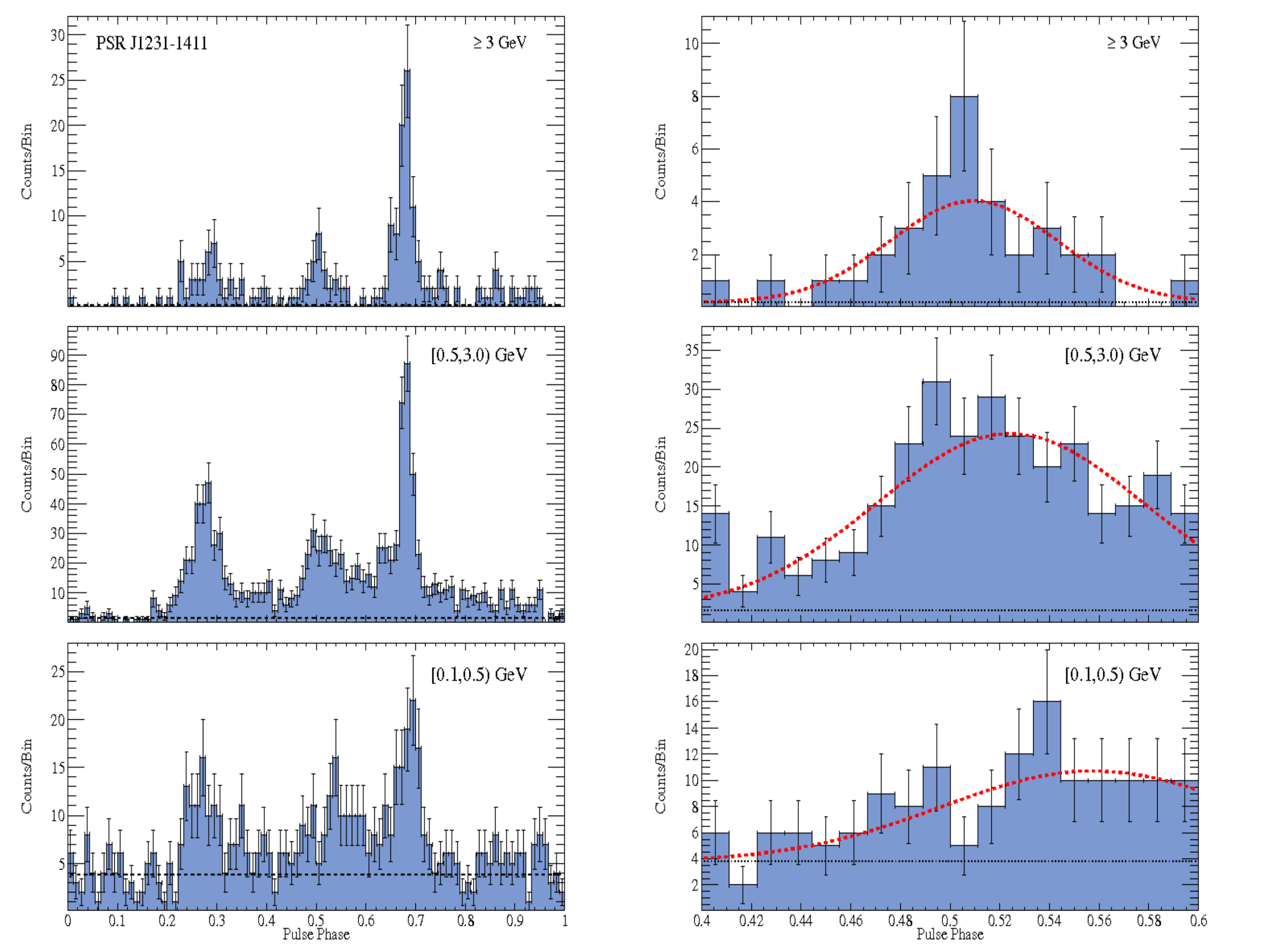}
\end{center}
\small\normalsize
\begin{quote}
\caption[Gamma-ray light curves of PSR J1231$-$1411 in different energy bands]{Gamma-ray light curves of PSR J1231$-$1411 in different energy bands (as labeled).  The left plot shows the full phase while the right plot shows a zoom in on the middle peak which has been fit with a Gaussian plus a constant background estimated using \emph{gtsrcprob}.\label{ch7J1231Ene}}
\end{quote}
\end{figure}

The gamma-ray light curves of this MSP in different energy bands are shown in the left side of Fig.~\ref{ch7J1231Ene}.  Note that, similar to some non-recycled pulsars, the second peak grows in strength relative to the first with increasing energy but the position of both peaks remains stable.  However, the right side of Fig.~\ref{ch7J1231Ene} demonstrates that the middle peak appears to move to earlier phases with increasing energy.

Note that a fit to this minor peak using a Gaussian shape on top of the \emph{gtsrcprob} derived background estimate in each energy band did yield different peak positions.  However, these positions are consistent with one value (reduced $\chi^{2}$ of 1.3) and thus the energy dependence can not be conclusively claimed.  Assuming that the energy dependence is real, it is probable that the emission which constitutes this minor peak originates in a different part of the magnetosphere than the main peaks as \citet{Du11} claim is true for the Vela pulsar.

\small\normalsize

Given the aforementioned model deficiencies, the only way to assess the viability of the best-fit models is through comparison with other observations, namely radio polarization measurements.  The simulation code described in Chapter 5 can be used to predict the polarization angle swings for a given geometry and beam model, these can be compared with observations without the need for RVM fits.  The power of such comparisons is well demonstrated in Section~\ref{ch7J1744} where the alTPC/OG fits to PSR J1744$-$1134 have comparable or better likelihood values than the PSPC fit but the observed polarization properties do not support the hypothesis of caustic radio emission.  Polarization measurements would be similarly helpful for the case of PSR J2214+3000 (see Section~\ref{ch7J2214}).

MSP spectral properties are much less dependent on the assumed emission model. The assumption of CR gamma rays is common to many models and agrees with observation which suggests that those results should be robust.  The reported $L_{\gamma}$ and $\eta_{\gamma}$ values do make use of the predicted f$_{\Omega}$ values but these are typically of order 1 and thus the arguments made in Chapter 8 regarding the $L_{\gamma}$ versus $\dot{E}$ should be little affected if the underlying models used for the fitting are wrong.

\section{Conclusions}\label{ch7conc}
The radio and gamma-ray light curves of nineteen MSPs have been fit with geometric emission models using a maximum likelihood technique which allows for statistical constraints on the emission geometries in these systems to be made.  In some cases, the extent of the emission regions has been constrained to within 10\% of the light cylinder radius.

The best-fit ($\alpha$,$\zeta$) values tend to agree with constraints from radio and X-ray observations with some particular exceptions.  It is clear that, for some MSPs, the single, hollow-cone radio beam is insufficient and use of either multiple cones and/or a core beam is warranted.  Additionally, as will be discussed in more detail in the next chapter, there is evidence that the emission altitude of \citet{KG03} is too low for MSPs.

While radio polarization measurements for many of the MSPs analyzed in this chapter do in fact exist, they do not often lead to constraining predictions on $\alpha$ and $\zeta$ using the standard RVM fitting procedures.  However, the simulations described in Chapter 5 can also produce the expected polarization properties for a given viewing geometry independent of the RVM.  Therefore, future studies could further test the maximum likelihood viewing geometries by comparing the observed and predicted polarization profiles.

The best-fit $\zeta$ values roughly follow a uniform, angular distribution, as expected for a random distribution of pulsar spin axes.  This distribution will be analyzed in more detail in Chapter 8.

The best-fit $\alpha$ values seem to favor all angles with roughly equal probability which may be in line with the predicted, recycled pulsar evolution.  However, uncertainties in the efficiency of the recycling process complicate the process further and it may require comparison with population synthesis models to fully address this issue.
\renewcommand{\thechapter}{8}
\chapter{\bf Discussion}\label{ch8}
A major goal of the study presented here is to statistically differentiate between emission models for LAT detected MSPs, whenever possible, using maximum likelihood.  As discussed in Chapter 6, the likelihood value depends strongly on the gamma-ray background estimate, with $-\log(\mathcal L)$ changing by 5-6 for changes in the background level of 1-2 counts per bin.  The potential systematic bias introduced by this fact must be taken into account when comparing different fits for the same MSP.

\begin{figure}
\begin{center}
\includegraphics[width=1.\textwidth]{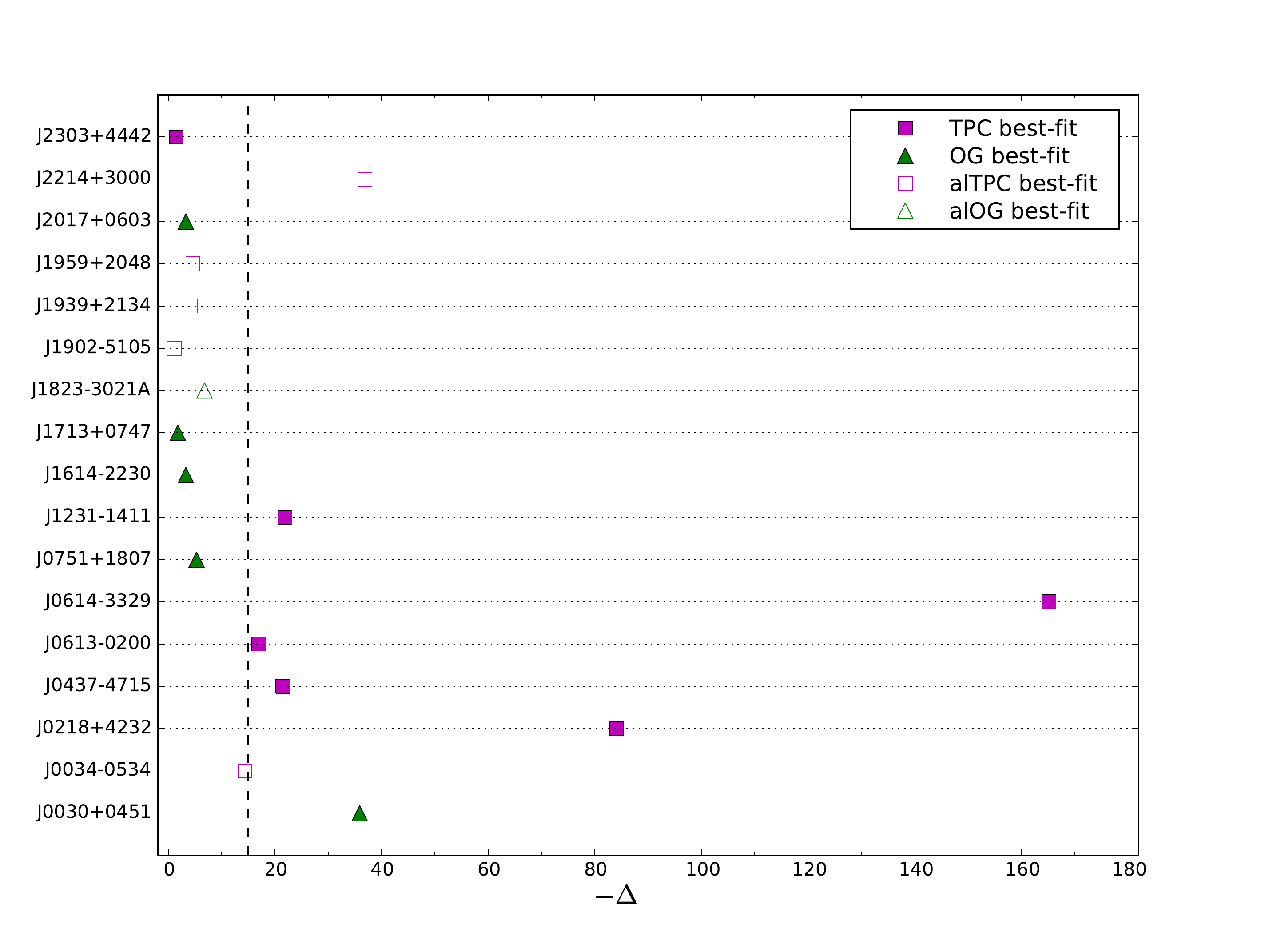}
\end{center}
\small\normalsize
\begin{quote}
\caption[Differences in $-\log(\mathcal L)$ for TPC and OG fits]{Differences in $-\log(\mathcal L)$ for the MSPs fit with TPC and OG type models.  The vertical dashed line denotes the threshold for preferring one model over another. \label{ch8LL}}
\end{quote}
\end{figure}

As such, when comparing two fits a conservative requirement of $-\Delta\log(\mathcal L)$ $\geq$ 15 is required for one model to be preferred over another.  This results in a total of nine MSPs for which a particular emission model can be selected.  Fig.~\ref{ch8LL} shows the $-\Delta\log(\mathcal L)$ values for those MSPs fit with TPC and OG type models and indicates which model is preferred.  It is interesting to note that even with a less conservative requirement of $-\Delta\log(\mathcal L)\ \geq$ 10 an emission model can be preferred for only one additional MSP (J0034$-$0534).

\small\normalsize

Note that for PSRs J1744$-$1134 and J2124$-$3358 while no fits with standard TPC and OG models have been presented these were investigated and found to have values of $-\log(\mathcal L)$ which were $\gtrsim$ 100 higher than the PSPC values.  For J1744$-$1134 fits have been presented with the alTPC and alOG models (see Section~\ref{ch7J1744}) which have likelihood values comparable to that of the PSPC fit but the viability of these models is questionable given the polarization characterstics of the source.

For the purposes of this chapter it will be necessary to select only one fit for each MSP, regardless of whether or not $-\Delta\log(\mathcal L)$ is above the threshold of 15.  As such, strong conclusions can not always be drawn from the data and only general trends will be discussed.  There are, however, some cases for which the results are independent of which model is chosen and thus the results are more robust.

Overall, the TPC models are found to be more strongly preferred than the OG models.  This could be due to issues with the background estimates as OG models will be systematically disfavored if this level is set too low.  However, note that TPC models are overwhelmingly preferred for $-\Delta\log(\mathcal L)\ \geq$ 15.

It should be noted that \citet{Venter09} found that MSP gamma-ray light curves were not well fit by OG models with infinitely thin emission layers.  While the study presented here did allow for non-zero width emission layers the current gap-width resolution resulted in this only being a viable possibility for OG gaps with a width of 0.10.  OG models with such large accelerating gap widths are not found to fit the observed MSP light curves well and, thus, the best-fit OG solutions all have zero width emission layers.  Note that the limits placed on the size of the emission layer in this thesis exclude the OG model \citet{Venter09} used for most MSPs with $r_{ovc}^{min}$ = 0.95 and $r_{ovc}^{max}$ = 1.0.

A preference for TPC models is contrary to the findings of \citet{RW10} and \citet{WR11} that the OG models are preferred for non-recycled pulsars. The former authors fit five of the \emph{EGRET} pulsars (excluding the Crab) using a modified $\chi^{2}$ statistic while the latter simulated a large population of gamma-ray pulsars and compared the observed and simulated properties from different emission models.  Both studies have specific differences, beyond modeling only non-recycled pulsars, in their implementations of pulsar emission models which could affect their results compared to those presented in Chapter 7.

The TPC model of \citet{RW10} adheres strictly to the details of the original \citet{DR03} paper limiting the emission to cylindrical distances of 0.75 R$_{\rm LC}$ and only following the emission to R$_{\max}^{\gamma}\ =\ \rm R_{LC}$.  Additionally, they do not use a fully illuminated gap, instead assuming that the emission comes from an infinitely-thin emitting layer on the upper boundary of field lines with $r_{ovc}\ \leq 1$.  In addition to the Deutsch field geometry used in this study \citet{RW10} also explored a pseudo force-free geometry which assumes the presence of charges enforcing the co-rotation of the magnetosphere but does not model the currents.

When evaluating simulated light curve properties \citet{WR11} assumed a core beam for the radio emission which leads to different radio-to-gamma lags than a cone beam though the difference is less pronounced in longer period pulsars which have much smaller PCs.  While they use the same TPC model realization of \citet{RW10} they also include versions which follow the emission out to cylindrical distances of 0.95 R$_{\rm LC}$ (as is done for this study) with emission from only the upper gap boundary and with the gap fully illuminated.

\section{Viewing Angle Distribution}\label{ch8zeta}
\begin{figure}[h]
\begin{center}
\includegraphics[width=.7\textwidth]{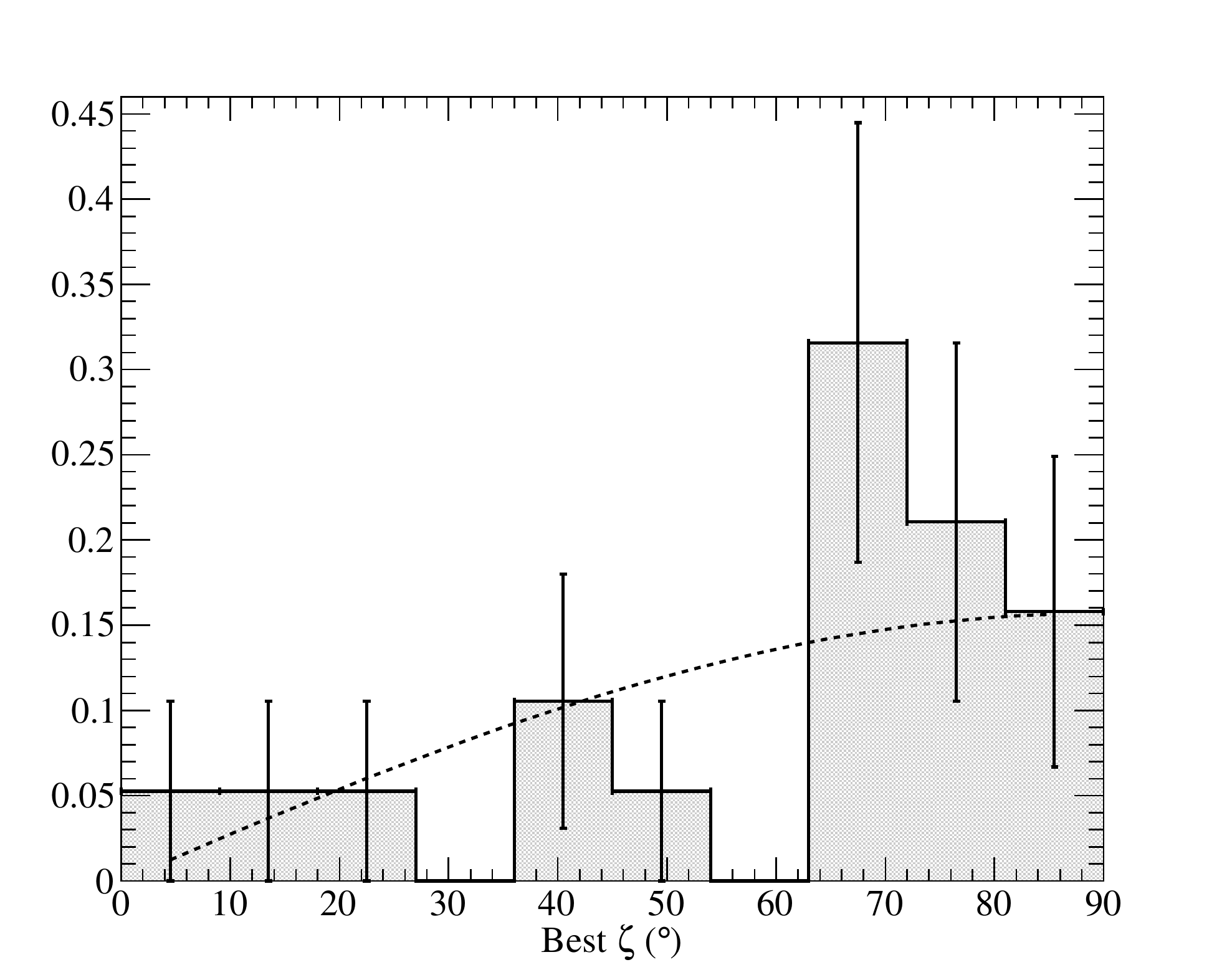}
\end{center}
\small\normalsize
\begin{quote}
\caption[Normalized distribution of preferred $\zeta$ values]{Normalized distribution of best-fit $\zeta$ values which are preferred by the likelihood.  Dashed black line shows the expectation of a uniform, angular distribution.\label{ch8bestzeta}}
\end{quote}
\end{figure}

In Chapter 7, it was noted that the best-fit viewing angles seemed to favor values near 90\DEG{} somewhat more than what is expected from just a uniform, angular distribution.  However, as shown in Fig.~\ref{ch8bestzeta}, when only one value of $\zeta$ is chosen for each MSP using the likelihood values it is clear that this is not the case.

\small\normalsize

A $\chi^{2}$ analysis comparing to the expected curve for a uniform angular distribution (assuming counting errors in each bin and accounting for normalizing the distribution) results in a reduced $\chi^{2}\ \equiv\ \chi^{2}/\nu\ =\ 4.37/8\ =\ 0.55$ which indicates good agreement.  Thus, it is most likely that the extra favoritism towards $\zeta\ \sim$ 90\DEG{} implied by Fig.~\ref{ch71Dzs} was due to low statistics and double counting of MSPs for which the TPC and OG models predict similar geometries.

\section{Need for Emission Below the NCS}\label{ch8NCSemit}
There are a number of MSPs fit with the TPC and OG models for which the predicted geometry is very similar, namely PSRs J0034$-$0534, J0613$-$0200, J0751+1807, J1614$-$2230, J1713+0747, J1823$-$3021A, J1939+2134, J1959+2048, J2017+0603, and J2302+4442.  This suggests that the predicted geometries are fairly stable to uncertainties in the models and polarization measurements for such MSPs can strongly test both emission models.  However, as MSP polarization angle swings are not typically well fit by the RVM model producing predicted polarization curves which include relativistic effects is a logical next step.

One reason the TPC and OG model finds similar geometries for these MSPs may be that the observed emission is from only one pole.  However, of the MSPs noted above the phase plots in Appendix~\ref{appA} suggest that the observed emission of J0751+1807, J1823$-$3021A, and J1939+2134 is from two poles and casts doubt on this explanation.

However, another conclusion which can be drawn is that not all of the emission below the null charge surface is necessary to explain the data.  This might suggest an OG model in which the gap is allowed to extend some distance below the NCS.

Several authors have demonstrated that the inner gap boundary of a realistic OG model which is disturbed from vacuum conditions will not be at the NCS (e.g., Takata et al. 2004, Hirotani 2005,2006).  In these OG models there are two competing currents of electrons considered, one coming from the stellar surface and one entering from the outer boundary.

\citet{Takata04} found that the inner boundary of the OG gap was at the NCS in the vacuum case and if the two currents were equal.  However, if the current of electrons from the outer boundary is greater, the inner boundary can be pushed towards the stellar surface though there are suggestions that such regions can not significantly contribute to the HE flux \citep{Wang11}.  Note that \citet{Takata04} demonstrated this when the electron current from the surface was zero (their ``case 4'').  Additionally, they assume that the flow of electrons coming in through the outer boundary exists but it is not clear from where the charges originate.

Assuming the existence of such an electron current coming in through the outer boundary of the gap, this may explain the preference for TPC models and the many cases where TPC and OG models predict similar geometries if MSPs are unable to produce a large flow of electrons across the inner bap boundary.  Depending on the relative strengths of the electron currents a given MSP may appear more or less TPC-like resulting in similar best-fit geometries and little statistical difference between the two models.

\section{MSP Gamma-ray Efficiency and Luminosity}\label{ch8effLum}
As discussed in Chapter 2, the efficiency with which a pulsar converts rotational energy into gamma rays is expected to be proportional to the gap voltage.  For non-recycled pulsars with $\dot{E}\ \gtrsim\ 10^{35}$ erg $s^{-1}$ this suggests $L_{\gamma}\ \propto\ \dot{E}^{1/2}$ and thus the efficiency $\eta_{\gamma}\ \equiv\ L_{\gamma}/\dot{E}\ \propto\ \dot{E}^{-1/2}$.  This dependence predicts $>$ 100\% efficiency near a spin-down power of a few times $10^{34}$ erg s$^{-1}$; thus, some transition in the gamma-ray emission properties must occur for pulsars with measured $\dot{E}$ below $\sim10^{34}$ erg s$^{-1}$.

\begin{figure}[h]
\begin{center}
\includegraphics[width=.7\textwidth]{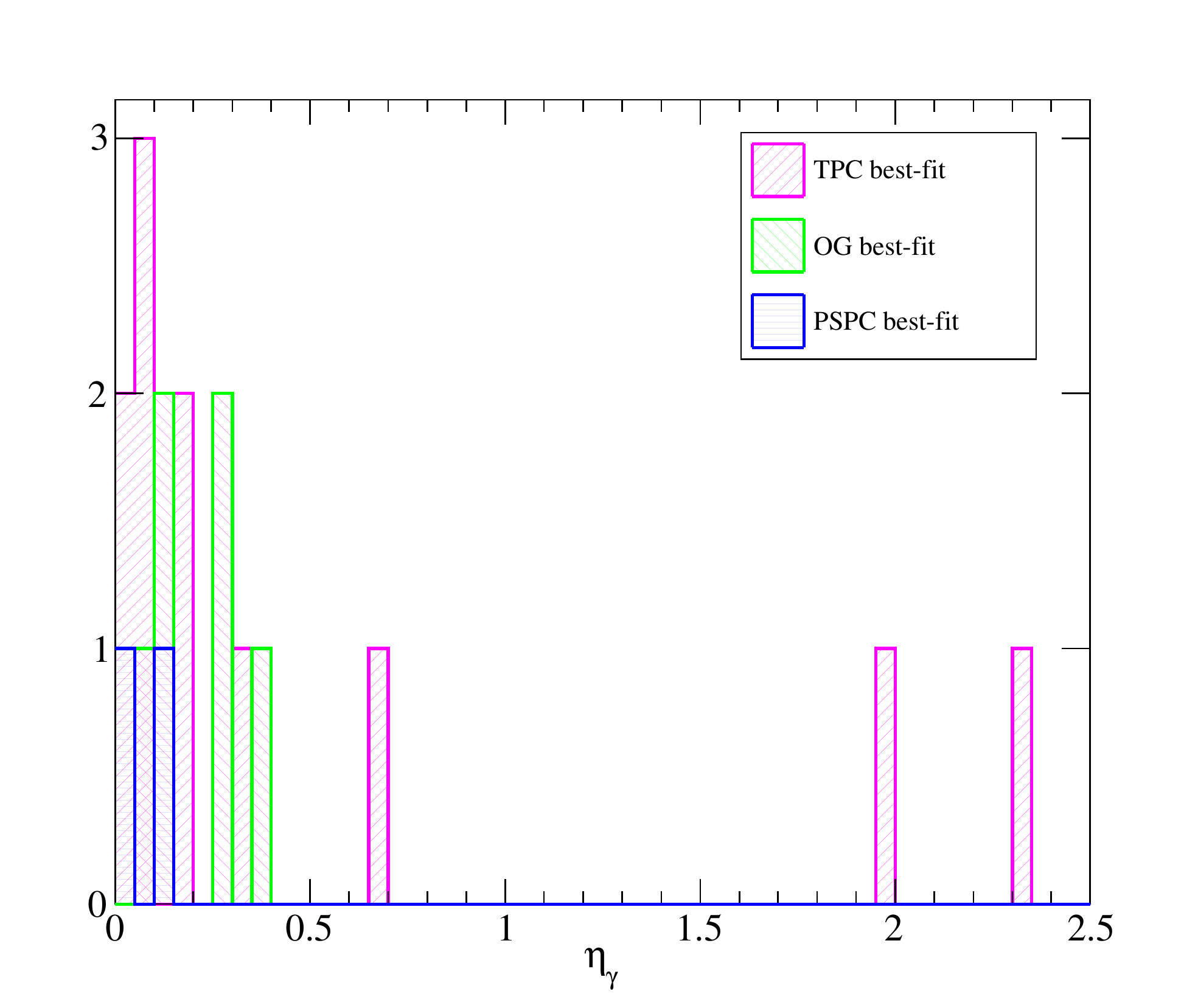}
\end{center}
\small\normalsize
\begin{quote}
\caption[Gamma-ray efficiencies using predicted beaming factors favored by the likelihood]{Calculated efficiencies for each of the 19 MSPs modeled here using the predicted f$_{\Omega}$ from the model with the lowest $-\log{\mathcal L}$ regardless of whether or not the difference is significant.  TPC models are shown in pink, OG in green, and PSPC in blue.\label{ch8besteff}}
\end{quote}
\end{figure}

Note that most of the gamma-ray MSPs presented here have measured $\dot{E}$ values near the transition point and thus might be expected to have efficiencies near 100\%.  However, as can be seen in Fig.~\ref{ch8besteff}, this is not the case.  While the efficiencies cluster near 10-20\%, as opposed to efficiencies on the order of a few percent for non-recycled pulsars \citep{AbdoPSRcat}, only three have values above 50\%.  Note that the two MSPs with efficiencies in excess of 100\% are newly discovered and only have DM-derived distances which may be overestimated.

\small\normalsize

\begin{figure}[h]
\begin{center}
\includegraphics[width=.7\textwidth]{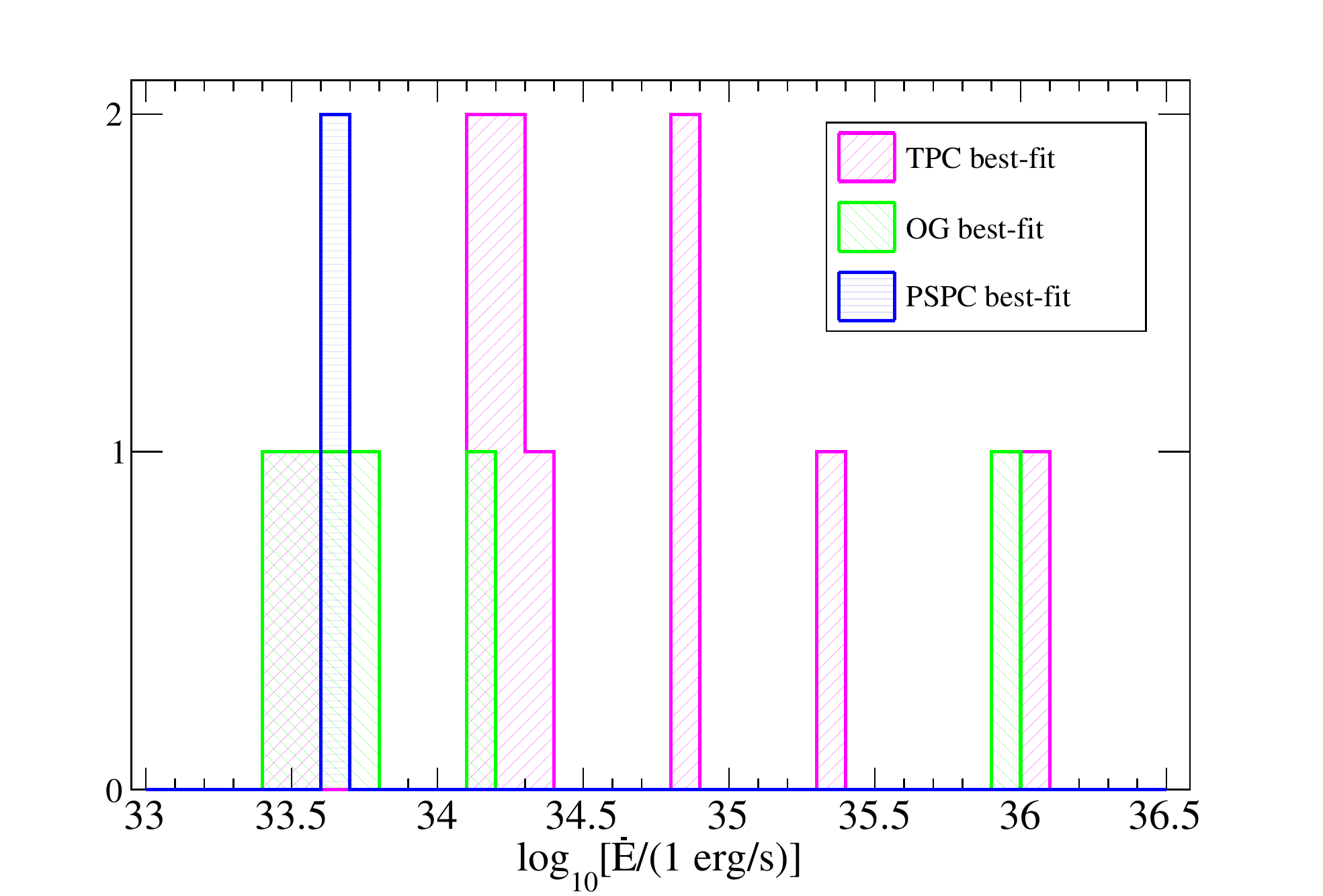}
\end{center}
\small\normalsize
\begin{quote}
\caption[Distribution of $\dot{E}$ separated by preferred model]{Measured $\dot{E}$ values for the 19 MSPs modeled here with color coding indicating the model with the lowest $-\log{\mathcal L}$ regardless of whether or not the difference is significant.  TPC models are shown in pink, OG in green, and PSPC in blue.  Note that the one outlier OG point corresponds to PSR J1823$-$3021A which is best-fit by the alOG model.\label{ch8bestedot}}
\end{quote}
\end{figure}

Note that MSPs for which the OG models are favored (though not necessarily by a significant amount) preferentially populate the lower range of $\dot{E}$ values (see Fig.~\ref{ch8bestedot}).  The fact that these MSPs have relatively low efficiencies may suggest that higher-order magnetic multipoles are important, this will be discussed in more detail later.

\small\normalsize

\begin{figure}
\begin{center}
\includegraphics[width=1.\textwidth]{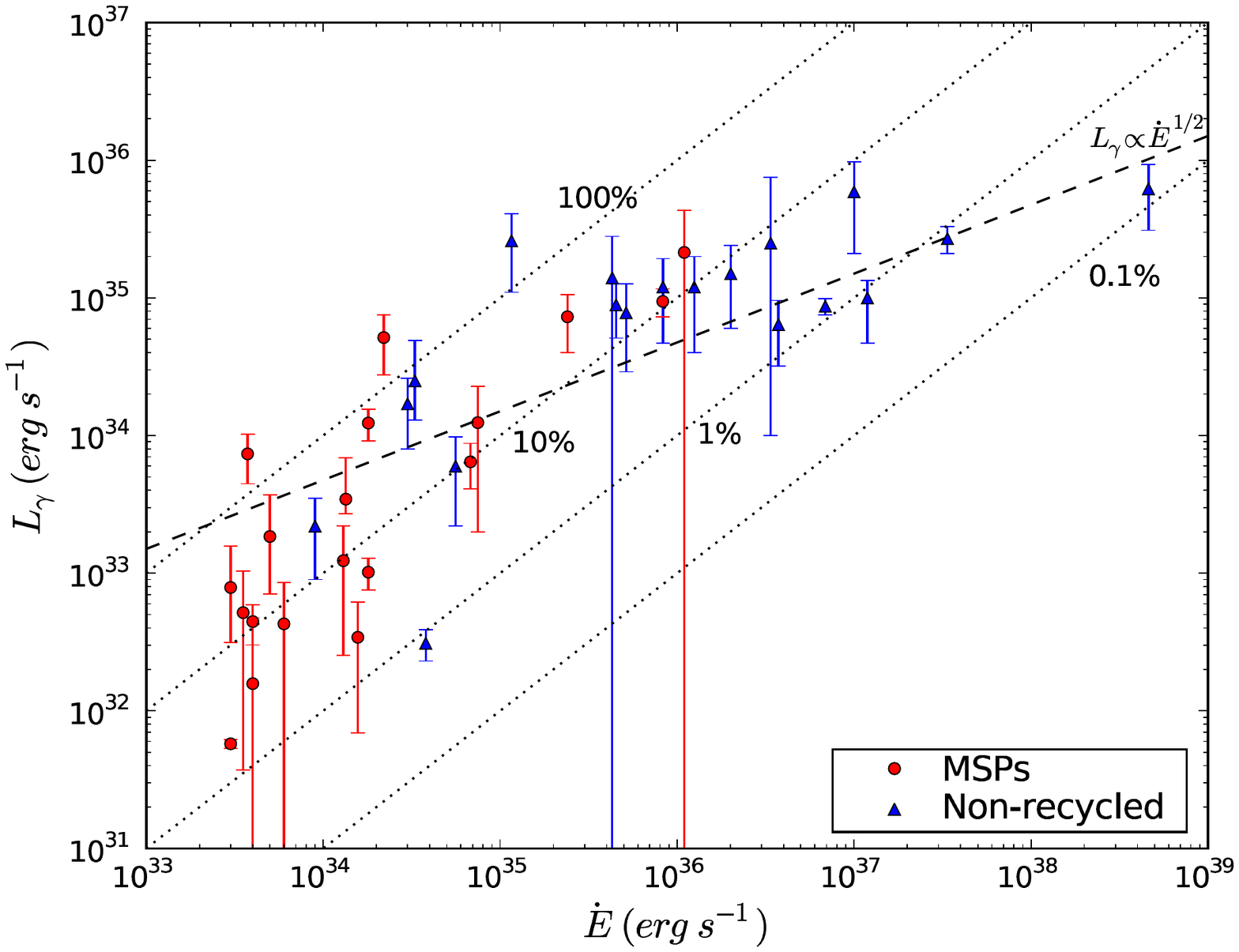}
\end{center}
\small\normalsize
\begin{quote}
\caption[$L_{\gamma}$ vs. $\dot{E}$ for known gamma-ray pulsars]{$L_{\gamma}$ versus $\dot{E}$ for known gamma-ray MSPs (red circles) and non-recycled pulsars (blue triangles).  The non-recycled pulsars are taken from the sample of \citet{AbdoPSRcat}.  Only those pulsars with specified error bars, and not just a range of $L_{\gamma}$, are included which results in a comparable sample.  Note that the non-recycled sample assumes f$_{\Omega}$ = 1 while the MSPs use the values estimated from the best-fit models (the uncertainties on f$_{\Omega}$ have not been propagated to the error bars shown here).  The dotted lines demarcate bands of constant efficiency.  A line of $L_{\gamma}\propto\dot{E}$ (dashed) is drawn for instructive purposes, note that this is not a fit. \label{ch8LgEdall}}
\end{quote}
\end{figure}

Using the first forty-six gamma-ray pulsars from which HE pulsations were firmly detected with the LAT in six months of sky survey, \citet{AbdoPSRcat} noted weak evidence for the expected change in the $L_{\gamma}$ versus $\dot{E}$ trend.  However, their sample only included eight MSPs and a handful of non-recycled pulsars with low $\dot{E}$.  With the addition of eleven gamma-ray MSPs the break in the $L_{\gamma}$ vs. $\dot{E}$ trend is much more obvious (see Fig.~\ref{ch8LgEdall}).

\small\normalsize

Fig.~\ref{ch8LgEdall} is improved over that in \citet{AbdoPSRcat} (their Fig. 6) by the inclusion of more statistics (i.e. MSPs) in the range $\lesssim\ 10^{34}$ erg s$^{-1}$.  Note that more non-recycled gamma-ray pulsars have been detected since the first LAT pulsar catalog was published but this study will remain constrained to those published results in as far as non-recycled pulsars are concerned.

The two samples now have sizeable overlap from $\dot{E}\sim10^{34}$-$10^{36}$ which allows for a qualitative assessment of the emission mechanisms at work in the two populations to be made.  As first noted by \citet{AbdoMSPpop}, it is clear that the emission mechanisms in gamma-ray MSPs and non-recycled pulsars are the same.

Those MSPs with $\dot{E}\ \gtrsim\ 10^{35}$ erg s$^{-1}$ (roughly 5) have measured values of $L_{\gamma}$ very similar to the non-recycled pulsars with similar $\dot{E}$ values and follow the general $\dot{E}^{1/2}$ trend nicely.  Additionally, $L_{\gamma}$ values for those non-recycled pulsars with $\dot{E}$ below a few times $10^{34}$ tend to cluster with values from MSPs in that same range with efficiencies around 10\%.  These facts suggest that what is observed in Fig.~\ref{ch8LgEdall} from high to low $\dot{E}$ is an efficiency evolution in the mechanism powering the gamma-ray pulsars.

\begin{figure}[h]
\begin{center}
\includegraphics[width=.75\textwidth]{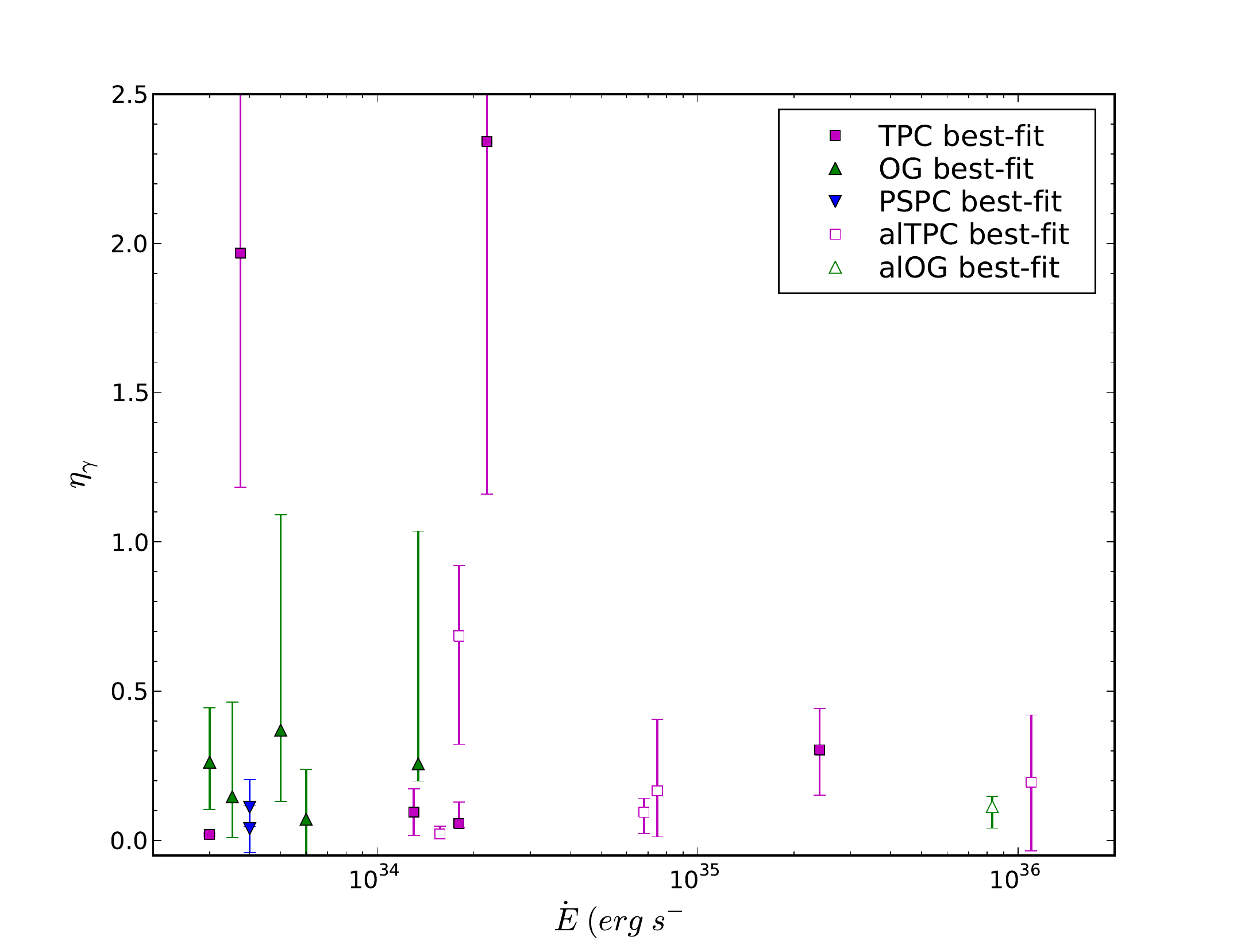}
\end{center}
\small\normalsize
\begin{quote}
\caption[Gamma-ray efficiency vs. $\dot{E}$ separated into preferred models]{Measured efficiency vs. $\dot{E}$ values for the 19 MSPs modeled here with color coding indicating the model with the lowest $-\log{\mathcal L}$ regardless of whether or not the difference is significant.  TPC models are shown in pink, OG in green, and PSPC in blue.  Luminosities were corrected with the f$_{\Omega}$ values predicted from the best-fit models, error bars reflect the confidence range in those beaming factors.\label{ch8effved}}
\end{quote}
\end{figure}

Note that while the change in the $L_{\gamma}$ versus $\dot{E}$ trend is visible with only the MSPs, without the non-recycled pulsars it would not be significant.  In fact, as shown in Fig.~\ref{ch8effved} it could be argued that gamma-ray MSPs follow $L_{\gamma}\ \approx\ 0.10\dot{E}$ as the MSPs do not extend to high enough $\dot{E}$ to fully demonstrate the decrease in efficiency which is clear in Fig.~\ref{ch8LgEdall} with the non-recycled pulsars.  Thus, to fully understand HE pulsar emission both recycled and non-recycled gamma-ray pulsars must be considered.

\small\normalsize

The change in $L_{\gamma}$ is expected to be due to gap saturation in which the accelerating voltage has reached a maximum.  While this is true for both SG and OG models, the predicted consequences are quite different.  In both models saturation means that the accelerating voltage can not be increased by simply widening the gap.  For SG models the gap can continue to grow but the efficiency is approximately constant with $L_{\gamma}\ \propto\ \dot{E}$.  For OG models this the gap occupies the whole open volume above the NCS at saturation and for $\dot{E}$ below this point gamma-ray production turns off.

For SG models, \citet{HMZ02} predicted that the $L_{\gamma}$ break should occur at 1.4$\times10^{34}\rm P^{-1/7} \mathnormal B_{12}^{-2/7}$ erg s$^{-1}$.  For the nineteen gamma-ray MSPs considered here the break occurs in the range 5.7-22$\times10^{34}$ erg s$^{-1}$.  Below this break the luminosity should be proportional to $\dot{E}$ with efficiencies as high as 10\%.

\begin{figure}[h]
\begin{center}
\includegraphics[width=.75\textwidth]{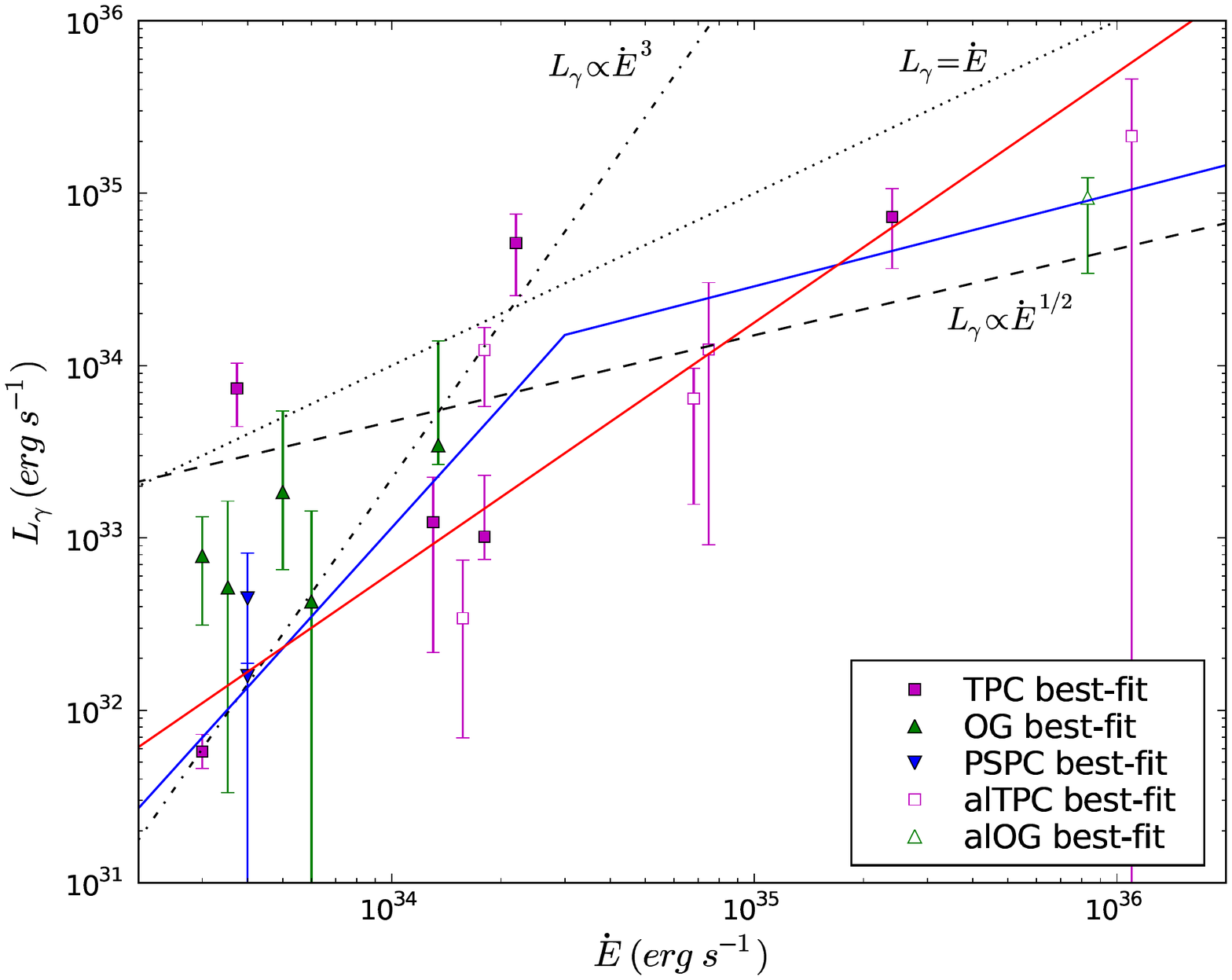}
\end{center}
\small\normalsize
\begin{quote}
\caption[$L_{\gamma}$ vs. $\dot{E}$ for MSPs only]{$L_{\gamma}$ versus $\dot{E}$ for the nineteen MSPs presented in this study.  The dotted line represents 100\% efficiency, the dashed line is the same, demonstrative line of $L_{\gamma}\propto\dot{E}^{1/2}$ as in Fig.~\ref{ch8LgEdall}.  The dash-dot line ($L_{\gamma}\propto\dot{E}^{3}$) is purely an instructive attempt to highlight the approximate location of the expected transition.  The red solid line is a power law fit to all of the MSPs while the blue solid line is the combination of two different power law fits above and below the transition.\label{ch8MSPlgedfit}}
\end{quote}
\end{figure}

The estimates of \citet{HMZ02} agree well with the data in Fig.~\ref{ch8MSPlgedfit} but it should be noted that they were derived assuming that 50\% of the primary electron beam power is converted into gamma rays and thus there is some uncertainty in these values.  In particular, if the electrons are in the CR reaction limit the primary beam efficiency should be 100\% implying MSP efficiencies $>$ 10\%.


For OG models \citet{Zhang04} predict that the $L_{\gamma}$ break should be due to saturation of the gap voltage which results in 100\% efficiency.  They predict this will occur at $1.5\times10^{34}(\rm P)^{1/3}$ erg s$^{-1}$.  For the nineteen gamma-ray MSPs considered here the break values range from 1.7-2.7$\times10^{33}$ erg s$^{-1}$.  While all of these MSPs are above these estimated break values they are all very close and one would expect the efficiencies to cluster nearer 100\% efficiency as opposed to 10\%.

Even when the confidence ranges in the beaming factors are taken into account (see Fig.~\ref{ch8MSPlgedfit}) the majority of MSPs favor efficiencies $\lesssim$ 50\%.  Additionally, there is one non-recycled pulsar in the sample of \citet{AbdoPSRcat} with a very low spin down power.  PSR J1741$-$2054 has $\dot{E}$ = 0.9$\times10^{34}$ erg s$^{-1}$ but a spin period of 0.414 s implies a critical spin down value of 1.12$\times10^{34}$ erg s$^{-1}$ using the formula of \citet{Zhang04}.  Therefore, this pulsar should not be active in gamma rays (assuming the OG model of Zhang et al., 2004) unless some non-dipolar field structure is important.  \citet{AbdoPSRcat} assumed f$_{\Omega}$ = 1 but a beaming factor of 4 is possible with outer-magnetospheric models which would increase the derived efficiency; however, this requires a very particular geometry and thus speaks against the model of \citet{Zhang04}.

\small\normalsize

The statistics of \citet{AbdoPSRcat} were too poor to attempt any model discrimination based on the transition location.  With an increased pulsar sample (particularly in the low $\dot{E}$ range) it is worthwhile to revisit this prospect.  Attempts to fit the observed trend to a smoothly broken power law using just the nineteen MSPs presented in this study proved unsuccessful.  A large part of the difficulty may come from a lack of data in the high $\dot{E}$ range which could be improved by the inclusion of the non-recycled pulsar sample of \citet{AbdoPSRcat}; however, such an analysis would not be a fair comparison as the luminosities are from less data with an earlier IRF version and all assume f$_{\Omega}$ = 1.

The $L_{\gamma}$ values for all the MSPs were fit to a single power law (solid red line in Fig.~\ref{ch8MSPlgedfit}) resulting in the relation $L_{\gamma}\ =\ 10^{-16.5}\dot{E}^{1.45}$.  Next the MSPs were divided into two groups, those above and below a trial transition value of $\dot{E}_{t}$, and fit to power laws separately.  Values of $\dot{E}_{t}$ ranging from $3\times10^{33}$ to $7\times10^{34}$ erg s$^{-1}$ were tested but none of the fits resulted in more than marginal improvement over the single powerlaw fit.

However, when matching the best-fit power laws above and below the $\dot{E}_{t}$ values most intersected far from the corresponding $\dot{E}_{t}$.  Keeping only those solutions which matched near the assumed $\dot{E}_{t}$ the best transition value was chosen to be $\dot{E}_{t}\ =\ 2\times10^{34}$ erg s$^{-1}$ for which the two solutions converged at $3\times10^{34}$ erg s$^{-1}$.  This solution is shown as the solid blue line in Fig.~\ref{ch8MSPlgedfit}.  Below the transition $L_{\gamma}\ =\ 10^{-46.16}\dot{E}^{2.33}$ and above it $L_{\gamma}\ =\ 10^{15.56}\dot{E}^{0.54}$.  It should be stressed that this expression is not from a rigorous fit and thus no strong conclusions can be drawn from it.

With more sources above the implied transition (i.e., non-recycled gamma-ray pulsars) it may be possible to fit the entire population (above and below some $\dot{E}_{t}$) simultaneously.  With just the MSPs neither emission model can be ruled out.  Note that $\dot{E}_{t}\ \approx 3\times10^{34}$ erg s$^{-1}$ does agree well with the predicted values using the equation of \citet{HMZ02}.  Additionally, this $\dot{E}_{t}$ is roughly an order of magnitude above the values predicted using the equation of \citet{Zhang04}.

\section{Nature of the Phase-aligned MSPs}\label{ch8Align}
While the previous sections have argued in favor of gamma-ray MSPs and non-recycled pulsars emitting via, essentially, the same mechanisms, there are some clear differences between the two populations.  In particular, one striking difference is the number of pulsars in each class which display near alignment of their radio and gamma-ray profiles.  Among non-recycled gamma-ray pulsars, only the Crab displays this remarkable trait.  However, there are currently six gamma-ray MSPs which show evidence for phase-alignment.

The first three to be detected (J00340$-$0534, J1939+2134, and J1959+2048) all have spin periods less than 2 ms.  Combined with the relatively low spin period of the Crab (compared to other non-recycled gamma-ray pulsars) this initially suggested that phase-alignment might be restricted to the fastest members of a given pulsar class.  However, three additional MSPs have phase-aligned profiles including one with a spin period of 5.44 ms and one with 3.12 ms.  Therefore, the secret to phase-alignment remains unclear.

Given the findings of \citet{Ravi10}, which suggest that pulsar radio emission should also be modeled as extended fan-like beams, the manifestation of phase-alignment could be simply a side effect of co-located emission regions as has been argued for some MSPs (e.g., Abdo et al. 2010d, Guillemot et al. 2011, and Venter et al. 2011).  However, if that is the case then the question still remains of what leads to this arrangement in some pulsars and not in others.

The altitude-limited models described in Chapter 5 can produce offsets similar to those seen in the MSPs for which the radio was modeled as a hollow-cone beam if the radio emission is constrained to be $\lesssim0.7\ \rm R_{LC}$.  It is not clear if such restrictions allow the models to reproduce all of the features observed in radio MSP profiles and even if that is the case there is still no clear answer as to what controls the location of the radio emission zone.    As such, a careful comparison between the properties of MSPs with and without aligned profiles is in order.

\begin{figure}[h]
\begin{center}
\includegraphics[width=.75\textwidth]{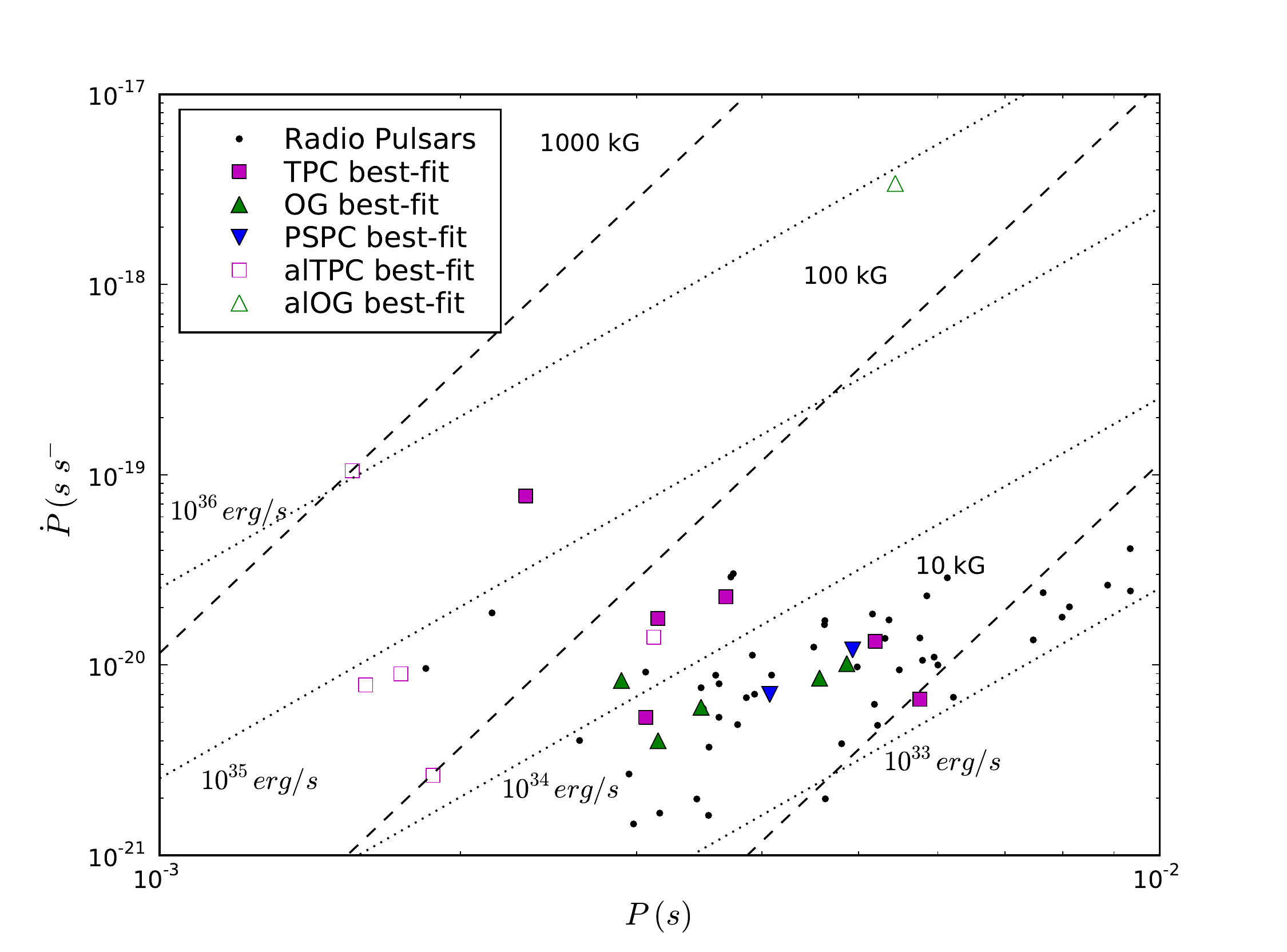}
\end{center}
\small\normalsize
\begin{quote}
\caption[$\dot{\rm P}$-P distribution for MSPs only]{Distribution of $\dot{\rm P}$ versus P for field MSPs from the ATNF catalog with gamma-ray MSPs labeled by best-fit model.  Magenta squares are for TPC (filled) and alTPC (open) models.  Green triangles are for OG (filled) and alOG (open) models.  Blue triangles are for PSPC models.  Lines of constant $B_{\rm LC}$ (dashed) and $\dot{E}$ (dotted) are shown.\label{ch8PdotPbest}}
\end{quote}
\end{figure}

\begin{figure}[h]
\begin{center}
\includegraphics[width=.75\textwidth]{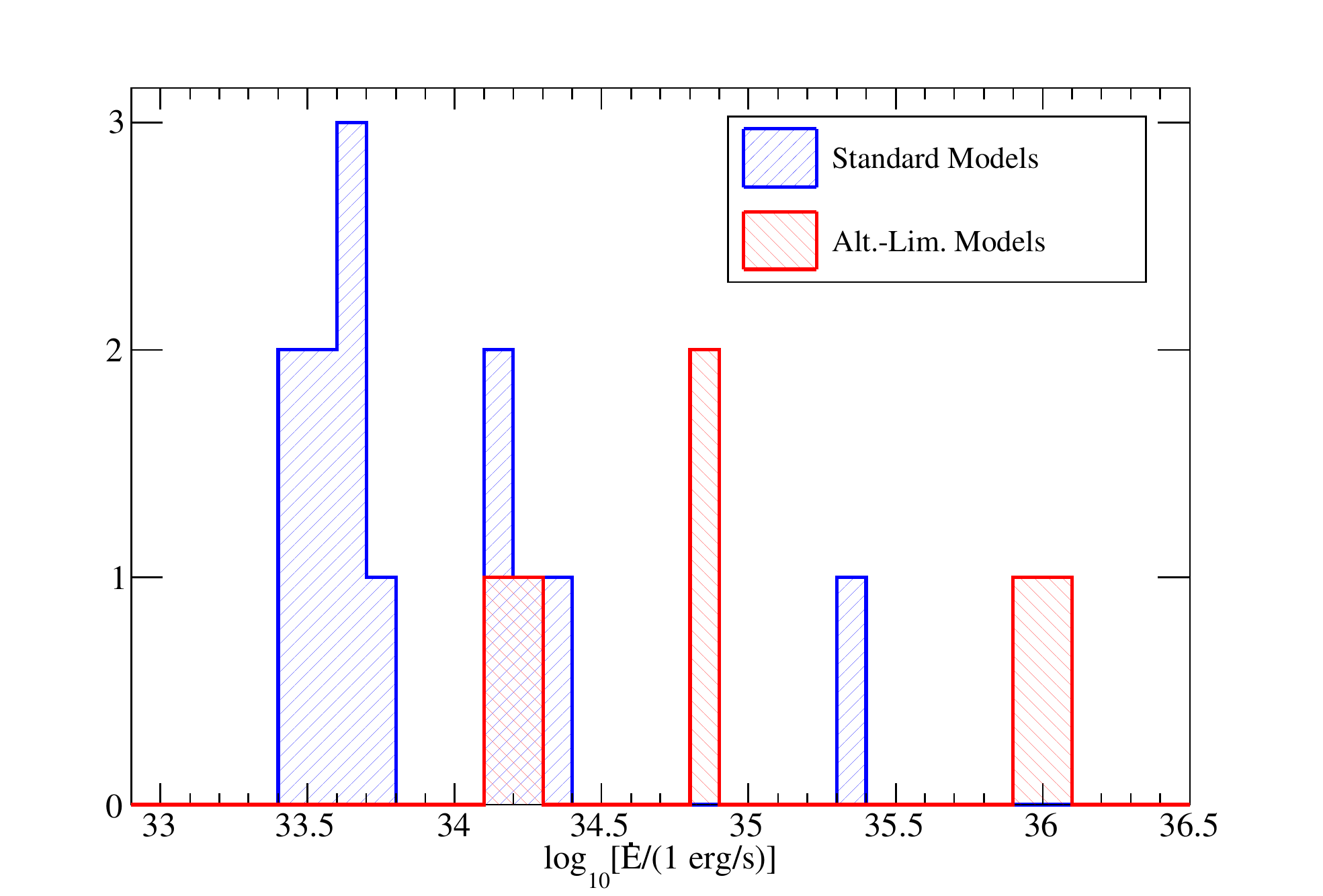}
\end{center}
\small\normalsize
\begin{quote}
\caption[$\dot{E}$ distribution separated into MSPs with and without aligned profiles]{Distribution of $\dot{E}$ values separated into MSPs best-fit by altitude-limited (red) and standard models (blue), this last encompasses PSPC as well as TPC and OG models.\label{ch8compEdot}}
\end{quote}
\end{figure}

\begin{figure}[h]
\begin{center}
\includegraphics[width=.75\textwidth]{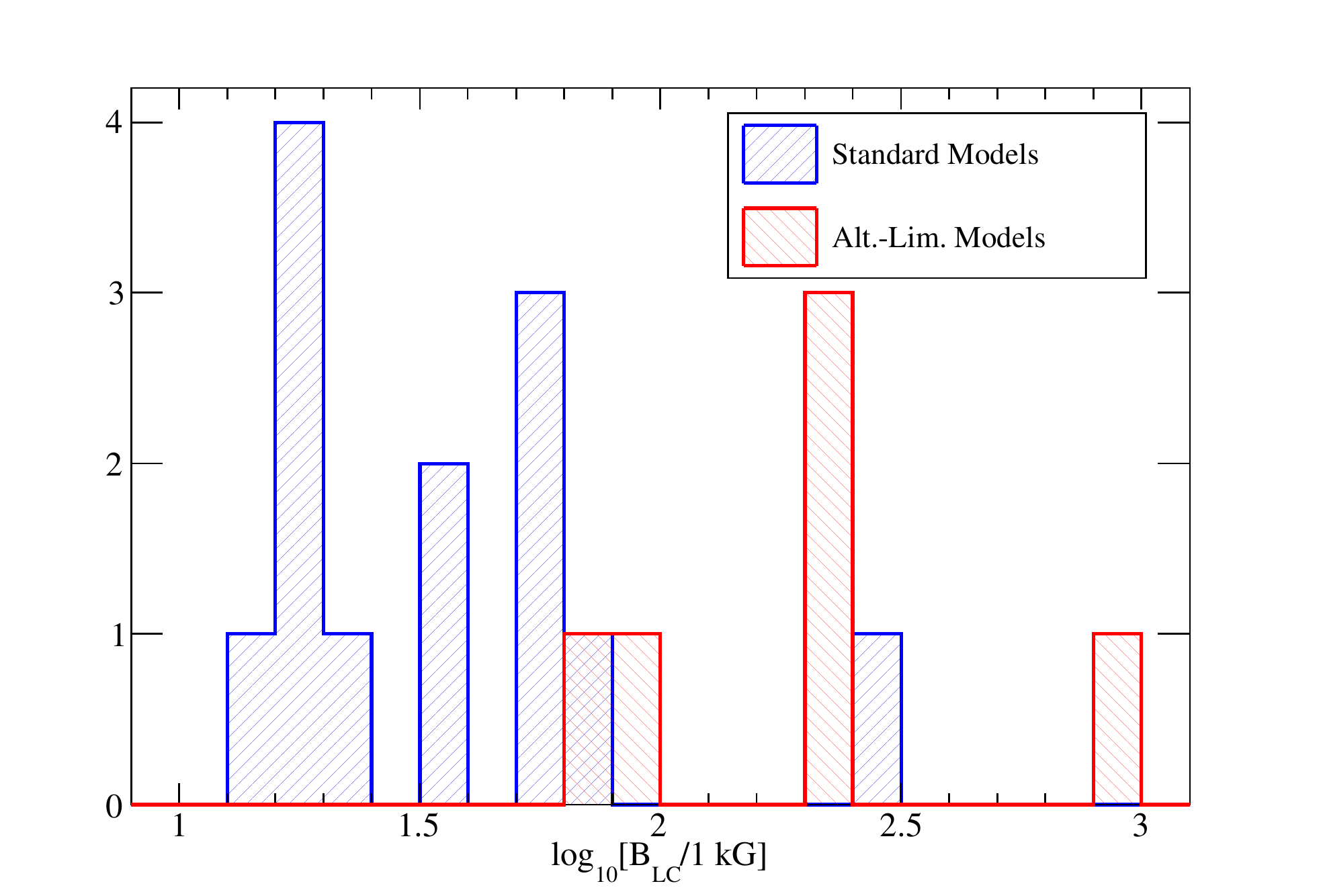}
\end{center}
\small\normalsize
\begin{quote}
\caption[$B_{\rm LC}$ distribution separated into MSPs with and without aligned profiles]{Distribution of $B_{\rm LC}$ values separated into MSPs best-fit by altitude-limited (red) and standard models (blue), this last encompasses PSPC as well as TPC and OG models.\label{ch8compBlc}}
\end{quote}
\end{figure}

As shown in Fig.~\ref{ch8PdotPbest} the phase-aligned MSPs (open symbols) do not obviously occupy a special place in $\dot{\rm P}$-P space.  While they do have some of the highest values of $\dot{E}$ and $B_{\rm LC}$ (see Figs.~\ref{ch8compEdot} and~\ref{ch8compBlc}), there are other MSPs without aligned profiles which have similar values.

\small\normalsize

\small\normalsize

\small\normalsize

\begin{figure}[h]
\begin{center}
\includegraphics[width=.75\textwidth]{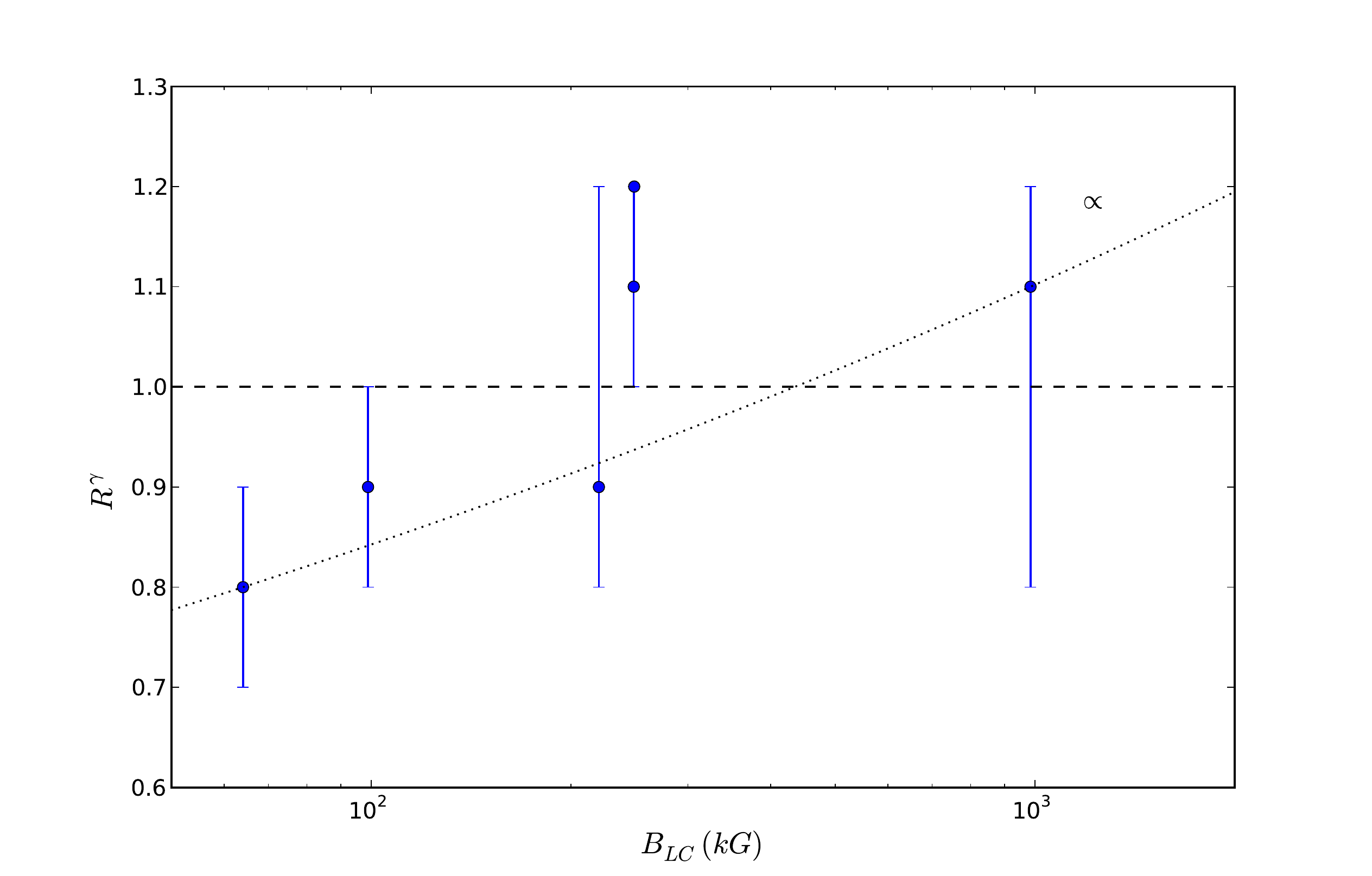}
\end{center}
\small\normalsize
\begin{quote}
\caption[R$_{max}^{\gamma}$ vs. $B_{\rm LC}$ for the phase-aligned MSPs using values from preferred models.]{Best-fit values of R$_{max}^{\gamma}$ vs. $B_{\rm LC}$.  Error bars are from profile likelihood.  Dashed line corresponds to the average value of 1.0 R$_{\rm LC}$.  Dotted line is derived by connecting the first and last points in log-log space.\label{ch8gmaxBlc}}
\end{quote}
\end{figure}

To investigate whether or not $B_{\rm LC}$ does in fact have some influence over the gamma-ray emission region the best-fit R$_{max}^{\gamma}$ versus $B_{\rm LC}$ are shown in Fig.~\ref{ch8gmaxBlc}.  While there is an implied trend, suggesting that the emission extends to higher radial distances for MSPs with stronger $B_{\rm LC}$, the current level of uncertainties suggest that the values are consistent with a constant value of 1.0 R$_{\rm LC}$.

\small\normalsize

The dotted line in Fig.~\ref{ch8gmaxBlc} is derived by connecting the first and last points with a line in log-log space.  This gives R$_{max}^{\gamma}\propto B_{\rm LC}^{0.112}$, with an exponent very close to zero further supporting a constant value.

\begin{figure}[h]
\begin{center}
\includegraphics[width=.75\textwidth]{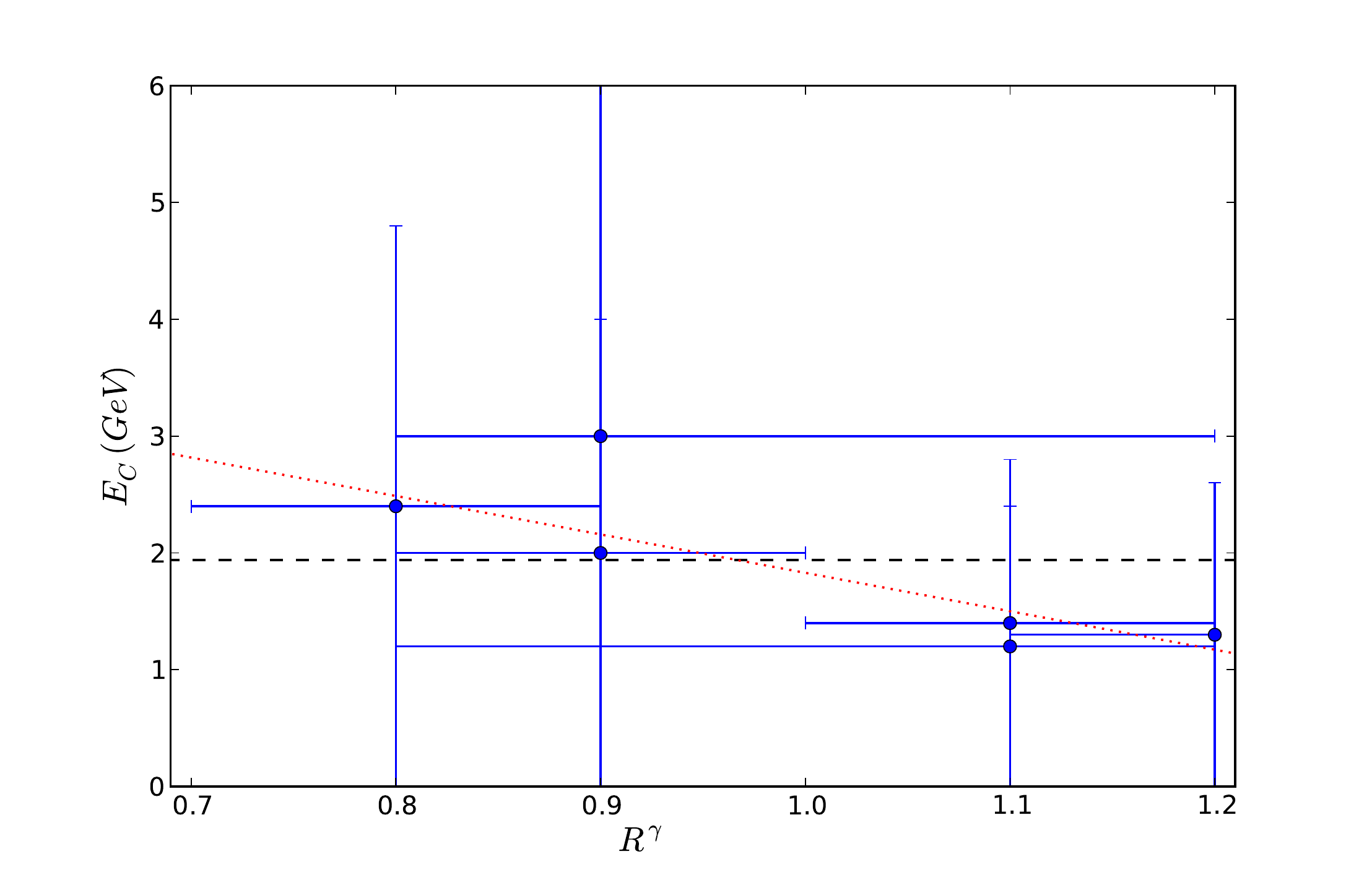}
\end{center}
\small\normalsize
\begin{quote}
\caption[$E_{C}$ versus R$_{max}^{\gamma}$ for the phase-aligned MSPs using values from preferred models.]{Best-fit values of $E_{C}$ versus R$_{max}^{\gamma}$.  Error bars are on $E_{C}$ are from spectral analyses and those on R$_{max}^{\gamma}$ from the profile likelihood described in Chapter 6.  Dotted red line is a fit using just the errors on $E_{C}$ while the dashed line is the weighted average $E_{C}$.\label{ch8EcvsRm}}
\end{quote}
\end{figure}

As noted in Chapters 1 and 4, the CR cutoff energy depends on the local radius of curvature at the point of emission.  Thus, it is interesting to examine the trend of measured cutoff energies from Table~\ref{ch7Vitals} with R$_{max}^{\gamma}$ as shown in Fig.~\ref{ch8EcvsRm}.  Note that this can only meaningfully be done for MSPs fit with alTPC/OG models as R$_{max}^{\gamma}$ is held fixed at 1.2 R$_{\rm LC}$ for the standard TPC/OG models in this thesis.

\small\normalsize

The points in Fig.~\ref{ch8EcvsRm} do seem to imply a systematic trend of decreasing cutoff energy with increasing R$_{max}^{\gamma}$ but such a claim can not be justified with the present uncertainties.  Considering only the errors on $E_{C}$ the points are consistent with a constant value of $E_{C}\ =\ 1.9\pm0.2$ GeV.  This suggests that even when the emission continues out to higher altitudes the bulk of the emission comes from the same part of the magnetosphere with approximately the same local radius of curvature.  This is in line with the findings of \citet{Venter11} that changing R$_{max}^{\gamma}$ affects the shape of minor features in the light curves but not the main peak structure.

\section{Concerning MSP Radio Emission Models}\label{ch8Radio}
The discovery of a growing number of phase-aligned MSPs presents further challenges to the traditional radius-to-frequency mapping (RFM) model in which each frequency is emitted at a single altitude with higher frequencies closer to the stellar surface.  The RFM has been shown to work well for non-recycled pulsars \citep{hdbk} but there have been prior indications that it does not hold for MSPs.  For instance, the observed peak separation in the radio profile of PSR J1939+2134 suggested a near-orthogonal configuration but the sharpness of the peaks and little frequency dependence in the widths did not fit in with the RFM \citep{TS90}.  However, such sharp peaks are naturally explained if the radio emission is caustic in origin.

\begin{figure}[h]
\begin{center}
\includegraphics[width=.7\textwidth]{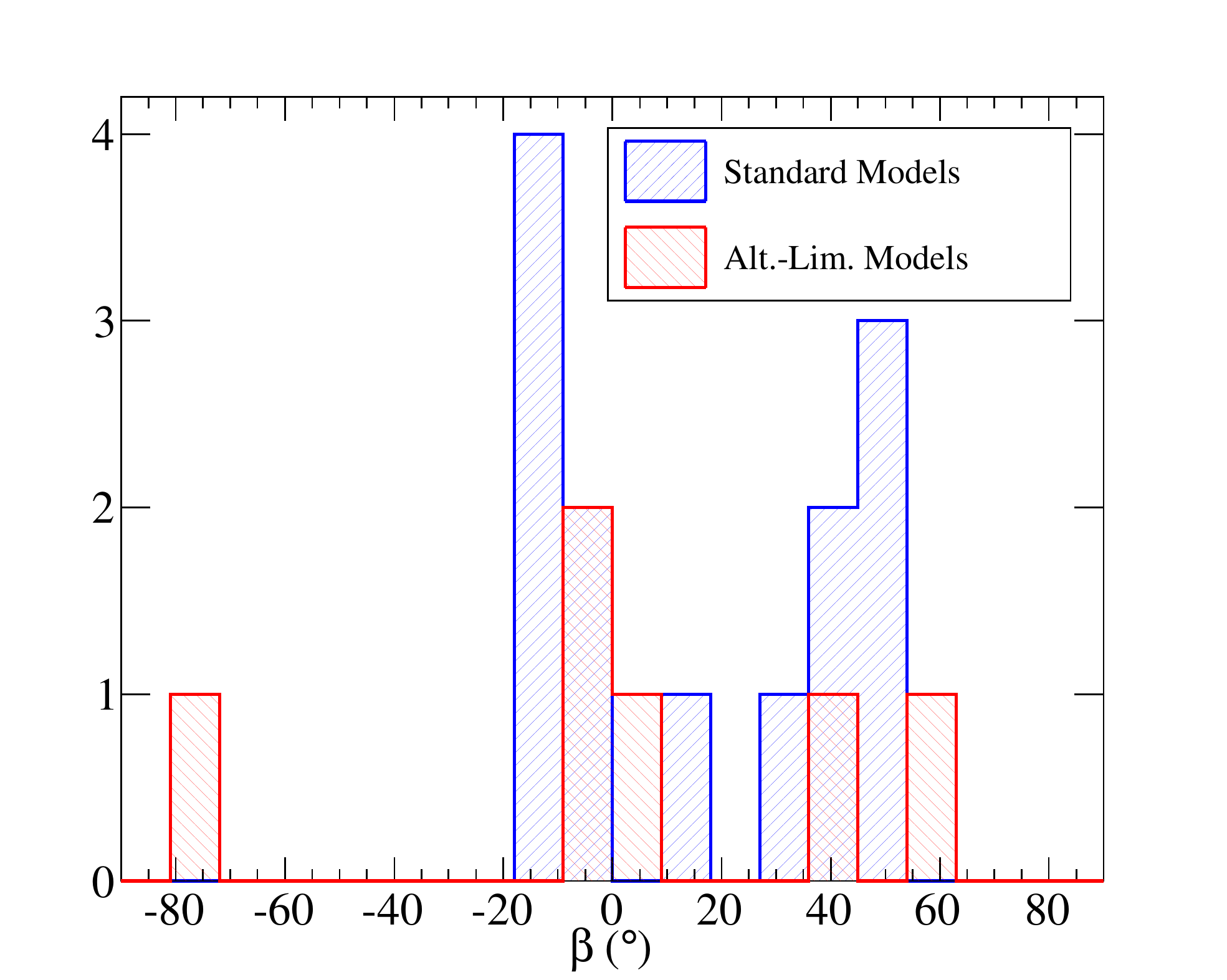}
\end{center}
\small\normalsize
\begin{quote}
\caption[Distribution of preferred $\beta$ values separated into standard and altitude-limited models]{Distribution of $\beta$ values preferred by the likelihood.  Values for MSPs fit with standard models are shown in blue while those fit with altitude-limited models are shown in red.\label{ch8bestbeta}}
\end{quote}
\end{figure}

Comparison of the predicted $\beta$ values, Fig.~\ref{ch8bestbeta}, for the models fit with hollow-cone beams (standard) and those fit with alTPC/OG models does not reveal any striking differences in the two populations.

\small\normalsize

With the exception of one outlier (PSR J2214+3000) the phase-aligned MSPs occupy the same range of $\beta$ values as the non-aligned MSPs.  For some of the phase-aligned MSPs the likelihood predicts that the radio emission regions should be very near the light cylinder which is in conflict which the findings of \citet{KG03} which predicts altitudes $\sim$ 30\% of the light cylinder for the shortest period MSPs.  With emission occuring at higher altitudes the distribution of $\beta$ values should cover a broader range, but with only six phase-aligned MSPs the statistics may be too low to demonstrate this.

Note that the extent of the blue histogram in Fig.~\ref{ch8bestbeta} is constrained by the size of the cone beam and the requirement that all of the MSPs in this study be both radio and gamma-ray loud.  The MSPs in this sample do seem to disfavor geometries with smaller $|\beta|$, perhaps arguing for larger cones.  Additionally, combining the observed gamma-ray and radio light curves of many of these MSPs does seem to argue for a modification to the emission height formula of \citet{KG03}.

\begin{figure}[h]
\begin{center}
\includegraphics[width=.75\textwidth]{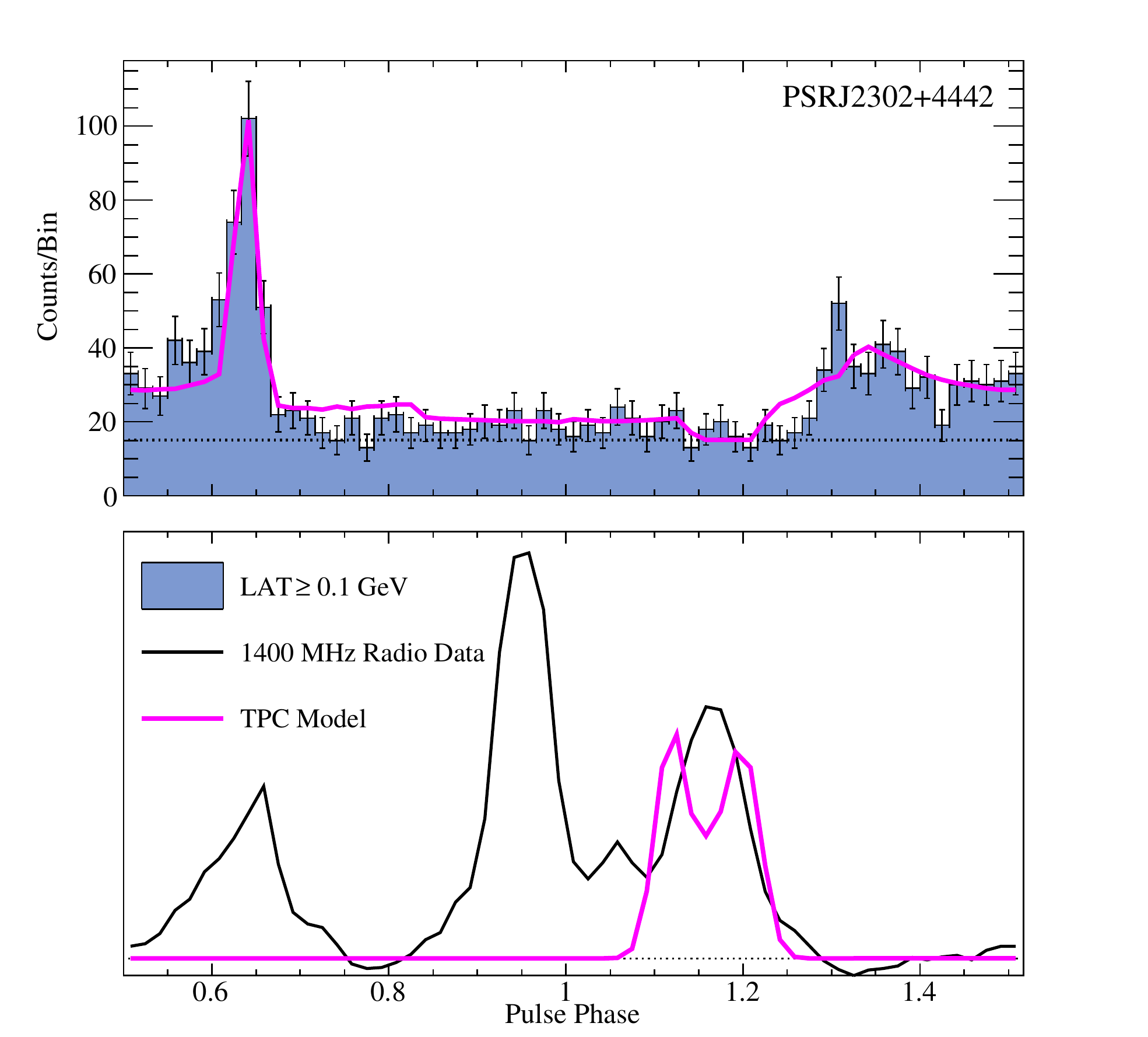}
\end{center}
\small\normalsize
\begin{quote}
\caption[Data and best-fit TPC light curves for PSR J2302+4442]{Best-fit TPC and data profiles for PSR J2302+4442, gamma-ray on top radio on bottom.  Note that the phase has been shifted by 0.5 in order to more prominently display the two-peaked radio structure on either side of phase 1.\label{ch8J2302TPC}}
\end{quote}
\end{figure}

PSR J2302+4442 demonstrates this point well.  This MSP has a complex profile with a main, two-peaked structure centered on phase 0 (1) which has the shape of a classical cone beam and the possibility of a weak core component.  The best-fit TPC geometry (see Fig.~\ref{ch8J2302TPC}) reproduces the gamma-ray light curve well and can find two radio peaks but they are not spaced far enough apart.

The marginalized confidence contours suggest a high-level of confidence in the best-fit geometry.  This suggests that if geometries exist which reproduce the observed radio profile more accurately they do not reproduce the gamma-ray light curve well.  This could be seen as an argument against the geometric, gamma-ray emission models used in this study; however, if the altitude of the cone beam was increased (implying that the radio models are at fault) then the same geometry should reproduce the observed radio profile while leaving the predicted gamma-ray light curve unchanged.

\small\normalsize

\begin{figure}[h]
\begin{center}
\includegraphics[width=.75\textwidth]{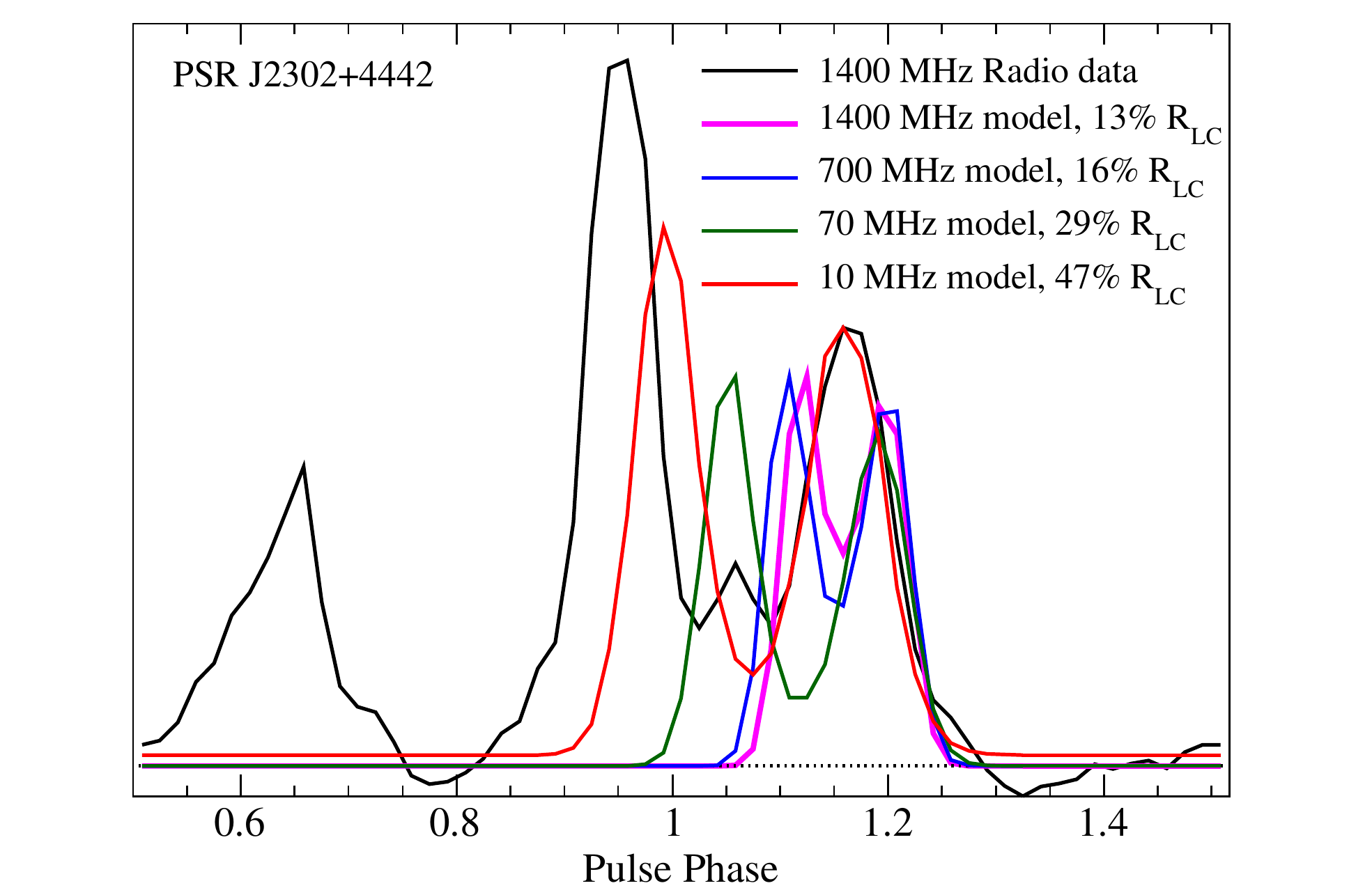}
\end{center}
\small\normalsize
\begin{quote}
\caption[Data and best-fit radio profiles with different simulated frequencies for PSR J2302+4442]{Observed radio profile, at 1400 MHz, for PSR J2302+4442 with model profiles with emission at different altitudes.  The pink curve corresponds to the best-fit model with emission height simulated assuming a frequency of 1400 MHz.  The blue, green and red curves are simulations for the same, best-fit geometry found by the MCMC but with the emission height calculated assuming frequencies of 700, 70, and 10 MHz, respectively.  The blue and green model curves were normalized to have the same maximum height as the pink curve while the red curve was normalized such that the peak near phase 1.2 matched the height of the same peak in the data.\label{ch8J2302comp}}
\end{quote}
\end{figure}

In order to demonstrate the effect of increasing the emission altitude of the cone beam, simulations with the same period (5.5 ms) were made for the best-fit $\alpha$ (59\DEG{}) but with lower frequencies (700, 70, and 10 Hz).  The light curves for each of the lower frequency simulations with the best-fit $\zeta$ value (46\DEG{}) are shown against the data and 1400 MHz model profiles in Fig.~\ref{ch8J2302comp}.

\small\normalsize

Note that not only does the separation of the model radio peaks increase as the emission altitude is increased but the same asymmetry observed in the data peak heights appears in the models.  These new simulations were not fit to the data and the same phase shift was used in order to demonstrate how, for the exact same gamma-ray model, the radio fit improves.

While the 10 MHz (red) model does do much better than the standard emission height the data suggests that the cone must move higher still, beyond 50\% of the light cylinder radius.  It should be noted that a singe hollow-cone beam is unlikely to acount for all of the observed features, there may be a weak core component between the two main peaks and the feature near phase 0.65 may suggest that emission from both poles is seen (which would necessitate a different geometry) or perhaps the presence of a second, wider cone which is only partially illuminated.  Polarization measurements will be instrumental in assessing the nature of each of these components.

In addition to MSPs with double peaked radio structures, there are also suggestions that the emission altitudes are too low in MSPs with single-peak radio profiles.  For many of these MSPs the best-fit geometry is found very far from the maximum of the cone.  This would suggest that the line of sight from Earth barely clips the beam and then the measured radio flux is much lower than the true value (i.e., for the radio beam f$_{\Omega}$ is considerably greater than 1).

In principle this is not a problem; however, such large f$_{\Omega}$ values would imply the existence of incredibly bright radio MSPs.  One is then left to wonder why such bright MSPs have not been discovered with their beams pointed more directly at Earth.  It should be noted that while polarization measurements of some MSPs with a single, strong radio peak do suggest core emission (thereby invalidating the previous argument) others do display the flat position angle swings characteristic of cone beams.

\section{Conclusions}\label{ch8concl}
All of gamma-ray astrophysics has experienced a resurgence and new areas have been opened with the launch of \Fermi{}.  Many more gamma-ray blazars are now known \citep{Abdo1LAC}; delayed HE emission from some gamma-ray bursts has been confirmed (e.g., Hurley et al., 1994 and Abdo et al., 2009i); new classes of gamma-ray pulsars are now well established (e.g., Abdo et al., 2009e,g); and \\completely unexpected sources have been discovered \citep{AbdoV407Cyg}.  Additionally, constraints from searches for dark matter signals and other new physics in the \FL{} data are driving theoretical developments (e.g., Abdo et al., 2009h and 2010a).

The field of gamma-ray pulsar science has benefited enormously from \FL{} observations, increasing the known population by more than an order of magnitude with approximately 2 years of sky survey.  MSPs have emerged as a numerous class of HE emitters via the detection of pulsed gamma rays from twenty thus far (including the MSPs presented here and PSR J2241$-$5236 announced by Keith et al., 2011); the discovery of $>$ 20 new radio MSPs in searches of unassociated LAT sources; and the detection of pulsar-like HE emission from several globular clusters (e.g., Abdo et al., 2009f and 2010j).

Understanding the exact details of the emission mechanisms at work in gamma-ray MSPs will involve detailed radiation models such as has been done for Vela \citep{Du11} and the Crab \citep{Harding08}.  However, geometric emission models are capable of predicting viewing geometries and can be especially powerful when paired with observations at other wavelengths.

When fitting just the gamma-ray light curves, it is easy to find geometries which reproduce the observations well for either TPC or OG models.  Thus, inclusion of constraints from other wavelengths (radio lag, X-ray torus fitting, etc.) is crucial when attempting to decide between different models.  The shapes of both the radio and gamma-ray light curves are expected to carry information about the magnetic field structure; thus, it is important to fit both light curves, whenever possible, in order to obtain a more complete picture of the magnetosphere.

In reality, the magnetosphere of a pulsar will be filled with charged particles and thus the field structure will deviate from the vacuum retarded dipole geometry which has been used in this thesis.  However, the magnetosphere can not be filled in a completely force-free manner as pulsars are known to accelerate particles to high energies.  Therefore, geometric modeling of gamma-ray pulsar light curves with both the vacuum and force-free magnetic field geometries is necessary in order to provide a metric for how far between the two cases a real pulsar magnetosphere is.

Joint fitting of gamma-ray and radio profiles provides a strong test of not only the gamma-ray but also the radio emission models.  There is evidence for radio emission originating much further out in the magnetosphere than previously thought, in agreement with the studies of \citet{Ravi10}.  Caustic modeling of the radio profiles of phase-aligned MSPs has specific predictions concerning the polarization characteristics which, to first order, are in good agreement with observations.

Of the nineteen MSPs considered in this study only two seem to have pair starved magnetospheres while the other seventeen are well fit using models with narrow accelerating gaps similar to what has been found for non-recycled pulsars.  Six of the known gamma-ray MSPs demonstrate phase-aligned pulse profiles, as opposed to one for non-recycled pulsars.  These six pulsars tend to populate the high-end of the MSP $\dot{E}$ and $B_{\rm LC}$ ranges.

While the likelihood analysis only allows for a decision between the TPC and OG models for seven MSPs it should be noted that TPC models tend to be preferred.  This is in contrast to findings for non-recycled pulsars (Romani \& Watters, 2010 and Watters \& Romani, 2011) though the two may be reconciled if modified OG models with emission below the NCS are invoked and/or if differences between model implementations are considered as discussed at the beginning of this chapter.

The predominance of gamma-ray MSPs displaying narrow peaked profiles suggests that some mechanism moves the CR pair-creation death line below previous predictions.  This further suggests that previously discarded, older, non-recycled pulsars near the old death line may merit further investigation as gamma-ray sources.

The two main possibilities which have been considered are higher-order magnetic multipoles and offset dipoles.  While multipoles are likely to be present it is not clear how strongly they control the observed MSP characteristics, especially if the emission originates, predominantly, in the outer magnetosphere.  Offset dipoles provide a natural explanation but it has not yet been demonstrated if such models can reproduce the observed gamma-ray light curves (Harding \& Muslimov, 2011 and Harding et al., 2011).

It is clear that the emission mechanisms in non-recycled gamma-ray pulsars and MSPs are the same (assumed to be CR in the radiation-reaction limit), though there seems to be an inherent efficiency transition near a spin-down power of a few times $10^{34}$ erg s$^{-1}$.  This transition can not be due to an inability of low $\dot{E}$ pulsars to effectively screen the accelerating field as most observed MSP gamma-ray light curves are indicative of narrow gaps which requires a screened field.  However, following the offset dipole findings of \citet{HM11}, this transition may be due to low $\dot{E}$ pulsars being only partially screened.  As the gamma-ray pulsar $L_{\gamma}$-$\dot{E}$ plot continues to be populated the nature of this transition should become more clear and serve as a further test of possible emission models.

Understanding this transition will be crucial for proper population synthesis models which attempt to assess the MSP contribution to the diffuse gamma-ray backgrounds (e.g., Takata et al., 2011).  These studies must fold this transition into the simulated $\dot{E}$ distribution in order to make realistic estimates and should investigate the possibility of higher altitude, and thus larger beam, radio emission when addressing the question of radio-quiet gamma-ray MSPs.

All told, this is a very exciting time in pulsar physics with important discoveries continuing to be made and technological advancements promising even more to come.
\appendix
\renewcommand{\thechapter}{A}

\chapter{MCMC Plots for all MSPs}\label{appA}
This appendix collects the best-fit light curve, marginalized confidence contour, and simulated emission plots for all nineteen MSPs considered in this thesis.  Additionally, this appendix will collect basic historical information such as radio and gamma-ray discovery papers.

\section{PSR J0030+0451}\label{appAJ0030}
PSR J0030+0451 is an isolated, 4.8655 ms pulsar first detected in the radio by \citet{Lommen00}.  Gamma-ray pulsations were first reported by \citet{AbdoJ0030} and later by \citet{AbdoMSPpop} and \citet{AbdoPSRcat}.  The gamma-ray light curve of this MSP were previously modeled by \citet{Venter09} using geometric OG and TPC models with a hollow-cone beam radio model.

\begin{figure}
\begin{center}
\includegraphics[width=0.75\textwidth]{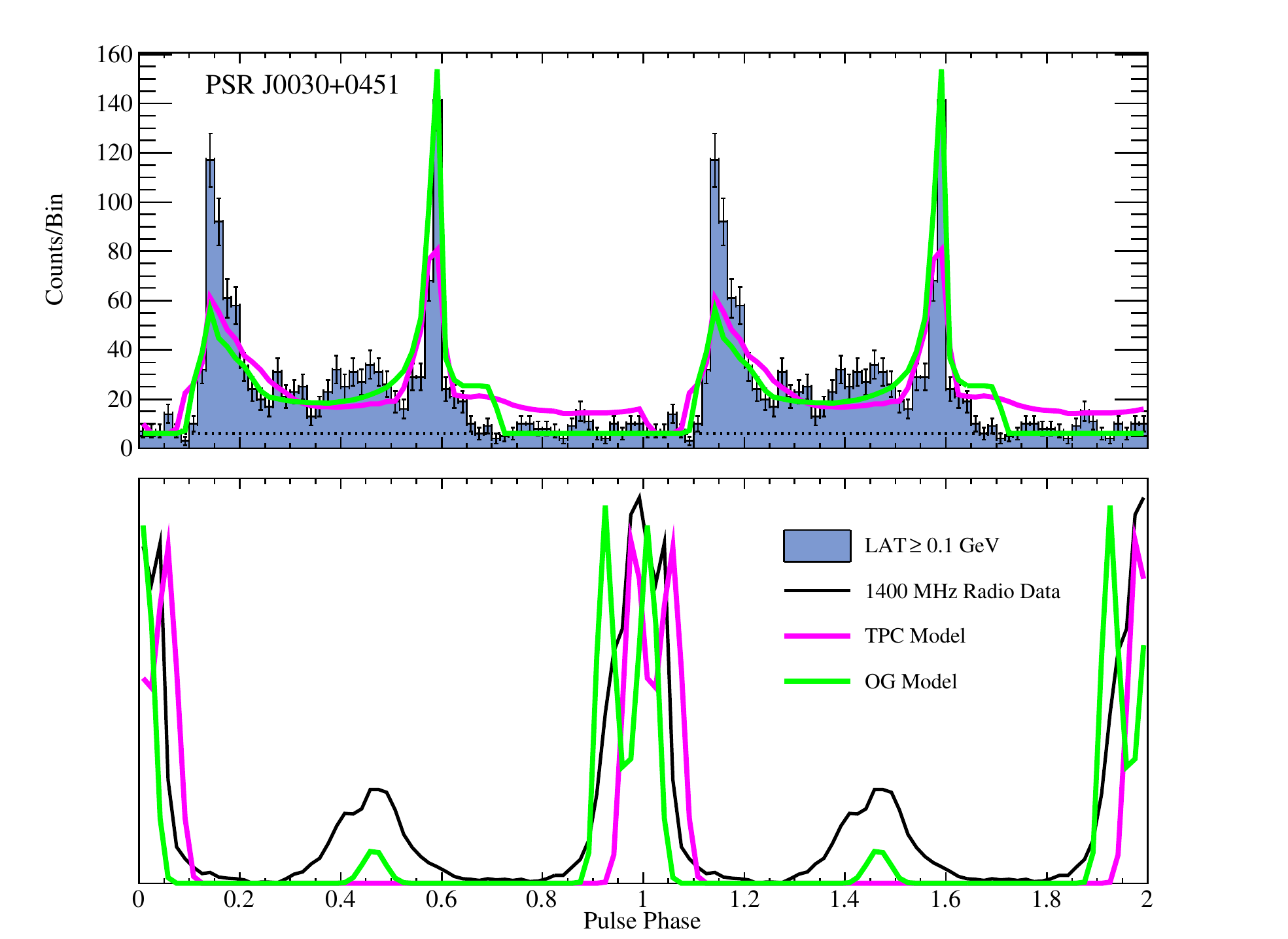}
\end{center}
\small\normalsize
\begin{quote}
\caption[Data and best-fit light curves for PSR J0030+0451]{Best-fit gamma-ray and radio light curves for PSR J0030+0451 using the TPC and OG models.\label{appAJ0030LCs}}
\end{quote}
\end{figure}

The best-fit gamma-ray and radio light curves are shown in Fig.~\ref{appAJ0030LCs}.  The gamma-ray light curve was fit with the TPC and OG models.  The radio profile was fit with a hollow-cone beam model.  These light curve fits have used 1400 MHz Nan\c{c}ay radio profile.

\small\normalsize

\begin{figure}
\begin{center}
\includegraphics[width=0.75\textwidth]{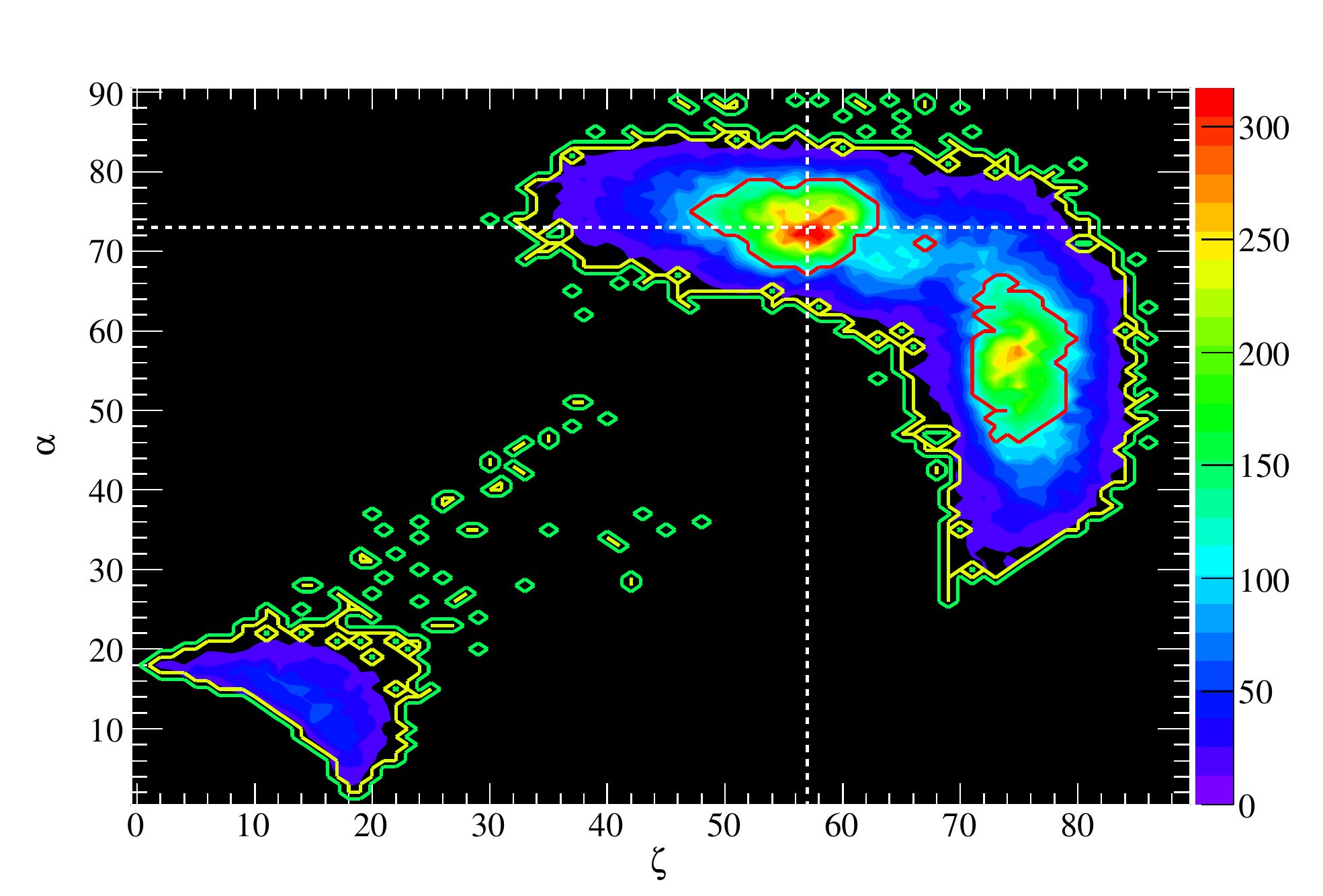}
\end{center}
\small\normalsize
\begin{quote}
\caption[Best-fit TPC contours for PSR J0030+0451]{Marginalized confidence contours for PSR J0030+0451 for the TPC model.  The color scale represents the number of entries in each $(\alpha,\zeta)$ bin.\label{appAJ0030TPCcont}}
\end{quote}
\end{figure}

The marginalized $\alpha$-$\zeta$ confidence contours corresponding to the TPC fit are shown in Fig.~\ref{appAJ0030TPCcont}, the best-fit geometry is indicated by the horizontal and vertical dashed, white lines.

\small\normalsize

\begin{figure}
\begin{center}
\includegraphics[width=0.75\textwidth]{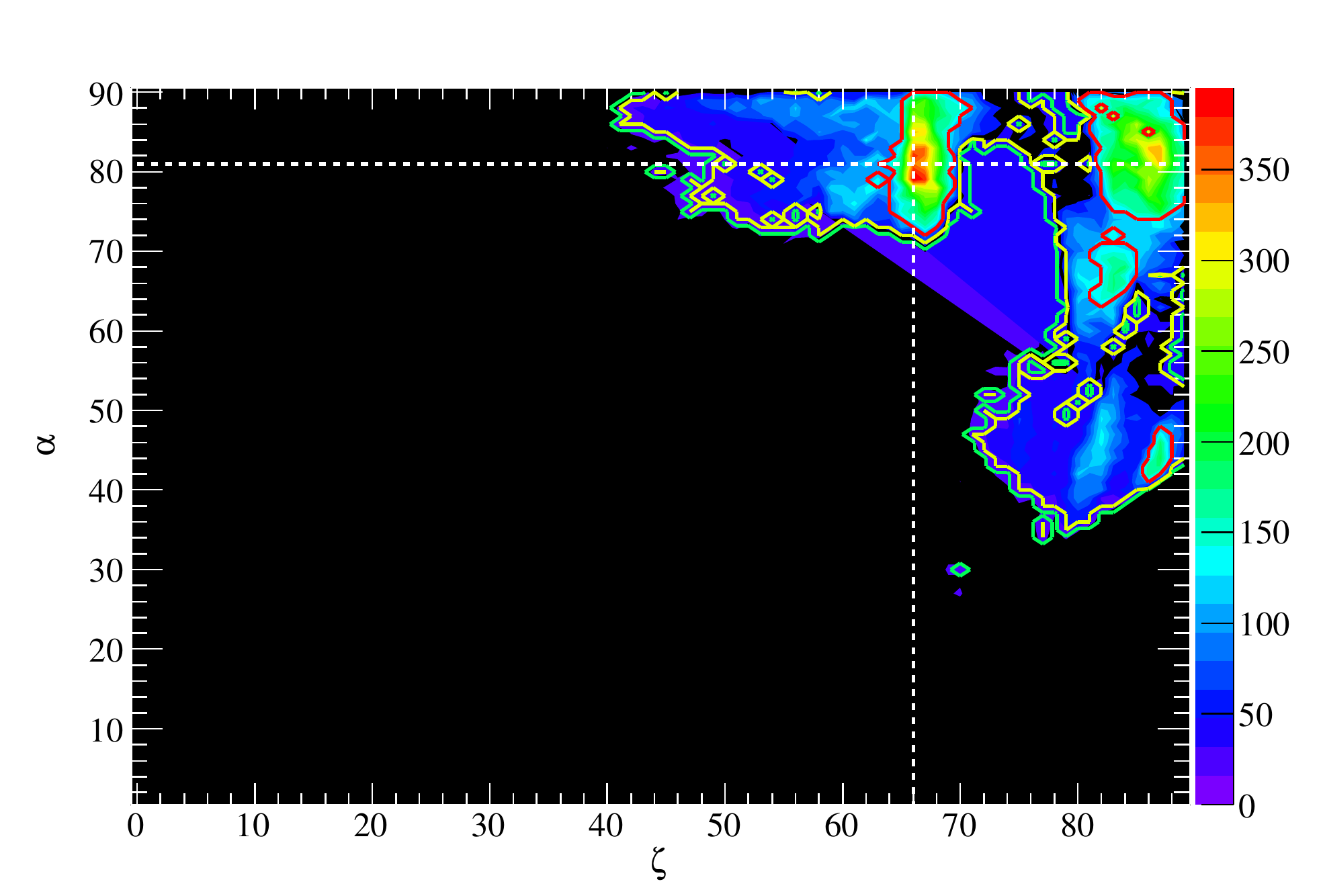}
\end{center}
\small\normalsize
\begin{quote}
\caption[Best-fit OG contours for PSR J0030+0451]{Marginalized confidence contours for PSR J0030+0451 for the OG model.\label{appAJ0030OGcont}}
\end{quote}
\end{figure}

The marginalized $\alpha$-$\zeta$ confidence contours corresponding to the OG fit are shown in Fig.~\ref{appAJ0030OGcont}, the best-fit geometry is indicated by the horizontal and vertical dashed, white lines.

\small\normalsize

\begin{figure}
\begin{center}
\includegraphics[width=0.75\textwidth]{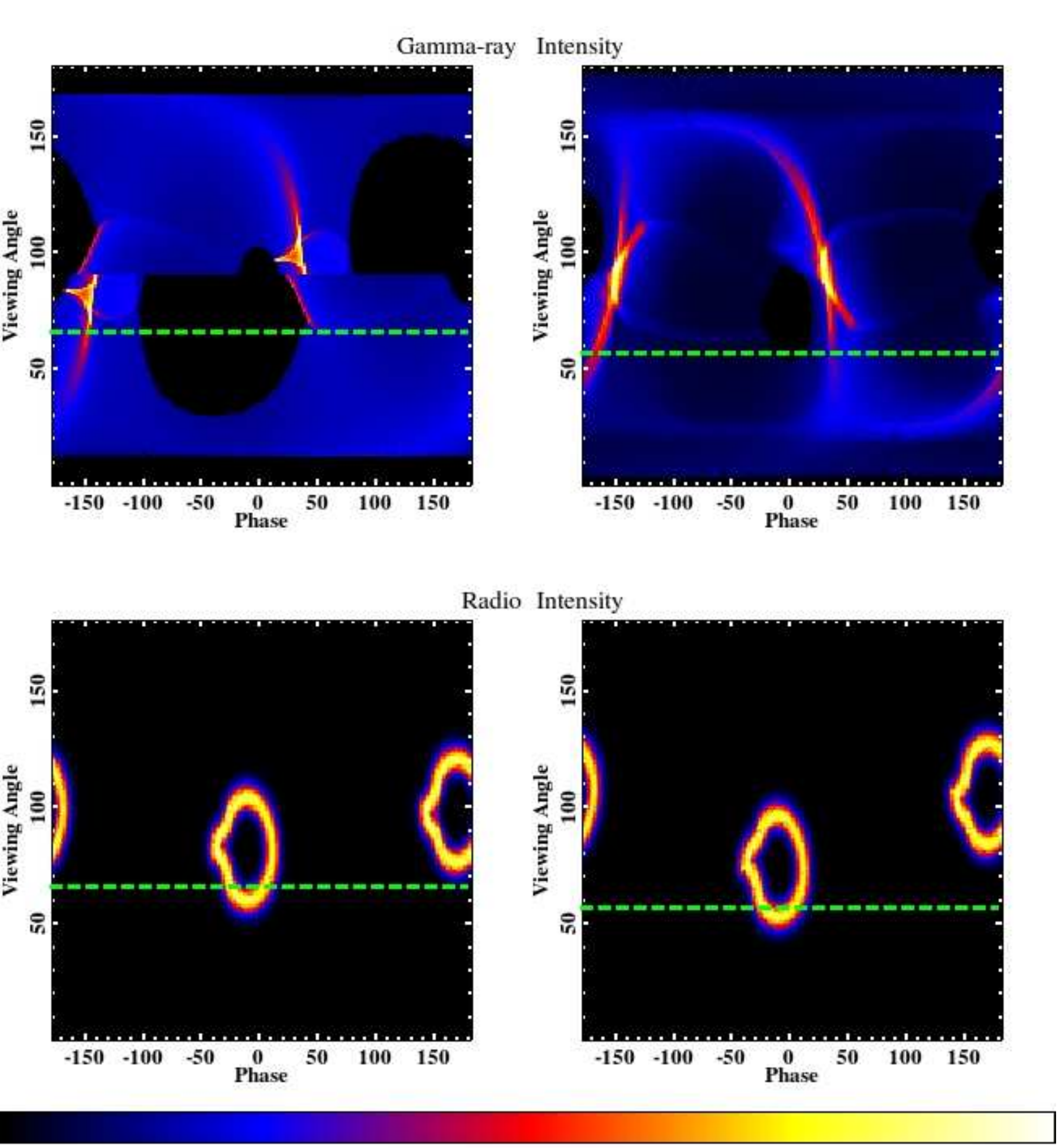}
\end{center}
\small\normalsize
\begin{quote}
\caption[Best-fit phase plots of simulated emission for PSR J0030+0451]{Distribution of simulated emission as a function of viewing angle and pulse phase for models used to fit PSR J0030+0451.  The best-fit $\zeta$ values are indicated by the dashed green lines.  The top plots correspond to the gamma-ray phase plots while the bottom are for the radio.  The left plots correspond to fits with the OG model with TPC plots on the right.  Intensity units are arbitrary increasing in value from black to white as indicated by the color bar.  The color scale in the top-left plot is square root in order to bring out fainter features.\label{appAJ0030PhPlt}}
\end{quote}
\end{figure}

Plots of simulated emission corresponding to the best-fit models are shown in Fig.~\ref{appAJ0030PhPlt}, OG models are on the left and TPC on the right, gamma-ray models are on the top and radio on the bottom.

\small\normalsize

\section{PSR J0034$-$0534}\label{appAJ0034}
PSR J0034$-$0534 is a 1.8772 ms pulsar in a 1.6 d orbit with a low-mass white dwarf companion \citet{Bailes94}.  Gamma-ray pulsations from this MSP were first reported by \citet{AbdoJ0034} who also introduced the altitude-limited models (based on models used by Harding et al., 2008 to model the emission from the Crab pulsar) to explain the near phase-aligned profiles.  An analysis of the off-peak region by \citet{PWNcat} revealed that there is evidence for magnetospheric emission through nearly the entire pulse.  The MCMC likelihood analysis described in this thesis was also applied to this MSP by \citet{Venter11} using the same data span as \citet{PWNcat}.

\begin{figure}
\begin{center}
\includegraphics[width=0.75\textwidth]{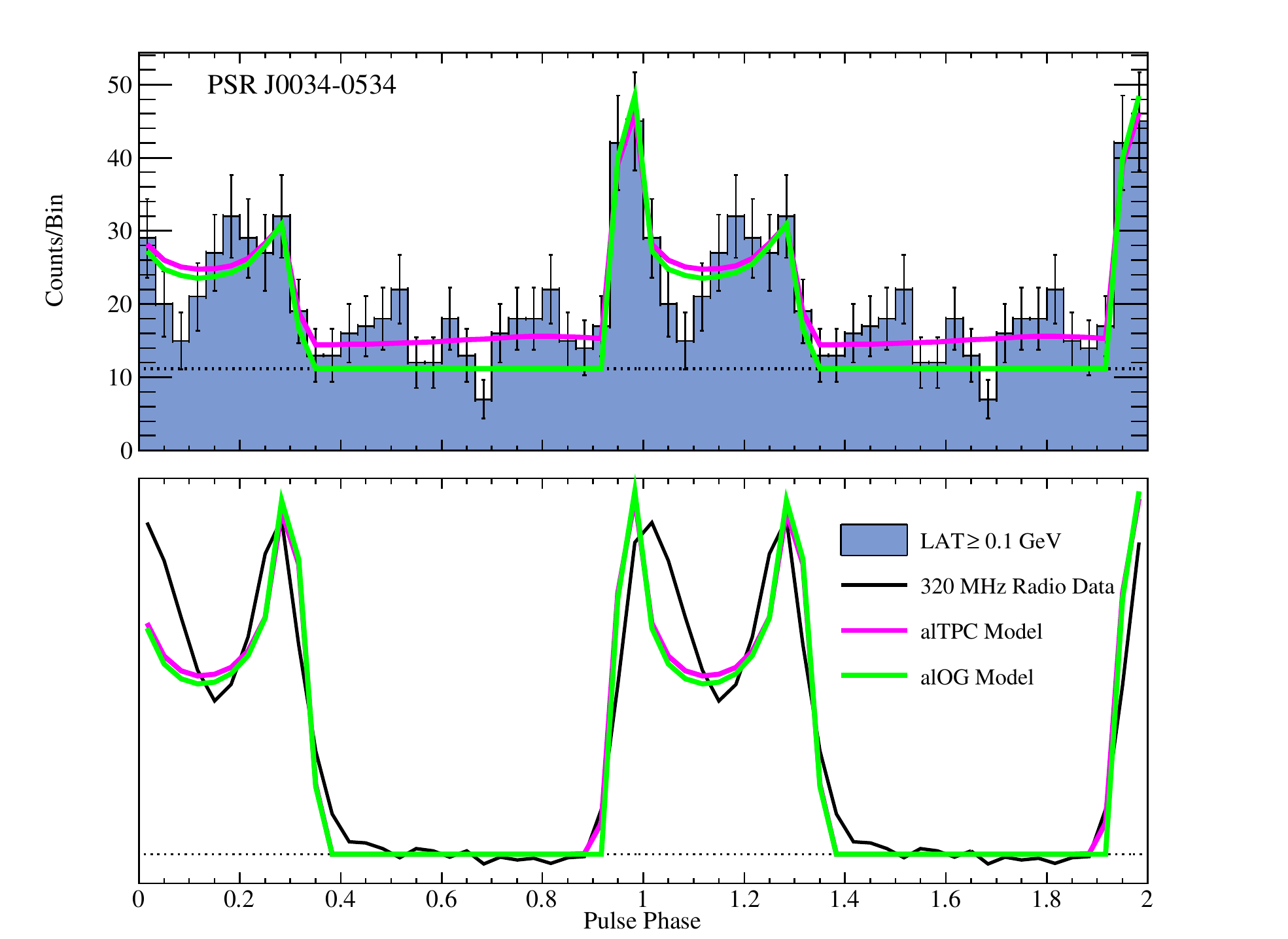}
\end{center}
\small\normalsize
\begin{quote}
\caption[Data and best-fit light curves for PSR J0034$-$0534]{Best-fit gamma-ray and radio light curves for PSR J0034$-$0534 using the alTPC and alOG models.\label{appAJ0034LCs}}
\end{quote}
\end{figure}

The best-fit light curves are shown in Fig.~\ref{appAJ0034LCs}.  The gamma-ray and radio profiles have both been fit with the alTPC and alOG models.  These light curve fits have used the 324 MHz Westerbork radio profile.

\small\normalsize

\begin{figure}
\begin{center}
\includegraphics[width=0.75\textwidth]{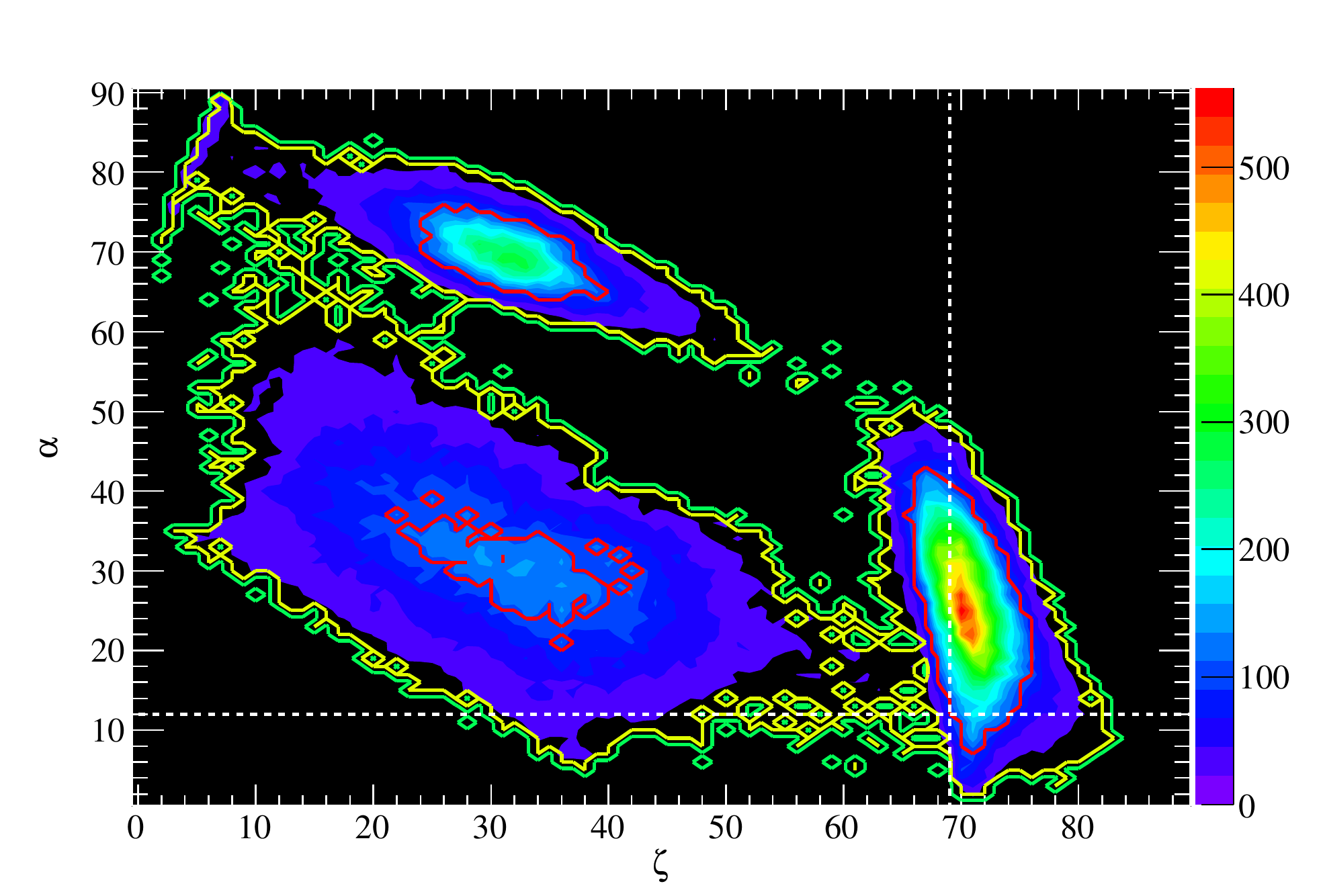}
\end{center}
\small\normalsize
\begin{quote}
\caption[Best-fit alTPC contours for PSR J0034$-$0534]{Marginalized confidence contours for PSR J0034$-$0534 for the alTPC model.\label{appAJ0034TPCcont}}
\end{quote}
\end{figure}

The marginalized $\alpha$-$\zeta$ confidence contours corresponding to the alTPC fit are shown in Fig.~\ref{appAJ0034TPCcont}, the best-fit geometry is indicated by the vertical and horizontal, white dashed lines.

\small\normalsize

\begin{figure}
\begin{center}
\includegraphics[width=0.75\textwidth]{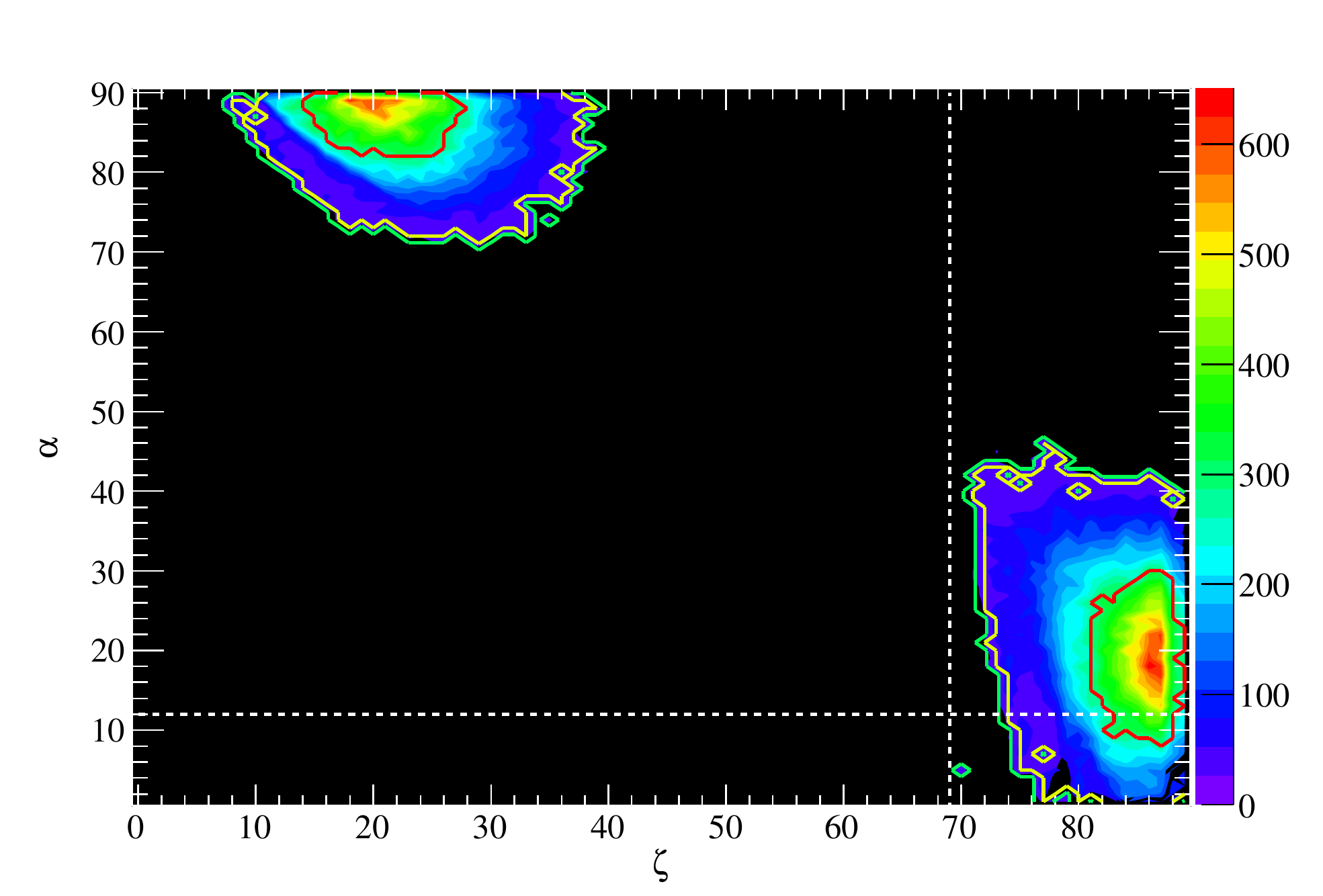}
\end{center}
\small\normalsize
\begin{quote}
\caption[Best-fit alOG contours for PSR J0034$-$0534]{Marginalized confidence contours for PSR J0034$-$0534 for the alOG model.\label{appAJ0034OGcont}}
\end{quote}
\end{figure}

The marginalized $\alpha$-$\zeta$ confidence contours corresponding to the alOG fit are shown in Fig.~\ref{appAJ0034OGcont}, the best-fit geometry is indicated by the vertical and horizontal, white dashed lines.

\small\normalsize

\begin{figure}
\begin{center}
\includegraphics[width=0.75\textwidth]{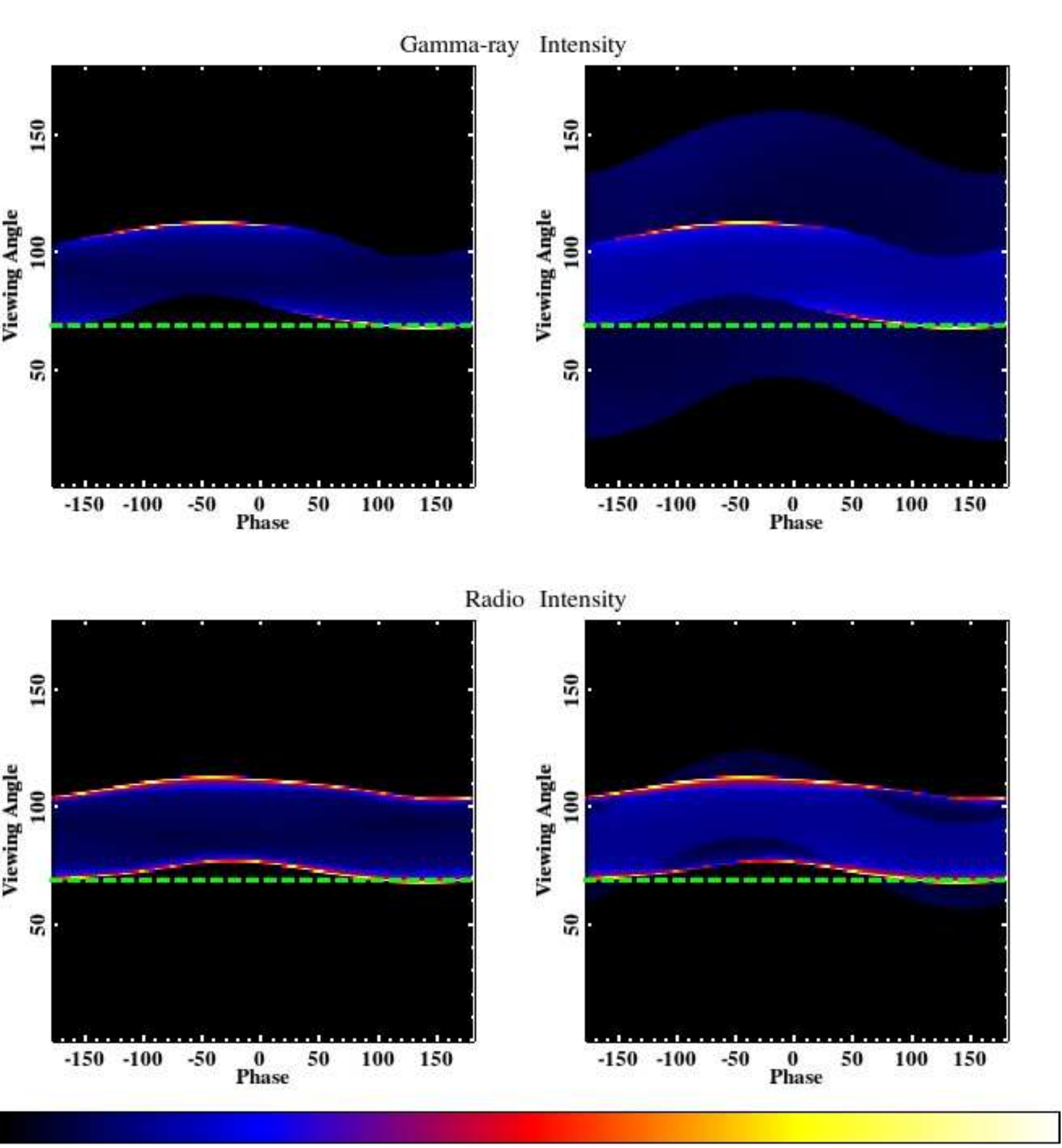}
\end{center}
\small\normalsize
\begin{quote}
\caption[Best-fit phase plots of simulated emission for PSR J0034$-$0534]{Distribution of simulated emission as a function of viewing angle and pulse phase for models used to fit PSR J0034$-$0534.  The best-fit $\zeta$ values are indicated by the dashed green lines.  The top plots correspond to the gamma-ray phase plots while the bottom are for the radio.  The left plots correspond to fits with the alOG model with alTPC plots on the right.\label{appAJ0034PhPlt}}
\end{quote}
\end{figure}

Plots of simulated emission corresponding to the best-fit models are shown in Fig.~\ref{appAJ0034PhPlt}, alOG models are on the left and alTPC on the right, gamma-ray models are on the top and radio on the bottom. 

\small\normalsize

\section{PSR J0218+4232}\label{appAJ0218}
PSR J0218+4232 is a 2.3231 ms  pulsar in a 2 day orbit with a companion of mass $\gtrsim$0.16 M$_{\odot}$ and was first discovered in the radio by \citet{Navarro95}.  A marginal pulsed detection of this MSP using \emph{EGRET} data by \citet{Kuiper00} and was later confirmed with the LAT (Abdo et al. 2009g and 2010c).  The gamma-ray light curve of this MSP was modeled by \citet{Venter09} using geometric OG and TPC models with a hollow-cone beam radio model.

\begin{figure}
\begin{center}
\includegraphics[width=0.75\textwidth]{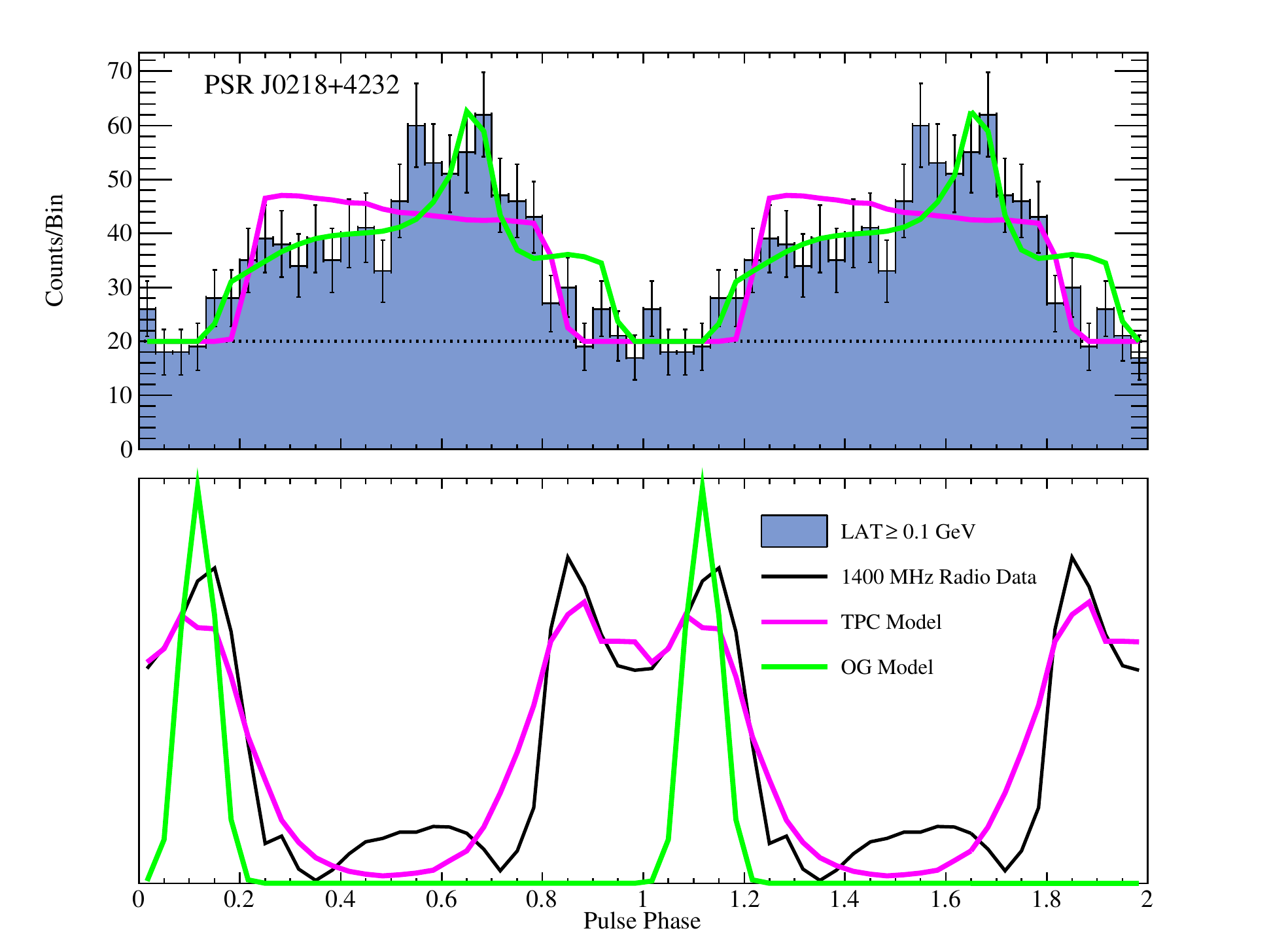}
\end{center}
\small\normalsize
\begin{quote}
\caption[Data and best-fit light curves for PSR J0218+4232]{Best-fit gamma-ray and radio light curves for PSR J0218+4232 using the TPC and OG models.\label{appAJ0218LCs}}
\end{quote}
\end{figure}

The best-fit gamma-ray and radio light curves are shown in Fig.~\ref{appAJ0218LCs}.  The gamma-ray light curve has been fit with the OG and TPC models.  The radio profile has been fit with a hollow-cone beam model.  These light curve fits have used the 1400 MHz Nan\c{cay} radio profile.

\small\normalsize

\begin{figure}
\begin{center}
\includegraphics[width=0.75\textwidth]{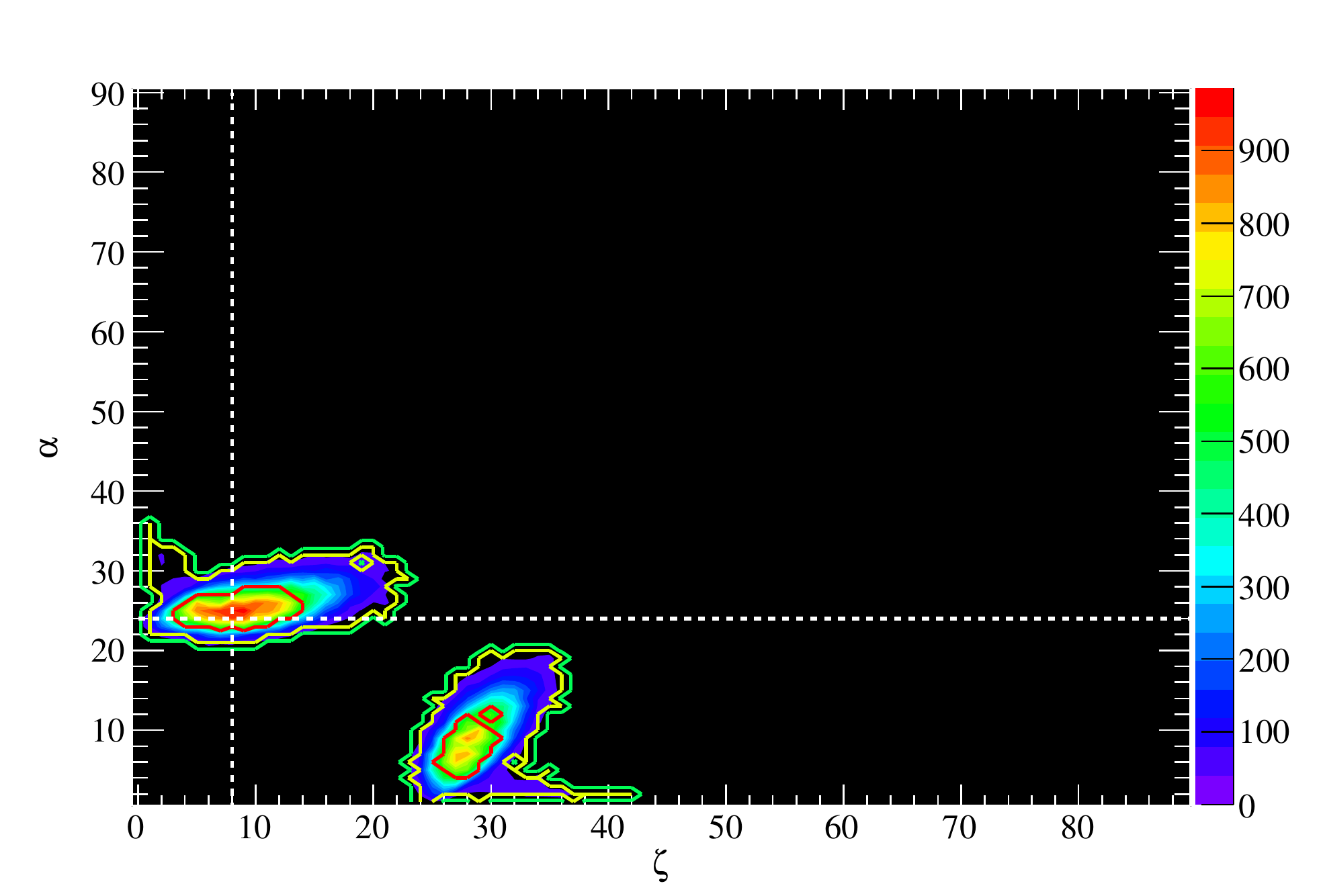}
\end{center}
\small\normalsize
\begin{quote}
\caption[Best-fit TPC contours for PSR J0218+4232]{Marginalized confidence contours for PSR J0218+4232 for the TPC model.\label{appAJ0218TPCcont}}
\end{quote}
\end{figure}

The marginalized $\alpha$-$\zeta$ confidence contours corresponding to the TPC fit are shown in Fig.~\ref{appAJ0218TPCcont}, the best-fit geometry is indicated by the vertical and horizontal dashed, white lines.

\small\normalsize

\begin{figure}
\begin{center}
\includegraphics[width=0.75\textwidth]{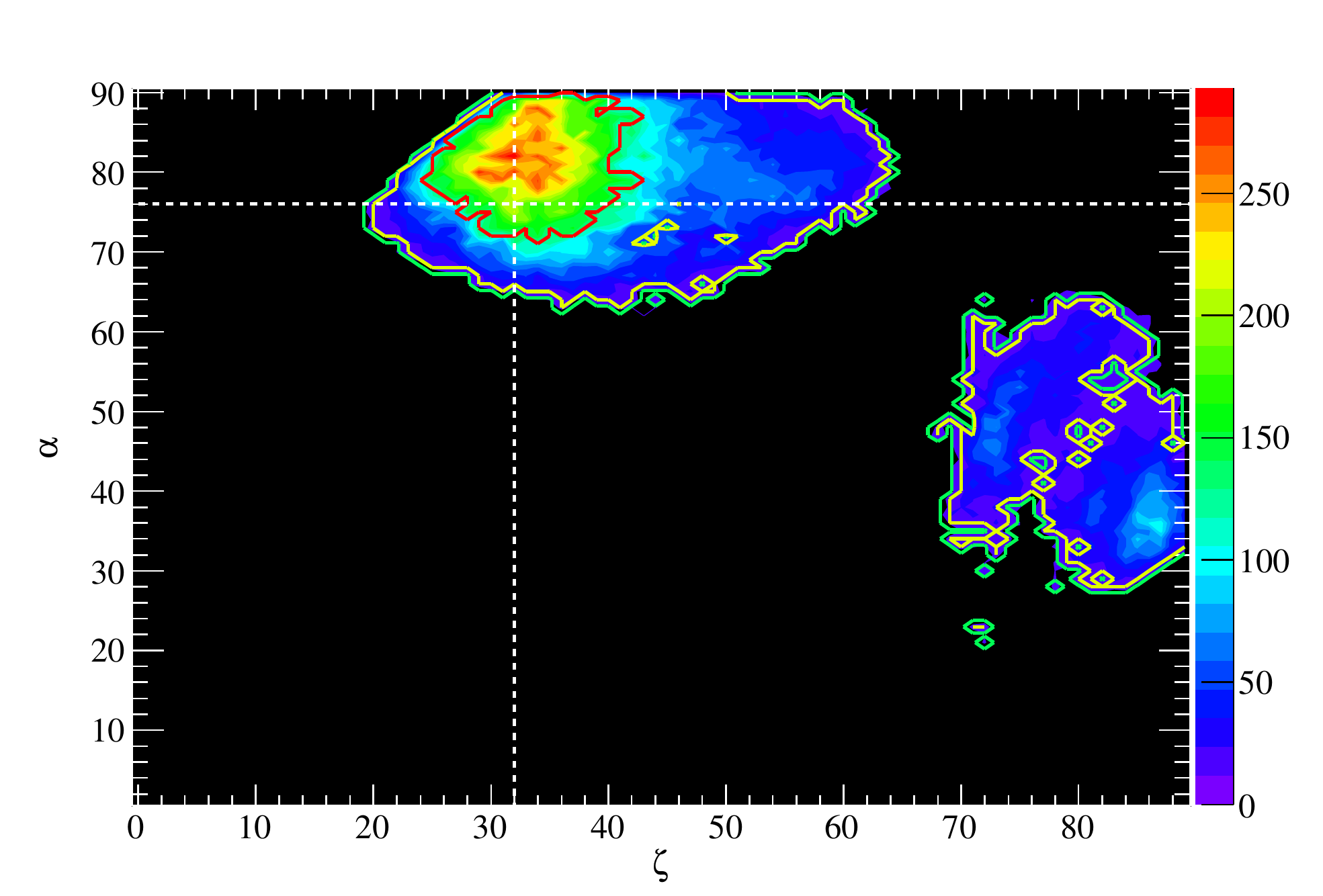}
\end{center}
\small\normalsize
\begin{quote}
\caption[Best-fit OG contours for PSR J0218+4232]{Marginalized confidence contours for PSR J0218+4232 for the OG model.\label{appAJ0218OGcont}}
\end{quote}
\end{figure}

The marginalized $\alpha$-$\zeta$ confidence contours corresponding to the OG fit are shown in Fig.~\ref{appAJ0218OGcont}, the best-fit geometry is indicated by the vertical and horizontal dashed, white lines.

\small\normalsize

\begin{figure}
\begin{center}
\includegraphics[width=0.75\textwidth]{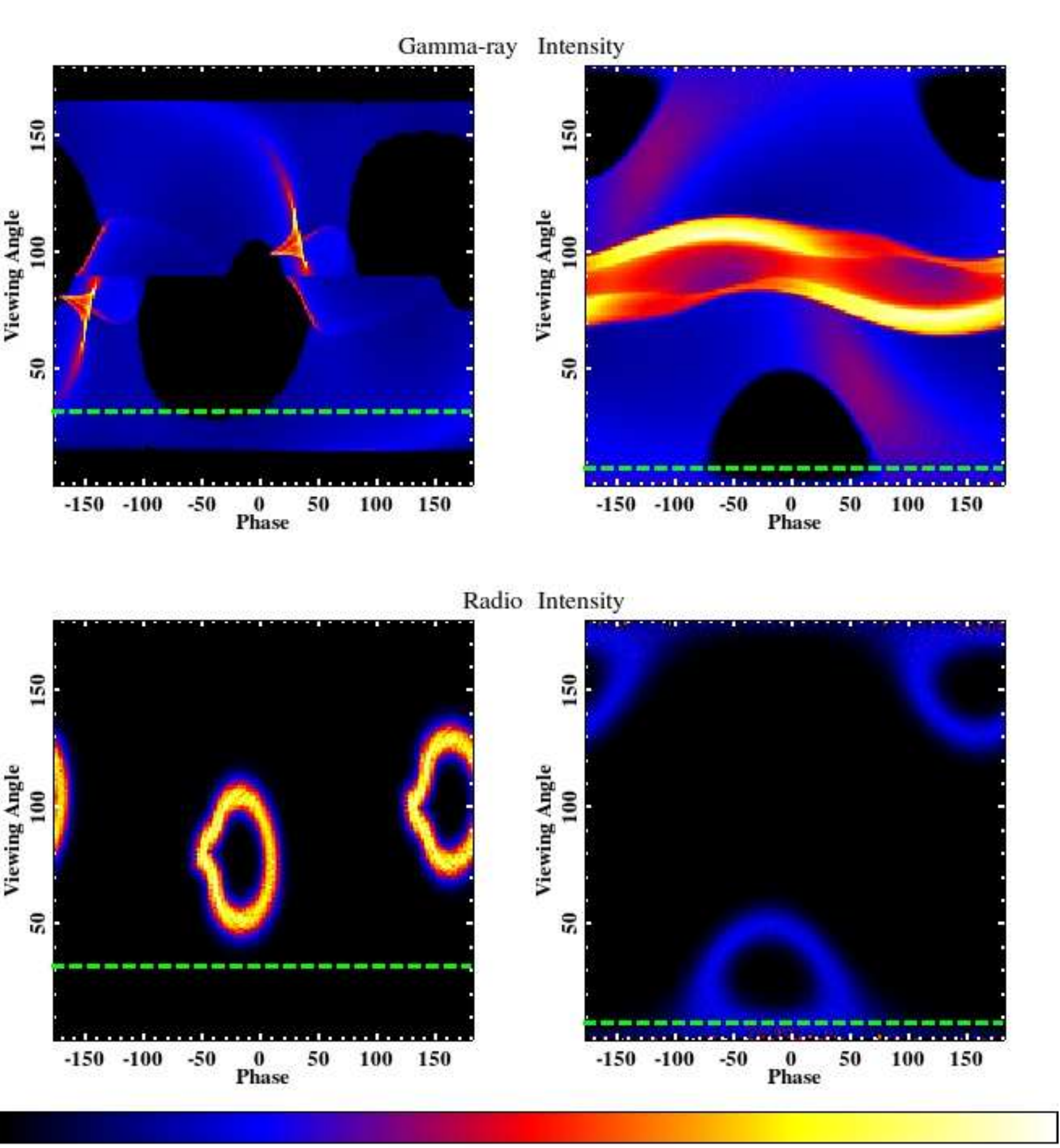}
\end{center}
\small\normalsize
\begin{quote}
\caption[Best-fit phase plots of simulated emission for PSR J0218+4232]{Distribution of simulated emission as a function of viewing angle and pulse phase for models used to fit PSR J0218+4232.  The best-fit $\zeta$ values are indicated by the dashed green lines.  The top plots correspond to the gamma-ray phase plots while the bottom are for the radio.  The left plots correspond to fits with the OG model with TPC plots on the right.  The color scale in the top-left plot is square root in order to bring out fainter features.\label{appAJ0218PhPlt}}
\end{quote}
\end{figure}

Plots of simulated emission corresponding to the best-fit models are shown in Fig.~\ref{appAJ0218PhPlt}, the OG models are on the left and TPC on the right, the gamma-ray models are on the top and radio on the bottom.

\small\normalsize

\section{PSR J0437$-$4715}\label{appAJ0437}
PSR J0437$-$4715 is a 5.7575 ms pulsar in a 5.7 d orbit with a $\sim$0.2 M$_{\odot}$ companion and was first discovered in the radio by \citet{Johnston93}.  Gamma-ray pulsations from this MSP were first reported by \citet{AbdoMSPpop} and later by \citet{AbdoPSRcat}.  The gamma-ray light curve of this MSP was modeled by \citet{Venter09} using geometric TPC and OG models with a hollow-cone beam radio model.

\begin{figure}
\begin{center}
\includegraphics[width=0.75\textwidth]{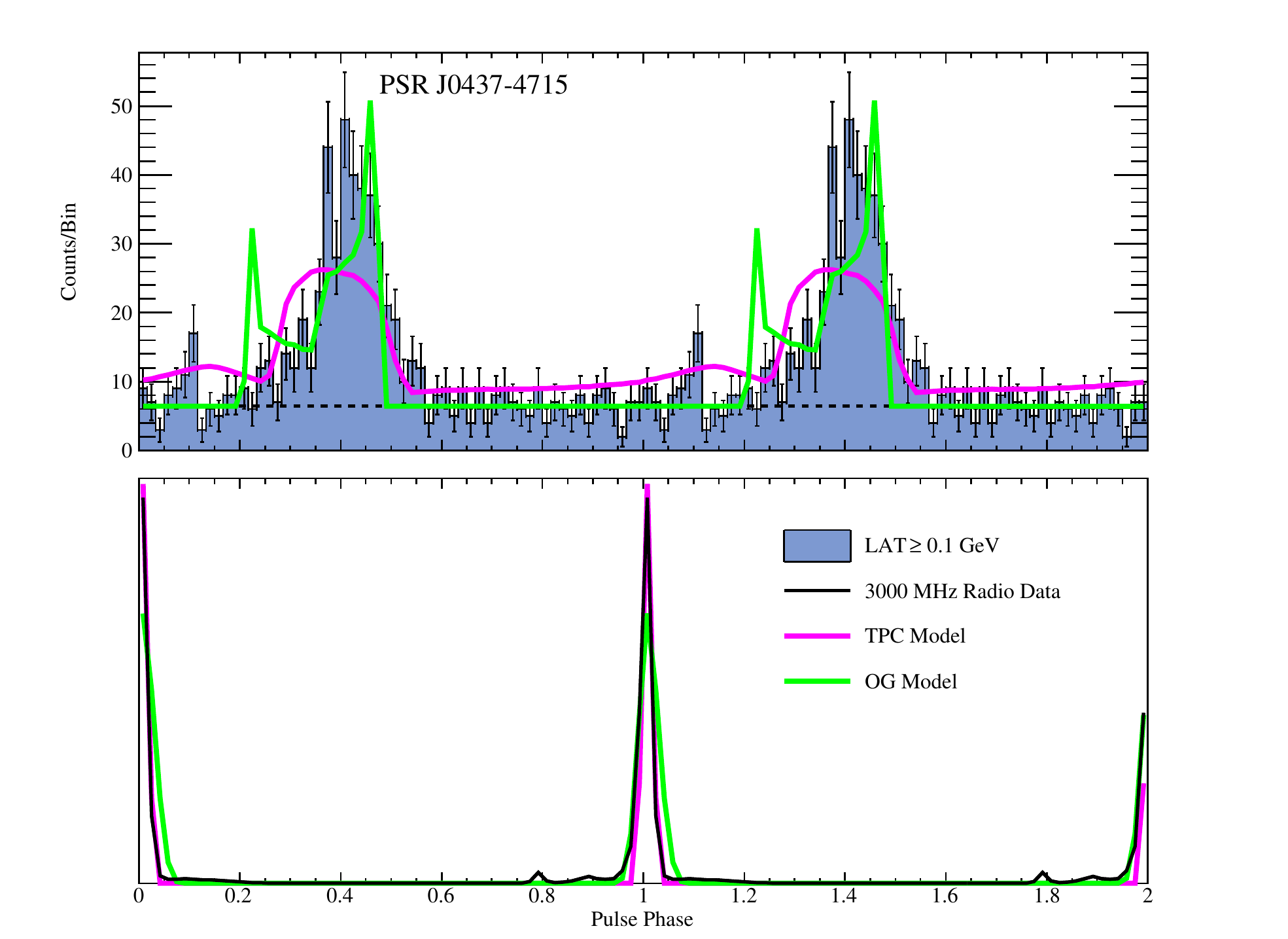}
\end{center}
\small\normalsize
\begin{quote}
\caption[Data and best-fit light curves for PSR J0437$-$4715]{Best-fit gamma-ray and radio light curves for PSR J0437$-$4715 using the TPC and OG models.\label{appAJ0437LCs}}
\end{quote}
\end{figure}

The best-fit gamma-ray and radio light curves are shown in Fig.~\ref{appAJ0437LCs}.  The gamma-ray light curve was fit with OG and TPC models.  The radio profile was fit with a hollow-cone beam model.  These light curve fits have used the 3000 MHz Parkes radio profile.

\small\normalsize

\begin{figure}
\begin{center}
\includegraphics[width=0.75\textwidth]{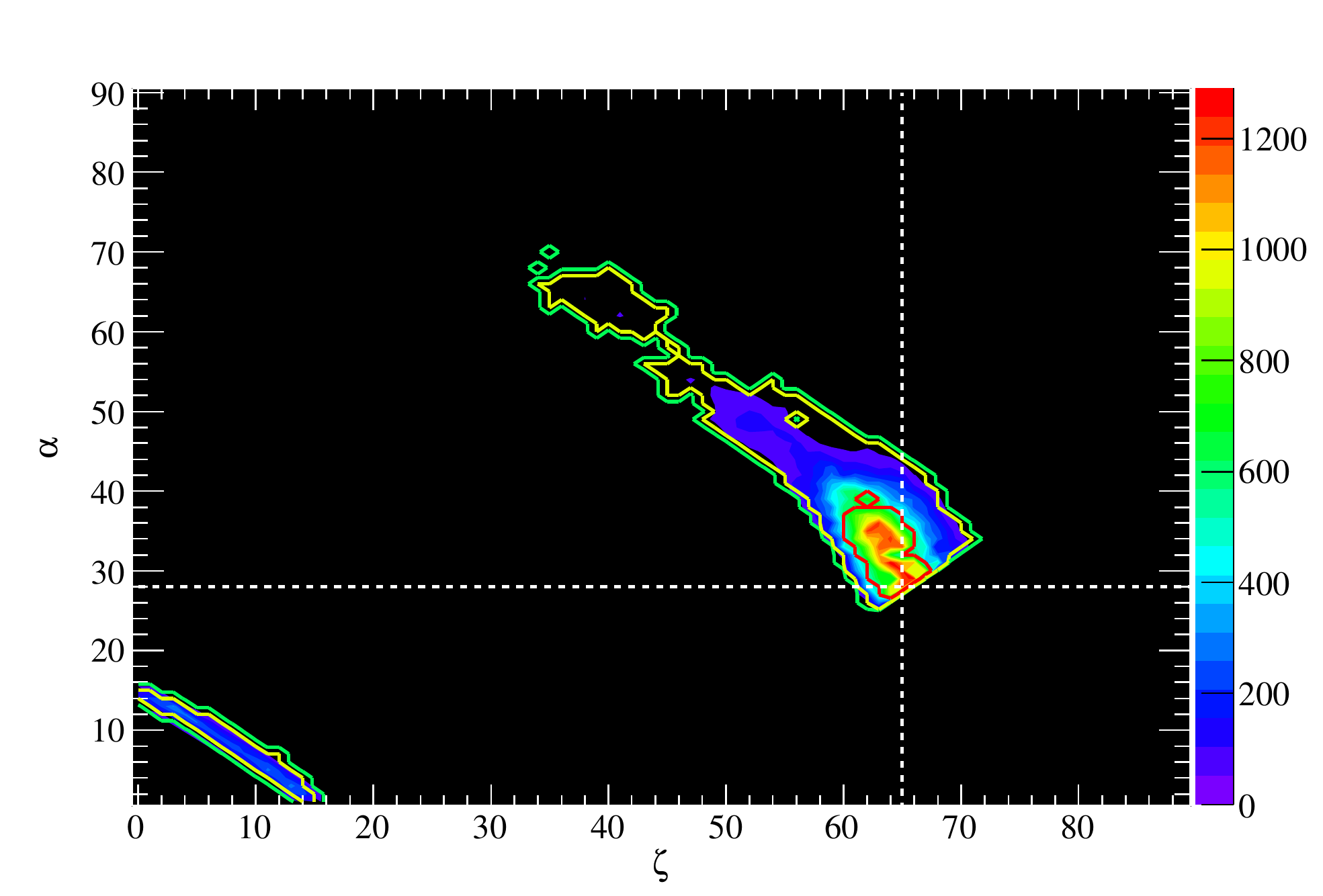}
\end{center}
\small\normalsize
\begin{quote}
\caption[Best-fit TPC contours for PSR J0437$-$4715]{Marginalized confidence contours for PSR J0437$-$4715 for the TPC model.\label{appAJ0437TPCcont}}
\end{quote}
\end{figure}

The marginalized $\alpha$-$\zeta$ confidence contours corresponding to the TPC fit are shown in Fig.~\ref{appAJ0437TPCcont}, the best-fit geometry is indicated by the vertical and horizontal dashed, white lines.

\small\normalsize

\begin{figure}
\begin{center}
\includegraphics[width=0.75\textwidth]{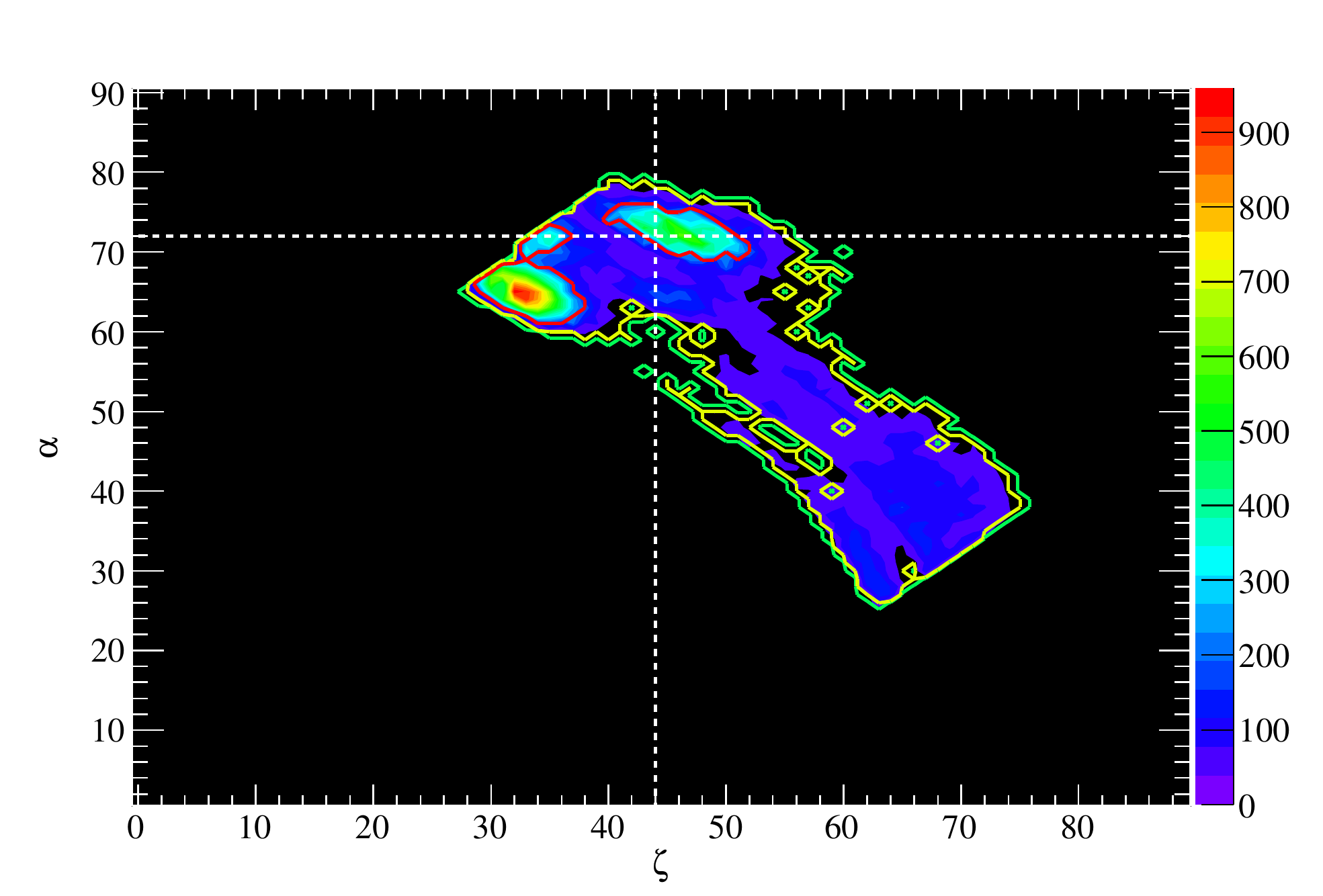}
\end{center}
\small\normalsize
\begin{quote}
\caption[Best-fit OG contours for PSR J0437$-$4715]{Marginalized confidence contours for PSR J0437$-$4715 for the OG model.\label{appAJ0437OGcont}}
\end{quote}
\end{figure}

The marginalized $\alpha$-$\zeta$ confidence contours corresponding to the OG fit are shown in Fig.~\ref{appAJ0437OGcont}, the best-fit geometry is indicated by the vertical and horizontal dashed, white lines.

\small\normalsize

\begin{figure}
\begin{center}
\includegraphics[width=0.75\textwidth]{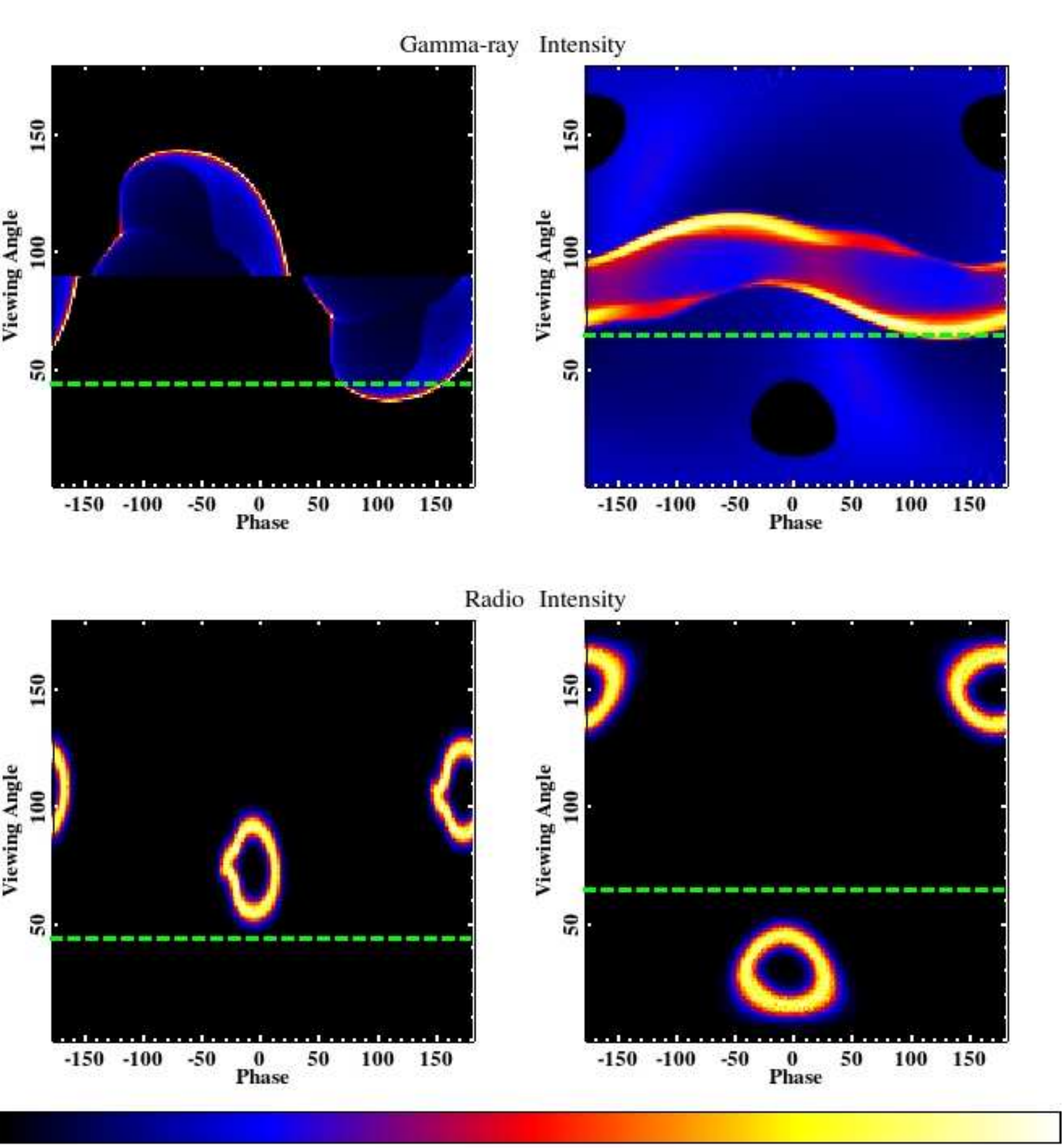}
\end{center}
\small\normalsize
\begin{quote}
\caption[Best-fit phase plots of simulated emission for PSR J0437$-$4715]{Distribution of simulated emission as a function of viewing angle and pulse phase for models used to fit PSR J0437$-$4715.  The best-fit $\zeta$ values are indicated by the dashed green lines.  The top plots correspond to the gamma-ray phase plots while the bottom are for the radio.  The left plots correspond to fits with the OG model with TPC plots on the right.\label{appAJ0437PhPlt}}
\end{quote}
\end{figure}

Plots of simulated emission corresponding to the best-fit models are shown in Fig.~\ref{appAJ0437PhPlt}, OG models are on the left and TPC on the right, gamma-ray models are on the top and radio on the bottom.

\small\normalsize

\section{PSR J0613$-$0200}\label{appAJ0613}
PSR J0613$-$0200 is a 3.0618 ms pulsar in a 1.2 d orbit with a low-mass companion ($\gtrsim$0.13 M$_{\odot}$) and was discovered in the radio by \citet{Lorimer95}.  Gamma-ray pulsations from this MSP were first reported by \citet{AbdoMSPpop} and later by \citet{AbdoPSRcat}.  The gamma-ray light curve was modeled by \citet{Venter09} using geometric OG and TPC models with a hollow-cone beam radio model.

\begin{figure}
\begin{center}
\includegraphics[width=0.75\textwidth]{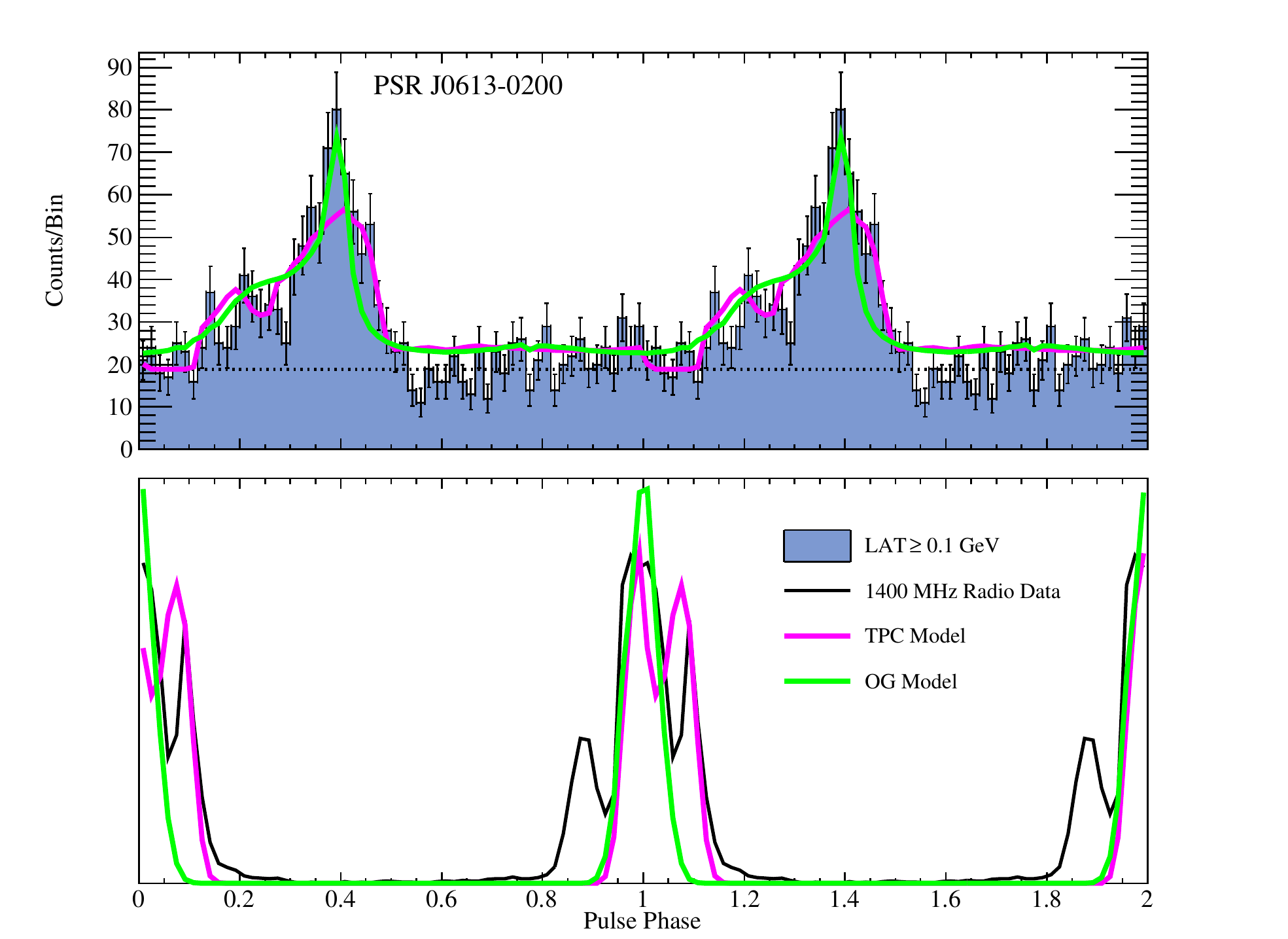}
\end{center}
\small\normalsize
\begin{quote}
\caption[Data and best-fit light curves for PSR J0613$-$0200]{Best-fit gamma-ray and radio light curves for PSR J0613$-$0200 using the TPC and OG models.\label{appAJ0613LCs}}
\end{quote}
\end{figure}

The best-fit gamma-ray and radio light curves are shown in Fig.~\ref{appAJ0613LCs}.  The gamma-ray light curve has been fit with the TPC and OG models.  The radio profile has been fit with a hollow-cone beam model.  These light curve fits have used the 1400 MHz Nan\c{cay} radio profile.

\small\normalsize

\begin{figure}
\begin{center}
\includegraphics[width=0.75\textwidth]{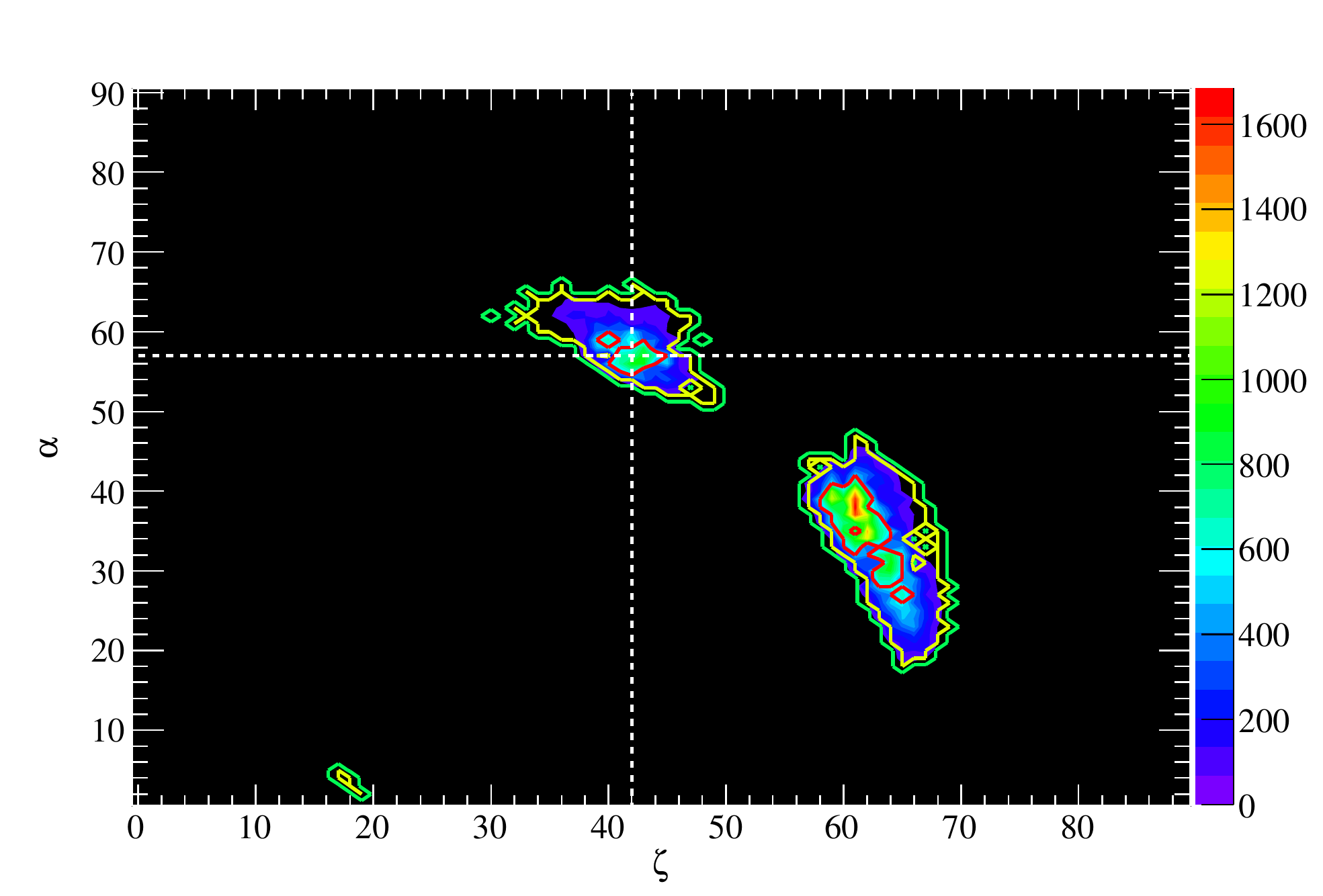}
\end{center}
\small\normalsize
\begin{quote}
\caption[Best-fit TPC contours for PSR J0613$-$0200]{Marginalized confidence contours for PSR J0613$-$0200 for the TPC model.\label{appAJ0613TPCcont}}
\end{quote}
\end{figure}

The marginalized $\alpha$-$\zeta$ confidence contours corresponding to the TPC fit are shown in Fig.~\ref{appAJ0613TPCcont}, the best-fit geometry is indicated by the vertical and horizontal dashed, white lines.

\small\normalsize

\begin{figure}
\begin{center}
\includegraphics[width=0.75\textwidth]{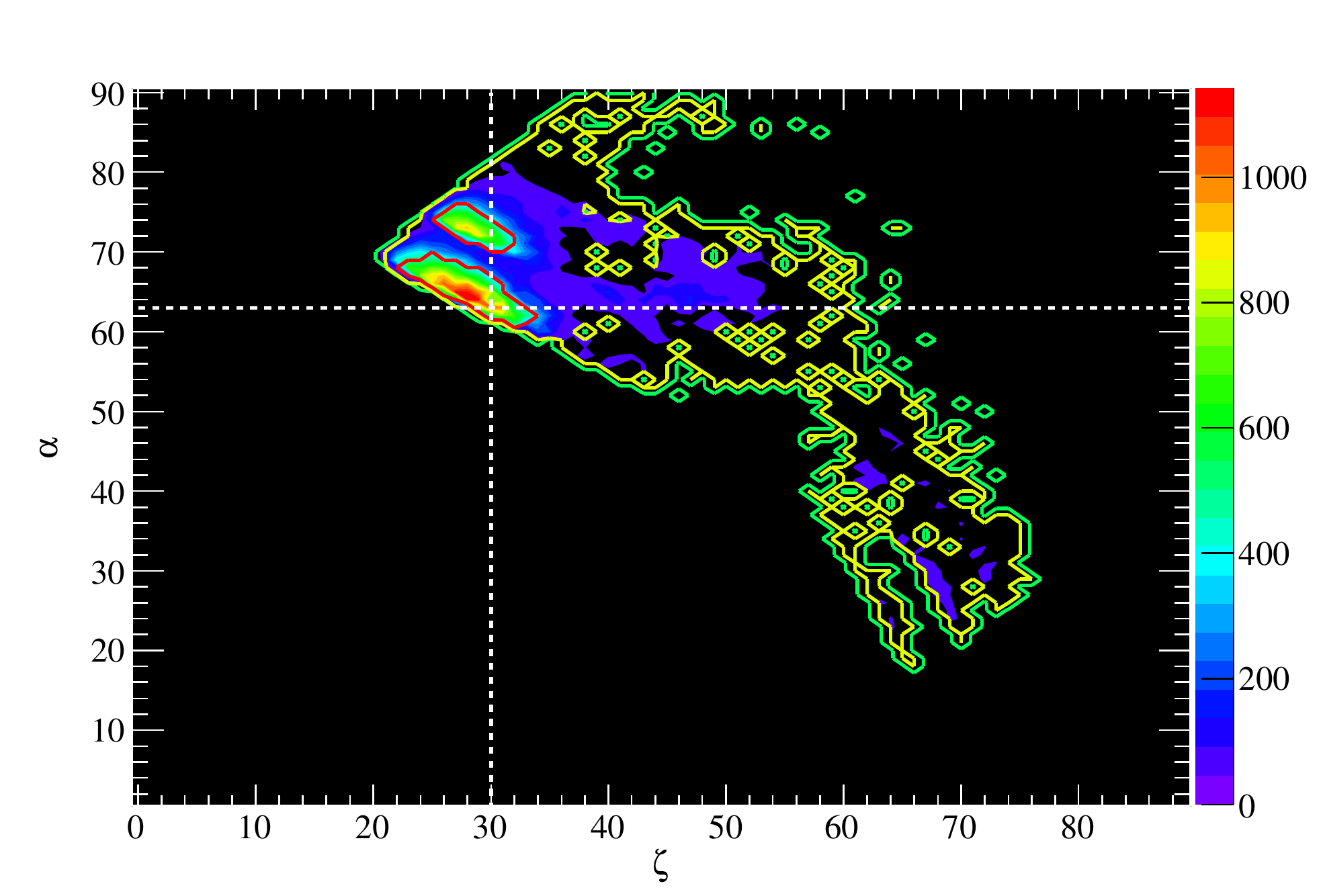}
\end{center}
\small\normalsize
\begin{quote}
\caption[Best-fit OG contours for PSR J0613$-$0200]{Marginalized confidence contours for PSR J0613$-$0200 for the OG model.\label{appAJ0613OGcont}}
\end{quote}
\end{figure}

The marginalized $\alpha$-$\zeta$ confidence contours corresponding to the OG fit are shown in Fig.~\ref{appAJ0613OGcont}, the best-fit geometry is indicated by the vertical and horizontal dashed, white lines.

\small\normalsize

\begin{figure}
\begin{center}
\includegraphics[width=0.75\textwidth]{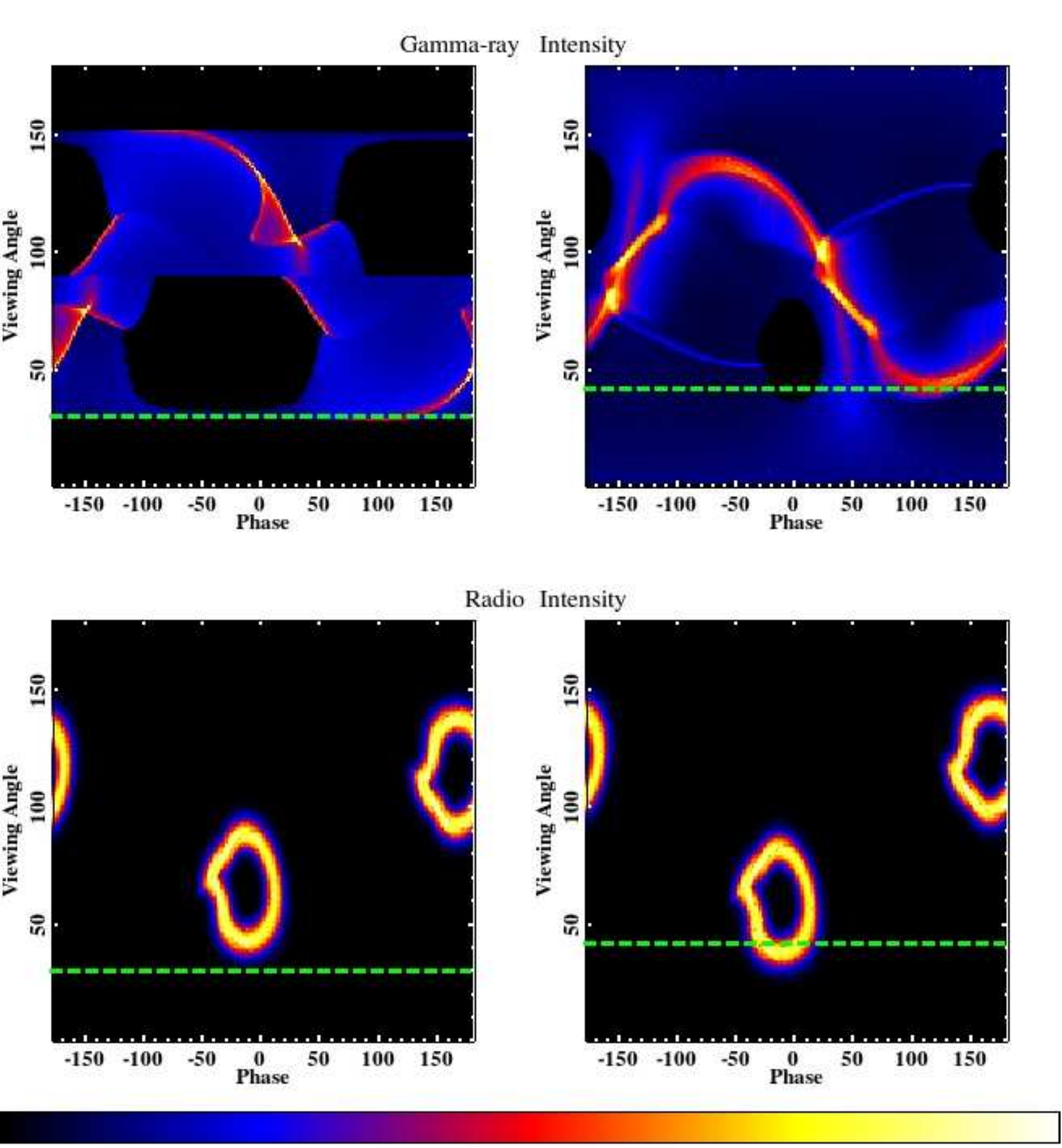}
\end{center}
\small\normalsize
\begin{quote}
\caption[Best-fit phase plots of simulated emission for PSR J0613$-$0200]{Distribution of simulated emission as a function of viewing angle and pulse phase for models used to fit PSR J0613$-$0200.  The best-fit $\zeta$ values are indicated by the dashed green lines.  The top plots correspond to the gamma-ray phase plots while the bottom are for the radio.  The left plots correspond to fits with the OG model with TPC plots on the right.  The color scale in the top-left plot is square root in order to bring out fainter features.\label{appAJ0613PhPlt}}
\end{quote}
\end{figure}

Plots of simulated emission corresponding to the best-fit models are shown in Fig.~\ref{appAJ0613PhPlt}, OG models are on the left and TPC on the right, gamma-ray models are on the top and radio on the bottom.

\small\normalsize

\section{PSR J0614$-$3329}\label{appAJ0614}
PSR J0614$-$3329 is a 3.1487 ms pulsar in a 53.6 d orbit with a low-mass companion ($\gtrsim$0.28 M$_{\odot}$).  This MSP was discovered in targeted radio observations of unassociated LAT sources with pulsar-like characteristics and seen to pulse in gamma rays soon after \citep{Ransom11}.

\begin{figure}
\begin{center}
\includegraphics[width=0.75\textwidth]{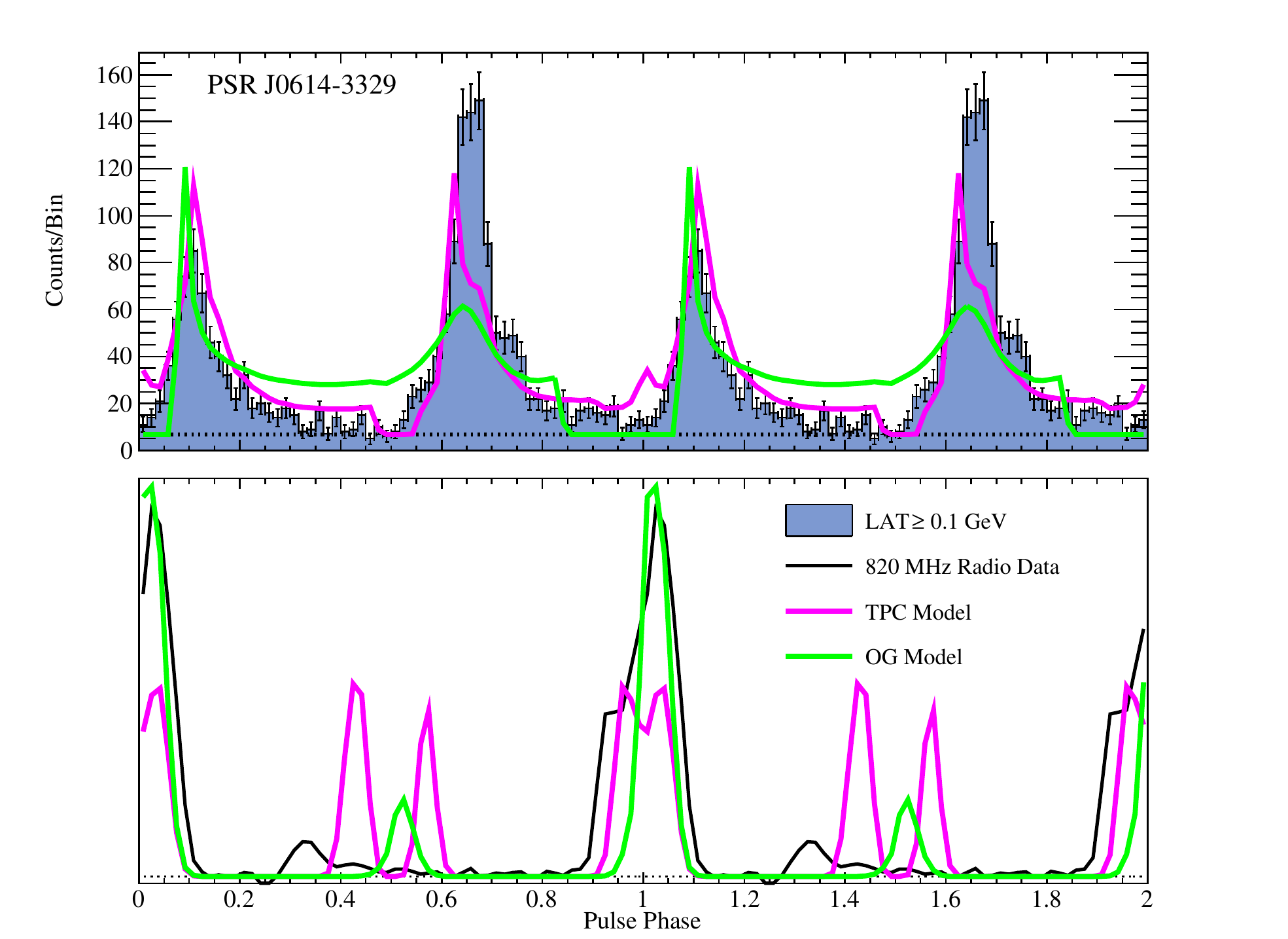}
\end{center}
\small\normalsize
\begin{quote}
\caption[Data and best-fit light curves for PSR J0614$-$3329]{Best-fit gamma-ray and radio light curves for PSR J0614$-$3329 using the TPC and OG models.\label{appAJ0614LCs}}
\end{quote}
\end{figure}

The best-fit gamma-ray and radio light curves are shown in Fig.~\ref{appAJ0614LCs}.  The gamma-ray light curve has been fit with TPC and OG models.  The radio profile has been fit with a hollow-cone beam model.  These light curve fits have used the 820 MHz Greenbank radio profile.

\begin{figure}
\begin{center}
\includegraphics[width=0.75\textwidth]{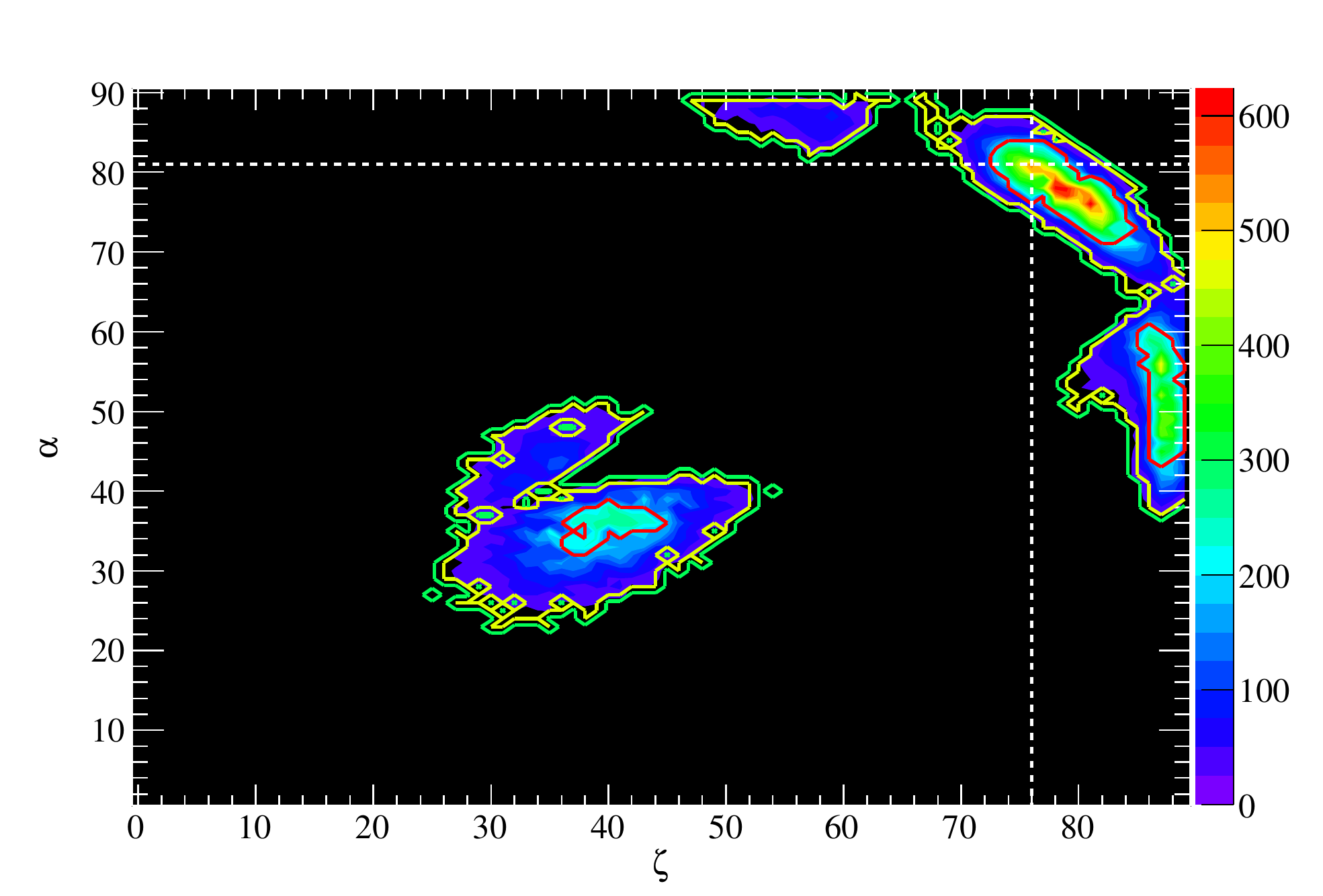}
\end{center}
\small\normalsize
\begin{quote}
\caption[Best-fit TPC contours for PSR J0614$-$3329]{Marginalized confidence contours for PSR J0614$-$3329 for the TPC model.\label{appAJ0614TPCcont}}
\end{quote}
\end{figure}

The marginalized $\alpha$-$\zeta$ confidence contours corresponding to the TPC fit are shown in Fig.~\ref{appAJ0614TPCcont}, the best-fit geometry is indicated by the vertical and horizontal dashed, white lines.

\begin{figure}
\begin{center}
\includegraphics[width=0.75\textwidth]{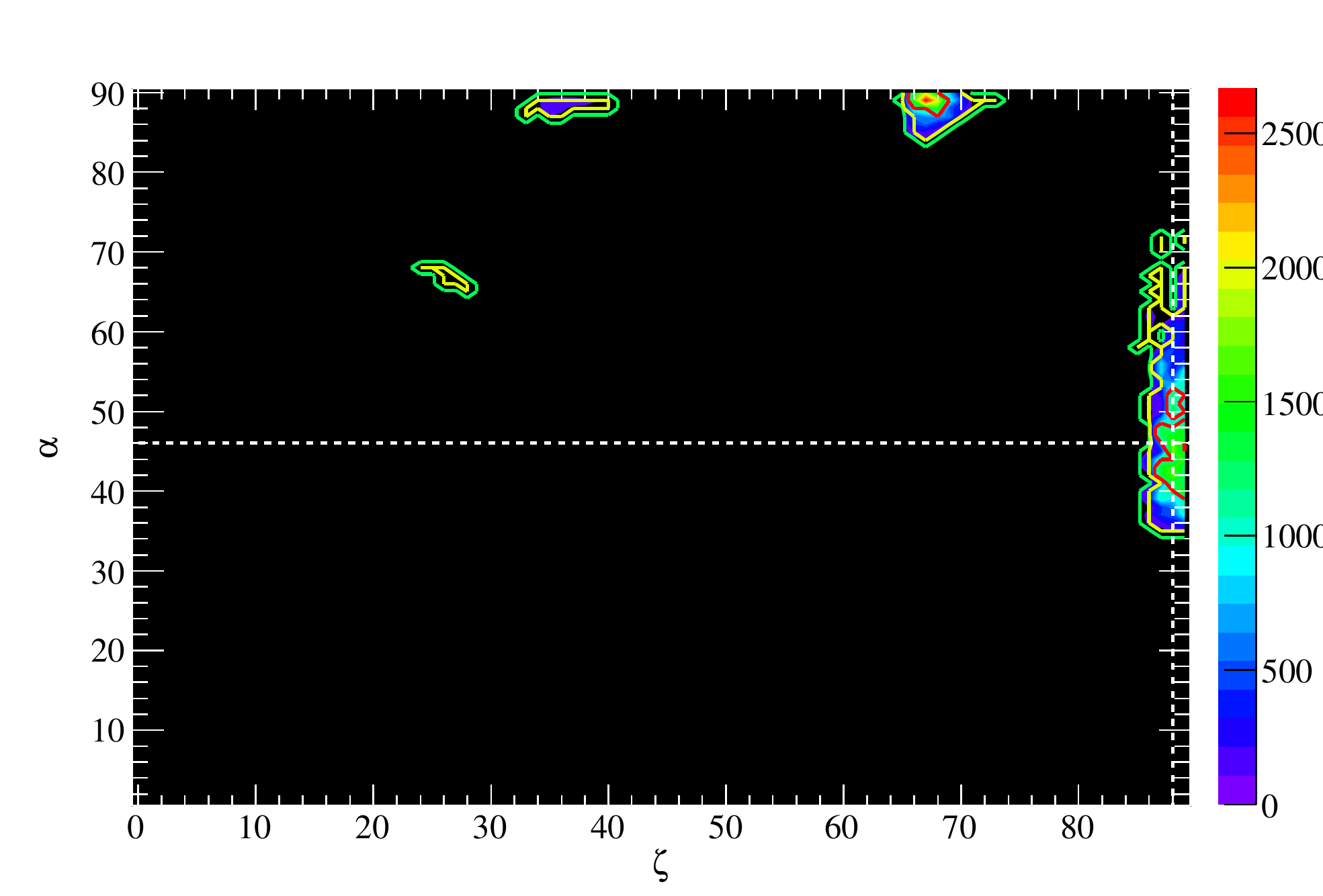}
\end{center}
\small\normalsize
\begin{quote}
\caption[Best-fit OG contours for PSR J0614$-$3329]{Marginalized confidence contours for PSR J0614$-$3329 for the OG model.\label{appAJ0614OGcont}}
\end{quote}
\end{figure}

The marginalized $\alpha$-$\zeta$ confidence contours corresponding to the OG fit are shown in Fig.~\ref{appAJ0614OGcont}, the best-fit geometry is indicated by the vertical and horizontal dashed, white lines.

\small\normalsize

\small\normalsize

\small\normalsize

\begin{figure}
\begin{center}
\includegraphics[width=0.75\textwidth]{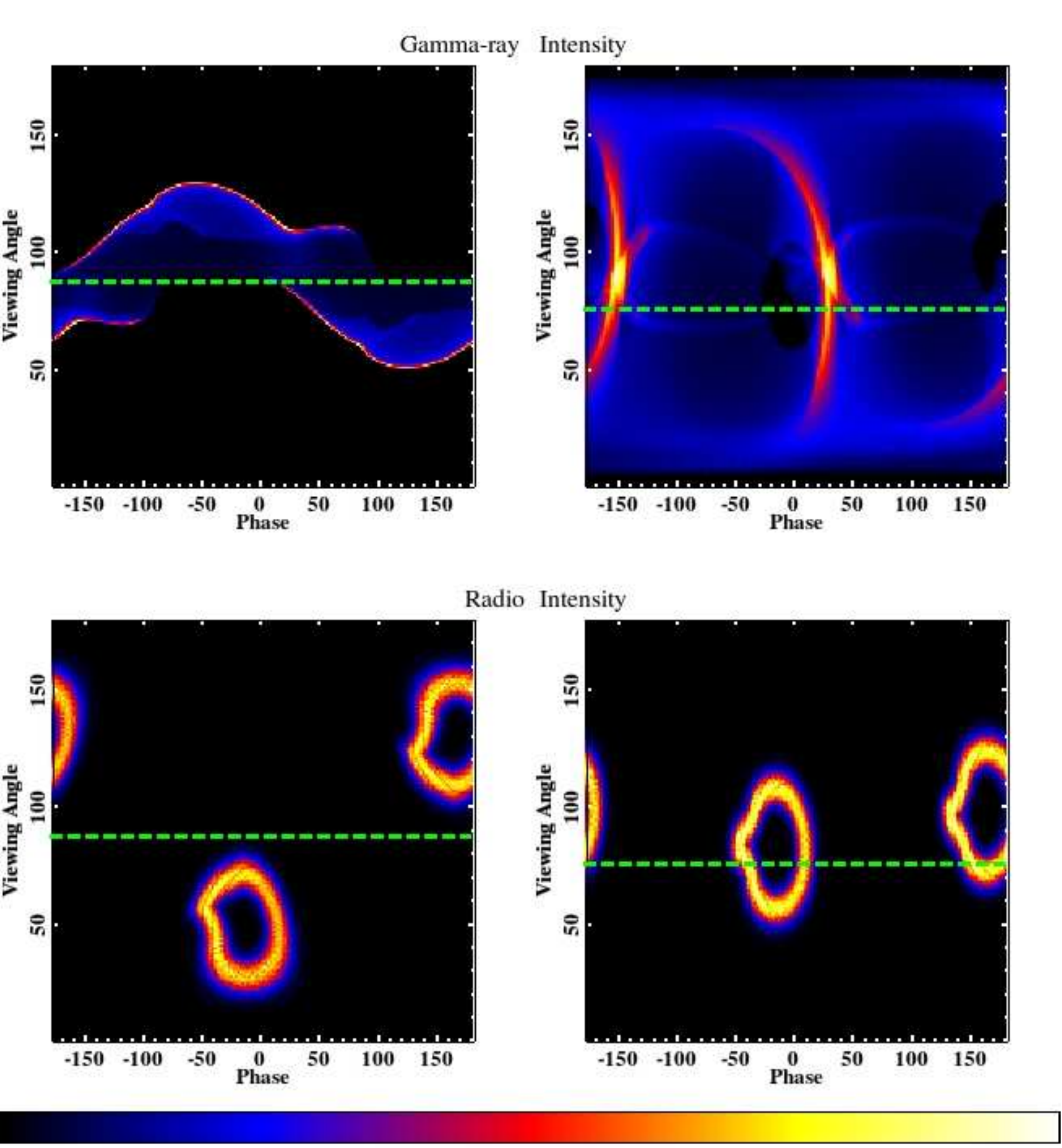}
\end{center}
\small\normalsize
\begin{quote}
\caption[Best-fit phase plots of simulated emission for PSR J0614$-$3329]{Distribution of simulated emission as a function of viewing angle and pulse phase for models used to fit PSR J0614$-$3329.  The best-fit $\zeta$ values are indicated by the dashed green lines.  The top plots correspond to the gamma-ray phase plots while the bottom are for the radio.  The left plots correspond to fits with the OG model with TPC plots on the right.\label{appAJ0614PhPlt}}
\end{quote}
\end{figure}

Plots of simulated emission corresponding to the best-fit models are shown in Fig.~\ref{appAJ0614PhPlt}, OG models are on the left and TPC on the right, gamma-ray models are on the top and radio on the bottom.

\small\normalsize

\section{PSR J0751+1807}\label{appAJ0751}
PSR J0751+1807 is a 3.4788 ms pulsar in a 6.3 hr orbit with a companion of mass between 0.12 and 0.6 M$_{\odot}$ and was first discovered in the radio by \citet{Lundgren95}.  Gamma-ray pulsations from this MSP were first reported by \citet{AbdoMSPpop} and later by \citet{AbdoPSRcat}.  The gamma-ray light curve of this MSP was modeled by \citet{Venter09} using geometric TPC and OG models with a hollow-cone beam radio model.

\begin{figure}
\begin{center}
\includegraphics[width=0.75\textwidth]{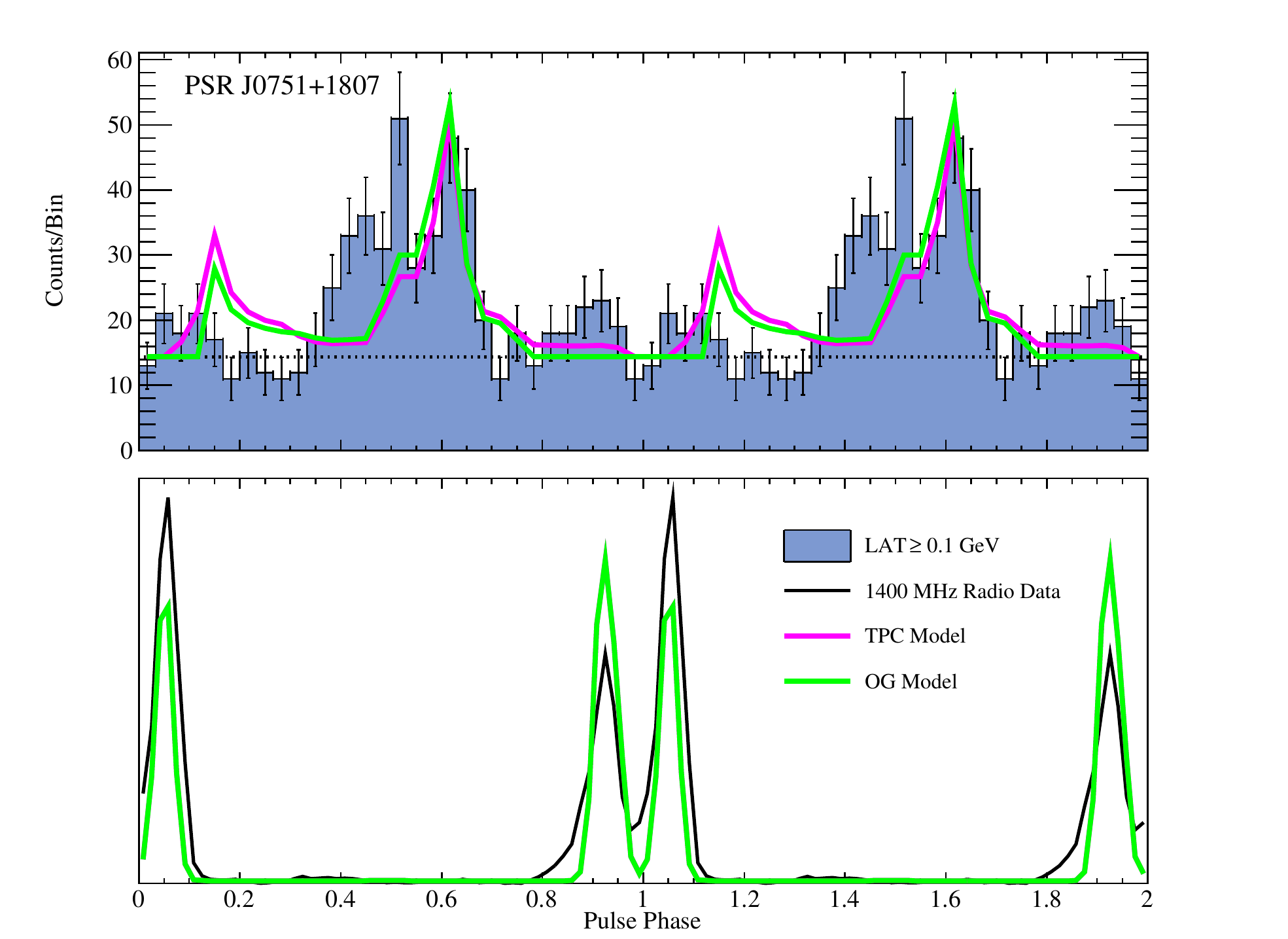}
\end{center}
\small\normalsize
\begin{quote}
\caption[Data and best-fit light curves for PSR J0751+1807]{Best-fit gamma-ray and radio light curves for PSR J0751+1807 using the TPC and OG models.\label{appAJ0751LCs}}
\end{quote}
\end{figure}

The best-fit gamma-ray and radio light curves are shown in Fig.~\ref{appAJ0751LCs}.  The gamma-ray light curve has been fit with TPC and OG models.  The radio profile has been fit with a hollow-cone beam model.  These light curve fits have used the 1400 MHz Nan\c{cay} radio profile.

\begin{figure}
\begin{center}
\includegraphics[width=0.75\textwidth]{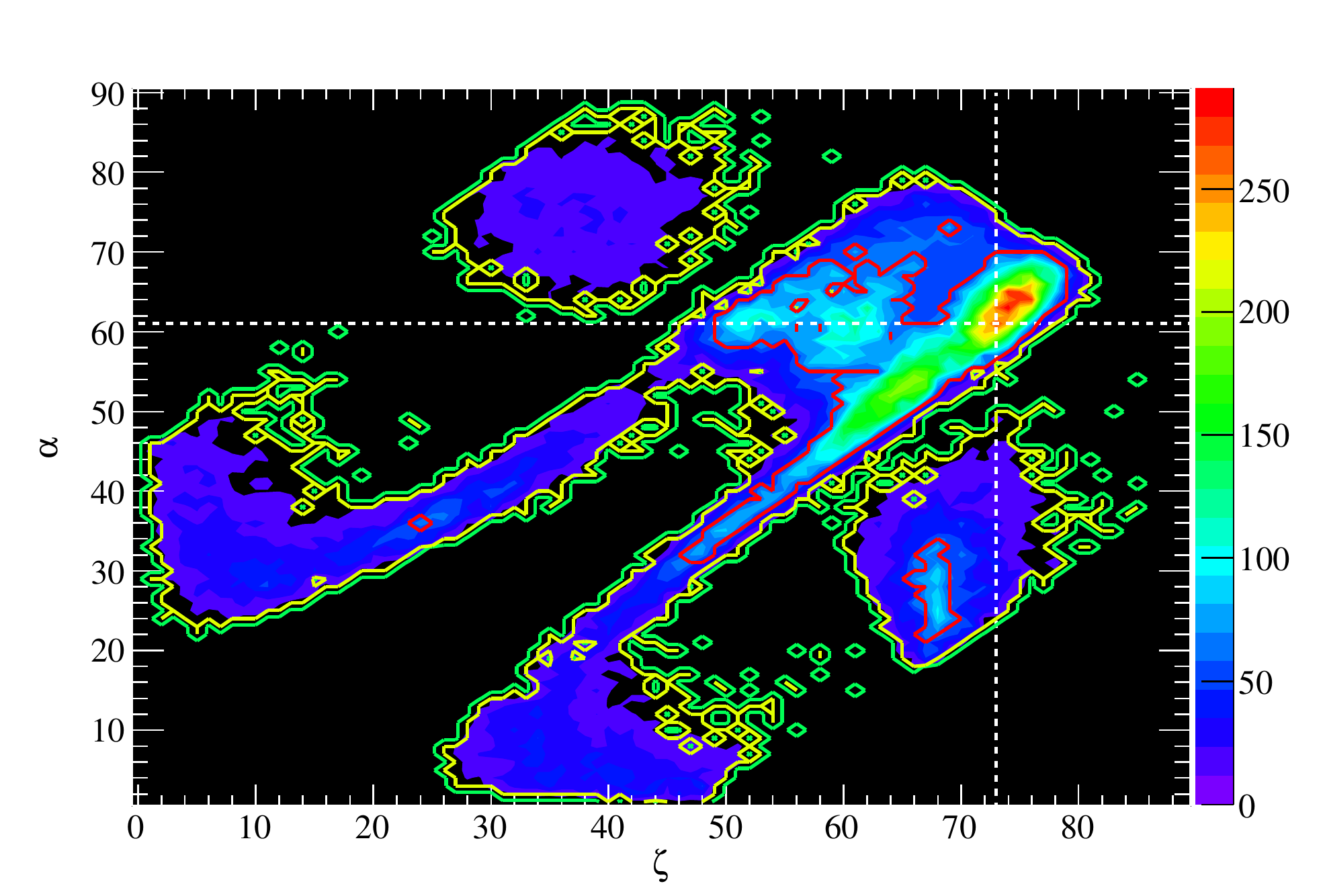}
\end{center}
\small\normalsize
\begin{quote}
\caption[Best-fit TPC contours for PSR J0751+1807]{Marginalized confidence contours for PSR J0751+1807 for the TPC model.\label{appAJ0751TPCcont}}
\end{quote}
\end{figure}

The marginalized $\alpha$-$\zeta$ confidence contours corresponding to the TPC fit are shown in Fig.~\ref{appAJ0751TPCcont}, the best-fit geometry is indicated by the vertical and horizontal dashed, white lines.

\begin{figure}
\begin{center}
\includegraphics[width=0.75\textwidth]{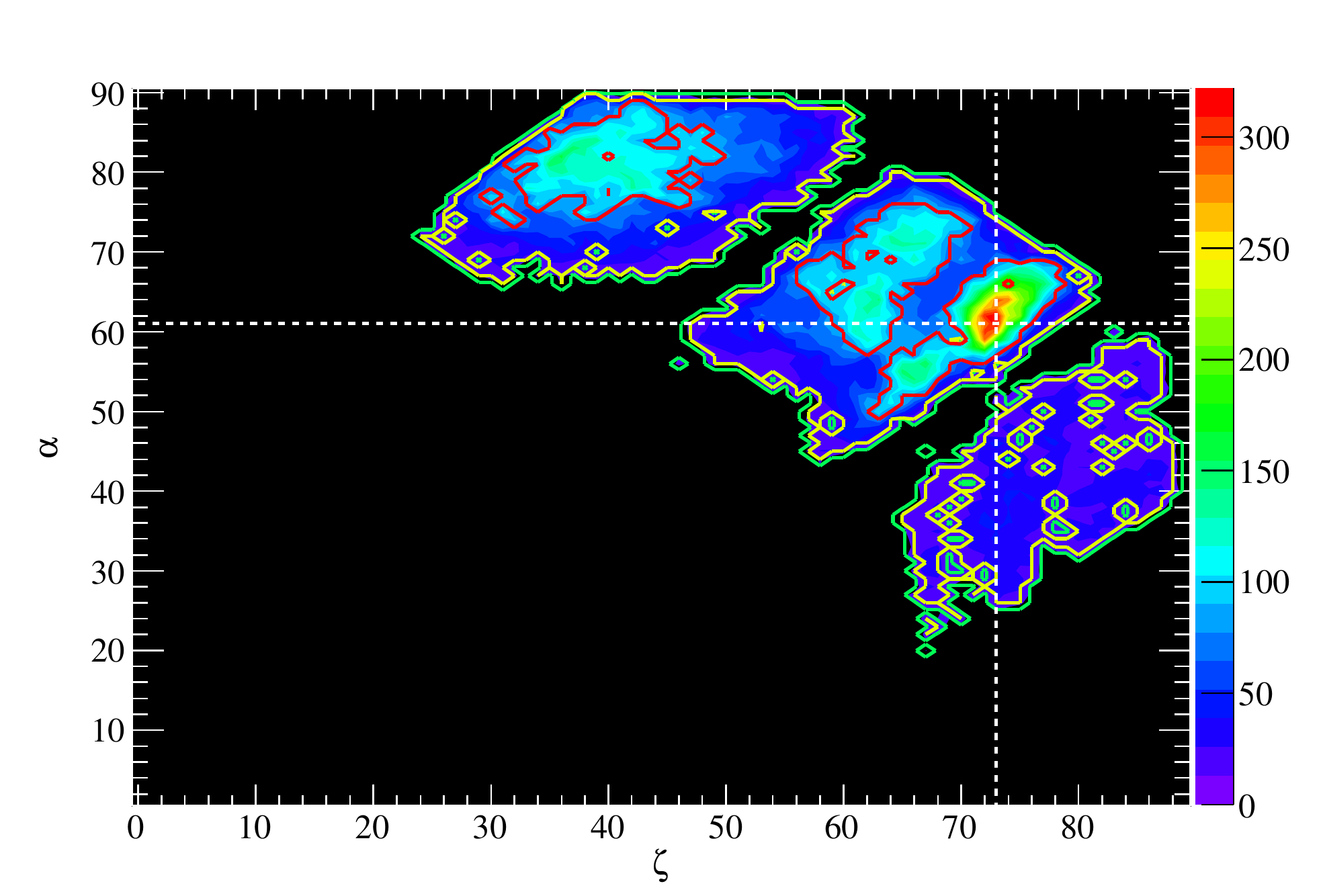}
\end{center}
\small\normalsize
\begin{quote}
\caption[Best-fit OG contours for PSR J0751+1807]{Marginalized confidence contours for PSR J0751+1807 for the OG model.\label{appAJ0751OGcont}}
\end{quote}
\end{figure}

The marginalized $\alpha$-$\zeta$ confidence contours corresponding to the OG fit are shown in Fig.~\ref{appAJ0751OGcont}, the best-fit geometry is indicated by the vertical and horizontal dashed, white lines.

\small\normalsize

\small\normalsize

\small\normalsize

\begin{figure}
\begin{center}
\includegraphics[width=0.75\textwidth]{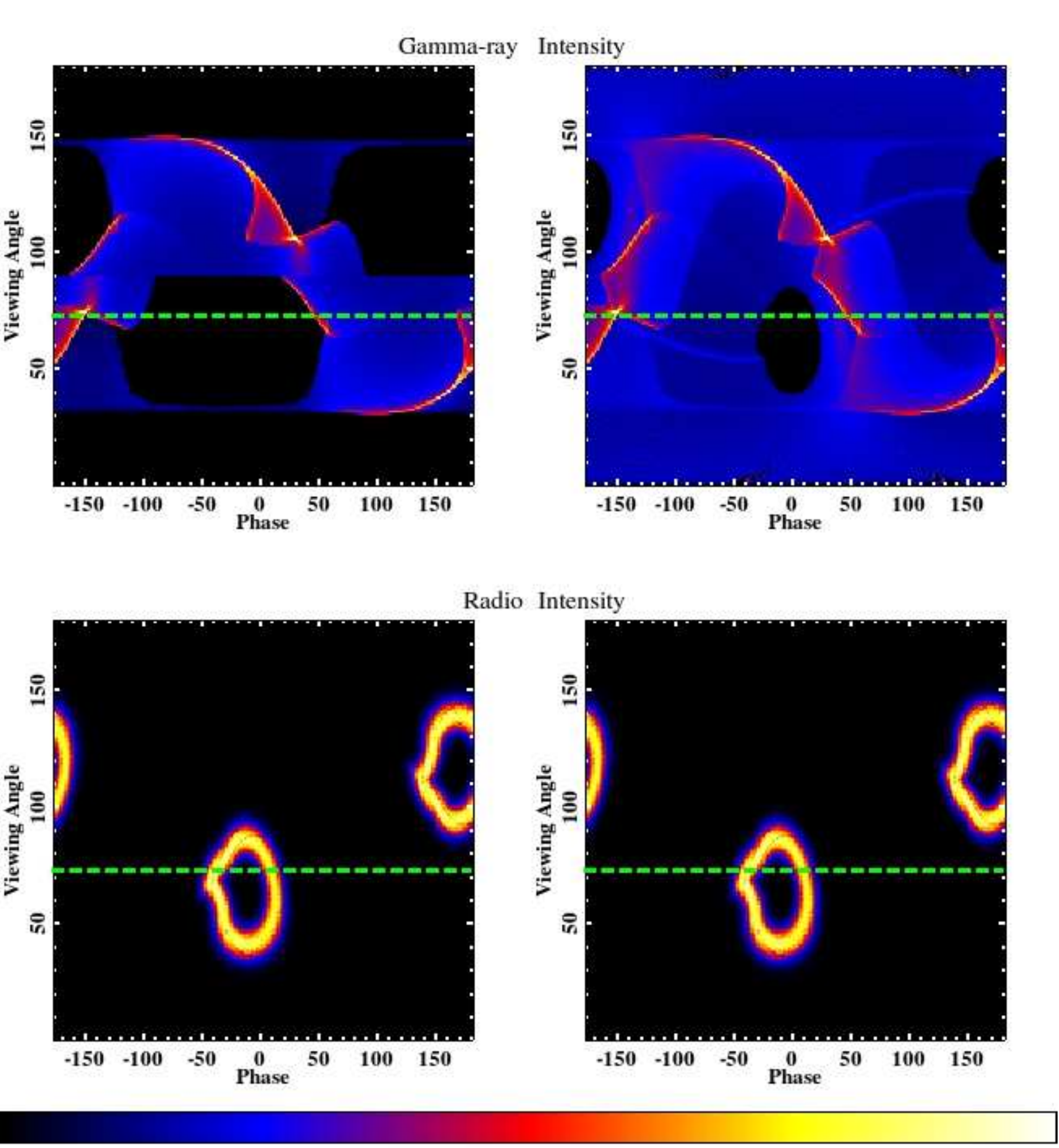}
\end{center}
\small\normalsize
\begin{quote}
\caption[Best-fit phase plots of simulated emission for PSR J0751+1807]{Distribution of simulated emission as a function of viewing angle and pulse phase for models used to fit PSR J0751+1807.  The best-fit $\zeta$ values are indicated by the dashed green lines.  The top plots correspond to the gamma-ray phase plots while the bottom are for the radio.  The left plots correspond to fits with the OG model with TPC plots on the right.  The color scales in the top plots are square root in order to bring out fainter features.\label{appAJ0751PhPlt}}
\end{quote}
\end{figure}

Plots of simulated emission corresponding to the best-fit models are shown in Fig.~\ref{appAJ0751PhPlt}, OG models are on the left and TPC on the right, gamma-ray models are on the top and radio on the bottom.

\small\normalsize

\section{PSR J1231$-$1411}\label{appAJ1231}
PSR J1231$-$1411 is a 3.6839 ms pulsar in a 1.9 d orbit with a low-mass companion ($\gtrsim$0.19 M$_{\odot}$).  This MSP was discovered in targeted radio observations of unassociated LAT sources with pulsar-like characteristics and seen to pulse in gamma rays soon after \citep{Ransom11}.

\begin{figure}
\begin{center}
\includegraphics[width=0.75\textwidth]{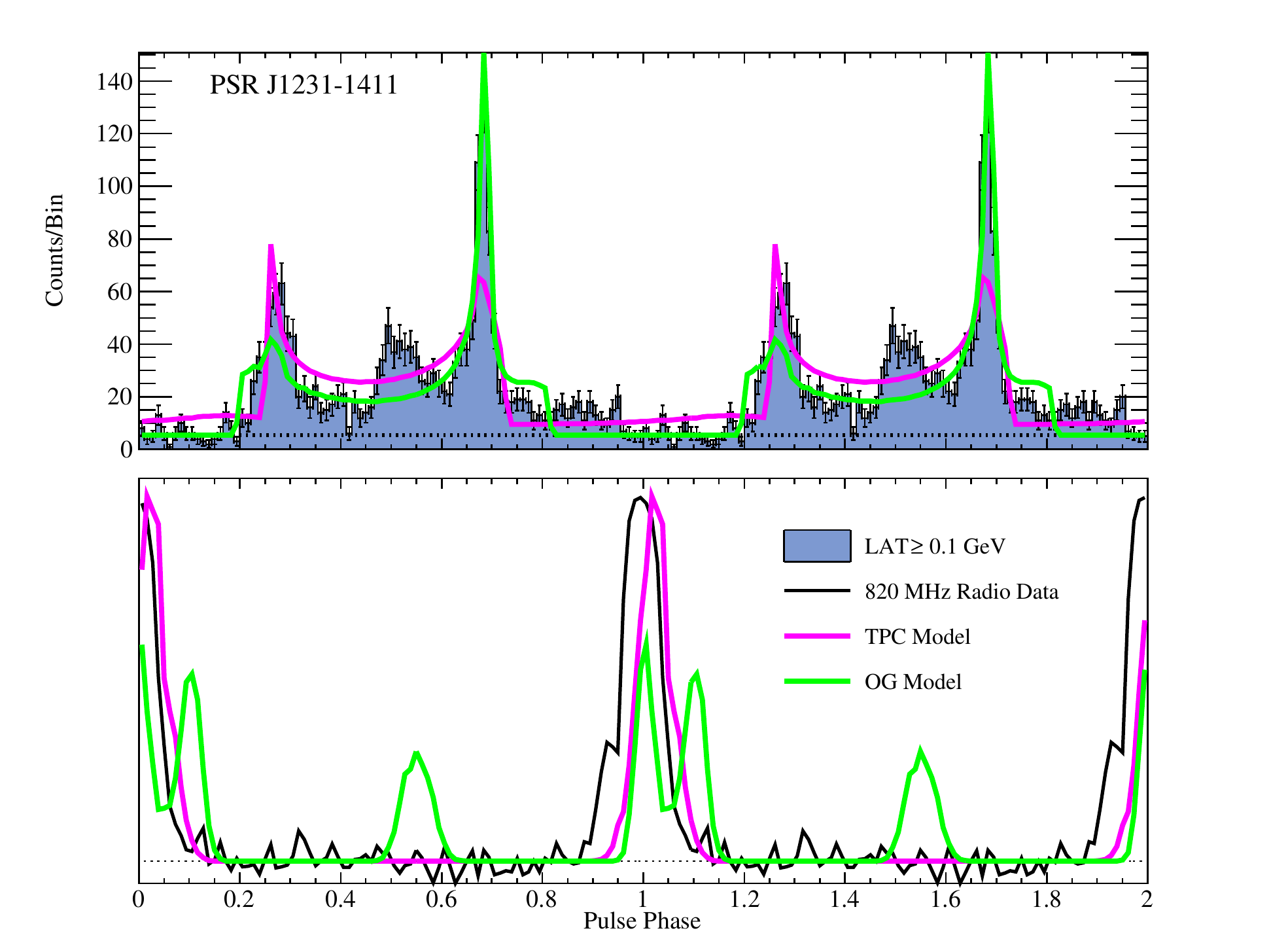}
\end{center}
\small\normalsize
\begin{quote}
\caption[Data and best-fit light curves for PSR J1231$-$1411]{Best-fit gamma-ray and radio light curves for PSR J1231-1411 using the TPC and OG models.\label{appAJ1231LCs}}
\end{quote}
\end{figure}

The best-fit gamma-ray and radio light curves are shown in Fig.~\ref{appAJ1231LCs}.  The gamma-ray light curve has been fit with TPC and OG models.  The radio profile has been fit with a hollow-cone beam model.  These light curve fits have used the 820 MHz Greenbank radio profile.

\small\normalsize

\begin{figure}
\begin{center}
\includegraphics[width=0.75\textwidth]{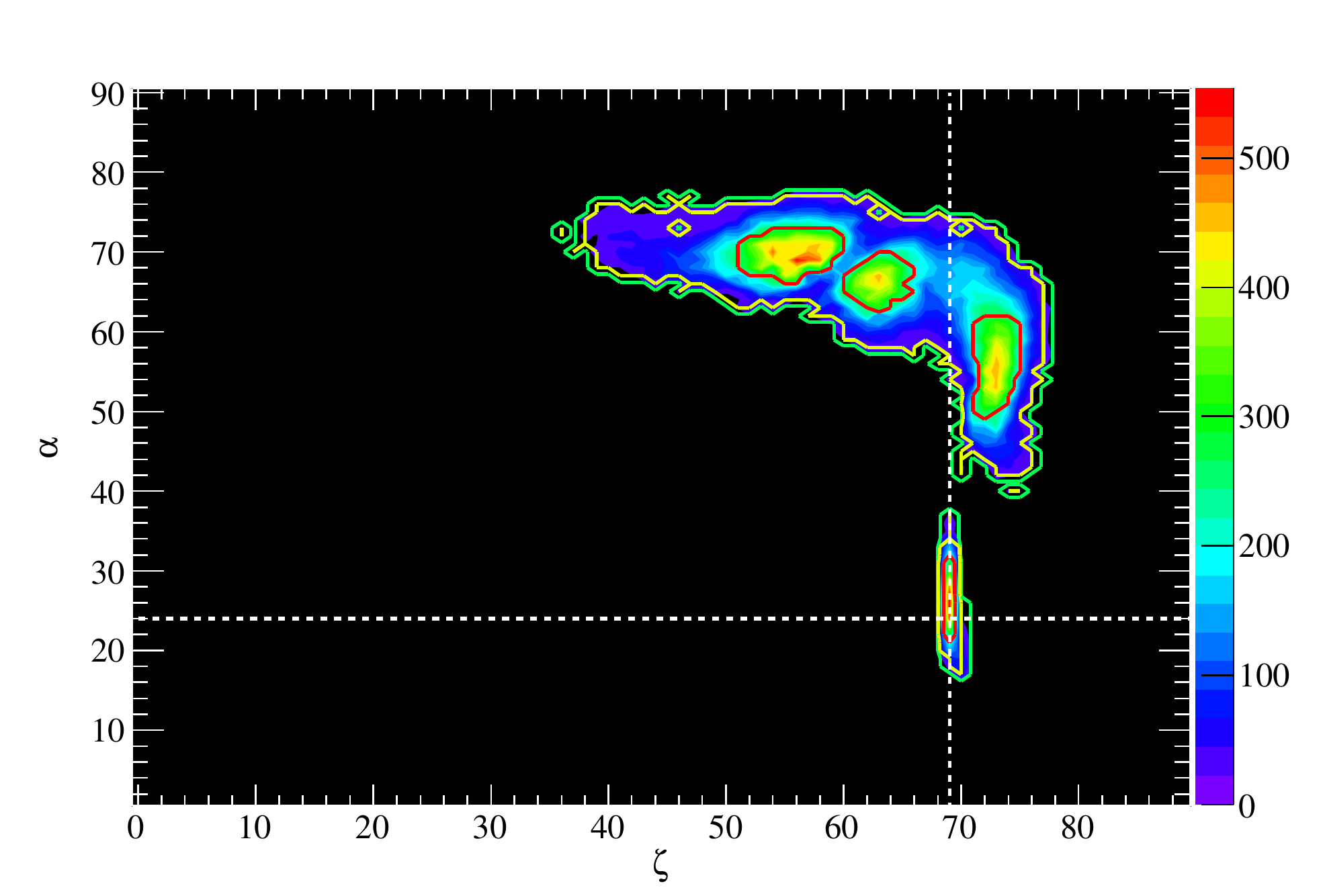}
\end{center}
\small\normalsize
\begin{quote}
\caption[Best-fit TPC contours for PSR J1231$-$1411]{Marginalized confidence contours for PSR J1231$-$1411 for the TPC model.\label{appAJ1231TPCcont}}
\end{quote}
\end{figure}

The marginalized $\alpha$-$\zeta$ confidence contours corresponding to the TPC fit are shown in Fig.~\ref{appAJ1231TPCcont}, the best-fit geometry is indicated by the vertical and horizontal dashed, white lines.

\begin{figure}
\begin{center}
\includegraphics[width=0.75\textwidth]{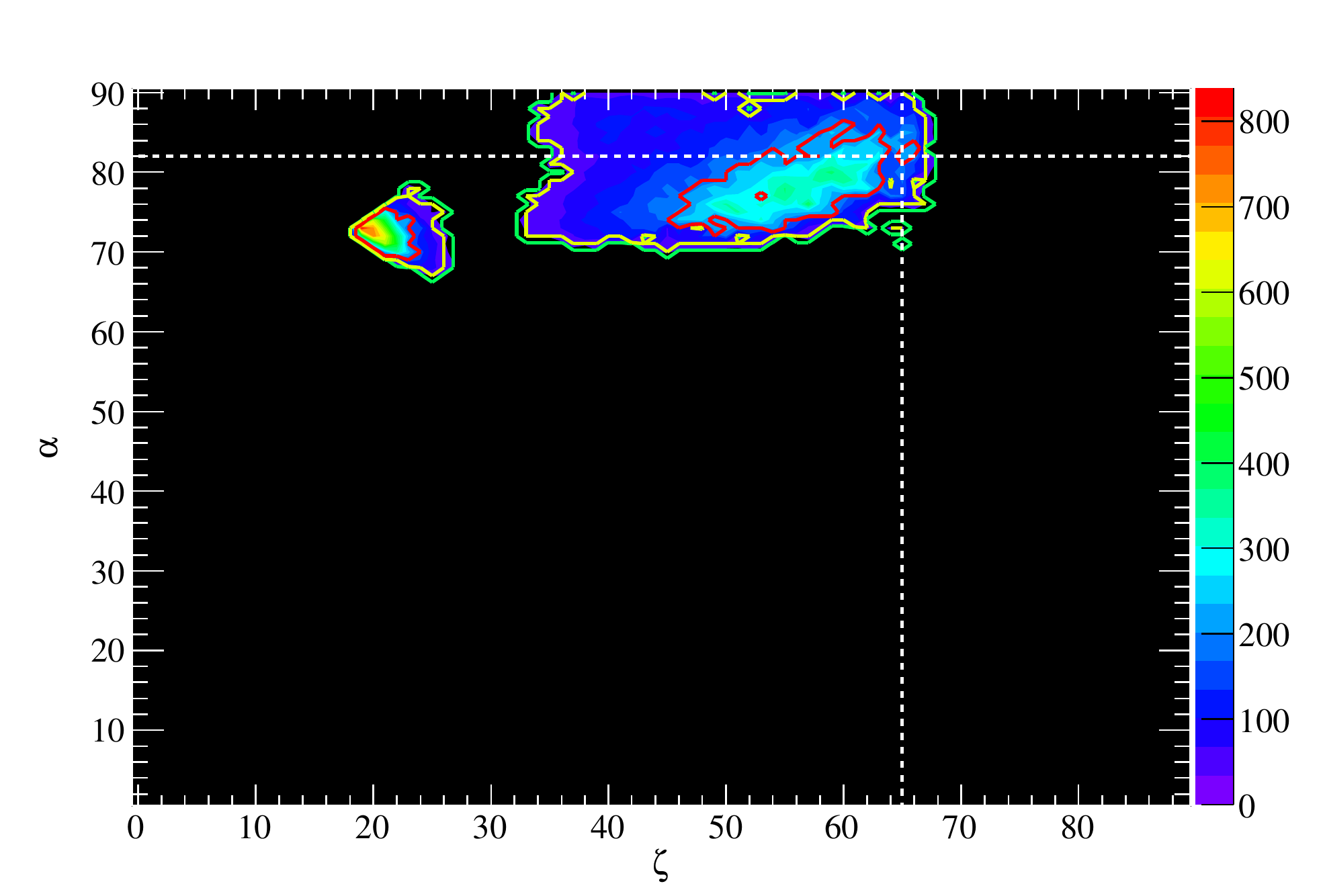}
\end{center}
\small\normalsize
\begin{quote}
\caption[Best-fit OG contours for PSR J1231$-$1411]{Marginalized confidence contours for PSR J1231$-$1411 for the OG model.\label{appAJ1231OGcont}}
\end{quote}
\end{figure}

The marginalized $\alpha$-$\zeta$ confidence contours corresponding to the OG fit are shown in Fig.~\ref{appAJ1231OGcont}, the best-fit geometry is indicated by the vertical and horizontal dashed, white lines.

\begin{figure}
\begin{center}
\includegraphics[width=0.75\textwidth]{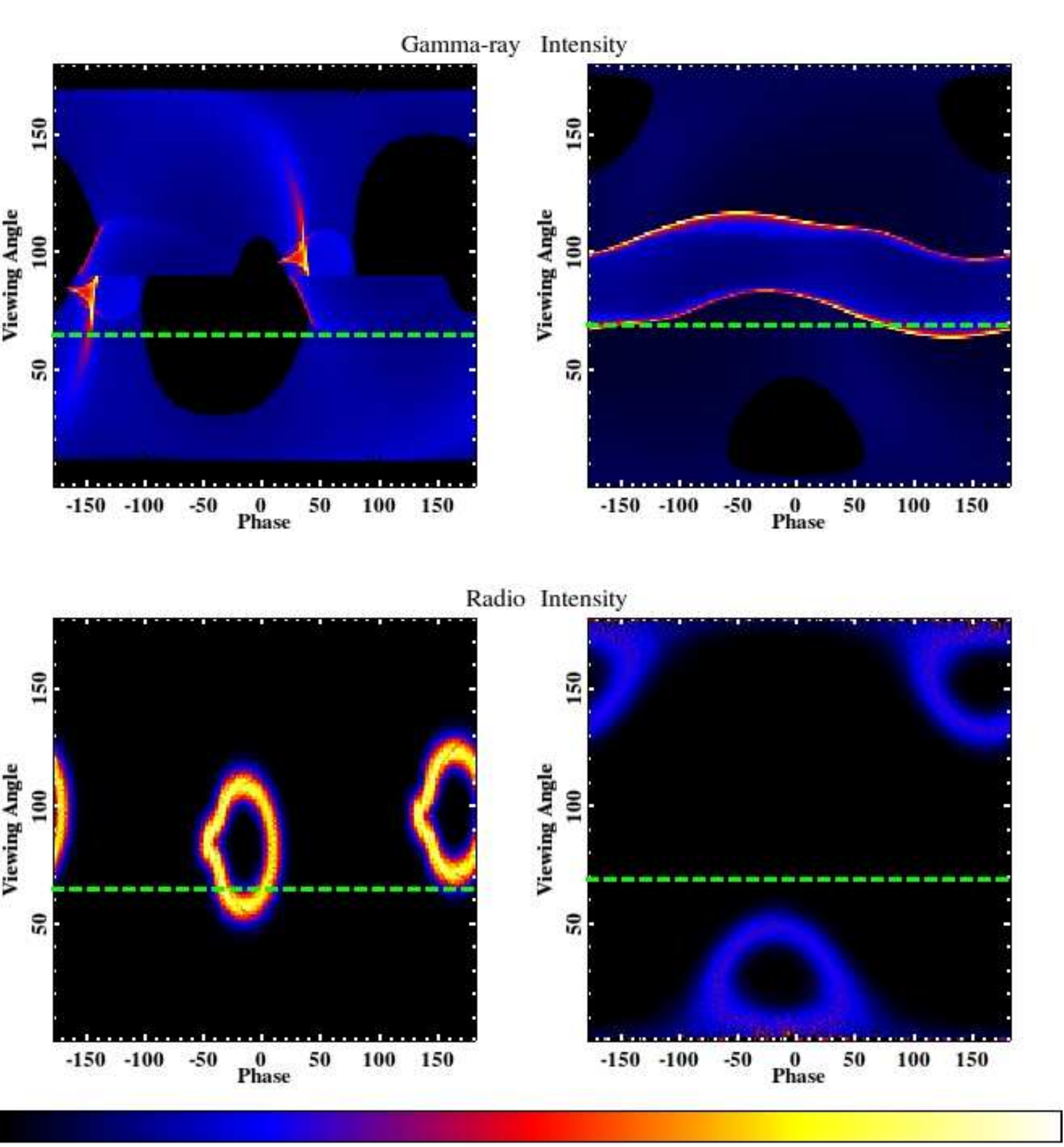}
\end{center}
\small\normalsize
\begin{quote}
\caption[Best-fit phase plots of simulated emission for PSR J1231$-$1411]{Distribution of simulated emission as a function of viewing angle and pulse phase for models used to fit PSR J1231$-$1411.  The best-fit $\zeta$ values are indicated by the dashed green lines.  The top plots correspond to the gamma-ray phase plots while the bottom are for the radio.  The left plots correspond to fits with the OG model with TPC plots on the right.  The color scale in the top-left plot is square root in order to bring out fainter features.\label{appAJ1231PhPlt}}
\end{quote}
\end{figure}

Plots of simulated emission corresponding to the best-fit models are shown in Fig.~\ref{appAJ1231PhPlt}, OG models are on the left and TPC on the right, gamma-ray models are on the top and radio on the bottom.

\section{PSR J1614$-$2230}\label{appAJ1614}
PSR J1614$-$2230 is a 3.1510 ms pulsar in a 8.7 d orbit with a 0.5 M$_{\odot}$ companion and was first discovered in the radio by \citet{Crawford06}.  Gamma-ray pulsations were first reported from this MSP by \citet{AbdoMSPpop} and later by \citet{AbdoPSRcat}.  Recently, radio observations have used the Shapiro delay to measure the mass of this pulsar to be 1.97$\pm$0.04 M$_{\odot}$ \citep{Demorest10}.  The gamma-ray light curve of this MSP was previously modeled by \citet{Venter09} using TPC and OG models with a hollow-cone beam radio model.

\small\normalsize

\small\normalsize

\small\normalsize

\begin{figure}
\begin{center}
\includegraphics[width=0.75\textwidth]{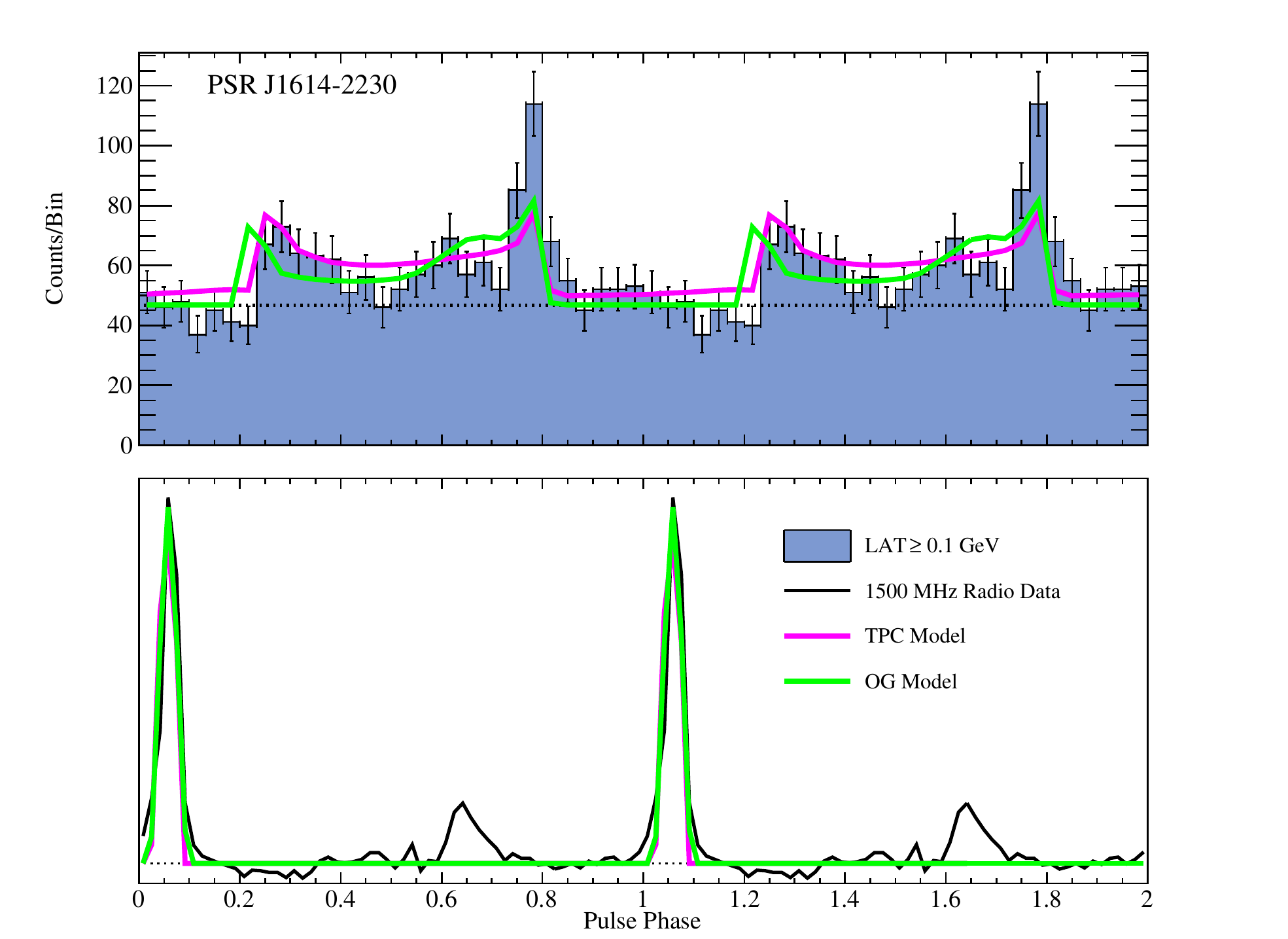}
\end{center}
\small\normalsize
\begin{quote}
\caption[Data and best-fit light curves for PSR J1614$-$2230]{Best-fit gamma-ray and radio light curves for PSR J1614$-$2230 using the TPC and OG models.\label{appAJ1614LCs}}
\end{quote}
\end{figure}

The best-fit gamma-ray and radio light curves are shown in Fig.~\ref{appAJ1614LCs}.  The gamma-ray light curve has been fit with TPC and OG models.  The radio profile has been fit with a hollow-cone beam model.  These light curve fits have used the 1500 MHz Greenbank radio profile.

\begin{figure}
\begin{center}
\includegraphics[width=0.75\textwidth]{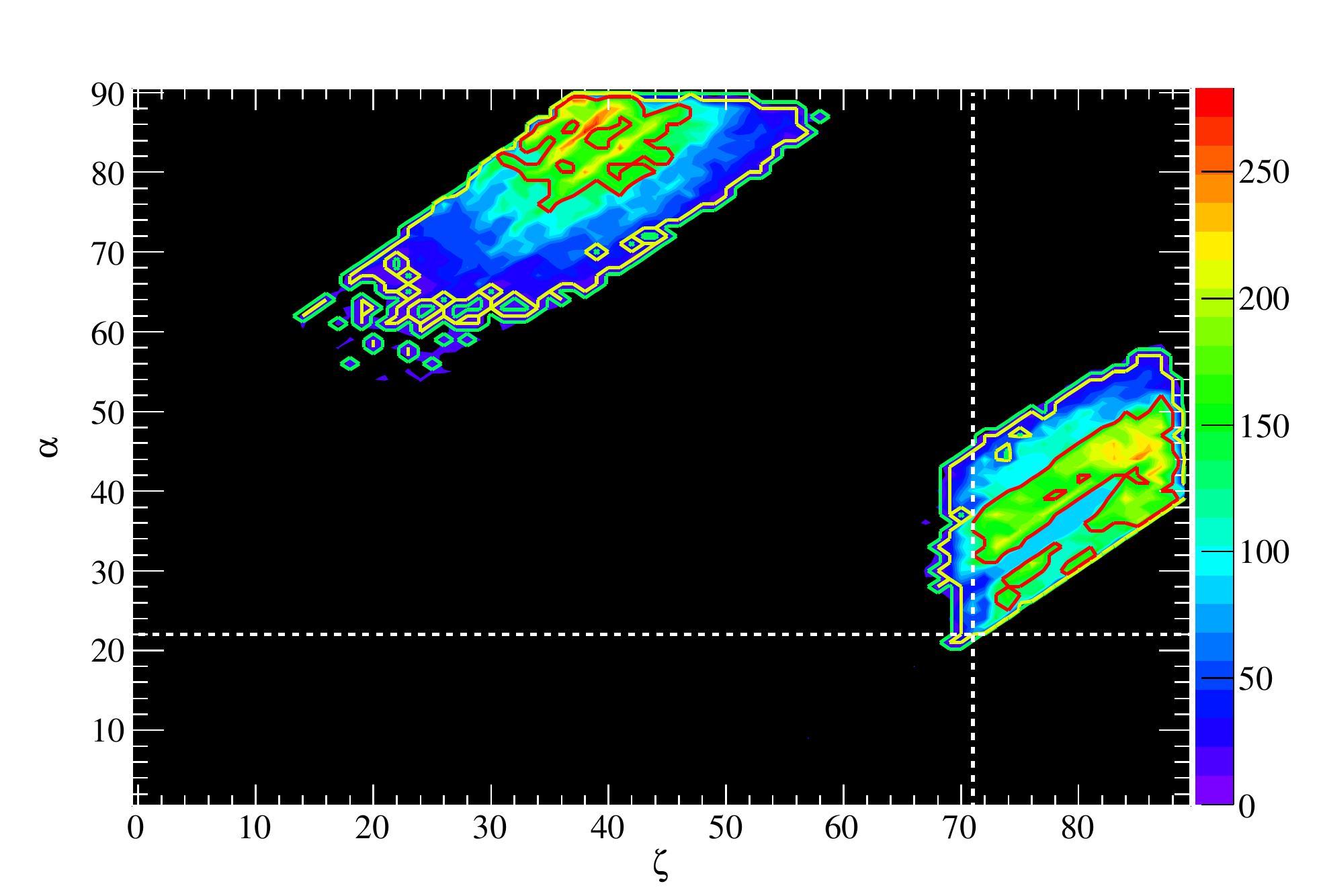}
\end{center}
\small\normalsize
\begin{quote}
\caption[Best-fit TPC contours for PSR J1614$-$2230]{Marginalized confidence contours for PSR J1614$-$2230 for the TPC model.\label{appAJ1614TPCcont}}
\end{quote}
\end{figure}

The marginalized $\alpha$-$\zeta$ confidence contours corresponding to the TPC fit are shown in Fig.~\ref{appAJ1614TPCcont}, the best-fit geometry is indicated by the vertical and horizontal dashed, white lines.

\begin{figure}
\begin{center}
\includegraphics[width=0.75\textwidth]{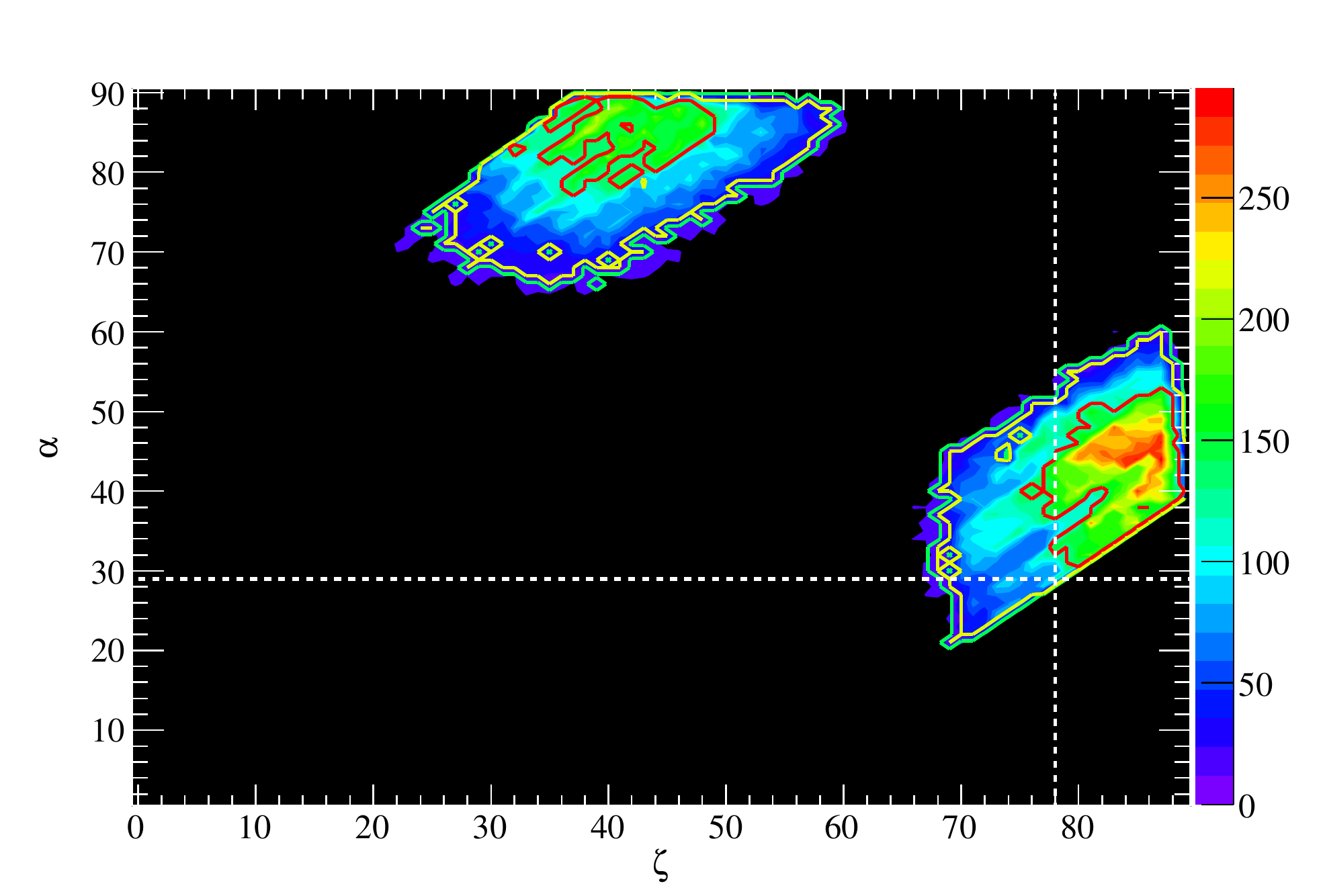}
\end{center}
\small\normalsize
\begin{quote}
\caption[Best-fit OG contours for PSR J1614$-$2230]{Marginalized confidence contours for PSR J1614$-$2230 for the OG model.\label{appAJ1614OGcont}}
\end{quote}
\end{figure}

The marginalized $\alpha$-$\zeta$ confidence contours corresponding to the OG fit are shown in Fig.~\ref{appAJ1614OGcont}, the best-fit geometry is indicated by the vertical and horizontal dashed, white lines.

\small\normalsize

\small\normalsize

\small\normalsize

\begin{figure}
\begin{center}
\includegraphics[width=0.75\textwidth]{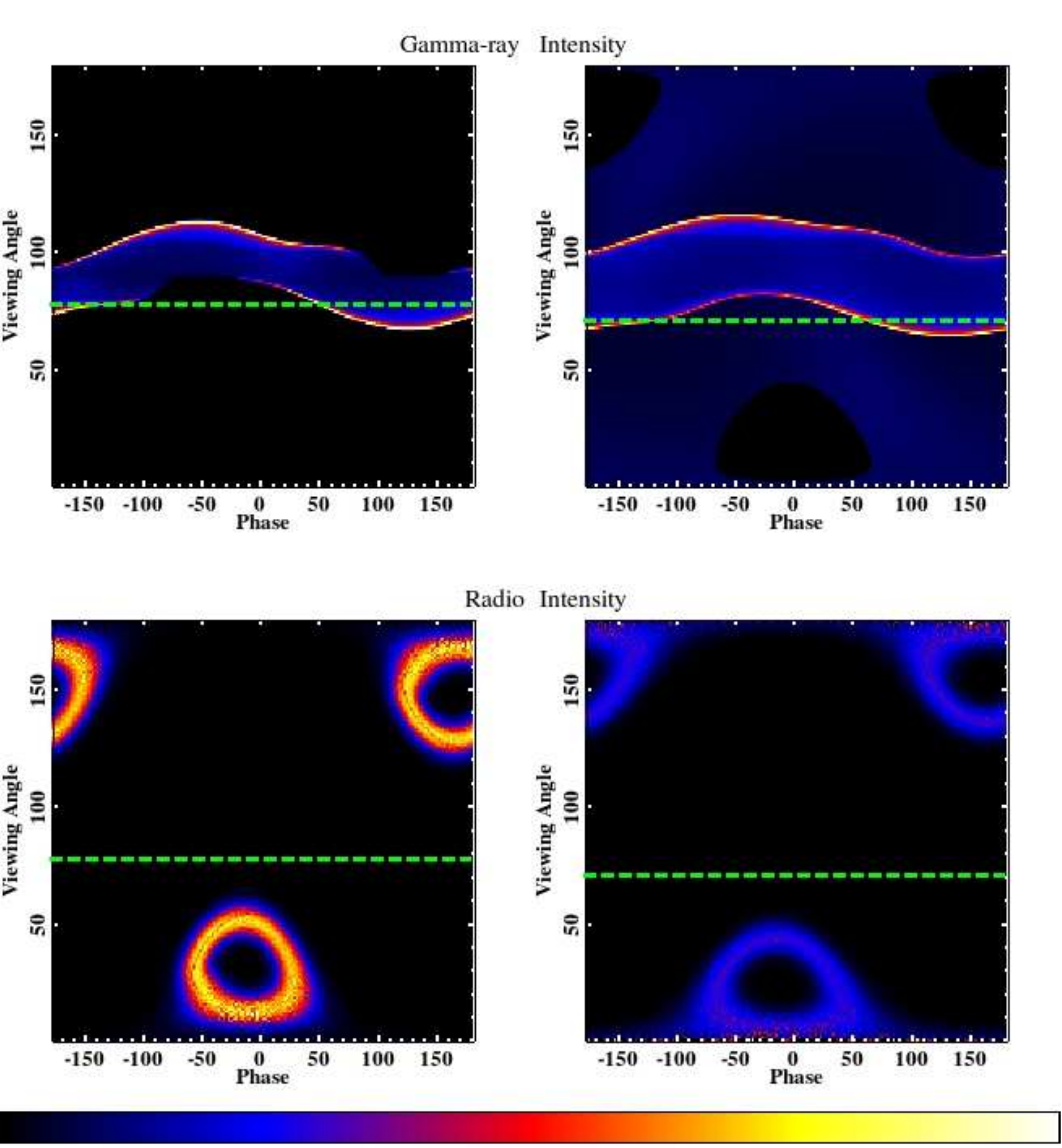}
\end{center}
\small\normalsize
\begin{quote}
\caption[Best-fit phase plots of simulated emission for PSR J1614$-$2230]{Distribution of simulated emission as a function of viewing angle and pulse phase for models used to fit PSR J1614$-$2230.  The best-fit $\zeta$ values are indicated by the dashed green lines.  The top plots correspond to the gamma-ray phase plots while the bottom are for the radio.  The left plots correspond to fits with the OG model with TPC plots on the right.\label{appAJ1614PhPlt}}
\end{quote}
\end{figure}

Plots of simulated emission corresponding to the best-fit models are shown in Fig.~\ref{appAJ1614PhPlt}, OG models are on the left and TPC on the right, gamma-ray models are on the top and radio on the bottom.

\small\normalsize

\section{PSR J1713+0747}\label{appAJ1713}
PSR J1713+0747 is a 4.5700 ms pulsar in a 67.8 d orbit with low-mass companion ($>$0.28 M$_{\odot}$) and was first discovered in the radio by \citet{Foster93}.  Gamma-ray pulsations from this MSP have not been announced prior to this thesis.

\begin{figure}
\begin{center}
\includegraphics[width=0.75\textwidth]{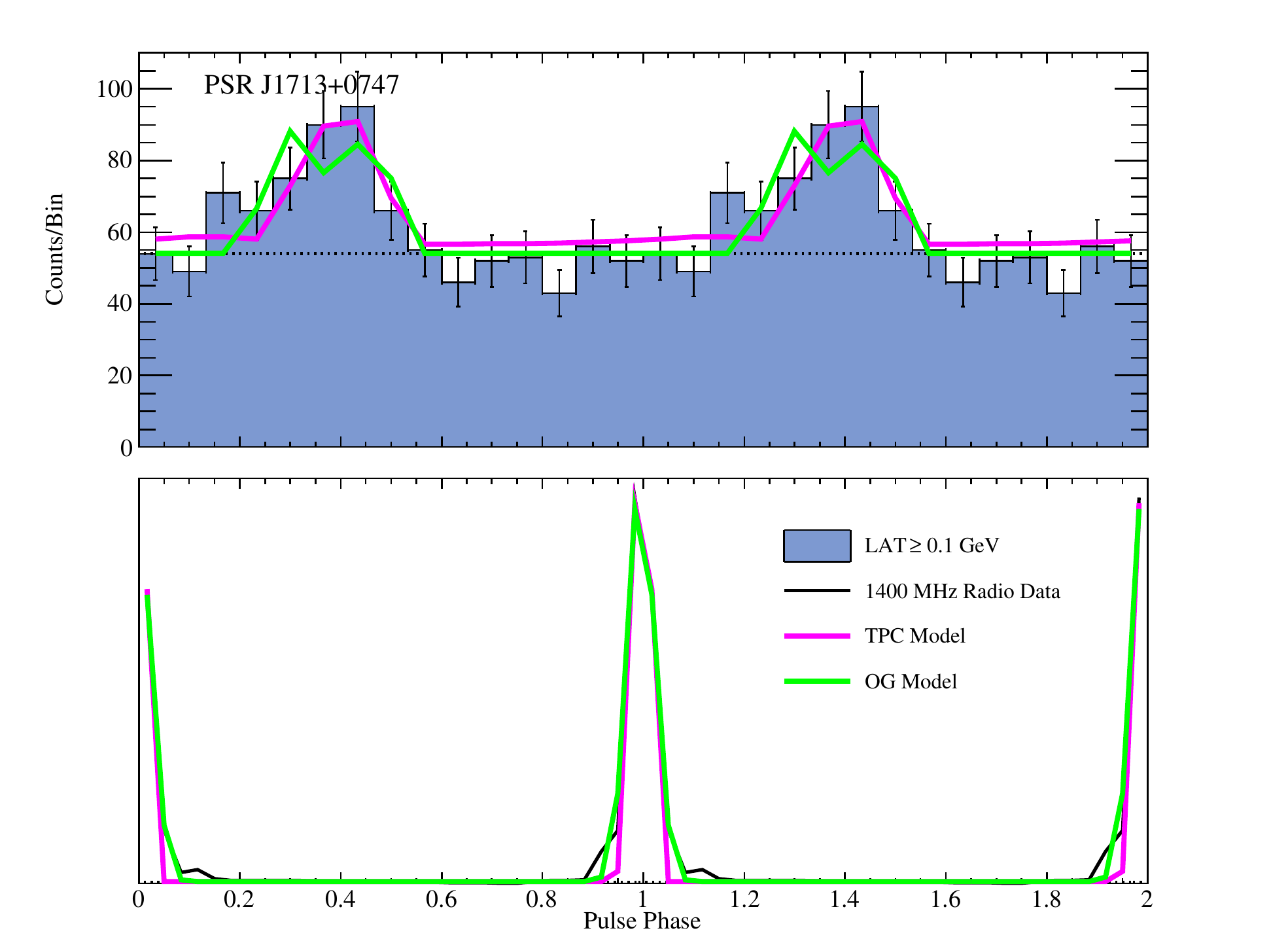}
\end{center}
\small\normalsize
\begin{quote}
\caption[Data and best-fit light curves for PSR J1713+0747]{Best-fit gamma-ray and radio light curves for PSR J1713+0747 using the TPC and OG models.\label{appAJ1713LCs}}
\end{quote}
\end{figure}

The best-fit gamma-ray and radio light curves are shown in Fig.~\ref{appAJ1713LCs}.  The gamma-ray light curve has been fit with TPC and OG models.  The radio profile has been fit with a hollow-cone beam model.  These light curve fits have used the 1400 MHz Jodrell Bank radio profile.

\begin{figure}
\begin{center}
\includegraphics[width=0.75\textwidth]{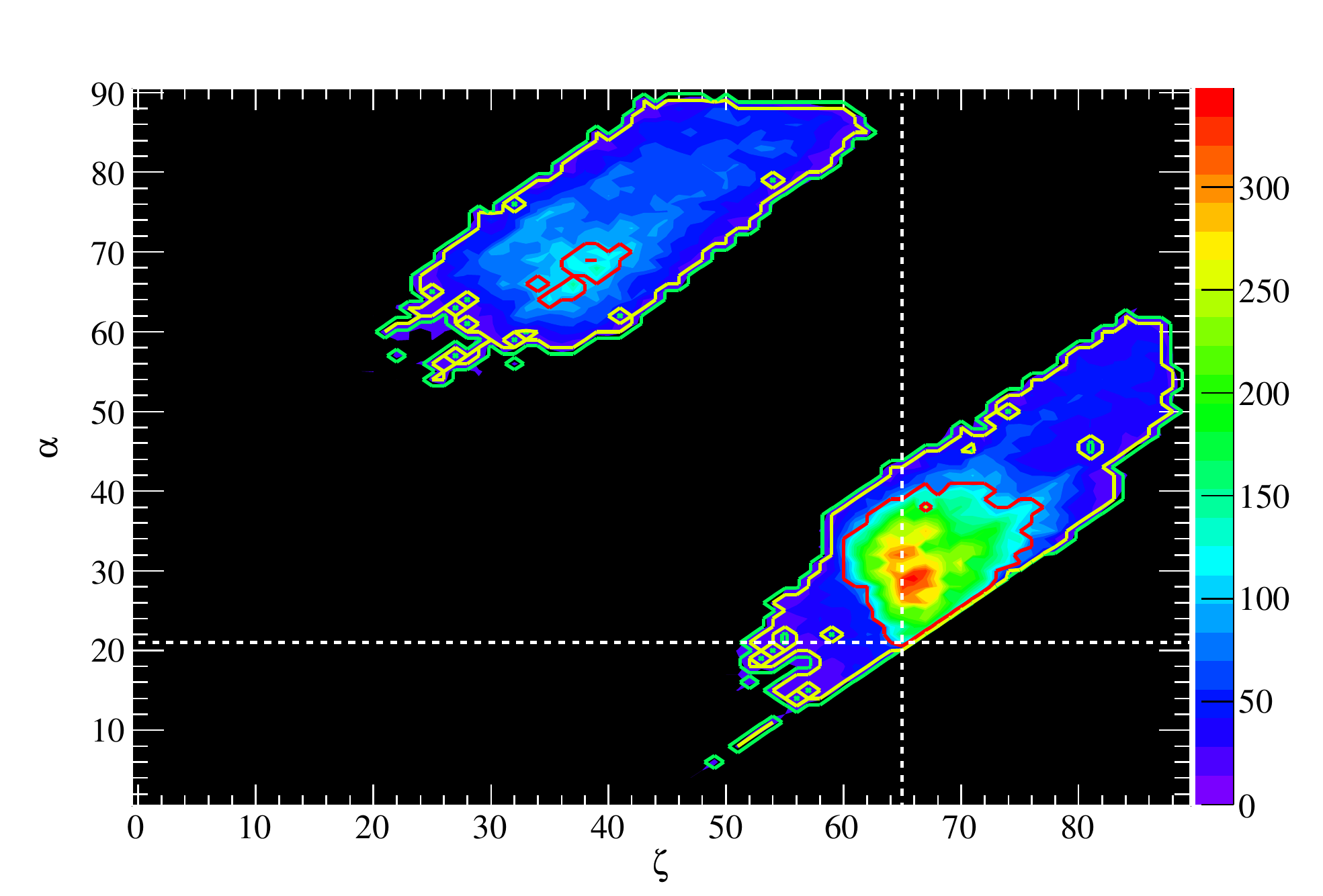}
\end{center}
\small\normalsize
\begin{quote}
\caption[Best-fit TPC contours for PSR J1713+0747]{Marginalized confidence contours for PSR J1713+0747 for the TPC model.\label{appAJ1713TPCcont}}
\end{quote}
\end{figure}

The marginalized $\alpha$-$\zeta$ confidence contours corresponding to the TPC fit are shown in Fig.~\ref{appAJ1713TPCcont}, the best-fit geometry is indicated by the vertical and horizontal dashed, white lines.

\begin{figure}
\begin{center}
\includegraphics[width=0.75\textwidth]{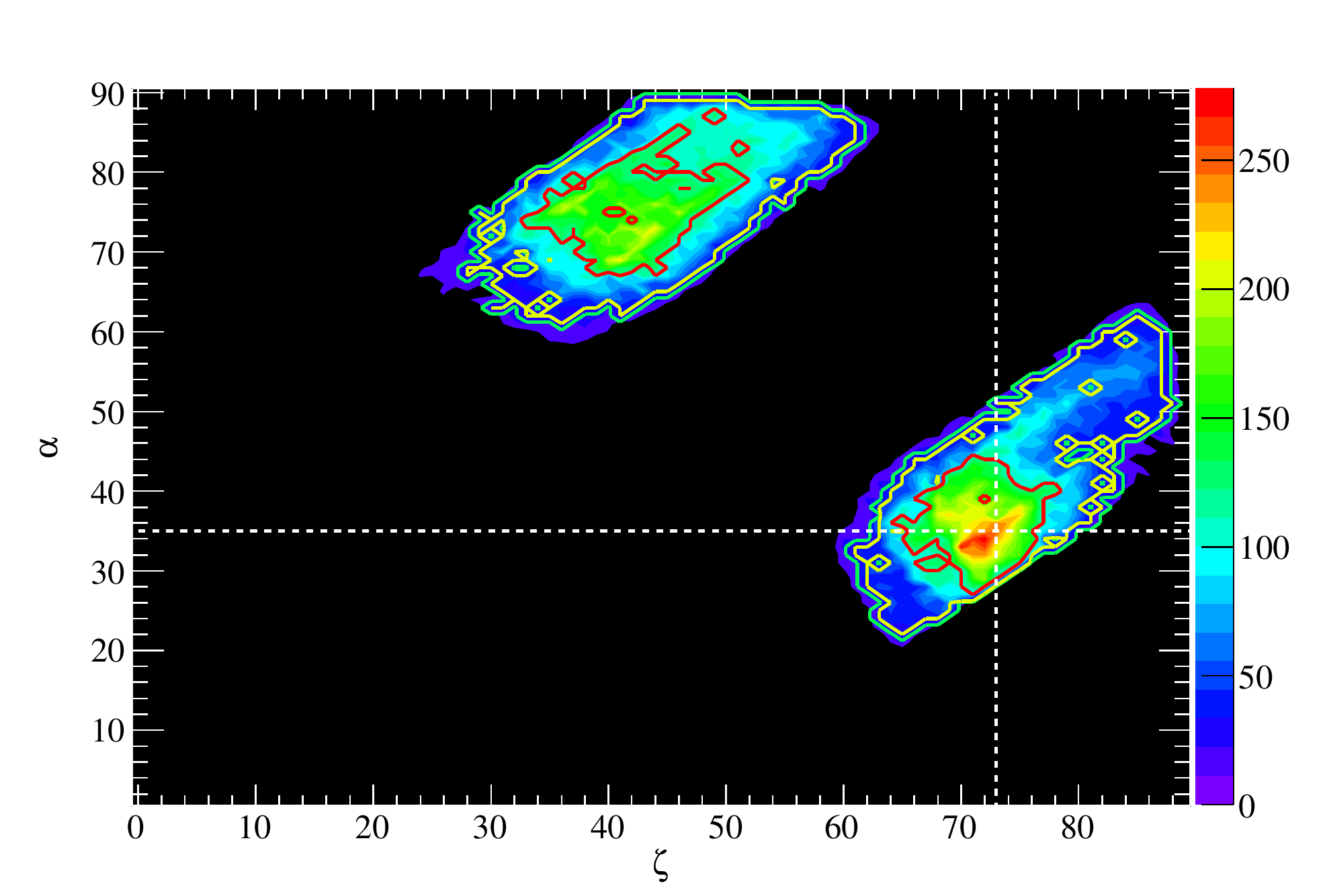}
\end{center}
\small\normalsize
\begin{quote}
\caption[Best-fit OG contours for PSR J1713+0747]{Marginalized confidence contours for PSR J1713+0747 for the OG model.\label{appAJ1713OGcont}}
\end{quote}
\end{figure}

The marginalized $\alpha$-$\zeta$ confidence contours corresponding to the OG fit are shown in Fig.~\ref{appAJ1713OGcont}, the best-fit geometry is indicated by the vertical and horizontal dashed, white lines.

\small\normalsize

\small\normalsize

\small\normalsize

\begin{figure}
\begin{center}
\includegraphics[width=0.75\textwidth]{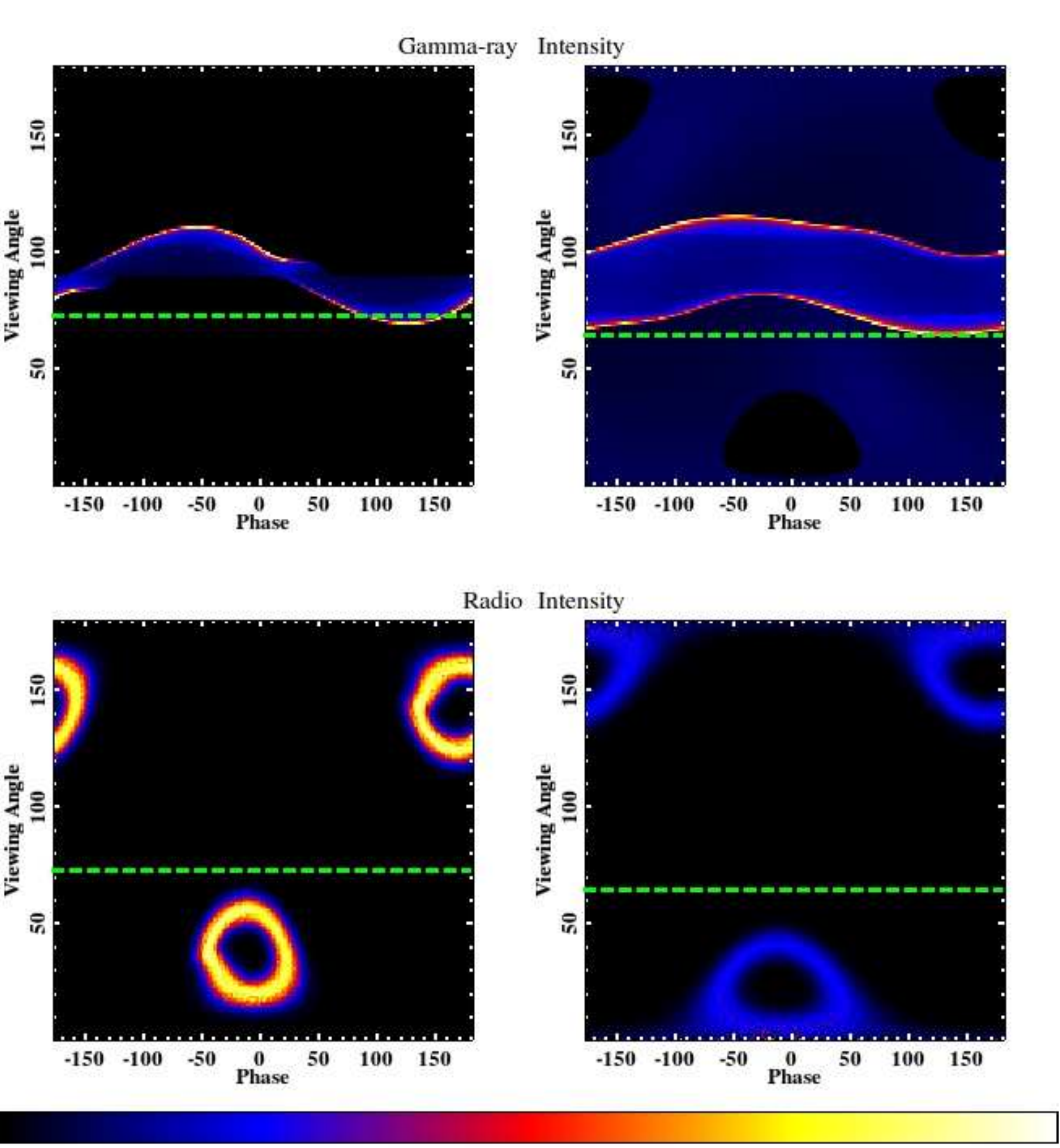}
\end{center}
\small\normalsize
\begin{quote}
\caption[Best-fit phase plots of simulated emission for PSR J1713+0747]{Distribution of simulated emission as a function of viewing angle and pulse phase for models used to fit PSR J1713+0747.  The best-fit $\zeta$ values are indicated by the dashed green lines.  The top plots correspond to the gamma-ray phase plots while the bottom are for the radio.  The left plots correspond to fits with the OG model with TPC plots on the right.\label{appAJ1713PhPlt}}
\end{quote}
\end{figure}

Plots of simulated emission corresponding to the best-fit models are shown in Fig.~\ref{appAJ1713PhPlt}, OG models are on the left and TPC on the right, gamma-ray models are on the top and radio on the bottom.

\small\normalsize

\section{PSR J1744$-$1134}\label{appAJ1744}
PSR J1744$-$1134 is a 4.0745 isolated pulsar first discovered in the radio by \citet{Bailes97}.  Gamma-ray pulsations from this MSP were first reported by \citet{AbdoMSPpop} and later by \citet{AbdoPSRcat}.  The gamma-ray light curve of this MSP was modeled by \citet{Venter09} using a geometric PSPC model with a hollow-cone beam radio model.

\begin{figure}
\begin{center}
\includegraphics[width=0.75\textwidth]{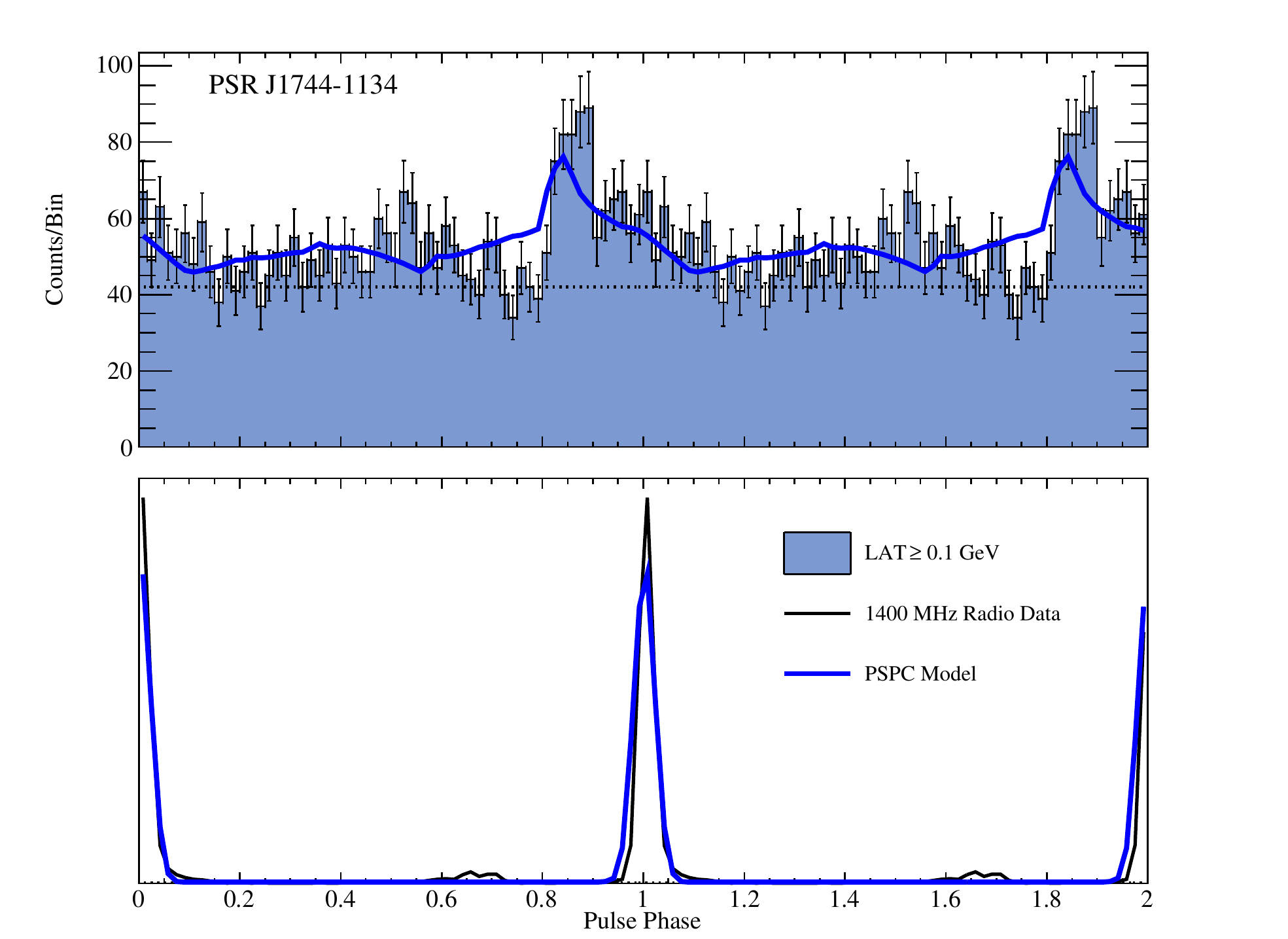}
\end{center}
\small\normalsize
\begin{quote}
\caption[Data and best-fit light curves for PSR J1744$-$1134]{Best-fit gamma-ray and radio light curves for PSR J1744$-$1134 using the PSPC model.\label{appAJ1744LCs}}
\end{quote}
\end{figure}

The best fit gamma-ray and radio light curves are shown in Fig.~\ref{appAJ1744LCs}.  The gamma-ray light curve of this MSP was fit with a PSPC model.  The radio profile was fit with a hollow-cone beam model.  These light curve fits have used the 1400 MHz Nan\c{cay} radio profile.

\small\normalsize

\begin{figure}
\begin{center}
\includegraphics[width=0.75\textwidth]{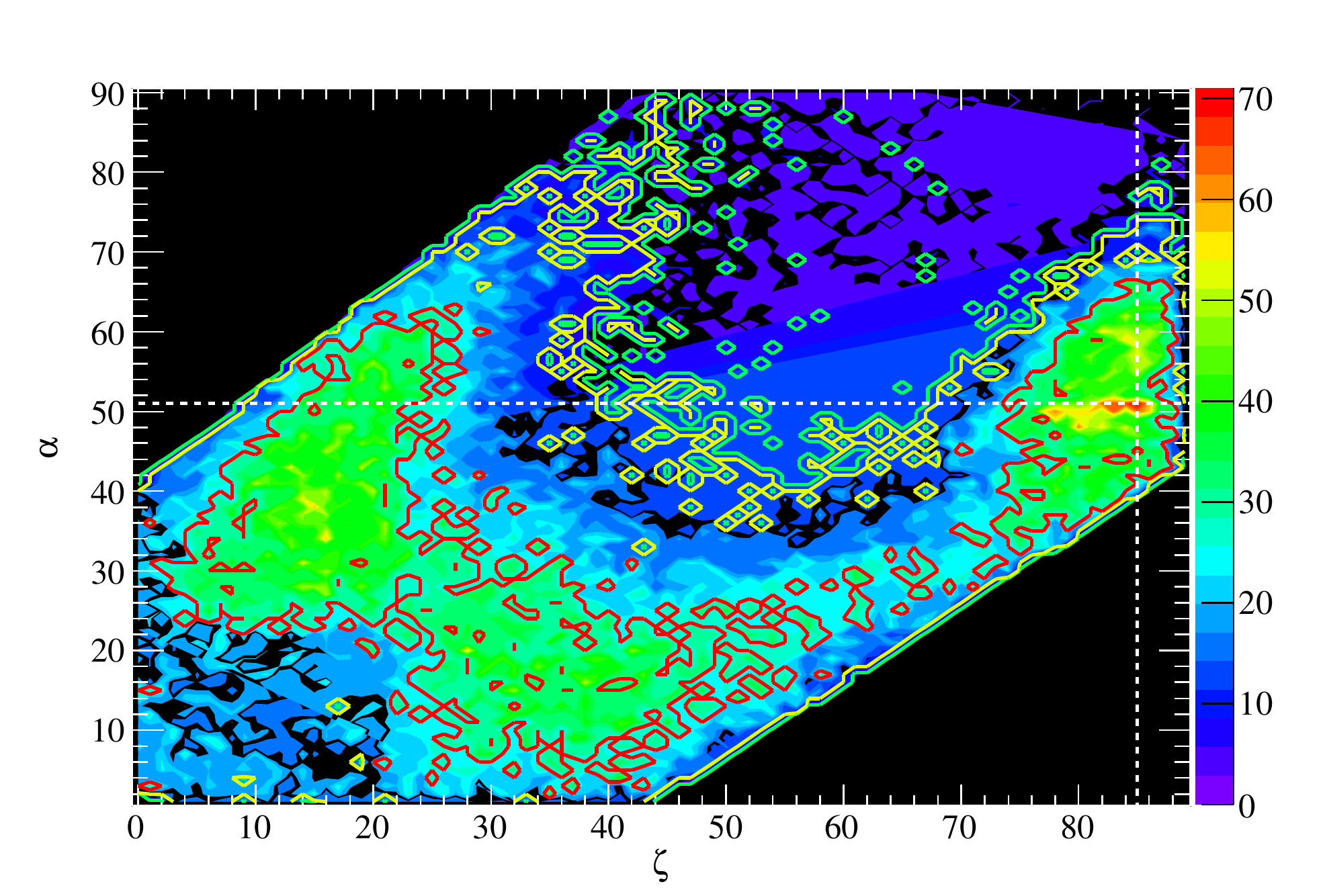}
\end{center}
\small\normalsize
\begin{quote}
\caption[Best-fit PSPC contours for PSR J1744$-$1134]{Marginalized confidence contours for PSR J1744$-$1134 for the PSPC model.\label{appAJ1744PSPCcont}}
\end{quote}
\end{figure}

The marginalized $\alpha$-$\zeta$ confidence contours corresponding to the PSPC fit are shown in Fig.~\ref{appAJ1744PSPCcont}, the best-fit geometry is indicated by the vertical and horizontal dashed, white lines.

\small\normalsize

\begin{figure}
\begin{center}
\includegraphics[width=0.5\textwidth]{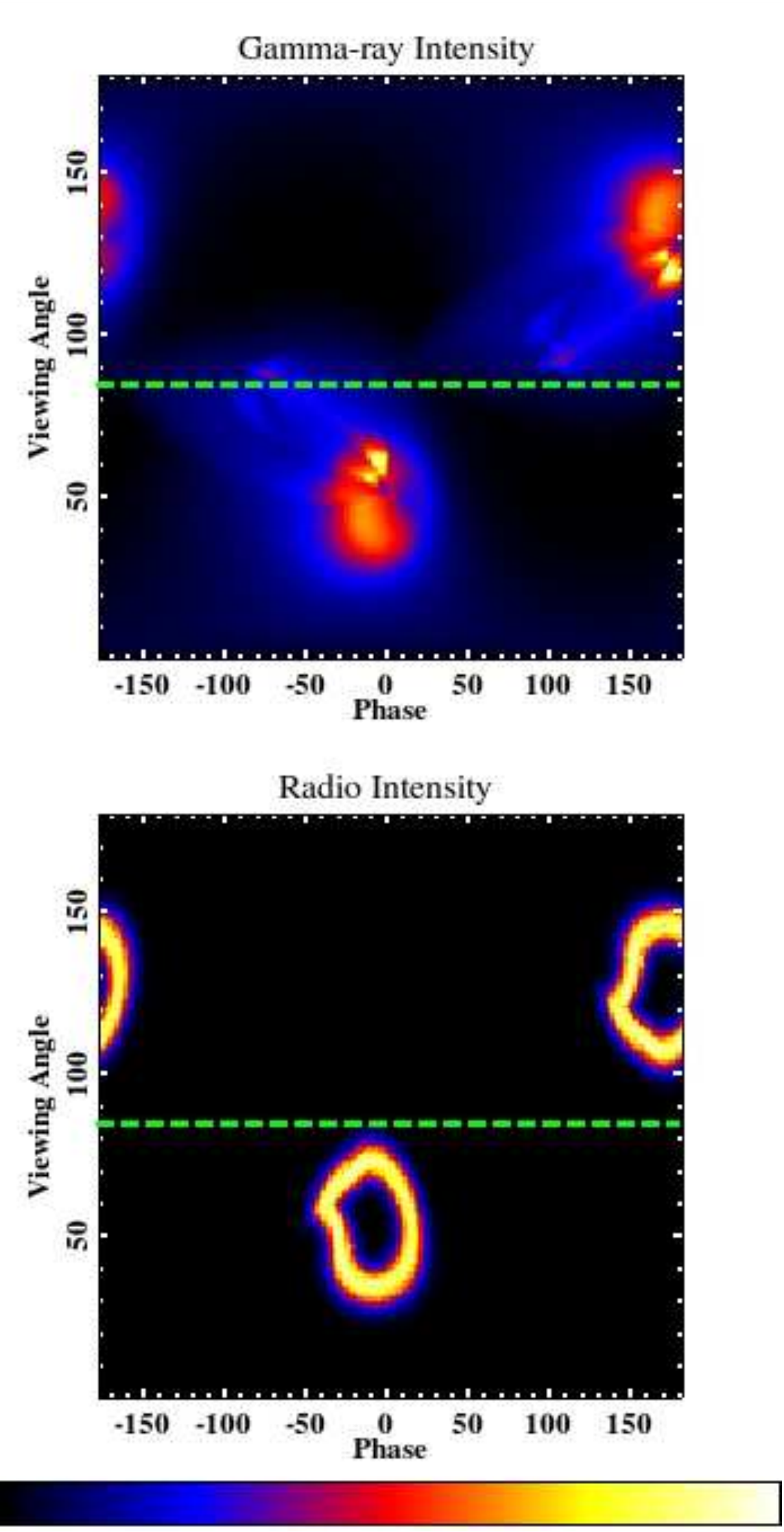}
\end{center}
\small\normalsize
\begin{quote}
\caption[Best-fit phase plots of simulated emission for PSR J1744$-$1134]{Distribution of simulated emission as a function of viewing angle and pulse phase for models used to fit PSR J1744$-$1134 with the PSPC model.  The best-fit $\zeta$ value is indicated by the dashed green lines.  The top plot corresponds to the gamma-ray phase plot while the bottom is for the radio.\label{appAJ1744PhPlt}}
\end{quote}
\end{figure}

Plots of simulated emission corresponding to the best-fit models are shown in Fig.~\ref{appAJ1744PhPlt}, the gamma-ray model is on the top and radio on the bottom.

\small\normalsize

\section{PSR J1823$-$3021A}\label{appAJ1823}
PSR J1823$-$3021A is a 5.4400 isolated pulsar.  This MSP is located in the globular cluster NGC6624 and was first discovered in the radio by \citet{Biggs94}.  Details of the gamma-ray pulsation and spectral fits will be reported by \citet{Freire11} who also use the MCMC likelihood technique described in this thesis to fit the gamma-ray and radio profiles with alTPC and alOG models.

\begin{figure}
\begin{center}
\includegraphics[width=0.75\textwidth]{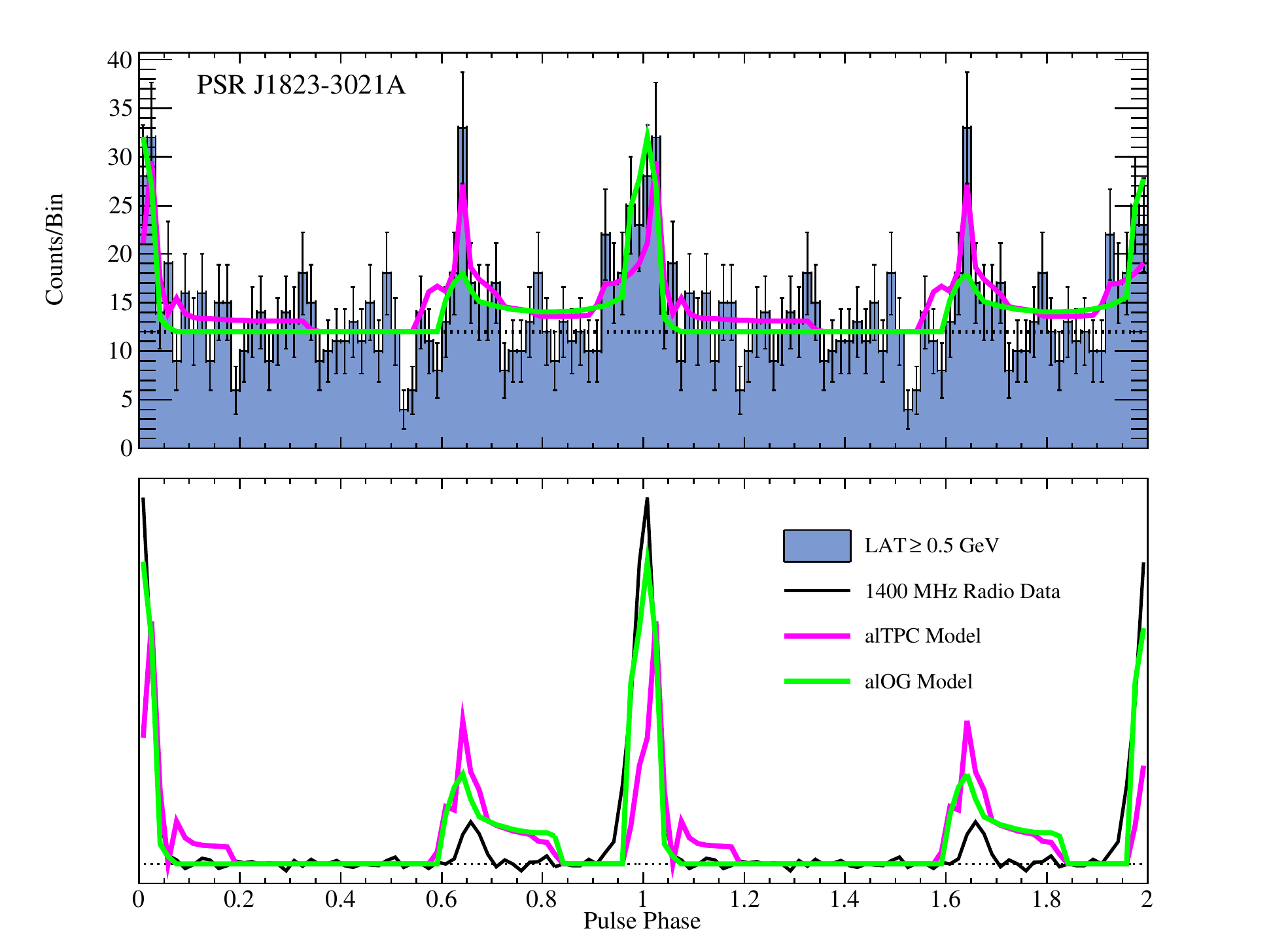}
\end{center}
\small\normalsize
\begin{quote}
\caption[Data and best-fit light curves for PSR J1823$-$3021A]{Best-fit gamma-ray and radio light curves for PSR J1823$-$3021A using the alTPC and alOG models.\label{appAJ1823LCs}}
\end{quote}
\end{figure}

The best-fit gamma-ray and radio light curves are shown in Fig.~\ref{appAJ1823LCs}.  The gamma-ray and radio profiles have been modeled with the alOG and alTPC models.  These light curve fits have used the 1400 MHz Nan\c{cay} radio profile.

\begin{figure}
\begin{center}
\includegraphics[width=0.75\textwidth]{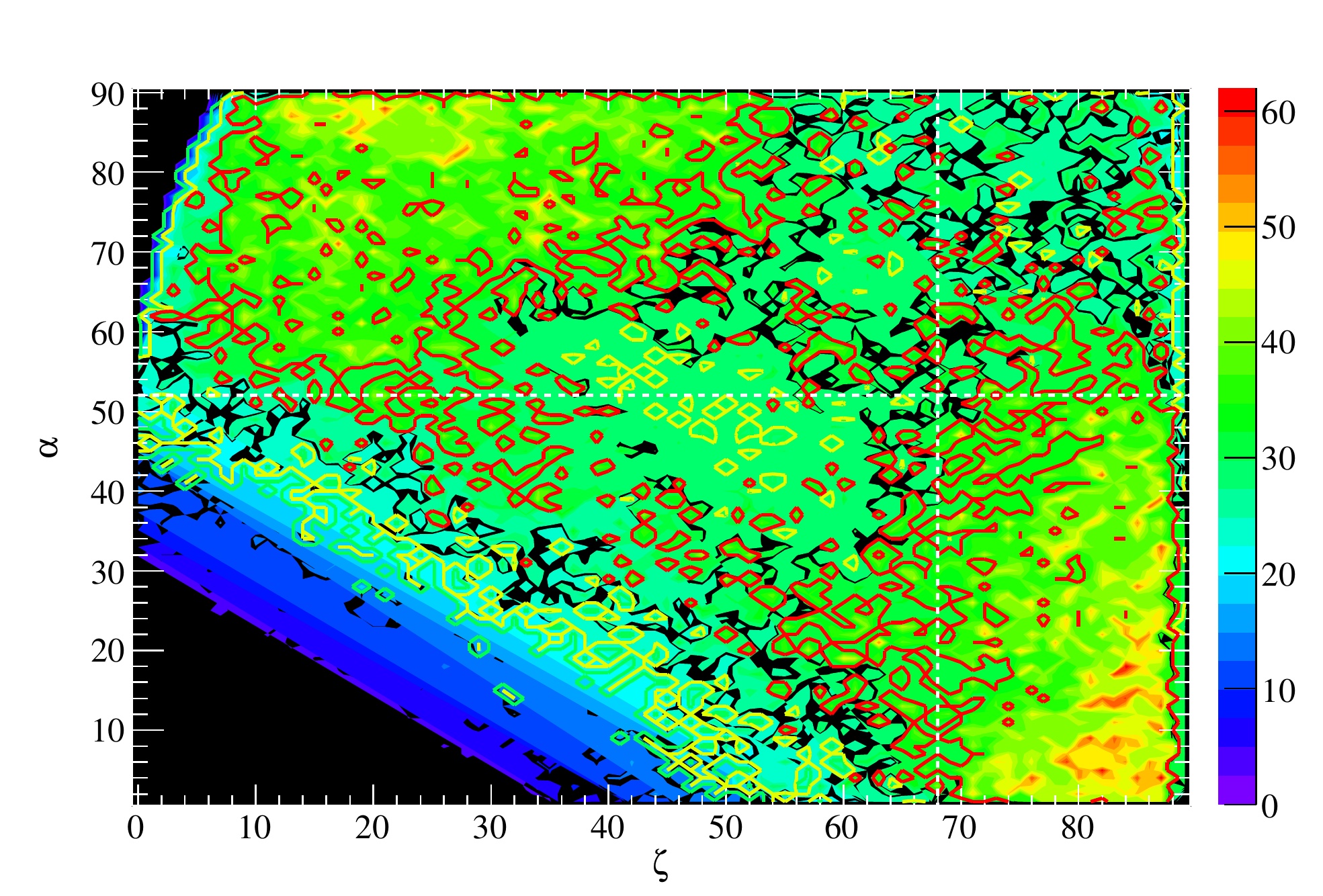}
\end{center}
\small\normalsize
\begin{quote}
\caption[Best-fit alTPC contours for PSR J1823$-$3021A]{Marginalized confidence contours for PSR J1823$-$3021A for the alTPC model.\label{appAJ1823TPCcont}}
\end{quote}
\end{figure}

The marginalized $\alpha$-$\zeta$ confidence contours corresponding to the alTPC fit are shown in Fig.~\ref{appAJ1823TPCcont}, the best-fit geometry is indicated by the vertical and horizontal dashed, white lines.

\begin{figure}
\begin{center}
\includegraphics[width=0.75\textwidth]{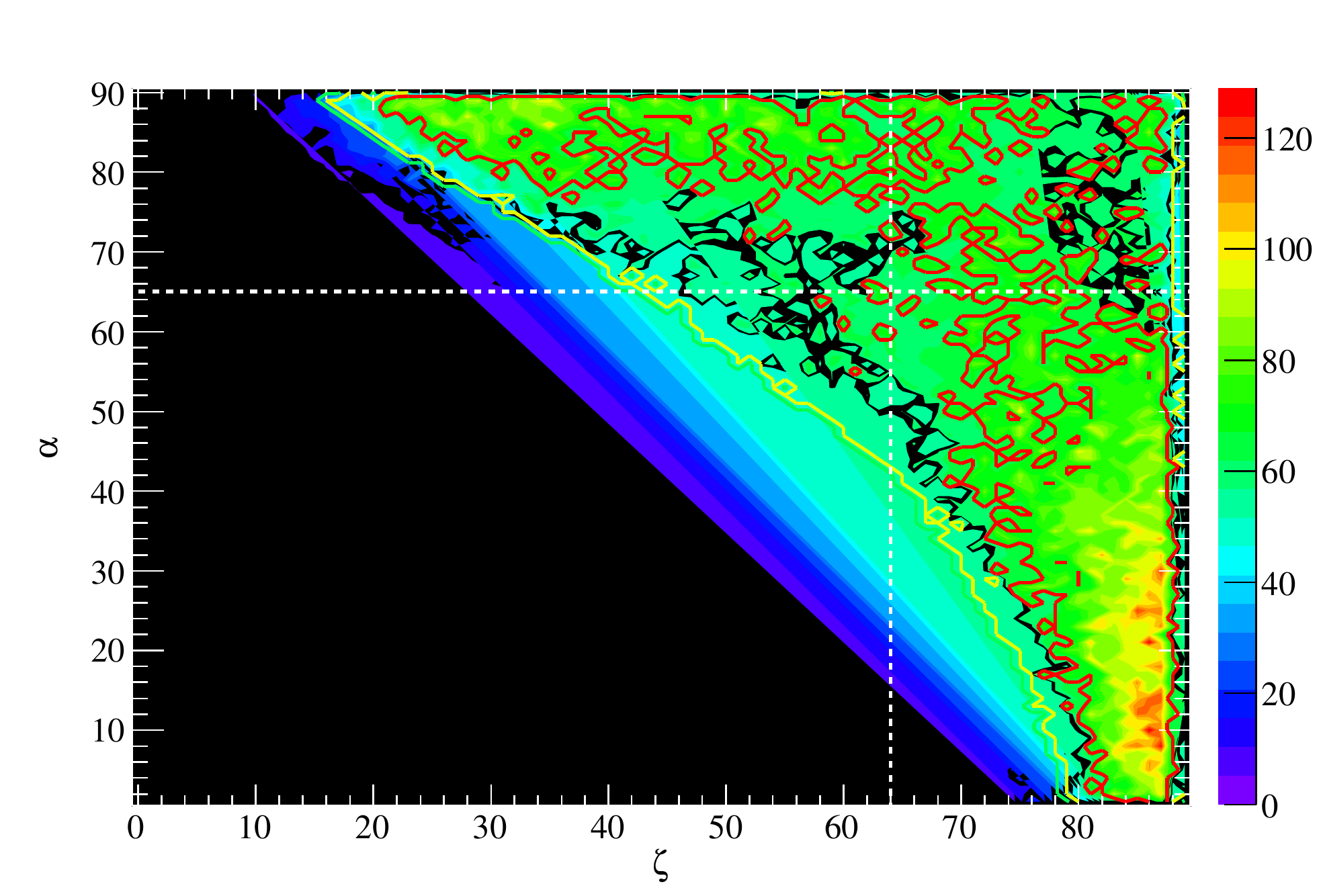}
\end{center}
\small\normalsize
\begin{quote}
\caption[Best-fit alOG contours for PSR J1823$-$3021A]{Marginalized confidence contours for PSR J1823$-$3021A for the alOG model.\label{appAJ1823OGcont}}
\end{quote}
\end{figure}

The marginalized $\alpha$-$\zeta$ confidence contours corresponding to the alOG fit are shown in Fig.~\ref{appAJ1823OGcont}, the best-fit geometry is indicated by the vertical and horizontal dashed, white lines.

\small\normalsize

\small\normalsize

\small\normalsize

\begin{figure}
\begin{center}
\includegraphics[width=0.75\textwidth]{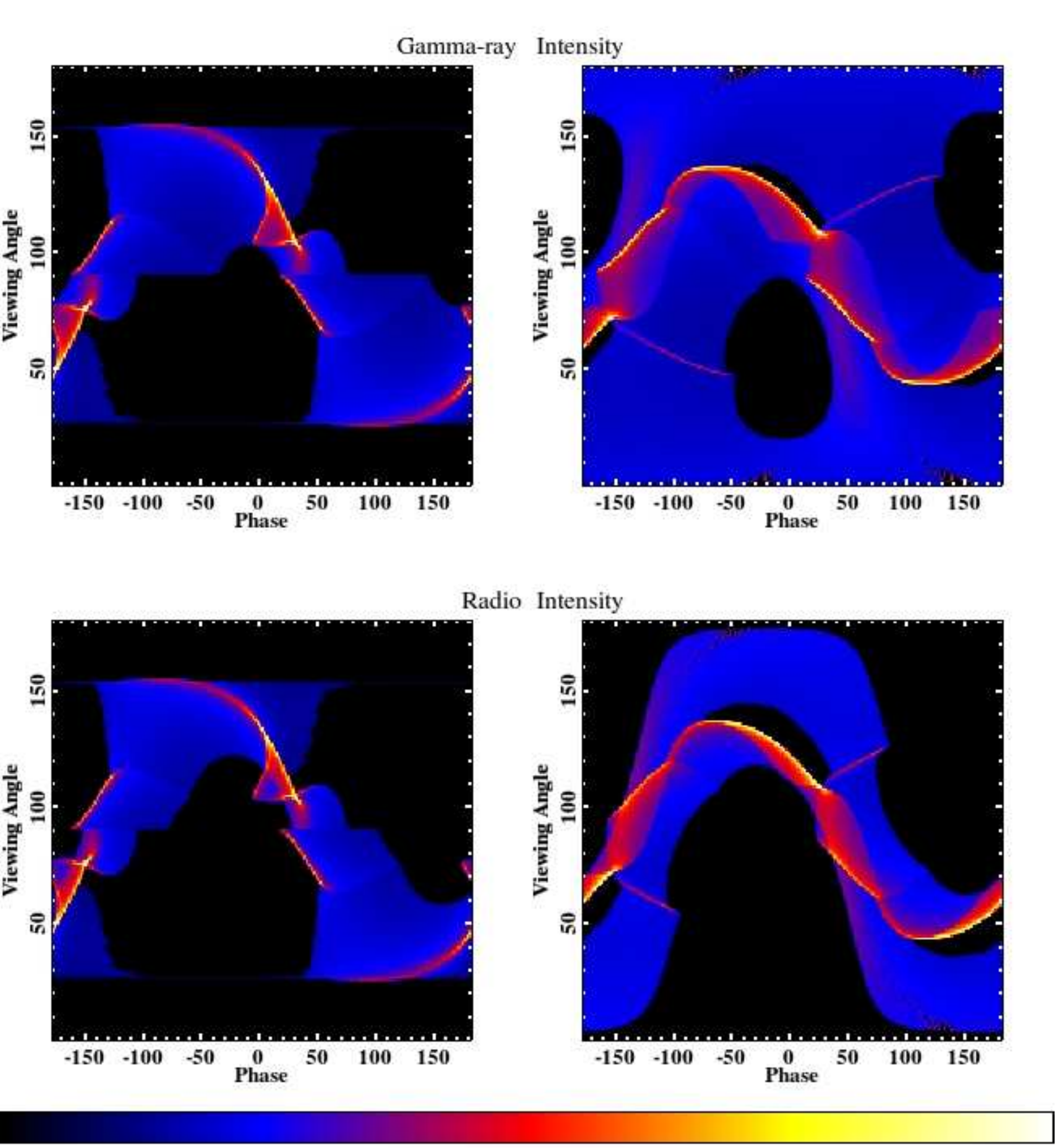}
\end{center}
\small\normalsize
\begin{quote}
\caption[Best-fit phase plots of simulated emission for PSR J1823$-$3021A]{Distribution of simulated emission as a function of viewing angle and pulse phase for models used to fit PSR J1823$-$3021A.  The best-fit $\zeta$ values are indicated by the dashed green lines.  The top plots correspond to the gamma-ray phase plots while the bottom are for the radio.  The left plots correspond to fits with the alOG model with alTPC plots on the right.  The color scales are square root in order to bring out fainter features.\label{appAJ1823PhPlt}}
\end{quote}
\end{figure}

Plots of simulated emission corresponding to the best-fit models are shown in Fig.~\ref{appAJ1823PhPlt}, alOG models are on the left and alTPC on the right, gamma-ray models are on the top and radio on the bottom.

\small\normalsize

\section{PSR J1902$-$5105}\label{appAJ1902}
PSR J1902$-$5105 is a 1.7424 ms pulsar in a 2 d orbit with a $\sim$0.2 M$_{\odot}$ companion.  This MSP was discovered in targeted radio observations of unassociated LAT sources with pulsar-like characteristics and seen to pulse in gamma rays soon after.  The radio and gamma-ray discoveries will be discussed in a forthcoming paper \citep{Camilo11}, the timing solution and radio light curve were used in this thesis courtesy of F. Camilo.

\begin{figure}
\begin{center}
\includegraphics[width=0.75\textwidth]{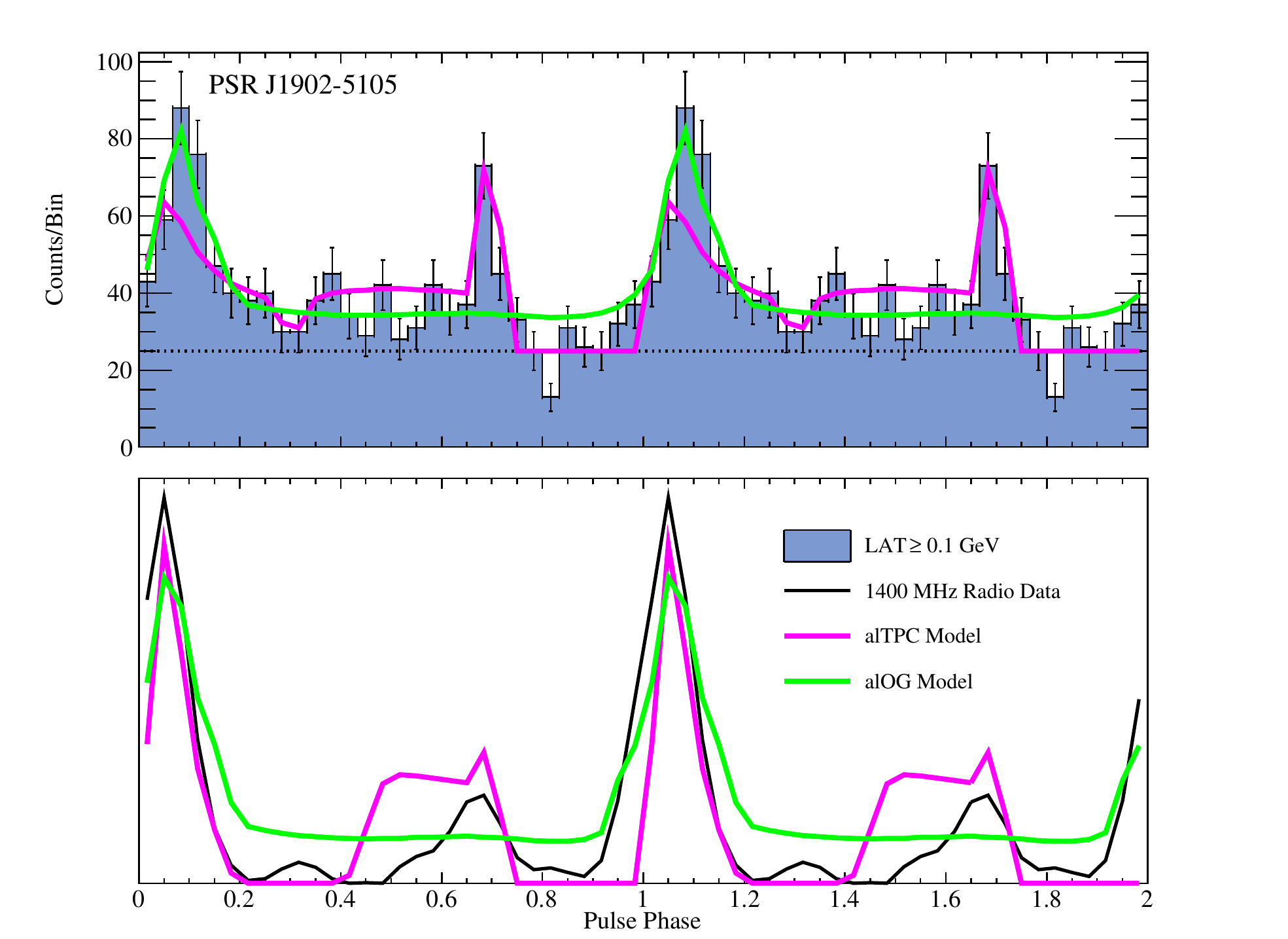}
\end{center}
\small\normalsize
\begin{quote}
\caption[Data and best-fit light curves for PSR J1902$-$5105]{Best-fit gamma-ray and radio light curves for PSR J1902$-$5105 using the alTPC and alOG models.\label{appAJ1902LCs}}
\end{quote}
\end{figure}

The best-fit gamma-ray and radio light curves are shown in Fig.~\ref{appAJ1902LCs}.  The gamma-ray and radio light curves were fit with the alTPC and alOG models. These light curve fits have used the 1400 MHz Parkes radio profile.

\begin{figure}
\begin{center}
\includegraphics[width=0.75\textwidth]{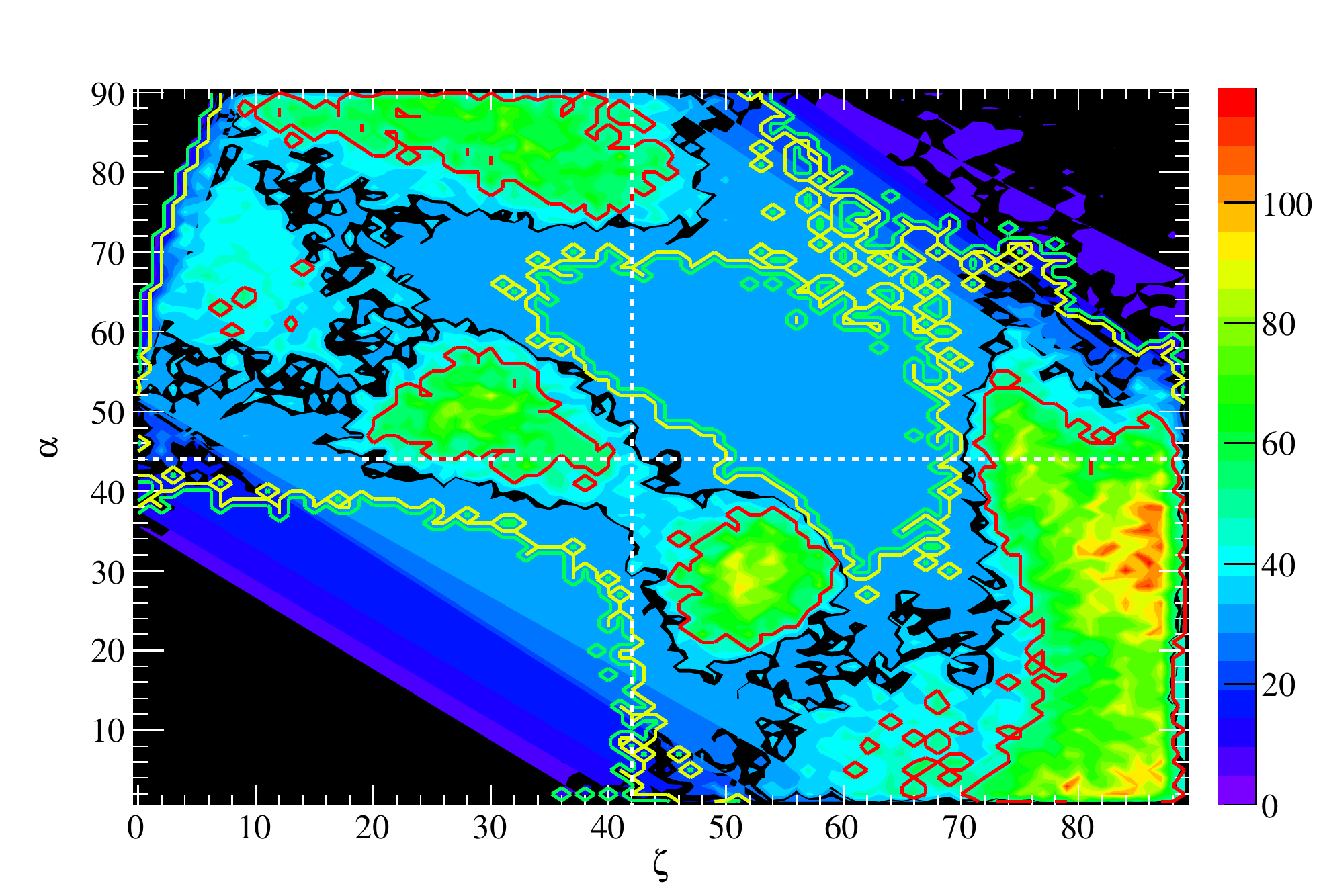}
\end{center}
\small\normalsize
\begin{quote}
\caption[Best-fit alTPC contours for PSR J1902$-$5105]{Marginalized confidence contours for PSR J1902$-$5105 for the alTPC model.\label{appAJ1902TPCcont}}
\end{quote}
\end{figure}

The marginalized $\alpha$-$\zeta$ confidence contours corresponding to the alTPC fit is shown in Fig.~\ref{appAJ1902TPCcont}, the best-fit geometry is indicated by the vertical and horizontal dashed, white lines.

\begin{figure}
\begin{center}
\includegraphics[width=0.75\textwidth]{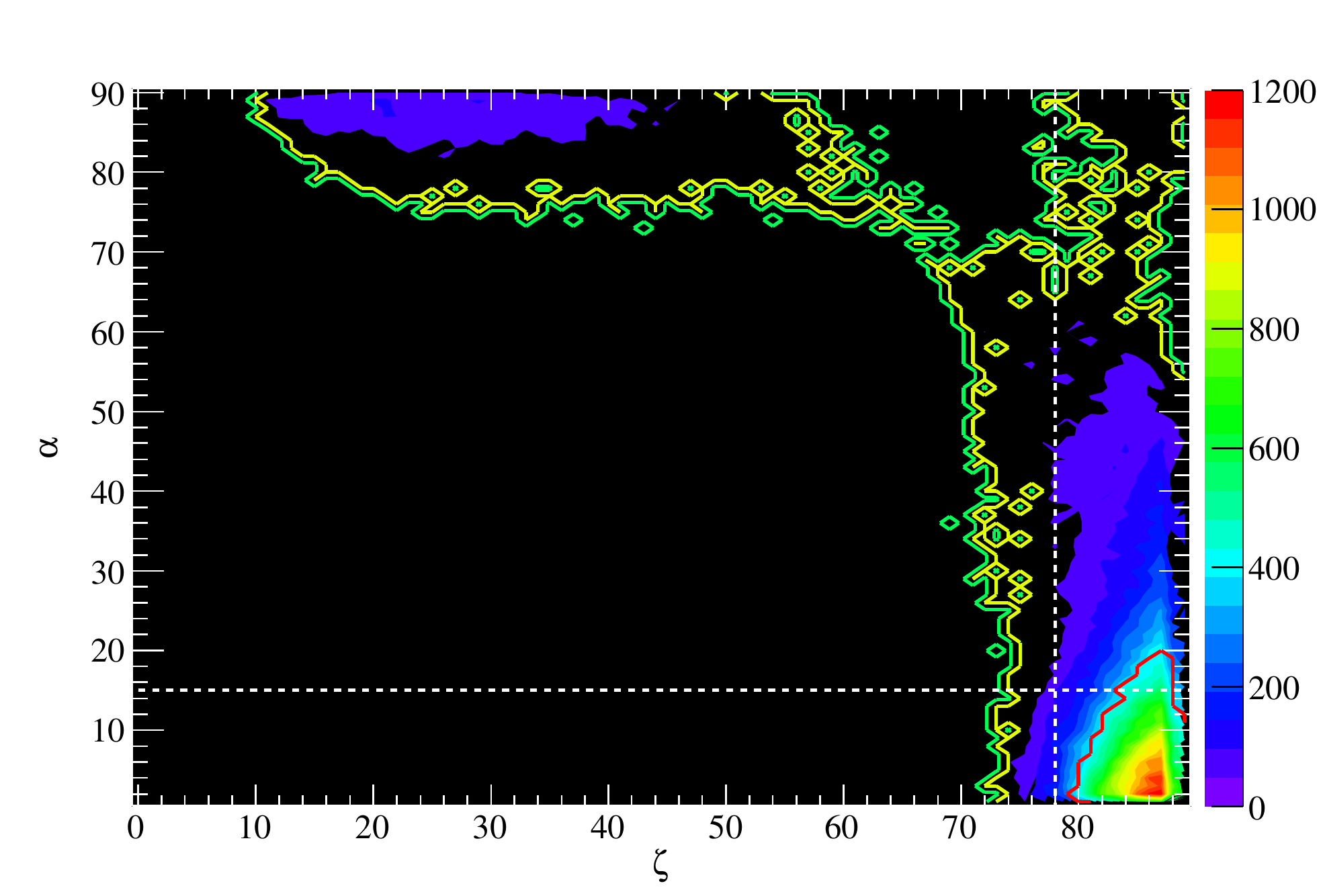}
\end{center}
\small\normalsize
\begin{quote}
\caption[Best-fit alOG contours for PSR J1902$-$5105]{Marginalized confidence contours for PSR J1902$-$5105 for the alOG model.\label{appAJ1902OGcont}}
\end{quote}
\end{figure}

The marginalized $\alpha$-$\zeta$ confidence contours corresponding to the alOG fit are shown in Fig.~\ref{appAJ1902OGcont}, the best-fit geometry is indicated by the vertical and horizontal dashed, white lines.

\small\normalsize

\small\normalsize

\small\normalsize

\begin{figure}
\begin{center}
\includegraphics[width=0.75\textwidth]{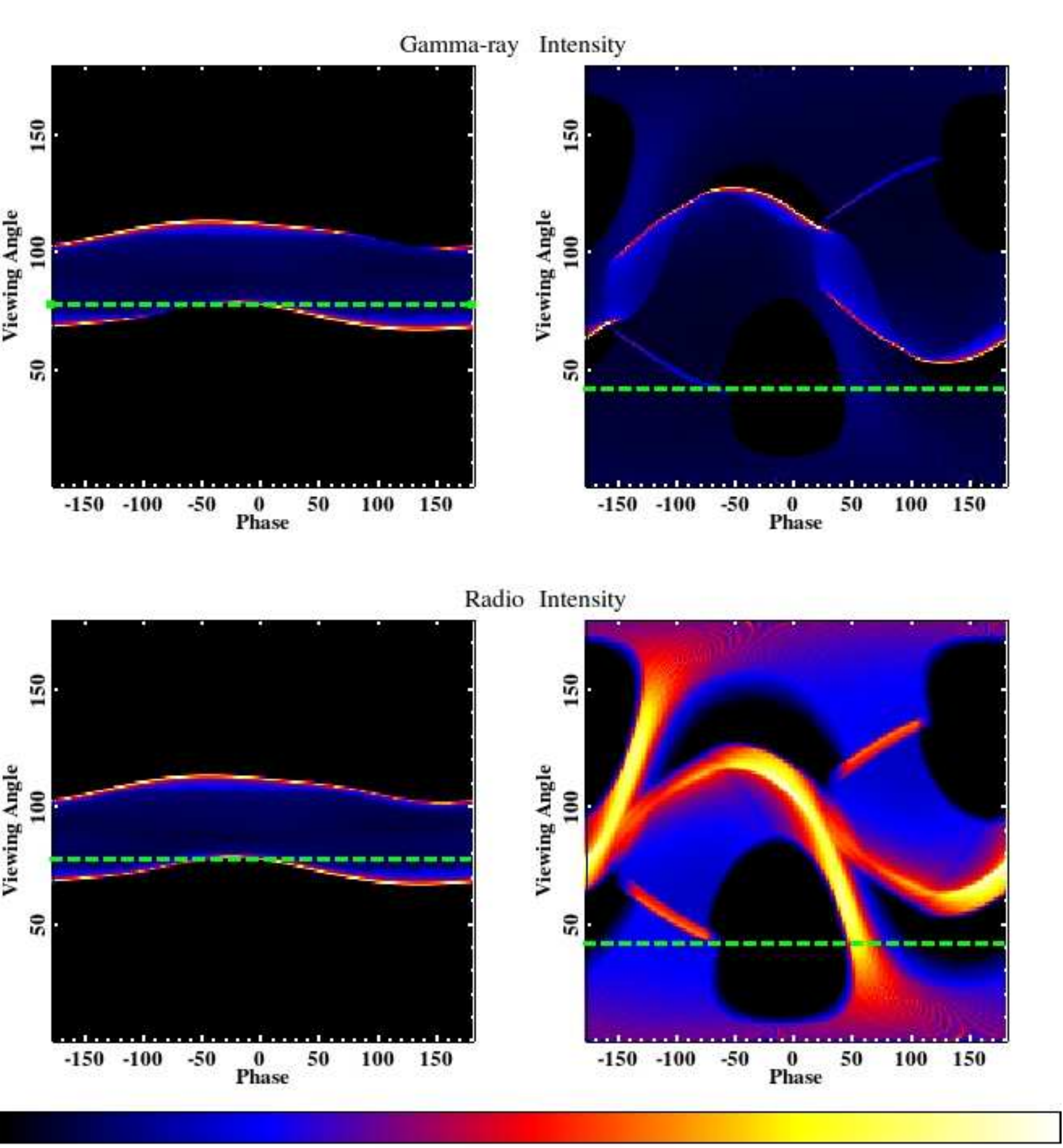}
\end{center}
\small\normalsize
\begin{quote}
\caption[Best-fit phase plots of simulated emission for PSR J1902$-$5105]{Distribution of simulated emission as a function of viewing angle and pulse phase for models used to fit PSR J1902$-$5105.  The best-fit $\zeta$ values are indicated by the dashed green lines.  The top plots correspond to the gamma-ray phase plots while the bottom are for the radio.  The left plots correspond to fits with the alOG model with alTPC plots on the right.\label{appAJ1902PhPlt}}
\end{quote}
\end{figure}

Plots of simulated emission corresponding to the best-fit models are shown in Fig.~\ref{appAJ1902PhPlt}, alOG models are on the left and alTPC on the right, gamma-ray models are on the top and radio on the bottom.

\small\normalsize

\section{PSR J1939+2134}\label{appAJ1939}
PSR J1939+2134 (a.k.a PSR B1937+21) was the first MSP ever discovered by \citet{Backer82}.  This is an isolated MSP with a 1.5578 ms period.  Gamma-ray pulsations will be reported in a forthcoming publication \citep{Guillemot11}.  They also fit the gamma-ray and radio profiles of this MSP with the alOG ad alTPC models using the MCMC likelihood technique described in this thesis.

\begin{figure}
\begin{center}
\includegraphics[width=0.75\textwidth]{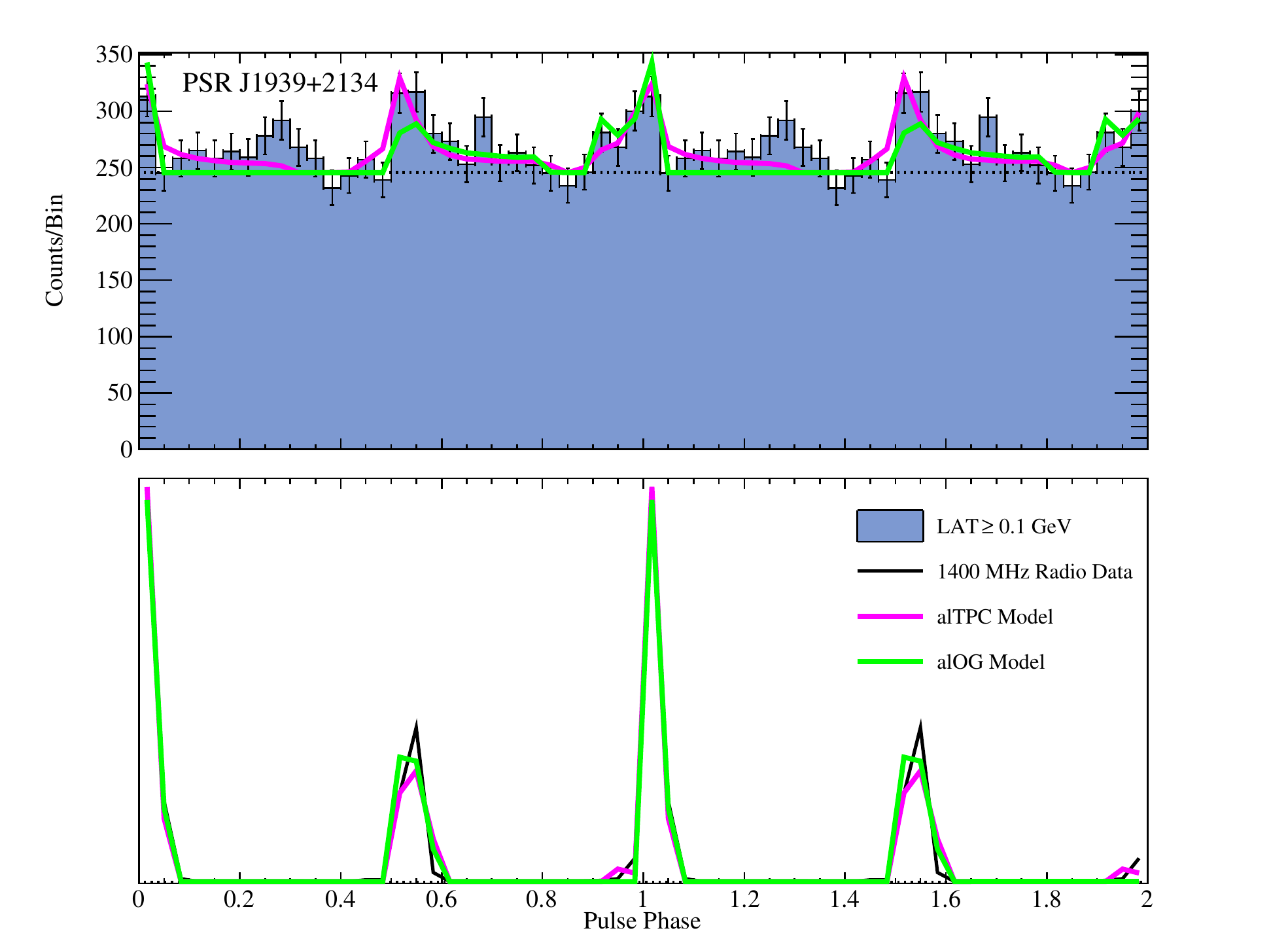}
\end{center}
\small\normalsize
\begin{quote}
\caption[Data and best-fit light curves for PSR J1939+2134]{Best-fit gamma-ray and radio light curves for PSR J1939+2134 using the alTPC and alOG models.\label{appAJ1939LCs}}
\end{quote}
\end{figure}

The best-fit gamma-ray and radio light curves are shown in Fig.~\ref{appAJ1939LCs}.  The gamma-ray and radio light curves have been fit using the alTPC and alOG models.  These light curve fits have used the 1400 MHz Nan\c{cay} radio profile.

\begin{figure}
\begin{center}
\includegraphics[width=0.75\textwidth]{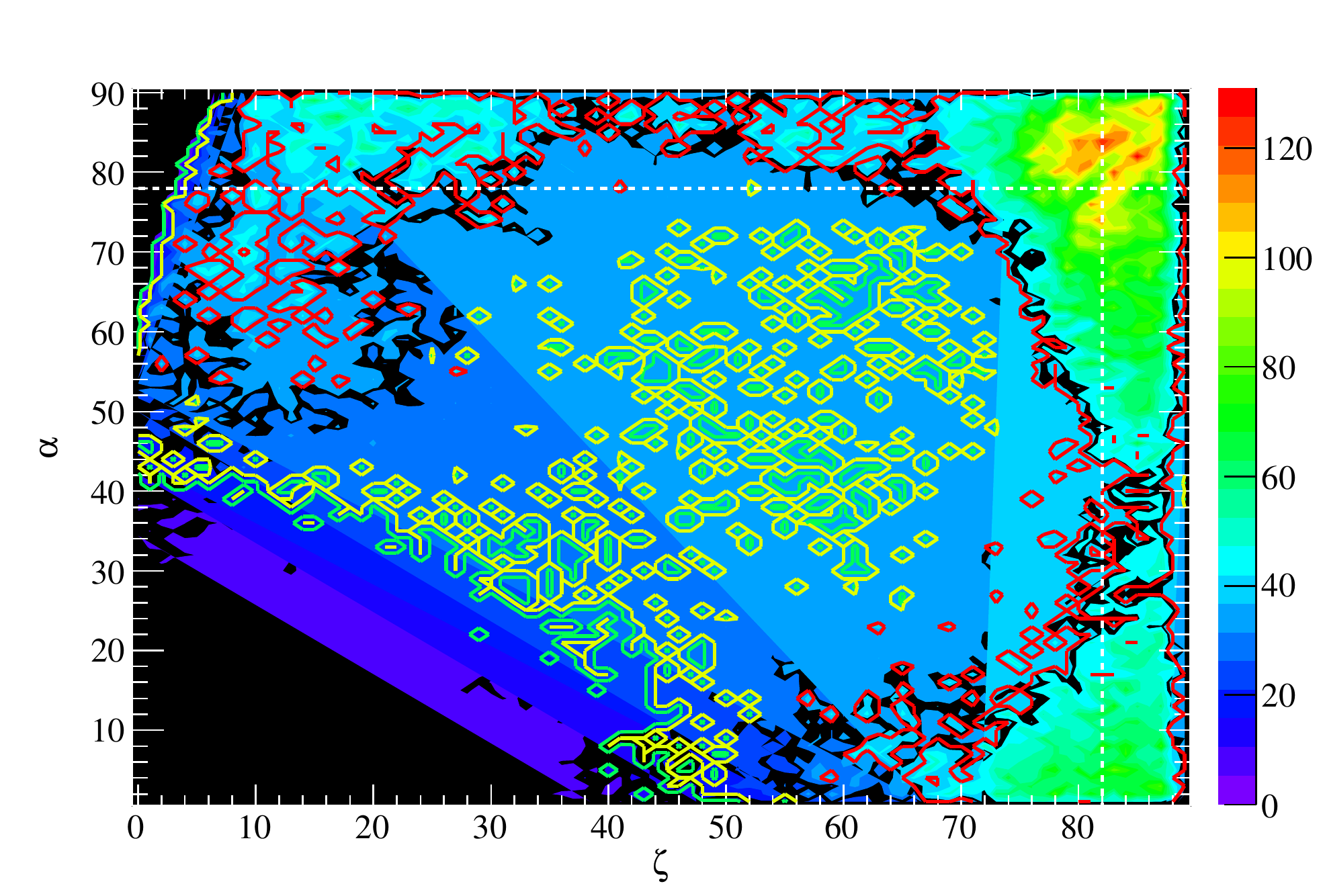}
\end{center}
\small\normalsize
\begin{quote}
\caption[Best-fit alTPC contours for PSR J1939+2134]{Marginalized confidence contours for PSR J1939+2134 for the alTPC model.\label{appAJ1939TPCcont}}
\end{quote}
\end{figure}

The marginalized $\alpha$-$\zeta$ confidence contours corresponding to the alTPC fit are shown in Fig.~\ref{appAJ1939TPCcont}, the best-fit geometry is indicated by the vertical and horizontal dashed, white lines.

\begin{figure}
\begin{center}
\includegraphics[width=0.75\textwidth]{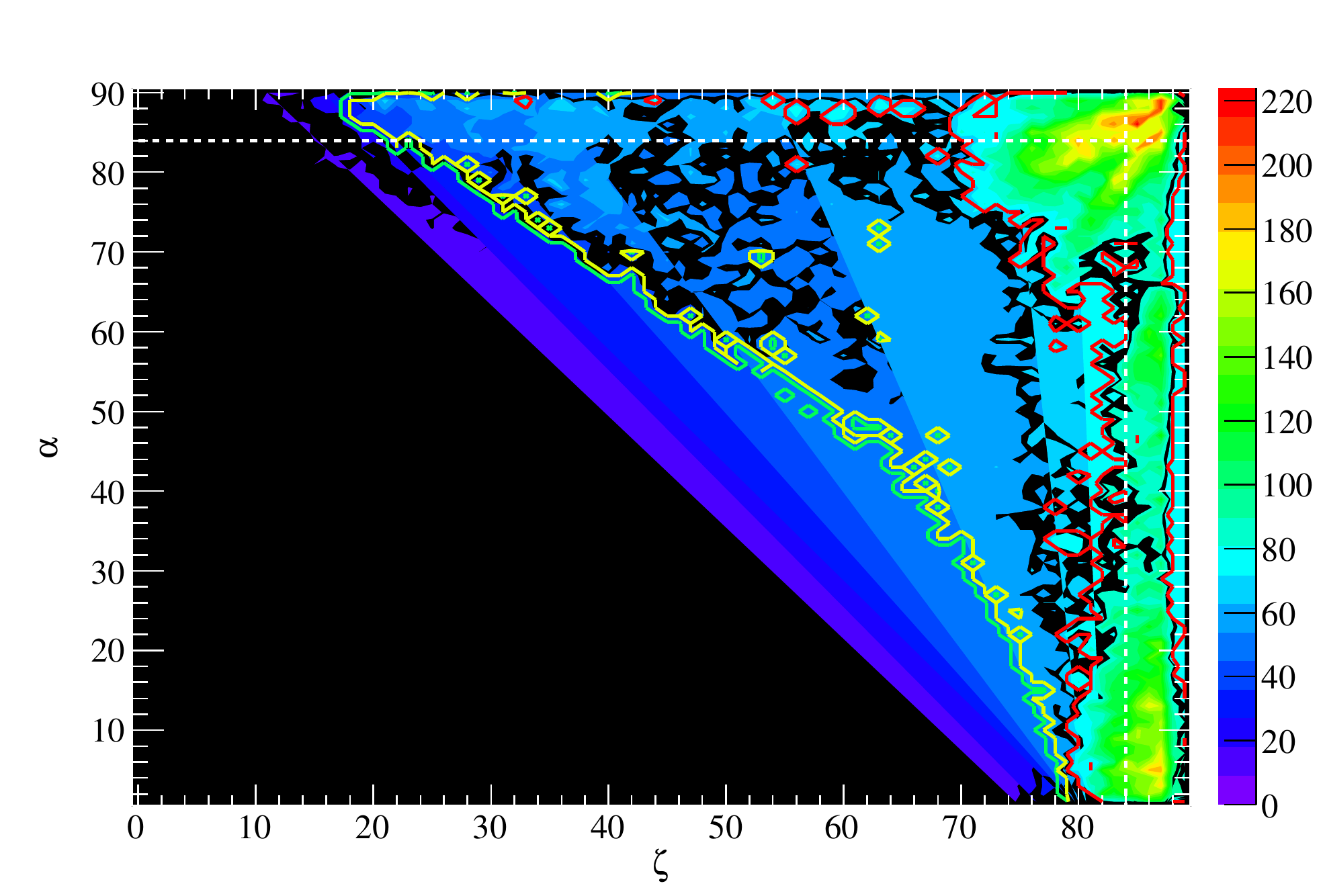}
\end{center}
\small\normalsize
\begin{quote}
\caption[Best-fit alOG contours for PSR J1939+2134]{Marginalized confidence contours for PSR J1939+2134 for the alOG model.\label{appAJ1939OGcont}}
\end{quote}
\end{figure}

The marginalized $\alpha$-$\zeta$ confidence contours corresponding to the alOG fit are shown in Fig.~\ref{appAJ1939OGcont}, the best-fit geometry is indicated by the vertical and horizontal dashed, white lines.

\small\normalsize

\small\normalsize

\small\normalsize

\begin{figure}
\begin{center}
\includegraphics[width=0.75\textwidth]{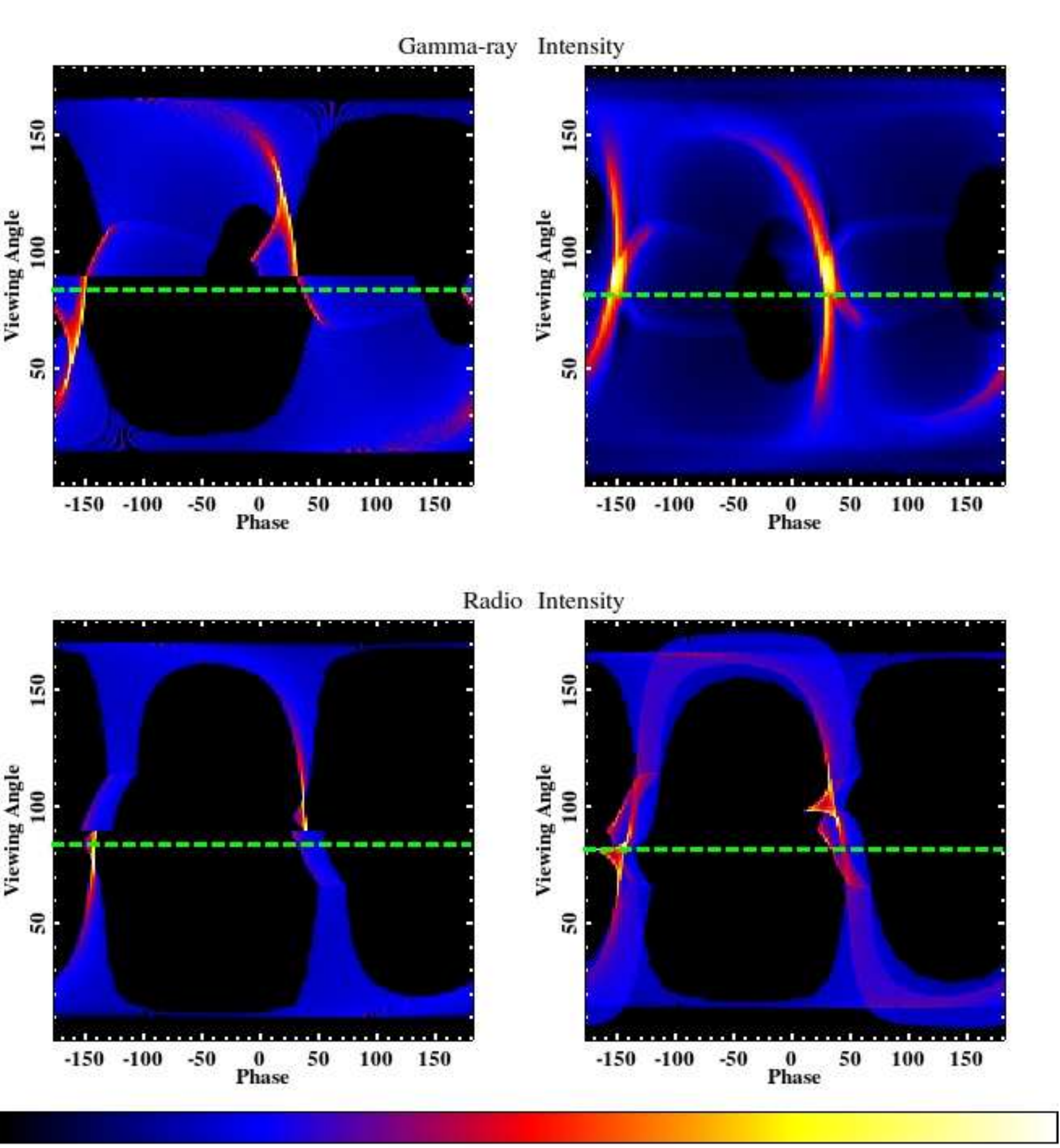}
\end{center}
\small\normalsize
\begin{quote}
\caption[Best-fit phase plots of simulated emission for PSR J1939+2134]{Distribution of simulated emission as a function of viewing angle and pulse phase for models used to fit PSR J1939+2134.  The best-fit $\zeta$ values are indicated by the dashed green lines.  The top plots correspond to the gamma-ray phase plots while the bottom are for the radio.  The left plots correspond to fits with the alOG model with alTPC plots on the right.  The color scales in the top-left and bottom plots are square root in order to bring out fainter features.\label{appAJ1939PhPlt}}
\end{quote}
\end{figure}

Plots of simulated emission corresponding to the best-fit models are shown in Fig.~\ref{appAJ1939PhPlt}, alOG models are on the left and alTPC on the right, gamma-ray models are on the top and radio on the bottom.

\small\normalsize

\section{PSR J1959+2048}\label{appAJ1959}
PSR J1959+2048 was the fist black-widow pulsar ever discovered \citep{Fruchter88}.  This is a 1.6074 ms pulsar in a 9.2 hr orbit with a 0.022 M$_{\odot}$ companion.  Gamma-ray pulsations will be reported in a forthcoming publication \citep{Guillemot11}.  They also fit the gamma-ray and radio light curves with alOG and alTPC models using the MCMC likelihood technique described in this thesis.

\begin{figure}
\begin{center}
\includegraphics[width=0.75\textwidth]{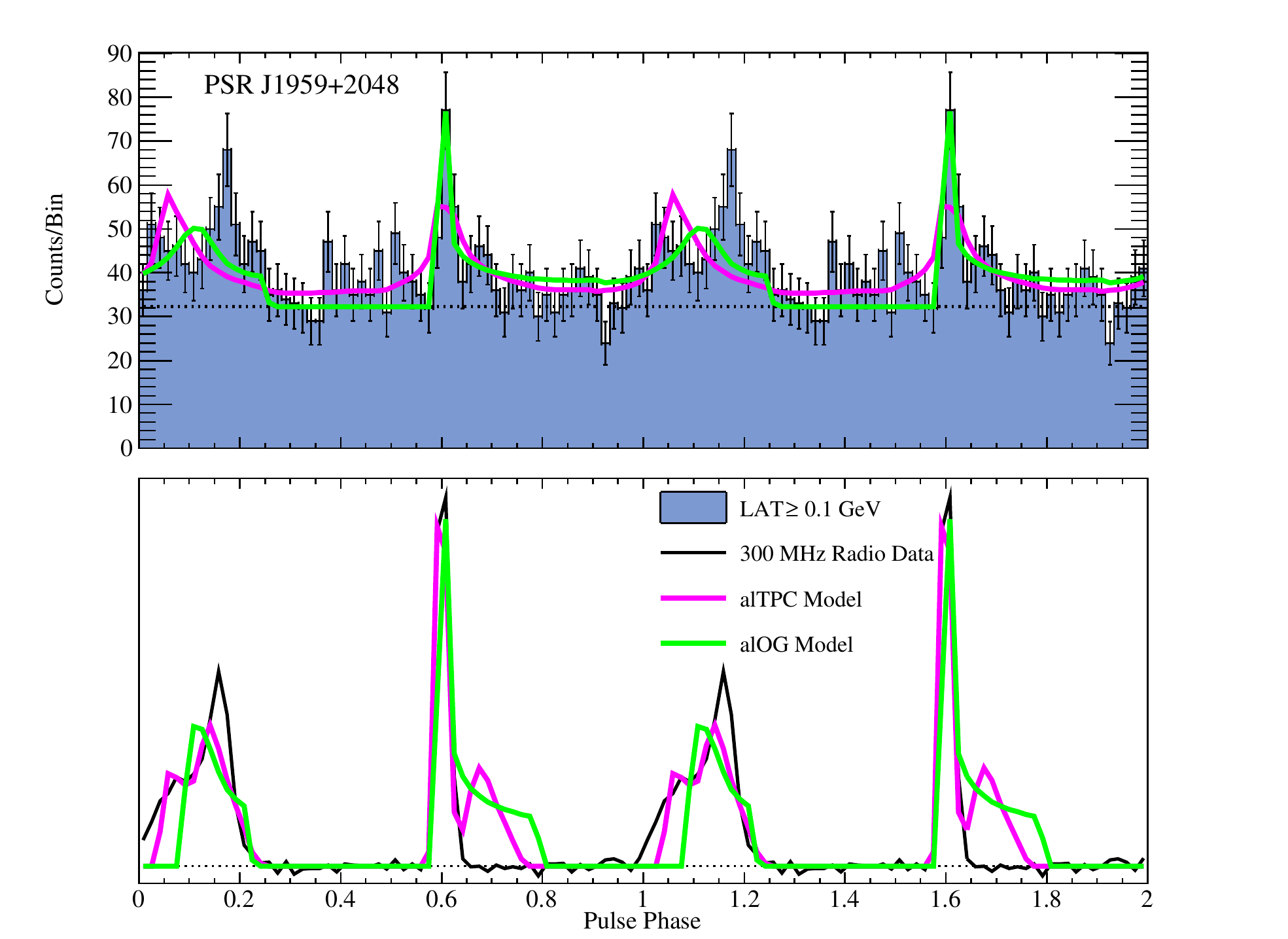}
\end{center}
\small\normalsize
\begin{quote}
\caption[Data and best-fit light curves for PSR J1959+2048]{Best-fit gamma-ray and radio light curves for PSR J1959+2048 using the alTPC and alOG models.\label{appAJ1959LCs}}
\end{quote}
\end{figure}

The best-fit gamma-ray and radio light curves are shown in Fig.~\ref{appAJ1959LCs}.  The gamma-ray and radio light curves have been fit with the alTPC and alOG models.  These light curve fits have used the 300 MHz Westerbork radio profile.

\begin{figure}
\begin{center}
\includegraphics[width=0.75\textwidth]{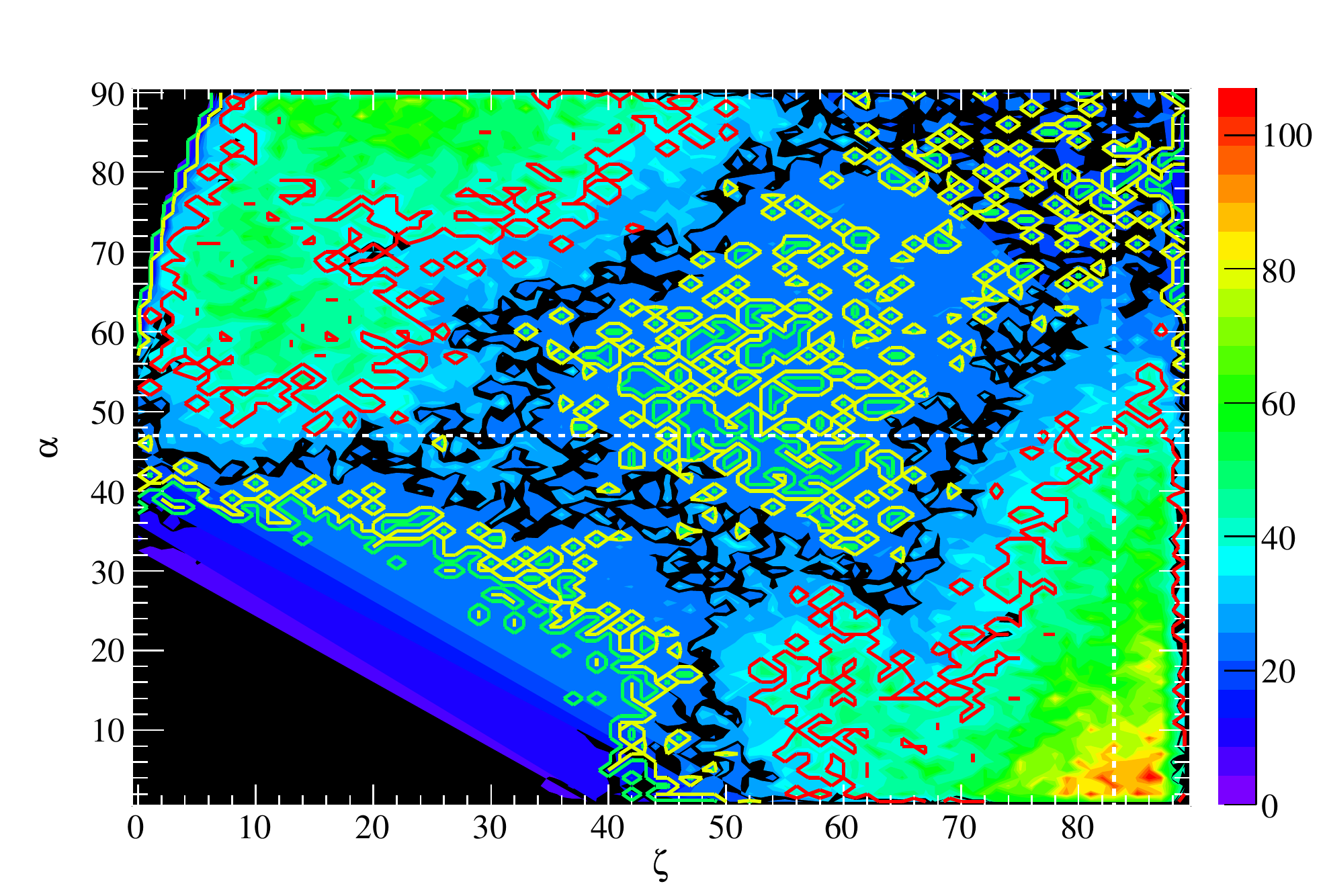}
\end{center}
\small\normalsize
\begin{quote}
\caption[Best-fit alTPC contours for PSR J1959+2048]{Marginalized confidence contours for PSR J1959+2048 for the alTPC model.\label{appAJ1959TPCcont}}
\end{quote}
\end{figure}

The marginalized $\alpha$-$\zeta$ confidence contours corresponding to the alTPC fit are shown in Fig.~\ref{appAJ1959TPCcont}, the best-fit viewing geometry is indicated by the vertical and horizontal dashed, white lines.

\begin{figure}
\begin{center}
\includegraphics[width=0.75\textwidth]{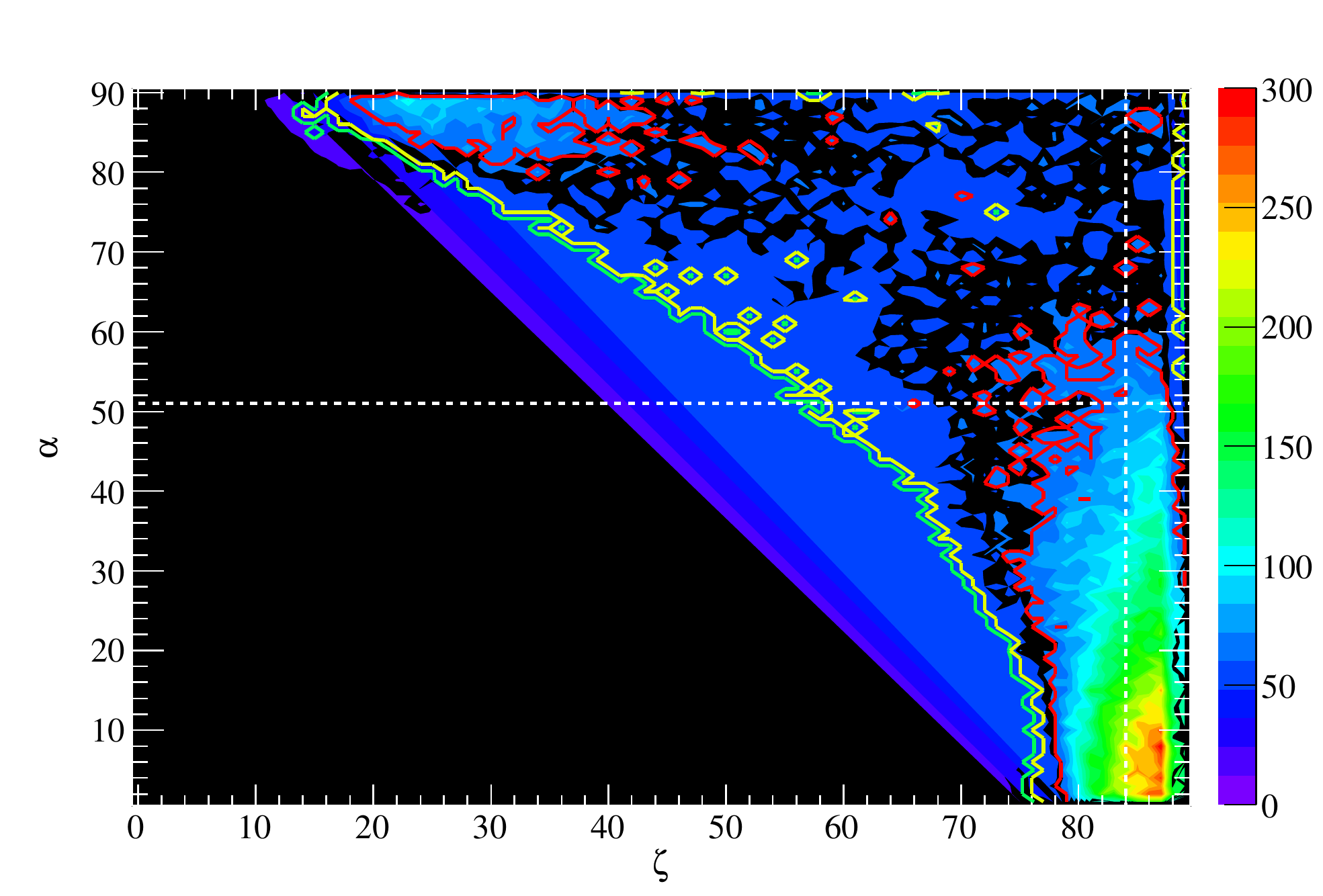}
\end{center}
\small\normalsize
\begin{quote}
\caption[Best-fit alOG contours for PSR J1959+2048]{Marginalized confidence contours for PSR J1959+2048 for the alOG model.\label{appAJ1959OGcont}}
\end{quote}
\end{figure}

The marginalized $\alpha$-$\zeta$ confidence contours corresponding to the alOG fit are shown in Fig.~\ref{appAJ1959OGcont}, the best-fit viewing geometry is indicated by the vertical and horizontal dashed, white lines.

\begin{figure}
\begin{center}
\includegraphics[width=0.75\textwidth]{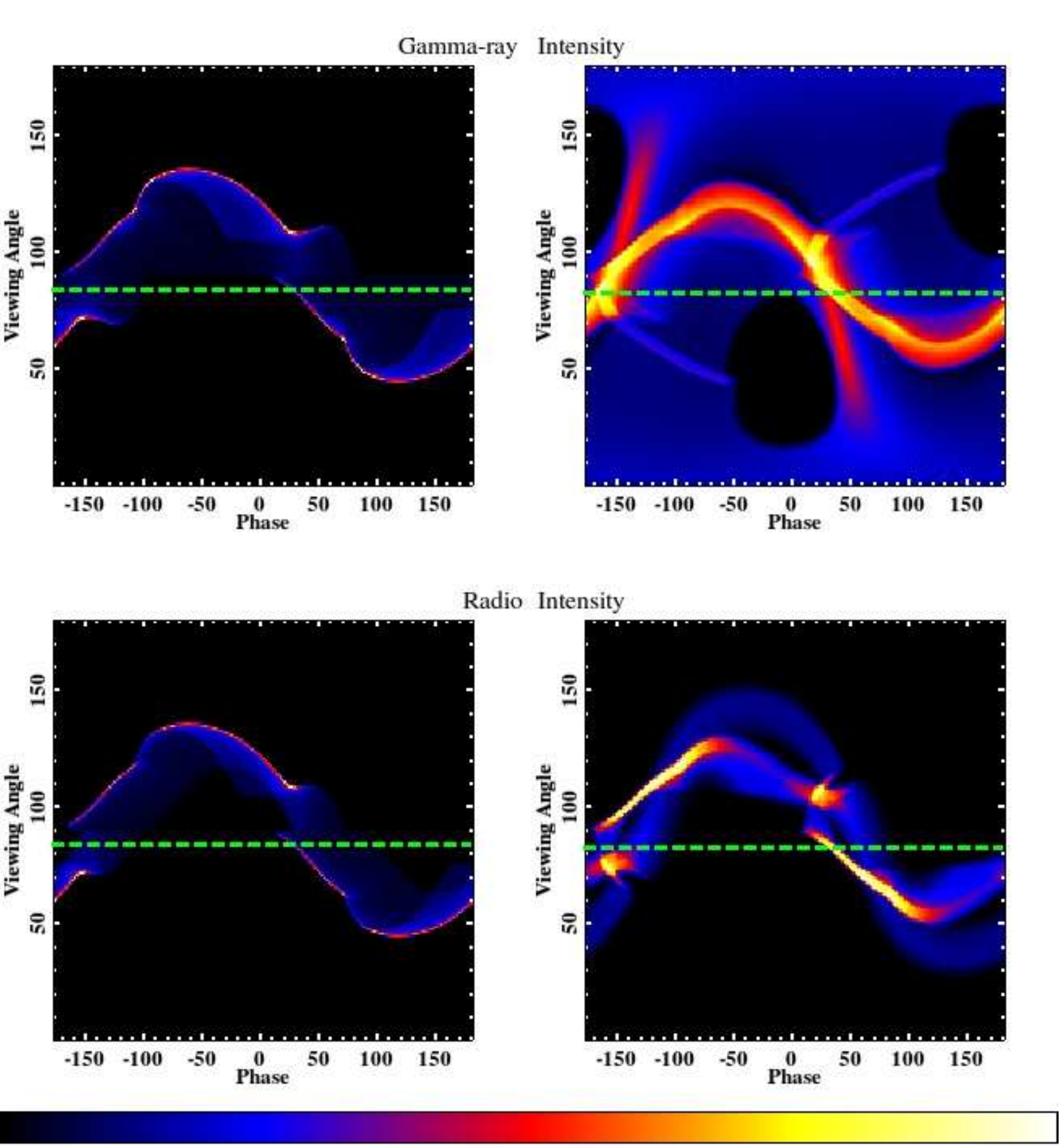}
\end{center}
\small\normalsize
\begin{quote}
\caption[Best-fit phase plots of simulated emission for PSR J1959+2048]{Distribution of simulated emission as a function of viewing angle and pulse phase for models used to fit PSR J1959+2048.  The best-fit $\zeta$ values are indicated by the dashed green lines.  The top plots correspond to the gamma-ray phase plots while the bottom are for the radio.  The left plots correspond to fits with the OG model with TPC plots on the right.\label{appAJ1959PhPlt}}
\end{quote}
\end{figure}

Plots of simulated emission corresponding to the best-fit models are shown in Fig.~\ref{appAJ1959PhPlt}, alOG models are on the left and alTPC on the right, gamma-ray models are on the top and radio on the bottom.

\small\normalsize

\small\normalsize

\small\normalsize

\small\normalsize

\section{PSR J2017+0603}\label{appAJ2017}
PSR J2017+0603 is a 2.8962 ms pulsar in a 2.2 d orbit with a low-mass companion ($\geq$0.18 M$_{\odot}$).  This MSP was discovered in targeted radio observations of unassociated LAT sources with pulsar-like characteristics and seen to pulse in gamma rays soon after \citep{Cognard11}.  They also fit the gamma-ray and radio light curves using the MCMC likelihood technique described in this thesis.  Geometric TPC and OG models were used for the gamma-ray light curve with a hollow-cone beam radio model.

\begin{figure}
\begin{center}
\includegraphics[width=0.75\textwidth]{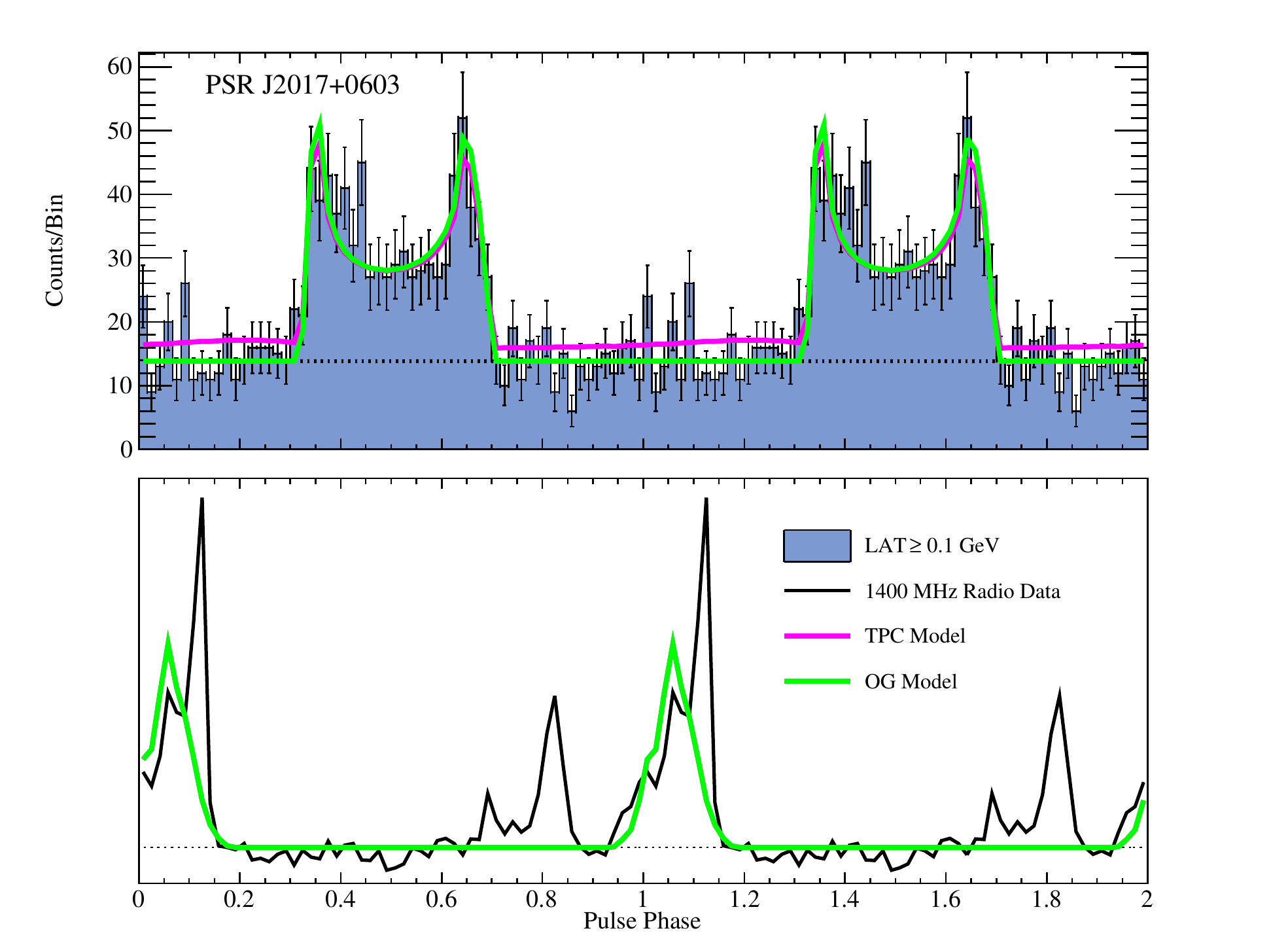}
\end{center}
\small\normalsize
\begin{quote}
\caption[Data and best-fit light curves for PSR J2017+0603]{Best-fit gamma-ray and radio light curves for PSR J2017+0603 using the TPC and OG models.\label{appAJ2017LCs}}
\end{quote}
\end{figure}

The best-fit gamma-ray and radio light curves are shown in Fig.~\ref{appAJ2017LCs}.  The gamma-ray light curve has been fit with TPC and OG models.  The radio profile has been fit with a hollow-cone beam model.  These light curve fits have used the 1400 MHz Nan\c{cay} radio profile.

\begin{figure}
\begin{center}
\includegraphics[width=0.75\textwidth]{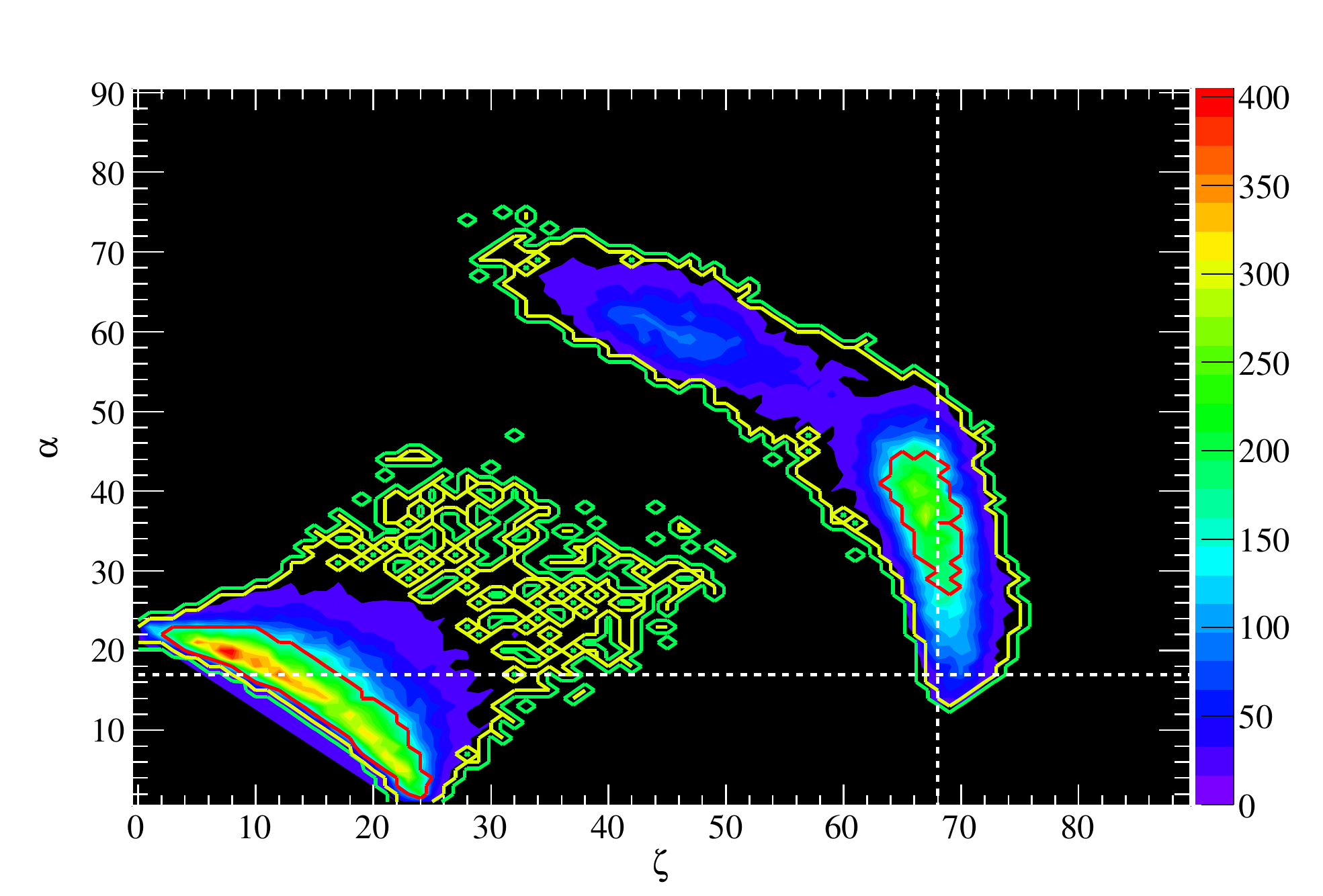}
\end{center}
\small\normalsize
\begin{quote}
\caption[Best-fit TPC contours for PSR J2017+0603]{Marginalized confidence contours for PSR J2017+0603 for the TPC model.\label{appAJ2017TPCcont}}
\end{quote}
\end{figure}

The marginalized $\alpha$-$\zeta$ confidence contours corresponding to the TPC fit are shown in Fig.~\ref{appAJ2017TPCcont}, the best-fit geometry is indicated by the vertical and horizontal dashed, white lines.

\begin{figure}
\begin{center}
\includegraphics[width=0.75\textwidth]{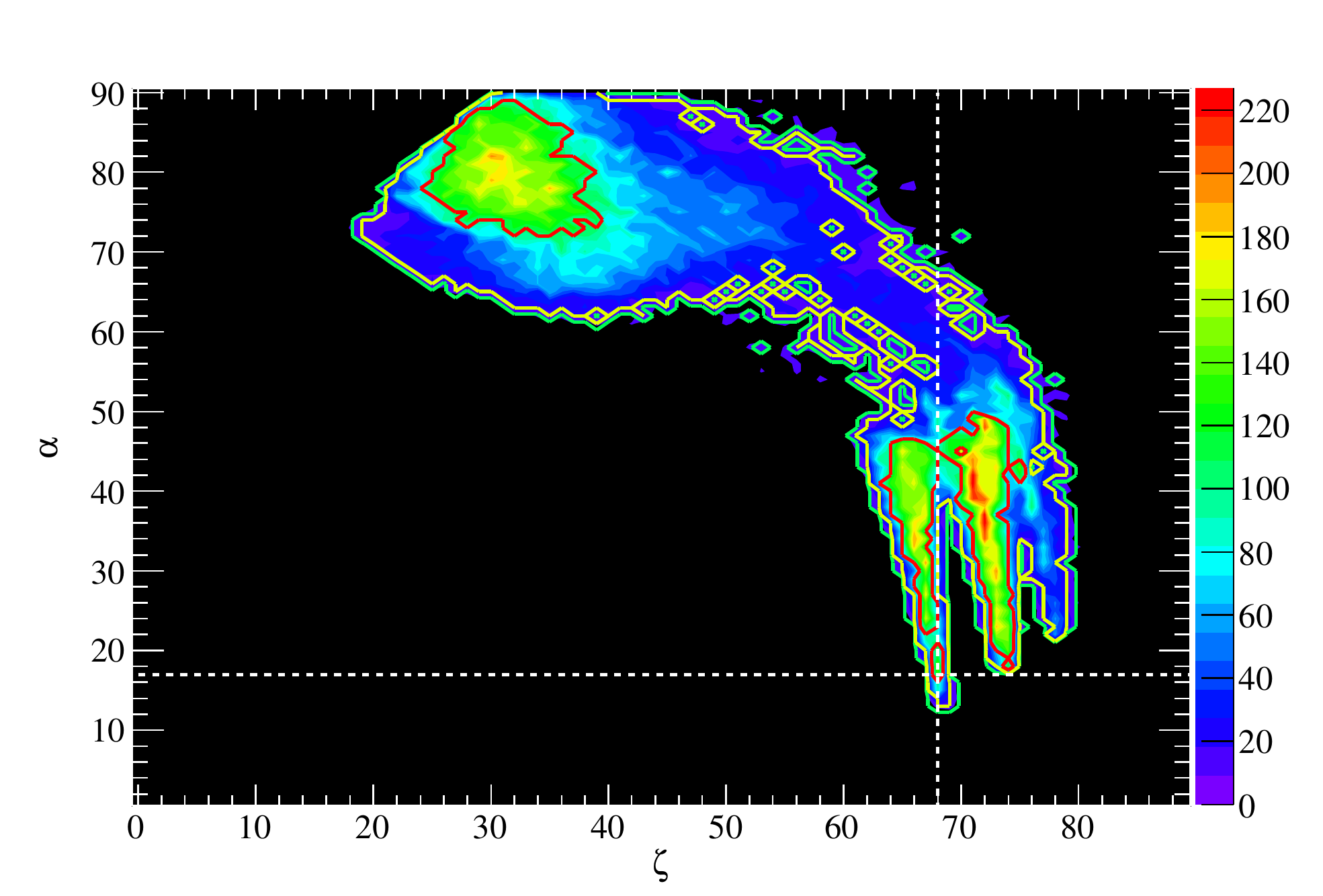}
\end{center}
\small\normalsize
\begin{quote}
\caption[Best-fit OG contours for PSR J2017+0603]{Marginalized confidence contours for PSR J2017+0603 for the OG model.\label{appAJ2017OGcont}}
\end{quote}
\end{figure}

The marginalized $\alpha$-$\zeta$ confidence contours corresponding to the OG fit are shown in Fig.~\ref{appAJ2017OGcont}, the best-fit geometry is indicated by the vertical and horizontal dashed, white lines.

\small\normalsize

\small\normalsize

\small\normalsize

\begin{figure}
\begin{center}
\includegraphics[width=0.75\textwidth]{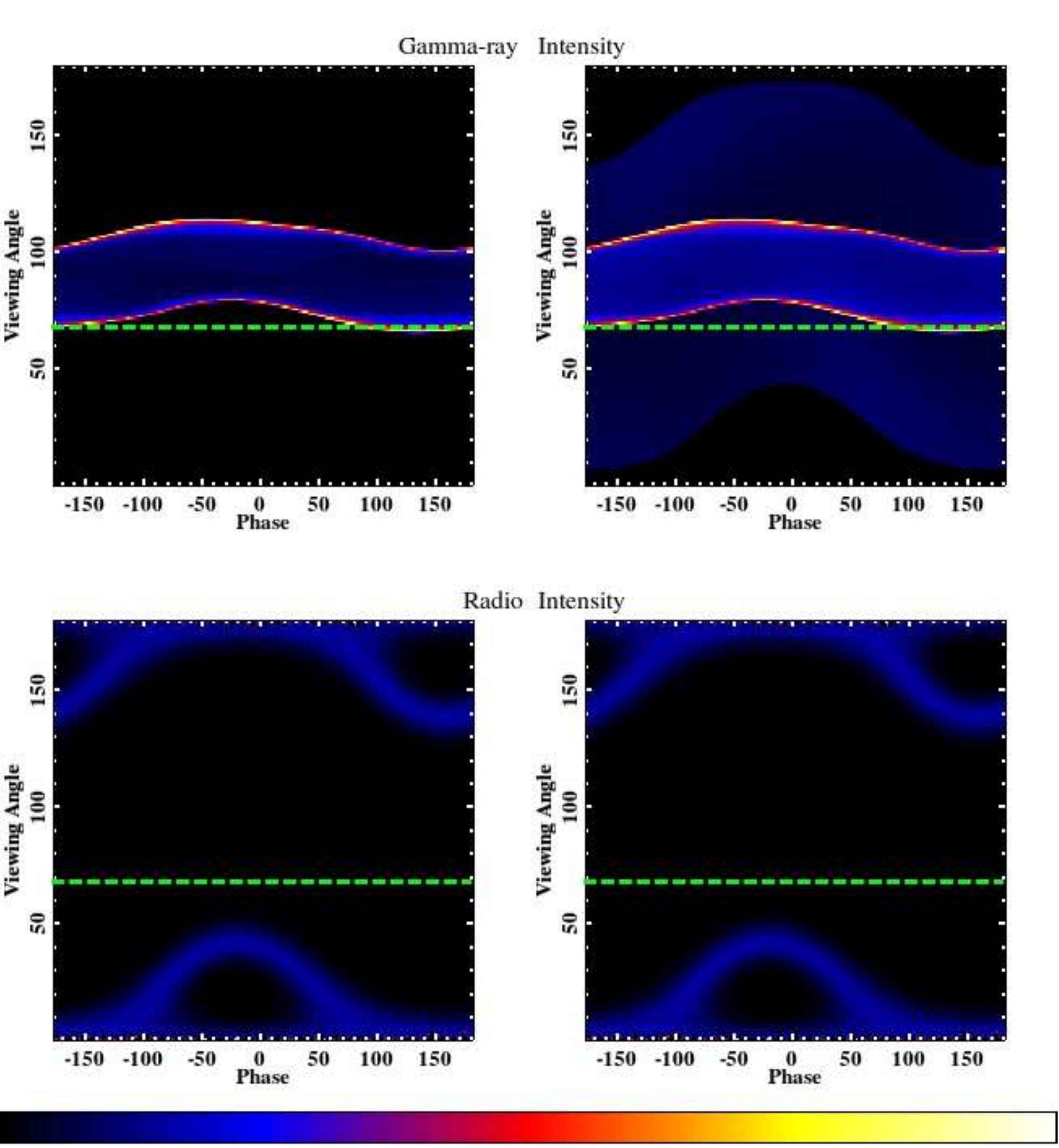}
\end{center}
\small\normalsize
\begin{quote}
\caption[Best-fit phase plots of simulated emission for PSR J2017+0603]{Distribution of simulated emission as a function of viewing angle and pulse phase for models used to fit PSR J2017+0603.  The best-fit $\zeta$ values are indicated by the dashed green lines.  The top plots correspond to the gamma-ray phase plots while the bottom are for the radio.  The left plots correspond to fits with the OG model with TPC plots on the right.\label{appAJ2017PhPlt}}
\end{quote}
\end{figure}

Plots of simulated emission corresponding to the best-fit models are shown in Fig.~\ref{appAJ2017PhPlt}, OG models are on the left and TPC on the right, gamma-ray models are on the top and radio on the bottom.

\small\normalsize

\section{PSR J2124$-$3358}\label{appAJ2124}
PSR J2124$-$3358 is a 4.9311 ms, isolated pulsar first discovered in the radio by \citet{Bailes97}.  Gamma-ray pulsations were first reported from this MSP by \citet{AbdoMSPpop} and later by \citet{AbdoPSRcat}.  The gamma-ray light curve of this MSP was modeled by \citet{Venter09} using a PSPC model with hollow-cone beam radio model.

\begin{figure}
\begin{center}
\includegraphics[width=0.75\textwidth]{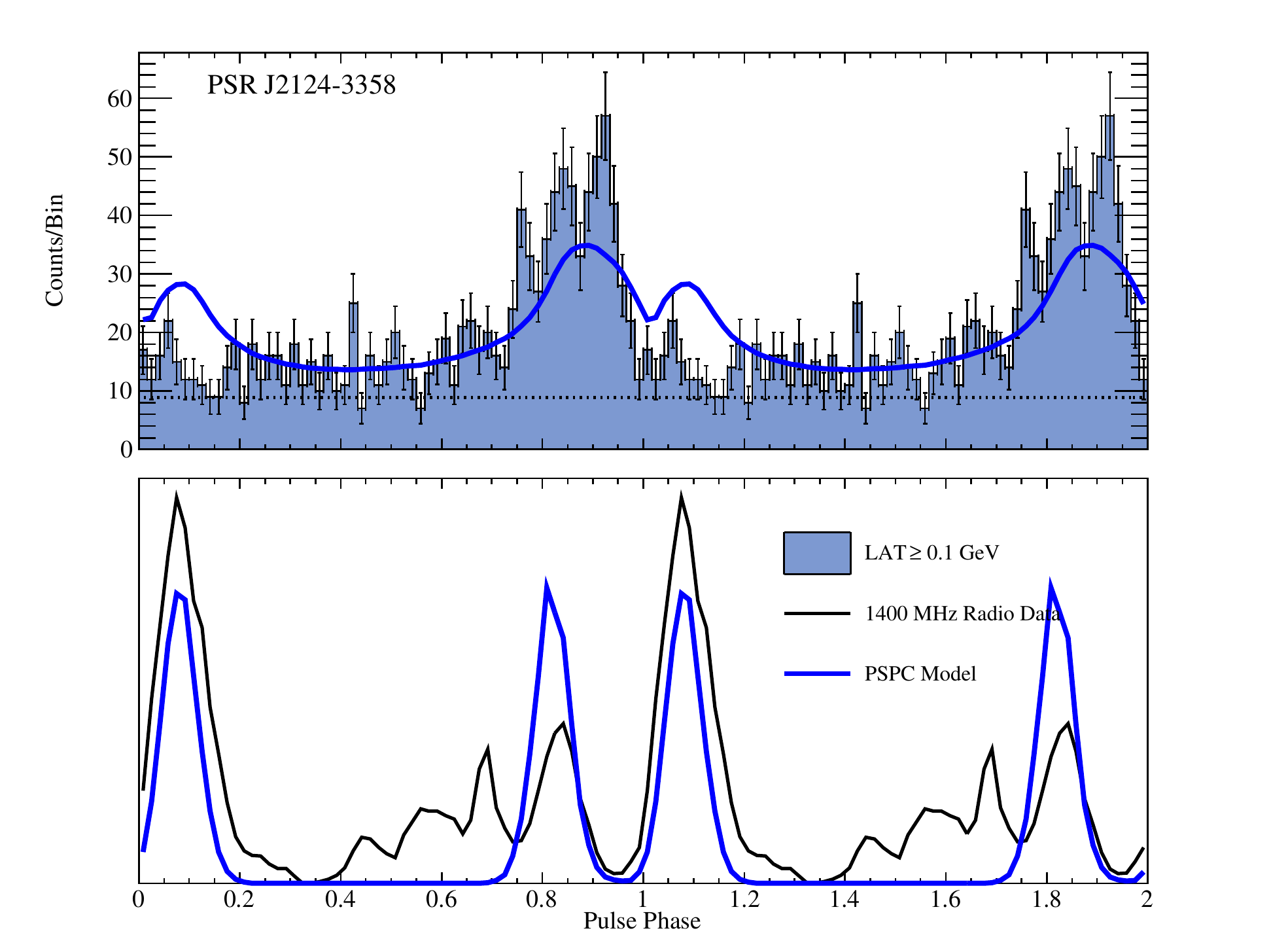}
\end{center}
\small\normalsize
\begin{quote}
\caption[Data and best-fit light curves for PSR J2124$-$3358]{Best-fit gamma-ray and radio light curves for PSR J2124$-$3358 using the PSPC model.\label{appAJ2124LCs}}
\end{quote}
\end{figure}

The best-fit gamma-ray and radio light curves are shown in Fig.~\ref{appAJ2124LCs}.  The gamma-ray light curve has been fit with a PSPC model.  The radio profile has been fit with a hollow-cone beam model.  These light curve fits have used the 1400 MHz Nan\c{cay} radio profile.

\begin{figure}
\begin{center}
\includegraphics[width=0.75\textwidth]{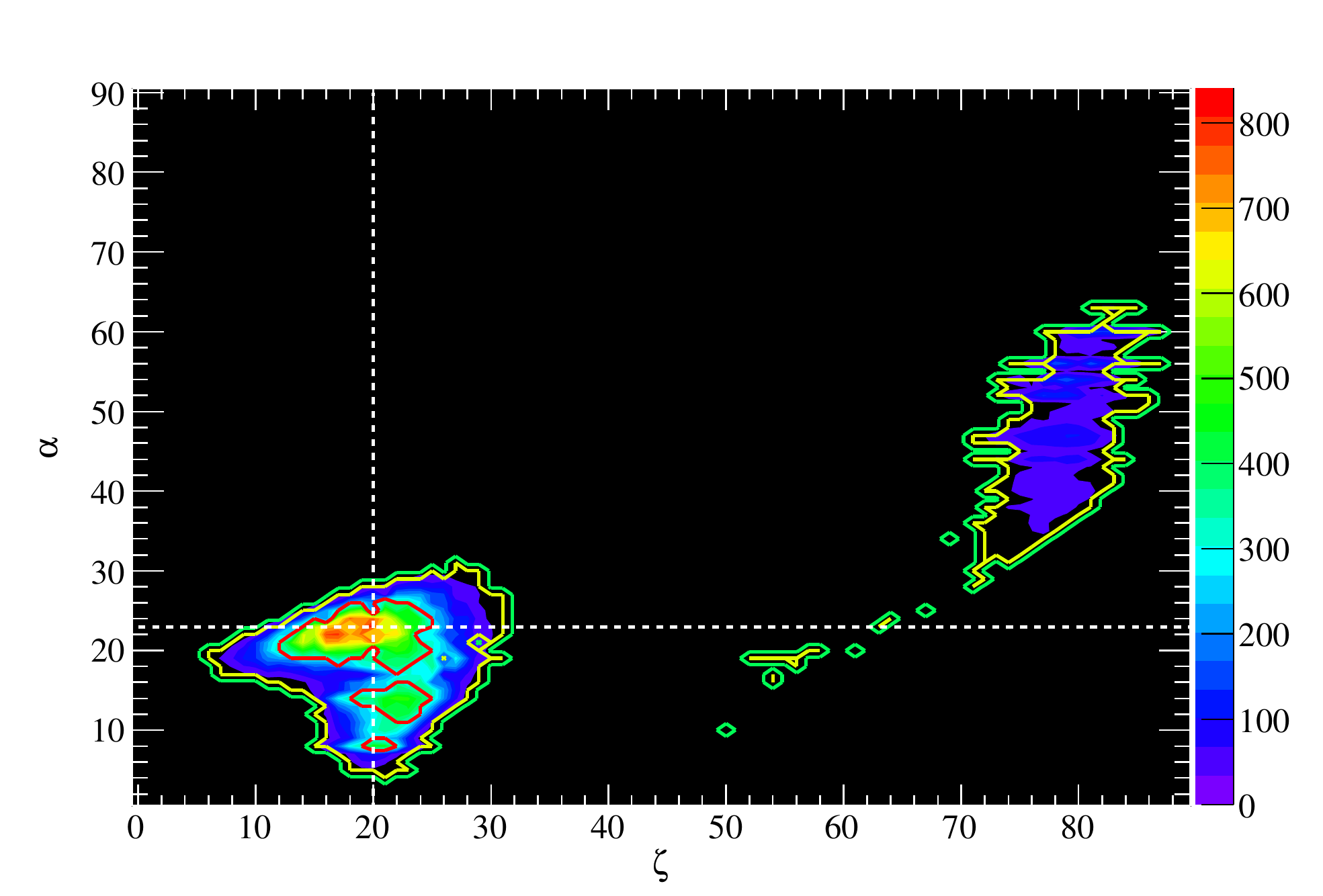}
\end{center}
\small\normalsize
\begin{quote}
\caption[Best-fit PSPC contours for PSR J2124$-$3358]{Marginalized confidence contours for PSR J2124$-$3358 for the PSPC model.\label{appAJ2124PSPCcont}}
\end{quote}
\end{figure}

The marginalized $\alpha$-$\zeta$ confidence contours corresponding to the PSPC fit are shown in Fig.~\ref{appAJ2124PSPCcont}, the best-fit geometry is indicated by the vertical and horizontal dashed, white lines.

\small\normalsize

\small\normalsize

\begin{figure}
\begin{center}
\includegraphics[width=0.5\textwidth]{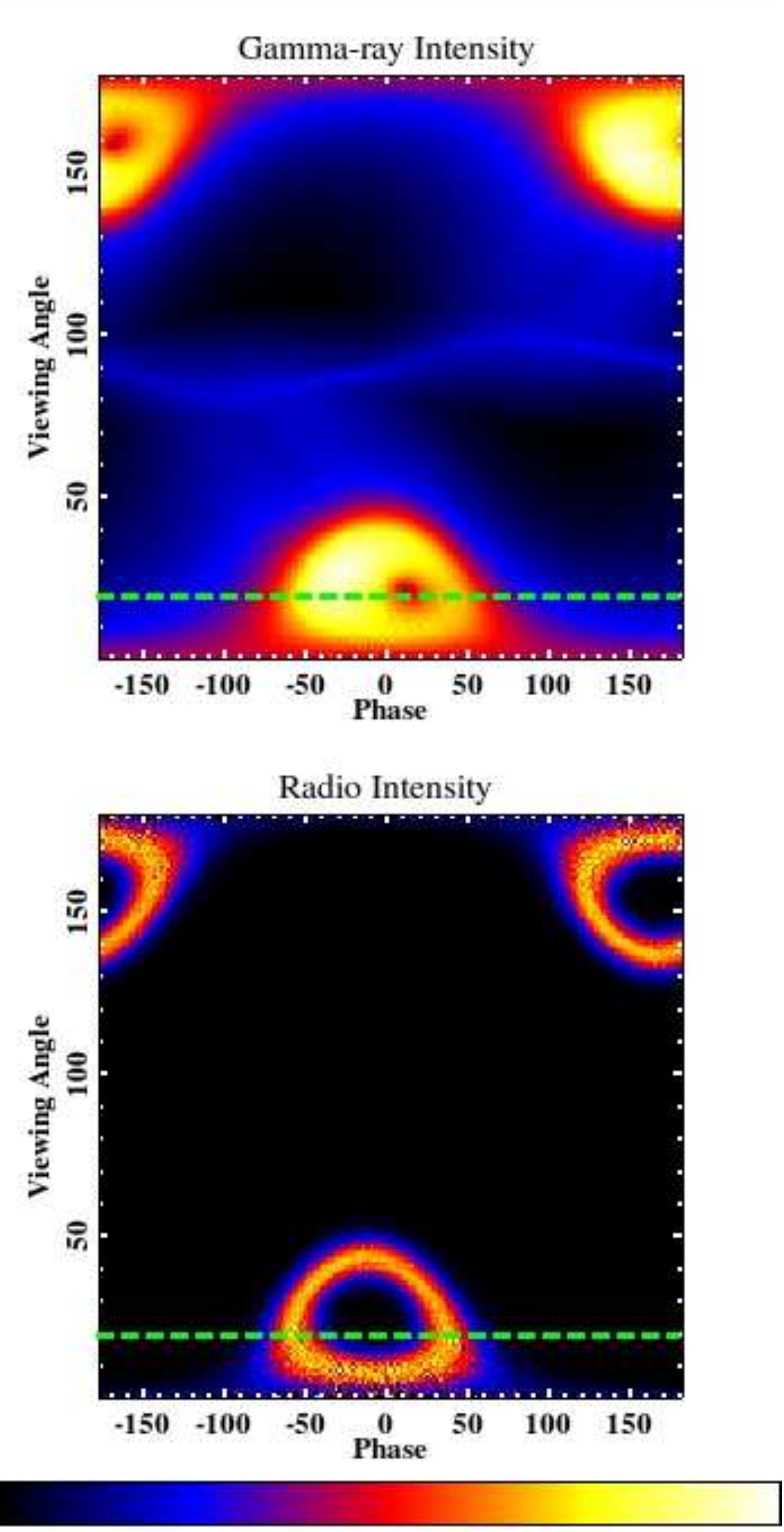}
\end{center}
\small\normalsize
\begin{quote}
\caption[Best-fit phase plots of simulated emission for PSR J2124$-$3358]{Distribution of simulated emission as a function of viewing angle and pulse phase for models used to fit PSR J2124$-$3358 with the PSPC model.  The best-fit $\zeta$ value is indicated by the dashed green lines.  The top plot corresponds to the gamma-ray phase plot while the bottom is for the radio.\label{appAJ2124PhPlt}}
\end{quote}
\end{figure}

Plots of simulated emission corresponding to the best-fit models are shown in Fig.~\ref{appAJ2124PhPlt}, gamma-ray models are on the top and radio on the bottom.

\small\normalsize

\section{PSR J2214+3000}\label{appAJ2214}
PSR J2214+3000 is a 3.1192 ms pulsar in a 0.4 d orbit with a low-mass companion ($\geq$0.014 M$_{\odot}$).  This MSP was discovered in targeted radio observations of unassociated LAT sources with pulsar-like characteristics and seen to pulse in gamma rays soon after \citep{Ransom11}.

\begin{figure}
\begin{center}
\includegraphics[width=0.75\textwidth]{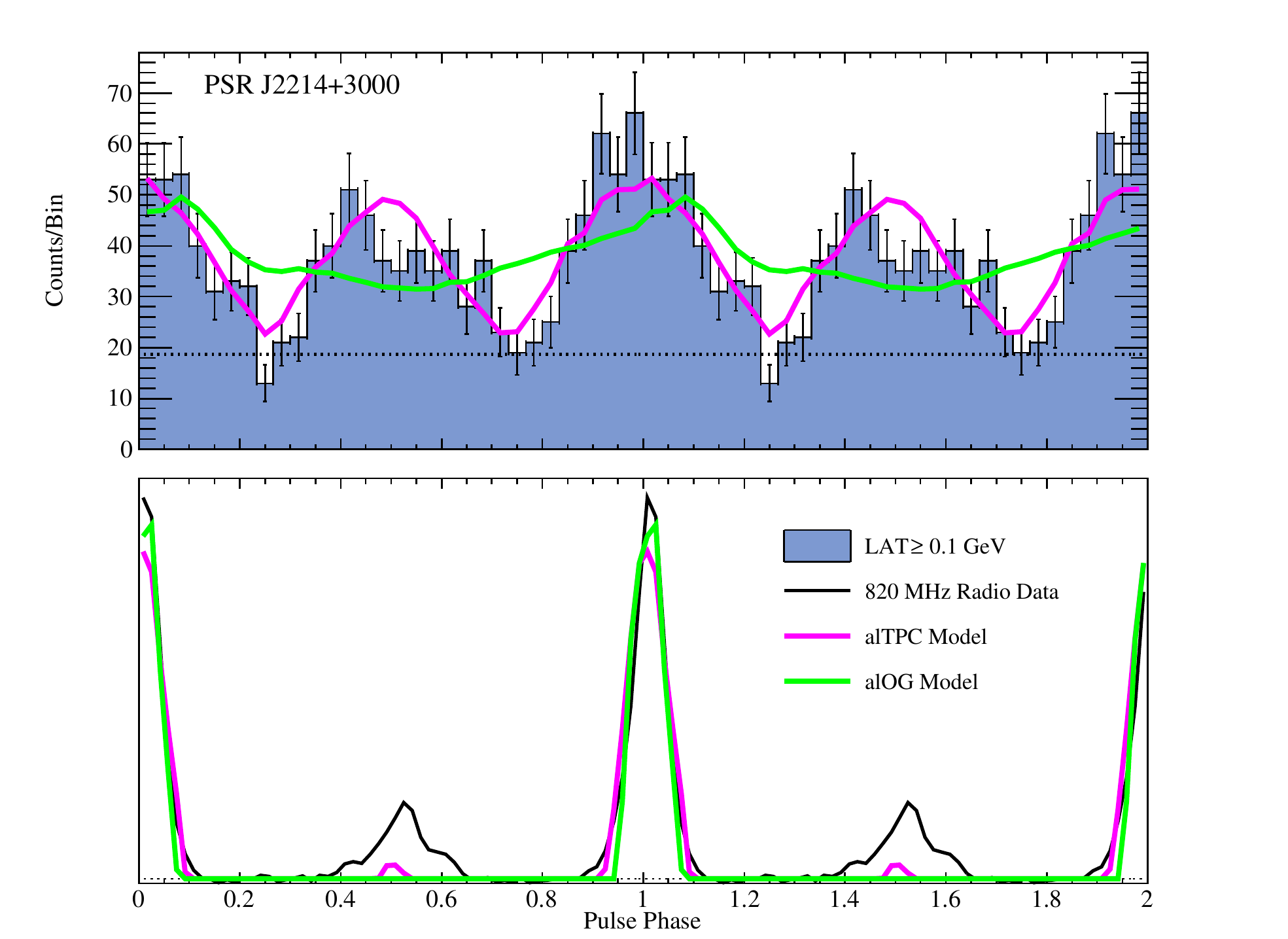}
\end{center}
\small\normalsize
\begin{quote}
\caption[Data and best-fit light curves for PSR J2214+3000]{Best-fit gamma-ray and radio light curves for PSR J2214+3000 using the alTPC and alOG models.\label{appAJ2214LCs}}
\end{quote}
\end{figure}

The best-fit gamma-ray and radio light curves are shown in Fig.~\ref{appAJ2214LCs}.  The gamma-ray and radio profiles have been fit with the alTPC and alOG models.  These light curve fits have used the 820 Greenbank radio profile.

\small\normalsize

\begin{figure}
\begin{center}
\includegraphics[width=0.75\textwidth]{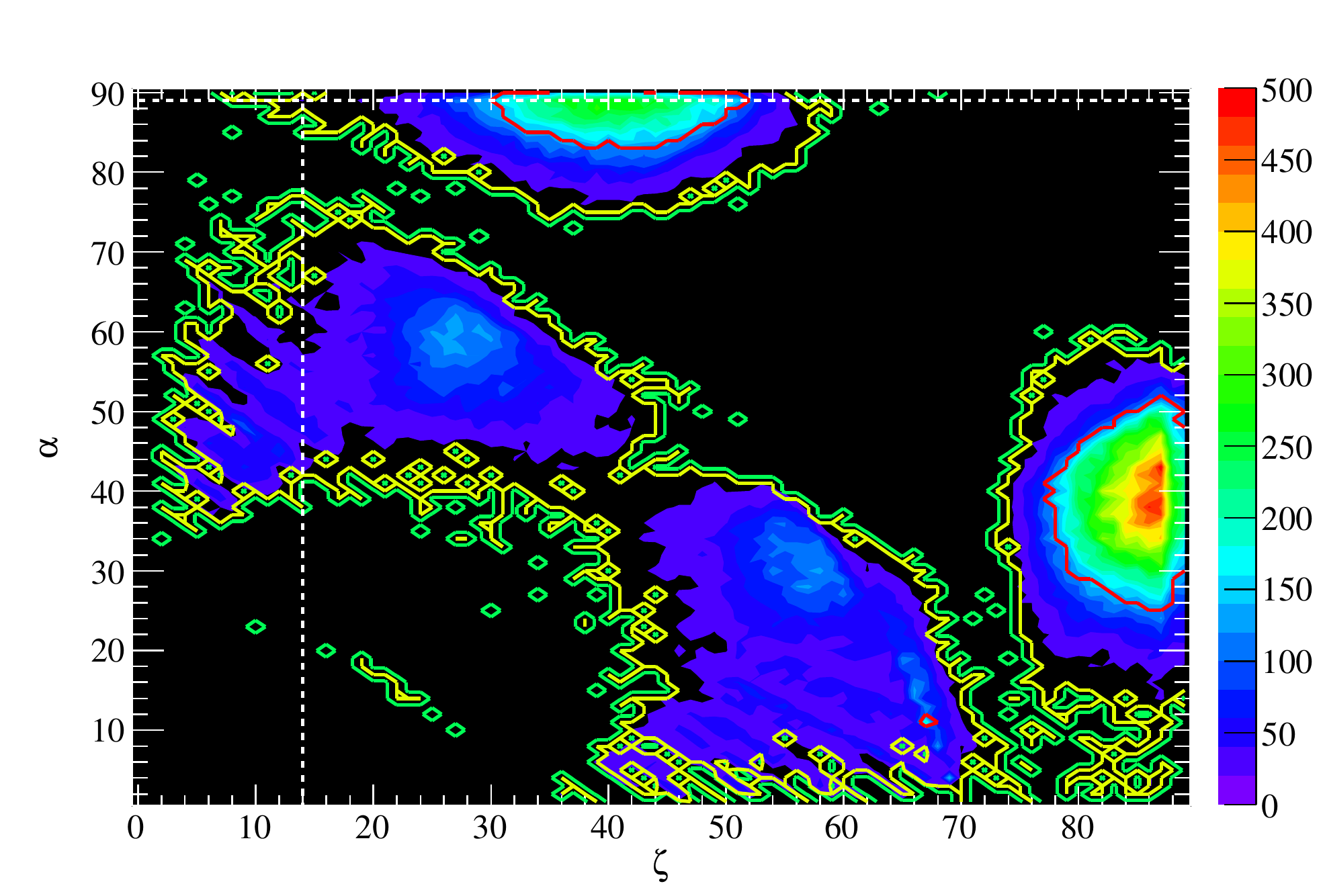}
\end{center}
\small\normalsize
\begin{quote}
\caption[Best-fit alTPC contours for PSR J2214+3000]{Marginalized confidence contours for PSR J2214+3000 for the alTPC model.\label{appAJ2214TPCcont}}
\end{quote}
\end{figure}

The marginalized $\alpha$-$\zeta$ confidence contours corresponding to the alTPC fit are shown in Fig.~\ref{appAJ2214TPCcont}, the best-fit geometry is indicated by the vertical and horizontal dashed, white lines.

\small\normalsize

\begin{figure}
\begin{center}
\includegraphics[width=0.75\textwidth]{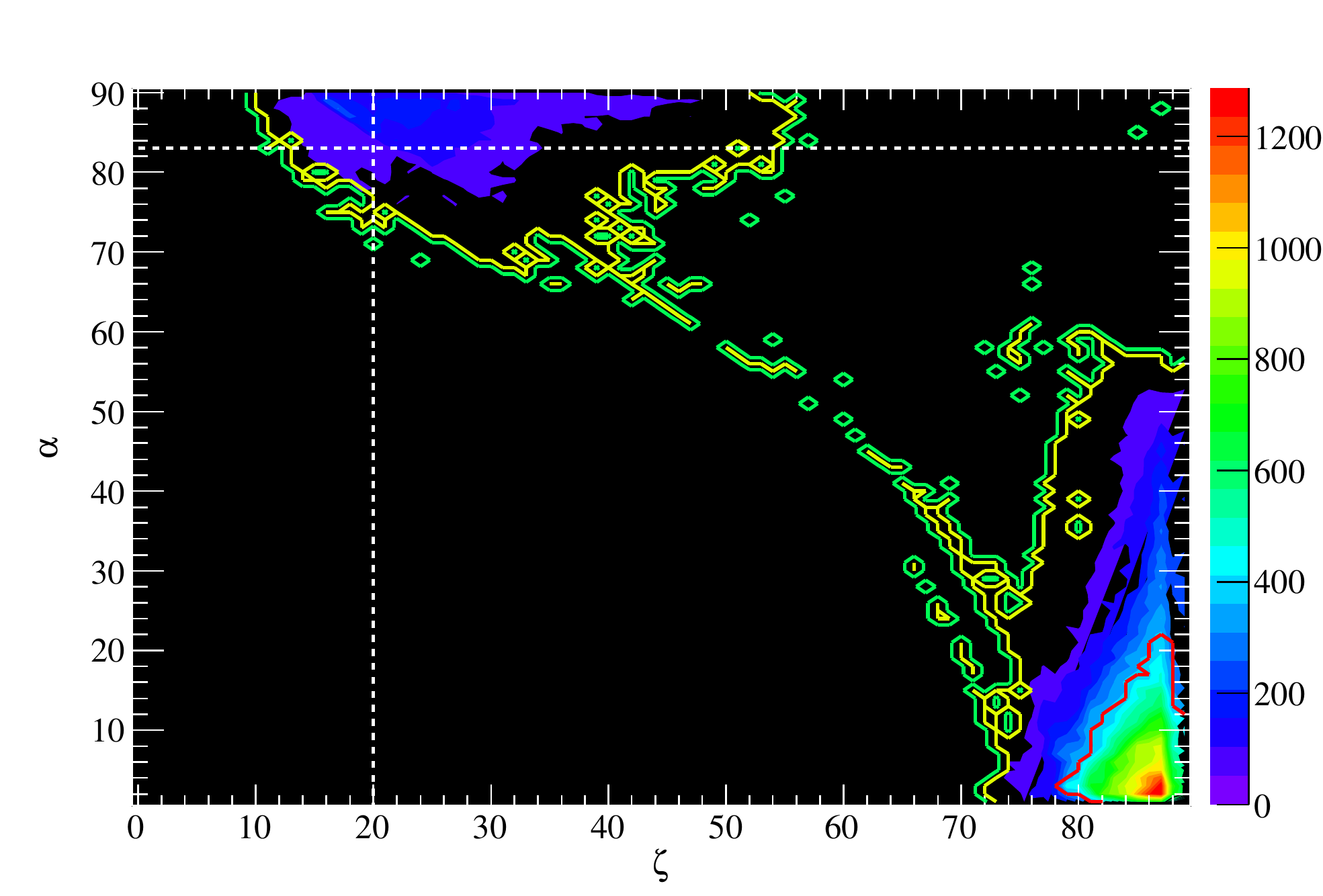}
\end{center}
\small\normalsize
\begin{quote}
\caption[Best-fit alOG contours for PSR J2214+3000]{Marginalized confidence contours for PSR J2214+3000 for the alOG model.\label{appAJ2214OGcont}}
\end{quote}
\end{figure}

The marginalized $\alpha$-$\zeta$ confidence contours corresponding to the alOG fit is shown in Fig.~\ref{appAJ2214OGcont}, the best-fit geometry is indicated by the vertical and horizontal dashed, white lines.

\small\normalsize

\begin{figure}
\begin{center}
\includegraphics[width=0.75\textwidth]{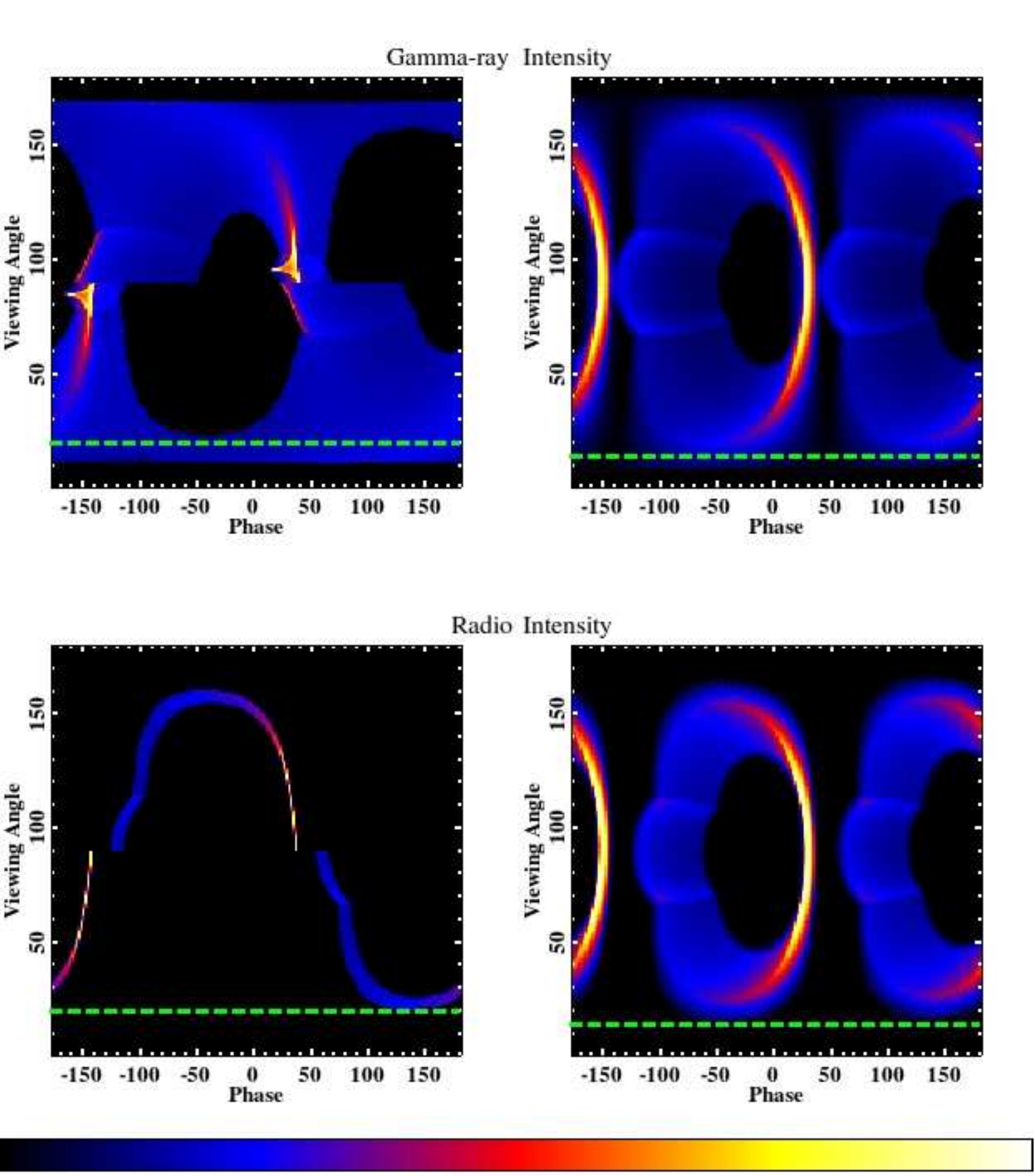}
\end{center}
\small\normalsize
\begin{quote}
\caption[Best-fit phase plots of simulated emission for PSR J2214+3000]{Distribution of simulated emission as a function of viewing angle and pulse phase for models used to fit PSR J2214+3000.  The best-fit $\zeta$ values are indicated by the dashed green lines.  The top plots correspond to the gamma-ray phase plots while the bottom are for the radio.  The left plots correspond to fits with the OG model with TPC plots on the right.  The color scale in the top-left plot is square root in order to bring out fainter features.\label{appAJ2214PhPlt}}
\end{quote}
\end{figure}

Plots of simulated emission corresponding to the best-fit models are shown in Fig.~\ref{appAJ2214PhPlt}, alOG models are on the left and alTPC on the right, gamma-ray models are on the top and radio on the bottom.

\small\normalsize

\section{PSR J2302+4442}\label{appAJ2302}
PSR J2302+4442 is a 5.1932 ms pulsar in a 51.4 d orbit with a 0.30 M$_{\odot}$ companion.  This MSP was discovered in targeted radio observations of unassociated LAT sources with pulsar-like characteristics and seen to pulse in gamma rays soon after \citep{Cognard11}.  They also fit the gamma-ray and radio light curves with the MCMC likelihood technique described in this thesis.  Geometric TPC and OG models were used for the gamma-ray light curve with a hollow-cone beam radio model.

\begin{figure}
\begin{center}
\includegraphics[width=0.75\textwidth]{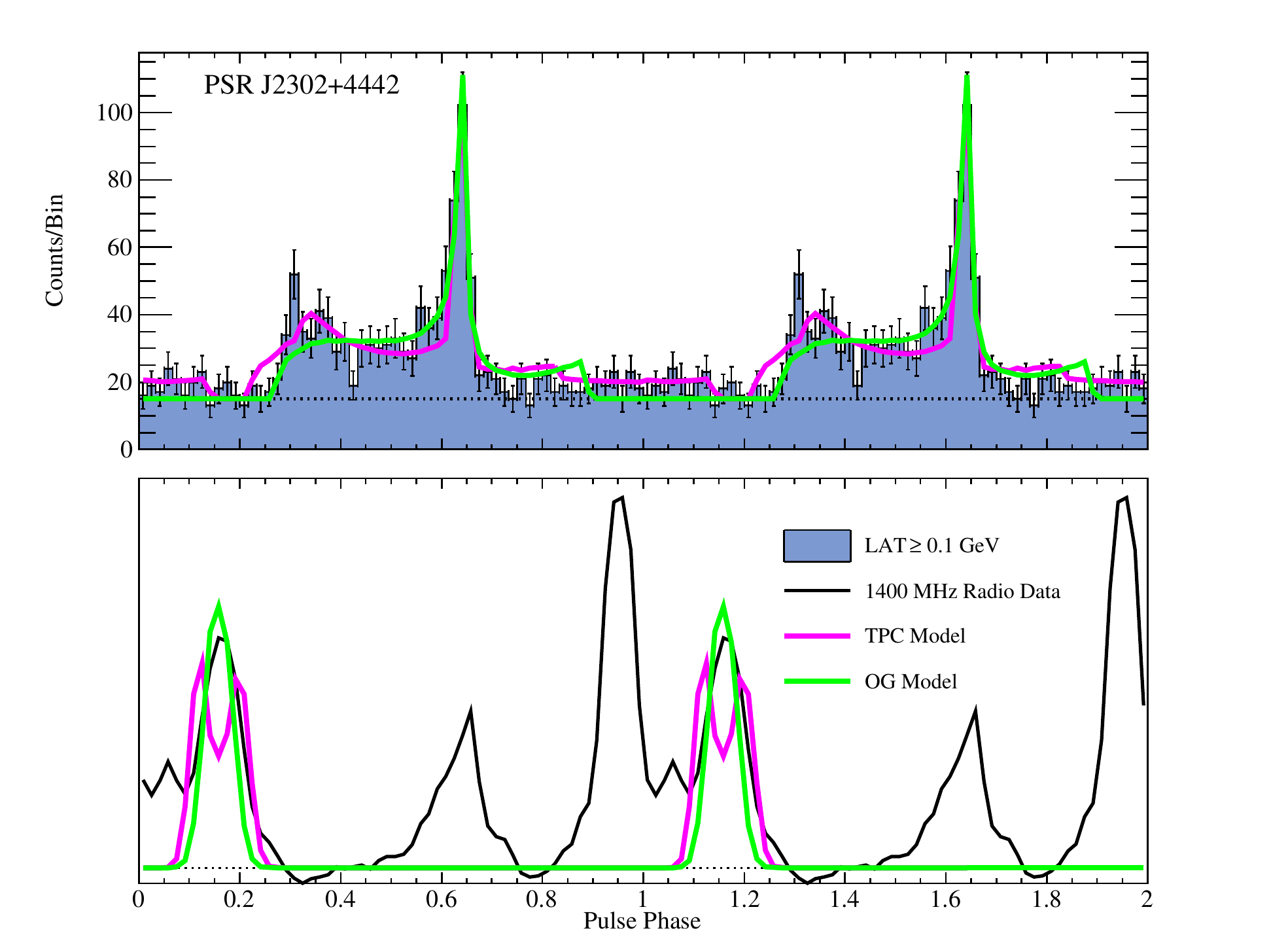}
\end{center}
\small\normalsize
\begin{quote}
\caption[Data and best-fit light curves for PSR J2302+4442]{Best-fit gamma-ray and radio light curves for PSR J2302+4442 using the TPC and OG models.\label{appAJ2302LCs}}
\end{quote}
\end{figure}

The best-fit gamma-ray and radio light curves are shown in Fig.~\ref{appAJ2302LCs}.  The gamma-ray light curve was fit with TPC and OG models.  The radio profile was fit with a hollow-cone beam model.  These light curve fits have used the 1400 MHz Nan\c{cay} radio profile.

\begin{figure}
\begin{center}
\includegraphics[width=0.75\textwidth]{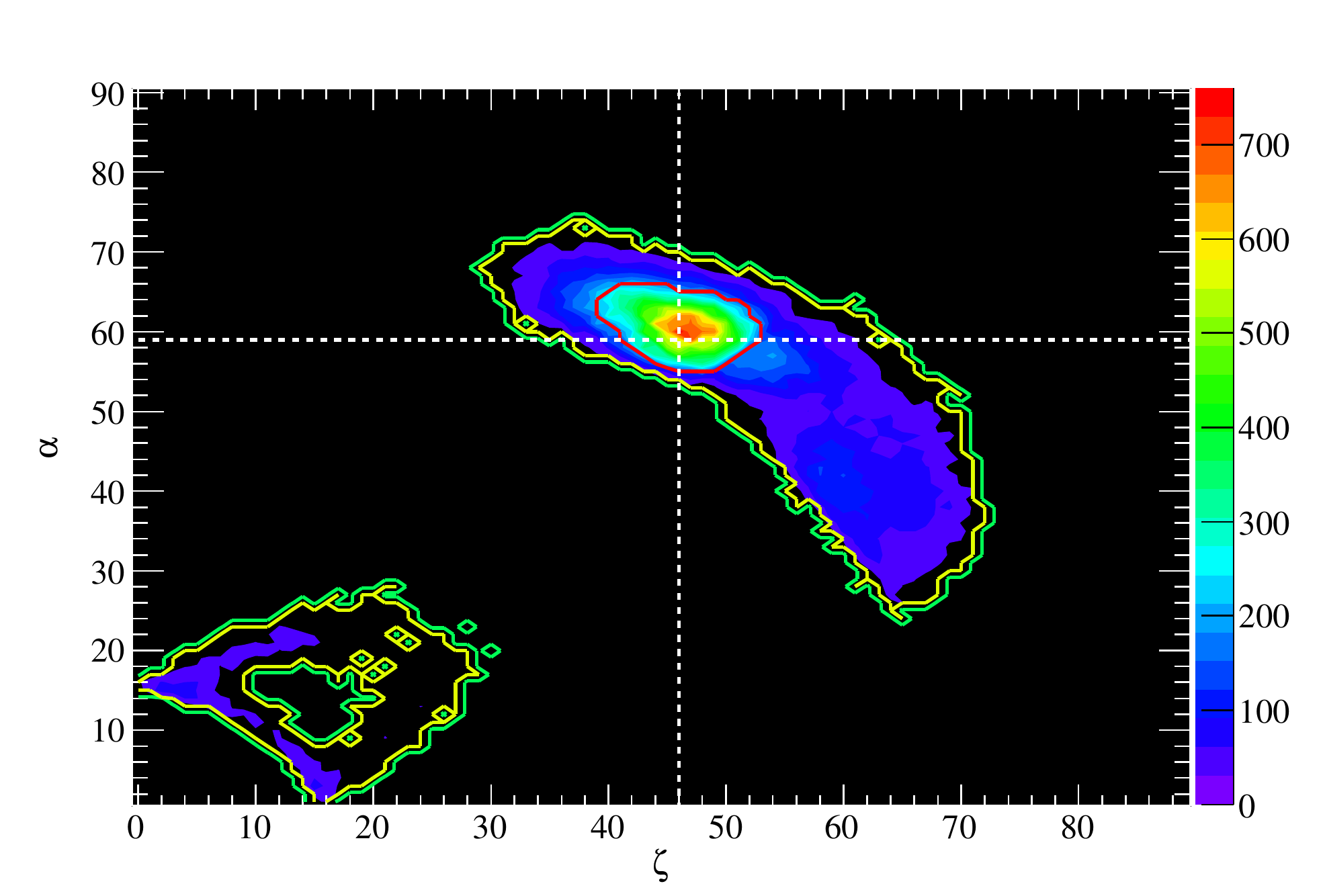}
\end{center}
\small\normalsize
\begin{quote}
\caption[Best-fit TPC contours for PSR J2302+4442]{Marginalized confidence contours for PSR J2302+4442 for the TPC model.\label{appAJ2302TPCcont}}
\end{quote}
\end{figure}

The marginalized $\alpha$-$\zeta$ confidence contours corresponding to the TPC fit are shown in Fig.~\ref{appAJ2302TPCcont}, the best-fit geometry is indicated by the vertical and horizontal dashed, white lines.

\begin{figure}
\begin{center}
\includegraphics[width=0.75\textwidth]{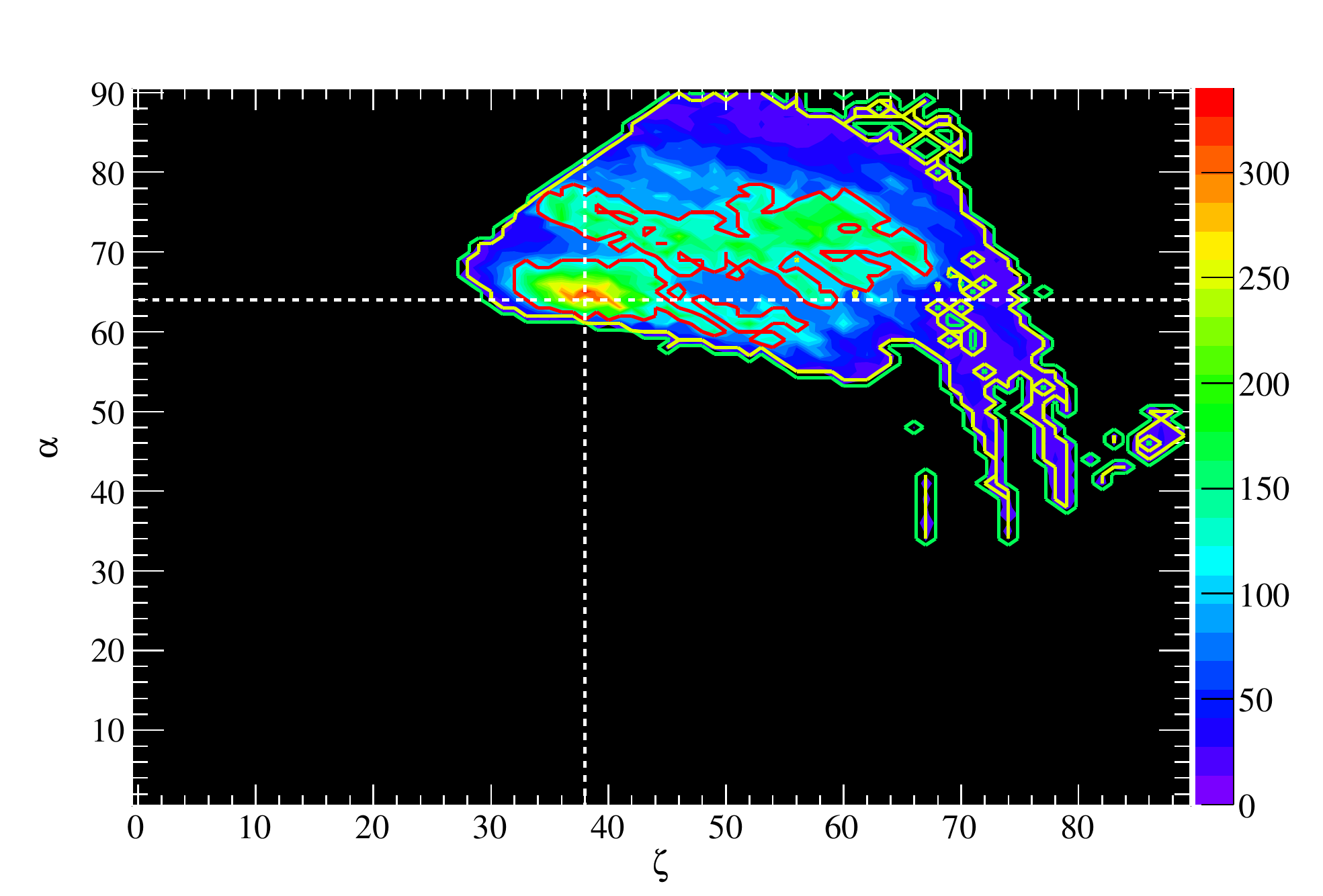}
\end{center}
\small\normalsize
\begin{quote}
\caption[Best-fit OG contours for PSR J2302+4442]{Marginalized confidence contours for PSR J2302+4442 for the OG model.\label{appAJ2302OGcont}}
\end{quote}
\end{figure}

The marginalized $\alpha$-$\zeta$ confidence contours corresponding to the OG fit are shown in Fig.~\ref{appAJ2302OGcont}, the best-fit geometry is indicated by the vertical and horizontal dashed, white lines.

\begin{figure}
\begin{center}
\includegraphics[width=0.75\textwidth]{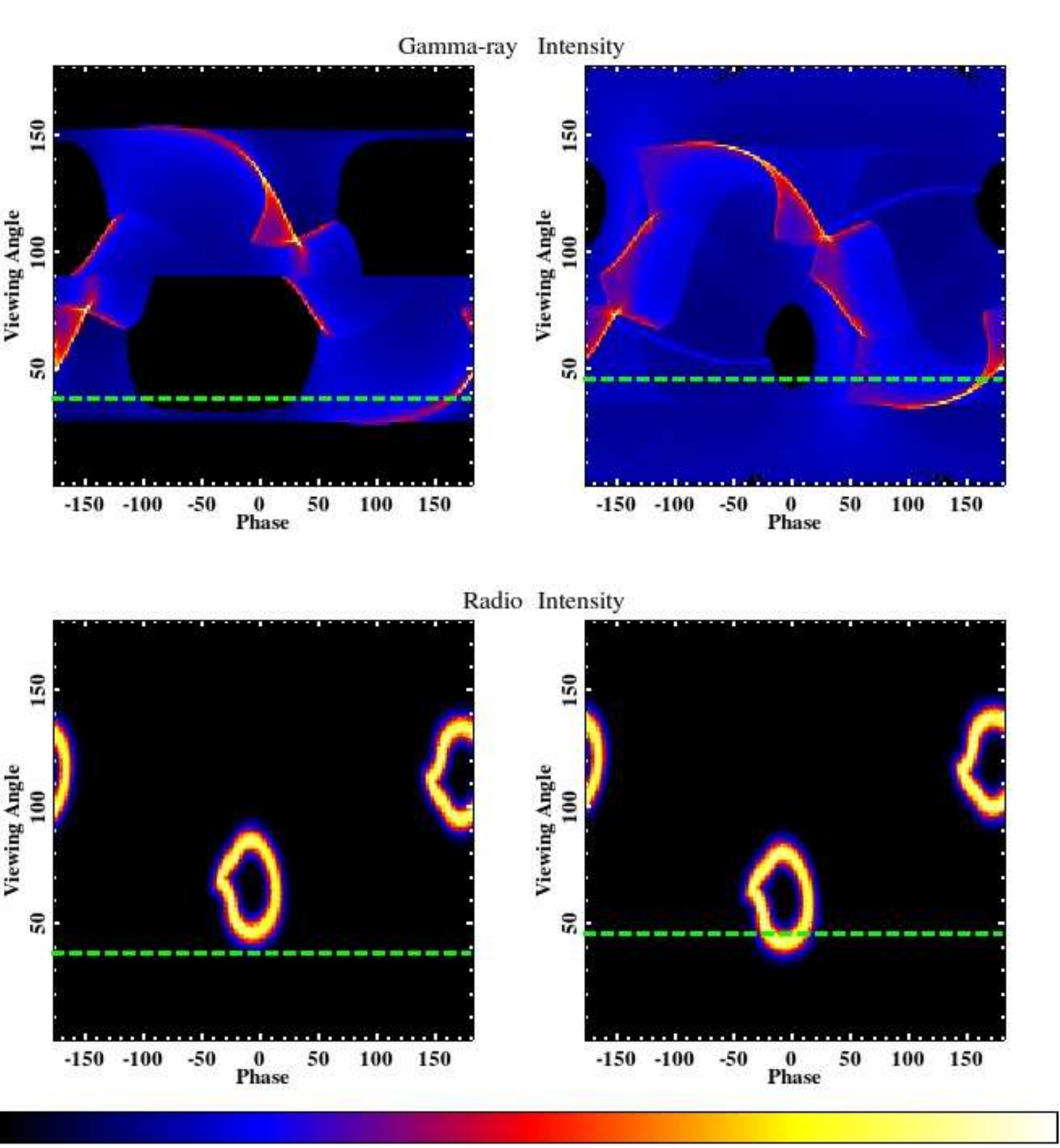}
\end{center}
\small\normalsize
\begin{quote}
\caption[Best-fit phase plots of simulated emission for PSR J2302+4442]{Distribution of simulated emission as a function of viewing angle and pulse phase for models used to fit PSR J2302+4442.  The best-fit $\zeta$ values are indicated by the dashed green lines.  The top plots correspond to the gamma-ray phase plots while the bottom are for the radio.  The left plots correspond to fits with the OG model with TPC plots on the right.  The color scales in the top plots are square root in order to bring out fainter features.\label{appAJ2302PhPlt}}
\end{quote}
\end{figure}

Plots of simulated emission corresponding to the best-fit models are shown in Fig.~\ref{appAJ2302PhPlt}, OG models are on the left and TPC on the right, gamma-ray models are on the top and radio on the bottom.

\small\normalsize

\small\normalsize

\small\normalsize

\small\normalsize
\appendix
\renewcommand{\thechapter}{B}

\chapter{Selected Formula Derivations}\label{appB}
This appendix contains a derivation of the relativistic aberration formula and outlines the manner in which the Lorentz transform of the retarded dipole field from the IOF to the CF is implemented in the \textbf{light curve simulation} code.  The former is not a new result nor is it particularly complex; however, a search of the literature did not return any papers in which this derivation is given.  While that is not to say that such references do not exist it seemed prudent to work through the derivation so that a more recent reference existed.  The latter is also not describing a new result, Lorentz transforms of electromagnetic fields are well known, but as this has been neglected in previous studies a description of how it is now included seems necessary for completeness.

\section{Relativistic Aberration}\label{appBab}
Following the geometric conventions described in Chapter 5, consider the point of emission \textbf{to be an} inertial frame of reference moving with three-velocity $\vec{\beta}_{\Omega}$ (the instantaneous co-rotation velocity in units of $c$, given by Eq.~\ref{ch5beta}) with respect to the IOF with corresponding Lorentz factor $\gamma\ =\ (1-\beta_{\Omega}^{2})^{-1/2}$.  Let the photon direction at the point of emission in the CF be $\hat{k}^{\prime}$.  To \textbf{calculate the IOF direction} ($\vec{k}$) it is helpful to separate $\vec{k}^{\prime}$ into components perpendicular and parallel to $\vec{\beta}_{\Omega}$ before applying the Lorentz transformations as given in Eqs.~\ref{appBkpar} and~\ref{appBkperp} (adapted from Eqs. 37, 38, and 39 of Chapter 2 in Ohanian 2001).

\begin{equation}\label{appBkpar}
\vec{k}_{\parallel}\ =\ \frac{\vec{k}_{\parallel}^{\prime}+\vec{\beta}_{\Omega}}{1+\vec{\beta}_{\Omega}\cdot\hat{k}^{\prime}}
\end{equation}

\begin{equation}\label{appBkperp}
\vec{k}_{\perp}\ =\ \gamma\vec{k}_{\perp}^{\prime}\bigg(1-\frac{\vec{\beta}_{\Omega}\cdot(\vec{k}_{\parallel}^{\prime}+\vec{\beta}_{\Omega})}{1+\vec{\beta}_{\Omega}\cdot\hat{k}^{\prime}}\bigg)
\end{equation}

Eq.~\ref{appBkperp} can be simplified by noting that $\vec{\beta}_{\Omega}\cdot\vec{k}_{\parallel}^{\prime}\ =\ \vec{\beta}_{\Omega}\cdot\hat{k}^{\prime}\ =\ \vec{\beta}_{\Omega}\cdot\hat{k}^{\prime}$.  By adding together the terms in parentheses, using a common denominator of $(1+\vec{\beta}_{\Omega}\cdot\hat{k}^{\prime})$, one arrives at Eq.~\ref{appBkperpmid} which is simplified using $(1-\beta_{\Omega}^{2})\ =\ \gamma^{-2}$.  Eq.~\ref{appBkperpmid} gives $\vec{k}_{\perp}$ in terms of the initial emission direction, the co-rotation velocity, and $\vec{k}_{\perp}^{\prime}$; however, it would be useful to simplify this further such that $\vec{k}_{\perp}^{\prime}$ did not explicitly appear.

\begin{equation}\label{appBkperpmid}
\vec{k}_{\perp}\ =\ \gamma\vec{k}_{\perp}^{\prime}\bigg(\frac{1-\beta_{\Omega}^{2}}{1+\vec{\beta}_{\Omega}\cdot\hat{k}^{\prime}}\bigg)\ =\ \frac{\vec{k}_{\perp}^{\prime}}{\gamma(1+\vec{\beta}_{\Omega}\cdot\hat{k}^{\prime})}
\end{equation}

It is useful to recognize that $(\vec{\beta}_{\Omega}\times\vec{k}^{\prime})\times\vec{\beta}_{\Omega}$ will be in the same direction as $\vec{k}_{\perp}^{\prime}$ but will have magnitude $\beta_{\Omega}^{2}k_{\perp}^{\prime}$.  Thus, $\vec{k}_{\perp}^{\prime}$ can be rewritten in terms of only the initial photon direction and the co-rotation velocity as given in Eq.~\ref{appBkprimeperp} (where the identity $(\vec{\beta}_{\Omega}\times\vec{k}^{\prime})\times\vec{\beta}_{\Omega}\ =\ -\vec{\beta}_{\Omega}\times(\vec{\beta}_{\Omega}\times\vec{k}^{\prime})$ and Lagrange's formula $\vec{a}\times(\vec{b}\times\vec{c})\ =\ (\vec{a}\cdot\vec{c})\vec{b}-(\vec{a}\cdot\vec{b})\vec{c}$ have been used).  Using Eq.~\ref{appBkprimeperp} in Eq.~\ref{appBkperpmid} leads to Eq.~\ref{appBkperpfin} which gives the component of the photon direction perpendicular to $\vec{\beta}_{\Omega}$ in the IOF in terms of the emitted direction and co-rotation velocity.

\begin{equation}\label{appBkprimeperp}
\vec{k}_{\perp}^{\prime}\ =\ \Big(\hat{k}^{\prime}-\frac{\vec{\beta}_{\Omega}\cdot\hat{k}^{\prime}}{\beta_{\Omega}^{2}}\vec{\beta}_{\Omega}\Big)
\end{equation}

\begin{equation}\label{appBkperpfin}
\vec{k}_{\perp}\ =\ \frac{1}{\gamma(1+\vec{\beta}_{\Omega}\cdot\hat{k}^{\prime})}\bigg(\hat{k}^{\prime}-\frac{\vec{\beta}_{\Omega}\cdot\hat{k}^{\prime}}{\beta_{\Omega}^{2}}\vec{\beta}_{\Omega}\bigg)
\end{equation}

Similarly, Eq.~\ref{appBkpar} can be simplified by recognizing that $\vec{\beta}_{\Omega}\cdot\vec{k}^{\prime}\ =\ \beta_{\Omega}k_{\parallel}^{\prime}$ and that $\vec{k}_{\parallel}^{\prime}$ is, by definition, along the same direction as $\vec{\beta}_{\Omega}$.  These considerations lead to Eq.~\ref{appBkprimepar} which can be used to cast Eq.~\ref{appBkpar} in terms of the emitted photon direction and co-rotation velocity as given in Eq.~\ref{appBkparfin}.

\begin{equation}\label{appBkprimepar}
\vec{k}_{\parallel}^{\prime}\ =\ \frac{\vec{\beta}_{\Omega}\cdot\hat{k}^{\prime}}{\beta_{\Omega}^{2}}\vec{\beta}_{\Omega}
\end{equation}

\begin{equation}\label{appBkparfin}
\vec{k}_{\parallel}\ =\ \bigg(\frac{\vec{\beta}_{\Omega}\cdot\hat{k}^{\prime}}{\beta_{\Omega}^{2}}+1\bigg)\frac{\vec{\beta}_{\Omega}}{1+\vec{\beta}_{\Omega}\cdot\hat{k}^{\prime}}
\end{equation}

The direction of the emitted photon in the IOF is obtained by combining Eqs.~\ref{appBkparfin} and~\ref{appBkperpfin} which returns,

\begin{equation}\label{appBabform}
\hat{k}\ =\ \frac{\hat{k}^{\prime}+\Big(\gamma+(\gamma-1)\frac{\vec{\beta}_{\Omega}\cdot\hat{k}^{\prime}}{\beta_{\Omega}^{2}}\Big)\vec{\beta}_{\Omega}}{\gamma(1+\vec{\beta}_{\Omega}\cdot\hat{k}^{\prime})},
\end{equation}

which is the relativistic aberration formula given in Eq.~\ref{ch5abform}.

\section{Implementation Of The Magnetic Field Lorentz Transformation}\label{appBLT}
As noted in Chapter 5, previous studies of gamma-ray pulsar light curves in the retarded vacuum dipole geometry (e.g.; Venter et al., 2009; Abdo et al., 2010d) \\assumed that the direction of the magnetic field (which determines the initial photon direction) in the CF was the same as that in the IOF.  However, \citet{BS10a} noted that this was incorrect and the magnetic field should be first transformed to the CF.  They demonstrated that neglecting this transform affected the simulated light curve shapes.

It should be noted that the authors of previous studies did not neglect the transformation because they were not aware of it; rather, the overall effect is of second order in $r_{n}$ and thus is of lesser importance.  However, given the results of \citet{BS10a} this has now been included but note that the affects on the light curve shapes are minimal.

The magnetic field in the IOF ($\vec{B}_{IOF}$) is assumed to have the Deutsch field geometry \citep{Deutsch55} as given in Chapter 5.  As discussed in Chapter 5, the emission is taken to occur at a time $t$ such that the IOF and CF axes are aligned.  Consider the point of emission ($x$, $y$, $z$) to be, instantaneously, an inertial frame with velocity $\vec{\beta}_{\Omega}$ with respect to the IOF, the magnetic field in the IOF can be transformed to the CF using Eqs.~\ref{appBbpar} and~\ref{appBbper}, where the parallel and perpendicular indices are referenced to $\vec{\beta}_{\Omega}$.
\begin{equation}\label{appBbpar}
\vec{B}^{CF}_{\parallel}\ =\ \vec{B}^{IOF}_{\parallel}
\end{equation}
\begin{equation}\label{appBbper}
\vec{B}^{CF}_{\perp}\ =\ \gamma\Big(\vec{B}^{IOF}_{\perp}-\vec{\beta}_{\Omega}\times\vec{E}^{IOF}\Big)
\end{equation}

Assuming force-free conditions $\vec{E}^{IOF}\ =\ -\vec{\beta}_{\Omega}\times\vec{B}^{IOF}$ and using Lagrange's formula Eq.~\ref{appBbper} can be reduced to,
\begin{equation}\label{appBbperRed}
\vec{B}^{CF}_{\perp}\ =\ \gamma^{-1}\vec{B}^{IOF},
\end{equation}
\noindent{}where $(1-\beta_{\Omega}^{2})\ =\ \gamma^{-1}$ has been used as well.

At any given point above the stellar surface the co-rotation velocity is $\vec{\beta}_{\Omega}\ =\ \vec{\Omega}\times\vec{r}\ =\ \rho/\rm R_{LC} \hat{n}_{1}$ with the usual cylindrical radius $\rho\ =\ \sqrt{x^{2}+y^{2}}$.  With $\vec{\Omega}\ =\ \Omega\hat{z}$, the unit vector $\hat{n}_{1}$ (which is parallel to $\vec{\beta}_{\Omega}$) will be in the $x$-$y$ plane as given by,
\begin{equation}\label{appBn1}
\hat{n}_{1}\ =\ \Big\langle-\frac{y}{\rm R_{LC}},\frac{\mathnormal x}{\rm R_{LC}},0\Big\rangle.
\end{equation}
\noindent{}An instantaneous coordinate system can be built at the emission point using the unit vectors,
\begin{equation}\label{appBn2}
\hat{n}_{2}\ =\ \langle0,0,1\rangle
\end{equation}
\noindent{}and,
\begin{equation}\label{appBn3}
\hat{n}_{3}\ =\ \Big\langle\frac{x}{\rm R_{LC}},\frac{\mathnormal y}{\rm R_{LC}},0\Big\rangle
\end{equation}
\noindent{}which are orthogonal to each other and $\hat{n}_{1}$.  The coordinate system defined by Eqs.~\ref{appBn1},~\ref{appBn2}, and~\ref{appBn3} can be used to calculate the components of $\vec{B}^{IOF}$ parallel and perpendicular to $\vec{\beta}_{\Omega}$ as given in Eqs.~\ref{appBB1},~\ref{appBB2}, and~\ref{appBB3}.
\begin{equation}\label{appBB1}
\vec{B}_{\parallel}^{IOF}\ =\ \Big(\vec{B}^{IOF}\cdot\hat{n}_{1}\Big)\hat{n}_{1}
\end{equation}
\begin{equation}\label{appBB2}
\vec{B}_{\perp2}\ =\ \Big(\vec{B}^{IOF}\cdot\hat{n}_{2}\Big)\hat{n}_{2}
\end{equation}
\begin{equation}\label{appBB3}
\vec{B}_{\perp3}\ =\ \Big(\vec{B}^{IOF}\cdot\hat{n}_{3}\Big)\hat{n}_{3}
\end{equation}

The magnetic field in the CF is thus obtained by combining the magnetic field components given in Eqs.~\ref{ch5Bretx},~\ref{ch5Brety}, and~\ref{ch5Bretz} with Eqs.~\ref{appBn1} through~\ref{appBB3} and then setting $\vec{B}^{CF}\ =\ \vec{B}_{\parallel}^{IOF}+\gamma^{-1}\Big(\vec{B}^{IOF}_{\perp2}+\vec{B}^{IOF}_{\perp3}\Big)$.

\renewcommand{\baselinestretch}{1}


\small\normalsize
\begin{thebibliography}{99}
\bibitem[Abdo et al.(2008)]{AbdoCTA1} Abdo, A.~A., et al. 2008, \emph{Science}, 322, 1218
\bibitem[Abdo et al.(2009a)]{AbdoJ1028} Abdo, A.~A., et al. 2009a, \ApJ{} \emph{Lett.}, 695, L72
\bibitem[Abdo et al.(2009b)]{AbdoVela} Abdo, A.~A., et al. 2009b, \ApJ{}, 696, 1084
\bibitem[Abdo et al.(2009c)]{AbdoBSL} Abdo, A.~A., et al. 2009c, \ApJ{} \emph{Suppl.}, 183, 46
\bibitem[Abdo et al.(2009d)]{AbdoJ0030} Abdo, A.~A., et al. 2009d, \ApJ{}, 699, 1171
\bibitem[Abdo et al.(2009e)]{AbdoBlndSrc} Abdo, A.~A.,et al. 2009e, \emph{Science}, 325, 840
\bibitem[Abdo et al.(2009f)]{Abdo47Tuc} Abdo, A.~A., et al. 2009f, \emph{Science}, 325, 845
\bibitem[Abdo et al.(2009g)]{AbdoMSPpop} Abdo, A.~A., et al. 2009g, \emph{Science}, 325, 848
\bibitem[Abdo et al.(2009h)]{AbdoLIVlim} Abdo, A.~A., et al. 2009h, \emph{Nature}, 462, 331
\bibitem[Abdo et al.(2009i)]{Abdo090902B} Abdo, A.~A., et al. 2009i, \ApJ{} \emph{Lett.}, 706, L138
\bibitem[Abdo et al.(2009j)]{AbdoDF} Abdo, A.~A., et al. 2009j, \ApJ{}, 700, 1059
\bibitem[Abdo et al.(2009k)]{AbdoOnorb} Abdo, A.~A., et al. 2009k, \AsPart{}, 32, 193
\bibitem[Abdo et al.(2009l)]{AbdoDiffuse} Abdo, A.~A., et al. 2009l, \emph{Phys. Rev. Lett.}, 103, 251101
\bibitem[Abdo et al.(2009m)]{AbdoAlbedo} Abdo, A.~A., et al. 2009m, \emph{Phys. Rev. D}, 80, 122004
\bibitem[Abdo et al.(2010a)]{AbdoDMLines} Abdo, A.~A., et al. 2010a, \emph{Phys, Rev. Lett.}, 104, 091302
\bibitem[Abdo et al.(2010b)]{Abdo1907} Abdo, A.~A., et al. 2010b, \ApJ{}, 711, 64
\bibitem[Abdo et al.(2010c)]{AbdoPSRcat} Abdo, A.~A., et al. 2010c, \ApJ{} \emph{Suppl.}, 187, 460
\bibitem[Abdo et al.(2010d)]{AbdoJ0034} Abdo, A.~A., et al. 2010d, \ApJ{}, 712, 957
\bibitem[Abdo et al.(2010e)]{AbdoVX} Abdo, A.~A., et al. 2010e, \ApJ{} 713, 146
\bibitem[Abdo et al.(2010f)]{AbdoVII} Abdo, A.~A., et al. 2010f, \ApJ{}, 713, 154
\bibitem[Abdo et al.(2010g)]{Abdo1LAC} Abdo, A.~A., et al. 2010g, \ApJ{}, 715, 429
\bibitem[Abdo et al.(2010h)]{Abdo1FGL} Abdo, A.~A., et al. 2010h, \ApJ{} \emph{Suppl.}, 188, 405
\bibitem[Abdo et al.(2010i)]{AbdoV407Cyg} Abdo, A.~A., et al. 2010i, \emph{Science}, 329, 817
\bibitem[Abdo et al.(2010j)]{AbdoGCpop} Abdo, A.~A., et al. 2010j, \AA{}, 524, A75
\bibitem[Abdo et al.(2001a)]{AbdoCrabFlare} Abdo, A.~A., et al. 2011a, \emph{Science}, 331, 739
\bibitem[Abdo et al.(2011b)]{Abdo2FGL} Abdo, A.~A., et al. 2011b, \emph{in preparation}
\bibitem[Ackermann et al.(2011)]{PWNcat} Ackermann,~M., et al. 2011, \ApJ{}, 726, 35
\bibitem[Agostinelli et al.(2003)]{Agostinelli03} Agostinelli,~S., et al. 2003, \NIMA{}, 506, 250
\bibitem[Aguilar et al.(2002)]{Aguilar02} Aguilar,~M., et al. 2002, \emph{Phys. Rep.}, 366, 331
\bibitem[Akerlof et al.(1992)]{Whipple} Akerlof, ~C.~W., et al., 1992, \emph{Proc. of The Compton Obs. Sci. Workshop}, NASA GSFC, 406
\bibitem[Albert et al.(2008)]{MAGICCRAB} Albert,~J., et al., 2008, \ApJ{}, 674, 1037
\bibitem[Allison et al.(2006)]{Allison06} Allison,~J., et al. 2006, \ITNS{}, 53, 270
\bibitem[Alpar et al.(1982)]{Alpar82} Alpar, M.~A., Cheng, A.~F., Ruderman, M.~A., \& Shaham,~J. 1982, \emph{Nature}, 300, 728 
\bibitem[Archibald et al.(2009)]{Archibald09} Archibald, Anne~M., et al. 2009, \emph{Science}, 324, 1411
\bibitem[Arons \& Scharlemann(1979)]{AS79} Arons,~J. \& Scharlemann, E.~T. 1979, \ApJ{}, 231, 854
\bibitem[Arons(1983)]{Arons83} Arons,~J. 1983, \ApJ{}, 266, 215
\bibitem[Arons(1996)]{Arons96} Arons,~J. 1996, \AA{} \emph{Suppl.}, 120, 49
\bibitem[Arzoumanian et al.(2002)]{Arzoumanian02} Arzoumanian,~Z., Chernoff, D.~F., \& Cordes, J.~M. 2002, \ApJ{}, 568, 289%
\bibitem[Ashworth et al.(1983)]{Ashworth83} Ashworth,~M., Lyne, A.~G., \& Smith, F.~G. 1983, \emph{Nature}, 301, 313
\bibitem[Atwood et al.(1994)]{Atwood94} Atwood, W.~B., et al. 1994, \NIMA{}, 342, 302
\bibitem[Atwood et al.(2006)]{Atwood06} Atwood, W.~B., Ziegler,~M., Johnson, R.~P., \& Baughman, B.~M. 2006, \ApJ{} \emph{Lett.}, 652, L49
\bibitem[Atwood et al.(2007)]{Atwood07} Atwood, W.~B., et al. 2007, \emph{Astropart. Phys.}, 28, 422
\bibitem[Atwood et al.(2009)]{Atwood09} Atwood, W. B., et al. 2009, \ApJ{}, 697, 1071
\bibitem[Baade \& Zwicky(1934a)]{BZ34a} Baade,~W. \& Zwicky,~F. 1934a, \emph{Proc. of the Nat. Acad. of Sci. of the U.S.A.}, 20, 259
\bibitem[Baade \& Zwicky(1934b)]{BZ34b} Baade,~W. \& Zwicky,~F. 193b, \emph{Phys. Rev.}, 46, 76
\bibitem[Backer et al.(1982)]{Backer82} Backer, D. C., et al. 1982, \emph{Nature}, 300, 615 
\bibitem[Backus et al.(1982)]{Backus82} Backus, P.~R., Taylor, J.~H., \& Damashek,~M. 1982, \ApJ{} \emph{Lett.}, 255, L63
\bibitem[Bai \& Spitkovsky(2010a)]{BS10a} Bai, X.-N. \& Spitkovsky,~A. 2010a, \ApJ{}, 715, 1270
\bibitem[Bai \& Spitkovsky(2010b)]{BS10b} Bai, X.-N. \& Spitkovsky,~A. 2010b, \ApJ{}, 715, 1282
\bibitem[Bailes et al.(1994)]{Bailes94} Bailes,~M., et al. 1994, \ApJ{} \emph{Lett.}, 425, L41
\bibitem[Bailes et al.(1997)]{Bailes97} Bailes,~M., et al. 1997, \ApJ{}, 481, 386
\bibitem[Baldini et al.(2006)]{Baldini06} Baldini,~L., et al. 2006, \ITNS{}, 53, 466
\bibitem[Baring(2004)]{Baring04} Baring, M.~G.2004, \emph{Adv. in Space Research}, 33, 552
\bibitem[Bertsch et al.(1992)]{Bertsch92} Bertsch, D.~L., et al. 1992, \emph{Nature}, 357, 306
\bibitem[Biggs et al.(1994)]{Biggs94} Biggs, J.~D., et al. 1994, \MNRAS{}, 267, 125
\bibitem[Blaskiewicz et al.(1991)]{Blaskiewicz91} Blaskiewicz,~M., Cordes, J.~M., \& Wasserman,~I. 1991, \ApJ{}, 370, 643
\bibitem[Bogdanov et al.(2007)]{Bogdanov07} Bogdanov,~S., Rybicki, G.~B., Grindlay, J.~E. 2007, \ApJ{}, 670, 668
\bibitem[Bogdanov et al.(2008)]{Bogdanov08} Bogdanov,~S., Grindlay, J.~E., \& Rybicki, G.~B. 2008, \ApJ{} 689, 407
\bibitem[Brisken et al.(2002)]{Brisken02} Brisken, W.~F., Benson, J.~M., Goss, W.~M., \& Thorsett, S.~E. 2002, \ApJ{}, 571, 906
\bibitem[Browning et al.(1971)]{Browning71} Browning,~R., Ramsden,~D., \& Wright, P.~J. 1971, \emph{Nature}, 232, 99
\bibitem[Byrd et al.(1995)]{Byrd95} Byrd, R.~H., Lu,~P., \& Nocedal,~J. 1995, \emph{SIAM Journ. on Sci. and Stat. Computing}, 16,1190
\bibitem[Cameron(1959)]{Cameron59} Cameron, A.~G.~W. 1959, \ApJ{}, 130, 884
\bibitem[Camilo et al.(1994)]{Camilo94} Camilo,~F., Foster, R.~S., \& Wolszczan,~A. 1994, \ApJ{} \emph{Lett.}, 437, L39
\bibitem[Camilo et al.(2009)]{Camilo09} Camilo,~F., et al. 2009, \ApJ{}, 705, 1
\bibitem[Camilo et al.(2011)]{Camilo11} Camilo,~F., et al. 2011, \emph{in prep}
\bibitem[Carroll \& Ostlie(1996)]{BOB} Carroll, B.~W. \& Ostlie, D.~A. 1996, \emph{An Introduction to Modern Astrophysics}, (USA: Addison-Wesley Pub. Co., Inc.)
\bibitem[\c{C}elik et al.(2010)]{Celik11} \c{C}elik,~\"{O}., Johnson, T.~J., \& The \FL{} Collab. 2011, \emph{Proc. of the Pulsar Conf. 2010:  Radio Pulsars: a key to Unlock the Secrets of the Universe}, ed. M.~Burgay, (USA: AIP)
\bibitem[CXO(2011)]{CXO} \emph{Chanda X-ray Observatory} Multimedia Coloring Space\\ $\langle$http://chandra.harvard.edu/art/color/colorspace.html$\rangle$
\bibitem[Chandrasekhar(1931a)]{Chandra31a} Chandrasekhar,~S. 1931a, \MNRAS{}, 91, 456
\bibitem[Chandrasekhar(1931b)]{Chandra31b} Chandrasekhar,~S. 1931b, \ApJ{}, 74, 81
\bibitem[Chandrasekhar(1935)]{Chandra35} Chandrasekhar,~S. 1935, \MNRAS{}, 95, 207
\bibitem[Cheng et al.(1986a)]{Cheng86a} Cheng, K.~S., Ho,~C., \& Ruderman,~M. 1986a, \ApJ{}, 300, 500
\bibitem[Cheng et al.(1986b)]{Cheng86b} Cheng, K.~S., Ho,~C., \& Ruderman,~M. 1986b, \ApJ{}, 300, 522
\bibitem[Cheng et al.(2000)]{Cheng00} Cheng, K.~S., Ho,~C., \& Ruderman,~M. 2000, \ApJ{}, 537, 964
\bibitem[Chiang(2002a)]{Chiang0} Chiang,~J. 2002a, \emph{Internal LAT Document} 
\bibitem[Chiang(2002b)]{Chiang1} Chiang,~J. 2002b, \emph{Internal LAT Document}
\bibitem[Chiang(2002c)]{Chiang2} Chiang,~J. 2002c, \emph{Internal LAT Document}
\bibitem[Chiang(2002d)]{Chiang3} Chiang,~J. 2002d, \emph{Internal LAT Document}
\bibitem[Chiang(2010)]{Chiang10} Chiang,~J. 2010, \emph{Presentation at the LAT Collab. Weekly Analysis Meeting}, 14 May
\bibitem[Cognard et al.(2011)]{Cognard11} Cognard, I., et al. 2011, \ApJ{}, \emph{in press} (arXiv:1102.4192)
\bibitem[Colin et al.(2009)]{MAGIC} Colin,~P., et al. for the MAGIC Collab. 2009, \emph{Proc. of the 31st ICRC}, Lodz, Poland (arXiv:0907.0960)
\bibitem[Comella et al.(1969)]{Comella69} Comella, J.~M., Craft, H.~D. jr., Lovelace, R.~V.~E., Sutton, J.~M., \& Tyler, G.~L. 1969, \emph{Nature}, 221, 453
\bibitem[Contopolous et al.(1999)]{Contopolous99} Contopolous,~I., Kazanas,~D., \& Fendt,~C. 1999, \ApJ{}, 511, 351
\bibitem[Cordes \& Stinebring(1984)]{CS84} Cordes, J.~M. \& Stinebring, D.~R. 1984, \ApJ{} \emph{Lett.}, 277, L53
\bibitem[Cordes \& Lazio(2002)]{NE2001} Cordes, J.~M. \& Lazio, T.~J.~W. 2002, arXiv:astro-ph/0207156
\bibitem[Crawford et al.(2006)]{Crawford06} Crawford,~F., et al. 2006, \ApJ{}, 652, 1499
\bibitem[Deutsch(1955)]{Deutsch55} Deutsch, A.~J. 1955, \emph{Annales d'Astrophysique}, 18, 1
\bibitem[do Couto e Silva et al.(2001)]{Couto01} do Couto e Silva,~E., et al. 2001, \NIMA{}, 474, 19
\bibitem[Daugherty \& Harding(1982)]{DH82} Daugherty, J.~K. \& Harding, A.~K. 1982, \ApJ{}, 252, 337
\bibitem[Daugherty \& Harding(1994)]{DH94} Daugherty, J.~K. \& Harding, A.~K. 1994, \ApJ{}, 429, 325
\bibitem[Daugherty \& Harding(1996)]{DH96} Daugherty, J.~K. \& Harding, A.~K. 1996, \ApJ{}, 458, 278
\bibitem[De Jager et al.(1989)]{deJager89} De Jager, O.~C., Raubenheimer, B.~C., \& Swanepoel, J.~W.~H. 1989, \AA{}, 221, 180
\bibitem[Derdeyn et al.(1972)]{SAS2} Derdeyn, S.~M., Ehrmann, C.~H., Fichtel, C.~E., Kniffen, D.~A., \& Ross, R.~W. 1972, \emph{Nucl. Instr. and Meth.}, 98, 557
\bibitem[Demorest et al.(2010)]{Demorest10} Demorest, P.~B., Pennucci,~T., Ransom, S.~M., Roberts, M.~S.~E., \& Hessels, J.~W.~T 2010, \emph{Nature}, 467, 1081
\bibitem[Du et al.(2011)]{Du11} Du, Y.~J., Han, J.~L., Qiao, G.~J., \& Chou, C.~K. 2011, \ApJ{}, \emph{in press} (arXiv:1102.2476)
\bibitem[Dyks \& Rudak(2003)]{DR03} Dyks,~J. \& Rudak,~B. 2003, \ApJ{}, 598, 1201 
\bibitem[Dyks et al.(2004)]{Dyks04} Dyks,~J., Harding, A.~K., \& Rudak,~B. 2004, \ApJ{}, 606, 1125
\bibitem[Dyks \& Harding(2004)]{DH04} Dyks,~J. \& Harding A.~K. 2004, \ApJ{}, 614, 869
\bibitem[Eddington(1931)]{Eddington31} Eddington, Sir A.~S. 1931, \MNRAS{}, 91, 444
\bibitem[Eddington(1933)]{Eddington33} Eddington, Sir A.~S. 1933, \MNRAS{}, 93, 320
\bibitem[Eddington(1935)]{Eddington35} Eddington, Sir A.~S. 1935, \MNRAS{}, 95, 194
\bibitem[Erber(1966)]{Erber66} Erber,~T. 1966, \emph{Rev. of Modern Phys.}, 38, 626
\bibitem[Fermi(2011)]{Fermi} \emph{Fermi} LAT Performance Page\\$\langle$http://www-glast.slac.stanford.edu/software/IS/glast\_lat\_performance.htm$\rangle$
\bibitem[Fermi(2011)]{FermiST} \emph{Fermi} Science Support Center Analysis Threads\\$\langle$http://fermi.gsfc.nasa.gov/ssc/data/analysis/scitools/$\rangle$
\bibitem[Fermi(2011)]{FermiUser} \emph{Fermi} Science Support Center User Contributed Software\\$\langle$http://fermi.gsfc.nasa.gov/ssc/data/analysis/user/
\bibitem[Ferreira et al.(2004)]{Ferreira04} Ferreira,~O., et al. 2004, \emph{Nucl. Instr. Meth. Phys. Res. A}, 530, 323
\bibitem[Fich et al.(1989)]{Fich89} Fich,~M., Blitz,~L., \& Start, A.~A. 1989, \ApJ{}, 342, 272
\bibitem[Fierro et al.(1993)]{Fierro93} Fierro, J.~M., et al. 1993, \ApJ{} \emph{Lett.}, 413, L27
\bibitem[Fierro et al.(1995)]{Fierro95} Fierro, J.~M., et al. 1995, \ApJ{}, 447, 807 
\bibitem[Fierro et al.(1998)]{Fierro98} Fierro, J.~M., et al. 1998, \ApJ{}, 494, 734
\bibitem[Finzi \& Wolf(1969)]{FW69} Finzi,~A. \& Wolf, R.~A. 1969, \ApJ{} \emph{Lett.}, 155, L107
\bibitem[Foster et al.(1993)]{Foster93} Foster, R.~S., Wolszczan,~A., \& Camilo,~F. 1993, \ApJ{} \emph{Lett.}, 410, L91
\bibitem[Fowler(1926)]{Fowler26} Fowler, R.~H. 1926, \MNRAS{}, 87 114
\bibitem[Friere et al.(2011)]{Freire11} Friere, P.~C.~C., et al. 2011, \emph{in preparation}
\bibitem[Fruchter et al.(1988)]{Fruchter88} Fruchter, A.~S., Stinebring, D.~R., \& Taylor, J.~H. 1988, \emph{Nature}, 333, 237
\bibitem[Fr\"{u}hwirth et al.(2000)]{Fruhwirth00} Fr\"{u}hwirth,~R., Regler,~M., Bock, R.~K., Grote,~H., \& Notz,~D. 2000, \emph{Data Analysis Techniques for High-Energy Physics} (2nd ed.; Cambridge: Cambridge Univ. Press)
\bibitem[Guillemot(2009)]{Guillemot09} Guillemot,~L. 2009, Ph.D. Thesis, Universit\'{e} de Bordeaux I, Centre d'Etudes Nucleaires de Bordeaux-Gradignan, France (arXiv:0910.4707)
\bibitem[Guillemot et al.(2011)]{Guillemot11} Guillemot,~L., et al. 2011, \emph{in preparation}
\bibitem[Gelman \& Rubin(1992)]{GR92} Gelman,~A., \& Rubin,~D. 1992, \emph{Stat. Sci.}, 7, 457
\bibitem[Gil \& Krawczyk(1997)]{GK97} Gil,~J. \& Krawczyk,~A. 1997, \MNRAS{}, 285, 561
\bibitem[Gold(1968)]{Gold68} Gold,~T. 1968, \emph{Nature}, 218, 731
\bibitem[Gold(1969)]{Gold69} Gold,~T. 1969, \emph{Nature}, 221, 25
\bibitem[Goldreich \& Julian(1969)]{GJ69} Goldreich,~P. \& Julian, W.~H. 1969, \ApJ{}, 157, 869
\bibitem[Gonthier et al.(2004)]{Gonthier04} Gonthier, P.~L., Van Gilder,~R., \& Harding, A.~K. 2004, \ApJ{}, 604, 775
\bibitem[Gonthier et al.(2006)]{Gonthier06} Gonthier, P.~L., et al. 2006, \emph{Chinese Journ. of Astron. and Astrophys.}, 6, 97
\bibitem[Gonz\`{a}lez(2008)]{HAWC} Gonz\`{a}lez, M.~M. for the HAWC Collab. 2008, \emph{Proc. of the 30th ICRC}, Merida, Mexico
\bibitem[Guan et al.(2006)]{Guan06} Guan,~Y., et al. 2006, \emph{Statist. Comput.}, 16, 193
\bibitem[Guan \& Krone(2007)]{GK07} Guan,~Y. \& Krone, S.~M. 2007, \emph{Annals of Appl. Prob.}, 17, 284
\bibitem[Gunn \& Ostriker(1969)]{GO69} Gunn, J.~E. \& Ostriker, J.~P. 1969, \emph{Nature}, 221, 454
\bibitem[Haino et al.(2004)]{Haino04} Haino,~S., et al. 2004, \emph{Phys. Lett. B}, 594, 35
\bibitem[Halpern \& Holt(1992)]{HH92} Halpern, J.~P. \& Holt, S.~S. 1992, \emph{Nature}, 357, 222
\bibitem[Halpern et al.(2008)]{Halpern08} Halpern, J.~P., et al. 2008, \ApJ{} \emph{Lett.}, 688, L33
\bibitem[Hardee(1977)]{Hardee77} Hardee, P.~E. 1977, \ApJ{}, 216, 873
\bibitem[Harding et al.(1978)]{HTE78} Harding, A.~K., Tademaru,~E., \& Esposito, L.~W. 1978, \ApJ{}, 225, 226
\bibitem[Harding(1981)]{Harding81} Harding, A.~K. 1981, \ApJ{}, 245, 267
\bibitem[Harding \& Muslimov(1998)]{HM98} Harding, A.~K. \& Muslimov, A.~G. 1998, \ApJ{}, 508, 328
\bibitem[Harding \& Muslimov(2002)]{HM02} Harding, A.~K. \& Muslimov, A.~G. 2002, \ApJ{}, 568, 862
\bibitem[Harding et al.(2002)]{HMZ02} Harding, A.~K., Muslimov, A.~G., \& Zhang,~B. 2002, \ApJ{}, 576, 366
\bibitem[Harding et al.(2005)]{Harding05} Harding, A.~K., Usov, V.~V., \& Muslimov, A.~G. 2005, \ApJ{}, 622, 531
\bibitem[Harding \& Lai(2006)]{Harding06} Harding, A.~K. \& Lai,~D. 2006, \emph{Reports on Progress in Physics}, 69, 2631
\bibitem[Harding et al.(2008)]{Harding08} Harding, A.~K., Stern, J.~V., Dyks,~J., \& Frackowiak,~M. 2008, \ApJ{}, 680, 1378
\bibitem[Harding(2009)]{Harding09} Harding, A.~K. 2009, \emph{High-energy Emission from the Polar Cap and Slot Gap}, \emph{Neutron Stars and Pulsars}, ed. W.~Becker, Astrophys. and Space Sci. Library, 357, 521
\bibitem[Harding \& Muslimov(2011)]{HM11} Harding, A.~K. \& Muslimov, A.~G. 2011, \ApJ{} \emph{Lett.}, 726, L10
\bibitem[Harding et al.(2011)]{Harding11} Harding, A.~K., DeCesar,~M., \& Miller, M.~C. 2011, \emph{Gamma-ray Pulsar Light Curves in Offset Polar Cap Geometry}, abstract submitted to the 2011 \emph{Fermi} Symposium
\bibitem[Hartman et al.(1999)]{Hartman99} Hartman, R.~C., et al. 1999, \ApJ{} \emph{Supplements}, 123, 79
\bibitem[Hastings(1970)]{Hastings70} Hastings, W.~K. 1970, \emph{Biometrika}, 57, 97
\bibitem[Hewish et al.(1968)]{HB68} Hewish,~A., Bell, S.~J., Pilkington, J.~D.~H., Scott, P.~F., \& Collins, R.~A. 1968, \emph{Nature}, 217, 709
\bibitem[Hinton(2004)]{HESS} Hinton, J.~A. for the HESS Collab. 2004, \emph{New Astron. Rev.}, 48, 331
\bibitem[Hirotani(2005)]{Hirotani05} Hirotani,~K. 2005, \emph{Adv. in Space Res.}, 35, 1085
\bibitem[Hirotani(2006)]{Hirotani06} Hirotani,~K. 2006, \ApJ{}, 652, 1475
\bibitem[Hobbs et al.(2006)]{TEMPO2} Hobbs, G.~B., Edwards, R.~T., \& Manchester, R.~N. 2006, \MNRAS{}, 369, 655
\bibitem[Holder(2006)]{VERITAS} Holder,~J. for the VERITAS Collab. 2006, \emph{Proc. of Science with New Gen. of High Energy Gamma-ray Experiments}, Elba
\bibitem[Hurley et al.(1994)]{Hurley94} Hurley,~K., et al. 1994, \emph{Nature}, 372, 652
\bibitem[Jackson(1999)]{Jackson} Jackson, J.~D. 1999, \emph{Classical Electrodynamics} (3rd ed.; USA: John Wiley \& Sons, Inc.)
\bibitem[Johnson et al.(2001)]{Johnson01} Johnson, W.~N., et al. 2001, \ITNS{}, 48, 1182
\bibitem[Johnston et al.(1993)]{Johnston93} Johnston,~S., et al. 1993, \emph{Nature}, 361, 613
\bibitem[Johnston \& Weisberg(2006)]{JW06} Johnston,~S., \& Weisberg, J.~M. 2006, \MNRAS{}, 368, 1856
\bibitem[Jones(1976)]{Jones76} Jones, P.~B. 1976, \emph{Astrophys. and Space Sci.}, 45, 369
\bibitem[Jones(1998)]{Jones98} Jones, B.~B. 1998, Ph. D. Thesis, Stanford University, Menlo Park, California, USA (astro-ph/0202088)
\bibitem[Kalman(1960)]{Kalman60} Kalman, R~.E. 1960, \emph{Transactions of the ASME--Journal of Basic Engineering}, 82, 35
\bibitem[Kanbach et al.(1994)]{Kanbach94} Kanbach,~G., et al. 1994, \AA{}, 289, 855
\bibitem[Keith et al.(2008)]{Keith08} Keith, M.~J., et al. 2008, \MNRAS{}, 389 1881
\bibitem[Keith et al.(2009)]{Keith09} Keith, M.~J., et al. 2009, \MNRAS{}, 393, 623
\bibitem[Keith et al.(2011)]{Keith11} Keith, M.~J., et al. 2011, \MNRAS{}, \emph{in press} (arXiv:1102.0648)
\bibitem[Kerr(2010)]{Kerr10} Kerr,~M. 2010, Ph. D. Thesis, University of Washington, Seattle, Washington, U.S.A. (arXiv:1101.6072)
\bibitem[Kijak \& Gil(2003)]{KG03} Kijak,~J. \& Gil,~J. 2003, \AA{}, 397, 969
\bibitem[Kraushaar \& Clark(1962)]{EXOXI} Kraushaar, W.~L. \& Clark, G.~W. 1962, \emph{Phys. Rev. Lett.}, 8, 106
\bibitem[Kuiper et al.(2000)]{Kuiper00} Kuiper,~L., et al. 2000, \AA{}, 359, 615 
\bibitem[Large et al.(1968)]{Large68} Large, M.~I., Vaughan, A.~E., \& Mills, B.~Y. 1968, \emph{Nature}, 220, 340
\bibitem[Latimer \& Prakash(2004)]{LP04} Lattimer, J.~M. \& Prakash,~M. 2004, \emph{Science}, 304, 536
\bibitem[Lommen et al.(2000)]{Lommen00} Lommen, A.~N., et al. 2000, \ApJ{}, 545, 1007
\bibitem[Lorimer et al.(1995)]{Lorimer95} Lorimer, D.~R., et al. 1995, \ApJ{}, 439, 933
\bibitem[Lorimer \& Kramer(2004)]{hdbk} Lorimer, D.~R. \& Kramer,~M. 2004, \emph{Handbook of Pulsar Astronomy} (Cambridge: Cambridge Univ. Press)
\bibitem[Lundgren et al.(1995)]{Lundgren95} Lundgren, S.~C., Zepka, A.~F., \& Cordes, J.~M. 1995, \ApJ{}, 453, 419
\bibitem[Lyne \& Smith(1968)]{LS68} Lyne, A.~G. \& Smith, F.~G. 1968, \emph{Nature}, 218, 124
\bibitem[Lyne \& Manchester(1988)]{LM88} Lyne, A.~G. \& Manchester, R~.N. 1988, \MNRAS{}, 234, 477
\bibitem[Malyshev et al.(2010)]{Malyshev10} Malyshev,~D., Cholis,~I., \& Gelfand, J.~D. 2010, \ApJ{}, 722, 1939
\bibitem[Manchester \& Johnston(1995)]{MJ95} Manchester, R.~N. \& Johnston,~S. 1995, \ApJ{} \emph{Lett.}, 441, L65
\bibitem[Manchester \& Han(2004)]{ManHan04} Manchester, R.~N. \& Han, J.~L. 2004, \ApJ{}, 609, 354
\bibitem[Manchester(2005)]{Manchester05} Manchester, R.~N. 2005, \emph{Astrophys. \& Space Sci.}, 297, 101
\bibitem[Manchester et al.(2005)]{ATNF} Manchester, R.~N., Hobbs, G.~B., Teoh,~A., \& Hobbs,~M. 2005, \emph{Astronom. Journal}, 129, 1993
\bibitem[Manchester(2011)]{Manchester11} Manchester, R.~N. 2011, \emph{Proc. of the Pulsar Conf. 2010:  Radio Pulsars: a key to Unlock the Secrets of the Universe}, ed. M.~Burgay, (USA: AIP)
\bibitem[Marinari \& Parisi(1992)]{MP92} Marinari,~E. \& Parisi,~G. 1992, \emph{Europhys. Lett.}, 19, 451 
\bibitem[Markov(1906)]{Markov06} Markov, A.~A. 1906, \emph{Izv. Fiz.-math. obsch. pri Kaz. Univ.}, 15, 135.
\bibitem[Mattox et al.(1996)]{Mattox96} Mattox, J.~R., et al. 1996, \ApJ{}, 461, 396
\bibitem[Meegan et al.(2009)]{Meegan09} Meegan,~C., et al. 2009, \ApJ{}, 702, 791
\bibitem[Mestel(1971)]{Mestel71} Mestel,~L. 1971, \emph{Nature}, 233, 149
\bibitem[Mestel \& Pryce(1992)]{MestelPryce92} Mestel,~L. \& Pryce, M.~H.~L. 1992, \MNRAS{}, 254, 344
\bibitem[Metropolis et al.(1953)]{Metropolis53} Metropolis,~N., et al. 1953, \emph{Journal of Chem. Phys.}, 21, 1087
\bibitem[Michel(1974)]{Michel74} Michel, F.~C. 1974, \ApJ{}, 187, 585
\bibitem[Michel(1969)]{Michel69} Michel, F.~W. 1969, \ApJ{}, 158, 727
\bibitem[Mizuno et al.(2004)]{Mizuno04} Mizuno,~T., et al. 2004, \ApJ{}, 614, 1113
\bibitem[Moiseev et al.(2004)]{Moiseev04} Moiseev, A.~A., et al. 2004, \AsPart{}, 22, 275
\bibitem[Moiseev et al.(2007)]{Moiseev07} Moiseev, A.~A., et al. 2007, \AsPart{}, 27, 339
\bibitem[Morini(1983)]{Morini83} Morini,~M. 1983, \MNRAS{}, 202, 495
\bibitem[Morrison(1958)]{Morrison58} Morrison,~P. 1958, \emph{Il Nuovo Cimento}, 7, 858
\bibitem[Muslimov \& Tsygan(1992)]{MT92} Muslimov, A.~G. \& Tsygan, A.~I. 1992, \MNRAS{}, 255, 61
\bibitem[Muslimov \& Harding(1997)]{MH97} Muslimov, A.~G. \& Harding, A.~K. 1997, \ApJ{}, 485, 735
\bibitem[Muslimov \& Harding(2003)]{MH03} Muslimov, A.~G. \& Harding, A.~K. 2003, \ApJ{}, 588, 430
\bibitem[Muslimov \& Harding(2004a)]{MH04a} Muslimov, A.~G. \& Harding, A.~K. 2004a, \ApJ{}, 606, 1143
\bibitem[Muslimov \& Harding(2004b)]{MH04b} Muslimov, A.~G. \& Harding, A.~K. 2004b, \ApJ{}, 617, 471
\bibitem[Navarro et al.(1995)]{Navarro95} Navarro,~J., et al. 1995, \ApJ{} \emph{Lett.}, 455, L55
\bibitem[Navarro et al.(1997)]{Navarro97} Navarro,~J., Manchester, R.~N., Sandhu, J.~S., Kulkarni, S.~R., \& Bailes,~M. 1997, \ApJ{}, 486, 1019
\bibitem[Ng \& Romai(2008)]{NR08} Ng, C.-Y. \& Romani, R.~W. 2008, \ApJ{}, 673, 411
\bibitem[Ohanian(2001)]{Ohanian01} Ohanian, H.~C. 2001, \emph{Special Relativity: A Modern Introduction} (1st ed.; USA: Physics Curr. \& Instr., Inc.)
\bibitem[Oppenheimer \& Volkoff(1939)]{OV39} Oppenheimer, J.~R. \& Volkoff, G.~M. 1939, \emph{Phys. Rev}, 55, 374
\bibitem[Ord et al.(2004)]{Ord04} Ord, S.~M., van Straten,~W., Hotan, A.~W., \& Bailes,~M. 2004, \MNRAS{}, 352, 804
\bibitem[Pacini(1968)]{Pacini68} Pacini,~F. 1968, \emph{Nature}, 219, 145
\bibitem[Pellizzoni et al.(2009)]{Pellizzoni09} Pellizzoni, A., et al. 2009, \ApJ{}, 695, L115 
\bibitem[Pierbattista(2011)]{Pierbattista} Pierbattista,~M. 2011, Ph. D. Thesis, Universit\'{e} Paris 7 - Diderot, France
\bibitem[Pierbattista et al.(2011)]{Pierbattista11} Pierbattista,~M., Grenier, I.~A., Harding, A.~K., \& Gonthier, P.~L. 2011, \emph{Proc. of the Pulsar Conf. 2010:  Radio Pulsars: a key to Unlock the Secrets of the Universe}, ed. M.~Burgay, (USA: AIP)
\bibitem[Radhakrishnan \& Cooke(1969)]{RVM} Radhakrishnan,~V. \& Cooke, D.~J. 1969, \emph{Astrophys. Lett.}, 3, 225
\bibitem[Ramanamurthy et al.(1995)]{Ramanamurthy95} Ramanamurthy, P.~V., et al. 1995, \ApJ{} \emph{Lett.}, 447, L109
\bibitem[Rando et al.(2009)]{Rando09} Rando, R. 2009, \emph{Proceedings of the 31$^{st}$ ICRC} (arXiv:0907.0626) 
\bibitem[Rankin(1993)]{Rankin93} Rankin, J.~M. 1993, \ApJ{}, 405, 285
\bibitem[Ransom et al.(2011)]{Ransom11} Ransom, S.~M., et al. 2011, \ApJ{}, 727, 16
\bibitem[Ravi et al.(2010)]{Ravi10} Ravi,~V., Manchester, R.~N., \& Hobbs,~G. 2010, \ApJ{} \emph{Lett.}, 716, L85
\bibitem[Ray et al.(2011)]{Ray11} Ray, P. S., et al. 2011, \ApJ{}, \emph{submitted} (arXiv:1011.2468)
\bibitem[Razzano et al.(2009)]{Razzano09} Razzano,~M., et al. 2009, \emph{Astropart. Phys.}, 32, 1
\bibitem[Reifenstein et al.(1969)]{Reifenstein69} Reifenstein, E.~C.~III, Brundage, W.~D., \& Staelin, D.~H. 1969, \emph{Phys. Rev Lett.}, 311
\bibitem[Romani(1990)]{Romani90} Romani, R.~W. 1990, \emph{Nature}, 347, 741
\bibitem[Romani \& Yadigaroglu(1995)]{RY95} Romani, R.~W. \& Yadigaroglu, I.-A. 1995, \ApJ{}, 438, 314
\bibitem[Romani \& Watters(2011)]{RW10} Romani, R.~W. \& Watters, K.~P. 2010, \ApJ{}, 714, 810
\bibitem[ROOT(2011)]{ROOT} ROOT Website $\langle$http://root.cern.ch/drupal/$\rangle$
\bibitem[Roth et al.(2011)]{Roth11} Roth,~M., et al. 2011, \emph{in preparation}
\bibitem[Ruderman \& Sutherland(1975)]{RS75} Ruderman, M.~A. \& Sutherland, P.~G. 1975, \ApJ{}, 196, 51
\bibitem[Ruderman(1991)]{Ruderman91} Ruderman,~M. 1991, \ApJ{}, 366, 261
\bibitem[Rybicki \& Lightman(1979)]{RL79} Rybicki, G.~B. \& Lightman, A.~P. 1979, \emph{Radiative Processes in Astrophysics}, (1st ed.; New York: John Wiley \& Sons, Inc.)
\bibitem[Saz Parkinson et al.(2010)]{SazPark10} Saz, Parkinson, P.~M., et al. 2010, \ApJ{}, 725, 571
\bibitem[Scipy(2011)]{Scipy} Scipy Documentation $\langle$http://docs.scipy.org/doc/$\rangle$
\bibitem[Shapiro(1964)]{Shapiro64} Shapiro, I.~I. 1964, \emph{Phys. Rev. Lett.}, 13, 789
\bibitem[Shitov(1983)]{Shitov83} Shitov, Yu.~P. 1983, \emph{Soviet Astron.}, 27, 314
\bibitem[Shitov(1985)]{Shitov85} Shitov, Yu.~P. 1985, \emph{Soviet Astron.}, 29, 33
\bibitem[Shklovskii(1970)]{Shklovskii70} Shklovskii, I.~S. 1970, \emph{Soviet Astron.}, 13, 562 
\bibitem[Smith et al.(2006)]{Smith06} Smith, D.~A., Grove, J.~E., \& Dumora,~D. 2006, \emph{An End-to-End test of GLAST LAT Absolute Timing}, LAT-TD-08777-03
\bibitem[Smith et al.(2007)]{Smith07} Smith, D.~A., Grove, J.~E., Dumora~D., Sandora, D.~P., \& Siskind, E.~J. 2007, \emph{A First Look at GLAST LAT Absolute Timing}, update of LAT-TD-08777-03
\bibitem[Smith et al.(2008)]{Smith08} Smith, D.~A., et al. 2008, \AA{}, 492, 923
\bibitem[Sokolov \& Ternov(1968)]{ST68} Sokolov, A.~A. \& Ternov, I.~M. 1968, \emph{Synchrotron Radiation}, 1st ed (New York: Akademi-Verlag)
\bibitem[Srinivasan(1990)]{Srinivasan90} Srinivasan,~G. 1990, \emph{Adv. in Space Res.}, 10, 167
\bibitem[Staelin \& Reifenstein(1968)]{SR68} Staelin, D.~H. \& Reifenstein, E.~C.~III 1968 ,\emph{Science}, 162, 1481
\bibitem[Stairs et al.(1999)]{Stairs99} Stairs, I.~H., Thorsett, S.~E., \& Camilo,~F. 1999, \ApJ{} \emph{Suppl.}, 123, 627
\bibitem[Stappers et al.(2003)]{Stappers03} Stappers, B.~W., Gaensler, B.~M., Kaspi, V.~M., van der Klis,~M., \& Lewin, W.~H.~G. 2003, \emph{Science}, 299, 1372
\bibitem[Stecker et al.(2008)]{Stecker08} Stecker, F.~W., Hunter, S.~D., \& Kniffen, D.~A. 2008, \emph{Astropart. Phys.}, 29, 25
\bibitem[Stinebring(1983)]{Stinebring83} Stinebring, D.~R. 1983, \emph{Nature}, 302, 690
\bibitem[Stinebring \& Cordes(1983)]{SC83} Stinebring, D.~R. \& Cordes, J.~M. 1983, \emph{Nature}, 306, 349
\bibitem[Strong et al.(2004)]{Strong04} Strong, A.~W., Moskalenko, I.~V., \& Reimer,~O. 2004, \ApJ, 613, 962
\bibitem[Sturrock(1971)]{Sturrock71} Sturrock, P.~A. 1971, \ApJ{}, 164, 529
\bibitem[Story et al.(2007)]{Story07} Story, S.~A., Gonthier, P.~L., \& Harding, A.~K. 2007, \ApJ{}, 671, 713
\bibitem[Sullivan et al.(2001)]{Milagro} Sullivan,~G., et al. 2001, \emph{Proc. of the 27th Int. Comsic-Ray Conf.}, Hamburg, 2773
\bibitem[Swanenburg et al.(1981)]{Swanenburg81} Swanenburg, B.~N., et al. 1981, \ApJ{} \emph{Lett.}, 243, L69
\bibitem[Takata et al.(2004)]{Takata04} Takata,~J., Shibata,~S., \& Hirotani,~K. 2004, \MNRAS{}, 354, 1120
\bibitem[Takata, et al.(2011)]{Takata11} Takata,~J., Wang,~Y., \& Cheng, K.~S. 2011, \MNRAS{}, \emph{in press} (arXiv:1102.2746)
\bibitem[Tavani et al.(2009)]{Tavani09} Tavani,~M., et al. 2009, \AA{}, 502, 995
\bibitem[Taylor \& Weisberg(1982)]{TW82} Taylor, J.~H., \& Weisberg, J.~M. 1982, \ApJ{}, 253, 908
\bibitem[TEMPO2(2011)]{T2} TEMPO2 Download Page $\langle$http://tempo2.sourceforge.net/$\rangle$
\bibitem[Thompson et al.(1974)]{DJT74} Thompson, D.~J., et al. 1974, \ApJ{} \emph{Lett.}, 190, L51
\bibitem[Thompson et al.(1975)]{DJT75} Thompson, D.~J., et al. 1975, \ApJ{} \emph{Lett.}, 200, L79
\bibitem[Thompson et al.(1992)]{DJT92} Thompson, D.~J., et al. 1992, \emph{Nature}, 359, 615
\bibitem[Thompson et al.(1993)]{DJT93} Thompson, D.~J., et al. 1993, \ApJ{} \emph{Suppl.}, 86, 629
\bibitem[Thompson et al.(2002)]{DJT02} Thompson, D.~J., et al. 2002, \ITNS{}, 49, 1898
\bibitem[Thompson(2004)]{DJT04} Thompson, D.~J. 2004, \emph{Cosmic Gamma-ray Sources}, ed. K.~S. Cheng \& G.~E. Romero (Dordrecht: Kluwer), 149
\bibitem[Thorsett \& Stinebring(1990)]{TS90} Thorsett, S.~E. \& Stinebring, D.~R. 1990, \ApJ{}, 361, 644
\bibitem[Timokhin(2006)]{Timokhin06} Timokhin, A.~N. 2006, \MNRAS{}, 368, 1055
\bibitem[Usov(1983)]{Usov83} Usov, V.~V. 1983, \emph{Nature}, 305, 409
\bibitem[van den Heuvel(1984)]{vdH84} van den Heuvel, E.~P.~J. 1984, \emph{Journ. of Astrophys. and Astron.}, 5, 209
\bibitem[Venter(2008)]{Venter08} Venter,~C. 2008, Ph. D. Thesis, North-West University, Potchefstroom Campus, South Africa
\bibitem[Venter et al.(2009)]{Venter09} Venter,~C., Harding, A.~K., \& Guillemot,~L. 2009, \ApJ{}, 707, 800 
\bibitem[Venter \& De Jager(2010)]{Venter10} Venter,~C. \& De Jager, O.~C. 2010, \ApJ{}, 725, 1903
\bibitem[Venter et al.(2011)]{Venter11} Venter,~C., Harding, A.~K., \& Johnson, T.~J. 2011, \emph{submitted}
\bibitem[Verde et al.(2003)]{Verde03} Verde,~L., et al. 2003, \ApJ{}, 148, 195
\bibitem[Wang et al.(2010)]{Wang10} Wang,~Y., Takata,~J., \& Cheng, K.~S. 2010, \ApJ{}, 720, 178
\bibitem[Wang et al.(2011)]{Wang11} Wang,~Y, Takata,~J., \& Cheng, K.~S. 2011, \MNRAS{}, \emph{in press} (arXiv:1102.4474)
\bibitem[Watters et al.(2009)]{Watters09} Watters, K.~P., et al. 2009, \ApJ{}, 695, 1289
\bibitem[Watters \& Romani(2011)]{WR11} Watters, K.~P. \& Romani, R.~W. 2011, \ApJ{}, 727, 123
\bibitem[Wilks(1938)]{Wilks38} Wilks, S.~S. 1938, \emph{Ann. Math. Statist.}, 9, 60
\bibitem[Wijnands \& van der Klis(1998)]{WK98} Wijnands,~R. \& van der Klis,~M. 1998, \emph{Nature}, 394, 344
\bibitem[Xilouris et al.(1998)]{Xilouris98} Xilouris, K.~M., et al. 1998, \ApJ{}, 501, 286
\bibitem[Yadigaroglu(1997)]{Y97} Yadigaroglu, I.-A. 1997, Ph. D. Thesis, Stanford University, Menlo Park, California, USA
\bibitem[Yan et al.(2011)]{Yan11} Yan, W.~M., et al. 2011, \MNRAS{}, \emph{in press} (arXiv:1102.2274)
\bibitem[Young et al.(2010)]{Young10} Young, M.~D.~T., Chan, L.~S., Burman, R.~R., \& Blair, D.~G. 2010, \MNRAS{}, 402, 1317
\bibitem[Zeissig \& Richards(1969)]{ZR69} Zeissig, G.~A. \& Richards, D.~W. 1969, \emph{Nature}, 222, 150
\bibitem[Zhang \& Cheng(2003)]{ZC03} Zhang,~L. \& Cheng, K.~S. 2003, \AA{}, 398, 639
\bibitem[Zhang et al.(2004)]{Zhang04} Zhang,~L., Cheng, K.~S., Jiang, Z.~J., \& Leung,~P. 2004, \ApJ{}, 604, 317
\bibitem[Zhang et al.(2007)]{Zhang07} Zhang,~L., Fang,~J., \& Chen, S.~B. 2007, \ApJ{}, 666, 1165
\bibitem[Zhu et al.(1997)]{Zhu97} Zhu,~C., Byrd, R.~H., \& Nocedal,~J. 1997, \emph{ACM Trans. on Math. Software}, 23, 550
\end{thebibliography}
\end{document}